\documentclass[11pt,a4paper,openany]{book}
\usepackage{amsmath,amssymb}
\usepackage{bm}
\usepackage[dvipdfmx]{graphicx}
\usepackage{ascmac}
\usepackage{fancyhdr}
\usepackage{fancybox}
\usepackage{color}

\usepackage{url}
\usepackage{multicol}
\usepackage{multirow}
\usepackage{threeparttable}

\definecolor{violet}{rgb}{0.4,0.0,0.4}
\definecolor{dblue}{rgb}{0.0,0.0,0.5}

\usepackage[
    backend=biber,
    style=authoryear,
    natbib=true,
    sorting=nyt,
    giveninits=true, 
    uniquename=init,
    doi=false, isbn=false, url=false
    ]{biblatex}

\usepackage{journals}
\usepackage{referencelist}

\defbibheading{subbibliography}[References]{\subsection*{#1}}

\title{\bf GREX-PLUS Science Book v2}
\author{GREX-PLUS Science Team}

\begin{document}

\maketitle

\section*{Preface}
\label{sec:executivesummary}

GREX-PLUS (Galaxy Reionization EXplorer and PLanetary Universe Spectrometer) is a mission candidate for a JAXA strategic L-class mission to be launched in the 2030s.
Its primary science goals are two-fold: galaxy formation and evolution, and planetary system formation and evolution. 
The GREX-PLUS spacecraft will carry a telescope with a 1 m primary mirror aperture cooled down to 50 K.
The two science instruments will be onboard: a wide-field camera in the 2--8 $\mu$m wavelength band and a high-resolution spectrometer with a wavelength resolution of 30,000 in the 10--18 $\mu$m band.
The GREX-PLUS wide-field camera aims to detect the first generation of galaxies at redshift $z>15$.
The GREX-PLUS high-resolution spectrometer aims to identify the location of the water ``snowline'' in protoplanetary disks. 
Both instruments will provide unique datasets for a broad range of scientific topics, including galaxy mass assembly, the origin of supermassive blackholes, infrared background radiation, molecular spectroscopy in the interstellar medium, transit spectroscopy of exoplanet atmospheres, planetary atmospheres in the Solar System, and so on.

This document is the second version of a collection of scientific themes that can be achieved with GREX-PLUS.
Each section in Chapters~\ref{chap:extragalactic} and \ref{chap:galacticplanetary} is based on presentations at several GREX-PLUS Science Workshops listed below:

\begin{itemize}
    \item G-REX$+$ Science Workshop held on 24-25 March 2021 at Waseda University\footnote{\url{http://www.obsap.phys.waseda.ac.jp/grex-hrs-sws-210324-25.html}}
    
    \item GREX-PLUS Science Workshop held on 24-25 March 2022 at Waseda University\footnote{\url{http://www.obsap.phys.waseda.ac.jp/grex-plus-sws-220324-25.html}}

    \item Early Universe Revealed by JWST and Future Prospects for Wide Field Surveys with GREX-PLUS held on 11-12 February 2025 at Waseda University Brussels Office\footnote{\url{http://www.obsap.phys.waseda.ac.jp/grex-plus-brussels-ws-20250211012.html}}

    \item GREX-PLUS Science Workshop held on 17-18 November 2025 at Waseda University\footnote{\url{http://www.obsap.phys.waseda.ac.jp/grex-plus-sws-20251117-18.html}}

    \item GREX-PLUS Science Workshop held on 23-24 November 2025 at Bologna University
    
\end{itemize}

\clearpage

\section*{Authors}

\begin{multicols}{2}

\noindent
Baba, Shunsuke (ISAS/JAXA) \S\ref{sec:AGNmolecularoutflow} \& \S\ref{sec:IGMmoleculargas}

\noindent
Belli, Sirio (Universit{\`a} di Bologna) \S\ref{sec:first-quiescent}

\noindent
Benotto, Pietro (INAF - Osservatorio Astronomico di Padova / Universit{\`a} di Bologna)

\noindent
Delvecchio, Ivan (INAF – Osservatorio di Astrofisica e Scienza dello Spazio di Bologna) \S\ref{sec:environment} \& \S\ref{sec:first_LRDs}

\noindent
Fudamoto, Yoshinobu (Chiba University) \S\ref{sec:submmgalaxies}

\noindent
Fujii, Yuka (NAOJ) \S\ref{sec:exoplanetatmosphere}

\noindent
Gouda, Naoteru (NAOJ) \S\ref{sec:galaxycenter}

\noindent
Harikane, Yuichi (University of Tokyo) \S\ref{sec:veryhighz} \& \S\ref{sec:first_LRDs}

\noindent
Hirahara, Yasuhiro (Nagoya University) \S\ref{sec:interstellarmolecules}

\noindent 
Inoue, Akio K.\ (Waseda University) \S\ref{chap:overview}, \S\ref{sec:veryhighz}, \S\ref{sec:cosmicinfraredbackground}, \& \S\ref{chap:synergywithotherprojects}

\noindent
Inoue, Yoshiyuki (University of Osaka) \S\ref{sec:smbh_coa}

\noindent
Kawahara, Hajime (ISAS/JAXA / University of Tokyo) \S\ref{sec:bd}

\noindent
Kawamuro, Taiki (University of Osaka) \S\ref{sec:smbh_coa}

\noindent
Kawashima, Yui (Kyoto University) \S\ref{sec:exoplanetatmosphere} \& \S\ref{sec:bd}

\noindent
Kimmig, Lucas (Ludwig-Maximilians-University / University of Nottingham) \S\ref{sec:protocluster}

\noindent
Kodama, Tadayuki (Tohoku University) \S\ref{sec:lssandmassassembly}

\noindent
Kokubo, Mitsuru (NAOJ) \S\ref{sec:agn_timedomain}


\noindent
Kurokawa, Hiroyuki (University of Tokyo) \S\ref{sec:PS_theoreticalperspective}

\noindent
Kusakabe, Katsunori (University of Osaka) \S\ref{sec:smbh_coa}

\noindent
Matsumoto, Kosei (Ghent University) \S\ref{sec: Attenuation curves in high-redshift galaxies}

\noindent
Matsunaga, Noriyuki (University of Tokyo) \S\ref{sec:AGBdust}

\noindent
Matsuo, Taro (University of Osaka) \S\ref{sec:exoplanetatmosphere}

\noindent
Matsuoka, Yoshiki (Ehime University) \S\ref{sec:highzquasars}

\noindent
Matsuura, Shuji (Kwansei Gakuin University) \S\ref{sec:cosmicinfraredbackground}


\noindent
Misawa, Toru (Shinshu University) \S\ref{sec:IGMmoleculargas}

\noindent
Miyazaki, Shota (ISAS/JAXA) \S\ref{sec:bd}

\noindent
Moretti, Alessia (INAF - Osservatorio Astronomico di Padova) \S\ref{sec:environment}

\noindent
Morihana, Kumiko (NAOJ) \S\ref{sec:galacticplane}

\noindent
Moriya, Takashi (NAOJ) \S\ref{sec:highzsupernovae}

\noindent
Nagamine, Kentaro (University of Osaka / University of Tokyo / University of Nevada Las Vegas) \S\ref{sec:EG_theoreticalperspective}

\noindent
Nakajima, Kimihiko (Kanazawa University) \S\ref{sec:empg}

\noindent
Nomura, Hideko (NAOJ) \S\ref{sec:protoplanetarydisks}

\noindent
Notsu, Shota (University of Tokyo) \S\ref{sec:protoplanetarydisks}

\noindent
Ootsubo, Takafumi (University of Occupational and Environmental Health) \S\ref{sec:icysmallbodies}

\noindent
Ohno, Kazumasa (NAOJ) \S\ref{sec:exoplanetatmosphere}

\noindent
Peluso, Giorgia (INAF – Osservatorio di Astrofisica e Scienza dello Spazio di Bologna) \S\ref{sec:environment}

\noindent
Poggianti, Bianca M. (INAF - Osservatorio Astronomico di Padova) \S\ref{sec:environment}

\noindent
Radovich, Mario (INAF - Osservatorio Astronomico di Padova) \S\ref{sec:environment}

\noindent
Rodighiero, Giulia (University of Padova) \S\ref{sec: Attenuation curves in high-redshift galaxies}

\noindent
Sagawa, Hideo (Kyoto Sangyo University) \S\ref{sec:solarsystemplanet}

\noindent
Shimasaku, Kazuhiro (University of Tokyo) \S\ref{sec:first_LRDs}

\noindent
Shimonishi, Takashi (Niigata University) \S\ref{sec:magellanicclouds}

\noindent
Tadaki, Ken-ichi (Hokkai-Gakuen University) \S\ref{sec:submmgalaxies}

\noindent
Takahashi, Kosuke (Tohoku University) \S\ref{sec:lssandmassassembly}

\noindent
Takami, Michihiro (ASIAA) \S\ref{sec:starformingregions}

\noindent
Tan, Shuya (JAMSTEC) \S\ref{sec:icysmallbodies}

\noindent
Tanaka, Takumi (University of Tokyo) \S\ref{sec:first_LRDs}

\noindent
Terai, Tsuyoshi (NAOJ) \S\ref{sec:icysmallbodies}

\noindent
Toba, Yoshiki (Ritsumeikan University) \S\ref{sec:dustyagns}

\noindent
Tripodi, Roberta (INAF - Astronomical Observatory of Rome) \S\ref{sec:first_LRDs}


\noindent
Valentino, Francesco (DAWN / Technical University of Denmark) \S\ref{sec:first-quiescent}

\noindent
Vulcani, Benedetta (INAF - Osservatorio Astronomico di Padova) \S\ref{sec:environment}

\noindent
Yano, Taihei (NAOJ) \S\ref{sec:galaxycenter}

\noindent
Yasui, Chikako (NAOJ) \S\ref{sec:starformingregions}

\noindent
Zibetti, Stefano (INAF - Arcetri Astrophysical Observatory) \S\ref{sec:stellarmass}

\end{multicols}

\tableofcontents

\clearpage


\chapter{Overview of GREX-PLUS}
\label{chap:overview}


\begin{refsection}[1_overviewofproject/overviewofproject.bib]

\section{Introduction}

GREX-PLUS (Galaxy Reionization EXplorer and PLanetary Universe Spectrometer) is a mission concept for a space telescope with a 1 m aperture and a temperature of 50 K, equipped with a wide-field camera in the 2--8 $\mu$m wavelength band and an optional high-resolution spectrometer with a wavelength resolution of 30,000 at 10--18 $\mu$m, as a JAXA strategic L-class mission to be launched in the 2030s (Table~\ref{tab:GPbasedesign}).
It aims to revolutionize research on the formation and evolution of galaxies and planetary systems by achieving high sensitivity, which is impossible from the ground.

The main goals of GREX-PLUS in galaxy formation and evolution are to discover rare bright ``first galaxies'' in the earliest epoch of the Universe (at a cosmic age of less than a few hundred million years) and to observe ``building blocks'' with one-hundredth the mass of a galaxy over 95\% of the history of the Universe (after a cosmic age of several hundred million years). 
For this purpose, the GREX-PLUS wide-field camera will perform super wide-field surveys in the 2--8 $\mu$m wavelength band, which are 10--100 times deeper (i.e. more sensitive) for the same area or 100--1000 times wider for the same depth than previous surveys conducted by the Spitzer Space Telescope.
The image data obtained from the GREX-PLUS surveys will also be used to search for the most distant supernova explosions, to search for massive blackhole objects in the most distant Universe or those heavily obscured by dust, to measure infrared background radiation, and to perform a census in the Galactic plane and time-domain astronomy, and will become legacy data in the history of astronomy.

The main goals of GREX-PLUS in planetary system formation and evolution are to determine the location of the water ``snowline'' in protoplanetary disks in the Galaxy and to understand the planet formation process, such as the segregation of rocky planets and gas giant planets.
Furthermore, GREX-PLUS explores planetary biosphere research to understand planetary atmospheric structures, the origin of surface oceans on rocky planets, and synthesis processes of organic materials (and eventually the origin of life), by observing organic molecules in protoplanetary disks and planetary atmospheres inside and outside the Solar System. 
For this purpose, the GREX-PLUS high-resolution spectrometer will open a new window to interstellar molecular spectroscopy with its extremely unique capability of a wavelength resolving power of 30,000 from space in the 10--18 $\mu$m wavelength band, which is referred to as the ``fingerprint region'' of molecular spectroscopy.

GREX-PLUS is a project that leverages the collective efforts of the Group of Optical and Infrared Astronomers (GOPIRA) in Japan by combining cryogenic space telescope technology and planetary science in the mid-infrared band, developed in SPICA \citep{2018PASA...35...30R}, a former ESA and JAXA joint mission candidate, and the world-leading super wide-field imaging surveys of distant galaxies conducted with the Subaru Telescope.
There was a former JAXA strategic L-class mission candidate, WISH \citep{2012SPIE.8442E..1AY}, which was a mission concept to expand highly successful optical wide-field imaging surveys conducted with the Subaru Telescope into near-infrared surveys from space.
The WISH working group (WG) under the ISAS/JAXA studied a passively cooled telescope with a 1.5 m aperture and a wide-field camera covering 1--5 $\mu$m wavelengths from 2008 to 2015.
Based on the research heritage, the Galaxy and Reionization EXplorer (G-REX) project was started in January 2020.
SPICA was a space infrared observatory project studied for about 20 years, aiming at mid- and far-infrared observations using a space telescope cooled to cryogenic temperatures.
Following the cancellation of SPICA, we developed the GREX-PLUS mission concept. This concept improved the cooling performance and reliability of the G-REX telescope by leveraging the most advanced space telescope cooling technology available at the time, which was developed through the SPICA study. It also enhanced scientific capabilities by installing SPICA's mid-infrared high-resolution spectrograph. 
In this way, GREX-PLUS is designed to achieve an important component of SPICA's highly evaluated science.

The Japanese optical-infrared astronomy community, GOPIRA, has published its Roadmap 2025 \citep{GopiraRoadmap2025}, in which GREX-PLUS is positioned as one of the highest-priority projects in the 2030s.
GOPIRA aims to shift its focus toward space telescope missions after the Thirty Meter Telescope (TMT).
GOPIRA's plan for space telescope missions is to first realize JASMINE, the Japan Astrometry Satellite Mission for Infrared Exploration \citep{2024PASJ...76..386K}, at the beginning of the 2030s, followed by the launch of GREX-PLUS in the mid-2030s. 
Finally, the plan is to participate in NASA's Habitable Worlds Observatory (HWO) in the 2040s.
This step-by-step development plan for these three space telescope missions paves the way for the future era of large-scale space telescopes.
Therefore, GREX-PLUS plays a central role in the GOPIRA roadmap 2025.


\begin{table}
    \label{tab:GPbasedesign}
    \begin{center}
    \caption{GREX-PLUS baseline design.}
    \begin{tabular}{|l|l|}
    \hline
    Telescope & $\phi1.0$ m, 50 K, diffraction limit at 4 $\mu$m \\
    \hline
    Wide-Field Camera (WFC) & 1,792 arcmin$^2$ divided into 5 bands in 2--8 $\mu$m \\
    \hline
    High Resolution Spectrometer (HRS) & Resolving power $R=30,000$ in 10--18 $\mu$m \\
    \hline
    Lifetime & 5 years \\
    \hline
    Orbit & Sun-Earth L2 Halo \\
    \hline
    Launch & 2030s by JAXA's H3 launch vehicle \\
    \hline
    \end{tabular}
    \end{center}
\end{table}

\section{Scientific Significance}
\label{sec:scientificsignificance}

\begin{figure}
 \begin{center}
  \includegraphics[width=15cm]{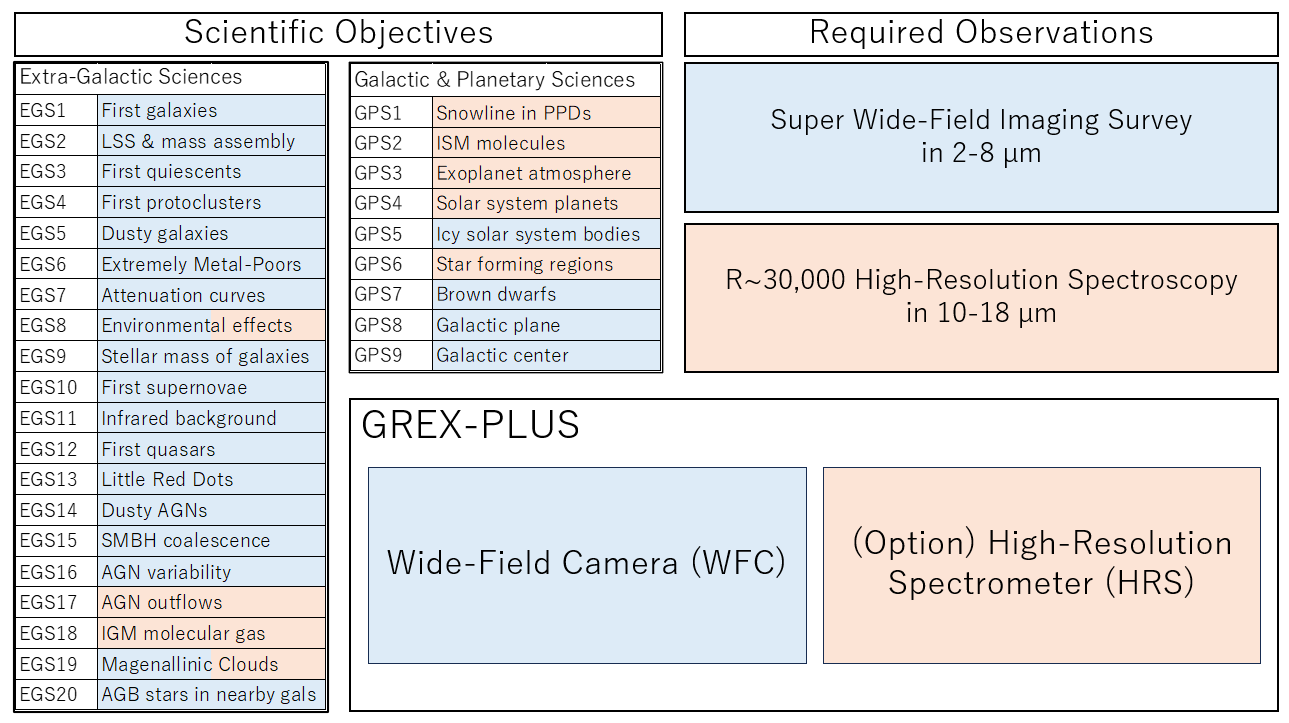}
 \end{center}
 \caption{Scientific objectives, required observations, and instruments of GREX-PLUS.
 \label{fig:SG_RG_INS}}
\end{figure}

The scientific goals in astronomy and astrophysics of ISAS/JAXA are ``to understand the origin of space and matter in the Universe'' and ``to explore the possibility of life in the Universe.''
GREX-PLUS is a project to tackle both of these goals directly. 
In this section, we describe the scientific significance of Extra-Galactic Sciences (EGS) and Galactic and Planetary Sciences (GPS) that GREX-PLUS aims to achieve, corresponding to these two major goals, respectively. 
Figure~\ref{fig:SG_RG_INS} shows the relation between the individual science cases needed to achieve the two major goals and the two GREX-PLUS instruments required for them. 
The following provides a brief overview of these scientific cases. 
More detailed descriptions will be presented in Chapters~\ref{chap:extragalactic} and \ref{chap:galacticplanetary}.

Before presenting a short review of the science cases, we summarize the significance and impact of GREX-PLUS observations.
GREX-PLUS provides super wide-field imaging data with high sensitivity and angular resolution in the 2--8 $\mu$m wavelength band and unique mid-infrared high-resolution spectroscopic data with a wavelength resolution of 30,000, all of which will eventually be released to the world as archival data and become part of the precious intellectual property of humanity. 
The data provided by both instruments will be unique and unparalleled and will be utilized in all fields of astronomy for a long time.
Analyzing archival data from new perspectives and with analysis methods that we do not have today will lead to various revolutionary discoveries in the future. 
This includes, of course, the discovery of unknown objects and phenomena. 
In order to promote such discoveries, we believe that developing an effective method of releasing archival data will also have a very high scientific value.

\subsection{Extra-Galactic Sciences (EGS)}

The primary goal of EGS with GREX-PLUS is to discover ``first galaxies'' and to reveal the formation processes of galaxies (EGS1; Section~\ref{sec:veryhighz}).
According to the current concordance structure formation theory based on cold dark matter, the ``first galaxies'' are thought to have formed at a cosmic age of less than 300 million years and a redshift of more than 15. 
The scientific value of proving this hypothesis by direct detection of ``first galaxies'' is extremely high. 
Such young galaxies can be detected in ultraviolet radiation in the source rest-frame, but at such a high redshift, infrared observations at wavelengths longer than 2 $\mu$m are required due to the significant redshift caused by the expansion of the Universe. 
In addition, since ``first galaxies'' are rare and faint, a sufficiently wide and deep imaging survey is required to detect them.

The secondary goal is to detect ``building blocks'' one-hundredth the mass of our Galaxy over 95\% of the history of the Universe and to determine how they merged, grew, and formed mature galaxies such as our Galaxy (EGS2; Section~\ref{sec:lssandmassassembly}). 
This is one of the ultimate goals of galactic astronomy, which began about 100 years ago at the beginning of the 20th century, and is of great scientific value. 
To achieve this goal, it is necessary to observe optical radiation in the source rest-frame up to a redshift of about 8, and infrared observations at wavelengths of 2 $\mu$m or longer are again required due to the redshift. 
In addition, deep imaging with a wide field of view is essential.

Galaxies evolve into a quiescent phase with less star formation activity.
The representative population of such quiescent galaxies is elliptical galaxies in the present-day Universe.
When and how elliptical galaxies were formed is a fundamental and long-standing question in extragalactic astronomy.
GREX-PLUS imaging surveys at wavelengths longer than 2 $\mu$m are essential to discover the cosmic first quiescent galaxies and to solve this question (EGS3; Section~\ref{sec:first-quiescent}).
In addition, protoclusters of galaxies in the early Universe are important objects for examining structure formation theory (EGS4; Section~\ref{sec:protocluster}).

Distant galaxies bright at submillimeter wavelengths, called submm galaxies, are a relevant galaxy population bridging quasars and dusty blackholes (EGS5; Section~\ref{sec:submmgalaxies}).
It is very important to conduct an imaging survey of distant galaxies with extremely low elemental abundances, which are a signature of primitive chemical evolutionary phases of the system (EGS6; Section~\ref{sec:empg}).
Dust attenuation curves of high-$z$ galaxies are also an important ingredient in analyzing the spectral energy distributions of galaxies (EGS7; Section~\ref{sec: Attenuation curves in high-redshift galaxies}).
Environmental effects on galaxy evolution are another very long-standing question in extragalactic astronomy (EGS8; Section~\ref{sec:environment}).
Stellar mass is a fundamental quantity of galaxies for understanding their evolution (EGS9; Section~\ref{sec:stellarmass}).

The discovery of the most distant supernova explosions elucidates various properties of first stars (EGS10; Section~\ref{sec:highzsupernovae}).
The rarity of such cosmic first supernovae requires super wide-field and sufficiently deep imaging surveys at wavelengths longer than 2 $\mu$m, which are possible only with GREX-PLUS.
Signs of first stars and blackholes can be imprinted on the infrared background radiation (EGS11; Section~\ref{sec:cosmicinfraredbackground}).
Discovering such signatures is also of high scientific value.

Revealing the origin of supermassive blackholes and galaxy-blackhole ``coevolution'' also has extremely high scientific value (EGS12-17). 
Galaxy-blackhole ``coevolution'' is a relatively new concept proposed about 20 years ago, in which a galaxy and a supermassive blackhole at the center of the galaxy mutually influence each other in their evolution.
It is a topic that has attracted a very high level of interest in recent years, and its high scientific value is indisputable.
To achieve this goal, we need to search for the most distant quasars in the rest-frame ultraviolet radiation (EGS12; Section~\ref{sec:highzquasars}).
A new AGN population, the so-called Little Red Dots (LRDs) discovered with JWST, is also a very important target because they are likely to be rapidly ``forming'' supermassive blackholes (EGS13; Section~\ref{sec:first_LRDs}).
The search for dusty supermassive blackholes in the rest-frame optical to near-infrared radiation (EGS14; Section~\ref{sec:dustyagns}) is another fundamental study to understand AGN evolution. 
For these topics, infrared imaging observations in the 2--8 $\mu$m wavelength range with a sufficiently wide field-of-view and depth are required.

The merging of blackholes can be an important evolutionary path to supermassive blackholes.
Binary blackholes are gravitational wave sources and one of the central targets of multi-messenger astronomy, including newly born gravitational wave astronomy.
GREX-PLUS imaging surveys contribute to finding dual AGNs as binary supermassive blackhole candidates (EGS15; Section~\ref{sec:smbh_coa}).
GREX-PLUS surveys also significantly contribute to time-domain astronomy through observations of AGN variability (EGS16; Section~\ref{sec:agn_timedomain}).

A census imaging survey of young stars in the Magellanic Clouds (EGS17; Section~\ref{sec:magellanicclouds}) is also very important for understanding star formation processes at lower elemental abundances than those in the interstellar medium in our Galaxy, which is relevant to star formation in young galaxies in the distant Universe. 
A survey of dust suppliers such as Asymptotic Giant Branch (AGB) stars in the local Universe is also realized with GREX-PLUS (EGS20; Section~\ref{sec:AGBdust}).
These goals also require wide-field deep imaging surveys at wavelengths longer than 2 $\mu$m.

In addition, high-resolution spectroscopy in the 10--18 $\mu$m wavelength range allows us to perform various unique extragalactic sciences, including observations of molecular gas outflows from supermassive blackholes that are essential for revealing their feedback processes (EGS17; Section~\ref{sec:AGNmolecularoutflow}), observations of molecular gas in the intergalactic medium and cosmological experiments (EGS18; Section~\ref{sec:IGMmoleculargas}), and spectroscopy of massive young stars in the Magellanic Clouds (EGS19; Section~\ref{sec:magellanicclouds}).

\subsection{Galactic and Planetary Sciences (GPS)}

The first goal of GPS with GREX-PLUS is to identify the position of the water ``snowline'' near the equatorial plane of protoplanetary disks (GPS1; Section~\ref{sec:protoplanetarydisks}).
For the theory of planetary system formation, ``snowlines'' play an important role in determining the segregation of rocky planets and gas giant planets.
It is also essential information for understanding the water supply process to the surfaces of rocky planets, which elucidates the origin of the Earth's ocean.
Although previous observations have not succeeded in resolving the water ``snowline'' location in the equatorial plane of the disks, there is a new idea to observe a water emission line at a wavelength of 17.8 $\mu$m with a wavelength resolution of 30,000 and to velocity-resolve the Keplerian motion of water molecules, eventually identifying the ``snowline'' position.
To realize this experiment for a statistical sample of $\sim100$ protoplanetary disks, for the first time, is of great scientific value.

The second goal is to elucidate the interstellar synthesis process of organic molecules and other materials that may eventually evolve into life (GPS2; Section~\ref{sec:interstellarmolecules}). 
Understanding the evolutionary process from simple interstellar molecules to complex organic molecules that could be the original materials of life is of extremely high scientific value, as it addresses the ultimate question of the origin of life on Earth. 
For this purpose, it is necessary to realize high-resolution spectroscopy with a wavelength resolution of 30,000 in the 10--18 $\mu$m wavelength band, which is also called the ``fingerprint region'' of molecular spectroscopy.

The third goal is to examine the atmospheric properties of Solar System planets and extrasolar planets (GPS3-4; Sections~\ref{sec:solarsystemplanet}-\ref{sec:exoplanetatmosphere}).
Since rocky planetary surfaces are expected to be sites of life in the Universe, it is clear that understanding the atmospheric properties that determine planetary surface environments has high scientific value. 
For example, it is necessary to observe molecules that are difficult or impossible to observe from the ground, such as hydrogen molecular quadrupole emission lines at wavelengths of 12 $\mu$m and 17 $\mu$m, to obtain information that has never been obtained before. 
In this case, sufficiently high wavelength resolution is required to observe molecular emission lines in crowded wavelength bands and emission lines of rare molecules.

In addition, young stars and star-forming regions in the Galaxy are also very interesting targets for GREX-PLUS (GPS6; Section~\ref{sec:starformingregions}).
Velocity-resolved observations of the [Ne II] line at 12 $\mu$m with a high-resolution spectrometer enable us to identify the outflow motion of gas from protoplanetary disks and to constrain the lifetime of the disks.

GREX-PLUS imaging observations at wavelengths of 2--8 $\mu$m, especially around the 3 $\mu$m water ice feature, for the Solar System provide an excellent opportunity to survey icy small bodies (GPS5; Section~\ref{sec:icysmallbodies}).
The wide-field imaging data also contribute to surveying brown dwarfs (GPS7; Section~\ref{sec:bd}) in the Galaxy.
The Galactic plane (GPS8; Section~\ref{sec:galacticplane}) and Galactic center (GPS9; Section~\ref{sec:galaxycenter}) are also excellent targets for GREX-PLUS wide-field imaging surveys to understand the formation and evolution of our own Galaxy.

\section{GREX-PLUS Wide-Field Camera and Imaging Surveys}
\label{sec:camera}

\begin{figure}
 \begin{center}
  \includegraphics[width=14cm]{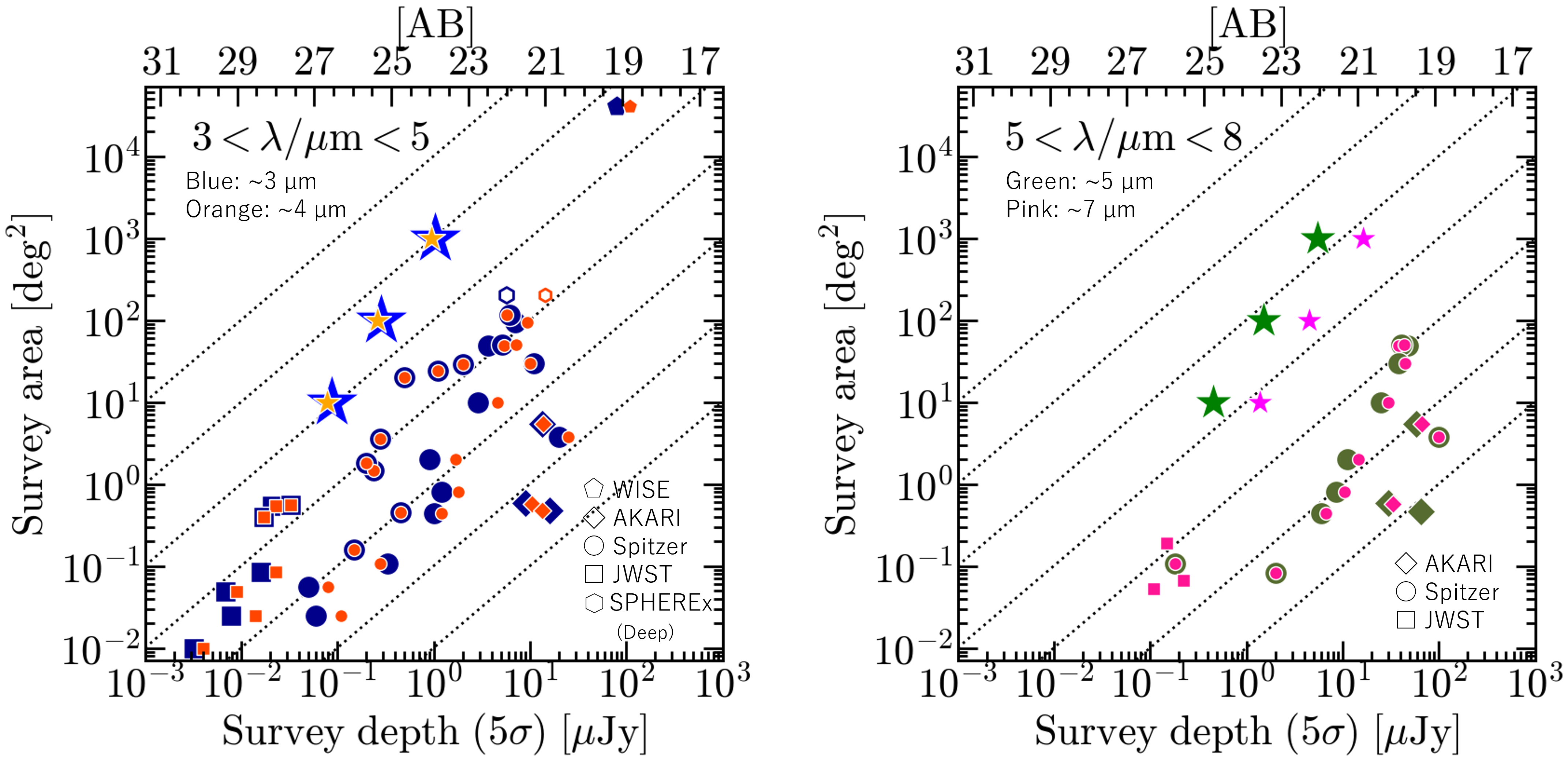}
 \end{center}
 \caption{Comparisons of survey areas and depths in the wavelength 3--5 $\mu$m bands (left) and 5--8 $\mu$m bands (right). The bluish (orange-ish) colors are $3$ (4) $\mu$m bands in the left panel, while the green-ish (magenta-ish) colors are $\sim5$ (7) $\mu$m bands in the right panel. The five-pointed stars are the GREX-PLUS survey parameters. The circles, diamonds, pentagons, and squares are Spitzer, AKARI, WISE, and JWST surveys, respectively. The open hexagons are SPHEREx deep surveys.
 \label{fig:surveypower}}
\end{figure}

To achieve a number of valuable science goals described briefly in the previous section and in detail in the following chapters, GREX-PLUS will be equipped with a wide-field camera and perform super wide-field imaging surveys in the 2--8 $\mu$m wavelength band.
Table~\ref{tab:GPcamera} summarizes the performance requirements for the GREX-PLUS Wide-Field Camera (WFC) to conduct the three types of imaging surveys listed in Table~\ref{tab:GPsurveys}.
As shown in Figure~\ref{fig:surveypower}, in the 3--4 $\mu$m and 5--8 $\mu$m wavelength bands, the three surveys of GREX-PLUS are about 10--100 times deeper in similar areas and about 100--1000 times wider at similar depths than the NASA/Spitzer Space Telescope's imaging surveys achieved over its 16 years of operation.
For comparison, Figure~\ref{fig:surveypower} also shows the AKARI imaging survey in the same wavelength band and the WISE all-sky survey depth. 
The survey parameters of surveys conducted with the NASA/James Webb Space Telescope (JWST) are also shown.
JWST is capable of reaching much deeper depths, but its narrow field-of-view limits the area it can cover to smaller than 1 deg$^2$.
The ongoing ESA/Euclid and the soon-to-start NASA/Nancy Grace Roman Space Telescope (hereafter referred to as Roman) super wide-field surveys are limited to wavelengths shorter than 2--2.3 $\mu$m and are not shown in Figure~\ref{fig:surveypower}.
NASA/SPHEREx is now conducting all-sky spectroscopic surveys in the wavelength range of 0.75--5 $\mu$m.
Figure~\ref{fig:surveypower} includes its planned Deep survey area and depth.

The imaging surveys with the Spitzer Space Telescope have produced more than 2,500 scientific papers, which include, for example, photometric data at the source rest-frame optical wavelengths of the most distant galaxies at that time.
The GREX-PLUS wide-field surveys surpass Spitzer's surveys by two to three orders of magnitude in depth and area, and completely revolutionize them.
The Spitzer Space Telescope has an angular resolution of 1.5 arcsec or worse at 4 $\mu$m wavelengths, while GREX-PLUS achieves an angular resolution of 1.0 arcsec at 4 $\mu$m thanks to a larger primary mirror aperture than that of the Spitzer Space Telescope.
This 1.5 times better angular resolution of GREX-PLUS is also a significant advantage for reaching deeper depths by overcoming the source confusion limit.
There is no other plan to realize deep and wide imaging surveys with good angular resolution in the 2--8 $\mu$m wavelength band that the GREX-PLUS WFC can achieve.
Therefore, the GREX-PLUS survey data will be an essential and fundamental legacy for almost all astronomical communities for a long time to come. 
Its scientific value will deserve the highest possible recognition.

\begin{table}
    \begin{center}
    \caption{GREX-PLUS Wide-Field Camera (WFC) performance requirements.}
    \label{tab:GPcamera}
    \begin{tabular}{|l|c|c|c|c|c|}
    \hline
     &  F232 & F303 & F397 & F520 & F680 \\
    \hline
    Central wavelength [$\mu$m] & 2.3 & 3.0 & 4.0 & 5.2 & 6.8 \\
    \hline
    Wavelength range [$\mu$m] & 2.0--2.6 & 2.6--3.4 & 3.4--4.5 & 4.5--5.9 & 5.9--7.7\\
    \hline
    Resolving power [$\lambda/\Delta\lambda$] & \multicolumn{5}{|c|}{3.7} \\
    \hline
    Pixel scale [arcsec pix$^{-1}$] & \multicolumn{5}{|c|}{0.48} \\
    \hline
    Field-of-view [arcmin$^2$] & 256 & 256 & 768 & 256 & 256 \\
    \hline
    Detector & \multicolumn{5}{|c|}{HgCdTe} \\
    \hline
    Sensitivity$^\dag$ [AB mag] & 23.4 & 23.5 & 23.0 & 21.7 & 20.5 \\
    \hline
    \end{tabular}
    \end{center}
    $^\dag$ 300 sec, $5\sigma$ for a point source, assuming a background intensity of 0.11 MJy str$^{-1}$ at a wavelength of 3 $\mu$m (three times higher than that in the North Ecliptic Pole).\\
\end{table}

\begin{table}
    \begin{center}
    \caption{GREX-PLUS wide-field survey plan.}
    \label{tab:GPsurveys}
    \begin{tabular}{|l|p{2cm}|p{2cm}|p{2cm}|}
    \hline
     & Deep & Medium & Wide \\
    \hline
    Area [deg$^2$] & 10 & 100 & 1000 \\
    \hline
    F232 [AB mag,~$5\sigma$] & 26.5 & 25.5 & 24 \\
    \hline
    F303 [AB mag,~$5\sigma$] & 26.5 & 25.5 & 24 \\
    \hline
    F397 [AB mag,~$5\sigma$] & 26.5 & 25.5 & 24 \\
    \hline
    F520 [AB mag,~$5\sigma$] & 24.5 & 23.5 & 22 \\
    \hline
    F680 [AB mag,~$5\sigma$] & 23.5 & 22.5 & 21 \\
    \hline
    \end{tabular}
    \end{center}
\end{table}

\section{GREX-PLUS High Resolution Spectrometer}
\label{sec:HRspectrometer}

\begin{figure}
 \begin{center}
  \includegraphics[height=8cm]{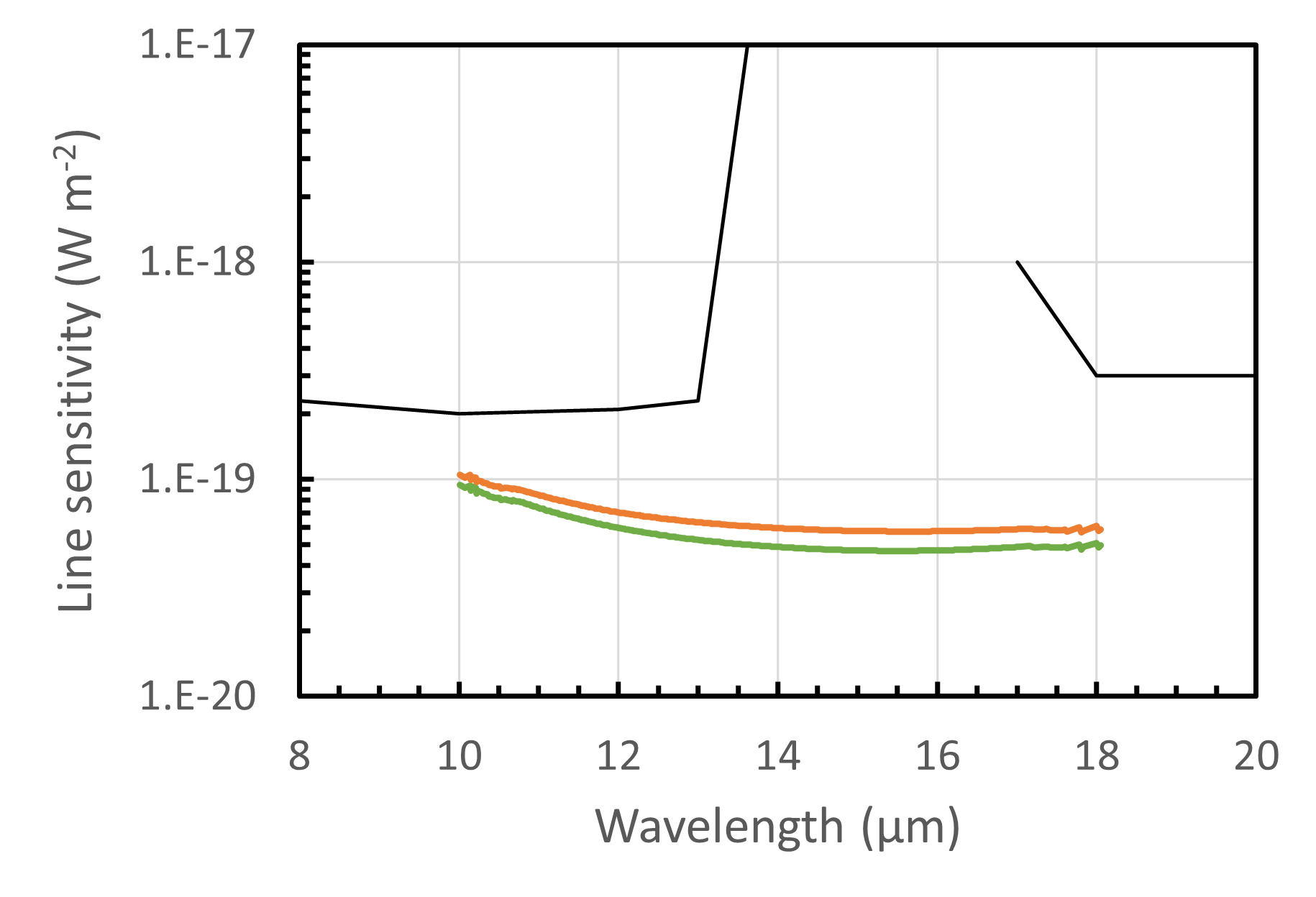}
 \end{center}
 \caption{A comparison of the line sensitivity (1 hr exposure, $5\sigma$) for unresolved narrow line emission between GREX-PLUS HRS ($\phi$ 1.0m, 50~K; low background case: green curve; high background case: orange curve; see Table~\ref{tab:HRparameters}) and a typical instrument on a ground-based 30 m-class telescope (black curves). The gap between 13.5 $\mu$m and 17 $\mu$m in the ground-based case is due to the very low transmission of the Earth's atmosphere.
 \label{fig:HRspec}}
\end{figure}

A number of highly valuable science cases, as described in Section~\ref{sec:scientificsignificance}, require a high-resolution spectrometer (HRS) with the instrument performance shown in Table~\ref{tab:HRparameters}.
GREX-PLUS is currently the only space telescope project aiming to achieve a high wavelength resolution of 30,000; JWST covers the same wavelength band, but its wavelength resolution is at most 3,000, which is one-tenth of that of GREX-PLUS.
On the other hand, ground-based large telescopes are planned to have even higher wavelength resolutions. 
For example, MICHI (Mid-Infrared Camera, High-disperser, and IFU; \citealt{2018cwla.conf...49M}), which is being considered as a second phase instrument for the 30-m telescope TMT, will have a wavelength resolution of 60,000 to 120,000.
However, the 10--18 $\mu$m wavelength band targeted by GREX-PLUS contains many transitions of interesting molecules such as water, carbon dioxide, and ammonia, which are also abundant in the Earth's atmosphere, making ground-based observations difficult due to atmospheric absorption. 
As a result, continuous coverage of the wavelength range from the ground is impossible, and the sensitivity is limited.
As shown in Figure~\ref{fig:HRspec}, GREX-PLUS has at least several times higher emission-line sensitivity than a TMT/MICHI-like instrument, thanks to the great advantage of observing from outside the Earth's atmosphere. 
Another major advantage of GREX-PLUS is continuous wavelength coverage, allowing GREX-PLUS to observe wavelength bands that ground-based telescopes cannot.
Compared to JWST, which covers the same wavelength band, GREX-PLUS has 10 times higher wavelength resolution. 
For example, interesting objects can be selected in advance with JWST's medium-resolution spectroscopy, and decisive results can be obtained with GREX-PLUS's high-resolution spectroscopy to develop very unique molecular spectroscopy. 
Specifically, there are many science cases requiring high velocity resolution, such as water ``snowlines'' in protoplanetary disks, exoplanetary atmospheres, and active galactic nuclei molecular outflows, and science cases in low-temperature regions, such as detection of various organic molecules in interstellar clouds and molecules in Solar System objects. 
No other project of this kind exists anywhere in the world, and it is easy to imagine that the scientific value of the data obtained with the GREX-PLUS HRS will continue to be of the highest standard for a long time.

Due to the challenges of procuring the detector and developing the key optical element, the immersion grating, the HRS is positioned as an optional instrument in the GREX-PLUS mission baseline.

\begin{table}
    \begin{center}
    \caption{GREX-PLUS high-resolution spectrometer performance.}
    \label{tab:HRparameters}
    \begin{tabular}{|l|l|}
    \hline
    Wavelength coverage [$\mu$m] & 10--18 \\
    \hline
    Resolving power [$\lambda/\Delta\lambda$] & $\sim30,000$ \\
    \hline
    Slit size [arcsec$^2$] & $9.8 \times 4.2$ \\
    \hline
    Pixel scale [arcsec pix$^{-1}$] & 1.8 \\
    \hline
    Detector & Si:As, 1k$\times$1k \\
    \hline
    Continuum sensitivity$^\dag$ [mJy, $5\sigma$, 1hr] & 6.0 / 7.2 \\
    \hline
    Line sensitivity$^\dag$ [10$^{-20}$ W m$^{-2}$, $5\sigma$, 1hr] & 4.9 / 5.9 \\
    \hline
    \end{tabular}
    \end{center}
    $^\dag$ Sensitivities under Zodiacal light intensities at a wavelength of 14 $\mu$m of 16 MJy str$^{-1}$ (low case) / 50 MJy str$^{-1}$ (high case).
\end{table}

\printbibliography[heading=subbibliography]
\end{refsection}

\chapter{Extra-Galactic Sciences}
\label{chap:extragalactic}


\begin{refsection}[2-1_EG_theoreticalperspective/EG_theoreticalperspective.bib]

\newcommand{\fesc}{f_{\rm esc}}
\newcommand{\Msun}{M_\odot}

\section{Theoretical Perspective}
\label{sec:EG_theoreticalperspective}

\noindent
\begin{flushright}
Kentaro Nagamine$^{1,2,3}$
\\
$^{1}$ The University of Osaka, $^{2}$ Kavli-IPMU (WPI), University of Tokyo, $^{3}$ University of Nevada Las Vegas
\end{flushright}
\vspace{0.5cm}


We now live in the era of precision cosmology, in which key cosmological parameters are constrained with accuracies better than 10\%.  
The energy budget of the Universe is dominated by dark matter ($\sim$30\%) and dark energy ($\sim$70\%) \citep{Planck18}, and the resulting best-fit framework is known as the concordance $\Lambda$ cold dark matter ($\Lambda$CDM) model.  Structure formation within the $\Lambda$CDM paradigm has been studied extensively over the past few decades, and substantial advances have been achieved on both theoretical and observational fronts.  Observations are increasingly unveiling the physical conditions of the early Universe through studies of the first galaxies and black holes, pushing the observational frontier to ever higher redshifts. 
The transformative impact of new observational capabilities was vividly demonstrated by the release of the James Webb Space Telescope's (JWST) first images in July 2022, marking a significant breakthrough in observational astronomy.

\begin{figure}[t]
    \centering
    \includegraphics[scale=0.45]{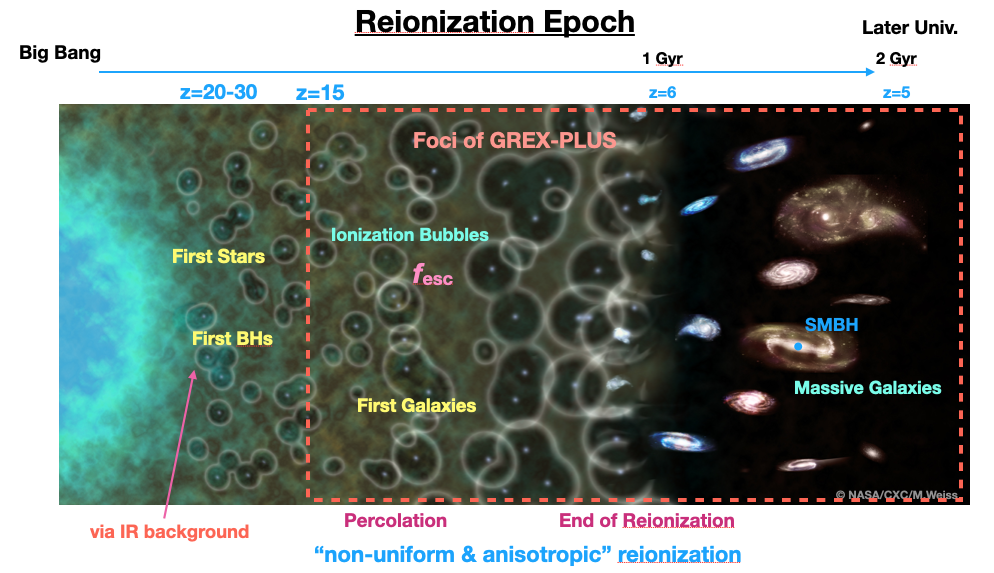}
    \caption{Structure formation in the $\Lambda$CDM cosmology and the scientific focus of GREX-PLUS.}
  \label{fig:cosmic}
\end{figure}

{\bf Figure~\ref{fig:cosmic}} summarizes the current picture of cosmic reionization within the $\Lambda$CDM framework.
The first stars and black holes are expected to form in mini-halos with masses $M_h \sim 10^{6}$--$10^{8}\,M_\odot$ at redshifts $z \approx 20$--30. These are followed by the emergence of the first galaxies at $z \approx 10$--20 in atomic-cooling halos with characteristic masses of $M_h \sim 10^{8}\,M_\odot$. Ionized regions subsequently develop around these sources, expand, and gradually percolate, leading to the completion of cosmic reionization by $z \sim 6$, as inferred from analyses of the Gunn--Peterson trough in quasar spectra \citep[e.g.,][]{Fan06a}. Cosmic reionization is therefore expected to be highly inhomogeneous and anisotropic, particularly during its early stages. A key physical quantity governing this process is the {\em escape fraction} of ionizing photons ($f_{\rm esc}$) from early galaxies, which regulates both the dominant sources of reionization and the temporal and spatial evolution of reionization. The relative contributions of faint galaxies and AGNs to the ionizing photon budget remain uncertain and are actively debated, as discussed below.

The central scientific theme of GREX-PLUS is understanding the origin of space and matter in the Universe. Within the context of galaxy formation and evolution, the specific primary science goals include the following: (i) discovering rare, bright first galaxies; (ii) observing the “building blocks” of Milky Way-type galaxies down to stellar masses of $M_\star \sim 10^{9}\,M_\odot$ out to $z \sim 8$; and (iii) searching for the most distant supernova (SN) explosions, uncovering massive black holes obscured by dust, and measuring the cosmic infrared background radiation.

Through these investigations, GREX-PLUS will critically test the standard cosmological model, theories of structure formation, and the baryonic physics governing galaxy formation in the early Universe. In particular, the mission will provide direct observational evidence for the emergence and assembly of the first massive galaxies. 
With these scientific objectives in mind, we now review the current status of cosmic reionization studies in detail. 

We begin by reviewing the standard methodology for modeling the history of cosmic reionization, starting with the evolution of the galaxy luminosity function (LF). Over the past several decades, observational efforts have made substantial progress in measuring the galaxy LF and its redshift evolution. 
\begin{figure}
    \centering
    \includegraphics[width=7cm]{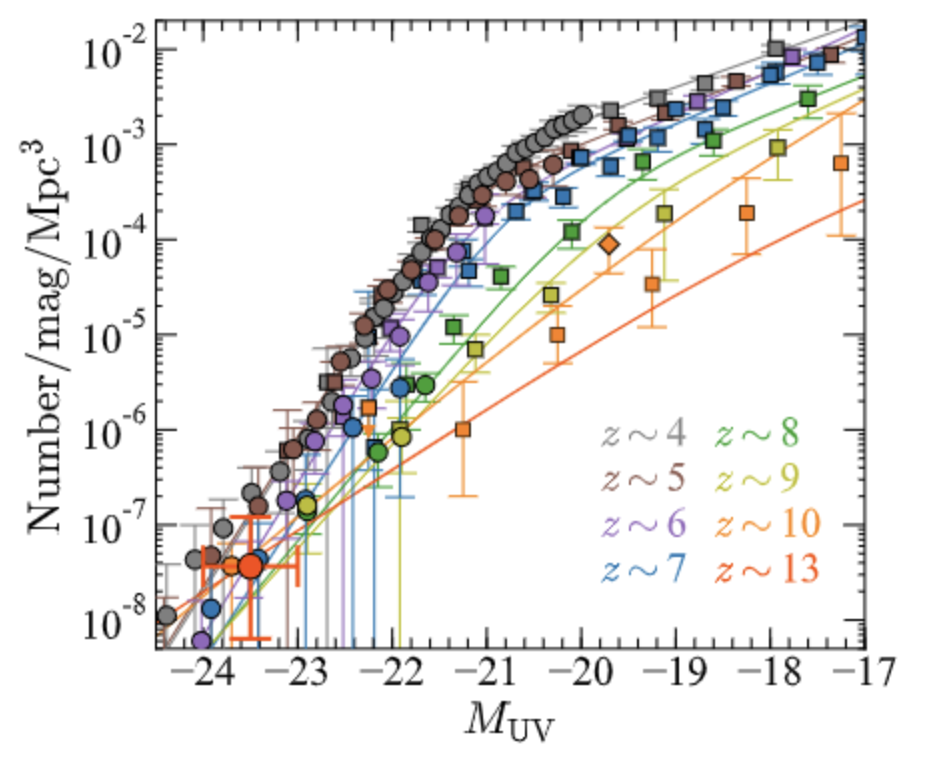}
        \includegraphics[width=8.5cm]{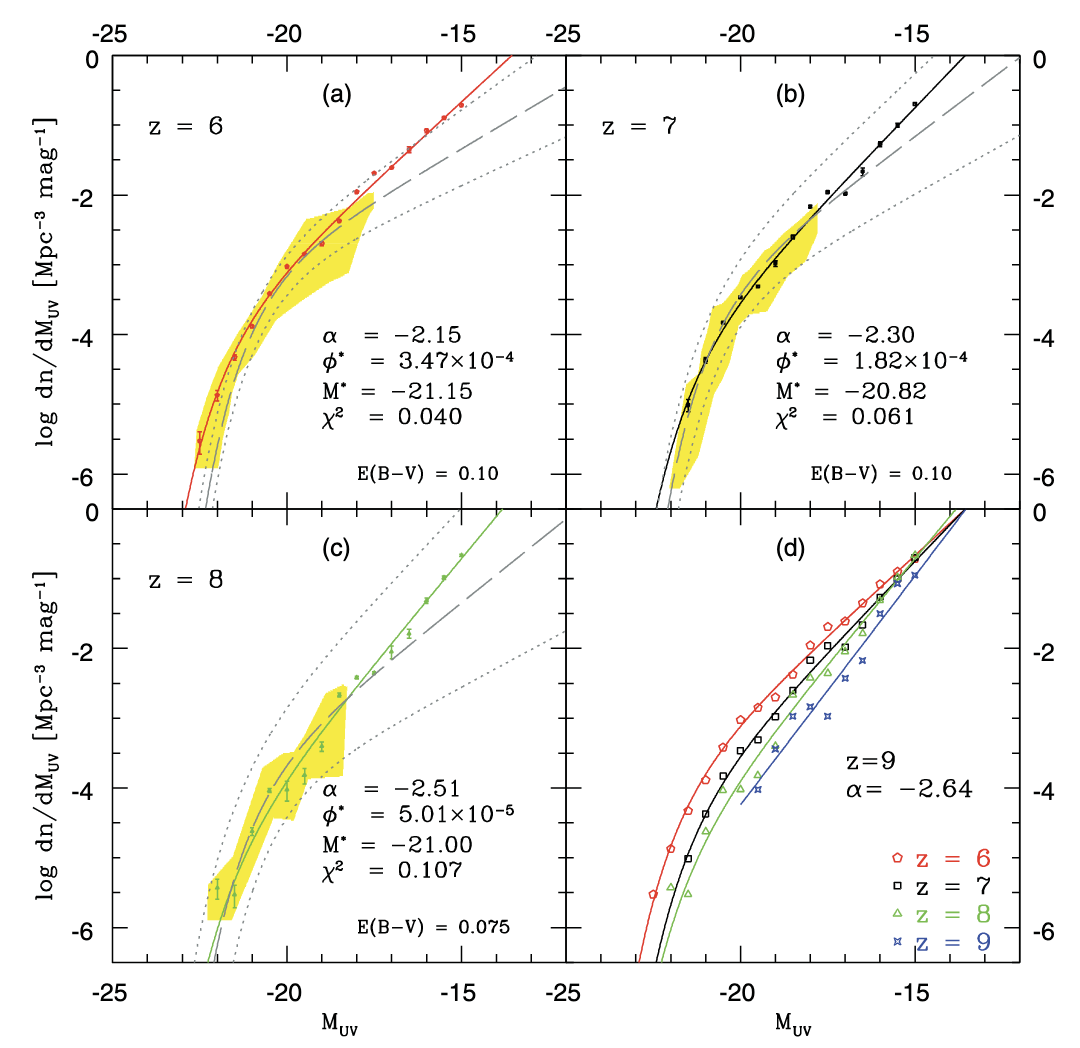}
    \caption{Examples of rest-frame UV luminosity functions of galaxies at $z=4$--13 from observations \citep[][left]{Harikane22} and cosmological hydrodynamic simulations \citep[][right]{Jaacks12b}.}
   \label{fig:LF}
\end{figure}
\citet{Harikane22} placed constraints on the bright end of the galaxy LF at $z \sim 13$, as shown in {\bf Figure~\ref{fig:LF}}. By integrating these LFs, one can derive the ultraviolet (UV) luminosity density of the Universe as a function of redshift. Observationally, the faint-end slope of the LF steepens from $\alpha \sim -1.2$ at $z = 0$ to $\alpha \sim -1.6$ at $z = 3$, and further to $\alpha \lesssim -2.0$ at $z \gtrsim 6$. This trend is in good agreement with earlier predictions from $\Lambda$CDM cosmological hydrodynamic simulations \citep{Nag04d,Nag04e,Night06,Jaacks12b}.

\begin{figure}[t]
    \centering
    \includegraphics[width=1.0 \textwidth]{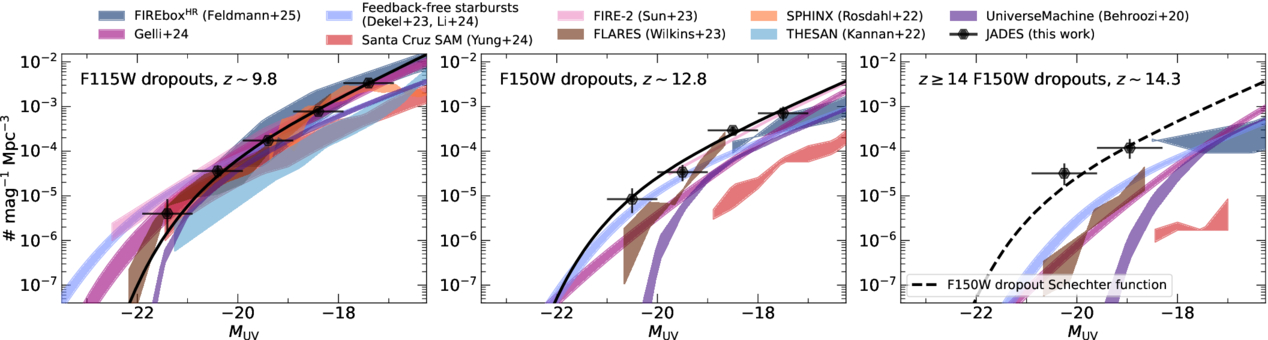}
    \caption{
    Rest-frame UVLFs from \citet{Whitler2025}, extending to $z \sim 16$ based on \textit{JWST} observations. The black hexagons show the binned observational data, and solid black lines indicate Schechter fits (with a dashed line for the $z \ge 14$ case, where no Schechter fit was performed). The observed LFs are consistent with some models at $z \sim 10$ but lie significantly above model predictions at $z \gtrsim 12$, suggesting more vigorous early star formation than anticipated.}
    \label{fig:WhitlerLF}
\end{figure}

\vspace{5mm}
\noindent
\textbf{Updated Constraints from \textit{JWST}.} \\
Recent deep imaging surveys with the \textit{JWST} have dramatically extended the empirical determination of the galaxy UVLF to $z \gtrsim 15$. 
In particular, \citet{Whitler2025} derived the UVLF over the redshift range $9 \lesssim z \lesssim 16$ using multiple JWST extragalactic fields, including JADES, CEERS, and PRIMER. 
{\bf Figure~\ref{fig:WhitlerLF}} shows the measured F115W, F150W, and $z \ge 14$ F150W dropout LFs, compared directly with a variety of theoretical models both before and after JWST. 

At $z \sim 10$, the observed galaxy LFs remain broadly consistent with several pre-JWST models
\citep[e.g.,][]{Kannan2022,Rosdahl2022,Wilkins2023}. 
However, at $z \gtrsim 12$, the observed number densities lie increasingly above all existing theoretical predictions across the full luminosity range. This indicates that star formation and galaxy assembly proceeded more efficiently in the first $\sim$300\,Myr of cosmic history than predicted by standard $\Lambda$CDM-based models. 
Post-JWST models that incorporate enhanced baryonic accretion and feedback physics \citep[e.g.,][]{Dekel2023,Gelli2024a,Feldmann2025,Yung2025} 
partially reconcile these discrepancies, but significant tension remains, particularly at the bright end ($M_{\rm UV} \lesssim -20$). 

Together with the earlier results of \citet{Harikane22} (Figure~\ref{fig:LF}), these JWST-based measurements highlight the rapid progress in constraining both the bright and faint ends of the UVLF and provide stringent new tests for models of early star formation, feedback, and reionization.


\vspace{3mm}
\noindent
\textbf{Implications for Cosmic Reionization.} \\
The elevated galaxy number densities inferred from JWST-based UVLFs at $z \gtrsim 10$--15 have direct and important implications for models of cosmic reionization. 
In particular, the higher normalization and steep faint-end slopes implied by these observations naturally increase the inferred cosmic star formation rate density (SFRD), $\rho_{\rm SFR}(z)$, relative to pre-JWST estimates based on extrapolations of \textit{HST} data. 
For fixed assumptions about the ionizing photon production efficiency, $\xi_{\rm ion}$, this would enhance the ionizing emissivity and favor an earlier onset of reionization.

At the same time, the tension between the observed UVLFs and theoretical predictions at $z \gtrsim 12$ highlights growing uncertainties in the efficiency of star formation, feedback, and baryon accretion in the earliest galaxies.
As a result, reionization models are now increasingly constrained by galaxy demographics, placing renewed emphasis on the escape fraction of ionizing photons, $\fesc$,
as a key regulating parameter: even modest changes in $\fesc$ for low-mass galaxies can significantly alter the timing and topology of reionization. 
As discussed below, combining these updated UVLF constraints with measurements of the neutral hydrogen fraction and the Thomson optical depth provides a powerful consistency check on reionization models and the underlying physics of early galaxy formation.

Assuming that the UV luminosity is dominated by massive stars with short lifetimes, the UV luminosity density can be converted into $\rho_{\rm SFR}(z)$ \citep[e.g.,][]{Madau96,Madau14,Robertson15}. The evolution of the ionized volume fraction, $Q_{\rm HII}$, is then governed by
\begin{equation}
    \dot{Q}_{\rm HII} = \frac{\dot{n}_{\rm ion}}{\langle n_{\rm H} \rangle} - \frac{Q_{\rm HII}}{t_{\rm rec}}, 
    \label{eq:QHII}
\end{equation}
where the first term represents the ionization rate and the second accounts for recombinations. The comoving production rate of ionizing photons is given by 
\begin{equation}
    \dot{n}_{\rm ion}(z) = \fesc\, \xi_{\rm ion}\, \rho_{\rm SFR}(z),  
\end{equation}
where $\xi_{\rm ion}$ denotes the ionizing photon production efficiency per unit SFR. 
$\fesc$ may, in principle, depend on galaxy properties such as SFR, halo mass, and redshift; however, it is often treated as a constant in simplified reionization models.

\begin{figure}
    \centering
    \includegraphics[width=8cm]{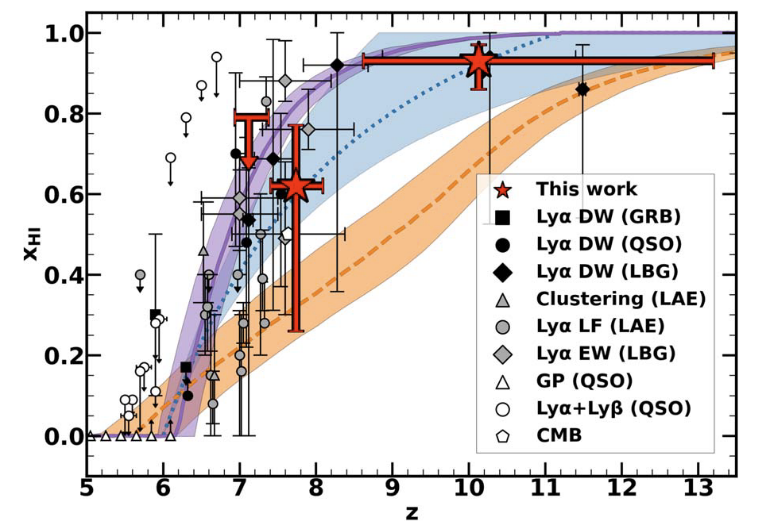}
    \includegraphics[width=7.6cm]{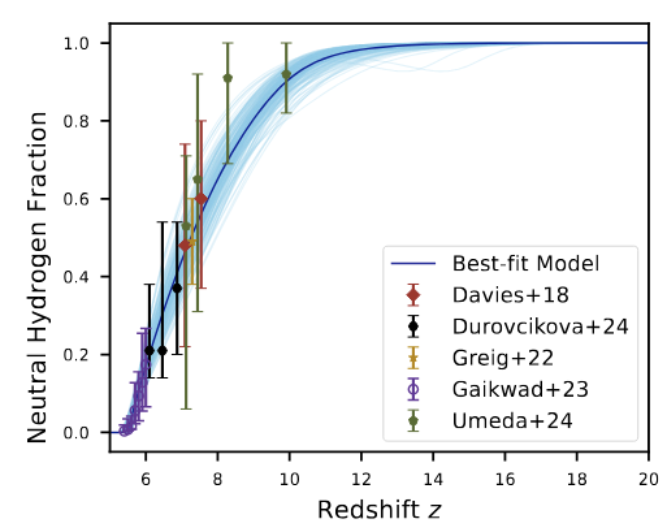}
    \caption{Volume-averaged neutral fraction as a function of redshift from \citet[][left]{Nakane2024} and \citet[][right]{Chakraborty2026}. 
    See the main text for more details. }
   \label{fig:XHII}
\end{figure}

Once $Q_{\rm HII}(z)$ is obtained from Eq.~(\ref{eq:QHII}), the volume-averaged neutral fraction can be computed as $x_{\rm HI} = 1 - Q_{\rm HII}$ as shown in Figure~\ref{fig:XHII}. 
The orange curve in the left panel corresponds to the {\em early reionization} scenario from \citet{Finkelstein19}, which assumes a higher $\fesc$ for low-mass halos together with a steep faint-end slope of the galaxy LF.  In this model, reionization is driven predominantly by low-mass galaxies and commences at earlier cosmic times.  
The purple curve in the left panel represents the {\em late reionization} scenario proposed by \citet{Naidu20}, which adopts a constant $\fesc$. In this case, more massive halos contribute more significantly at later times, leading to a delayed reionization history.
The red stars in Fig.~\ref{fig:XHII} (left panel) indicate estimates of neutral fraction inferred from the evolution of the Lyman-$\alpha$ emitter (LAE) LF by \citep{Nakane2024}.  



Semi-analytic models tuned to reproduce the JWST UVLFs over $6 \lesssim z \lesssim 15$ require a rapid increase in the SF efficiency and/or UV luminosity per unit stellar mass at $z \gtrsim 10$ relative to standard $\Lambda$CDM expectations \citep{Chakraborty2026}. 
When propagated into reionization calculations, such enhanced efficiencies naturally favor a substantial ionized fraction already by $z \gtrsim 8$ as shown in the right panel of Fig.~\ref{fig:XHII} 
unless the effective escaping ionizing emissivity is strongly regulated, for example by an evolving or halo-mass-dependent escape fraction.

\vspace{2mm}
\noindent
\textbf{Spectroscopic Observations and Ionized Bubbles.} \\
Recent spectroscopic observations with \textit{JWST} provide complementary, direct probes of the ionization state of the IGM at very high redshift.
In particular, the detection of Ly$\alpha$ emission from the galaxy JADES--GS--z13--1 at $z \simeq 13$ suggests the presence of a locally ionized region surrounding at least some early galaxies \citep{Witstok2025}.
At such redshifts, the volume-averaged neutral hydrogen fraction is expected to be high, and Ly$\alpha$ photons are strongly attenuated by the neutral IGM.  The observed Ly$\alpha$ emission therefore implies that ionized bubbles with non-negligible sizes must already exist at these early epochs, providing direct evidence for the patchy and inhomogeneous nature of reionization.

While current detections are based on small-number statistics and do not yet constrain the global reionization history, they provide important consistency checks on models of $x_{\rm HI}(z)$ and the topology of reionization.
In particular, the visibility of Ly$\alpha$ emission at $z \gtrsim 10$ disfavors scenarios in which reionization begins very late or proceeds exclusively through extremely small ionized regions.
As larger samples become available, Ly$\alpha$ observations will offer an increasingly powerful, independent constraint on the timing and spatial structure of reionization, complementary to LF measurements and CMB optical depth constraints.

Spectroscopic follow-up observations have revealed that many UV-bright galaxies at $z \gtrsim 8$--11 exhibit hard ionizing spectra and very young, often bursty SF histories, consistent with elevated ionizing photon production efficiencies, $\xi_{\rm ion}$ \citep[e.g.,][]{Nakane2024,Kokorev2025}.
Such systems can plausibly generate $\sim{\rm pMpc}$-scale ionized bubbles around individual sources already at $z \sim 7$--11 \citep{Napolitano2024}, reinforcing the picture that early reionization proceeds in a highly inhomogeneous manner.

At the same time, the combination of these high inferred efficiencies with the elevated galaxy number densities implied by JWST UVLFs introduces new tension in reionization models.
If moderate-to-high escape fractions commonly assumed in pre-JWST modeling are adopted, galaxies brighter than $M_{\rm UV} \lesssim -15$ can in principle complete reionization by $z \sim 8$, yielding a Thomson optical depth $\tau$ that is in tension with \textit{Planck} measurements and Ly$\alpha$ forest constraints. 
Including even fainter galaxies only exacerbates this so-called ``photon budget crisis'' \citep[e.g.,][]{Chakraborty2024,Cain2025}, highlighting the need for improved theoretical understanding of escape fractions, feedback, and the regulation of ionizing photon leakage in the
earliest galaxies.

\begin{figure}
    \centering
    \includegraphics[width=7.7cm]{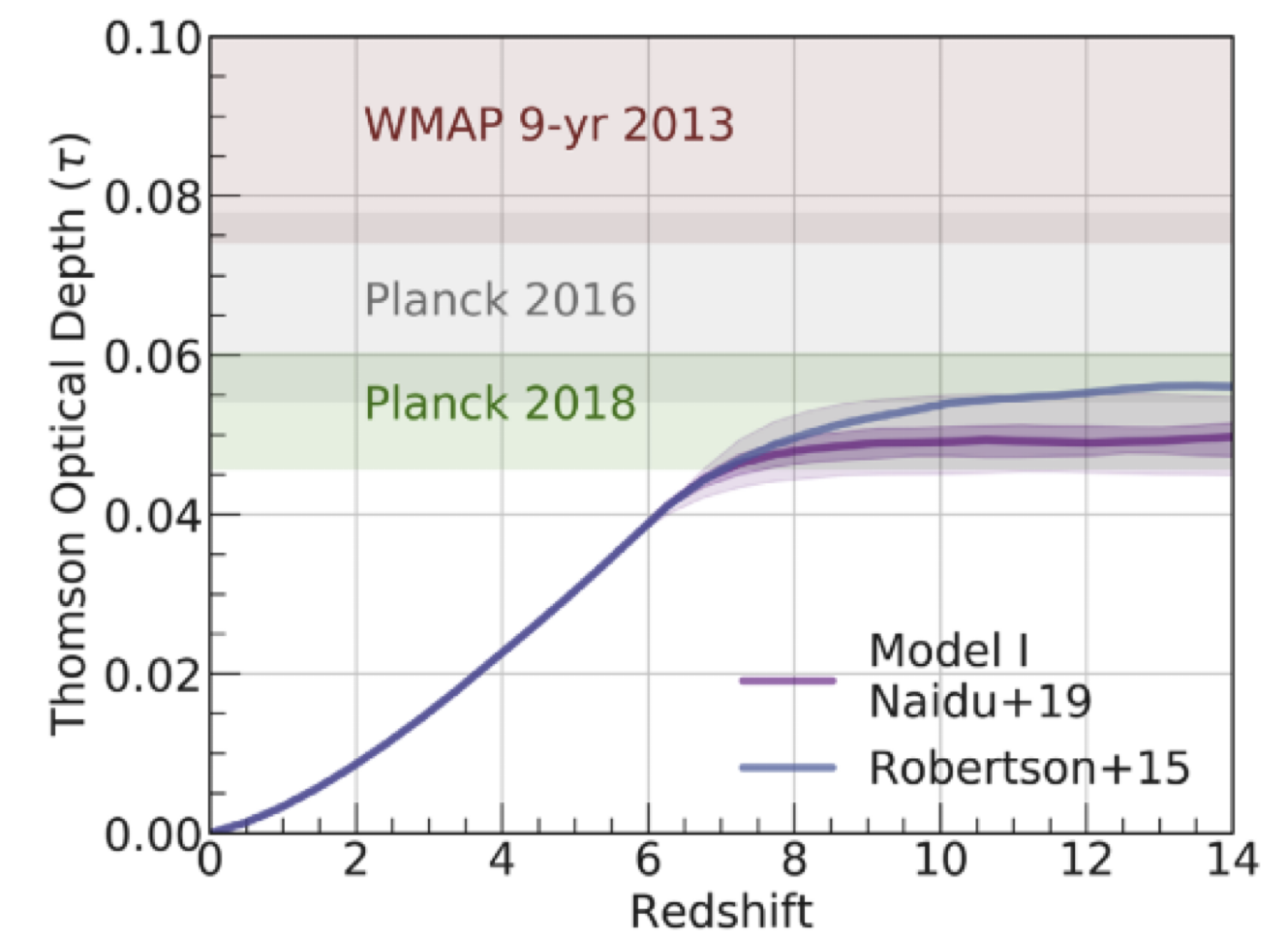}
     \includegraphics[width=7cm]{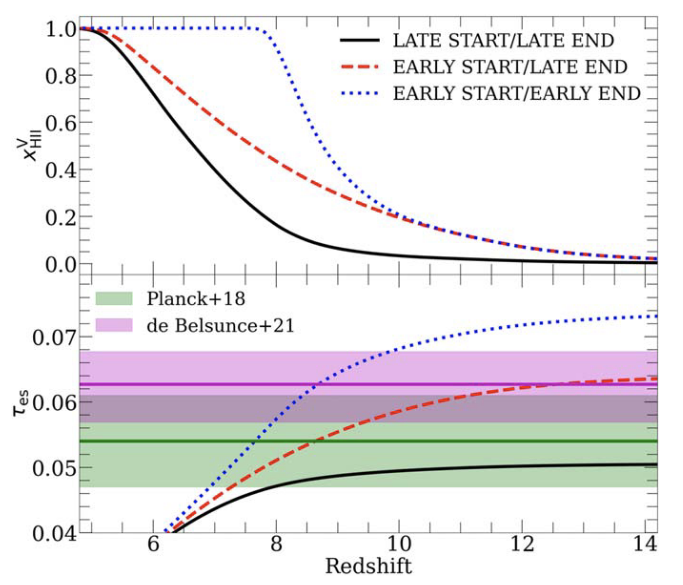}
    \caption{Thomson optical depth as a function of redshift from \citet[][{\it left}]{Naidu20} and \citet[][{\it right}]{Cain2025}. }
   \label{fig:tau}
\end{figure}

\vspace{3mm}
\noindent
\textbf{Thomson Optical Depth.} \\
An additional consistency check is provided by the cumulative Thomson scattering optical depth integrated up to redshift $z$, given by
\begin{equation}
    \tau(z) = c \langle n_H \rangle \sigma_T \int_0^z f_e Q_{\rm HII}(z^\prime) \frac{(1+z^\prime)^2}{H(z^\prime)}dz^\prime, 
\end{equation}
where $c$ is the speed of light, $\sigma_T$ is the Thomson scattering cross section, $f_e$ is the number of free electrons per hydrogen nucleus in the ionized IGM, and $H(z)$ is the Hubble parameter.
The resulting optical depth evolution is shown in Figure~\ref{fig:tau}, together with observational constraints from cosmic microwave background measurements by WMAP and Planck, which have progressively converged to lower values over successive data releases. 
Both the `early' and `late' reionization scenarios should satisfy this constraint; however, they differ in the redshift at which the optical depth approaches its asymptotic value of $\tau_e \simeq 0.05$. This distinction reflects differences in the timing and dominant sources of reionization in the two scenarios.

\vspace{3mm}
\noindent
\textbf{JWST Constraints Favoring a Late, Rapid Reionization}\\
Independent probes of the ionization state from Ly$\alpha$ emission and damping wings using JWST spectroscopy indicate that the IGM remains substantially neutral at $z \gtrsim 9$--10, pointing toward a relatively late and rapid reionization.

A large JWST census of Ly$\alpha$ emission in galaxies at $z\simeq 6.6$--13.2 finds a systematic decline of Ly$\alpha$ equivalent widths toward higher redshift for typical star-forming galaxies ($-22.5 \lesssim M_{\rm UV} \lesssim -17$). Likelihood analysis using Ly$\alpha$ EW distribution models calibrated on reionization simulations yields volume-averaged neutral fractions of $x_{\rm HI} < 0.79$ at $z\sim7$, and $x_{\rm HI}\simeq 0.93^{+0.04}_{-0.07}$ at $z\sim9$--13 \citep{Nakane2024}. 
A complementary Ly$\alpha$ luminosity function analysis over $z\sim5$--14 similarly infers $x_{\rm HI}\simeq 0.17^{+0.23}_{-0.16}$, $0.63^{+0.18}_{-0.28}$, $0.79^{+0.13}_{-0.21}$, and $0.88^{+0.11}_{-0.13}$ at $z\sim6$, 7, 8--9, and 10--14, respectively, favoring a late and sharp reionization with the main phase at $z\sim6$--7 \citep{Kageura2025}.

Damping-wing analyses of the smooth Ly$\alpha$ breaks in high-S/N NIRSpec prism spectra of $z\sim5.5$--13 galaxies support this picture. The median and variance of the continuum flux around the break decline with increasing redshift, consistent with smaller and rarer ionized regions and an increasingly neutral IGM. Fits to the highest-S/N spectra give $\overline{x}_{\rm HI}\approx 0.33^{+0.18}_{-0.27}$ at $z\approx 6.5$ and $\overline{x}_{\rm HI}\approx 0.64^{+0.17}_{-0.23}$ (or $>0.70$ excluding GNz11) at $z\approx 9.3$, implying that the IGM is significantly neutral at $z\gtrsim 9$ \citep{Mason2026}.

Radiative-transfer simulations comparing different reionization histories show that models that fully exploit the large JWST-inferred ionizing emissivity and therefore end reionization early ($z_{\rm end}\simeq 8$) are inconsistent with the $z<6$ end implied by the Ly$\alpha$ forest. When restricting to scenarios that end at $z_{\rm end}\simeq 5$, an ``early-start'' model beginning at $z\sim13$ requires up to an order-of-magnitude evolution in galaxy ionizing properties between $6\lesssim z \lesssim 12$, which may be in tension with the relatively mild evolution of $\xi_{\rm ion}$ inferred from JWST \citep{Cain2025}. Across a suite of observables (Ly$\alpha$ forest opacity, mean free path, IGM temperature, visibility of $z>8$ Ly$\alpha$ emitters, and the patchy kSZ signal), a ``late-start, late-end'' scenario with reionization occurring primarily between $z\sim9$ and $z\sim5$--6 is modestly preferred, although not yet decisively so \citep{Cain2025}.

\vspace{3mm}
\noindent
\textbf{Reconciling bright JWST galaxies with a late reionization}\\
Combining the JWST galaxy statistics with these IGM constraints suggests that the high ionizing photon production implied by the bright $z\gtrsim 10$ population must be strongly regulated at the level of the \emph{escaping} ionizing emissivity. Semi-analytic models that jointly fit the JWST/HST UVLFs, estimates of the neutral fraction, and CMB $\tau$ can reproduce a late-ending reionization history if the effective escaping ionizing efficiency
$\varepsilon_{\rm esc}\equiv f_{\rm esc}\,\xi_{\rm ion}$ is assumed to be halo-mass dependent, with higher Lyman-continuum leakage in low-mass galaxies and a population-averaged $\langle f_{\rm esc}\rangle \sim 0.1$--0.15 at $z\sim5$ for plausible $\xi_{\rm ion}$ \citep{Chakraborty2024,Chakraborty2026}. In this framework, the bright JWST galaxies at $z\gtrsim 10$ generate large local ionized bubbles but do not necessarily drive an \emph{early} completion of reionization.

Additional mechanisms have been proposed to reconcile the JWST-inferred stellar mass or UV luminosity densities with reionization constraints. Suppression of low-mass structure in warm dark matter models can reduce the number of low-mass galaxies and thus the total ionizing emissivity, allowing for high star-formation efficiencies in massive systems while still matching current reionization data \citep{Lin2024}. 

AGNs uncovered by JWST and other facilities may also contribute non-negligibly to the ionizing budget at $z\gtrsim5$. Revised AGN luminosity functions with relatively high space densities of faint AGNs indicate that active SMBHs could produce a large fraction of the required ionizing photons by $z\sim5$--6 \citep{Grazian2024}. However, reconciling the late, sharp reionization inferred from Ly$\alpha$ with an AGN-dominated scenario generally requires very high AGN duty cycles or escape fractions at $z\sim6$--7 \citep{Kageura2025}.

In the emerging picture, JWST's detection of numerous bright galaxies at $z\gtrsim 10$ pushes models toward an \emph{earlier onset} of reionization and large ionized regions around individual sources, while Ly$\alpha$-based neutral fraction measurements, Ly$\alpha$-forest constraints, and CMB $\tau$ favour a \emph{late and relatively rapid} global reionization, with the bulk of ionization occurring at $z\sim6$--7 and completion by $z\sim5$--6. Reconciling these requires strong redshift- and mass-dependent regulation of the escaping ionizing emissivity, and possibly a non-negligible but subdominant contribution from AGN.

\vspace{3mm}
\noindent
\textbf{Time-Dependent Escape Fraction and Burst-Driven Ionization}\\
In many semi-analytic reionization models, the escape fraction of ionizing photons, $f_{\rm esc}$, is treated as a time-averaged parameter. However, both recent \textit{JWST} observations and high-resolution radiation--hydrodynamic simulations indicate that ionizing escape is intrinsically time-dependent, anisotropic, and closely linked to the bursty nature of star formation in early galaxies.

Spectroscopic observations at $z \gtrsim 8$ reveal that many UV-bright galaxies host very young stellar populations with characteristic ages of only a few Myr and exhibit hard ionizing spectra consistent with elevated ionizing photon production efficiencies, $\xi_{\rm ion}$ \citep[e.g.,][]{Nakane2024,Kokorev2025,Mason2026}. 
In such systems, stellar feedback from massive stars and supernovae acts on short dynamical timescales, generating superbubbles that can puncture the interstellar medium and create low-column-density channels. As a consequence, $f_{\rm esc}$ is expected to vary strongly in time, typically peaking within $\sim 3$--$10$ Myr after the onset of a star-formation episode, when feedback-driven clearing becomes effective \citep[e.g.,][]{Kimm14,Ma2020}. 

Zoom-in simulations that resolve individual star-forming regions support this picture. The escape fraction can fluctuate by more than an order of magnitude over short timescales, tracking the formation and breakout of superbubbles \citep{Ma2020}. 
During embedded phases, newly formed clusters are surrounded by dense gas and $f_{\rm esc}$ remains low. After supernova-driven evacuation of the surrounding medium, ionizing photons can escape efficiently, often along preferential directions. The leakage is therefore highly anisotropic, and the angle-averaged $f_{\rm esc}$ depends sensitively on the relative timing between star formation, feedback, and gas inflow. Radiation--hydrodynamic simulations such as SPHINX and related works further demonstrate that binary stellar evolution can prolong ionizing photon production and enhance the effective escape fraction \citep{Rosdahl18,Katz21}.

This burst-driven behavior has important implications for interpreting the apparent tension between the large number densities of bright $z \gtrsim 10$ galaxies inferred from \textit{JWST} UV luminosity functions and the late, rapid reionization favored by Ly$\alpha$ and CMB constraints \citep[e.g.,][]{Nakane2024,Cain2025,Mason2026}. 
Individual galaxies may generate large local ionized bubbles during short-lived phases of enhanced ionizing emissivity without sustaining a high time-averaged escaping photon budget. In other words, the instantaneous ionizing output inferred from UV luminosities does not necessarily translate into a proportionally large global contribution to reionization unless the duty cycle of high-escape phases is substantial.

Consequently, the relevant quantity for reionization modeling is not simply a global mean $\langle f_{\rm esc} \rangle$, but the joint distribution of escape fraction, burst duty cycle, halo mass, and environment. Time variability introduces additional scatter in the ionizing emissivity,
$\dot{n}_{\rm ion} = f_{\rm esc}\,\xi_{\rm ion}\,\rho_{\rm SFR}$,
and may enhance the patchiness of ionized regions. Capturing this burst-driven, anisotropic leakage requires radiation--hydrodynamic simulations that self-consistently resolve the interplay between star formation, feedback, and gas dynamics across multiple spatial scales. We therefore turn next to recent advances in hydrodynamic simulations that attempt to predict $f_{\rm esc}$ from first principles.

\vspace{3mm}
\noindent
\textbf{Escape Fractions in Hydrodynamic Simulations} \\
Given that the escape fraction $f_{\rm esc}$ is highly time-variable and regulated by burst-driven feedback processes, it is essential to predict it theoretically using radiation--hydrodynamic simulations that resolve the complex coupling between stars, gas, and radiation. Semi-analytic models can explore parameter space efficiently, but only numerical simulations can self-consistently capture the multi-scale physics that governs the opening of low-density channels, the breakout of superbubbles, and the anisotropic leakage of ionizing photons.

Over the past two decades, numerous studies have investigated $f_{\rm esc}$ in early galaxies and its implications for cosmic reionization \citep[e.g.,][]{Cen03d,Razoumov06a,Gnedin08b,Wise09,Yajima17}. 
While we do not review each study in detail here, it has long been recognized that modeling SN feedback from first principles remains challenging, requiring subgrid prescriptions on spatial scales below $\sim 100\,{\rm pc}$. Despite these limitations, early simulation studies reached a broad consensus that $f_{\rm esc}$ tends to decrease with increasing halo mass, albeit with substantial scatter.
Galaxies residing in more massive halos are embedded in deeper gravitational potential wells and denser interstellar and circumgalactic gas, making it more difficult for ionizing photons to escape into the IGM.

\begin{figure}
    \centering
    \includegraphics[width=13cm]{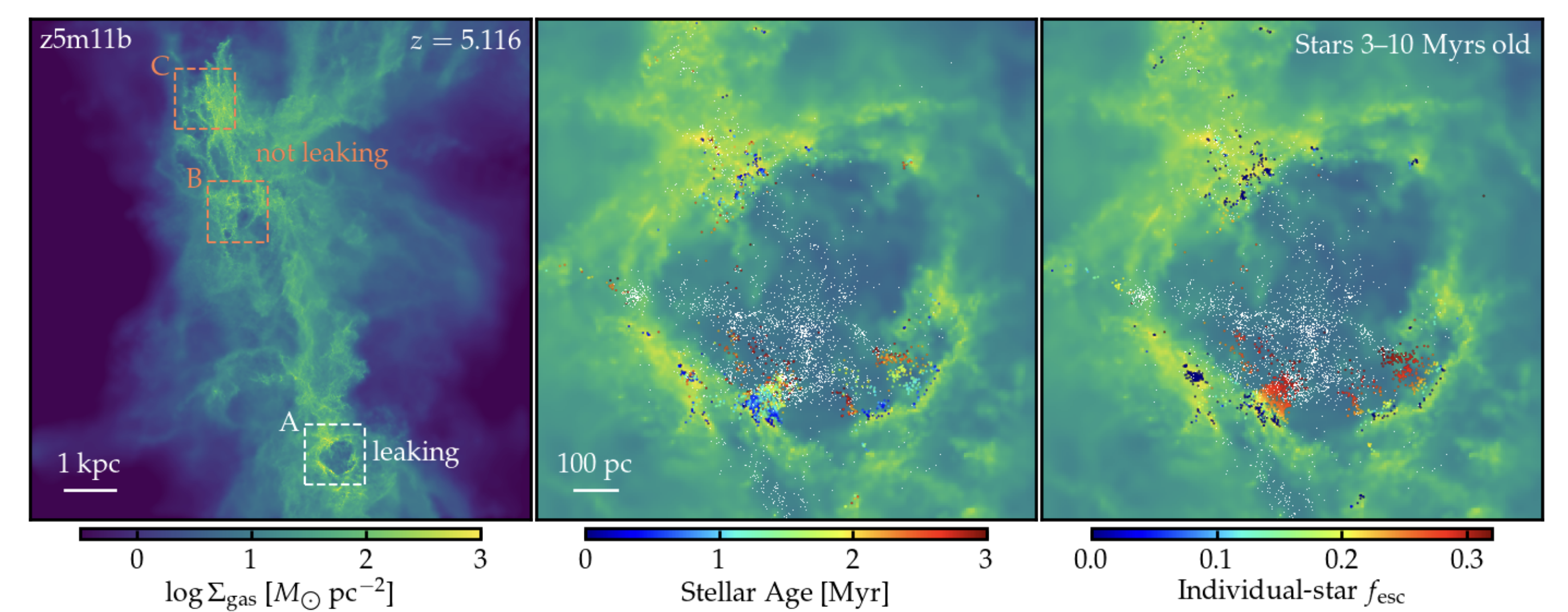}
    \caption{Zoom-in cosmological hydrodynamic simulation of high-redshift galaxies, resolving individual superbubbles with different escape fractions \citep{Ma2020}. The left panel shows the large-scale gaseous filament. The middle and right panels show the star particles overlaid on the gas density in region A of the left panel, color-coded by their stellar ages and $\fesc$, respectively.}
   \label{fig:Ma}
\end{figure}

Recently, advances in numerical techniques and computational resources have enabled parsec- and even sub-parsec-scale resolution in simulations of early galaxies, allowing the formation and evolution of individual superbubbles to be resolved explicitly. 
Cosmological zoom-in simulations now reach resolutions sufficient for capturing clustered star formation and its feedback-driven regulation of ionizing photon escape.

For example, \citet{Ma2020} carried out a cosmological zoom-in simulation using the GADGET-3 SPH code as part of the FIRE-2 project, achieving parsec-scale spatial resolution and a baryonic mass resolution of $\sim 100\,M_\odot$ (see Fig.~\ref{fig:Ma}). This resolution enabled the explicit modeling of clustered star formation and feedback-driven superbubbles that regulate the escape of ionizing photons. 
More recently, \citet{Hirai2025} performed a cosmological zoom-in simulation of a dwarf galaxy that resolves individual massive stars and follows chemical enrichment from single supernova events. By capturing stellar feedback at the level of individual massive stars, such simulations provide an unprecedented level of physical granularity, linking the timing and clustering of supernova explosions directly to the dynamical and chemical evolution of the
surrounding gas.

These cosmological zoom simulations consistently demonstrate that $f_{\rm esc}$ is highly time-dependent, with short-lived peaks associated with the breakout of superbubbles \citep{Ma2020,Katz21}. Ionizing leakage is therefore governed not solely by halo mass, but by the temporal sequencing and spatial clustering of star formation relative to gas clearing. As illustrated in Figure~\ref{fig:Ma}, star clusters embedded in dense gas exhibit low escape fractions, whereas those emerging from cleared channels can reach substantially higher values.


Complementary to these cosmological zoom studies, \citet{Hu19} achieved even finer spatial resolution in an isolated dwarf galaxy model, with a gas mass resolution of $1\,M_\odot$ and a spatial resolution of $0.3\,{\rm pc}$, enabling detailed modeling of individual superbubbles and the multiphase structure of the ISM. 
Related ISM-scale studies have further examined the interaction between supernova remnants and dense gas \citep{Girichidis16,Martizzi16,Gatto17,Kim18,El-badry19,Lancaster21a,Oku2022}.
Recent extensions have incorporated additional physics such as magnetic fields and cosmic rays, which can modify ISM porosity, regulate wind launching, and influence the duration and anisotropy of high-$f_{\rm esc}$ phases. 

Despite substantial progress across scales, predicting $f_{\rm esc}$ from first principles remains an open challenge. The escape fraction depends sensitively on unresolved small-scale clumping, burst duty cycles in low-mass halos, binary stellar evolution \citep{Rosdahl18,Katz21}, and the adopted treatment of photon propagation, including reduced speed-of-light approximations \citep{Rosdahl18}. Quantifying these systematic uncertainties and connecting stellar-scale feedback to large-scale reionization remains 
essential for interpreting cosmological radiation-hydrodynamic simulations and their predictions for reionization topology.

\vspace{3mm}
\noindent
\textbf{Cosmological Radiative Hydrodynamics Simulations}\\
To connect stellar-scale feedback to the global reionization history, radiation-hydrodynamic simulations must also operate on cosmological volumes. Significant progress has been made in scaling such simulations to $\sim (100\,{\rm cMpc})^3$ volumes while coupling star formation, stellar feedback, and radiative transfer on-the-fly.

\begin{figure}
    \centering
    \includegraphics[width=11cm]{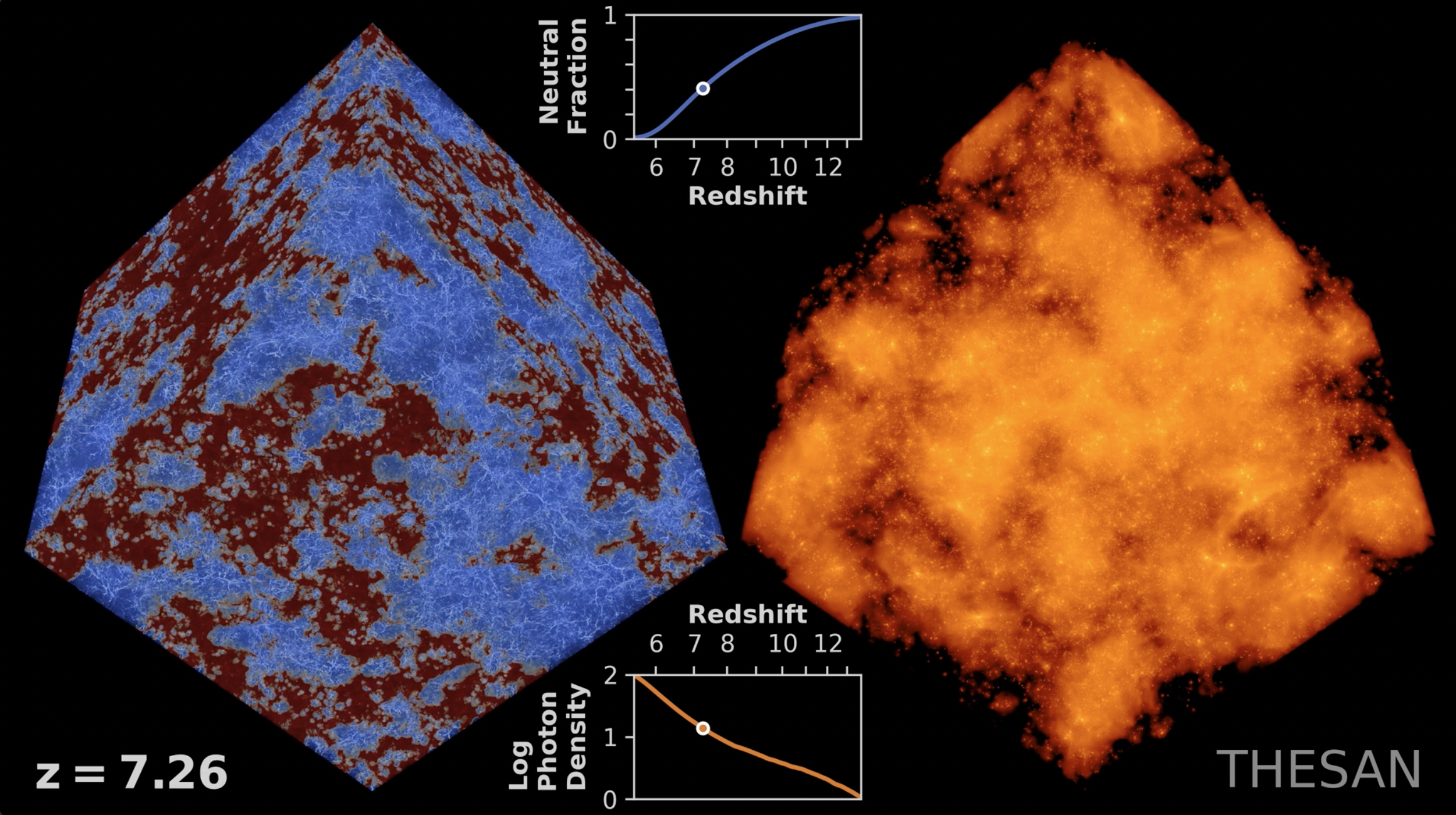}
    \caption{THESAN cosmological radiation hydrodynamic simulation. The left box shows the hydrogen neutral fraction, and the right box shows the ionizing radiation field.}
   \label{fig:THESAN}
\end{figure}

For example, the THESAN simulations \citep{Kannan22,Garaldi22,Yeh2023}
employ radiation-hydrodynamic calculations in a $(95.5\,{\rm cMpc})^3$ volume using the AREPO-RT moving-mesh code, incorporating star formation, stellar feedback, and radiative transfer self-consistently during the simulation (Figure~\ref{fig:THESAN}).
The left plot of Figure~\ref{fig:THESAN} shows the hydrogen neutral fraction at $z=7.26$, illustrating the patchy topology of reionization with large ionized bubbles surrounding overdense regions while neutral islands persist in the voids. The right panel shows the corresponding ionizing radiation field, whose spatial fluctuations trace the distribution of ionizing sources embedded in the cosmic web. The inset panels display the evolution of the volume-averaged neutral fraction and photon density as a function of redshift. 

Historically, performing cosmological simulations with all of these physical processes simultaneously was prohibitively expensive, and most studies relied on post-processing radiative transfer. THESAN builds upon the subgrid physics models developed for IllustrisTNG, treating $f_{\rm esc}$ as an effective subgrid parameter calibrated to reproduce global reionization observables.

The flagship THESAN-1 run achieves a maximum spatial resolution of $\sim 10$\,pc and produces a reionization history that lies between the previously discussed early and late scenarios.
Complementary to this effort, the SPHINX simulations \citep{Rosdahl18,Katz21} perform radiation-hydrodynamic calculations using the RAMSES-RT adaptive mesh refinement code, achieving a comparable maximum spatial resolution of $\sim 11$\,pc in smaller volumes of $5$--10\,cMpc.
Importantly, \citet{Rosdahl18} highlighted the critical role of binary stellar evolution, showing that including binary stars significantly enhances the ionizing photon budget of metal-poor stellar populations and increases the effective escape fraction by a factor of $\sim 3$ compared to single-star models.

\begin{figure}
    \centering
    \includegraphics[width=11cm]{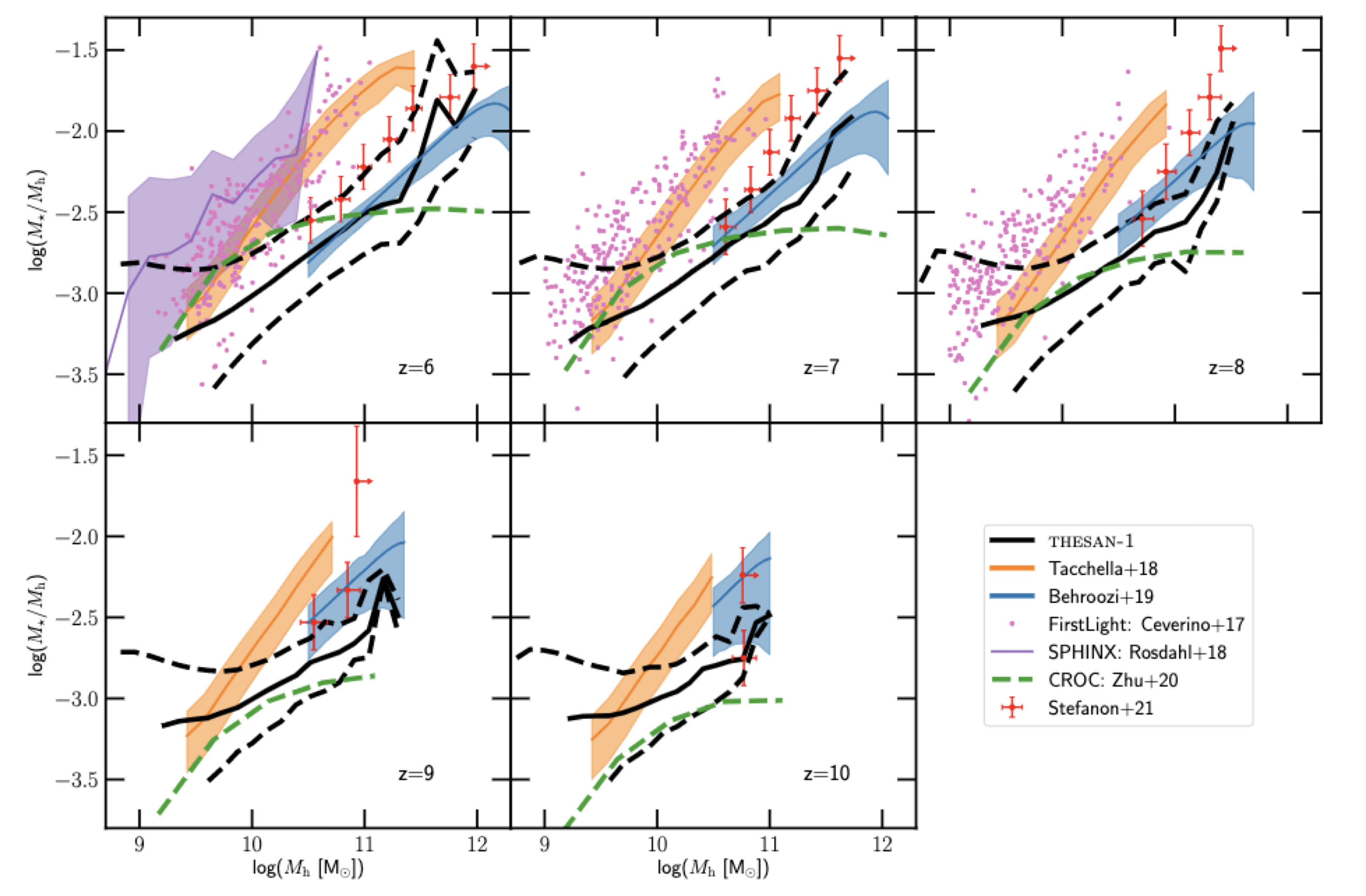}
    \caption{Comparison of SHMR from both simulations and observations \citep{Kannan22}. The blue line with shading is from the abundance matching method \citep{Behroozi19}, and the red data points with error bars are the observational estimates of \citet{Stefanon21}. }
   \label{fig:SHMR}
\end{figure}

\vspace{3mm}
\noindent
\textbf{Stellar-to-Halo Mass Ratio and CGM/IGM Tomography}\\
Beyond reproducing global reionization histories, simulations must also match the galaxy--halo connection inferred from observations.
In addition to constraints from the CMB optical depth and galaxy luminosity functions, measurements of the stellar-to-halo mass ratio (SHMR) have improved substantially over $z \sim 0$--10, as shown in Figure~\ref{fig:SHMR}. The SHMR exhibits a pronounced peak at $M_h \simeq 10^{12}\,M_\odot$, with declining efficiency toward both lower and higher halo masses, commonly interpreted as the cumulative effects of supernova feedback in low-mass halos and AGN feedback in high-mass systems. At $z \gtrsim 6$, however, current constraints probe the low-mass regime primarily, making this redshift range particularly sensitive to the treatment of stellar feedback and burst-driven regulation in simulations.
 
One can see that there are still some discrepancies even between the abundance-matching result of \citet{Behroozi19} and the observational estimate of \citet{Stefanon21}, which uses an abundance-matching technique to match the observed stellar mass function to the Bolshoi dark matter simulation.  This means that a consistency check between the latest galaxy observations and abundance-matching results remains to be performed.
Comparisons with other simulation data yield mixed results, and it is clear that we have not yet reached consensus on the SHMR at high redshift.  It is also noteworthy that \citet{Stefanon21} finds little evolution in SHMR from $z\sim 10$ to $z\sim 6$.  
See also \citet{Shuntov22} for comparisons of SHMR at lower redshifts.

\begin{figure}
    \centering
    \includegraphics[width=6.3cm]{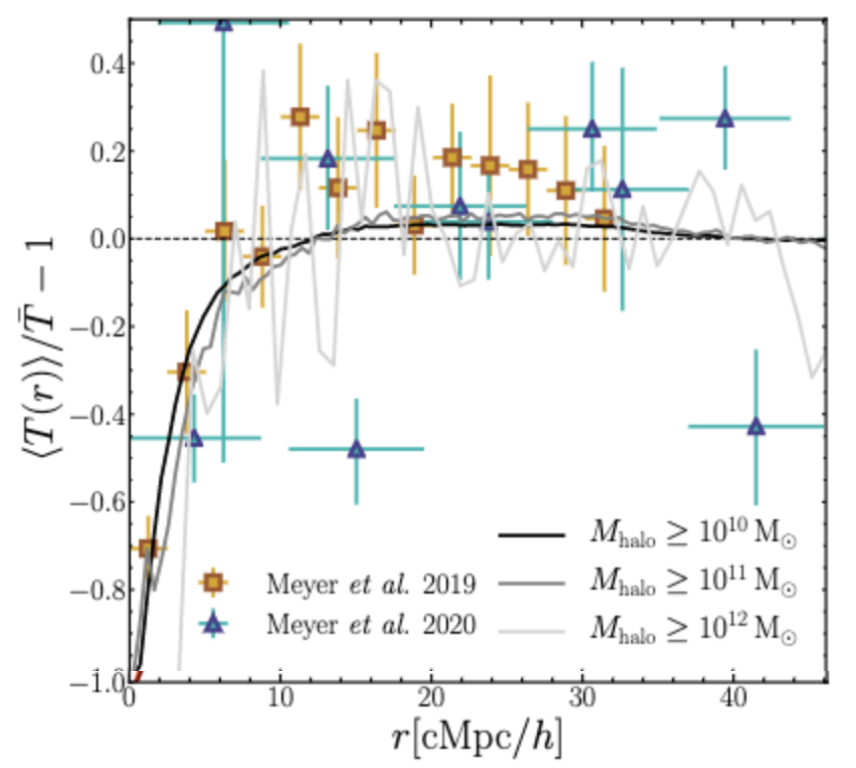}
    \includegraphics[width=7.5cm]{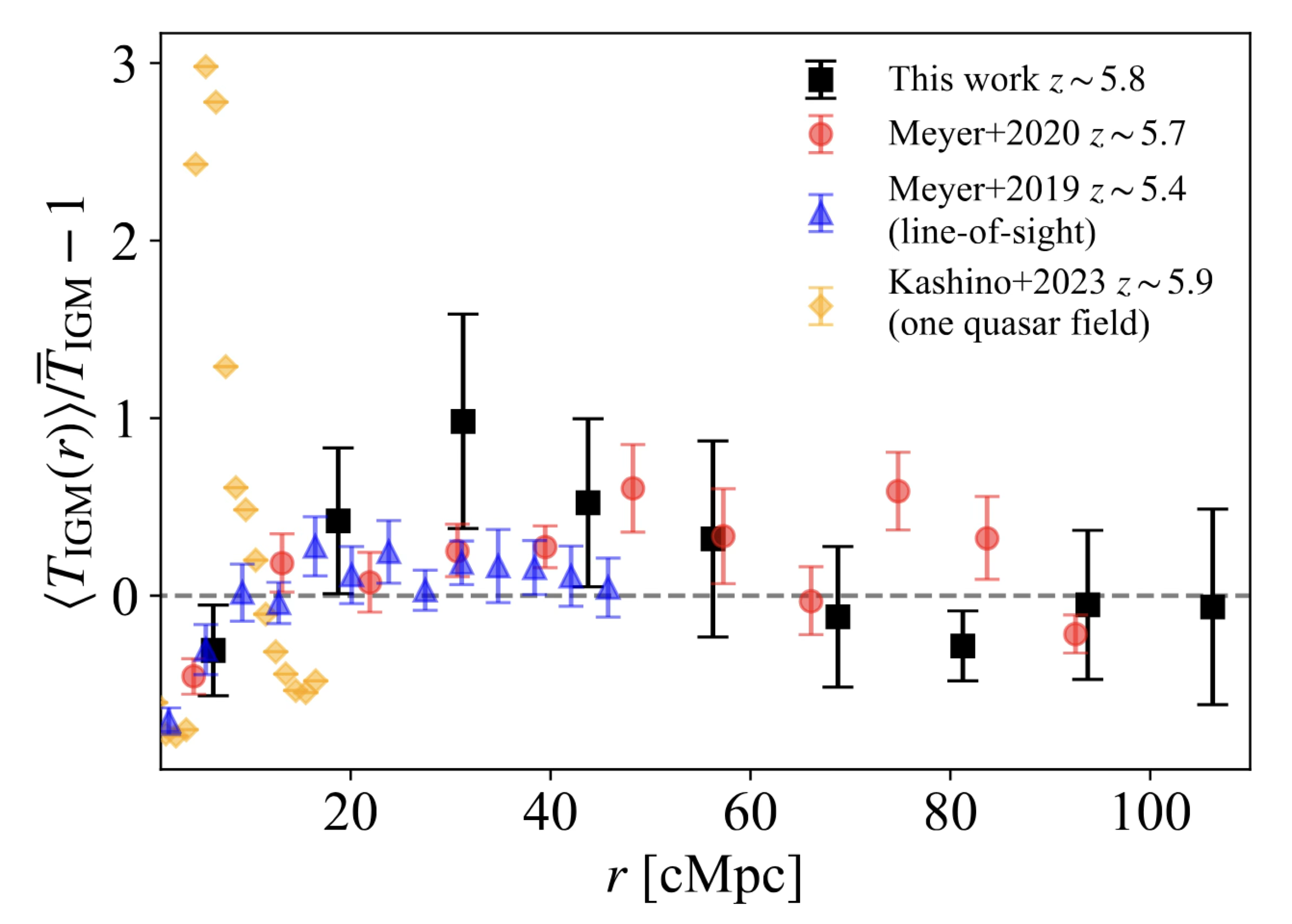}
    \caption{The galaxy--Ly$\alpha$ transmission cross-correlation as a probe of neutral hydrogen distribution around galaxies during the epoch of reionization. \textit{Panel (a):}~Average Ly$\alpha$ transmission relative to the mean, $\langle T(r) \rangle / \bar{T} - 1$, as a function of distance from galaxies in the THESAN-1 simulation at $z=5.5$
    \citep{Garaldi22}. Solid curves show predictions for different minimum
    host halo masses ($M_{\rm halo} \geq 10^{10}$, $10^{11}$, and
    $10^{12}\,M_\odot$), compared with observational data from
    \citet{Meyer19} (orange squares) and \citet{Meyer20} (blue triangles).
    \textit{Panel (b):}~The IGM transmission cross-correlation measured around
    {[O\,{\sc iii}]} emitters using JWST ASPIRE data across five quasar fields
    \citep[black squares at $z\sim5.8$]{Kakiichi2025}. 
    Excess transmission is detected at $\sim20$--40~cMpc, while absorption
    is seen at smaller scales due to gas overdensities around galaxies.}
   \label{fig:transmit}
\end{figure}

Finally, an even more stringent test for simulations would be the distribution of neutral hydrogen around galaxies as probed by the Ly$\alpha$ transmission (or the so-called flux decrement), a technique also known as CGM/IGM tomography \citep{Lee14a,Meyer19,Bosman20,Momose21a,Momose21b,Nag21,Garaldi22,Kakiichi2025}. The cross-correlations between galaxies, neutral hydrogen, and metal absorption lines will provide us with a unique opportunity to constrain the effects of feedback and the ionizing radiation field. 
Figure~\ref{fig:transmit}(a) shows the average Ly$\alpha$ transmission as a function of distance from galaxies, $\langle T(r) \rangle / \bar{T} - 1$, in the THESAN-1 radiation-hydrodynamic simulation at $z=5.5$
\citep{Garaldi22}. The simulation predicts that the Ly$\alpha$ transmission is suppressed near galaxies due to gas overdensities in their environments, and gradually recovers toward the mean IGM value at larger distances. The amplitude and radial extent of this suppression depend on the host halo mass, with more massive halos ($M_{\rm halo} \geq 10^{12}\,M_\odot$) producing deeper and more extended absorption troughs, which is also consistent with the results at $z\sim2$--3 by \citet{Nag21}. 
The THESAN predictions are compared with the observational data from \citet{Meyer19,Meyer20}, showing broad qualitative agreement. \\
\indent Figure~\ref{fig:transmit}(b) presents a complementary observational measurement from \citet{Kakiichi2025}, who used JWST ASPIRE data across five quasar fields to measure the galaxy--Ly$\alpha$ transmission cross-correlation around [{\sc O iii}] emitters at $z\sim5.8$. Their data reveal $2\sigma$ evidence for excess Ly$\alpha$ forest transmission at scales of $\sim 20$--40\,cMpc, indicating that [{\sc O iii}] emitters reside within large, highly ionized regions of the IGM. At smaller separations ($r \lesssim 10$\,cMpc), the transmission is preferentially suppressed, consistent with the gas overdensities seen in the THESAN simulation. Also shown is the earlier result from \citet{Kashino2023}, which is based on a single quasar field from the EIGER survey. More recently, \citet{Kashino2026} presented the full EIGER dataset comprising 948 [{\sc O iii}] emitters across six quasar sightlines, finding clear redshift evolution in the galaxy--IGM correlation: suppressed transmission in overdense regions at $z < 5.5$, transitioning to enhanced transmission at $5.7 < z < 6.15$, supporting an inside-out progression of reionization. Current observational errors remain dominated by cosmic variance, and future analyses incorporating additional QSO fields from JWST will significantly improve these constraints. 
The 2D correlation maps can also provide additional information on the gas dynamics (inflows/outflows) around galaxies \citep{Turner17,Chen20}.

\vspace{3mm}
\noindent
\textbf{Summary} \\
The escape fraction of ionizing photons remains a critical parameter for reionization studies, but it is increasingly clear that $f_{\rm esc}$ is neither constant nor purely mass-dependent. Rather, it reflects the complex, burst-driven coupling between star formation, feedback, gas dynamics, and radiation. 
Understanding this radiation--matter coupling across scales is one of the central theoretical challenges in first-galaxy formation physics and cosmic reionization.



Within this framework, the scientific return of GREX-PLUS can be assessed at three levels. 
At minimum, the mission will discover and characterize the brightest
first galaxies, establishing the existence and basic properties of the most massive systems at $z \gtrsim 10$. 
The ``nominal success" would be to significantly improve constraints on the UVLFs, stellar mass functions, and SHMR at $z > 6$, thereby anchoring the mapping between dark matter halos and baryonic growth. 
At the most ambitious level, the mission will connect galaxy properties to their ionizing impact on the surrounding IGM --- distinguishing between steady and burst-driven ionization scenarios, constraining the effective escaping ionizing emissivity, and providing decisive evidence for the timing, duration, and topology of cosmic reionization.

In this sense, GREX-PLUS will not only refine the reionization history but also directly constrain the baryon cycle and radiation--matter coupling that govern galaxy formation in the early Universe.

\printbibliography[heading=subbibliography]
\end{refsection}

\clearpage

\begin{refsection}[2-2_veryhighz/veryhighz.bib]

\section{First Galaxy and Cosmic Reionization}
\label{sec:veryhighz}

\noindent
\begin{flushright}
Yuichi Harikane$^{1}$, 
Akio K. Inoue$^{2}$
\\
$^{1}$ ICRR, University of Tokyo, 
$^{2}$ Waseda University 
\end{flushright}
\vspace{0.5cm}

\subsection{Scientific background and motivation}

Observations of the highest-redshift galaxies provide the strongest observational constraints on structure formation in the Universe.
In the concordance $\Lambda$ cold dark matter (CDM) structure formation model, the number density of bright, massive galaxies is smaller in the earlier Universe \citep[e.g.,][]{2004ApJ...610...45N,2004MNRAS.350..385N,2006MNRAS.366..705N,2012MNRAS.420.1606J}.
If there are more massive galaxies than predicted by the theory, it is either due to a lack of understanding of the baryonic physics involved in galaxy formation (gas cooling, heating, and star formation) or perhaps a flaw in the established theoretical model of structure formation.
It is thought that these early galaxies also emitted ultraviolet (UV) photons and were responsible for cosmic reionization.
Observations of the Universe at redshift $z=6$--7, when cosmic reionization was completed, advanced considerably with large telescopes; however, the beginning of cosmic reionization at $z>10$ was poorly explored before the launch of the James Webb Space Telescope (JWST) in 2021.

One of the few galaxies identified at $z>10$ before 2021 is GN-z11 \citep{2016ApJ...819..129O,2021NatAs...5..256J,2023A&A...677A..88B}.
A major surprise of GN-z11 is its remarkably high luminosity, $M_\mathrm{UV}=-21.5$ mag.
Given that it is not gravitationally lensed, GN-z11 is located in the brightest part of the rest-frame UV luminosity function.
Although the narrow field-of-view (FoV) of the Hubble Space Telescope (HST)/Wide Field Camera 3 (WFC3) in the near-infrared has limited the imaging survey areas to $<1\ \mathrm{deg^2}$, several studies using HST report very luminous Lyman break galaxy (LBG) candidates at $z\sim9-10$ more frequently than expectations from a Schechter-shaped luminosity function (e.g., \citealt{2018ApJ...867..150M,2022ApJ...928...52F,2024ApJ...961..209B,2023MNRAS.524.5454L}, see also \citealt{2022ApJ...927..236R}).

More statistically robust results have come from near-infrared imaging surveys covering a few square degrees with the Visible and Infrared Survey Telescope for Astronomy (VISTA) and UK Infrared Telescope (UKIRT) such as UltraVISTA \citep{2012A&A...544A.156M}, the UKIRT InfraRed Deep Sky Surveys (UKIDSS, \citealt{2007MNRAS.379.1599L}), and the VISTA Deep Extragalactic Observation (VIDEO) Survey \citep{2013MNRAS.428.1281J}.
These surveys have revealed that the UV luminosity functions at $z\sim9-10$ are more consistent with double power-law functions than with standard Schechter functions with an exponential cutoff at the bright end \citep[][]{2017ApJ...851...43S,2019ApJ...883...99S,2020MNRAS.493.2059B}.
Previous studies also report similar number density excesses beyond the Schechter function at $z\sim4-7$ \citep{2018PASJ...70S..10O,2018ApJ...863...63S,2020MNRAS.494.1771A,2022ApJS..259...20H}, implying little evolution of the number density of bright galaxies at $z\sim4-13$ \citep{2020MNRAS.493.2059B,2022ApJS..259...20H,2022ApJ...929....1H}.

In addition to these observations of bright galaxies at $z\sim9-13$, several studies independently suggest the presence of star-forming galaxies in the early Universe, even at $z\sim15$.
A candidate $z\sim12$ galaxy is photometrically identified in very deep {\it HST}/WFC3 images obtained in the Hubble Ultra Deep Field 2012 (UDF12) campaign \citep{2013ApJ...763L...7E}, and its redshift is confirmed with recent JWST observations \citep{2023NatAs...7..622C}.
An analysis of passive galaxy candidates at $z\sim6$ reports that their stellar population is dominated by old stars with ages of $\gtrsim700$ Myr, consistent with star-formation activity at $z>14$ \citep{2020ApJ...889..137M}.

\begin{figure}
    \centering
    \includegraphics[width=15cm]{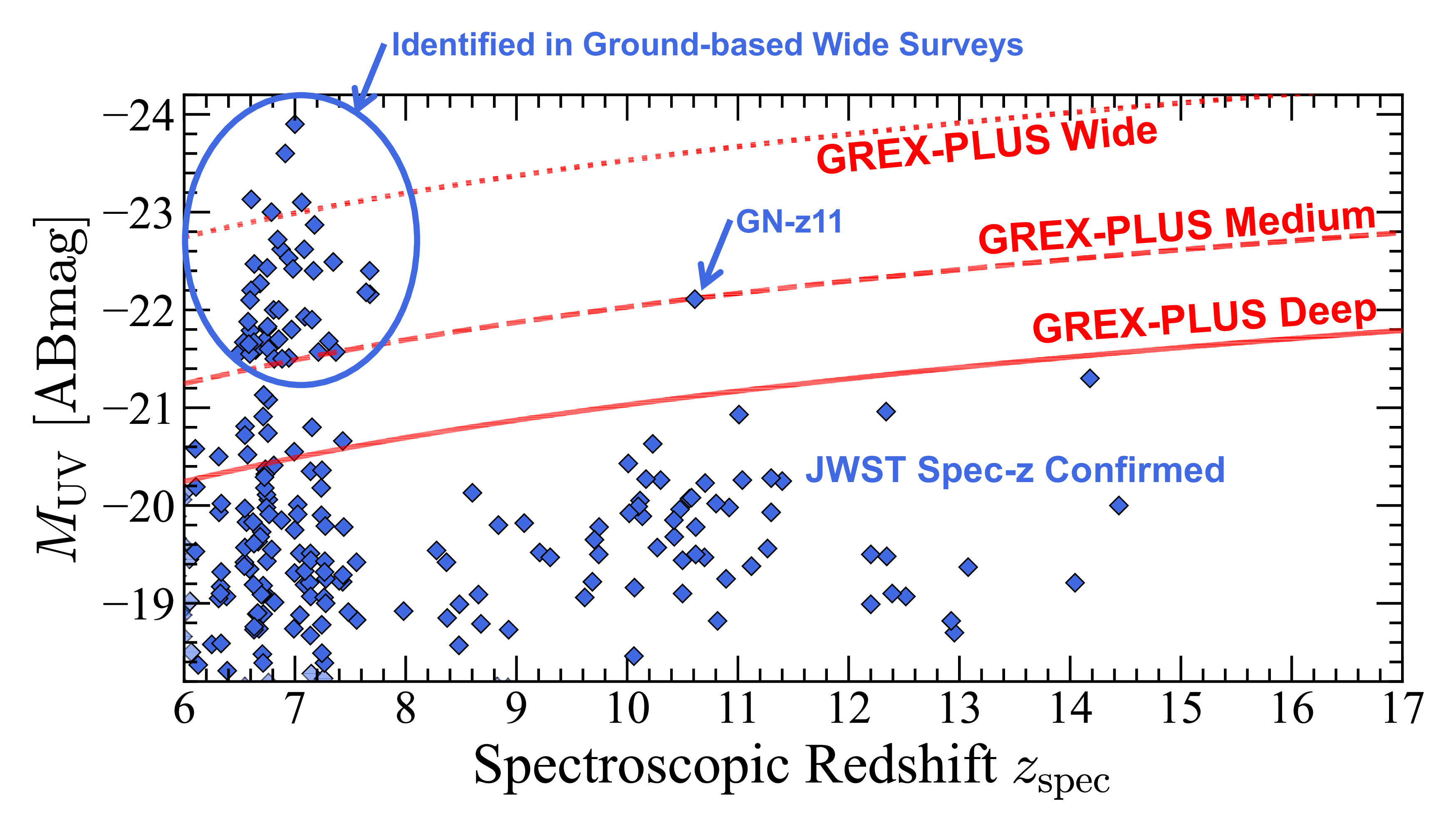}
    \caption{Absolute ultraviolet magnitude of galaxies as a function of redshift. The blue diamonds are spectroscopically confirmed galaxies compiled by \citet{2024ApJ...960...56H,2025ApJ...980..138H} and \citet{2025arXiv250821708R}. The red solid, dashed, and dotted curves are the 5$\sigma$ limiting magnitudes of the three GREX-PLUS imaging surveys: Deep, Medium, and Wide, respectively. Although the bright ($M_\mathrm{UV}<-21$ mag) galaxies have been identified at $z<8$ with ground-based wide-area surveys, no bright galaxies except for GN-z11 have been identified at $z>10$. GREX-PLUS surveys will probe this unexplored parameter space and search for bright galaxies in the early Universe.}
    \label{fig:GPandJWST-1}
\end{figure}

Finally, the JWST observations have identified and confirmed galaxies at $z
>10$, including JADES-GS-z14-0 \citep{2024Natur.633..318C,2025A&A...696A..87C,2025ApJ...988...19S} at $z=14.2$ and MoM-z14 at $z=14.4$ \citep{2026OJAp....956033N}.
These observations again suggest a high number density of bright galaxies even at $z\gtrsim12$ \citep[e.g.,][]{2022ApJ...940L..14N,2022ApJ...938L..15C,2022ApJ...940L..55F,2023MNRAS.518.6011D,2023ApJS..265....5H,2023MNRAS.523.1009B} (see Fig.~\ref{fig:GPandJWST-1}).
These studies indicate a higher number density of bright galaxies and more active star formation in the early Universe than previously thought.
Subsequent spectroscopic studies also confirm this trend of the overabundance of bright galaxies at $z>10$ \citep[e.g.,][]{2025ApJ...980..138H,2024ApJ...960...56H}.
Possible reasons for this high number density of bright galaxies include inefficient negative feedback in the galaxy formation process in massive halos in the early Universe, hidden AGN activity in these bright galaxies, a top-heavy initial mass function, and other unknown physical processes in massive galaxy formation, or perhaps a flaw in the theoretical model of structure formation (see discussion in, e.g., \citealt{2022ApJ...929....1H,2023ApJS..265....5H,2022MNRAS.514L...6P,2022ApJ...938L..10I,2023MNRAS.521..497M,2023MNRAS.522.3986F,2023ApJ...951L..40S}).
However, the number of bright galaxy candidates is limited, and the distinction between the Schechter function and the double power-law function still lacks statistical significance. 
Importantly, ongoing/upcoming wide-area surveys with space telescopes such as Euclid and Roman will observe only up to 1.8 and 2.3 $\mu$m, respectively, and are expected to identify bright galaxies up to $z=11$--13.
To search for bright galaxies at $z\gtrsim14$, when the first galaxies form, we need a wide-field survey covering $2$--5\,$\mu$m using the GREX-PLUS near-infrared wide-field camera.

\subsection{Required observations and expected results}

\begin{figure}
    \centering
    \includegraphics[width=15cm]{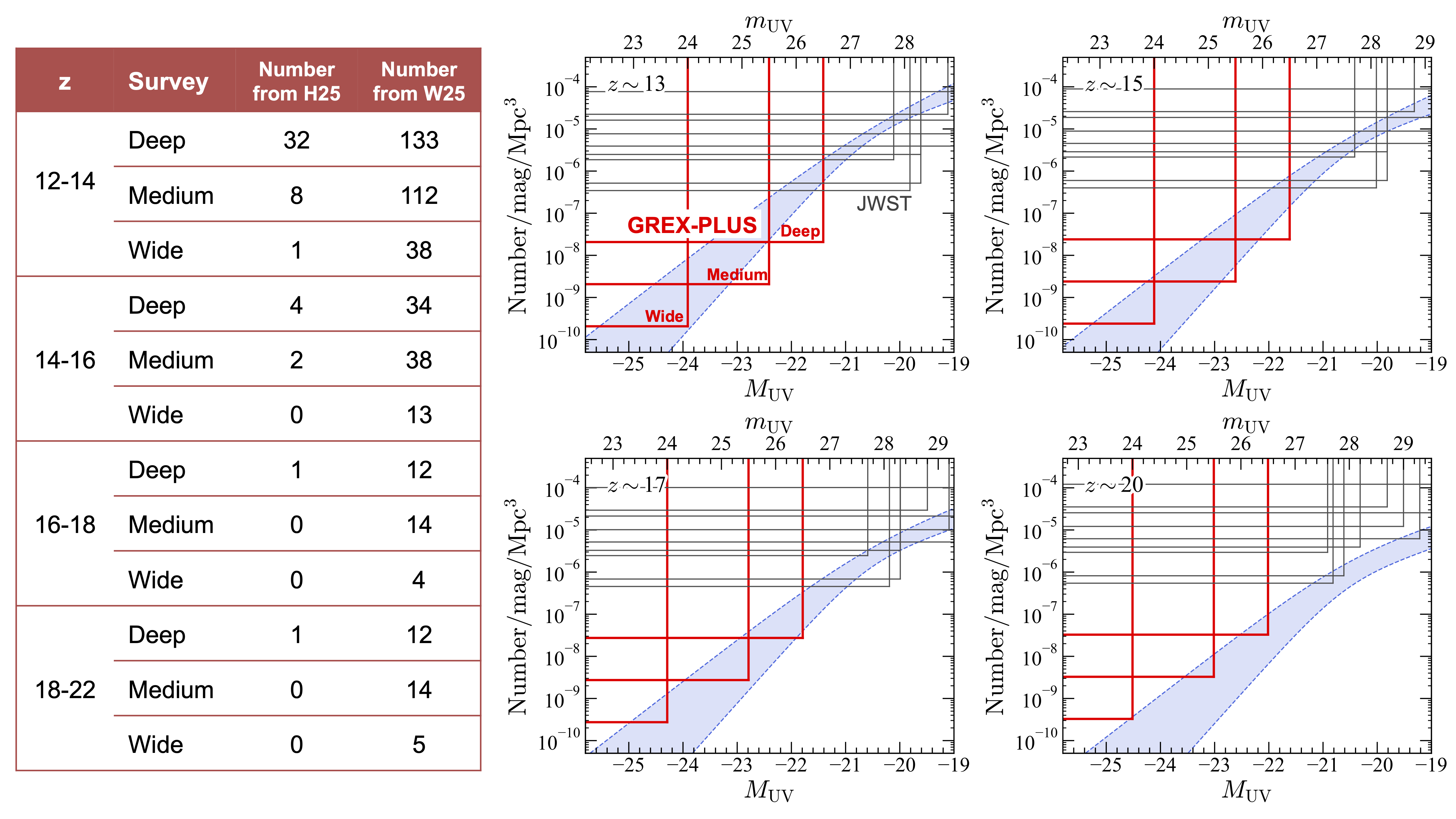}
    \caption{Expected numbers of galaxies detected in the GREX-PLUS imaging surveys (left table). In the right four panels, the blue shaded regions show the range of the ultraviolet luminosity functions constrained by JWST and their extrapolation to $z\sim20$ from \citet[][H25]{2025ApJ...980..138H} and \citet[][W25]{2025ApJ...992...63W}. The vertical and horizontal lines in the panels show the imaging survey sensitivity and area (or survey volume), respectively. The gray lines are ongoing JWST surveys, and the red lines are the three GREX-PLUS surveys.}
    \label{fig:GPandJWST-2}
\end{figure}

We consider three surveys aimed at finding galaxies in the early Universe at $z>15$ (Fig.~\ref{fig:GPandJWST-2}; Table~\ref{tab:fitstgal}).
First, due to Ly$\alpha$ scattering by neutral hydrogen in the intergalactic medium, photons from galaxies at $z>15$ will not be detected below an observed wavelength of 2 $\mu$m (Ly$\alpha$ break, or Gunn-Peterson Trough; \citealt{1965ApJ...142.1633G}).
Therefore, a wavelength of 2 $\mu$m or longer is required to observe galaxies at $z>15$.
In addition, in order to distinguish between the Ly$\alpha$ break of $z>15$ galaxies and the red colors of low-redshift galaxies (e.g., passive galaxies and/or dusty galaxies), two or more photometric data points are required at the rest-frame wavelengths of $0.15$--0.3 $\mu$m, longer than the wavelength of the Ly$\alpha$ break. 
GREX-PLUS is primarily designed to detect galaxies, not to spatially resolve them.
Therefore, the required spatial resolution is set to about 1 arcsec to separate high-redshift galaxies from surrounding foreground objects and avoid the confusion limit.
This resolution corresponds to $3$--4\,kpc at $z=10$ to 15, which is about the diameter of a bright high-redshift galaxy \citep{2015ApJS..219...15S,2024PASJ...76..219O,2025ApJ...991..222O,2023ApJ...951...72O}.
To achieve the 1 arcsec resolution at the observed wavelength of 5 $\mu$m, an aperture of $\sim$1.0 m is required, assuming the diffraction limit.
Regarding the depth of the search, in general, we need deeper surveys with a smaller survey area to detect high-redshift galaxies, whereas a larger area is required for shallower surveys.
Here, we consider a galaxy with a magnitude of $\sim26$ AB mag, whose emission lines can be detected and whose redshift can be spectroscopically determined in a few hours using ALMA, JWST, or a 30-m ground-based telescope.

We calculate the expected number of galaxies detected in the GREX-PLUS surveys using the UV luminosity functions in \citet{2025ApJ...980..138H} and \citet{2025ApJ...992...63W} by extrapolating their parameters to $z\sim20$.
Based on these calculations, a 10 (100) square-degree survey with a $5\sigma$ limiting magnitude of 26.5 (25.5) AB mag can detect tens to hundreds of galaxies with $z>15$ in the case of the double power-law function (Fig.~\ref{fig:GPandJWST-2}).
If extremely bright galaxies with $M_{\rm UV}\sim-24$ also exist in the earliest Universe, they will be efficiently detected by a $\sim24$ AB mag shallow but extremely wide area survey of 1000 square degrees.
Assuming the sensitivity of a cooled telescope with a 1.0 m primary mirror aperture, a focal plane field of view of 0.3 square degrees or more is required to realize such surveys in a total observing time of less than about one year. 
The deep survey fields of the Subaru Telescope (e.g., Hyper Suprime-Cam/HSC SSP Deep fields) and of Euclid and Roman are good candidates for GREX-PLUS survey fields, because the optical-to-near-infrared deep images up to 2 $\mu$m obtained by such telescopes are useful for removing foreground galaxies in the search for high-redshift galaxies.
Satellite orbits capable of observing these fields with sufficient frequency are necessary.
There is no requirement for observation cadence, time resolution, or pointing arrival time.

\subsection{Scientific goals}

We will conduct 10 square-degree, 26.5 AB mag-depth (5$\sigma$, point source) and 100 square-degree, 25.5 AB mag-depth (5$\sigma$, point source) surveys at wavelengths of 2$-$5\,$\mu$m and detect $>10$ galaxies with redshifts higher than those found by the JWST and Roman surveys.
This will provide stronger constraints on the physical processes of first galaxy formation and the onset of cosmic reionization.
We will also test $\Lambda$CDM structure formation models in the early Universe.

\begin{table}
    \begin{center}
    \caption{Required observational parameters.}\label{tab:fitstgal}
    \begin{tabular}{|l|p{9cm}|l|}
    \hline
     & Requirement & Remarks \\
    \hline
    Wavelength & 2--5 $\mu$m & \multirow{2}{*}{$a$} \\
    \cline{1-2}
    Spatial resolution & $\lesssim1$ arcsec & \\
    \hline
    Wavelength resolution & $\lambda/\Delta \lambda>3$ & $b$ \\
    \hline
    Field of view & 10 degree$^2$, 26.5 AB mag ($5\sigma$, point-source) & \multirow{2}{*}{$c$}\\
    \cline{1-1}
    Sensitivity & 100 degree$^2$, 25.5 AB mag ($5\sigma$, point-source) & \\
    \hline
    Observing field & Fields where deep imaging data at $\lambda<2$ $\mu$m are available. & $d$ \\
    \hline
    Observing cadence & N/A & \\
    \hline
    \end{tabular}
    \end{center}
    $^a$ Primary mirror $\phi\geq1.0$ m is required to achieve $\lesssim1$ arcsec at $\lambda=5$ $\mu$m for the diffraction limit.\\
    $^b$ Three or more bands are required for the color selection of very high-$z$ galaxies.\\
    $^c$ A $>0.3$ degree$^2$ field of view of a single pointing is required from the point-source sensitivity for a $\phi=1.0$ m telescope and an assumed amount of observing time.\\
    $^d$ For example, deep fields observed with Subaru and Roman.
\end{table}


\printbibliography[heading=subbibliography]
\end{refsection}

\clearpage

\begin{refsection}[2-3_lssandmassassembly/lssandmassassembly.bib]

\section{Large-Scale Structure and Galaxy Mass Assembly History}
\label{sec:lssandmassassembly}
\def\um{$\mu${\rm m}}
\noindent
\begin{flushright}
Kosuke Takahashi$^{1}$,
Tadayuki Kodama$^{1}$ 
\\
$^{1}$ Tohoku University, Japan
\end{flushright}
\vspace{0.5cm}

\subsection{Scientific background and motivation}
Exploring ultra-massive galaxies ($M_\ast\gtrsim10^{11}~\mathrm{M_\odot}$)
in the early Universe ($5<z<10$) yields crucial insights into the onset of star formation and the build-up of stellar mass, in a way complementary to the direct identification of the most distant galaxies at $z>14$ \citep[e.g.,][]{2024Natur.633..318C,2026OJAp....956033N}. The existence of such massive systems at high redshifts --- particularly quiescent galaxies --- implies that galaxy formation and quenching occurred much earlier and more rapidly than previously anticipated. This challenges the standard bottom-up structure formation scenario within the current cold dark matter (CDM) framework.

To date, discoveries such as a quiescent galaxy at $z=3.717$ with a stellar mass of $M_\ast=1.5\times10^{11}~\mathrm{M_\odot}$ \citep{2017Natur.544...71G} and another at $z=4.898$ with $M_\ast=9.9\times10^{10}~\mathrm{M_\odot}$ \citep{2025NatAs...9..280D} indicate the presence of rapid mass assembly and rapid quenching of massive galaxies at $z\gtrsim4$. To place tighter constraints on the timescales of early galaxy formation, it is essential to extend observations to even higher redshifts ($z>4.5$). Recently, the James Webb Space Telescope (JWST) has revealed a growing population of massive quiescent galaxies at $z>3$ through deep photometric and spectroscopic observations \citep[e.g.,][]{2023ApJ...947...20V}. However, JWST’s survey capability remains limited by its relatively small field of view (NIRCam: $2.2'\times2.2'\times2$ chips, MIRI imager: $1.23'\times1.88'$) and restricted filter set (number of available filters, especially medium-band filters and narrow-band filters) compared to ground-based telescopes.

GREX-PLUS is designed to cover a wide wavelength range from 2.0 \um\ to 7.7 \um. In particular, it provides unique access to wavelengths longer than 2.5 \um, which are inaccessible to ground-based observatories. 
This capability enables direct measurements of stellar emission at rest-frame wavelengths longer than 6000 \AA\ for galaxies out to redshifts of $z\sim10$, allowing robust determinations of their stellar masses.
Understanding the build-up of stellar mass in galaxies across cosmic time is a fundamental goal of galaxy evolution studies. 
However, galaxies with stellar masses exceeding $M_{\ast}\sim10^{10}~\mathrm{M_{\odot}}$ are intrinsically rare due to their low number densities, particularly at high redshift. 
As a result, wide-field surveys are essential to efficiently identify and characterize such massive systems.
\subsection{Required observations and expected results}
We propose three plans to achieve our science goals.
\begin{itemize}
\item Wide (1000 $\mathrm{deg.^2}$, 24.0 mag with $5\sigma$ in the NIR bands)

Find ultra-massive galaxies at $z\sim6$ 

\item Medium (100 $\mathrm{deg.^2}$, 25.5 mag with $5\sigma$ in the NIR bands)

Find ultra-massive galaxies at $z\lesssim8$ 

\item Deep (10 $\mathrm{deg.^2}$, 26.5 mag with $5\sigma$ in the NIR bands)

We will measure the stellar mass function and will be able to reach stellar masses of 
$M_{\ast} \sim 10^{9}\,\mathrm{M_{\odot}}$ for star-forming galaxies at 
$z \lesssim 10$. For quiescent galaxies, we can detect stellar light from 
systems with $M_{\ast} \sim 10^{10}\,\mathrm{M_{\odot}}$ out to $z = 8$.
\end{itemize}

Galaxies are commonly classified into star-forming galaxies (SFGs) and quiescent galaxies (QGs) according to their star-formation activity. 
These two populations provide complementary insights into galaxy growth and quenching processes, and we therefore propose distinct science cases for each.

\begin{figure}[h]
    \begin{center}
      \includegraphics[width=\linewidth]{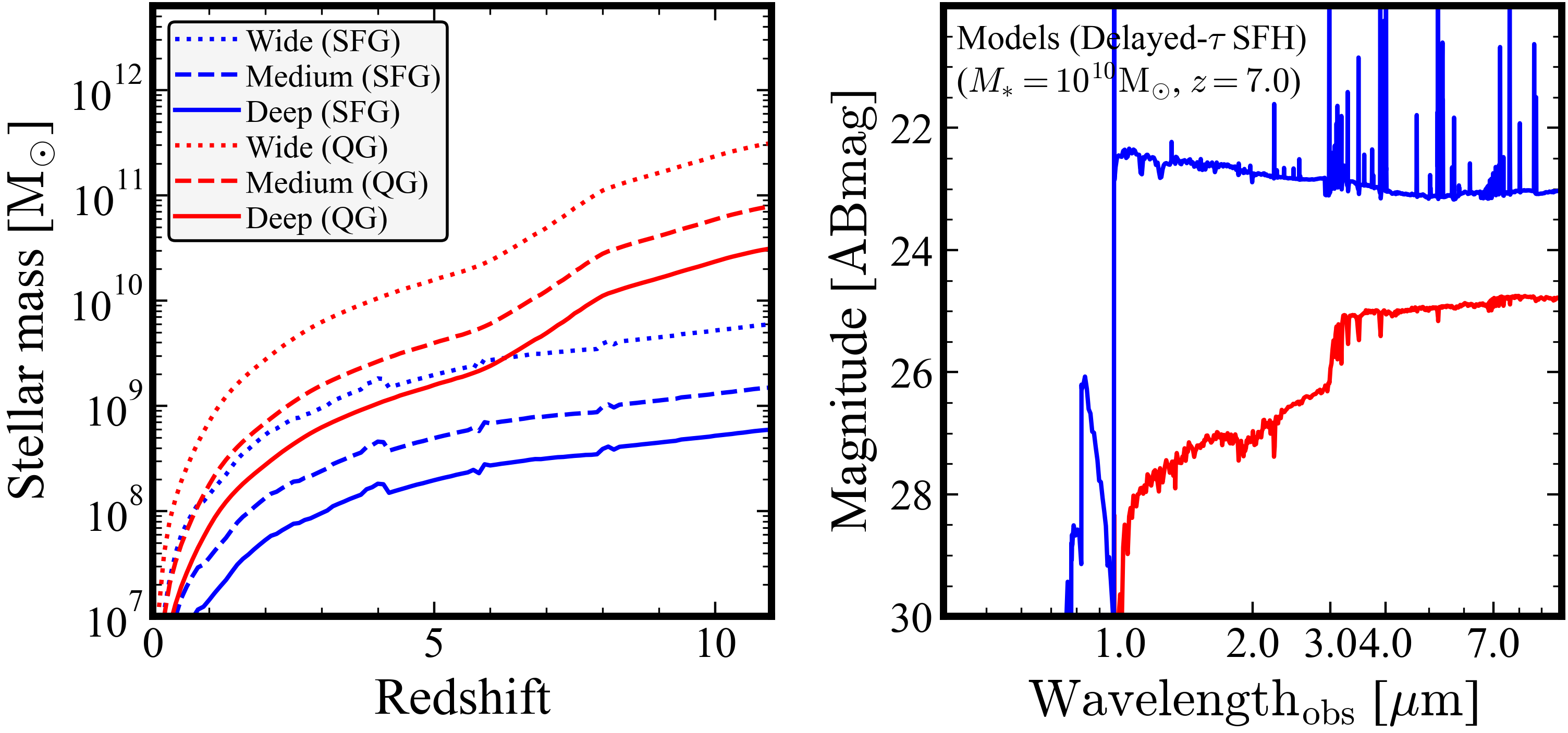}
    \end{center}
    \caption{({\it Left}) Limiting stellar mass of galaxies as a function of redshift with GREX-PLUS. For star-forming galaxies ({\it Blue}), we use the model flux in the F397 filter. In contrast, we use the F303 flux to calculate the mass limit for quiescent (Balmer-break) galaxies ({\it Red}). The first selection of quiescent galaxies at $z>5.5$ must be conducted by GREX-PLUS itself, capturing the Balmer break feature with two adjacent filters. The right panel shows the model SEDs for SFGs and QGs used to calculate the detection limits shown in the left panel. These models were created using the \texttt{CIGALE} SED fitting code \citep{2019A&A...622A.103B}.}
    \label{fig:grex_masslim_massive}
\end{figure}

For SFGs, GREX-PLUS will enable the detection of stellar emission from galaxies with stellar masses down to $M_{\ast}\sim10^{9}~\mathrm{M_{\odot}}$ out to $z\sim10$.
Large samples of such high-redshift SFGs are currently being identified through wide-area broadband imaging surveys with Subaru/HSC, Euclid, and the Roman Space Telescope, primarily using the Lyman-break technique. 
These surveys efficiently select high-redshift galaxies and provide measurements of rest-frame UV emission, from which star-formation rates and approximate stellar masses have been inferred.
However, stellar masses derived solely from rest-frame UV data are subject to large uncertainties, as UV emission primarily traces young, massive stars and is strongly affected by dust attenuation and recent star-formation history. 
Reliable stellar-mass estimates instead require observations at rest-frame optical wavelengths (longer than 6000 \AA), where the light is dominated by older stellar populations and more closely correlates with the total stellar mass.
To date, space-based facilities such as WISE, AKARI, and Spitzer, which observe at wavelengths longer than 3 \um\, have played a crucial role in probing this rest-frame optical emission at high redshift. 
Nevertheless, even the deepest Spitzer data have been insufficient for robustly measuring the stellar masses of SFGs at $z>6$ except for the most massive systems with $M_{\ast}\sim10^{10}~\mathrm{M_{\odot}}$. 
The medium-depth survey layer of GREX-PLUS will overcome this limitation by enabling the detection of intermediate-mass SFGs with $M_{\ast}\sim10^{9}~\mathrm{M_{\odot}}$ (Figure~\ref{fig:grex_masslim_massive}), thereby allowing us to directly trace the build-up of stellar mass in typical galaxies during the early stages of cosmic history at $z>6$.

For QGs, GREX-PLUS will be used to identify candidate systems based on their characteristic spectral features. 
In particular, we focus on the Balmer break, a prominent flux jump in the spectra that becomes evident in galaxies that have ceased star formation for approximately 0.1 Gyr or longer. 
By capturing this spectral break with adjacent broadband filters, GREX-PLUS enables efficient photometric selection of quiescent galaxy candidates at high redshift.
The stellar-mass detection limit for QGs is higher than that for SFGs, because identifying the Balmer break requires detecting the continuum emission on both sides of the break. 
Consequently, our survey primarily targets massive quiescent galaxies with $M_{\ast}\sim10^{10}~\mathrm{M_{\odot}}$ in the medium-depth layer (Figure~\ref{fig:grex_masslim_massive}). 
So far, only a few massive QGs at $z > 6$ have been reported \citep{2025NatAs...9.1541O, 2025ApJ...983...11W}, reflecting their extremely low number density. 
Detecting such rare systems therefore requires a wide-field survey covering on the order of 100 $\mathrm{deg.^2}$, which can only be achieved with GREX-PLUS. 
The discovery and characterization of massive quiescent galaxies at very high redshift thus represent key science goals uniquely enabled by GREX-PLUS.

\subsection{Medium-band filters}
We propose installing medium-band filters on GREX-PLUS to increase the photometric sampling. These filters can split the wavelength coverage of the broad-band filters and make differences in spectral shapes clearer.  

In previous work, the medium-band technique has been adopted by ZFOURGE \citep[e.g.,][]{2014ApJ...783L..14S, 2016ApJ...830...51S} in which the $J$ and $H$ bands were split into several medium bands to capture the Balmer break feature more precisely up to $z\sim4$, leading to the discovery of massive galaxies at $z=3$--4 \citep[e.g.,][]{2017Natur.544...71G, 2024Natur.628..277G, 2025ApJ...981...78N}.

\begin{figure}[h]
    \begin{center}
      \includegraphics[width=\linewidth]{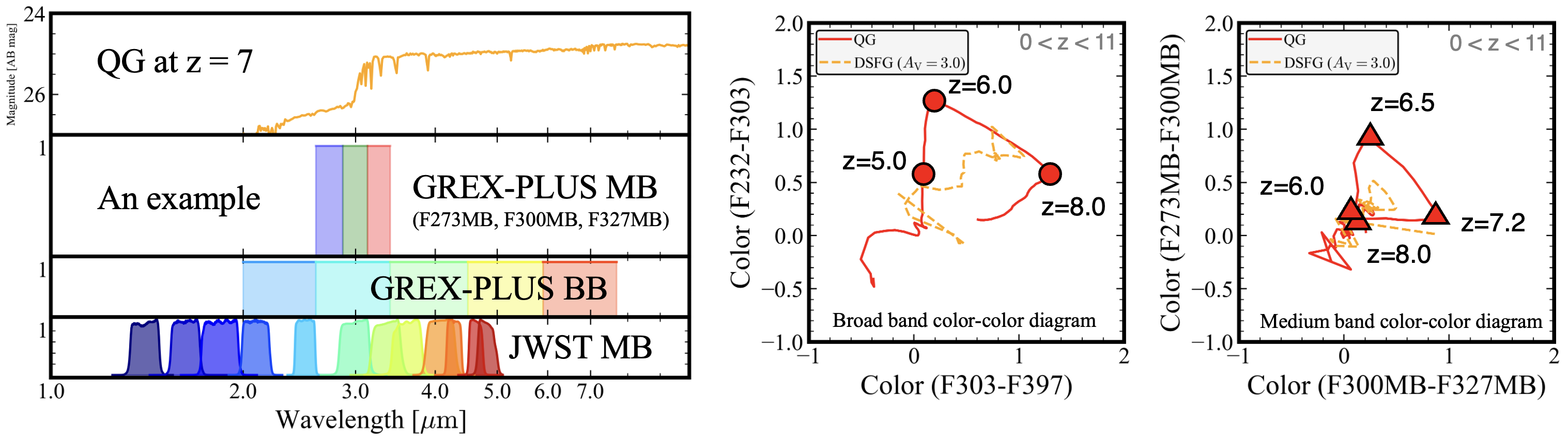}
    \end{center}
    \caption{The left panel shows the proposed medium-band filters for GREX-PLUS. The middle and right panels show color–color diagrams for the selection of Balmer-break galaxies (red) and dusty star-forming galaxies (orange) using broad-band and medium-band filters, respectively. Splitting the F303 filter into three medium bands makes the colors more sensitive to the redshifts of Balmer-break galaxies and improves their separation from lower-redshift dusty star-forming galaxies, enabling more robust candidate selection.}
    \label{fig:grex_mb}
\end{figure}

Figure~\ref{fig:grex_mb} illustrates a possible implementation of medium-band filters on GREX-PLUS. As an example, the F303 filter is divided into three medium-band filters. If these filters are available, the resulting medium-band colors would be more sensitive to the redshifts of Balmer-break galaxies compared with broad-band colors. In addition, the improved sampling of the SED would allow us to more easily distinguish Balmer-break galaxies from dusty star-forming galaxies. Therefore, the candidate selection would become significantly more robust.

\printbibliography[heading=subbibliography]
\end{refsection}

\clearpage

\begin{refsection}[2-4_First_Quiescent_Galaxies/first-quiescent.bib]

\section{The First Massive Quiescent Galaxies}
\label{sec:first-quiescent}

\noindent
\begin{flushright}
Francesco Valentino$^{1,2}$, 
Sirio Belli$^{3}$,
Tadayuki Kodama$^{4}$, 
Kosuke Takahashi$^{4}$. 
\\
$^{1}$ Cosmic Dawn Center (DAWN), Denmark, 
$^{2}$ DTU Space, Technical University of Denmark, Lyngby, Denmark,
$^{3}$ Università di Bologna, Italy
$^{4}$ Tohoku University, Japan
\end{flushright}
\vspace{0.5cm}

\subsection{Scientific background and motivation}


Deep imaging with the \emph{James Webb Space Telescope} (JWST) at wavelengths longer than $2\,\mu$m has revealed a substantial population of massive, quiescent galaxies at $z > 3$ \citep{Valentino2023, Carnall2023phot}.
Pronounced rest-frame optical breaks in JWST photometry indicate the existence of evolved stellar populations within the first two billion years of cosmic history, extending the onset of quiescence to unexpectedly early epochs. 
Spectroscopic follow-up with JWST has confirmed the presence of Balmer breaks, together with Balmer absorption lines, thus validating both the redshift and the quiescent nature of these systems out to $z \sim 7$ \citep{Carnall2023spec, Weibel2025}. 
This demonstrates that photometrically selected candidates are not dominated by dust-obscured star-forming contaminants, establishing quiescent galaxies as a genuine component of the
high-redshift population.

The inferred number densities and stellar masses of early quiescent galaxies challenge pre-JWST theoretical expectations, prompting the latest generation of models to rapidly adapt in order to reproduce the emergence of this population \citep{DeLucia2024, Lagos2024, Kimmig2025}.
Physically, the existence of these galaxies implies a formation pathway characterized by rapid stellar mass assembly followed by efficient quenching. Independent support for such rapid shutdown mechanisms comes from the widespread detection of massive neutral gas outflows at $z \sim 2$--3 \citep{Belli2024, DEugenio2024, Davies2024}, with emerging evidence that similar processes may already operate at higher redshift \citep{Wu2025, Valentino2025}. 
These outflows are capable of removing or heating large fractions of the interstellar medium, providing a plausible route to quenching on short timescales.
In contrast, the role of the environment remains uncertain. While environmental quenching is well established at lower redshift, its relevance at $z > 3$ is unclear, mostly due to the difficulty of systematically and reliably mapping the overdensities on the large scales typical of protoclusters at $z>3$.

The presence of massive, quiescent galaxies at such early times also has important
implications for cosmology and structure formation. Their abundance constrains the
efficiency and timing of baryonic mass assembly, the growth of massive dark matter halos,
and the regulation of star formation by feedback. In several cases, inferred stellar
masses approach the expected baryon budgets of their host halos, placing tension on
standard assumptions about star formation efficiencies at high redshift \citep{Glazebrook2024, Carnall2024, DeGraaff2025}.

Moreover, massive galaxy formation is expected to be accelerated in massive host haloes or higher-density regions due to biased galaxy formation \citep{Kimmig2025}.
This is because such overdense regions must have originated from regions with higher density fluctuations on large scales (cluster scales) in the early Universe,
and the density fluctuations of massive galaxy-scale haloes therein are also higher due to the bottom-up effect of the density fields of larger-scale overdensities.
Furthermore, the quenching process of massive galaxies would also be accelerated in overdense regions due to additional environmental effects such as ram-pressure stripping,
galaxy-galaxy interactions, and AGN activation. Therefore, we would expect to see such preferred environments where we find very massive quiescent galaxies.

Together, these developments strongly motivate a systematic investigation of the physical mechanisms,
timescales, and environmental context responsible for the emergence of quiescent galaxies
in the early Universe by conducting very wide-field observations that JWST cannot realize. 

\subsection{Required observations and expected results}


Currently, a systematic exploration of the first quiescent galaxies is limited by survey volume rather than depth. JWST
provides unprecedented sensitivity at near- and mid-infrared wavelengths and can efficiently
identify quiescent systems out to $z \sim 7$, but its field of view remains limited. As a
result, current surveys probe relatively small cosmic volumes, restricting the discovery
space for the rarest and most massive quiescent galaxies at high redshift. For example, JWST
surveys covering $\sim800~\mathrm{arcmin}^2$ yield only $\sim50$ quiescent galaxies with
$M_\star > 10^{11}\,M_\odot$ at $z>2$ \citep{Baker2025}, leading to large uncertainties from cosmic variance
and preventing robust measurements of the high-mass tail of the stellar mass function or
galaxy clustering.
Moreover, we also aim to test the scenario of accelerated formation of massive galaxies in high-density environments
as discussed above, and thus we need to split the galaxies into several environmental bins to quantify such effects.
This requires much wider-field observations to cover representative large-scale structures of the cosmic web.
These limitations propagate into comparisons with cosmological
simulations, which can reproduce observed abundances through different feedback
implementations but make divergent predictions for star formation histories, mass
distributions, and environmental dependence that cannot yet be distinguished
observationally.

\begin{figure}
    \centering
    \includegraphics[width=\linewidth]{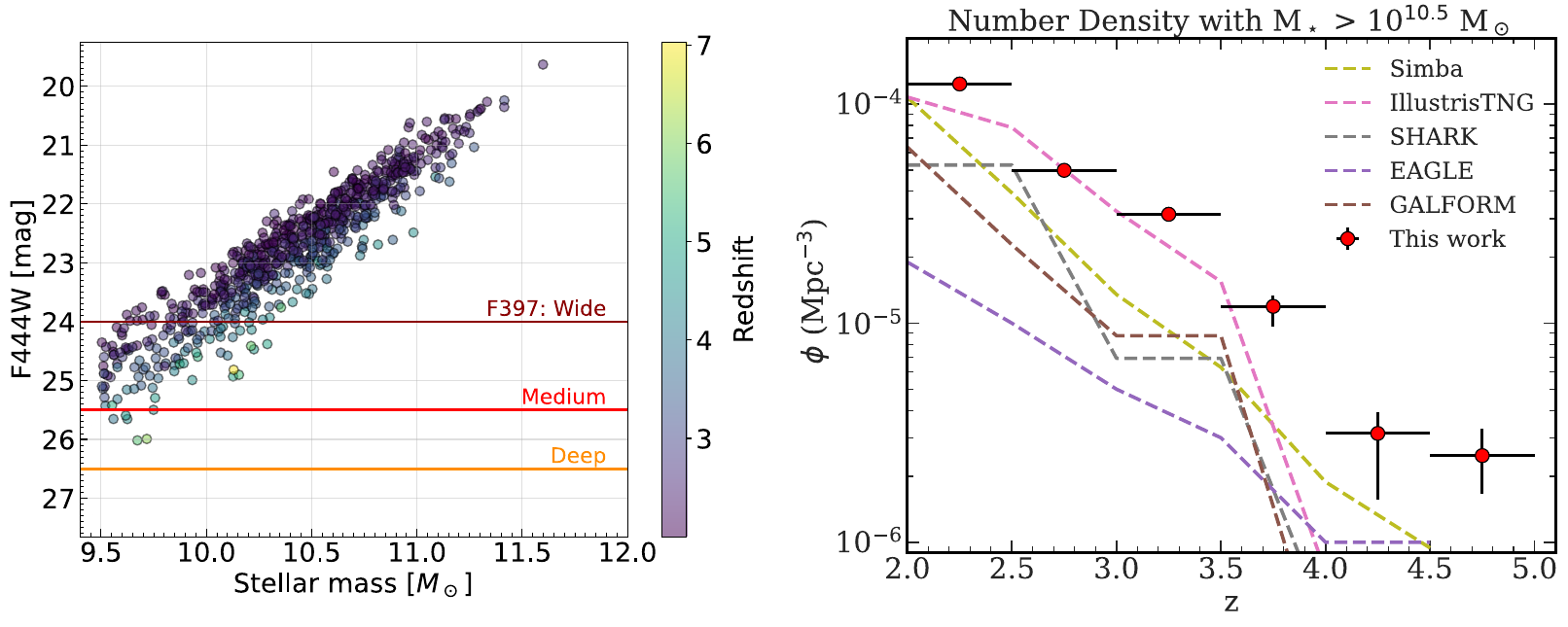}
    \caption{Left: Stellar mass and JWST/NIRCam F444W magnitude distribution for a sample of 700 quiescent galaxies at $z=2$--7. The circles are color-coded according to their redshift. The planned limits of the three-tier GREX-PLUS survey are 
    marked in red. Right: Comoving number densities for $M_{\star}>10^{10.5}\,M_\odot$ as a function of redshift and a comparison with a handful of widely used models and simulations. Panels adapted from \cite{Baker2025}.}
    \label{fig:mass_f444w}
\end{figure}

The wide-field infrared imaging capabilities of GREX-PLUS will directly address these limitations by combining long-wavelength coverage with orders-of-magnitude larger survey areas. Figure~\ref{fig:mass_f444w} showcases the opportunities that GREX-PLUS will open.
Considering the 700 quiescent galaxies with $M_\star \gtrsim 10^{9.5}\,M_\odot$ at $z=2$--7 detected by JWST/NIRCam over $\approx 800$ sq. arcmin in the literature compilation by \cite{Baker2025} as a reference, GREX-PLUS will be able to detect the vast majority in the medium- and deep-tier layers of the planned survey with the F397W filter. This mass limit corresponds to $\sim1$~dex below the knee of the stellar mass function at $z\sim3$. GREX-PLUS will detect all quiescent galaxies above $M_\star^* \sim 10^{10.5}\,M_\odot$ in all tiers of the planned surveys. Accounting for the area covered by the deep survey (10 sq. deg), we expect to detect $\sim3000$ massive quiescent galaxies at $z\sim3$--5, and 10--100$\times$ more in the medium and wide surveys (100 and 1000 sq. deg). 
These areas correspond to $10^8$--$10^{10}$ cMpc$^3$ per
$\Delta z=1$ and can fully cover representative volumes of the Universe. 

Combined with coverage of the $2$--7\,$\mu$m range and in synergy with existing or forthcoming facilities
spanning the rest-frame UV and optical wavelengths (Euclid, Rubin), GREX-PLUS will 
thus uniquely allow for robust measurements of the stellar mass function of quiescent 
galaxies across both the low- and high-mass ends at $z>3$, resolving current discrepancies in number densities
and placing constraints on $\Lambda$CDM cosmology and early structure formation,
particularly through the abundance of systems with $M_\star > 10^{11}\,M_\odot$.

With such large sample statistics and coverage of wide contiguous patches of the sky, GREX-PLUS
will also provide statistically meaningful measurements of clustering and environmental dependence
for quiescent galaxies at $z>3$, which are currently unfeasible. This will enable 
tests of whether early quenching is linked to halo growth, dense environments, or primarily internal 
feedback processes, and will provide a powerful tool to match early quiescent galaxies with their low-redshift descendants.

GREX-PLUS observations at long wavelengths ($\sim4$--7\,$\mu$m) will reduce contamination by dusty star-forming systems, thus yielding reliable photometric selection and stellar mass estimates.
These data will also probe, for the first time in a statistically robust manner, the incidence of obscured AGN in the first quiescent systems.
Additional benefits would come from the availability of one or more medium-band filters, which are sensitive to the presence of strong emission lines. Medium-band filters would allow better discrimination between the population of quiescent galaxies and dusty, very high-redshift, or unusual line-emitting contaminants. Moreover, they would enable an estimate of the equivalent width (EW) of key emission lines ([OIII]+H$\beta$, H$\alpha$+[NII]), placing constraints on residual star formation or possible emission from AGN---a common occurrence among 
distant quiescent galaxies ($\rm EW([OIII]\lambda\lambda4959,5007)\sim10$\,\AA\ on average for recent post-starburst systems 
from JWST/NIRSpec observations; \citealt{Ito2025,Bugiani2025}).

Finally, GREX-PLUS will provide efficient identification of the most extreme and informative systems for
spectroscopic follow-up with JWST and the upcoming Extremely Large Telescope, enabling detailed studies of star formation histories, outflows, chemical enrichment, and stellar populations.

In this context, GREX-PLUS is uniquely positioned to overcome the current limitations set
by low-number statistics and to provide the definitive observational samples required to
test models of early galaxy assembly and quenching.

\subsection{Scientific goals}

GREX-PLUS will detect $10^3$--$10^5$ massive quiescent galaxies at $z>3$, enabling a statistically robust analysis for the first time. The main goals will be measurements of the mass function and the clustering of quiescent galaxies at $3 < z < 7$, and the characterization of their environment. Moreover, the GREX-PLUS sample will provide the best targets for spectroscopic follow-up with JWST and the next generation of large ground-based telescopes.
Massive quiescent galaxies are extreme objects that offer unique probes of theoretical models, and GREX-PLUS observations will enable substantial progress in our understanding of cosmology, galaxy formation, and galaxy quenching.

\begin{table}
    \begin{center}
    \caption{Required observational parameters.}\label{tab:firstgals}
    \begin{tabular}{|l|p{9cm}|l|}
    \hline
     & Requirement & Remarks \\
    \hline
    Wavelength & 2--8 $\mu$m & \multirow{2}{*}{} \\
    \cline{1-2}
    Spatial resolution & $<2$ arcsec & \\
    \hline
    Wavelength resolution & $\lambda/\Delta \lambda>3$ & $a$ \\
    \hline
    Field of view & 100 degree$^2$, 25.5 ABmag ($5\sigma$, point-source) & \multirow{2}{*}{}\\
    \cline{1-1}
    Sensitivity & 1000 degree$^2$, 24 ABmag ($5\sigma$, point-source) & \\
    \hline
    Observing field & Fields where deep imaging data at $\lambda_{\rm obs}<2\,{\rm \mu m}$ are available. & $b$ \\
    \hline
    Observing cadence & N/A & \\
    \hline
    \end{tabular}
    \end{center}
    $^a$ Additional medium-band filters would improve the selection of quiescent galaxies.\\
    $^b$ For example, deep fields observed with Euclid or Roman.
\end{table}

\printbibliography[heading=subbibliography]
\end{refsection}

\clearpage

\begin{refsection}[2-5_protoclusterdetection/protocluster.bib]

\section{Identifying the First Protoclusters}
\label{sec:protocluster}

\noindent
\begin{flushright}
Lucas Kimmig$^{1,2}$
\\
$^{1}$ Ludwig-Maximilians-University, Munich, Germany, \\
$^{2}$ University of Nottingham, Nottingham, England
\end{flushright}
\vspace{0.5cm}

\subsection{Scientific background and motivation}


Our understanding of the earliest collapsed structures has been revolutionized with the advent of the \textit{James Webb Space Telescope} (JWST). The first galaxies have been found to evolve on far shorter timescales than previously assumed, undergoing rapid growth of their stellar mass and, in some cases, a shut-down of their star formation already at redshifts $z>4$ \citep{harikane:2022,carnall:2023}. Observations of these earliest and most massive ``lighthouse" galaxies have already pushed models of galaxy formation forward, finding increased burstiness and a quicker growth of the central supermassive black holes for the first galaxies as compared to the present day \citep{nanayakkara:2025,kimmig:2025,weller:2025}. 

Beyond internal processes, the properties of galaxies such as their ages and ongoing star formation are known to be linked with their host environment at the present day $z\sim0$, with strong observed differences between the field and galaxies residing in galaxy clusters \citep{dressler:1980,postman:1984}. The role of the environment, however, has remained elusive at high redshifts, with some observations finding multiple quenched galaxies in overdensities \citep{kakimoto:2024,jin:2024} while others find galaxies in the central regions of overdensities to be particularly star forming \citep{wang:2025}. \citet{kimmig:2025} showed that their cosmological simulations well capture the emergence of both quiescent and star-forming galaxies, and found that quiescent galaxies tend to reside in local underdensities as compared to same-mass star-forming counterparts at $z>3$, implying a reversal of the morphology-density relation seen at $z=0$. Beyond direct measures of overdensity and underdensity, the large-scale flow field of the gas falling into the earliest galaxies is found to have a profound impact on their quenching \citep{kimmig:2025,fortune:2025}.

Answering whether the environment enhances or suppresses star formation in galaxies, or whether it has an impact at all, is thus a key open question in the field of galaxy formation physics. Doing so first requires the correct identification of early protocluster regions. In this context, a particularly crucial limitation of the JWST is the small field of view of $\approx9.7\mathrm{arcmin}^2$. For comparison, the infall region of a Coma-mass galaxy cluster, $M_\mathrm{vir}=10^{15}\mathrm{M}_\odot$, has a total extent of around $D\sim30\,\mathrm{cMpc}$, or equivalently $\sim100\,\mathrm{arcmin}^2$ at $z=5$ \citep{remus:2023}. This is without accounting for their extreme rarity, as we discuss later. Consequently, identifying protoclusters and galaxies residing in their infall regions, which may follow a biased assembly, requires prohibitive amounts of observational time with the JWST.

\subsection{Required observations and expected results}

The GREX-PLUS telescope will directly fill this gap in observational facilities by combining a wide-field imager with coverage of the infrared beyond 2~microns. The field of view of $1800\,\mathrm{arcmin}^2$ is 200~times larger than that of the JWST, and captures 20~times the extent of a typical protocluster infall region in one pointing. This is critical for candidate protoclusters, because distinguishing those that actually become galaxy clusters by $z=0$ from those that represent local but not global overdensities is not possible simply from their stellar mass alone. Instead, protocluster identification requires accurate stellar-mass data over entire regions. 

\begin{figure}
    \centering
    \includegraphics[width=\linewidth]{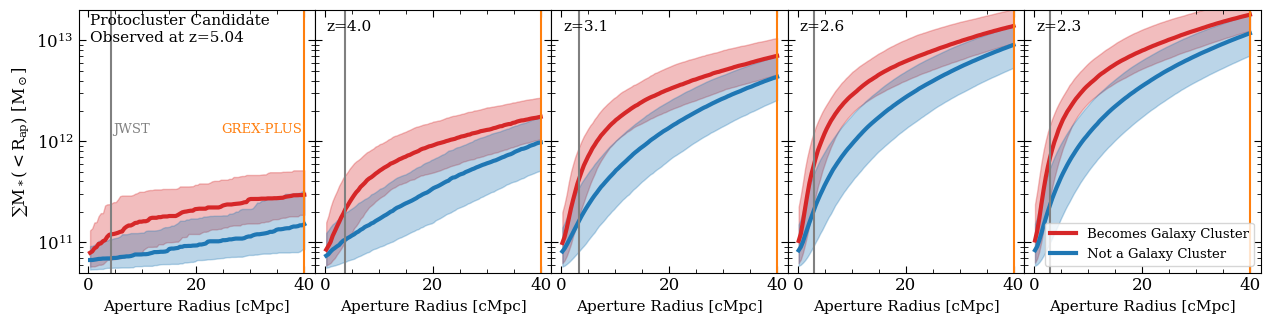}
    \includegraphics[width=\linewidth]{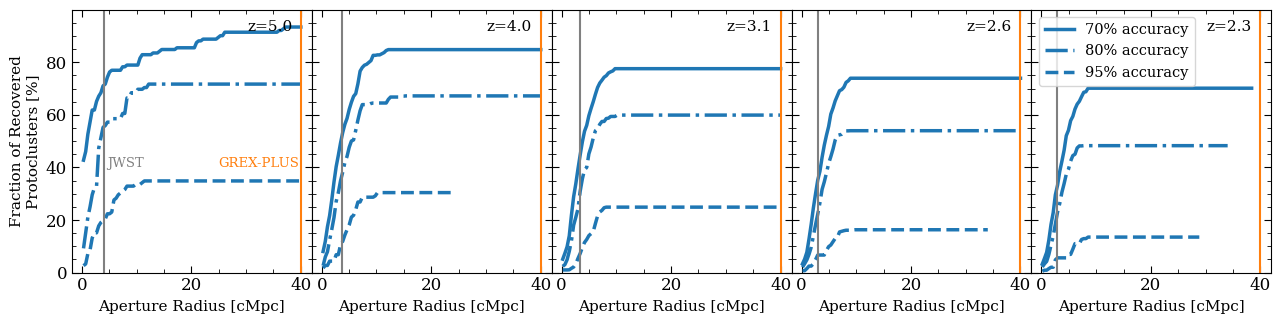}
    \caption{Top row: For protocluster candidate galaxies with $M_*>5\times10^{10}\mathrm{M}_\odot$ from $z=5$ (left) to $z=2.3$ (right), we plot the summed stellar mass of surrounding galaxies with $M_*>10^{10}\mathrm{M}_\odot$ in a projected aperture with a depth of $\pm750\mathrm{km/s}$, plotted as a function of aperture radius. Candidates that become galaxy clusters by $z=0$ (red line) have a strongly enhanced amount of stellar mass in their surroundings compared to those that do not become galaxy clusters (blue line). Bottom row: The fraction of protoclusters which can be recovered with a prediction accuracy of $70\%$, $80\%$, and $95\%$ (solid, dash-dotted, and dashed blue lines) as a function of aperture radius.}
    \label{fig:proto}
\end{figure}

We show this by drawing on simulated systems from the large-volume cosmological hydrodynamical \textit{Magneticum Pathfinder} Box2/hr simulation \citep{dolag:2025}, and consider the environment around protocluster candidate massive galaxies with $M_*\geq 5\times10^{10}\mathrm{M}_\odot$. The top row of Fig.~\ref{fig:proto} demonstrates that for small apertures, comparable to the field of view of the JWST, the distribution of stellar masses in true and false protoclusters strongly overlaps. 

An observational campaign of~$100\mathrm{deg}^2$ with sufficient depth to sample galaxy stellar masses down to $M_*\geq10^{10}\mathrm{M}_\odot$ would not only increase the number of observed galaxies at $z>4$ by almost two orders of magnitude relative to prior surveys such as UltraVISTA \citep{mccracken:2012}, but would also significantly increase the expected number of true protocluster systems between $z=4$--5. For a redshift range of $z=4$--5 (with a comoving depth of $L\sim600\,\mathrm{cMpc}$) with comoving angular diameter distance $D_\mathrm{A}=9000\,\mathrm{cMpc}$, one square degree corresponds to a comoving volume of $V = (9000\,\mathrm{cMpc}\cdot\pi/180)^2\cdot 600\,\mathrm{cMpc}\approx 1.5\times 10^{7}\, \mathrm{cMpc}^3$. For the halo mass function given by \citet{tinker:2008}, this results in~100 protocluster regions for galaxy clusters of mass $M_\mathrm{vir}=10^{14}\mathrm{M}_\odot$, and~1.5~regions for the most massive clusters with $M_\mathrm{vir}=10^{15}\mathrm{M}_\odot$. 

\subsection{Scientific goals}

Sampling the large-scale environment with galaxies serving as the lighthouses of the underlying dark matter distribution opens a window to witness the assembly of the cosmic web at the earliest times in the Universe. GREX-PLUS will, for the first time, combine a super-wide-field imager with coverage of the 2--8 $\mu$m wavelength range, equivalent to the rest-frame optical for $3<z<15$ and thus crucial for measuring accurate stellar masses. Thus, it is uniquely positioned for the tight identification of true protoclusters that become galaxy clusters by $z=0$ (as opposed to only local overdensities). A full $100\,\mathrm{deg}^2$ survey will, at $z=4$--5, capture $10^4$~progenitor regions of $M_\mathrm{vir}\sim10^{14}\mathrm{M}_\odot$ galaxy clusters, as well as $100$~regions that turn into the most massive galaxy clusters seen today, such as Coma with $M_\mathrm{vir}\sim10^{15}\mathrm{M}_\odot$. This will enable direct tests of whether the impact of environment on the formation and evolution of galaxies occurs already at the earliest times, and if so, whether it suppresses or enhances star formation, with important consequences for our understanding of cosmic structure assembly. 

\begin{table}
    \begin{center}
    \caption{Required observational parameters.}\label{tab:proto}
    \begin{tabular}{|l|p{9cm}|l|}
    \hline
     & Requirement & Remarks \\
    \hline
    Wavelength & 2--8 $\mu$m & \multirow{2}{*}{} \\
    \cline{1-1}
    Spatial resolution & $<1$ arcsec & $a$\\
    \hline
    Wavelength resolution & $\lambda/\Delta \lambda>3$ & $b$ \\
    \hline
    Field of view & 100 degree$^2$ & $c$\\
    \cline{1-1}
    Sensitivity & 25 AB mag ($5\sigma$, point-source) & \\
    \hline
    Observing field & Overlap with deep JWST fields preferable &  \\
    \hline
    Observing cadence & N/A & \\
    \hline
    \end{tabular}
    \end{center}
    $^a$ For accurate stellar masses, including the separation of merging galaxy pairs.\\
    $^b$ Medium-band filters would be useful as well, and a redshift accuracy of $\Delta z\sim0.05\sim30\mathrm{cMpc}$.\\
    $^c$ At least, for a minimum of~100 protocluster regions of $10^{15}\mathrm{M}_\odot$ galaxy clusters at $z=4$--5.
\end{table}

\printbibliography[heading=subbibliography]
\end{refsection}

\clearpage

\begin{refsection}[2-6_submmgalaxies/submmgalaxies.bib]

\section{Dusty Star-Forming Galaxies}
\label{sec:submmgalaxies}

\noindent
\begin{flushright}
Ken-ichi Tadaki$^{1}$,
Yoshinobu Fudamoto$^{2}$
\\
$^{1}$ Hokkai-Gakuen University,
$^{2}$ Chiba University
\end{flushright}
\vspace{0.5cm}
\newcommand{\YF}[1]{\textcolor{blue}{#1}}

\subsection{Scientific background and motivation}

\begin{figure}[tbh]
\centering
\includegraphics[width=.95\linewidth]{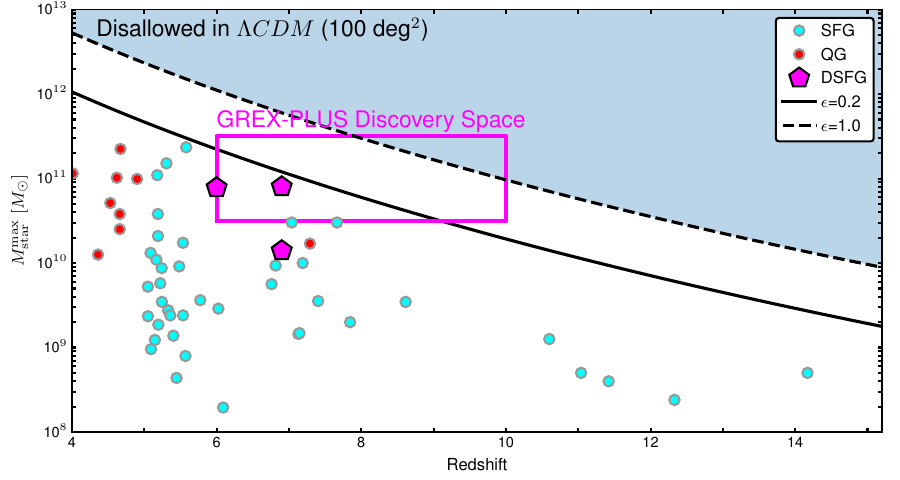}\ \ 
\caption{
Stellar masses of the most massive galaxies expected in a 100 deg$^2$ survey field (\citealt{2013A&C.....3...23M}), compared with galaxies actually discovered by JWST and other facilities (DSFG:\citealt{2022ApJ...933..242Z,2023A&A...671A.105A}, QG: \citealt{2019ApJ...885L..34T,2023Natur.619..716C,2024MNRAS.534..325C,2024ApJ...963...49K,2025NatAs...9..280D,2025ApJ...983...11W,2025ApJ...978...90A,2025MNRAS.537.3453B}, SFG: \citealt{2023ApJ...952...74T,2023Natur.622..707A,2024Natur.635..311X,2025NatAs...9..155Z,2025NatAs...9..729H,2025ApJ...988...19S}).
\label{fig:Mmax_100deg2}
}
\end{figure}

Understanding how the most massive galaxies were assembled provides key insights into the physical mechanisms that governed the earliest phases of galaxy formation, including the emergence of bright star-forming galaxies (SFGs) at $z=10$--14 (Figure~\ref{fig:Mmax_100deg2}). 
The star formation history of nearby quiescent galaxies (QGs) indicates that more massive galaxies formed earlier in the Universe and on shorter timescales, referred to as \textit{downsizing} of galaxy formation \citep{2010MNRAS.404.1775T}. 
Near-infrared spectroscopic observations of massive QGs at $z=4$--5 support the scenario that they were formed through intense star formation at $z>5$, even out to $z>7$--8, such as SFRs of $\sim200$--$800~M_\odot~{\rm yr}^{-1}$, with durations of a few to several hundred million years \citep[e.g.][]{2024MNRAS.534..325C,2025NatAs...9..280D}. 
The identification of massive QGs and the prediction of their highly star-forming progenitors are surprising, as such progenitors are extremely rare or even absent in rest-frame UV observations.
Alternatively, we could be missing these progenitors because such galaxies are expected to host large stellar masses (e.g., $\log(M_\star/M_\odot) > 10.5$) and substantial dust reservoirs as a result of their rapid buildup.
Dusty star-forming galaxies (DSFGs) at $z>6$ are therefore strong candidates for the progenitors of these massive QGs at the highest redshifts, as their high SFRs naturally reproduce the short, intense formation episodes implied by downsizing.
We define DSFGs primarily as galaxies with observed 870\,$\mu$m flux densities exceeding 2~mJy (corresponding to SFR $\sim200~M_\odot~{\rm yr}^{-1}$).

While ALMA observations have identified many DSFGs at $z=4$--6, only three DSFGs are currently known at $z>6$: SPT0311$-$58 at $z=6.9$ \citep{2018Natur.553...51M}, HFLS3 at $z=6.3$ \citep{2013Natur.496..329R}, and G09-83808 at $z=6.0$ \citep{2018NatAs...2...56Z}. 
Two of them, SPT0311$-$58 and HFLS3, are exceptionally bright sources with 870\,$\mu$m flux densities exceeding 15 mJy (SFR $>1000~M_\odot~{\rm yr}^{-1}$), whereas G09$-$83808 is intrinsically $\sim$4 mJy at 870~$\mu$m (SFR $\sim400~M_\odot~{\rm yr}^{-1}$) after correcting for its strong gravitational lensing magnification ($\mu=8$--9).  
Theoretical models predict that the surface number density of unlensed $\sim4$ mJy DSFGs at $z=6$--7 is as low as $\sim0.3~{\rm deg}^{-2}$ and increases to $\sim2~{\rm deg}^{-2}$ when $\sim2$ mJy sources are included \citep{2020ApJ...891..135P}.  
However, these number densities remain poorly constrained by current observations.  
Identifying DSFGs at $z>7$ and constructing a statistical sample of DSFGs at $z=4$--7 are therefore essential for understanding when and how the most massive galaxies formed in the early Universe.


\subsection{Required observations and expected results}

DSFGs are typically massive, with stellar masses of $\log(M_\star/M_\odot) > 10.5$ \citep{2022ApJ...933..242Z}, resulting in bright continuum emission at long wavelengths ($\sim$25 AB mag at $4~\mu$m).  
However, they remain faint at shorter wavelengths ($\lambda_{\rm obs}<2~\mu$m) due to strong dust extinction and are sometimes completely undetected even in HST $1.4$--1.6 $\mu$m images \citep{2019Natur.572..211W,2021Natur.597..489F}.
As a consequence, DSFGs can be efficiently characterized by their red colors at $\lambda_{\rm obs}\sim2$--4\,$\mu$m.  
JWST observations have demonstrated that a red 2--4 $\mu$m source detected in dust continuum maps ($S_{870}\sim2$ mJy) corresponds to a DSFG at $z=5$--6, although the number density of such DSFGs remains highly uncertain due to the small field of view of current JWST observations \citep{2023ApJ...943L...9Z}.
These results show that wide-field imaging at $\lambda_{\rm obs}\sim2$--4\,$\mu$m is a powerful strategy for assembling a large sample of DSFGs at $z=6$--7 and for discovering DSFGs at $z>7$.  
A Roman survey alone is not ideal for this purpose because it does not cover wavelengths longer than $2~\mu$m, where DSFGs become significantly brighter.  
Coordinated surveys with Roman ($\sim$27 AB mag at $2~\mu$m) and GREX-PLUS ($\sim$25 AB mag at $4~\mu$m) will therefore be highly effective for identifying DSFG candidates at $z>6$.  

A red color at 2--4 $\mu$m, however, can also arise from a Balmer break produced by old stellar populations, not only from heavy dust attenuation.  
Photometry with the GREX-PLUS/F520 filter is useful for breaking this age--dust degeneracy, because DSFGs at $z>6$ are even brighter at $5~\mu$m, corresponding to the rest-frame $i$ band.  
Including $\sim$5 $\mu$m photometry will also enable more accurate stellar mass estimates for DSFGs at $z>6$, as it is less affected by dust extinction.  
\citet{2020ApJ...889..137M} find that roughly half of the red 2--4 $\mu$m sources are detected in ALMA $870~\mu$m continuum maps, supporting that many of them are DSFGs.  
If deep $5~\mu$m imaging is unavailable, deep submillimeter follow-up observations targeting the $870~\mu$m dust continuum provide a robust alternative for distinguishing DSFGs from QGs, as QGs lack significant infrared emission.
Once red 2--4 $\mu$m sources are identified as DSFGs, ALMA observations of the [C~{\sc ii}] $158\,{\rm \mu m}$ emission line can provide secure spectroscopic redshifts.  
Because the [C~{\sc ii}] line is bright in DSFGs even at $z>6$ ($\sim$0.75~Jy km s$^{-1}$ for a 2 mJy source; \citealt{2018NatAs...2...56Z}), only $\sim$4 minutes of ALMA integration are required for a 5$\sigma$ detection.  
With the ALMA Wideband Sensitivity Upgrade \citep{2023pcsf.conf..304C}, such redshift searches and dust continuum observations will become even more efficient. 
Therefore, wide-area surveys with GREX-PLUS will enable the construction of the first statistical sample of DSFGs at $z>6$.


\subsection{Scientific goals}

The minimum goal is to identify $>100$ DSFGs at $z>6$ to robustly measure the redshift evolution of their number densities.  
Achieving this requires a wide survey covering $\sim$100 deg$^2$ to depths of 27 AB mag at $2~\mu$m and 25 AB mag at $4~\mu$m, assuming the theoretical predictions of \citet{2020ApJ...891..135P}.  
We also aim to constrain the stellar masses and halo masses of DSFGs through SED fitting and clustering analyses, respectively.  
By comparing their number densities, stellar masses, and halo masses with those of massive QGs at similar redshifts, we will be able to establish the evolutionary connection from DSFGs to quiescent systems.  
A further key objective is to discover DSFGs at $z>7$ (Figure~\ref{fig:Mmax_100deg2}), enabling us to determine when the most massive galaxies first emerged in the early Universe.

\begin{table}[tbh]
    \label{tab:smgrequirement}
    \begin{center}
    \caption{Required observational parameters.}
    \begin{tabular}{|l|p{9cm}|l|}
    \hline
     & Requirement & Remarks \\
    \hline
    Wavelength & 2--5 $\mu$m &  \\
    \hline
    Spatial resolution & $<2$ arcsec & \\
    \hline
    Wavelength resolution & N/A &  \\
    \hline
    Field of view & 100 degree$^2$, 27 AB mag at $2~\mu$m ($5\sigma$, point source) & \multirow{3}{*}{$a$}\\

    Sensitivity & 100 degree$^2$, 25 AB mag at $4~\mu$m ($5\sigma$, point source) & \\

    \hline
    Observing field & Fields where deep imaging data at $\lambda<2~\mu$m are available. &  \\
    \hline
    Observing cadence & N/A & \\
    \hline
    \end{tabular}
    \end{center}
    $^a$ When deep $2~\mu$m data are available from a Roman survey, we require $4.4~\mu$m and $8~\mu$m data for the GREX-PLUS survey.\\
\end{table}

\printbibliography[heading=subbibliography]
\end{refsection}

\clearpage

\begin{refsection}[2-7_empg/empg.bib]

\section{Extremely Metal-Poor Galaxies}
\label{sec:empg}

\noindent
\begin{flushright}
Kimihiko Nakajima$^{1}$
\\
$^{1}$ Kanazawa University
\end{flushright}
\vspace{0.5cm}

\subsection{Scientific background and motivation}

Galaxy mass is thought to be a fundamental quantity that governs the evolution of galaxies.
Theoretically, galaxy formation models based on $\Lambda$CDM cosmology predict that
galaxies grow through successive mergers of lower-mass objects and that
galaxy properties are largely determined by their masses through mass-dependent processes
at work in their evolution
(e.g., \citealt{blumenthal1984,davis1985,bardeen1986}).
Therefore, observations of the mass-dependence of galaxy properties back through cosmic time
are crucial for understanding how galaxies evolve to acquire present-day properties.
In particular, low-mass galaxies at high redshifts are interesting, since they are likely to be building blocks of massive galaxies seen in later epochs.

Another key property of galaxies is gas-phase metallicity.
This can be used to investigate complex physical processes that regulate galaxy evolution,
such as star-formation and explosion history, outflows of metal-enriched gas, and
infall of metal-poor intergalactic medium (IGM) gas.
Despite its complexity, a clear relationship between stellar mass and metallicity
has been recognized for local galaxies over four orders of magnitude in stellar mass
(e.g., \citealt{tremonti2004,AM2013}), where a galaxy with a larger mass has higher gas-phase metallicity.
A similar mass-metallicity relation has been observed for higher-redshift galaxies up to $z\sim 3.5$
with a tendency for galaxies to show lower metallicities at higher redshifts for a given stellar mass
(\citealt{MM2019} for a review).
More recently, early JWST spectroscopic data provide metallicity measurements for galaxies at $z=4$--10, suggesting the presence of a mass-metallicity relation with a small evolution from $z=2$--3 to $z=4$--10, albeit with a potential deficit of metallicity at $z>8$ (e.g., \citealt{nakajima2023}).
Such a clear relationship between mass and metallicity, and its secondary dependence on
star formation rate (e.g., \citealt{mannucci2010}), suggests that galaxies have
metallicity equilibrium conditions for the balance between star formation, gas outflows, and gas inflows
(e.g., \citealt{lilly2013}).
Several state-of-the-art cosmological simulations are able to reproduce the metallicity observations
in the local Universe and at redshifts up to $z\sim 2$--3,
but have dissimilar predictions for much higher redshifts in the early Universe,
such as the slope of the mass-metallicity relation, metallicities typically expected in the low-mass regime, 
and their evolution across cosmic time
(e.g., \citealt{ma2016,derossi2017,torrey2019,langan2020,ucci2023,wilkins2023,nakazato2023}).
Indeed, the recent JWST studies at $z=4$--10 (e.g., \citealt{nakajima2023}) suggest a possible inconsistency of the mass-metallicity relation with the existing simulations at the low-mass end ($\lesssim 10^8\,M_{\odot}$), although the sample size of the low-mass galaxies remains small. 
The potential break of the mass-metallicity-star formation rate relation at $z>8$ also needs to be examined with a larger sample. 
Determining the metallicity relations at high redshifts, including the low-mass regime, is crucial for constraining the models
and hence the physics affecting galaxy evolution across cosmic time.

In the coming decade, the exploration of galaxies in the early Universe will be driven by powerful infrared observing instruments, including the already operational JWST and Euclid, as well as upcoming missions like the Roman Space Telescope (Roman) and GREX-PLUS.
Using near- and mid-infrared multi-band imaging data, one can efficiently select high-redshift galaxy
candidates by detecting the Lyman break at a wavelength of $0.1216\times (1+z)\,\mu$m.
Galaxies selected in this way, called LBGs, will be followed up
spectroscopically with JWST, ALMA, and the next-generation extremely large telescopes
for studies of early galaxy evolution, including chemical enrichment.
However, LBGs are UV-selected galaxies and provide a UV-bright sample with modest stellar masses,
i.e. not ideal for a systematic sampling of low-mass galaxies.
In order to fully address early galaxy evolution, it is essential to explore galaxies in the low-mass,
low-metallicity regime as a representative population of galaxies in the early Universe.

\subsection{Required observations and expected results}
\begin{figure*}[htb]
  \centering
    \includegraphics[bb=0 0 879 284, width=0.99\textwidth]{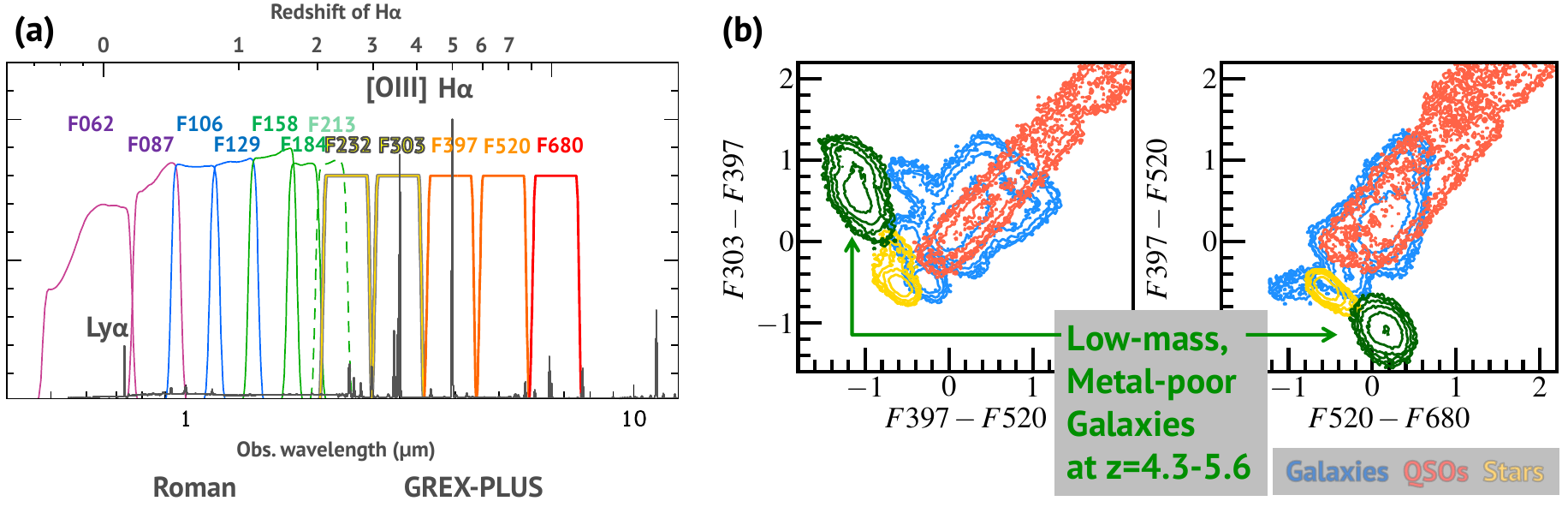}
    \caption{
      \textbf{(a:)} A modeled spectrum of a low-mass, metal-poor galaxy
      at $z=5$ (gray) along with the available filters of 
      Roman Space Telescope (up to $2\,\mu$m) and 
      GREX-PLUS (from $2$ to $8\,\mu$m based on the current plan).
      The infrared bands allow us to perform systematic searches of low-mass galaxies
      over $z=2$--6
      by capturing the characteristic intense emission lines such as H$\alpha$
      imprinted on the broadband photometric colors as shown in Panel (b).
      Combining this with Lyman-break / Ly$\alpha$ constraints provided by Roman, 
      such a low-mass galaxy search can be extended to $z=6$--8.
      \textbf{(b:)} 
      Simulated distributions of low-mass, metal-poor galaxies at $z=4.3$--5.6 (green) 
      on the color-color diagrams using the GREX-PLUS bands,
      ensuring their efficient selection and discrimination from other populations 
      of evolved / different redshift galaxies (blue), QSOs (orange), and Galactic stars (yellow)
      solely using the photometric data.
      \label{fig:empg_spectrum_z5}
    }
\end{figure*}

We propose that GREX-PLUS, which covers from 2 to 8\,$\mu$m with five different photometric bands in a seamless way, 
can uniquely provide a systematic sample of low-mass, metal-poor galaxies
over the redshift range of $z=2$--6 (and beyond with Roman) 
by exploiting the rest-frame optical intense emission lines.
Figure \ref{fig:empg_spectrum_z5} presents a model spectrum of low-mass, metal-poor galaxies at $z=5$
associated with intense emission lines such as
the Hydrogen Balmer lines (H$\alpha$, H$\beta$, etc) and the metal lines (e.g., [OIII]$\lambda\lambda5007,4959$)
in the rest-frame optical wavelength regime.
The presence of such intense emission lines would boost the broadband photometry of GREX-PLUS, producing characteristic photometric colors.
Using the low-mass, metal-poor galaxy templates and pseudo-observations,
Panel (b) illustrates that we can isolate candidates of such low-mass, metal-poor galaxies at high redshifts
from other populations (evolved / different redshift galaxies, QSOs, and Galactic stars)
efficiently using the GREX-PLUS photometric bands.
Note that two bands longer than H$\alpha$ are ideally needed to remove strong [OIII] emitting objects at slightly higher redshifts by confirming that there are no strong emission lines in the longer bands.
This novel technique was originally adopted by \citet{kojima2020} for a local ($z<0.03$) extremely metal-poor galaxy search with Subaru/Hyper Suprime-Cam's wide and deep imaging data. Developing a machine learning classifier, \citet{kojima2020} successfully constructed a metal-poor galaxy sample in the local Universe. 
\citet{nishigaki2023} extended the technique toward higher redshifts at $z=4$--5 using early JWST photometric data, and demonstrated its novel application at high redshifts, albeit with a limited survey volume.
We will extend the technique with GREX-PLUS (and with Roman) to perform a more systematic sampling of low-mass, metal-poor galaxies at high redshifts.

For a reliable classification, two of the most intense emission lines, H$\alpha$ and [OIII]$\lambda 5007$,
must fall into two adjacent photometric bands rather than a single band at any given redshift. 
The current conceptual design, with the five photometric bands tentatively shown in Figure \ref{fig:empg_spectrum_z5},
perfectly meets this criterion.
Using the GREX-PLUS band set alone, this method allows us to search for low-mass, metal-poor galaxies across two distinct redshift ranges: from $z=3.1$ to 4.1 and from $z=4.3$ to 5.6.
By combining this with Roman $\sim 1\,\mu$m data, which provides additional redshift constraints by capturing rest-frame UV spectroscopic signatures like the Lyman break and/or Ly$\alpha$, this method can be extended to higher redshifts, up to $z=8$. At these redshifts, a single band redward of H$\alpha$ is sufficient to capture the stellar continuum without redshift ambiguity.
Lower-redshift metal-poor galaxies can also be explored from $z=2.2$ to 2.8 by combining GREX-PLUS and Roman. Therefore, while this science case exhibits strong synergy with Roman, it cannot be achieved at $z>2$ using Roman alone.

Table \ref{tab:empg_expectations} lists the estimated number of galaxies expected to be found in each redshift range across different survey depths and areas. To derive these estimates, we assume (i) the number density of metal-poor galaxies explored at $z=0$ \citep{kojima2020}, (ii) the redshift evolution of the mass-metallicity relation \citep{torrey2019,sanders2021}, and (iii) the redshift evolution of the stellar mass function \citep{baldry2012,song2016,davidzon2017,stefanon2021}.
We only count galaxies whose stellar continuum redward of H$\alpha$ is detectable
in the corresponding GREX-PLUS photometric bands.
Our estimates indicate that a survey design covering the widest area with a modest depth yields the largest sample of low-mass, metal-poor galaxies across all redshift bins: approximately 4k galaxies each at $z\sim 2.5$, 3k galaxies at $z\sim 3.5$, $\sim 800$ galaxies at $z\sim 5$, and $\sim 100$ galaxies at $z\sim 7$. This yield at $z\sim 7$ represents the minimum sample size necessary to systematically examine their detailed properties at the highest redshifts.
This requirement for a large survey volume underscores the unique necessity of GREX-PLUS in the JWST era. Because extremely metal-poor galaxies possess an intrinsically low number density, finding a statistically significant sample requires sweeping wide areas of the sky. While JWST offers unprecedented sensitivity, its narrow field of view makes it highly inefficient for mapping such rare populations. In contrast, the wide-field survey capability of GREX-PLUS is highly advantageous and essential for overcoming this low number density.
Another approach to finding these rare systems involves serendipitous near-infrared Integral Field Unit observations targeting Ly$\alpha$ emission. This method has successfully identified low-mass galaxies up to $z=6-7$ \citep{vanzella2020,vanzella2023} that have been confirmed to be extremely metal-poor \citep{nakajima2026}. However, because Ly$\alpha$ emission is frequently attenuated by the neutral intergalactic medium during the reionization epoch, it is not a universally reliable tracer. Consequently, targeting Balmer emission lines combined with the absence of strong metal lines provides a more robust and advantageous selection method.
To implement this strategy, Table \ref{tab:empg_requirements} summarizes the required observational parameters. We strongly advocate for maintaining the current Wide survey design for this science case, ideally with greater depth at $\lambda>5\,\mu$m to probe even lower-mass systems at $z>5$.

\subsection{Scientific goals}
Building a valuable sample of low-mass galaxies at $z=2$--8 provided by GREX-PLUS + Roman,
and comparing it with the LBG sample, 
we can pursue key scientific goals, including the following three:
\begin{itemize}
  \vspace{-1.mm}
  \setlength{\itemsep}{0cm}
  \setlength{\parskip}{0.25em}
\item[$\bullet$] Determining the mass-metallicity relation covering the low-mass regime to discuss relevant physical processes.
\item[$\bullet$] Characterizing other ISM properties and the nature of the ionizing spectrum in early galaxies to discuss stellar populations and ionizing properties.
\item[$\bullet$] Quantifying the dark matter halo properties of low-mass galaxies through clustering analysis to discuss their descendants.
\end{itemize}
For the first two objectives, we need spectroscopic follow-up observations ideally in the GREX-PLUS wavelength regime to observe the rest-frame optical spectra. JWST is expected to play an essential role in this effort (as demonstrated by, e.g., \citealt{nakajima2026}). `21If not available, we can use the photometric color excesses to infer the strengths of [OIII]+H$\beta$ and H$\alpha$ to examine the nebular properties (e.g., \citealt{stark2013,smit2014,roberts-borsani2016,bouwens2016,nishigaki2023}).
The strong optical emission lines are well-calibrated to estimate gas-phase metallicities (e.g., \citealt{nakajima2022} for the latest updates).
Moreover, we can use extremely large telescopes such as TMT to examine the nebular properties with rest-frame UV spectra (e.g., \citealt{nakajima2018}).

As listed in the second item, we can address another important question of the nature of the ionizing spectrum
with detailed spectroscopic follow-up observations.
Galaxies hosting the first generation of stars (PopIII stars) are expected to be found between
the end of the dark ages ($z=20$--30) and the epoch of reionization ($z\sim 7$), and are thought
to play an important role in reionizing the Universe and in subsequent structure formation.
Low-mass, metal-poor galaxies explored with the above method may contain such a primordial structure.
We can use the spectroscopic diagnostics as proposed by, e.g., \citet{schaerer2003,inoue2011,NM2022}
to explore galaxies hosting PopIII stars and distinguish them from another important class of primeval objects: 
direct collapse black holes (DCBHs), using the rest-frame optical and UV spectra.

Finally, we can directly examine the link between such low-mass, metal-poor galaxies in the early Universe and the evolved galaxies found in later epochs via the dark matter halo properties. Such investigations will be of particular importance for assessing whether they are likely building blocks of galaxies in later epochs, including our Milky Way.

\begin{table}[htb]
  \begin{center}
    \caption{Expected numbers of low-mass, metal-poor galaxies with the current survey plans.}
    \label{tab:empg_expectations}
    \renewcommand{\arraystretch}{1.3}
    \begin{tabular}{|l|c|c|c|c|c|c|c|c|c|c|c|c|}
      \hline
      Redshift & \multicolumn{3}{c|}{2.2--2.8} & \multicolumn{3}{c|}{3.1--4.1} & \multicolumn{3}{c|}{4.3--5.6} & \multicolumn{3}{c|}{6.0--7.7$^{(\dag)}$} \\
      \hline
      Survey Type & D & M & W & D & M & W & D & M & W & D & M & W \\
      Depth (AB) & 26.5 & 25.5 & 24.0 & 26.5 & 25.5 & 24.0 & 24.5 & 23.5 & 22.0 & 23.5 & 22.5 & 21.0 \\
      Area (deg$^{2}$) & 10 & 100 & 1000 & 10 & 100 & 1000 & 10 & 100 & 1000 & 10 & 100 & 1000 \\
      \hline
      $N^o$ of galaxies & 100 & 700 & 4k & 100 & 700 & 3k & 20 & 200 & 800 & 10 & 80 & 100 \\
      \hline
    \end{tabular}
  \end{center}
  $\dag$ The highest-redshift search requires an additional redshift constraint such as Lyman break and Ly$\alpha$ provided by Roman.
\end{table}

\begin{table}[htb]
  \begin{center}
    \caption{Required observational parameters.}
    \label{tab:empg_requirements}
    \begin{tabular}{|l|p{9cm}|l|}
      \hline
      & Requirement & Remarks \\
      \hline
      Wavelength & 2--8 $\mu$m & \multirow{2}{*}{$a$} \\
      \cline{1-2}
      Spatial resolution & $<1$ arcsec & \\
      \hline
      Wavelength resolution & $\lambda/\Delta \lambda=3$--$4$ & $b$ \\
      \hline
      Field of view & 1000 degree$^2$, F232=23.4, F303=23.5, F397=23.0, & \multirow{2}{*}{$c$}\\
      \cline{1-1}
      Sensitivity & F520=21.7, F680=20.5 AB mag ($5\sigma$, point source) & \\
      \hline
      Observing field & Fields where deep imaging data at $\lambda<2\,\mu$m are available. & $d$ \\
      \hline
      Observing cadence & N/A & \\
      \hline
    \end{tabular}
  \end{center}
  $^a$ The wavelength coverage at $\lambda>5\,\mu$m is especially unique and crucial for high-redshift galaxy searches. \\
  $^b$ H$\alpha$ and [OIII]$\lambda 5007$ should fall into two adjacent photometric bands, and not into a single band at any redshift ($\lambda=(5007+6563)/2\times(1+z)$\,\AA, $\Delta\lambda=(6563-5007)\times(1+z)$\,\AA).\\
  $^c$ The Wide survey plan in the current conceptual design will result in the largest sample of low-mass, metal-poor galaxies at high redshifts. With deeper observations at $\lambda>5\,\mu$m, we would be able to probe much lower-mass systems at $z>5$.\\
  $^d$ Deep fields observed with Roman would be ideal.
\end{table}

\printbibliography[heading=subbibliography]
\end{refsection}

\clearpage

\begin{refsection}[2-8_highzattenuationcurves/highz_attenuation.bib]

\section{Attenuation curves in high-redshift galaxies}
\label{sec: Attenuation curves in high-redshift galaxies}

\noindent
\begin{flushright}
Kosei Matsumoto$^{1}$, 
Giulia Rodighiero$^{2}$
\\
$^{1}$ Ghent University, 
$^{2}$ University of Padova 
\end{flushright}
\vspace{0.5cm}

\subsection{Scientific background and motivation}
Dust is a fundamental component of the ISM that plays a critical role in galaxy evolution. It regulates the thermal and chemical state of the ISM, promotes molecular hydrogen formation, and strongly modifies the observed SEDs of galaxies by absorbing and scattering UV and optical light and re-emitting it in the infrared. As a result, dust directly affects key physical quantities inferred from observations, such as stellar mass, star formation rate, and dust mass \citep{Conroy2013}. Understanding the nature and evolution of dust is therefore essential for interpreting galaxy observations across cosmic time.

The dust attenuation curve, which describes the wavelength dependence of dust attenuation in galaxies, provides a powerful diagnostic of dust properties but is influenced by radiative processes within galaxies. Unlike extinction curves, attenuation curves encode not only intrinsic dust grain properties but also the relative spatial distribution of stars and dust (i.e., star-dust geometry), various stellar populations within galaxies, and scattering processes. In recent years, observations with JWST have begun to reveal attenuation curves in high-redshift galaxies by applying SED fitting techniques that allow for flexible attenuation laws \citep[e.g.,][]{Salim2018, Li2008}, suggesting that galaxies at higher redshift ($z>3$) exhibit systematically flatter attenuation curves at a fixed optical attenuation ($A_V$) than those observed in the local Universe \citep{Markov2025_Nat,Markov2025, Shivaei2025}. These findings point to potentially fundamental differences in dust properties or ISM conditions in early galaxies.

However, current observational constraints remain limited. JWST samples of attenuation curves at high redshift are still small, and thus, how attenuation curves evolve over cosmic time is still uncertain \citep[][]{Shivaei2025}. 
In addition, the wavelength coverage is often insufficient to robustly constrain the full shape of the attenuation curve and stellar populations within galaxies. As a result, it is observationally challenging to disentangle the physical origin of flat attenuation curves. Several explanations have been proposed, including (i) young stellar populations dominating the observed UV-optical light \citep[][]{Narayanan2018, Sommovigo2025}, (ii) compact dust geometries distinct from nearby galaxies \citep{Matsumoto2026}, and (iii) intrinsically different dust grain size distributions \citep{Shivaei2025}. At present, these scenarios remain highly degenerate, and no observational framework exists to distinguish between them.

GREX-PLUS offers a unique opportunity to overcome these limitations at higher redshifts. Prior to its launch, wide-field surveys conducted by the Roman Space Telescope and LSST are expected to provide deep photometric data over the wavelength range $0.3$--2 $\mu\mathrm{m}$ down to AB magnitudes of $\sim 27.5$. 
When combined with Roman and LSST data, GREX-PLUS will extend the photometric coverage continuously from $0.3$ to $8$ $\mathrm{\mu}$m, enabling robust measurements of attenuation curves across wide survey areas and over a wide redshift range ($z=2.5$--12). This broad wavelength coverage will reduce degeneracies in the SED fitting, allowing simultaneous constraints on the photometric redshift, dust attenuation curves, stellar ages, stellar masses, and star formation histories \citep[][]{Ilbert2006, Brammer2008}. In particular, for galaxies at $z<5$, GREX-PLUS will provide the full rest-frame UV-to-near-infrared SED coverage in the rest-frame wavelength range of $0.09$--1.0 $\mathrm{\mu}$m, which is essential for accurately characterizing attenuation curves and stellar populations.

Moreover, the full scientific potential of GREX-PLUS will be realized through synergy with follow-up observations. High-resolution integral-field spectroscopy with ELT/HARMONI will constrain the spatial distribution and ages of stellar populations, while ALMA observations will directly probe the spatial distribution and mass of dust \citep[][]{Hamed2023}. Together, these datasets will allow us to disentangle the relative roles of dust grain properties and star-dust geometry in shaping attenuation curves. This multi-facility approach will enable, for the first time, a physically grounded interpretation of attenuation curve evolution across cosmic time.

By providing statistically robust attenuation curves for galaxies from cosmic noon to the epoch of reionization, GREX-PLUS will place better constraints on the origin of flatter attenuation curves in the early Universe. These measurements will fundamentally improve our understanding of dust evolution, star formation, and the interpretation of galaxy SEDs at high redshift, establishing a critical legacy dataset for studies of galaxy evolution in the coming decade.

\subsection{Required observations and expected results}
\begin{figure*}[t!]
  \centering
    \includegraphics[width=0.99\textwidth]{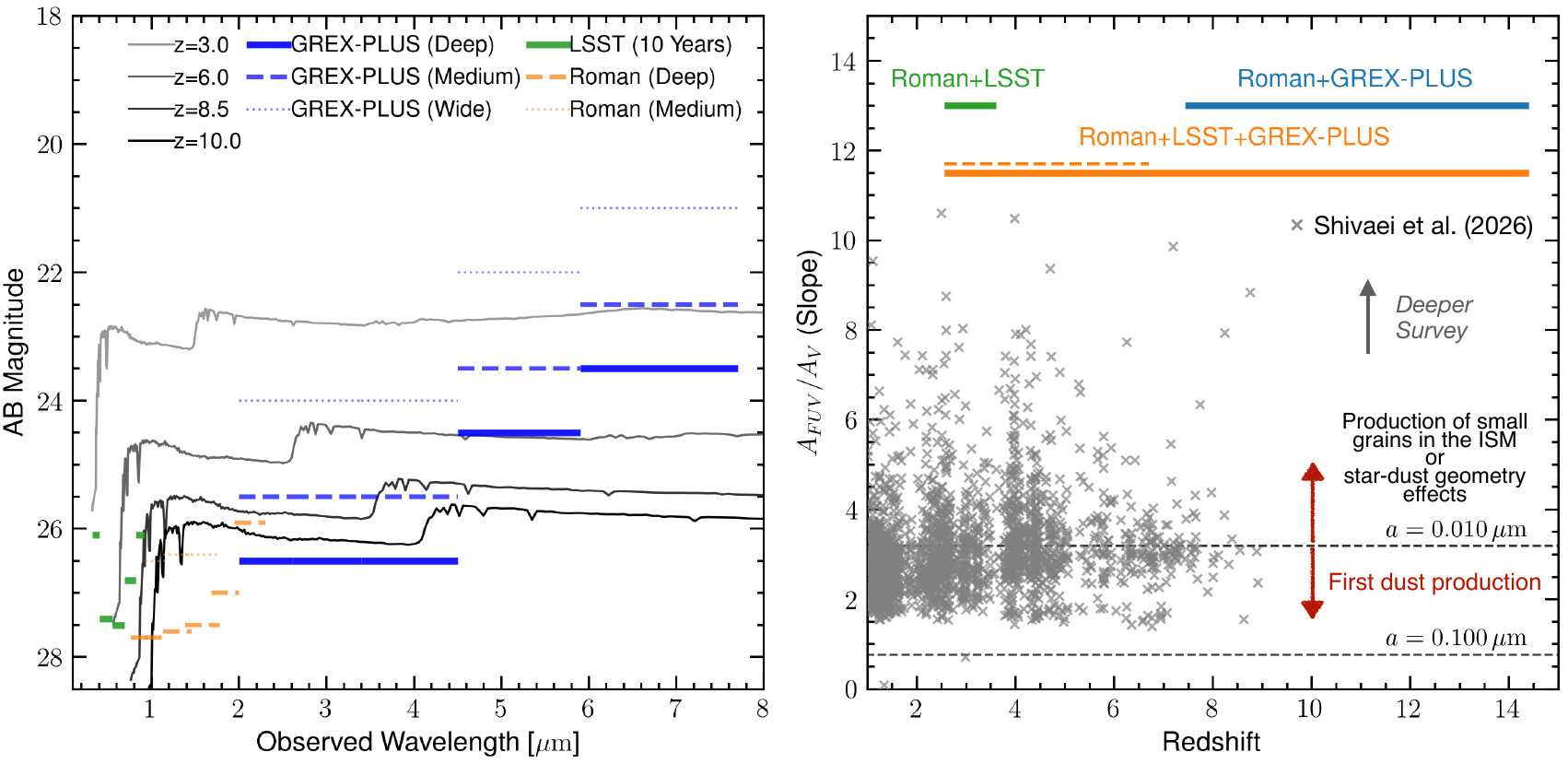}
    \caption{
      \textit{Left:} A mock dust-attenuated stellar spectrum of a star-forming galaxy with a stellar mass of $4\times10^{10} \ \mathrm{M_\odot}$ and $A_V=0.3$ at various redshifts is shown in black tones, together with the sensitivity limits of the available filters of the Roman Space Telescope, LSST, and GREX-PLUS (plotted in orange, green, and blue, respectively).
      \textit{Right:} The redshift evolution of the slopes of the attenuation curves for JWST-detected galaxies \citep[gray crosses; ][]{Shivaei2025}. The orange, green, and blue solid lines represent the redshift ranges accessible through the joint observations of the Roman Space Telescope, LSST, and GREX-PLUS, assuming a rest-frame wavelength coverage from $0.09$ to $0.5$ $\mathrm{\mu m}$. The orange dashed line shows the corresponding redshift range with the rest-frame wavelength coverage extended up to $1.0$ $\mathrm{\mu m}$. This highlights that GREX-PLUS is crucial for obtaining attenuation curves of galaxies at $z>4$ and the complete stellar continuum measurements for galaxies in the range $z=2.5$--6.5.
      The horizontal black dashed lines indicate the estimated slopes of the extinction curves for graphite grains with certain radii \citep[$a=0.01$ and $0.1$ $\mathrm{\mu m}$;][]{Draine2007}. If the slopes of the observed galaxies are below $3.2$, dust in these galaxies may be primarily composed of large grains formed in supernovae \citep[][]{Matsumoto2026}. Detecting such a shallower slope at $z\sim10$ is key for probing the onset of dust formation by the first supernovae.
      \label{fig:attenuation_curve}
    }
\end{figure*}

Combined with data from LSST and the Roman Space Telescope, the wide-field imaging surveys conducted with GREX-PLUS will enable the construction of a comprehensive catalog of galaxy attenuation curves over a broad range of redshifts. To achieve this goal, we require multi-band photometric coverage over the rest-frame wavelength range of $0.09$--0.5 $\mathrm{\mu}$m. This wavelength coverage allows us to capture key spectral features, including the Lyman break at $912$~\AA{}, the Balmer 4000~\AA{} break, and the optical continuum (see the left panel of Fig.~\ref{fig:attenuation_curve}). The UV-to-optical slope of attenuation curves ($A_{FUV}/A_V$) is characterized.
As discussed in Section~\ref{sec:veryhighz}, robust discrimination between Ly$\alpha$ break galaxies and low-redshift contaminants with intrinsically red colors (e.g., passive or dusty galaxies) requires at least two photometric data points at rest-frame wavelengths of $0.15$--0.3 $\mathrm{\mu}$m, redward of the Ly$\alpha$ break. Although spatially resolved imaging is not required for this science goal, we impose a spatial resolution better than $1^{\prime\prime}$ to mitigate source confusion caused by multiple galaxies falling within a single beam.
Based on these requirements, the observational parameters of GREX-PLUS are summarized in Table~\ref{tab: highz attenuation curves observational parameters}. 

\begin{table}[h!]
    \begin{center}
    \caption{Required observational parameters.}\label{tab: highz attenuation curves observational parameters}
    \begin{tabular}{|l|p{9cm}|l|}
    \hline
     & Requirement & Remarks \\
    \hline
    Wavelength & 2--8 $\mathrm{\mu}$m & \multirow{2}{*}{$a$} \\
    \cline{1-2}
    Spatial resolution & $<1$ arcsec & \\
    \hline
    Wavelength resolution & $\lambda/\Delta \lambda>3$ & $b$ \\
    \hline
    Field of view & 10 degree$^2$, 26.5 AB mag ($5\sigma$, point source) & \multirow{2}{*}{$c$}\\
    \cline{1-1}
    Sensitivity & 100 degree$^2$, 25.5 AB mag ($5\sigma$, point source) & \\
    & 1000 degree$^2$, 24.5 AB mag ($5\sigma$, point source) & \\
    \hline
    Observing field & Fields where imaging data at $\lambda<2$ $\mathrm{\mu}$m are available. & $d$ \\
    \hline
    \end{tabular}
    \end{center}
    $^a$ A primary mirror with $\phi\geq1.0$ m is required to achieve $\lesssim1$ arcsec at $\lambda=5$ $\mathrm{\mu}$m for the diffraction limit.\\
    $^b$ Three or more bands are required for characterizing the galaxy SEDs of galaxies at $z=12$.\\
    $^c$ A $>0.3$ degree$^2$ field of view of a single pointing is required from the point source sensitivity for a $\phi=1.0$ m telescope and an assumed amount of observing time.\\
    $^d$ The wide and medium fields are designed to overlap with the Roman Medium Tier, whereas the deep fields are placed within the Roman Deep Tier.
\end{table}

We estimate the expected number of galaxies in each survey tier by adopting the UV luminosity function from \citet{Bouwens2021} over the redshift range $2 \leq z \leq 10$ and the double power-law parameterization from \citet{Whitler2025} at $z > 10$.
We assume a mock dust-attenuated stellar spectrum corresponding to a star-forming galaxy with a stellar mass of $4 \times 10^{10}\,\mathrm{M_\odot}$ and $A_V = 0.3$, as shown in the left panel of Fig.~\ref{fig:attenuation_curve}.
For each redshift bin, we estimate the number of observable galaxies with the rest-frame wavelength coverage between $0.09$ and $0.5\,\mathrm{\mu}$m. The redshifts of observed galaxies are identified via detection of the Lyman break within the observed bands. We require detections in all bands corresponding to the rest-frame wavelengths between $0.09$ and $0.5\,\mathrm{\mu}$m, excluding the Lyman break at $\lambda = 0.912\,\mathrm{\mu}$m.
The resulting number of galaxies for each survey tier and redshift bin is summarized in Table~\ref{tab: Expected number of detected galaxies}.
In addition, we estimate the number of galaxies observable over an extended rest-frame wavelength range of $0.09$--1.0\,$\mathrm{\mu}$m with GREX-PLUS, as listed in Table~\ref{tab: Expected number of detected galaxies (0.09-1.0)}.
Below, we outline the survey strategy, additional requirements, and expected results for each GREX-PLUS survey tier.

\begin{table}[h]
\begin{center}
\caption{Expected number of detected galaxies with the rest-frame wavelengths between $0.09$ and $0.5$ $\mathrm{\mu m}$ for each survey tier$^\dagger$.}
\label{tab: Expected number of detected galaxies}
\begin{tabular}{|c|c|c|c|c|c|c|c|c|}
\hline
$z$ 
& $z_{\rm range}$ 
& $t_{\rm age}^*$ [Gyr]
& Deep 
& Medium 
& Wide 
& Most limiting band$^{**}$\\
\hline
2.95$^a$ & 2.51--3.39 & 1.87--2.60 & 158,412 & 1,550,884 & 9,339,032 & F232/GREX-PLUS \\
4.22$^a$ & 3.39--5.05 & 1.16--1.87 & 211,394 & 1,225,299 & 425,269 & F232/GREX-PLUS \\
5.81$^a$ & 5.05--6.58 & 0.83--1.16 & 33,934 & 46,585 & 2,671 & F232/GREX-PLUS \\
7.27$^b$ & 6.58--7.97 & 0.64--0.83 & 4,607 & 3,848 & 19 & F232/GREX-PLUS$^{***}$ \\
8.54$^b$ & 7.97--9.11 & 0.54--0.64 & 10 & 4 & 0 & F520/GREX-PLUS \\
10.62$^b$ & 9.16--12.07 & 0.36--0.53 & 4 & 5 & 1 & F680/GREX-PLUS \\
\hline
\end{tabular}
\end{center}

$^\dagger$ The numbers of detected galaxies are estimated using UV luminosity functions from \citet{Bouwens2021} and \citet{Whitler2025} for $z \leq 10$ and $z > 10$, respectively, assuming a dust-attenuated stellar spectrum ($M_\ast = 4 \times 10^{10}\,\mathrm{M_\odot}$, $A_V = 0.3$) shown in the left panel of Fig.~\ref{fig:attenuation_curve}, and requiring detections across rest-frame wavelengths of $0.09$--0.5 $\mathrm{\mu}$m but excluding the Lyman break.  \\
$^a$ The combined observations of LSST, the Roman Space Telescope, and GREX-PLUS.\\
$^b$ The combined observations of Roman Space Telescope and GREX-PLUS.\\
$^*$ The cosmic age range corresponds to the associated redshift range.\\
$^{**}$ We list the band that most strongly limits the detection, showing that the counts are primarily constrained by long-wavelength sensitivity. \\
$^{***}$ Only for the Medium survey at $z=7.27$, the z band from the LSST determines the observational limit.\\
\end{table}
\begin{table}[h]
\begin{center}
\caption{The same as Table~\ref{tab: Expected number of detected galaxies}, but showing the number of detected galaxies with rest-frame wavelengths between $0.09$ and $1.0$ $\mathrm{\mu m}$.}
\label{tab: Expected number of detected galaxies (0.09-1.0)}
\begin{tabular}{|c|c|c|c|c|c|c|}
\hline
$z$ 
& $z_{\rm range}$ 
& $t_{\rm age}$ [Gyr]
& Deep 
& Medium 
& Wide
& Most limiting band\\
\hline
2.95 & 2.51--3.39 & 1.87--2.60 & 158,412 & 1,504,265 & 927,815 & F397/GREX-PLUS \\
4.22 & 3.39--5.05 & 1.16--1.87 & 22,202 & 86,772 & 36,438 & F680/GREX-PLUS \\
\hline
\end{tabular}
\end{center}
\end{table}

\subsubsection{Wide Survey}

The GREX-PLUS Wide survey will be synchronized with the LSST survey and the Roman Space Telescope Medium survey. The LSST will conduct a survey with a survey area of $18{,}000$~deg$^2$ for $10$ years using the $u$, $g$, $r$, $i$, $z$, and $y$ bands ($\lambda = 0.367$, $0.481$, $0.617$, $0.752$, $0.866$, and $0.911~\mu$m, respectively)\footnote{\url{https://rubinobservatory.org/for-scientists/rubin-101/key-numbers}}. The Roman Medium survey will provide imaging over $2{,}400$~deg$^2$ with the F106, F129, and F158 filters ($\lambda = 1.06$, $1.29$, and $1.58~\mu$m, respectively)\footnote{\url{https://roman-docs.stsci.edu/roman-community-defined-surveys/high-latitude-wide-area-survey}}.
By observing a $1{,}000$~deg$^2$ area that overlaps with both the LSST and Roman survey regions, GREX-PLUS will observe around $1 \times 10^{7}$ galaxies over the redshift range $2.5 < z < 12$. This represents a substantial improvement compared to the $\sim3{,}800$ galaxies at $z > 3$ analyzed in \citet{Shivaei2025}.

\subsubsection{Medium Survey}

By observing a $100$~deg$^2$ area overlapping with both the LSST and Roman survey regions, the GREX-PLUS Medium survey is expected to detect more than $3 \times 10^{6}$ galaxies over the redshift range $2.5 < z < 12$. This survey tier provides the best balance between survey area and depth, enabling statistical studies of attenuation properties of fainter galaxies at $z>3.3$. Remarkably, this survey offers the greatest capability for identifying a few dust-attenuated galaxies at $z\sim10$.

\subsubsection{Deep Survey}

The GREX-PLUS Deep survey will be synchronized with the Roman Space Telescope Deep surveys. The Roman Deep survey will provide imaging over $19.2$~deg$^2$ using a rich set of filters including the F087, F105, F129, F158, F184, and F213 ($\lambda = 0.87$, $1.06$, $1.29$, $1.58$, $1.84$, and $2.13~\mu$m, respectively).
By observing a $10$~deg$^2$ area overlapping with the Roman Deep survey fields, GREX-PLUS is expected to detect more than $4 \times 10^{5}$ galaxies over the redshift range $2.5 < z < 12$. Notably, most of these galaxies at $z>5$ are expected to be UV-faint systems that are not detectable in the medium survey. These faint populations are likely to include dusty galaxies with steeper attenuation curves in the early Universe.
These measurements will, in turn, place constraints on the timescales of the first star formation and resulting dust production in our Universe (see the right panel of Fig.~\ref{fig:attenuation_curve}). The detection of a shallower slope ($A_{FUV}/A_V<0.5$) is key for probing the first dust production by the first supernovae \citep{Matsumoto2026}.


\subsection{Scientific goals}
The scientific goals of this program are to establish a statistically robust and physically grounded understanding of dust attenuation curves in galaxies from cosmic noon to the epoch of reionization ($2.5 \leq z \leq 12$) and to identify the first dust production around $z\geq10$ (see the right panel of Fig.~\ref{fig:attenuation_curve} and Table~\ref{tab: Expected number of detected galaxies}). 
By uniquely combining the wide-area imaging capability of GREX-PLUS with complementary data from the Roman Space Telescope and LSST, this program will also provide contiguous photometric coverage from $0.3$ to $8$ $\mathrm{\mu}$m, which is essential for breaking long-standing age-dust degeneracies in SED fitting.
This broad wavelength coverage enables simultaneous and self-consistent constraints on photometric redshifts, dust attenuation curve shapes, stellar ages, stellar masses, and star formation histories for galaxies at $z<5$ (see Table~\ref{tab: Expected number of detected galaxies (0.09-1.0)}).
GREX-PLUS is indispensable for extending rest-frame UV-to-near-infrared coverage to high redshifts over wide survey areas, enabling robust measurements of attenuation curves for statistically significant galaxy samples. This represents a substantial increase in sample size compared to existing JWST observations, which are limited by survey area. The wide survey will deliver a statistically representative galaxy population over a large cosmic volume and a broad redshift range, allowing us to establish global trends in dust attenuation, the slope of attenuation curves, and their redshift evolution. The medium survey will extend these measurements toward fainter galaxies, improving constraints on attenuation properties as a function of stellar mass. The deep survey will further probe UV-faint galaxies that are preferentially low-mass, older, or more heavily dust-obscured, accessing regions of parameter space that are inaccessible in the wide survey.

Through synergy with follow-up observations from ALMA and ELT/HARMONI, the role of star-dust geometry in shaping the slope of attenuation curves will be directly investigated. Together, these measurements will address the physical origin of the systematically flatter attenuation curves observed in early galaxies, distinguishing between intrinsic variations in dust grain properties and changes in star-dust geometry. This program will establish the first comprehensive observational framework for interpreting the evolution of dust attenuation across cosmic time, firmly positioning GREX-PLUS as a cornerstone facility for studies of galaxy and dust evolution in the early Universe.

\printbibliography[heading=subbibliography]
\end{refsection}

\clearpage

\begin{refsection}[2-9_Environmental_effects/Env_eff.bib]

\section{Galaxies under Pressure: Environmental Transformations from Clusters to the Cosmic Web}
\label{sec:environment}

\noindent
\begin{flushright}
Benedetta Vulcani$^{1}$, Bianca M. Poggianti$^{1}$, Alessia Moretti$^{1}$,
Ivan Delvecchio$^{2}$, Giorgia Peluso$^{2}$, Pietro Benotto$^{1,3}$, Mario Radovich$^{1}$
\\
$^{1}$INAF--Osservatorio astronomico di Padova, Vicolo Osservatorio 5, I-35122 Padova, Italy
$^{2}$INAF--Osservatorio di Astrofisica e Scienza dello Spazio di Bologna, Via Gobetti 93/3, I-40129 Bologna, Italy
$^{3}$Dipartimento di Fisica e Astronomia ‘G. Galilei’, Università di Padova, Vicolo dell’Osservatorio 3, 35122 Padova, Italy
\end{flushright}
\vspace{0.5cm}

\subsection{Scientific background and motivation}

Environment plays a fundamental role in shaping the evolution of galaxies, influencing their star formation activity, gas content, morphology, and nuclear activity \citep[e.g.,][]{Oemler1974, Dressler1980, Blanton2005, Vulcani2010, Vulcani2023}. Over the past decades, it has become clear that galaxy evolution cannot be understood in isolation, but rather as the result of a complex interplay between internal processes and external mechanisms linked to the surrounding environment. These environmental effects operate across a wide range of scales, from the dense cores of massive galaxy clusters to groups, filaments, and the diffuse structures of the cosmic web.

Several physical mechanisms have been proposed and observationally confirmed as key drivers of environmentally driven galaxy transformation. These include ram pressure stripping \citep{Gunn1972}, which removes cold gas from galactic disks as galaxies move through a dense intracluster medium; gas starvation or strangulation \citep{Larson1980, Balogh2000}, caused by the removal of the hot gaseous halo that feeds star formation; high-speed gravitational encounters \citep[harassment,][]{Moore1996}; and galaxy--galaxy mergers \citep{Toomre1972, Barnes1992}. Among these, hydrodynamical processes appear to be particularly efficient in dense environments. Observations of nearby clusters show that a significant fraction of star-forming galaxies (approximately 30--40\% within 2 virial radii) exhibit clear signatures of ongoing or recent ram pressure stripping, such as one-sided gas tails, truncated gas disks, and undisturbed stellar kinematics \citep{Vulcani2022}. On the other hand, internal mechanisms also affect galaxy evolution. Among these, stellar and Active Galactic Nuclei (AGN) feedback are particularly strong, as they can heat, expel, or redistribute gas, profoundly altering the interstellar medium \citep[e.g.,][]{Veilleux2005, Erb2015, King2015}.

Integral-field spectroscopic surveys and multiwavelength studies have provided compelling evidence that gas stripping primarily affects the gaseous component of galaxies, while leaving the stellar structure largely intact, at least in the initial phases \citep{Fumagalli2014, Poggianti2017, Boselli2022}. These observations reveal extended tails of ionized, neutral, and molecular gas, often maintaining coherent rotation several kiloparsecs downstream from the galactic disk \citep{Poggianti2025, Luber2022, Moretti2020, Jachym2019}. Such features demonstrate that environmental processes can efficiently remove the fuel for star formation on short timescales, leading to rapid quenching \citep{Vulcani2020}. However, current observations are still limited in their ability to fully characterize the multiphase nature of the stripped interstellar medium and to disentangle the relative roles of shocks, star formation, and AGN activity in shaping the observed emission.

Mid-infrared observations are uniquely suited to address these open questions. The mid-infrared wavelength range hosts a rich set of diagnostic features, including forbidden emission lines of different ionization states (e.g. [Ne II], [Ne III], [Ne V], [Fe II], [S IV]), rotational lines of warm molecular hydrogen, polycyclic aromatic hydrocarbon (PAH) features, and dust continuum emission. These tracers probe both distinct gas phases and dust emission, providing a comprehensive view of the physical conditions in galaxy disks, nuclear regions, and stripped tails. However, such observations are extremely challenging from the ground due to atmospheric absorption, and existing space facilities are limited either in spectral resolution or in survey capability.

Beyond rich clusters, there is growing evidence that environmental processes also operate in lower-density regions, such as galaxy groups and filaments \citep{Castignani2022, Vulcani2021}. Recent studies suggest that gas stripping and enhancement mechanisms may already be effective in the cosmic web, where galaxies interact with tenuous intergalactic gas. Understanding how and where environmental transformation begins is crucial for building a unified picture of galaxy evolution across environments. In this context, GREX-PLUS offers a unique opportunity to extend detailed environmental studies from cluster cores to the outskirts of large-scale structures and the cosmic web, enabling us to quantify how dust properties, PAH emission, and gas excitation change across environments and galaxy masses, and to identify where environmental processes begin to dominate over internal feedback. This requires the combination of spatially resolved mid-infrared diagnostics and the large, statistically representative samples uniquely accessible with GREX-PLUS.


\subsection{Required observations and expected results}

\subsubsection{Wide-Field Camera Imaging (2--8 $\mu$m)}

\begin{figure}
    \centering
    \includegraphics[width=0.65\linewidth]{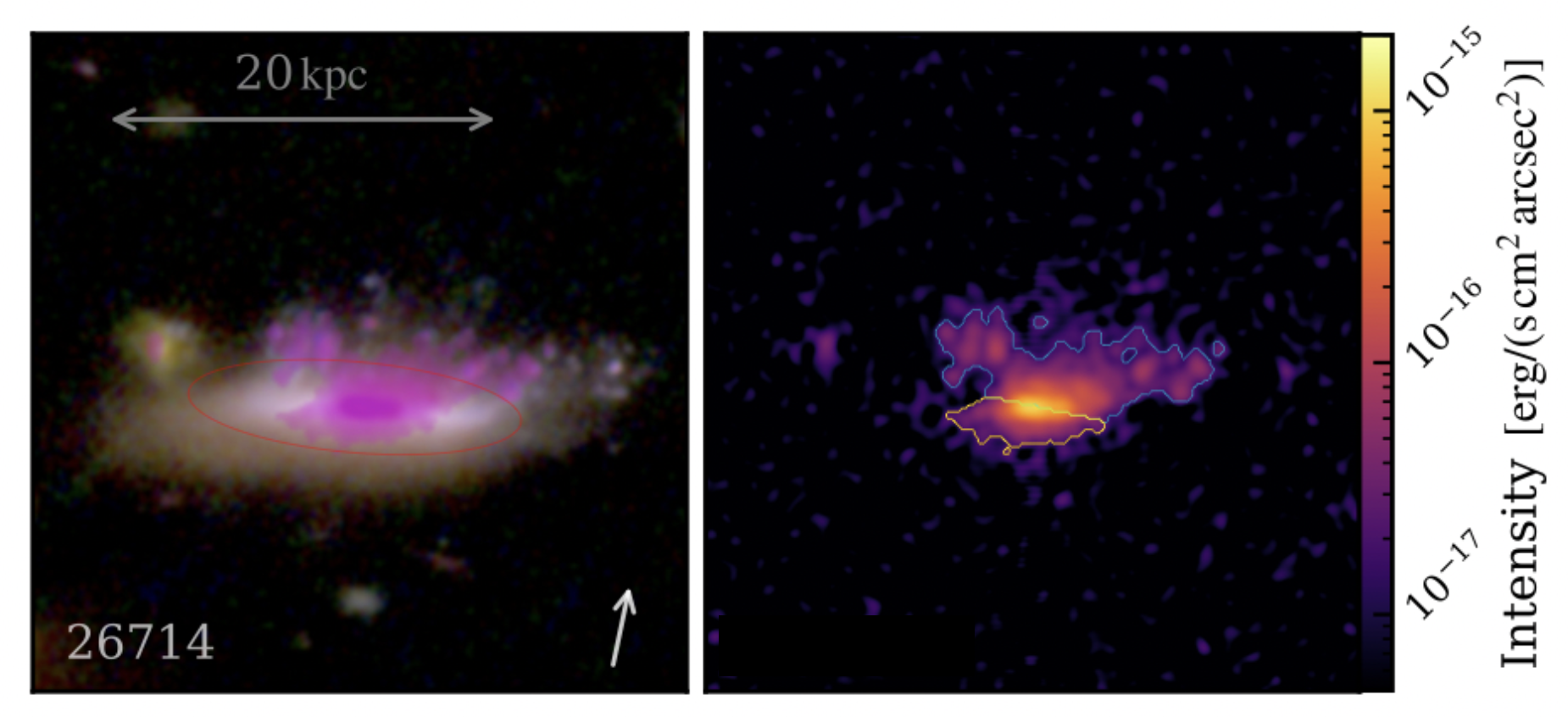}
    \caption{Example of a $3.3\,\mu$m PAH emission map for a galaxy undergoing ram pressure stripping in an intermediate-redshift cluster. An RGB image composite of F070W (blue), F115W (green), and
F200W (red) filters with the 5$\sigma$-significance PAH map of the galaxy (superimposed in purple) is shown on the left. The red ellipse indicates the
galaxy disk. The arrow indicates the RPS direction. The surface brightness map of the $3.3\,\mu$m PAH emission is shown on the right. The gold and blue
contours indicate the leading and trailing galactic regions with brightness greater than 5$\sigma_{bkg}$, respectively.  North is up, east is left. Taken from \cite{Benotto2025}.}
    \label{fig:PAH}
\end{figure}

The GREX-PLUS Wide-Field Camera (WFC) will provide deep, wide-area imaging in five bands covering the 2--8 $\mu$m wavelength range.  These observations are essential for tracing dust emission, PAH features, and stellar continuum in galaxies experiencing environmental interactions. Imaging observations will target representative samples of galaxies in nearby clusters, groups, and filamentary environments, including systems with known signatures of gas stripping. To maximize the scientific return of the WFC medium survey over $\sim$100 deg$^{2}$, the target fields will be selected to overlap with existing multi-wavelength surveys in the Southern Hemisphere. In particular, synergy with optical integral-field spectroscopy from MUSE (which has a comparable FoV) and molecular gas observations from ALMA will be crucial to link dust and PAH properties traced by GREX-PLUS to the distribution and kinematics of ionized and molecular gas. Existing datasets from surveys such as GASP, PHANGS--ALMA, and wide-area spectroscopic surveys of nearby clusters and groups provide an ideal foundation for this approach.

The WFC imaging will enable spatially resolved and quantitative studies of PAH emission and dust properties in galaxies undergoing environmental interactions. The availability of medium- and/or narrow-band filters would further enhance these capabilities, enabling improved constraints on PAH strength, spatial substructure, and variations across different galactic components. In particular, the $3.3\,\mu$m PAH feature (Fig.~\ref{fig:PAH}), accessible up to $z\sim1.2$, will be used as a tracer of recent star formation and as a sensitive probe of the PAH size distribution \citep[e.g.,][]{Shivaei2024, Lyu2025, Benotto2025}, while the $6.2\,\mu$m PAH feature, detectable up to $z\sim0.25$, will primarily provide constraints on the PAH ionization state in both galactic disks and stripped tails. 

In addition, the WFC wavelength coverage probes the mid-infrared continuum emission from warm dust with characteristic temperatures of $T\sim200$--400 K, providing constraints on dust column densities, geometry, and heating sources. Variations in the continuum shape can be used to identify differences in dust heating and processing between galaxy disks and stripped tails, and to distinguish between star-formation-dominated and AGN-dominated regions. These diagnostics offer a direct way to investigate how environmental mechanisms such as ram pressure stripping and shocks affect the dust component of galaxies.

Expected results include spatially resolved maps of PAH emission and dust spectral properties, quantitative measurements of star formation activity in disks and extraplanar regions, constraints on dust destruction and processing efficiencies, and a statistical characterization of how dust and PAH properties vary as a function of environment, from cluster cores to the cosmic web.

\subsubsection{High-Resolution Spectroscopy (10--18 $\mu$m)}

\begin{figure}
    \centering
    \includegraphics[width=0.5\linewidth]{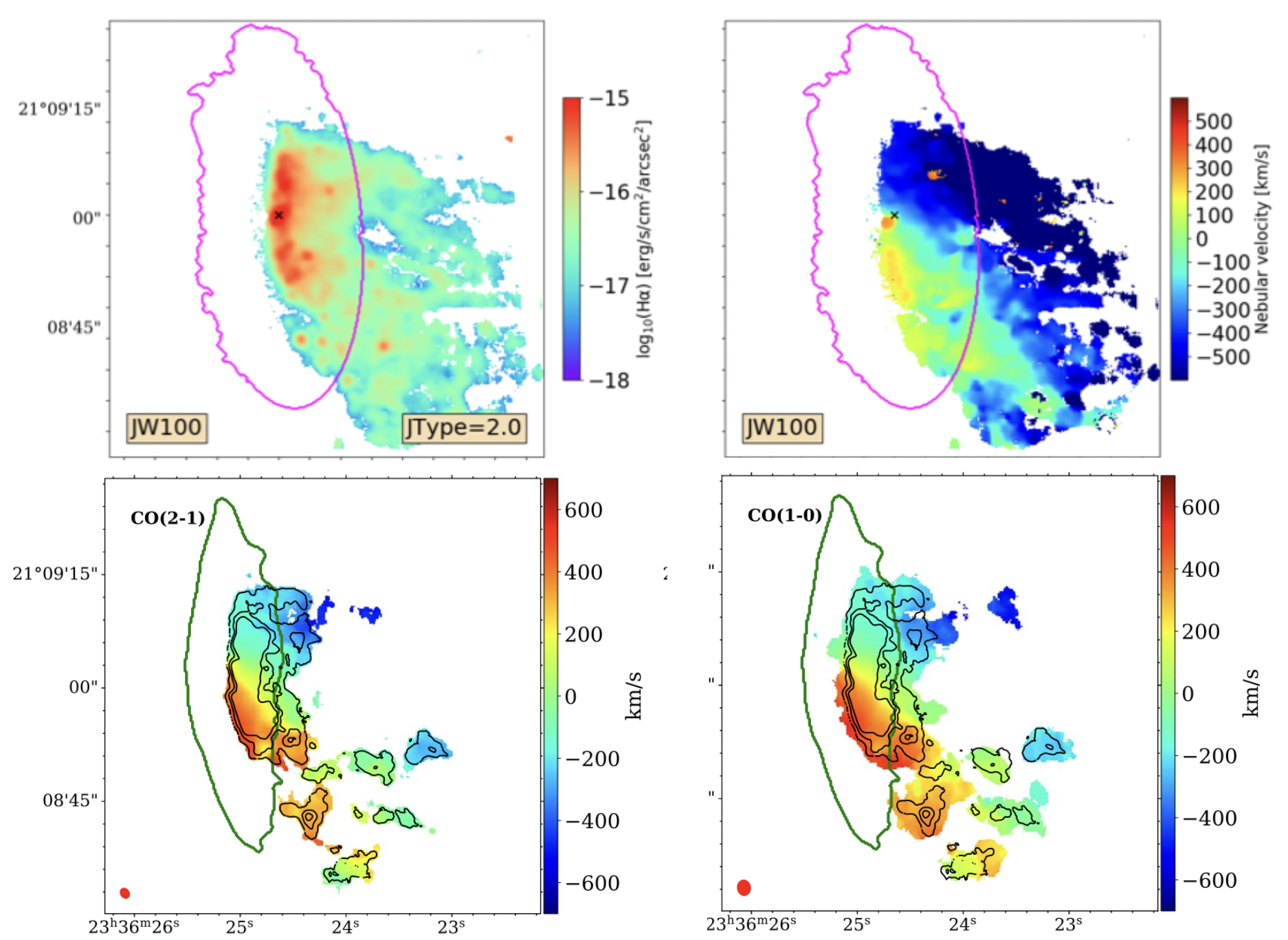}
    \caption{Example of a gas kinematics in a ram-pressure-stripped galaxy in different gas phases.  Top left: H$\alpha$ flux from MUSE, top right: ionized gas kinematics from MUSE; bottom left: molecular CO (2-1) gas kinematics from ALMA; bottom right: molecular CO (1-0) gas kinematics from ALMA. The pink and green lines delimit the stellar disk, computed in slightly different ways. Ionized gas information is taken from \cite{Poggianti2025}, molecular gas information is taken from \cite{Moretti2020_JW100}.}
    \label{fig:JW100}
\end{figure} 
The GREX-PLUS High-Resolution Spectrometer (HRS), with a resolving power of $R\sim30,000$ in the 10--18 $\mu$m range, will provide unprecedented access to the kinematics and excitation of the multiphase interstellar medium in environmentally affected galaxies. This spectral resolution corresponds to velocity resolutions on the order of 10 km s$^{-1}$,  allowing detailed separation of rotational motion, stripped components, inflows, and outflows. At $z=0$, the angular resolution of $\sim$1 arcsec is comparable to that achieved by state-of-the-art ground-based facilities, enabling direct and homogeneous comparisons between space-based mid-infrared spectroscopy and existing optical and millimeter data (Fig.~\ref{fig:JW100}).

For galaxies in the local Universe, key diagnostic lines include [Ne II] and [Ne III], which trace ionized gas associated with star formation and shocks; high-ionization lines such as [Ne V], which unambiguously signal AGN activity; [Fe II] lines, sensitive to shock excitation; and rotational lines of warm molecular hydrogen. By combining line ratios, line widths, and velocity fields, GREX-PLUS will enable robust discrimination between photoionization by young stars, shock heating induced by ram pressure stripping, and AGN-driven processes. 
For a local AGN with bolometric luminosity 
$\log L_{bol} \sim$44 \citep{Peluso2025}, the expected mid-infrared continuum flux density is of order a few mJy, while typical fine-structure line fluxes are expected to be of order $10^{-17}$--$10^{-16}$ W m$^{-2}$, enabling high signal-to-noise spectroscopy for representative samples of nearby galaxies and AGN. This will allow velocity measurements down to  $\sim$
10 km/s for the brightest systems, while still providing robust constraints on kinematics and excitation for larger statistical samples.
A resolving power of $R \geq 15\,000$ ($\Delta \lambda/\lambda \leq 7\times10^{-5}$) is required to disentangle the kinematics of ionized gas traced by mid-infrared fine-structure lines, while the HRS resolution of $R \sim 30\,000$ ($\Delta \lambda/\lambda \sim 3\times10^{-5}$) enables detailed measurements of velocity fields, line widths, and multiple dynamical components. Compared to current facilities such as JWST, GREX-PLUS will extend such analyses from targeted observations of individual systems to statistically representative samples across a wide range of environments.

These observations will allow us to measure velocity offsets between different gas phases, quantify turbulence and velocity dispersions in the brightest regions of stripped gas, and identify signatures of gas accretion or outflows in galaxy centers. While detailed kinematic mapping at the highest spectral resolution will be limited to relatively high surface-brightness regions (e.g. galaxy nuclei and bright clumps in stripped tails), integrated measurements will provide robust constraints on the kinematics of more diffuse gas. Ultimately, the combination of line luminosities and kinematics will provide estimates of mass and energy fluxes associated with environmental stripping and AGN feedback.

\subsection{Scientific goals}

The main goal of this program is to understand how environmental processes transform galaxies by reshaping their gaseous, dusty, and star-forming components across a wide range of environments, from massive clusters to the cosmic web. By exploiting the unique combination of wide-field mid-infrared imaging and high-resolution spectroscopy offered by GREX-PLUS, this science case aims to deliver a spatially and kinematically resolved characterization of the multiphase interstellar medium under environmental pressure, disentangling the relative roles of ram pressure stripping, shocks, star formation, and AGN activity in driving galaxy evolution.

\begin{table}
    \begin{center}
    \caption{Required observational parameters for PAH detection.}\label{tab:PAH}
    \begin{tabular}{|l|p{9cm}|l|}
    \hline
     & Requirement & Remarks \\
    \hline
    Wavelength & 3.2--8 $\mu$m & \multirow{2}{*}{$a$} \\
    \cline{1-2}
    Spatial resolution & $<1$ arcsec & \\
    \hline
    Field of view & 100 deg$^2$, 24.5 ABmag ($5\sigma$, point-source)  &\\
    \cline{1-1}
    Sensitivity &  & \\
    \hline
    \end{tabular}
    \end{center}
    $^a$ Depending on redshift.
\end{table}

\begin{table}
    \begin{center}
    \caption{Required observational parameters for kinematic and AGN characterization.}\label{tab:firstgals}
    \begin{tabular}{|l|p{9cm}|l|}
    \hline
     & Requirement & Remarks \\
    \hline
    Wavelength & 12--18 $\mu$m &  \\
    \cline{1-2}
    Spatial resolution & $<1$ arcsec & \\
    \hline
    Wavelength resolution & $\lambda/\Delta \lambda>15000$ & \\
    \hline
    Field of view & E.g., 100 deg$^2$, $F([\mathrm{Ne\,V}]\,14.3\,\mu\mathrm{m}) \sim 1 \times 10^{-17}\,\mathrm{W\,m^{-2}}$ & \multirow{2}{*}{$a$}\\
    \cline{1-1}
    Sensitivity &  & \\
    \hline
    \end{tabular}
    \end{center}
    $^a$ Obtained assuming a local AGN with 
$\log L_{\mathrm{bol}} \sim 44$ and the relation from \cite{Satyapal2007}.
\end{table}

\printbibliography[heading=subbibliography]
\end{refsection}

\clearpage

\begin{refsection}[2-10_lowzstellarmass/lowzstellarmass.bib]

\section{An accurate and unbiased mapping of stellar mass in the low-z Universe}
\label{sec:stellarmass}

\noindent
\begin{flushright}
Stefano Zibetti$^{1}$
\\
$^{1}$ INAF--Arcetri Astrophysical Observatory, Firenze, Italy
\end{flushright}
\vspace{0.5cm}

\newcommand{\mstar}{\ensuremath{M_\star}}
\newcommand{\mum}{\ensuremath{\mathrm{\mu m}}}

\subsection{Scientific background and motivation}

Stellar mass (\mstar) is a fundamental tracer, as well as a driver, of galaxy evolution. The evolution of the stellar mass function \citep[e.g.,][]{2023A&A...677A.184W} across cosmic time and environments, and the scaling relations that connect \mstar\ with the properties of the stellar populations \citep[SPs, e.g.,][]{2025A&A...703A...5M}, star-formation history (SFH), interstellar medium \citep[ISM, e.g.,][]{2004ApJ...613..898T}, and kinematics, both globally and spatially resolved \citep[e.g.,][]{2022MNRAS.512.1415Z}, represent key observational constraints for galaxy evolution models. Despite the advances of the last decades, reliable and unbiased estimates of \mstar\ at the level of $\lesssim 10\%$ remain a challenge: ideally one would need deep optical spectroscopy complemented with extended UV-VIS-NIR photometry to fully characterize the multiple stellar populations (SPs) that form a galaxy and their relative dust attenuation. This is feasible, although observationally expensive, for spatially unresolved estimates.

\begin{figure}
   \centering
   \includegraphics[width=7cm]{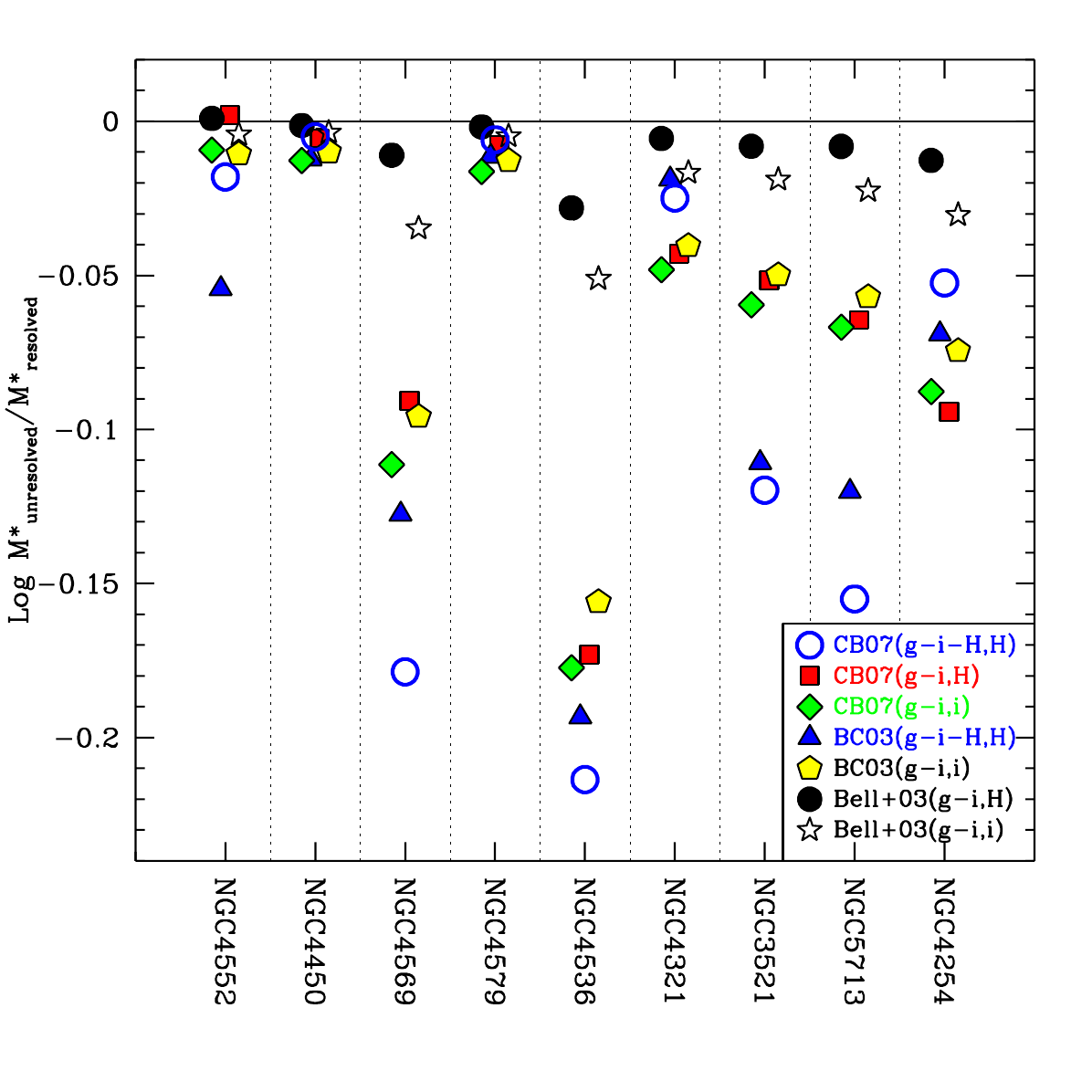}
   \caption{Logarithmic differences between total stellar mass estimates
  obtained from unresolved photometry and by integrating resolved
  maps, both based on the {\em same} method/SP models, for a sample of nearby galaxies with VIS and NIR imaging, spanning a range of morphologies, from ellipticals to late-type spirals (left to right). Different
  methods/models are shown with different symbols, as shown in the
  legend. Adapted from \citet{2009MNRAS.400.1181Z}.} 
   \label{fig:res_unres_Mstar}
\end{figure}

However, integrated observations suffer from systematic biases that cannot be corrected without spatial resolution. The mass-to-light ratio (M/L) varies dramatically within individual galaxies due to spatial variations in stellar age, metallicity, and dust attenuation. Younger, brighter stellar populations in spiral arms and star-forming regions systematically dominate the integrated light despite contributing a smaller fraction of the total stellar mass, leading to an ``outshining'' bias that can underestimate total \mstar\ by 0.1--0.15~dex (\citealt{2009MNRAS.400.1181Z}, see Fig. \ref{fig:res_unres_Mstar}). 

\begin{figure}
   \centering
   \includegraphics[width=7cm]{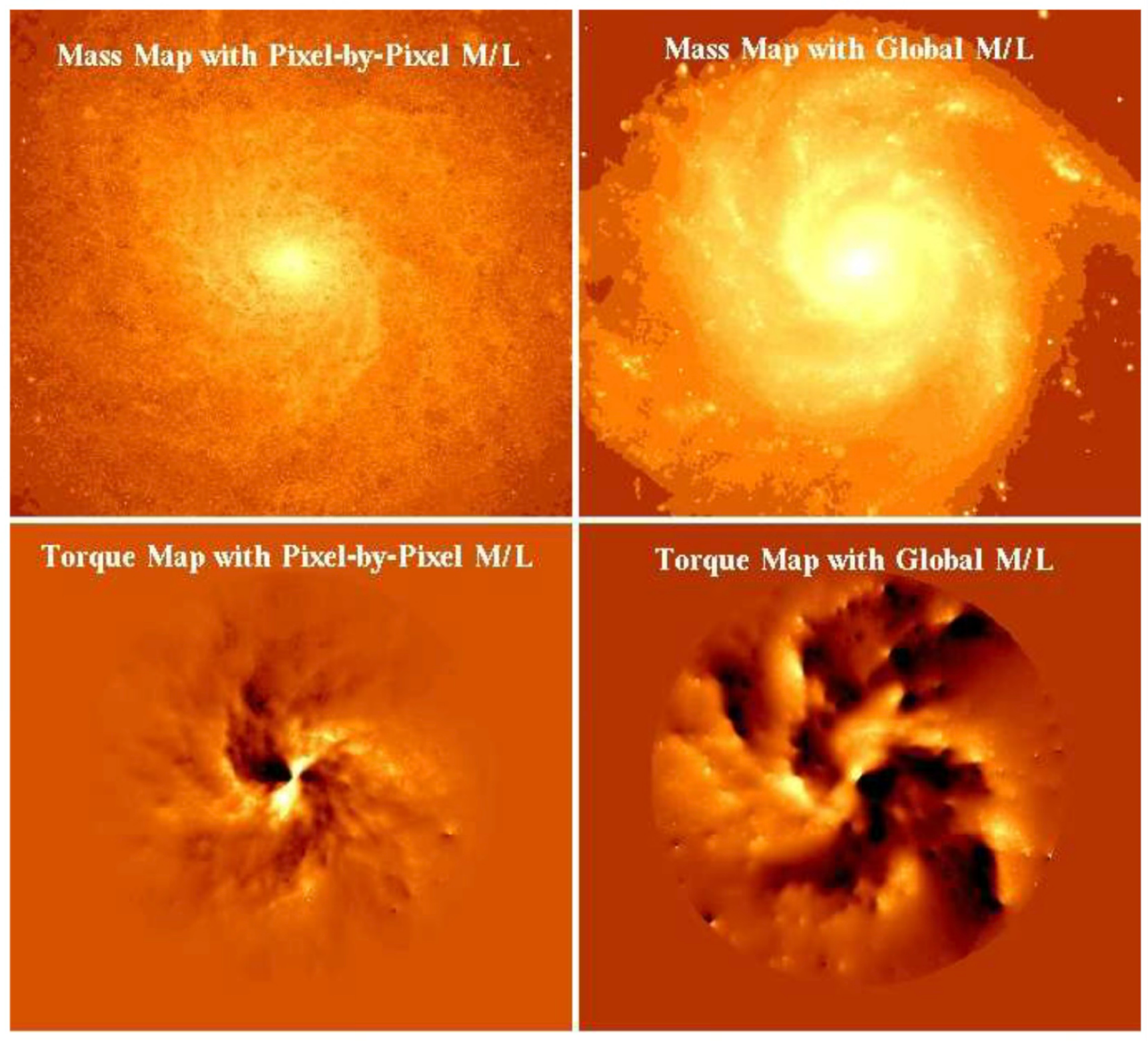}
   \caption{Mass maps of M101 ({\bf upper panels}) made from the pixel-by-pixel method of \citet{2009MNRAS.400.1181Z} using g- and i-band images from SDSS (left-hand panel) and from that using a single global M/L with an i-band image (right-hand panel). The torque maps for each respective mass image are shown in the {\bf bottom panels}. The global M/L methodology enhances the spiral arms and the torques associated with them. Adapted from \citet{2010MNRAS.407..163F}.} 
   \label{fig:mstar_torques}
\end{figure}

Moreover, accurate dynamical modeling---essential for constraining dark matter distributions and understanding secular evolution---requires knowledge of the stellar \textit{mass} distribution, not merely the light distribution. Distinguishing true mass overdensities from light overdensities in structures like spiral arms and bars is impossible without spatially resolved observations that properly account for local M/L variations (e.g., \citealt{2010MNRAS.407..163F}, also Fig. \ref{fig:mstar_torques}).

The near-infrared (NIR) at 1--3~\mum\ offers critical advantages: NIR light is dominated by evolved stellar populations (RGB and AGB stars) rather than main-sequence stars, yielding more stable M/L ratios---typically varying by only $\sim$0.2--0.3~dex across diverse SPs \citep{2001ApJ...550..212B}---compared to optical wavelengths where young, massive stars cause order-of-magnitude variations. Dust attenuation is significantly reduced, minimizing systematic uncertainties from uncertain dust geometries and attenuation laws. Furthermore, the 1--3~\mum\ range is free from contamination by polycyclic aromatic hydrocarbon (PAH) emission and hot dust continuum that plague longer wavelengths ($\lambda \gtrsim 3$~\mum), providing a clean window to the stellar photospheres \citep{2008MNRAS.388.1595D,2011MNRAS.417..812Z}.



Yet simply observing in a single NIR band is insufficient: sampling the full NIR SED with multi-band photometry (e.g., J, H, K) is essential to overcome extant uncertainties in SP models. In particular, NIR colors are sensitive to the contribution of thermally-pulsing AGB (TP-AGB) stars, one of the most uncertain phases in SPS models \citep[e.g.,][]{2005MNRAS.362..799M}; multi-band NIR observations of large samples spanning diverse SP parameters provide the only empirical benchmark for different SPS model prescriptions for evolved stars \citep[e.g.,][]{2010ApJ...722L..64K,2013MNRAS.428.1479Z}. Moreover, a well-sampled NIR SED enables clean extrapolation of the stellar continuum beyond 3~\mum, isolating PAH emission as a reliable SFR tracer \citep{2007ApJ...666..870C}. 

In summary, only spatially resolved, multi-band NIR observations of statistically significant samples can simultaneously address the outshining bias, SPS model systematics, and dust correction uncertainties.

\subsection{Required observations and expected results}

Euclid is currently surveying the full extragalactic sky ($\sim$14\,000~deg$^{2}$) and will provide exquisite Y, J, and H-band imaging (plus a broad visible filter) for low-$z$ galaxies, achieving a physical resolution $<1~(0.3)$~kpc at $z<0.05~(0.01)$. This would be ideally complemented by K-band ($\sim$2.2~\mum) imaging to complete the sampling of the stellar-dominated NIR SED. The planned 24~AB~mag ($5\sigma$) depth of the GREX-PLUS Wide survey matches well the Euclid imaging. To maximize constraints on SP properties, we focus on galaxies with available optical spectroscopy suitable for SP analysis ($R \gtrsim 2000$, S/N $> 10$--$20$). Requiring overlap with SDSS-I \citep{york00,sdss_DR7}, we reach a target density of 2--3.5 galaxies~deg$^{-2}$ at $0.01<z<0.05$ and 5--13 galaxies~deg$^{-2}$ at $0.05<z<0.10$ (the range reflecting the S/N$>20$ vs. S/N$>10$ cut). If $\gtrsim$50\% of the GREX-PLUS Wide survey footprint overlaps with SDSS, this yields a sample of a few thousand galaxies with sub-kpc resolution YJHK imaging plus integrated optical spectroscopy, spanning $\mstar \gtrsim 10^{9}$~M$_\odot$ and the full range of morphological types. On this sample, spatially resolved NIR SED analysis will enable us to:
\begin{itemize}
\item benchmark SPS models in evolutionary phases dominated by evolved NIR-bright stars (e.g.\ TP-AGB), leveraging sub-kpc resolution to isolate specific SFH phases;
\item derive state-of-the-art NIR-based stellar mass maps with minimal bias from dust attenuation and emission;
\item calibrate the bias affecting \mstar\ estimates based on integrated photometry;
\item assess the dynamical relevance of galactic structures such as spiral arms and bars, correcting for the luminosity bias affecting both optical and NIR imaging.
\end{itemize}
Leveraging the available optical spectroscopy will further refine the priors on NIR M/L, improving the accuracy of \mstar\ estimates.

The planned depth of GREX-PLUS observations will also be sufficient to detect low-surface-brightness stellar emission in galaxy outskirts---extended disks, envelopes, and tidal features of satellites. Combined with Euclid optical-NIR imaging, this will enable, for the first time, a census of stellar \emph{mass} in these structures.

Finally, GREX-PLUS imaging at $\lambda > 3$~\mum\ carries crucial information about PAH emission, whose flux---especially near $3.3$~\mum---must be disentangled from the underlying stellar continuum. Extrapolating the PAH-free NIR SED beyond 3~\mum\ isolates the PAH contribution, which is valuable in two ways: it provides clean PAH emission maps as tracers of hot dust and SFR across galaxies, and it calibrates recipes for correcting $3.6$~\mum\ photometry to the pure stellar component \citep[e.g.,][]{2012ApJ...744...17M}. To this end, the addition of intermediate-width ($W \sim 0.1$~\mum) filters---one centered on the 3.3~\mum\ feature and one or two on either side---would be crucial to cleanly isolate the two emission components.

\subsection{Scientific goals}

Leveraging the synergy with Euclid YJH imaging, GREX-PLUS will provide the community with deep, high-resolution stellar mass maps for a sizeable and representative sample of nearby galaxies. These data will allow a revision of the $z \sim 0$ stellar mass function---one of the most fundamental constraints for galaxy formation and evolution models and a key benchmark for higher-redshift observations. The resulting reference sample will calibrate photometric stellar mass relations applicable to much larger samples, enabling systematic control over parameters such as environment and gas properties. A systematic study of the stellar mass contrast in different galactic structures is paramount for evaluating internal dynamical evolution and for correctly estimating dark matter fractions; the resulting M/L calibrations will reduce degeneracies in dynamical studies of galaxies with detailed kinematic observations. Finally, clean PAH emission maps will enable systematic studies of how hot dust and SFR are distributed across galaxies.

\begin{table}
    \begin{center}
    \caption{Required observational parameters.}\label{tab:firstgals}
    \begin{tabular}{|l|p{9cm}|l|}
    \hline
     & Requirement & Remarks \\
    \hline
    Wavelength & 1.8--5 $\mu$m & {$a$}\\
    \hline
    Spatial resolution & $<1$ arcsec & \\
    \hline
    Wavelength resolution & $\lambda/\Delta \lambda>4$; photometry only & \\
    \hline
    Field of view & 1,000 deg$^2$ & \\
    Sensitivity & 24.0 AB mag ($5\sigma$, point source) & \\
    \hline
    Observing field & Maximize overlap with SDSS-I (Legacy) footprint & $b$ \\
    \hline
    Observing cadence & N/A & \\
    \hline
    \end{tabular}
    \end{center}
    $^a$ Classical broad bands: K, 3.6~\mum, 4.5~\mum, plus intermediate/narrow bands at 3.2, 3.3, and 3.4~\mum\\
    $^b$ Complementary optical spectroscopy for redshift and integrated SP analysis.
\end{table}

\printbibliography[heading=subbibliography]
\end{refsection}

\clearpage

\begin{refsection}[2-11_highzsupernovae/highzsupernovae.bib]

\section{High Redshift Supernovae}
\label{sec:highzsupernovae}

\noindent
\begin{flushright}
Takashi Moriya$^{1}$
\\
$^{1}$ NAOJ 
\end{flushright}
\vspace{0.5cm}

\subsection{Scientific background and motivation}
Massive stars play essential roles in the evolution of the early Universe. A significant fraction of the first stars in the Universe are predicted to be massive stars that explode as supernovae \citep[e.g.,][]{2015MNRAS.448..568H}. Supernovae provide energy and elements to the surrounding media and drive the evolution of the early Universe. Furthermore, massive stars are considered to be a major source of high-energy photons contributing to cosmic reionization. In order to know the exact contribution of massive stars to cosmic reionization, it is important to characterize the properties of massive stars in the early Universe. However, it is difficult to observe individual massive stars in the epoch of reionization ($z\gtrsim6$) even with JWST unless they are lensed (\citealt{2022ApJ...940L...1W}, see also a recent example of the JWST discovery of a lensed supernova at $z=5.13$, \citealt{2026arXiv260104156C}). Fortunately, some supernovae from massive stars such as superluminous supernovae are known to be extremely luminous. These luminous supernovae can be observed even if they appear at $z\gtrsim6$. In addition, massive stars in the early Universe may evolve and explode differently from those in the local Universe. For instance, metallicity is lower in the early Universe and thus mass loss from massive stars is also lower. Lower mass loss keeps the masses of massive stars higher throughout their evolution. A representative example of supernovae that may only exist in the early Universe is pair-instability supernovae, which are theoretical explosions of massive stars that retain core masses of more than 60~M$_\odot$ until the time of their explosion \citep{2002ApJ...567..532H}. Some pair-instability supernovae are predicted to be extremely luminous and can be observed even if they appear at $z\gtrsim6$ \citep{2011ApJ...734..102K}. Another possible class of luminous supernovae that may only exist in the early Universe comprises explosions of massive stars above 10,000~M$_\odot$. Such extremely massive stars are suggested to be a main source of supermassive black holes, but some of them may explode during explosive helium burning and become luminous supernovae that can be observed at very high redshifts \citep{2021MNRAS.503.1206M}.

In order to discover supernovae at $z\gtrsim6$, it is necessary to perform wide-field deep near-infrared transient surveys. Roman will be able to conduct such transient surveys and discover supernovae up to $z\simeq 7$ \citep{2022ApJ...925..211M}. The maximum redshift that Roman can reach is mainly determined by the maximum wavelength that Roman can observe ($\sim 2~\mu\mathrm{m}$). In order to discover supernovae from epochs when cosmic reionization is ongoing ($z\gtrsim7$), it is necessary to conduct deep and wide transient surveys at the longer wavelengths. Therefore, we propose to conduct high-redshift supernova surveys with the near-infrared wide-field camera on GREX-PLUS.

\subsection{Required observations and expected results}
\begin{figure}
    \centering
    \includegraphics[width=0.49\columnwidth]{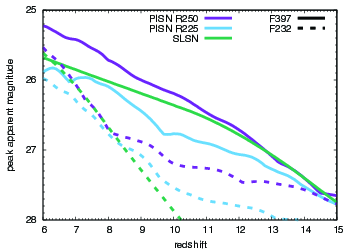}
    \includegraphics[width=0.49\columnwidth]{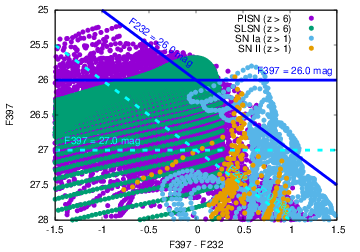}
    \caption{
    \textit{Left:} Expected peak brightness of pair-instability supernovae (PISN) and superluminous supernovae (SLSN) at high redshifts in the F232 (around $2~\mu\mathrm{m}$) and F397 (around $4~\mu\mathrm{m}$) bands. For pair-instability supernovae, we show the two brightest models (R250 and R225). By conducting supernova surveys that reach deeper than 26~AB~mag per epoch, we can discover supernovae that are more distant ($z\gtrsim7$) than those that can be found by Roman. \textit{Right:} Supernova color-magnitude diagram using the F397 and F232 bands. We show pair-instability supernovae at $z>6$, superluminous supernovae at $z>6$, Type~Ia supernovae at $z>1$ \citep{2007ApJ...663.1187H}, and Type~II supernovae at $z>1$ (the Nugent template from https://c3.lbl.gov/nugent/nugent\_templates.html). This shows that high-redshift supernovae can be separated from low-redshift ones through color.
    }
    \label{fig:moriya}
\end{figure}

Here, we investigate the observational strategy to discover pair-instability supernovae and superluminous supernovae at $z\gtrsim7$. First, we estimate their brightness around $2~\mu\mathrm{m}$ (F232 band) and around $4~\mu\mathrm{m}$ (F397 band) using pair-instability supernova models from \citet{2011ApJ...734..102K} and superluminous supernova templates from \citet{2022ApJ...925..211M}. In order to discover supernovae at $z\gtrsim7$, it is necessary to reach at least 26~AB~mag (Fig.~\ref{fig:moriya}). The two-band survey allows us to distinguish high-redshift supernovae and low-redshift supernovae even with a single-epoch observation (Fig.~\ref{fig:moriya}). Such a distinction would be important to efficiently discover rare high-redshift supernovae. It also implies that we need not rely solely on light curves to discover them and high-cadence observations are not necessarily required.

Based on the estimated light-curve properties, we conducted supernova survey simulations to estimate the expected numbers of high-redshift supernova discoveries by GREX-PLUS. We assume two-band surveys with the F232 and F397 bands. We assume a survey field of $10~\mathrm{deg^2}$ with a limiting magnitude of 26.0~AB~mag. We assume that the survey period is 5 or 6 years and the same field is observed repeatedly during the survey period. We count the number of supernovae that become brighter than the assumed limiting magnitude at least once in either band.

The number estimates are based on the cosmic star formation history of \citet{2015ApJ...802L..19R}. For pair-instability supernovae, we assume the Salpeter initial mass function to estimate their fiducial event rate. The stellar mass range is set to $0.1$--500~M$_\odot$ and the pair-instability supernova mass range is $150$--300~M$_\odot$. The superluminous supernova rate is estimated by extrapolating the local event rate ($30~\mathrm{Gpc^{-3}~yr^{-1}}$, \citealt{2013MNRAS.431..912Q}) based on the cosmic star formation history. In the case of pair-instability supernovae, we also adopted the event rate where the initial mass function is assumed to be flat above $100~M_\odot$ to show the possible case of a top-heavy initial mass function. We assumed the cadences of 0.5, 1.0, and 2.0~years. The survey period is set to 5~years for the 0.5 and 1.0~year cadence surveys, and it is set to 6~years for the 2.0~year cadence survey. A longer survey period is preferred to increase the discovery numbers.

Table~\ref{tab:highredshiftsupernovanumber} summarizes our simulation results. The numbers in parentheses are those for the case of the top-heavy initial mass function. In the $10~\mathrm{deg^2}$ survey simulations with a limiting magnitude of 26.0~AB~mag, we expect to discover around 10 pair-instability supernovae and around 2 superluminous supernovae at $z\gtrsim7$. The discovery numbers will increase if massive stars are preferentially formed at high redshifts.

Supernova surveys require repeated observations of the survey field. As previously discussed, the survey cadence does not need to be short in order to discover and identify high-redshift supernovae. Indeed, our survey simulations show that the expected discovery numbers do not change significantly when we assume the survey cadences of 0.5, 1.0, and 2.0~years. However, it is preferred to have one-year cadence so that we can confirm the appearance and disappearance of high-redshift supernova candidates.

To confirm high-redshift supernovae, spectroscopic observations of high-redshift supernova candidates would be ultimately required. The best candidates should be spectroscopically followed up by JWST or TMT, and their redshifts and supernova nature should be confirmed. Even if we fail to get supernova spectra, it is possible to obtain their distances by observing their host galaxies if they are bright enough. The required observational parameters are summarized in Table~\ref{tab:highredshiftsupernovarequirement}.

\subsection{Scientific goals}
By conducting supernova surveys with two bands in the $2$--5~$\mu\mathrm{m}$ wavelength range, we will discover supernovae that are at higher redshifts ($z\gtrsim7$) than those discovered by Roman. For example, in the $10~\mathrm{deg^2}$ deep field, a transient survey with a limiting magnitude of 26~AB~mag per epoch can be considered. The supernova surveys should be continued for the duration of the GREX-PLUS survey period (5~years) with a one-year cadence, i.e., the survey field needs to be observed every year for 5~years. The high-redshift supernovae from such surveys will enable us to reveal the supernova rates at $z\gtrsim7$. The supernova rates would allow us to estimate massive star formation rates at $z\gtrsim7$ and constrain the initial mass function in the early Universe, as well as the contribution of massive stars to cosmic reionization. In addition, we will systematically investigate the properties of supernovae such as pair-instability supernovae that are expected to exist mainly in the early Universe.

\begin{table}
    \begin{center}
    \caption{Expected numbers of high-redshift supernova discoveries by GREX-PLUS. The numbers in parentheses show the results for the top-heavy initial mass function.}
    \label{tab:highredshiftsupernovanumber}
    \begin{tabular}{|c|c|c|c|c|c|c|c|c|c|c|}
    \hline
     \multicolumn{11}{|c|}{$10~\mathrm{deg^2}$, 26~AB~mag per epoch} \\
    \hline
     & \multicolumn{5}{|c|}{Pair-instability supernovae} & \multicolumn{5}{|c|}{Superluminous supernovae} \\
    \hline
    Cadence & $z>6$ & $z>7$ & $z>8$ & $z>9$ & $z>10$ & $z>6$ & $z>7$ & $z>8$ & $z>9$ & $z>10$ \\
    \hline
    0.5~yr & 37 (270) & 12 (93) & 2.0 (14) & 0 (0) & 0 (0) & 6 & 2.7 & 1.4 & 0.6 & 0.3 \\
    \hline
    1.0~yr & 35 (253) & 11 (82) & 1.2 (9) & 0 (0) & 0 (0) & 4.5 & 2.2 & 1.0 & 0.5 & 0.2 \\
    \hline
    2.0~yr & 40 (295) & 12 (89) & 1.1 (8) & 0 (0) & 0 (0) & 3.2 & 1.6 & 0.7 & 0.3 & 0.1 \\
    \hline
    \end{tabular}
    \end{center}
\end{table}

\begin{table}
    \begin{center}
    \caption{Required observational parameters.}
    \label{tab:highredshiftsupernovarequirement}
    \begin{tabular}{|l|p{9cm}|l|}
    \hline
     & Requirement & Remarks \\
    \hline
    Wavelength & 2--5 $\mu$m &  \\
    \hline
    Spatial resolution & $<1$ arcsec & \\
    \hline
    Wavelength resolution & $\lambda/\Delta \lambda \geq 2$ & $a$ \\
    \hline
    Field of view & 10 deg$^2$, $> 26$ AB~mag ($5\sigma$, point source) per epoch & \multirow{2}{*}{}\\
    \cline{1-1}
    Sensitivity &  & \\
    \hline
    Observing field & Fields where deep imaging data at $\lambda<2$ $\mu$m are available. & $b$ \\
    \hline
    Observing cadence & Once per year & $c$ \\
    \hline
    \end{tabular}
    \end{center}
    $^a$ Having two-band information at around $2~\mu\mathrm{m}$ and $5~\mu\mathrm{m}$ helps distinguish high-redshift supernovae from low-redshift ones by color.\\
    $^b$ Host-galaxy information is useful to identify high-redshift supernova candidates. \\
    $^c$ The supernova survey should be continued for 5~years.   
\end{table}

\printbibliography[heading=subbibliography]
\end{refsection}

\clearpage

\begin{refsection}[2-12_cosmicinfraredbackground/cosmicinfraredbackground.bib]

\section{Cosmic Infrared Background}
\label{sec:cosmicinfraredbackground}

\noindent
\begin{flushright}
Shuji Matsuura$^{1}$, Akio K.\ Inoue$^{2}$
\\
$^{1}$ Kwansei Gakuin University \\
$^{2}$ Waseda University
\end{flushright}
\vspace{0.5cm}

\subsection{Scientific background and motivation}

\begin{figure}
    \centering
    \includegraphics[width=0.6\columnwidth]{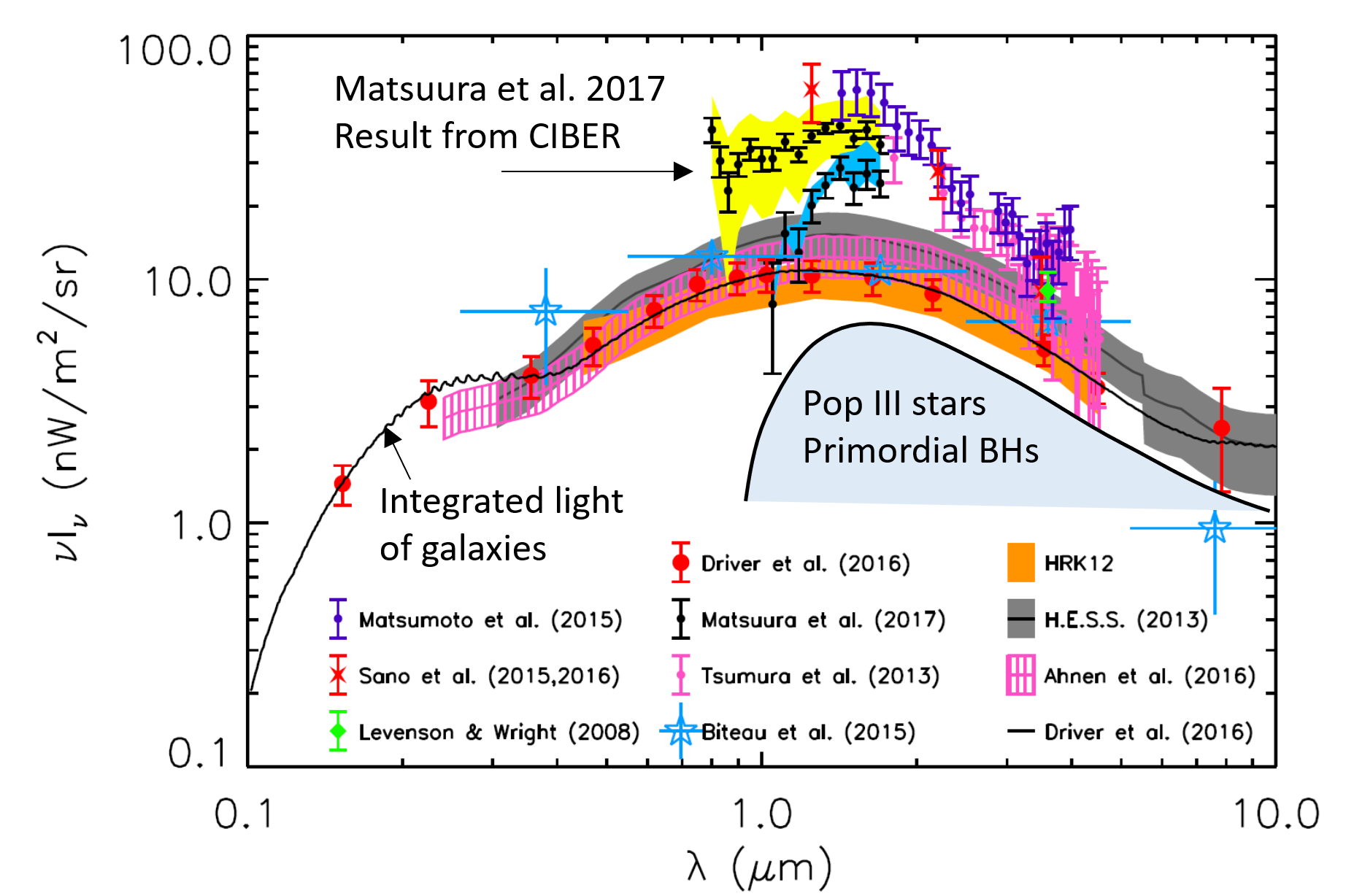}    
    \caption{A summary of background radiation observations. The solid black line and surrounding data show galaxy-integrated light. The gray shaded band is the upper limit to the background radiation from high-energy gamma-ray observations. Data points are the results of direct measurements of the background radiation and are several times higher than the galaxy-integrated light and the upper limit from the gamma-ray observations (adapted from \citealt{2018RvMP...90b5006K}).}
    \label{fig:EBL1}
\end{figure}

The first objects in the Universe (``first stars'' or ``Population III stars'') are expected to have formed around redshift 30.
They are thought to be massive stars with masses ranging from 10 to 1,000 solar masses centered at about 100 solar masses \citep{2014ApJ...781...60H,2015MNRAS.448..568H}.
Some of the first stars will eventually become blackholes. 
There may also be blackholes created by the direct gravitational collapse of primordial density fluctuations \citep{1971MNRAS.152...75H}.
We refer to all of these blackholes formed in the early Universe as primordial blackholes in this section. 
The first stars and primordial blackholes are too faint to detect individually.
However, their integrated light may be detectable as the cosmic infrared background (CIB) radiation.
It is known that the spectra of objects before the epoch of cosmic reionization (around redshift 6) have a break at shorter wavelengths than their Ly$\alpha$ emission line, which is called the Ly$\alpha$ break.
This characteristic spectral break may be imprinted in the CIB spectrum (Fig.~\ref{fig:EBL1}).
The CIB intensity measured by the rocket experiment, Cosmic Infrared Background Experiment (CIBER), is several times higher than the integrated light of galaxies, and understanding its origin is very important \citep{2017ApJ...839....7M}.
However, to measure the intensity of the CIB precisely, it is necessary to subtract the zodiacal light, which is more than 10 times brighter than the CIB, from the data.
Fortunately, the CIB and zodiacal light have different spectra, which are useful for the subtraction, but this is a major source of systematic errors because the zodiacal light dominates the measured signal.
On the other hand, there is an indirect measurement of the CIB from the observation of ultra-high-energy gamma rays.
Because gamma rays are limited in their mean free path by collisions and pair production with CIB photons, the maximum distance that gamma rays can reach the Earth is determined by the intensity of the CIB.
If we can observe gamma rays from distant objects, therefore, we can constrain the intensity of the CIB.
The CIB intensity estimated by this indirect method is consistent with the integrated light of galaxies and not with the direct CIB measurements \citep{2013A&A...550A...4H}.
However, we note that there are uncertainties caused by the assumed spectral shapes of gamma rays and the CIB.
On the other hand, the spatial fluctuation of the CIB has been detected by the analysis of images obtained with the Spitzer Space Telescope, in which individual galaxies are masked \citep{2005Natur.438...45K} and the AKARI satellite \citep{2011ApJ...742..124M}.
This indicates the existence of a CIB component with a different origin from the integrated light of galaxies.
In addition, a spatial correlation of fluctuations in the CIB and X-ray background radiation has been reported \citep{2013ApJ...769...68C}.
This is an interesting hint to consider the gas accretion onto blackholes as the origin of the CIB.
Another interpretation is that the CIB is the accumulation of stellar light from galactic halos \citep{2012Natur.490..514C}.
In summary, the origin of the CIB is still under debate, and solving this question remains an important scientific theme.

CIBER-2 \citep{2024SPIE13092E..0VM,2025ApJS..280...66Z}, a greatly improved version of the CIBER rocket experiment, has six wavelength bands covering optical wavelengths to 2 $\mu$m and tries to discriminate the contributions of primordial sources and galactic halo stars to the background fluctuations based on the spectral shapes.
However, it is important to observe at wavelengths above 2 $\mu$m to study the CIB originating from the first stars and primordial blackholes.
GREX-PLUS is equipped with a wide-field camera that covers the wavelength range from 2--8 $\mu$m to perform a wide-field imaging survey. 
Here, we propose to measure the intensity and fluctuations of the CIB using this camera.

\begin{figure}
    \centering
    \includegraphics[width=0.6\columnwidth]{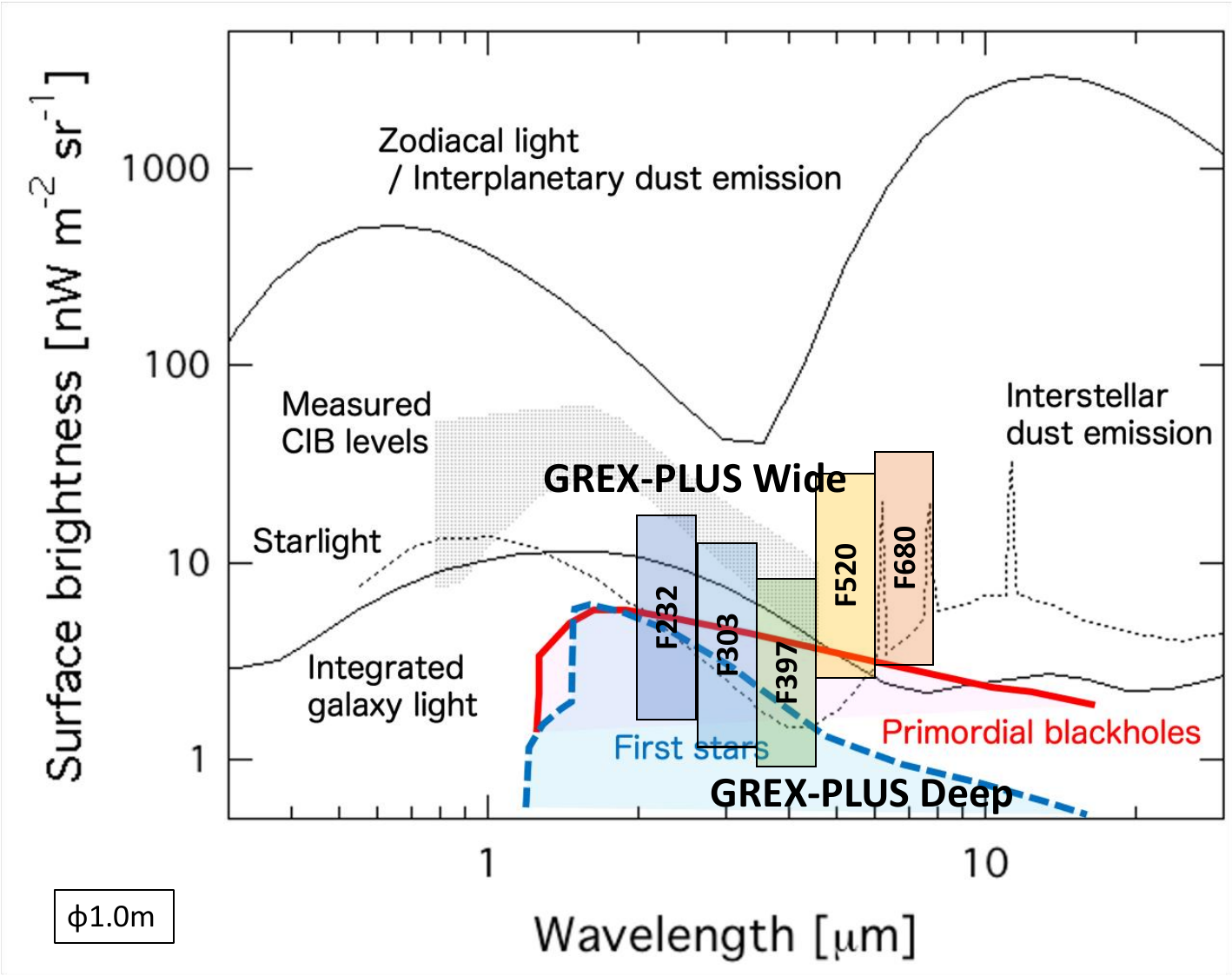}    
    \caption{
    A comparison of contributions from direct observations of the cosmic infrared background (CIB) \citep{2017ApJ...839....7M}, zodiacal light \citep{1998A&AS..127....1L}, Galactic interstellar dust emission \citep{1998A&AS..127....1L}, integrated galaxy light \citep{2017ApJ...839....7M}, and integrated radiation from the first stars and primordial blackholes \citep{2004MNRAS.351L..71C}. The expected sensitivities of the five bands of the GREX-PLUS imaging surveys are shown by the vertical bands. The lower and upper ends of the bands are sensitivities of the Deep and Wide surveys, respectively (see Table~\ref{tab:GPsurveys}). This is the sensitivity for detecting correlations on the smallest angular scale. For large-angle fluctuations and absolute intensity measurements, the sensitivity can be greatly improved by combining many pixels.
    }
    \label{fig:EBL2}
\end{figure}

\subsection{Required observations and expected results}

We aim to detect spatial fluctuations in the CIB beyond a degree-scale, where the effects of zodiacal light, Galactic radiation, and emission from foreground galaxies are considered to be small.
For this purpose, it is necessary to observe a region of several tens of square degrees or more.
It is almost impossible to obtain a contiguous field beyond a degree-scale field with JWST due to its narrow field of view (0.0027 square degrees in NIRCam).
Therefore, the GREX-PLUS super-wide-field surveys are essential for the CIB science.

As shown in Figure~\ref{fig:EBL2}, each component of the CIB has a different spectrum, so multi-color observations are needed to separate and extract them.
The most interesting contributions from the first stars and primordial blackholes to the CIB can be about 1--10 nW m$^{-2}$ sr$^{-1}$ in an optimistic estimate, and sensitivity at this level or better is needed.
We have examined the ability to constrain the CIB intensity with the GREX-PLUS Deep, Medium, and Wide imaging surveys that will perform near- and mid-infrared imaging over areas of more than 10 square degrees.
In Figure~\ref{fig:EBL2}, we show the CIB intensity limits converted from the $5\sigma$ point-source sensitivities for each band, assuming diffraction-limited imaging capability with a 1.0 m aperture telescope.
This is the CIB sensitivity on the smallest angular scale.
For larger angular scales, a significant improvement in sensitivity can be expected by stacking many pixels after point-source removal.
The sensitivity in F397, which is the highest, is sufficiently high to detect the expected intensity from the accumulation of the first stars and primordial blackholes.
Other bands are also sensitive enough to detect each component of the CIB and to separate and extract them.

There is no strict requirement for the choice of observing fields, and we can pursue the CIB science in the GREX-PLUS Deep, Medium, and Wide surveys.
If possible, it is desirable to include the fields previously observed with Spitzer and AKARI to compare the results.
Since the dominant foreground contribution comes from zodiacal light, it is also useful to evaluate its seasonal variation to better handle the zodiacal light subtraction.
Therefore, it is desirable to include monitoring observations near the North Ecliptic Pole (NEP), which can be easily and frequently observed from the satellite orbit assumed for GREX-PLUS.

\subsection{Scientific goals}

The science goal of this theme is to measure the spatial fluctuations of the CIB beyond a degree scale as an accumulation of primordial objects that are not resolved individually. 
The origin of the CIB will be identified by extracting the contribution of primordial objects from the CIB measurements based on the auto-correlation spectrum of the measured fluctuations, the cross-correlation spectrum among different wavelengths, and the spectral energy distribution.
Finally, theoretical models of primordial objects will be constrained based on the multi-wavelength cross-correlation analysis including the X-ray background radiation.

\begin{table}
    \label{tab:CIB}
    \begin{center}
    \caption{Required observational parameters.}
    \begin{tabular}{|l|p{9cm}|l|}
    \hline
     & Requirement & Remarks \\
    \hline
    Wavelength & 2--8 $\mu$m & \\
    \hline
    Spatial resolution & $<1$--2 arcsec & $a$ \\
    \hline
    Wavelength resolution & $\lambda/\Delta \lambda>3$ & \\
    \hline
    Field of view & More than several tens of square degrees, & \multirow{2}{*}{}\\
    \cline{1-1}
    Sensitivity & 1--10 nW m$^{-2}$ sr$^{-1}$ & \\
    \hline
    Observing field & There is no strict requirement, although it is desirable to include the fields observed with Spitzer and AKARI, for example, the North Ecliptic Pole (NEP). & \\
    \hline
    Observing cadence & It is desirable to include monitoring observations near the NEP to examine seasonal variation of zodiacal light. & \\
    \hline
    \end{tabular}
    \end{center}
    $^a$ Needed to avoid the confusion limit in each band.\\
\end{table}

\printbibliography[heading=subbibliography]
\end{refsection}

\clearpage

\begin{refsection}[2-13_highzquasars/highzquasars.bib]

\section{High Redshift Quasars}
\label{sec:highzquasars}

\noindent
\begin{flushright}
Yoshiki Matsuoka$^{1}$
\\
$^{1}$ Ehime University
\end{flushright}
\vspace{0.5cm}

\subsection{Scientific background and motivation}

Through extensive observations and theoretical works carried out over the past 30 years, we have come to realize that supermassive black holes (SMBHs), with masses ranging from $M_{\rm BH} \sim 10^6~{\rm M}_\odot$ to $10^{10}~{\rm M}_\odot$, are ubiquitous in the observable Universe. Almost all galaxies with evolved bulges have an SMBH at their nuclei, at least in the local Universe, and the SMBH mass is found to be roughly 0.1 \% of the bulge mass with a relatively small scatter. Such strong mass correlation may indicate that SMBHs have evolved with the host galaxies under close interactions, via yet unknown processes. Physical drivers behind this ``co-evolution" of galaxies and SMBHs have been a subject of extensive study, from both theoretical and observational perspectives \citep{2013ARA&A..51..511K}. 
This research field is also attracting attention, not only from academia but also from the general public, thanks to recent groundbreaking observations such as the gravitational wave detection from black hole mergers and imaging of SMBH shadows with the Event Horizon Telescope. 

Meanwhile, it is still not understood how such SMBHs are formed in the hierarchical structure formation. The strongest constraints come from observations of quasars in the early Universe -- the most distant quasar reported to date has a redshift of $z=7.64$ and the mass $M_{\rm BH} \sim 10^9~{\rm M}_\odot$. Even if we assume the classical maximum efficiency (the Eddington limit) of mass accretion, the progenitor mass would need to exceed $M_{\rm BH} \sim 10^4~{\rm M}_\odot$, which is difficult to achieve with the standard seeding models via Population III stars \citep[e.g.,][]{2021ApJ...907L...1W}. More exotic formation paths, such as the direct collapse of primordial gas clouds, have been proposed and actively discussed. To disentangle various theoretical scenarios, it is imperative to observe those high-$z$ quasars in greater detail to look for signatures of the preceding formation epochs. At the same time, we need to push to even higher redshifts at $z \geq 8$ and measure both statistical and individual properties, including the luminosity function and clustering of quasars, the stellar and gas contents of the host galaxies, dust obscuration, etc. 
Recent JWST discoveries of faint AGNs at very high redshifts, including a population known as Little Red Dots \citep[e.g.,][]{2024ApJ...963..129M}, have brought significant advances to this field, but they are mostly low-mass populations (if they are powered by SMBHs) and cannot provide as strong constraints as luminous quasars on the seeding mechanisms. A wide-field near-IR survey with GREX-PLUS will provide crucial pieces of information on when, where, and how the first quasars emerged and transitioned to SMBHs known in the low-$z$ Universe.

\subsection{Required observations and expected results}

\begin{figure}
 \centering
 \includegraphics[scale=0.3]{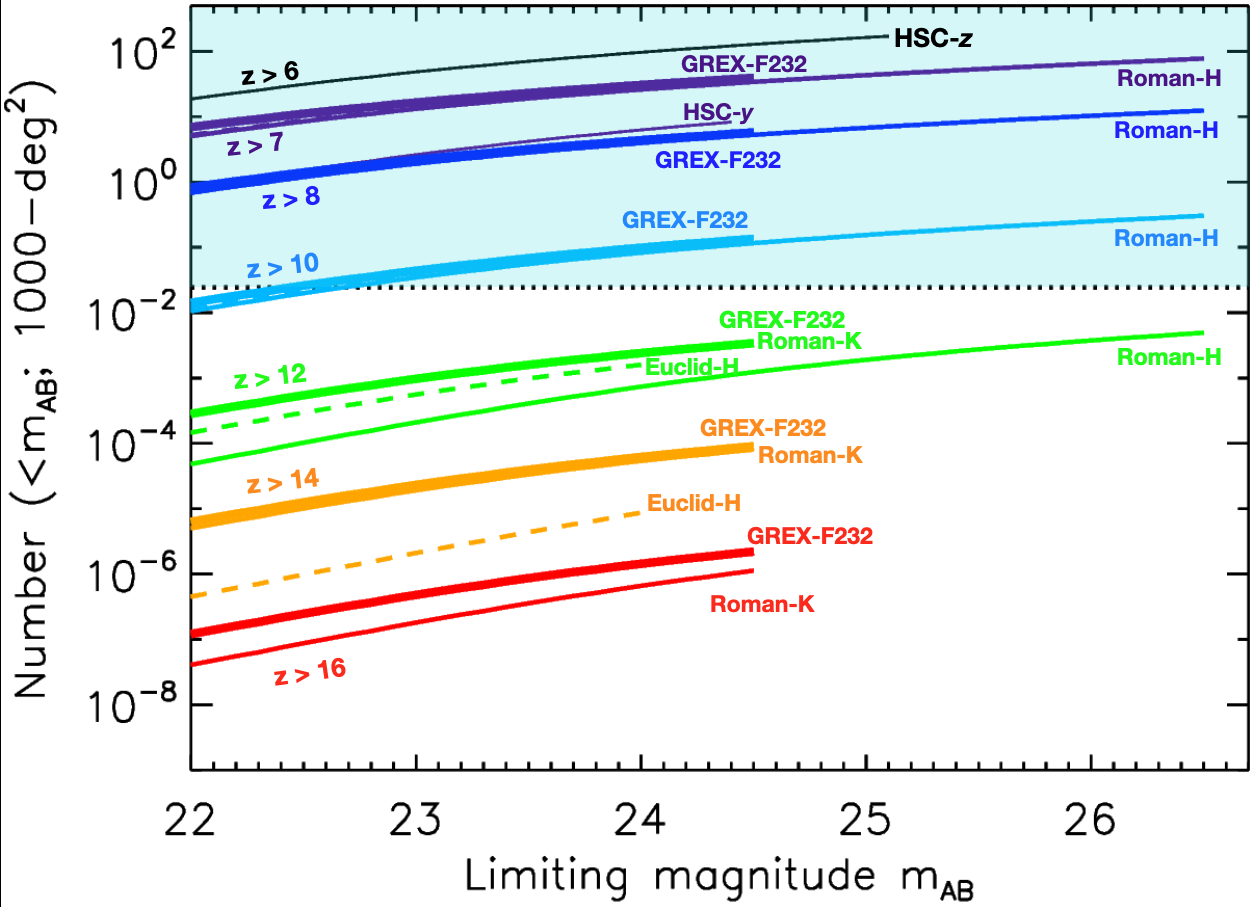}
 \caption{Expected surface density of quasars (per 1000 deg$^2$) for several cases of lower redshift cut, as a function of limiting magnitude. Four missions are considered here, i.e., Hyper Suprime-Cam (HSC) SSP survey (thin lines), {\it Euclid} (dashed lines; overlapping with other lines at $z>7$--10), {\it Roman Space Telescope} (medium lines), and GREX-PLUS (thick lines).
 The light-blue shading represents the surface densities providing $>$1 quasars in the whole sky; in other words, no quasar is expected on average in the lower part of this plot.}
 \label{fig:highzquasars}
\end{figure}

Figure~\ref{fig:highzquasars} presents the expected surface density of quasars (per 1000 deg$^2$) for several cases of lower redshift cut. Here, we assumed the quasar luminosity function measured at $z = 6$ extrapolated to higher redshifts, using the empirical relation of the number density $\phi \propto 10^{kz}$ with $k = -0.7$ \citep{2018ApJ...869..150M}. While this is the best possible estimate we can make at the moment, we caution that there is a very large uncertainty involved in the extrapolation. The present estimates suggest that {\it Euclid} is able to find quasars up to $z \sim 10$ with the planned 15,000 deg$^2$ survey down to $J \sim 24$ AB mag. The {\it Roman Space Telescope} plans to observe 2000 deg$^2$ down to $H \sim 26.5$ AB mag in the High Latitude Survey, which will also allow the detection of quasars up to $z \sim 10$. The additional $K$-band coverage could extend the highest redshift to $z \sim 15$, but the surface density may be too low to expect even a single quasar there. 

GREX-PLUS has an advantage over {\it Euclid} and {\it Roman} at the highest redshifts. However, Figure~\ref{fig:highzquasars} indicates that the situation is similar to the {\it Roman} $K$-band;  the expected number of quasars is only $\le 10^{-5}$ at $<24.5$ AB mag over 1000 deg$^2$, and thus the chance of finding a quasar at the relevant redshifts may be rather small. On the other hand, even a non-detection would provide a valuable piece of information on the evolution of quasar number densities. We have carried out a high-$z$ quasar survey with the imaging data obtained by the HSC SSP project, and discovered $\sim$200 quasars at $z \sim 6 - 7$ over $\sim$1000 deg$^2$, with the spectroscopic completeness nearly 100 \% down to $\sim$24.0 AB mag. Finding no quasar over a similar area at $z \sim 15$ would lead to the conclusion that the number densities evolve as $k < -0.4$ $(\phi \propto 10^{kz})$; in comparison, the measurements to date suggest $k = -0.5$ at $3 < z < 6$ and $k = -0.7$ at $5 < z < 6$ \citep{2016ApJ...833..222J}. For reference, a very flat slope of the evolution, $k = -0.1$ over $6 < z < 15$, would allow the discovery of a few quasars at $z \sim 15$ by probing 1000 deg$^2$ down to 25 AB mag (though we are not aware of any existing models that predict such slow evolution). Measurements of the declining number density toward high redshifts will provide useful constraints on the theoretical models of early quasar evolution, such as the seeding mechanisms of the first SMBHs and their obscured fraction. In the actual survey, it is desirable to target fields with deep images at shorter wavelengths from, e.g., Subaru and/or {\it Roman} observations, and use two GREX-PLUS filters to cover the redshifted Ly$\alpha$ line and continuum emission. We can use a classical two-color diagram to select candidates via dropout selection, and identify their spectroscopic nature with large ground-based or space telescopes.

Finally, let us consider how far we can go to hunt for the first quasars, from a pure sensitivity perspective. The current record holders of quasar redshifts are at $z \sim 7.5$, and have black hole masses $M_{\rm BH} \sim 10^9~{\rm M}_\odot$. Assuming the Eddington-limit accretion, the progenitor masses would be $M_{\rm BH} \sim 10^5$--$10^6~{\rm M}_\odot$ at $z \sim 15$ and $M_{\rm BH} \sim 10^4$--$10^5~{\rm M}_\odot$ at $z \sim 20$ -- within the mass range predicted by seeding scenarios via direct collapse black holes. The corresponding near-IR brightnesses are 29--31 AB mag at $z \sim 15$, assuming no dust extinction. If their positions are fixed by preceding observations, such sources could be interesting targets for very deep exposures with GREX-PLUS.

\subsection{Scientific goals}
With a 1000-deg$^2$-class survey at $2$--4 $\mu$m, we aim to obtain an upper limit on the rate of number density decline from $z \sim 6$ to 15. If the decline is much milder than currently measured at $z \le 6$, then we might be able to find a few quasars at $z \sim 15$. Comparison of the survey results with theoretical models will provide useful constraints on the properties of the earliest quasars, such as the formation of the first SMBHs and nuclear obscuration.

\begin{table}
    \label{tab:highzq}
    \begin{center}
    \caption{Required observational parameters.}
    \begin{tabular}{|l|p{9cm}|l|}
    \hline
     & Requirement & Remarks \\
    \hline
    Wavelength & 2--4 $\mu$m & $a$ \\
    \hline
    Spatial resolution & $<1$ arcsec & \\
    \hline
    Wavelength resolution & $\lambda/\Delta \lambda>3$ & $b$ \\
    \hline
    Field of view & $>$1000 deg$^2$& \\
    \hline
    Sensitivity & 25 AB mag ($5\sigma$, point source) & \\
    \hline
    Observing field & Fields where deep imaging data at $\lambda<2\,\mu$m are available. & $c$ \\
    \hline
    Observing cadence & N/A & \\
    \hline
    \end{tabular}
    \end{center}
    $^a$ We need to capture redshifted Ly$\alpha$ (2 $\mu$m at $z=15$) and continuum emission at longer wavelengths.\\
    $^b$ At least three bands are required for the color selection (can be reduced to two bands if the bluest band comes from previous [e.g., {\it Roman}] observations).\\
    $^c$ For example, deep fields observed with Subaru and/or {\it Roman}.
\end{table}

\printbibliography[heading=subbibliography]
\end{refsection}

\clearpage

\begin{refsection}[2-14_high-z-LRD/high-z-LRD.bib]

\section{Little Red Dots}
\label{sec:first_LRDs}

\noindent
\begin{flushright}
Takumi Tanaka$^{1}$, 
Yuichi Harikane$^{2}$, 
Kazuhiro Shimasaku$^{3}$, 
Roberta Tripodi$^{4}$,
Ivan Delvecchio$^{5}$
\\
$^{1}$ Kavli IPMU (WPI), UTIAS, The University of Tokyo,
$^{2}$ ICRR, The University of Tokyo\\
$^{3}$ The University of Tokyo
$^{4}$ INAF - Astronomical Observatory of Rome, Rome, Italy\\
$^{5}$ INAF - Astrophysics and Space Science Observatory, Bologna, Italy
\end{flushright}
\vspace{0.5cm}

\subsection{Scientific background and motivation}

It is widely accepted that supermassive black holes (SMBHs) with masses of $M_{\rm BH}\gtrsim10^6M_\odot$ reside at the centers of almost all galaxies in the local Universe. SMBHs are believed to play a crucial role in shaping the evolution of galaxies \citep[e.g.,][]{KormendyHo2013}, including our own Milky Way, and they also serve as unique laboratories for testing general relativity under strong gravitational fields, as well as key targets for next-generation gravitational-wave observations (also see Section\,\ref{sec:smbh_coa}).
Nevertheless, the origin of SMBHs---how seed BHs formed and grew to SMBHs---remains one of the most profound open questions in astrophysics \citep[e.g.,][]{Volonteri2010}. To investigate the origin of SMBHs, it is essential to identify and track the first generation of BHs in the early Universe. At high redshift, the key tracers for studying SMBHs are active galactic nuclei (AGN), which emit the energy released as mass accretes onto SMBHs.

Thanks to its unprecedented infrared sensitivity, since 2021, JWST has revealed that AGN are roughly an order of magnitude more abundant at high redshift ($z\sim6$) than previously thought \citep[e.g.,][]{Harikane2023_AGN, Kocevski2025_LRD}. Among these AGN, JWST has uncovered a new striking population dubbed \textit{\textbf{``Little Red Dots''}} (LRDs, e.g., \citealt{Matthee2024, Akins2025, Kocevski2025_LRD}), which feature extremely compact sizes ($\lesssim100\,{\rm pc}$), red optical and blue ultraviolet emission, resulting in a characteristic ``V-shaped'' spectral energy distribution (SED) with a turnover around the Balmer limit (rest-frame 3646\AA). Spectroscopic observations have found broad Balmer emission lines for nearly all LRDs \citep[e.g.,][]{Greene2024}, implying SMBH masses of $M_{\rm BH} \sim 10^{6\,\mathchar`-\,8}M_\odot$ assuming local scaling relations calibrated for type-I AGN.
Intriguingly, LRDs are completely elusive from X-ray, radio, and (sub)-mm observations \citep{Ananna2024, Akins2025, Casey2025}, despite their red colors.
In addition, some LRDs show extremely strong Balmer breaks that cannot be explained by stellar population alone \citep{deGraaff2025_Cliff, Naidu2025_BHstar}, and detailed observations of LRDs identified at lower redshifts revealed that the rest-optical components exhibit blackbody-like SEDs \citep{Lin2025_local}.
This has led to the hypothesis that LRDs are SMBHs embedded in dense gas envelopes \citep{Inayoshi2025_uv_opt, InayoshiMaiolino2025, Naidu2025_BHstar, Liu2025_break, Kido2025} representing the first growth phase just after the formation of seed BHs \citep{Inayoshi2025_uv_opt, Pacucci2025_first}, possibly with super-Eddington accretion \citep{Liu2025_break, Pacucci2024_superEdd, Inayoshi2025_superEdd}. These results suggest that LRDs are in a transition phase from theoretically predicted seed BHs into the well-studied non-LRD SMBHs, providing the long-awaited observational opportunity to study seed BHs.

\subsection{Required observations and expected results}
GREX-PLUS will uniquely provide valuable NIR-to-MIR wide-field deep imaging data in the 2030s.
In this section, we propose three science cases and observing strategies for LRDs that can be achieved only using GREX-PLUS.

\subsubsection{$\bullet$ Science Case 1: $z\gtrsim10$ LRD Exploration}
Uncovering seed BHs using LRDs requires us to \textbf{push the search for LRDs to even higher redshifts ($\bm{z \gtrsim 10}$), ultimately identifying the \textit{first LRDs}---that is, the first SMBHs in the Universe}.
Photometric selection of LRDs relies on their compactness and characteristic V-shaped SED. At $z\gtrsim10$, the Balmer break shifts to $\lambda_{\rm obs}\gtrsim4\,{\rm \mu m}$; thus, JWST/NIRCam, which can cover up to $\lambda_{\rm obs}\sim5\,{\rm \mu m}$, cannot distinguish between LRDs and non-LRD galaxies with strong emission lines or a strong Balmer break. As a result, it becomes challenging to robustly select LRDs at $z\gtrsim10$ with NIRCam photometry alone. To overcome this limitation, \citet{Tanaka2025_z10} developed a new color-selection technique that combines NIRCam and JWST/MIRI photometry (F770W), enabling the first robust identification of an LRD candidate. This discovery suggests that the evolution of LRDs' number density follows a log-normal distribution peaking at $z\sim5\,\mathchar`-\,6$ \citep{Inayoshi2025_num} extending up to $z\sim10$, and declining sharply toward lower redshifts \citep[e.g.,][]{Kocevski2025_LRD, Inayoshi2025_num, Ma2025_lowz}. Furthermore, they reported that the fraction of LRDs among the overall galaxy population increases toward higher redshifts. This trend highlights the significance of LRDs not only for understanding SMBH formation, but also for probing the early Universe.

\begin{figure}[!h]
    \centering
    \includegraphics[width=14cm]{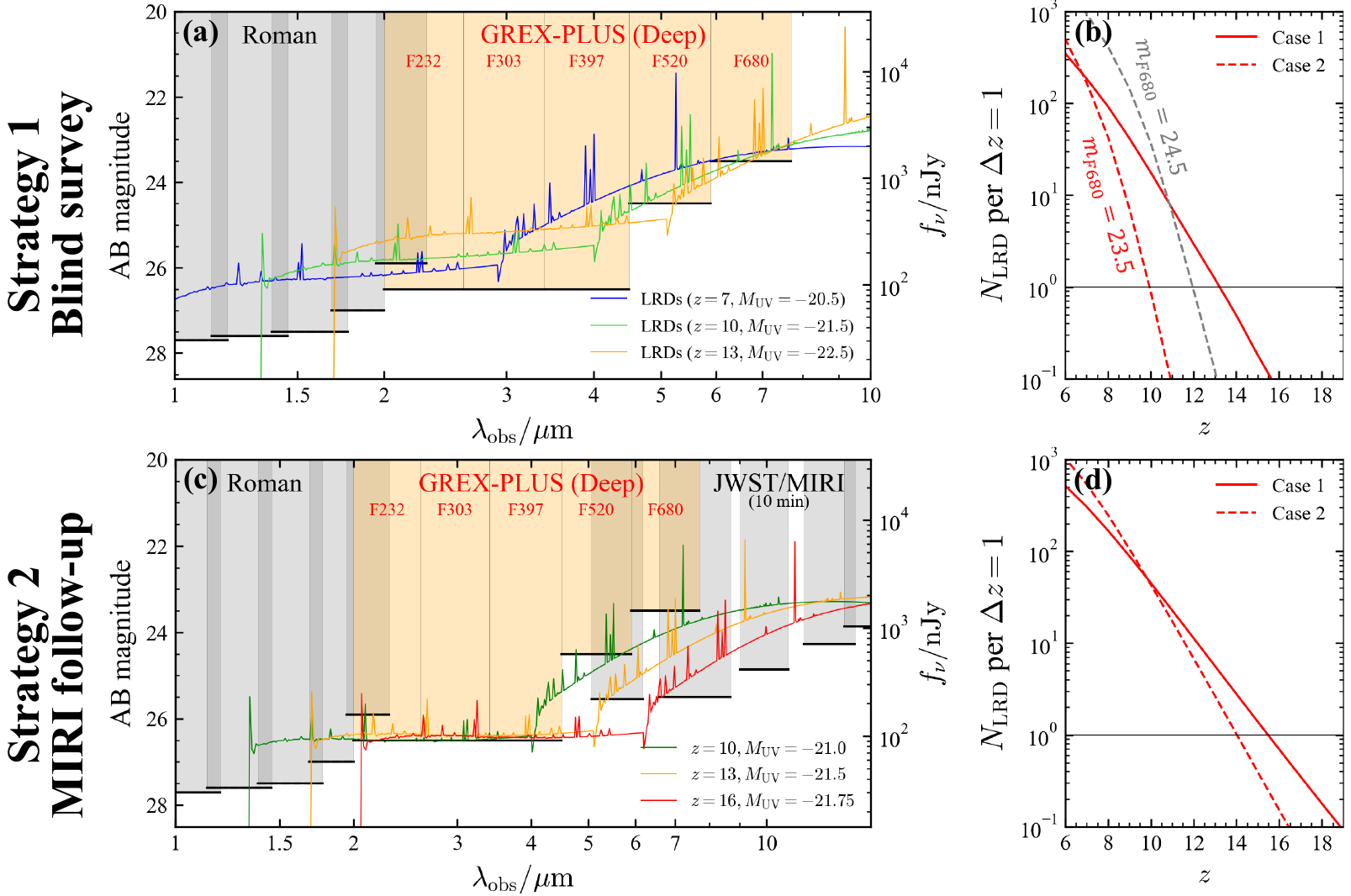}
    \caption{
    (a and c) SEDs for the detectable LRDs at each redshift with Strategies\,1 and 2, respectively.
    The survey depths of Roman (Deep tier of the High Latitude Wide Area Survey) and GREX-PLUS (Deep tier) are shown in gray and orange rectangles.
    In panel\,(c), we also plot the depth of JWST/MIRI 10-min-exposure imaging with gray rectangles.
    (b and d) The expected number of LRDs per $\Delta z = 1$ bin identified with Strategies\,1 and 2, respectively.
    Solid and dashed red lines indicate the expectation under Case\,1 and Case\,2 LFs, respectively.
    In panel\,(b), a gray dashed line shows the expectation under Case\,2 LF assuming a deeper F680 detection limit of $m_{\rm F680}=24.5$, rather than the current survey plan of $m_{\rm F680}=23.5$.
    }
    \label{fig:hzlrd_strategies}
\end{figure}

Nevertheless, the current conclusion is based on a single object, making the statistical robustness limited. Therefore, it is essential to enlarge the sample size and extend the search to an unexplored parameter space, including higher redshifts and a broad luminosity range. 
However, constructing a much larger $z>10$ LRD sample ($N>10$) is not expected with existing/planned NIRCam and MIRI survey data.
Thus, we need a joint near-IR + mid-IR survey in a parameter space complementary to that of JWST. GREX-PLUS can uniquely bridge this gap, providing the large survey volume and wavelength coverage necessary to select high-redshift LRDs.

To assess the feasibility of LRD searches at high redshift, we employ the LRD SED from \citet{Inayoshi2025_uv_opt} and explore two cases for the LRD UV LF: (i) {\bf Case 1}: A single power-law LF across the full UV luminosity range matching the observed LRD's number density at $z\sim7$ \citep{Kocevski2025_LRD}. (ii) {\bf Case 2}: A double power-law LF at $z\sim5$ in \citet{Ma2025_LFcut} with a cutoff on the luminous end. Based on the assumed SED template, we convert the 5100 \AA\ LF to a UV LF by assuming $M_{\mathrm{UV}} = M_{5100} + 1.9$. In both cases, we rescale the UV LF at each redshift based on the log-normal evolution \citep{Inayoshi2025_num}.
Building on these assumptions, we propose the following two complementary strategies to search for high-redshift LRDs with GREX-PLUS.

\begin{figure}[t]
    \centering
    \includegraphics[width=\linewidth]{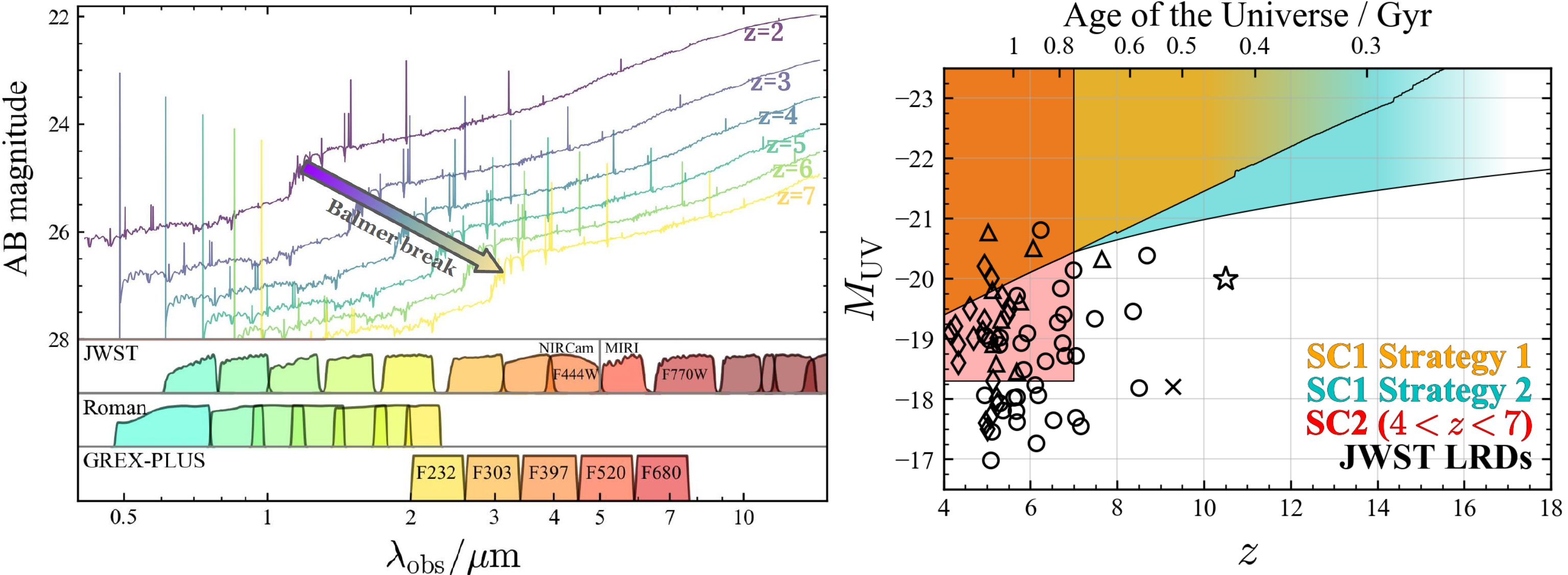}
    \caption{
    (Left): median SEDs of LRDs at $z=6.2$, $k$-corrected between $z=2$ and $z=7$. Filter transmission curves are shown at the bottom for JWST/NIRCam and MIRI, Roman/WFI, and GREX-PLUS Wide-Field Camera. 
    (Right): $M_{\rm UV}\,\mathchar`-\,z$ parameter space for LRDs. Orange and cyan shaded regions indicate the GREX-PLUS parameter space with Strategies\,1 and 2 in Science Case 1 ($z\gtrsim10$), respectively. Red shaded region indicates the parameter space for Science Case 2 ($4<z<7$). Black symbols mark some individual LRDs previously identified by JWST \citep{Kocevski2025_LRD, Lin2025_C3D, Greene2024, Matthee2024, Taylor2025_z9, Tanaka2025_z10}. GREX-PLUS will complement JWST in its ability to identify more luminous and distant LRDs at $z\gtrsim10$, as well as in probing more typical-luminosity LRDs at $4<z<7$ (down to $M_{\rm UV}\approx-18$). }
    \label{fig:lrd_sed_muv_z}
\end{figure}

\noindent \textbf{Strategy 1. Blind Survey:}
GREX-PLUS/F520 and F680 can trace the rest-optical component of the LRD SED at $z\lesssim13$ (Fig.\,\ref{fig:hzlrd_strategies}\,a); thus, we can blindly select LRDs at $z\lesssim13$ by combining Roman (Deep tier of the High Latitude Wide Area Survey, $\sim19\,{\rm deg^2}$) and GREX-PLUS (Deep tier, $\sim10\,{\rm deg^2}$) photometry.
As illustrated in Fig.\,\ref{fig:hzlrd_strategies}\,(a), F560 and F680 are the bottlenecks in this strategy, setting the detection limit for high-redshift LRDs.
Under Case\,1 LF, we can expect to find one LRD at $z\sim13\,\mathchar`-\,14$ and $\sim30$ LRDs at $z>10$, while Case\,2 LF predicts one LRD at $z\sim10\,\mathchar`-\,11$
The limiting redshifts under Case\,2 LF are comparable to the current highest redshift record found by JWST \citep{Tanaka2025_z10}.
However, note that GREX-PLUS excels in covering the more luminous side (Fig.\,\ref{fig:lrd_sed_muv_z}), complementary to JWST, and thus improves LF constraints. Also note that, under Case\,2 LF, deepening F680 by 1 magnitude with $\sim 6$ times longer exposure time (gray dashed line in Fig.\,\ref{fig:hzlrd_strategies}\,b) can increase the expected number of high-redshift LRDs by $\sim 50$ times.
Therefore, deepening F680 improves the high-redshift LRD search more effectively than simply expanding the survey area.

\noindent \textbf{Strategy 2. MIR Follow-Up Strategy:}
At $z \gtrsim13$, the Balmer break shifts to $\lambda_{\rm obs}\gtrsim6\,{\rm \mu m}$, making it impossible to distinguish LRDs from normal galaxies using GREX-PLUS alone. In this second strategy, we do not require detection in F520 and F680. Instead, we select high-$z$ candidates based on dropout selection from GREX-PLUS + Roman, following them up with MIRI imaging (Fig.\,\ref{fig:hzlrd_strategies}\,c).
Because GREX-PLUS preferentially detects brighter objects than JWST deep fields, the MIRI follow-up requires exposures of less than 10 min if they are LRDs.
Given the narrow field of view of MIRI imaging ($2.3\,{\rm arcmin^2}$), pre-selection from the GREX-PLUS Deep Survey ($10\,{\rm deg^2}$) yields a highly efficient identification strategy for $z \gtrsim 13$ LRDs.
With this strategy, we expect to detect LRDs up to $z\sim16\,\mathchar`-\,17$ and $z\sim14\,\mathchar`-\,15$ under Case\,1 and Case\,2 LFs, respectively (Fig.\,\ref{fig:hzlrd_strategies}\,d).
In addition, this strategy can probe fainter LRDs than Strategy\,1 even at $z\gtrsim7$ (Fig.\,\ref{fig:lrd_sed_muv_z}).

Objects identified through these two strategies remain photometric LRD candidates.
The main contamination source would be PAH-emitting galaxies at $z < 1$, where strong PAH features (e.g., $\lambda_{\rm rest}=6.2\,{\rm \mu m}$) enter F680 or MIRI filters and mimic the LRD rest-optical slope \citep{Tanaka2025_z10}.
Using medium-band filters to target their H$\alpha$, we can improve the reliability of photometric redshift estimates and LRD classification.
However, definitive confirmation of their nature still requires spectroscopic follow-up.
For LRDs as bright as $M_{\rm UV}\sim 21.75$ at $z\sim15$ (identified from Strategy\,2), NIRSpec/PRISM can detect the continuum around $2\,{\rm \mu m}$, just longer than the Lyman $\alpha$ break, at $3\sigma$ per spectral bin ($R\sim100$) with $\sim$30-min exposure. Furthermore, a MIRI/LRS $30$-min exposure can detect the H$\alpha$ at $5\sigma$.

\subsubsection{$\bullet$ Science Case 2: Statistical Sample of Typical-Luminosity LRDs at $4<z<7$}
While Science Case 1 aims to push the redshift frontier of LRD studies, it can only target the most luminous LRDs.
The identification of more ``typical''-luminosity LRDs is complementary to this approach and critically relies on the synergy between the Roman Deep survey and the GREX-PLUS Deep survey ($10\,\text{deg}^2$, 5$\sigma$ limit $m_{\rm AB}=26.5$ at $4\,{\rm \mu m}$). Fig.~\ref{fig:lrd_sed_muv_z} (left) displays the median SED obtained from NIRCam \& MIRI stacking of $>300$ LRDs over multiple JWST Legacy fields \citep{Akins2025, Delvecchio2025}. The same intrinsic SED is then $k$-corrected to various redshifts (in the range $2<z<7$). The respective Roman and GREX-PLUS 5$\sigma$ limits (see Fig.~\ref{fig:hzlrd_strategies}, a) will enable us to detect the median LRD SED up to $z\sim7$. This implies that a deep Roman \& GREX-PLUS survey is representative of the average NIRCam-selected LRD population. We also note that the redshift interval $4<z<7$ is a ``sweet spot'' that strongly purifies the photometric selection against (i) extended star-forming galaxies, which dominate at $z<4$; and (ii) emission-line contaminants, which are prevalent at $z>7$ since bright Balmer lines (e.g. ${\rm H \alpha}$) can boost the $4\,{\rm \mu m}$ fluxes mimicking the red LRD SED. Specifically, at $z>4$, Roman's angular resolution (PSF FWHM $\approx 0.\!\!^{\prime\prime}1$) efficiently removes extended sources \cite{Kocevski2025_LRD}, while at $z<7$ the emission-line contaminants drop to $<10\%$. In this respect, medium-band filters would further purify the LRD selection, thereby enabling a full characterization of LRDs in representative samples. A joint Roman \& GREX-PLUS deep survey over $10\,{\rm deg^2}$ would deliver $>20\times$ more LRD candidates at $4<z<7$ than currently available with NIRCam, and down to $M_{\rm UV}\approx-18.3$ (see Strategy 3 in Fig.~\ref{fig:lrd_sed_muv_z}, right). We estimate that such a novel LRD selection will be $>$50\% complete down to $M_{\rm UV}\approx-19$ (5,000 LRDs) and $>75\%$ complete down to $M_{\rm UV}\approx-19.5$ (3,000 LRDs).

\subsubsection{$\bullet$ Science Case 3: LRDs' Environments}

Characterizing the environments of LRDs provides a crucial test of their physical nature and evolutionary role. There is growing consensus that LRDs represent a specific phase in the evolution of SMBH host galaxies. In this framework, if LRDs are closely related to—or progenitors of—high-redshift quasars, they should preferentially reside in similarly overdense regions of the early Universe. Indeed, \citet{schindler2025} reported that an LRD at $z \sim 7$ is embedded within a massive overdensity comprising eight nearby galaxies. The inferred cross-correlation length is comparable to that measured for $z \sim 6$ quasars, lending support to a common underlying population. In contrast, \citet{escudero2025} found that several other LRDs inhabit environments that are no denser than those of typical high-redshift galaxies. This apparent discrepancy likely reflects current observational limitations, most notably the small sky coverage of JWST surveys, which are optimized for sensitivity rather than wide-area mapping. As a result, environmental studies of LRDs remain subject to significant cosmic variance.

In this context, leveraging the previous science cases, the GREX-PLUS survey will provide the wide-field coverage necessary to robustly characterize the large-scale environments of LRDs and systematically assess the presence and diversity of associated overdensities.
By combining Roman and GREX-PLUS, we will be able to probe not only LRDs themselves but also the surrounding high-redshift galaxies (see also Sections\,\ref{sec:lssandmassassembly}, \ref{sec:veryhighz}, \ref{sec:submmgalaxies}), enabling a simultaneous investigation of LRDs' environments by constraining the LRD-galaxy cross-correlation function. In this science case, as well, the introduction of medium-band filters is highly beneficial, enabling more robust photometric redshift estimation.

\subsection{Scientific goals}
By combining the GREX-PLUS Deep-tier imaging survey data with wide-area near-IR surveys such as Roman, we can blindly search for luminous LRDs at $z\lesssim13$ (Science Case 1, Strategy 1).
Furthermore, by following up high-redshift dropout candidates identified from the Roman + GREX-PLUS data with JWST/MIRI imaging, we can extend the search for luminous LRDs to $z\lesssim16$ (Science Case 1, Strategy 2).
Ultimately, the synergy between Roman and GREX-PLUS deep imaging will push the number of LRD detections by 20$\times$ even for the typical-luminosity LRD population at $4<z<7$ (Science Case 2). Increasing the sample size will enable a more statistically robust discussion of the environmental properties (Science Case 3).
These LRD searches with GREX-PLUS are highly complementary to JWST, enabling the exploration of previously uncharted regions of parameter space in the luminosity--redshift plane (Fig.\,\ref{fig:lrd_sed_muv_z}\,Right).
This will finally bridge the luminosity gap between the LRD and the quasar population at cosmic dawn, providing unprecedented constraints on the luminosity density of the global AGN population at $z>3$, and provide critical insights into how the first SMBHs formed and evolved in the early Universe ($z\gtrsim10$).
This immense discovery space will come entirely from a deep survey tier with Roman and GREX-PLUS. Subsequent spectroscopic follow-up will ideally require JWST, emphasizing the need for GREX-PLUS Deep survey during JWST's lifetime (during 2030s?).


\begin{table}
    \begin{center}
    \caption{Required observational parameters.}\label{tab:firstgals}
    \begin{tabular}{|l|p{9cm}|l|}
    \hline
     & Requirement & Remarks \\
    \hline
    \multirow{2}{*}{Wavelength} & 2--8\,$\mu$m (Strategy\,1) & \multirow{3}{*}{$a$} \\
    & 2--5\,$\mu$m (Strategy\,2) &\\
    \cline{1-2}
    Spatial resolution & $<1$ arcsec & \\
    \hline
    Wavelength resolution & $\lambda/\Delta \lambda>3$ & $b$ \\
    \hline
    Field of view & 10\,deg$^2$, 26.5 and 23.5 AB mag for 2--5 and 5--8\,${\rm \mu m}$, respectively (Strategy\,1) & \multirow{2}{*}{$c$}\\
    \cline{1-1}
    Sensitivity & 10\,deg$^2$, 26.5 AB mag for 2--5\,${\rm \mu m}$ (Strategy\,2) & \\
    \hline
    Observing field & Fields where deep imaging data at $\lambda_{\rm obs}<2\,{\rm \mu m}$ are available. & $d$ \\
    \hline
    Observing cadence & N/A & \\
    \hline
    \end{tabular}
    \end{center}
    $^a$ Primary mirror $\phi>1.2$ m is required to achieve $<1$ arcsec at $\lambda=5\,\mu$m for the diffraction limit.\\
    $^b$ This spectral resolution is required for the color selection of high-redshift LRDs.\\
    $^c$ Sensitivities are $5\sigma$ limiting magnitudes for point sources. In Strategy\,1, improving the depth at $\lambda_{\rm obs}=5$--$8\,{\rm \mu m}$ can efficiently improve the blind search for $z\gtrsim10$ LRDs compared to simply increasing the survey area with the same depth (Fig.\,\ref{fig:hzlrd_strategies}\,b).\\
    $^d$ For example, the Deep tier of the Roman High Latitude Wide Area Survey.
\end{table}

\printbibliography[heading=subbibliography]
\end{refsection}

\clearpage

\begin{refsection}[2-15_dustyagns/dustyagns.bib]

\section{Dust-Obscured Active Galactic Nuclei}
\label{sec:dustyagns}

\noindent
\begin{flushright}
Yoshiki Toba$^{1}$
\\
$^{1}$ Ritsumeikan University 
\end{flushright}
\vspace{0.5cm}

\subsection{Scientific background and motivation}
Dust-obscured active galactic nuclei (AGN) are thought to correspond to the maximum growth phase of both galaxies and supermassive black holes (SMBHs) in the context of their coevolution.
This makes them one of the ideal laboratories to investigate the physical mechanisms behind coevolution and AGN feedback \citep[e.g.,][]{2010MNRAS.407.1701N,2015PASJ...67...86T,2018MNRAS.478.3056B,2022ApJ...936..118Y,}.
Infrared (IR) satellites such as AKARI, Spitzer, and WISE have improved our understanding of dust-obscured AGN over a wide range of redshifts \citep[e.g.,][]{2018ARA&A..56..625H}. 
In particular, optically ``dark'' IR galaxies (that are faint or invisible at optical and/or near-IR wavelengths but bright in the IR and/or submillimeter) have recently been reported in several deep fields \citep[e.g.,][]{2019Natur.572..211W,2020A&A...640L...8U,2020ApJ...899...35T,2021Natur.597..489F,2022ApJ...926..155S}, and the advent of the JWST \citep{2006SSRv..123..485G} is increasing the number of discoveries of these objects \citep[e.g.,][]{2023MNRAS.522..449B,2023ApJ...946L..16P}.
Those optically dark IR galaxies are located at $1 < z < 7$ and are undetectable even by deep optical imaging surveys such as HST and Subaru HSC, meaning that they have been missed by previous optical surveys.
How many optically dark IR galaxies are there in the Universe?
How much do they contribute to the cosmic star formation and supermassive-black-hole mass accretion history?
Estimating the statistical properties (e.g., luminosity function and auto-correlation function) and physical properties (e.g., black hole mass accretion rates and star formation rates) of optically dark IR galaxies will provide an avenue for answering these questions.


\subsection{Required observations and expected results}
\begin{figure}
\centering
\includegraphics[width=0.8\textwidth]{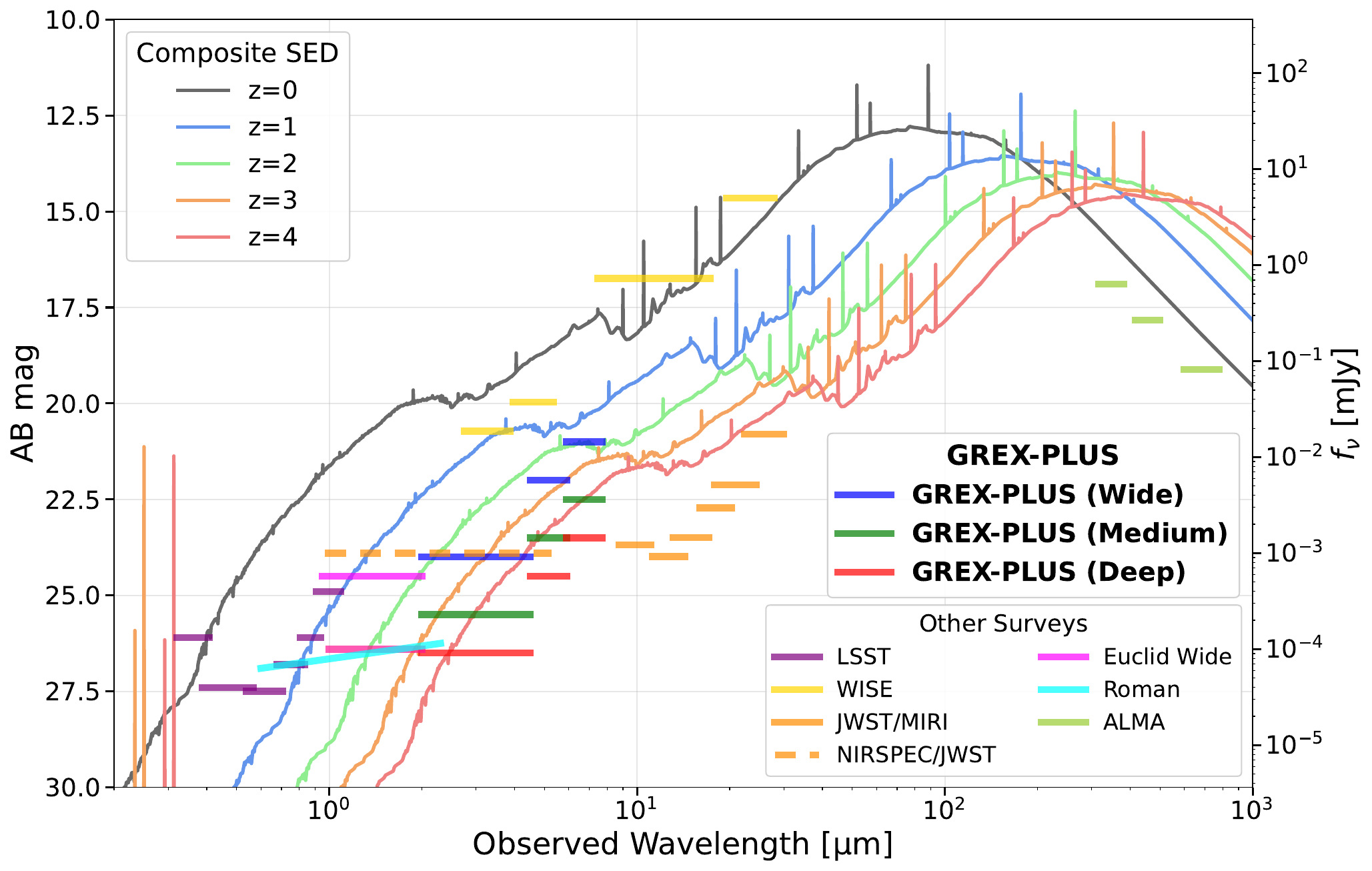}
\caption{Typical SEDs of optically dark IR galaxies found in the AKARI NEP \citep{2020ApJ...899...35T} as a function of redshift up to $z=4$. The detection limits of ongoing and forthcoming missions are overplotted. For MIRI, we assume a 10$\sigma$ limiting magnitude based on a 10 ksec observation.}
\label{SED}
\end{figure}

By the time GREX-PLUS becomes available, the combination of Euclid \citep[e.g.,][]{2022A&A...662A.112E} and Roman \citep[e.g.,][]{2019arXiv190205569A} with HSC and Rubin is expected to provide insight into the properties of optically dark IR galaxies, especially in relatively nearby ($z < 2$) sources. 
Hence, we propose a systematic search for optically dark IR galaxies at $z > 3$ by taking advantage of GREX-PLUS, a wide-area survey (100--1000 deg$^2$) with unique mid-IR bands (3.5--8 $\mu$m), which cannot be realized by Euclid and Roman.
According to a typical spectral energy distribution (SED) of optically dark IR galaxies found in the AKARI North Ecliptic Pole (NEP) survey \citep{2020ApJ...899...35T}, we can expect that a few tens of objects at $z > 3$ will be discovered, especially by the GREX-PLUS Medium survey with 100 deg$^2$ (Figure~\ref{SED}).
Those high-$z$ objects will not be able to be discovered even by Euclid and Roman due to their extremely large extinction.
We estimated the expected number of optically dark IR galaxies including such ``Roman-dropout'' galaxies to be discovered with GREX-PLUS based on their surface density and photometric redshifts reported in \cite{2020ApJ...899...35T}, as summarized in Table~\ref{tab:num}.
In particular, given that the number of optically dark galaxies with $z > 4$ has been limited to a few dozen so far, it is possible that only GREX-PLUS could construct a statistically robust sample.

\begin{table}[htb]
    \begin{center}
    \caption{Expected number of optically dark IR galaxies discovered by GREX-PLUS.}
    \label{tab:num}
    \begin{tabular}{ccrr}
    \hline\hline
     \multirow{2}{*}{Redshift} & \multirow{2}{*}{Expected surface density [deg$^{-2}$]} & \multicolumn{2}{c}{Expected number} \\
     \cline{3-4}
        &   &   Medium [100 deg$^2$]    &   Wide [1,000 deg$^2$] \\
        \hline
        $1 < z < 2$ &   5.6 &   560    &   5600    \\
        $2 < z < 3$ &   1.5 &    150    &    1500    \\
        $3 < z < 4$ &   2.1 &    210    &   ---      \\
        $z > 4$     &   0.2 &     20    &   ---        \\     
    \hline
    \end{tabular}
    \end{center} 
\end{table}
What are the properties of these ``Roman-dropout'' IR galaxies? 
We performed a physical parameter search from a SED point of view to satisfy the criteria for Roman dropouts using CIGALE \citep[Code Investigating GALaxy Emission;][]{2019A&A...622A.103B}, which is also a useful SED-template-generation tool.
We found that most of the objects satisfying the criteria for Roman dropouts have $3 < z < 5$, $A_{\rm V}$ (ISM) $>10$--100, and $L_{\rm IR}$(AGN)/$L_{\rm IR} > 0.8$, indicating that they may be very dusty AGN. 
This is observationally verifiable through follow-up observations with JWST (if it is still operating in the 2030s) and ALMA.
We will address the aforementioned question by calculating various physical quantities such as spectroscopic redshifts, star formation rates, and black hole masses.

Radio detections of optically dark IR galaxies will be examined to measure radio loudness and to search for radio-loud AGN harbored at the centers of optically dark IR galaxies. High radio loudness itself is strong evidence for an AGN. A high-$z$ radio-loud optically dark IR source was recently discovered \citep{2022MNRAS.512.4248E} and implies an abundance beyond expectations from the optically bright quasar population. Radio galaxies at $z>1$ show a dusty nature compared to the local radio galaxies \citep{2019ApJS..243...15T, 2020AJ....160...60Y}. We will explore the radio loudness of optically dark IR galaxies identified by GREX-PLUS with next-generation radio surveys by SKA and ngVLA.

\subsection{Scientific goals}
We propose a systematic search for optically dark IR galaxies, including dusty AGN (that are not detectable by Euclid, Roman, and LSST) with GREX-PLUS Wide and Medium surveys (Table~\ref{tab:dustyagn}).
The survey areas and 5$\sigma$ sensitivities at 3--8 $\mu$m for point sources in the Wide and Medium surveys are 1000 deg$^2$ and 21--24 AB mag and 100 deg$^2$ and 22.5--25.5 AB mag, respectively.
It would be ideal if our targets could be separated based on proper motion, since Galactic sources such as C-rich stars may be contaminated.
We will address the growth history of SMBHs and their host galaxies as a function of redshift by investigating the statistical properties (e.g., luminosity function and correlation function) and physical properties (SMBH mass accretion rate and star formation rate) of optically dark IR galaxies found in this project. 
We also aim to determine the contribution of dust-obscured AGN to cosmic star formation rate density, mass accretion rate density, and structure formation in the Universe.
\begin{table}
    \begin{center}
    \caption{Required observational parameters.}
    \label{tab:dustyagn}
    \begin{tabular}{|l|p{9cm}|l|}
    \hline
     & Requirement & Remarks \\
    \hline
    Wavelength & 3--8 $\mu$m & \multirow{2}{*}{$a$} \\
    \cline{1-2}
    Spatial resolution & $<2$ arcsec & \\
    \hline
    Wavelength resolution & $\lambda/\Delta \lambda>3$ & $b$ \\
    \hline
    Field of view & 1000 deg$^2$, 21--24 AB mag ($5\sigma$, point source) & \multirow{2}{*}{$c$}\\
    \cline{1-1}
    Sensitivity & 100 deg$^2$, 22.5--25.5 AB mag ($5\sigma$, point source) & \\
    \hline
    Observing field & Fields excluding the Galactic plane & $d$ \\
    \hline
    Observing cadence & $>$ 2 & $e$ \\
    \hline
    \end{tabular}
    \end{center}
    $^a$ F397, F520, and F680. Diffraction limit for a primary mirror with $\phi>1.0$ m and $\lambda=10\,\mu$m.\\
    $^b$ Three or more bands are required for the SED analysis.\\
    $^c$ A $>0.25$ deg$^2$ field of view of a single pointing is required for the point-source sensitivity for a $\phi=1.0$ m telescope and an assumed amount of observing time.\\
    $^d$ In particular, the overlap area with LSST, Euclid, and Roman is preferable.\\
    $^e$ It would be nice to be able to measure the proper motion to distinguish them from C-rich stars. 
\end{table}

\printbibliography[heading=subbibliography]
\end{refsection}

\clearpage

\begin{refsection}[2-16_SMBH_coalescence/SMBH_coalescence.bib]

\section{Supermassive Black Hole Coalescence}
\label{sec:smbh_coa}

\noindent
\begin{flushright}
Taiki Kawamuro$^{1}$, 
Yoshiyuki Inoue$^{1}$, 
Katsunori Kusakabe$^{1}$
\\
$^{1}$ The University of Osaka
\end{flushright}
\vspace{0.5cm}

\subsection{Scientific background and motivation}

Supermassive black holes (SMBHs) with masses of $10^{6}$--$10^{10}\,M_{\odot}$ are believed to reside ubiquitously in the centers of massive galaxies and exhibit tight correlations with the properties of their host bulges (e.g., $M_{\rm BH}$--$M_{\rm bulge}$ and $M_{\rm BH}$--$\sigma$ relations; \cite{Kormendy2013ARA&A..51..511K}). 
These empirical relations suggest that the growth of SMBHs and the assembly of galaxies are closely linked. 
Toward a complete understanding of this coevolution, a fundamental open question is how SMBHs have grown over cosmic time: what fraction of their mass has been assembled through ``secular processes'' that drive gas inward and ``galaxy mergers'' that trigger rapid accretion and result in SMBH coalescence (e.g., \cite{Alexander2012NewAR..56...93A}).

Galaxy mergers have been extensively studied as a key channel for SMBH growth. 
Hydrodynamical simulations have predicted that, as a merger progresses toward the final coalescence of galaxies and SMBHs, the SMBH accretion rate systematically increases, while the nuclear regions become progressively more deeply embedded in dense gas and dust \citep[e.g.,][]{Blecha2018MNRAS.478.3056B}. 
Observationally, multi-wavelength studies of interacting and merging galaxies have been performed as well \citep[e.g.,][]{Silverman2020ApJ...899..154S,Li2025ApJ...986..101L}. 
Various studies have revealed that dual AGNs become more luminous as their projected separation decreases \citep[e.g.,][]{Koss2012ApJ...746L..22K}, and that AGNs in late-stage mergers tend to be more heavily obscured, with an enhanced fraction of Compton-thick systems compared to isolated AGNs \citep[e.g.,][]{Ricci2017MNRAS.468.1273R}. 
Taken together, theory and observations suggest a luminous and extremely obscured phase of SMBH growth prior to coalescence.

In addition to the electromagnetic picture, a new window for quantifying the contribution of SMBH coalescence to SMBH growth has been opened by nanohertz (nHz) gravitational-wave (GW) observations. 
Recently, pulsar timing array (PTA) experiments have reported evidence for a stochastic GW background (SGWB) at nHz frequencies, with an amplitude and spectral shape that are broadly consistent with expectations from a cosmic population of SMBH binaries \citep[e.g.,][]{Agazie2023ApJ...951L...8A,Kusakabe2026ApJ...999..117K}. 
In principle, the SGWB encodes rich information about the SMBH coalescence history, such as the merger rate density, the typical masses of the binaries, and the efficiency of environmental hardening. 
In practice, however, PTA measurements alone cannot uniquely determine these quantities, because they are degenerate with each other. 

To break these degeneracies and translate the SGWB detection into a quantitative history of SMBH growth through mergers, independent electromagnetic constraints on the incidence and properties of dual and merging AGNs are indispensable \citep{Kusakabe2026ApJ...999..117K}.
For robust constraints, we need a large, homogeneous sample of dual AGNs over a wide range of redshift, luminosity, and SMBH mass, including heavily obscured systems. 
However, current dual-AGN samples are limited in several ways. 
Most are identified either via optical spectroscopy (e.g., double-peaked emission lines) or soft X-ray imaging, both of which are biased against the most obscured nuclei. 
In addition, luminous dual AGNs and late-stage mergers are intrinsically rare, so existing studies based on deep but narrow surveys or pointed follow-up observations are likely to miss a substantial fraction of this population. 
As a consequence, the dual-AGN fraction as a function of bolometric luminosity, SMBH mass, and redshift remains poorly constrained, particularly for the most obscured, luminous systems, highlighting the need for less biased and wide-field surveys.

GREX-PLUS is uniquely suited to overcome the limitations described above. 
Its wide-field camera (WFC) will perform imaging in the 2--8\,$\mu$m band with high sensitivity and good angular resolution \citep{2023arXiv230408104G}. 
At $z \sim 0$--3, these wavelengths probe the rest-frame mid-to-near-IR regime where dust extinction is modest and AGN-heated dust emission contributes significantly to the spectral energy distribution (SED), enabling the detection of dust-obscured AGNs that are faint or invisible in the optical band. 
In addition, GREX-PLUS will map hundreds to thousands of square degrees in the 2--8\,$\mu$m band, far exceeding the total area of past Spitzer/IRAC legacy surveys such as SWIRE ($\sim 50$~deg$^{2}$; e.g., \cite{Lonsdale2004ApJS..154...54L}), while reaching comparable or greater depth. 
Thus, the planned wide-area survey will make it possible to build statistically meaningful samples of rare, luminous dual and late-stage merging AGNs.
Finally, as an important characteristic of GREX-PLUS, the WFC point-spread function will achieve an angular resolution of $\lesssim 1$~arcsec and will allow us to resolve dual nuclei with projected separations of a few to several kpc across a wide redshift range.

In this science theme, we propose to use GREX-PLUS to perform a less biased census of dual AGNs with respect to obscuration and high luminosity across cosmic time.  
By identifying dual AGNs and estimating their luminosities, we will estimate the dual-AGN fraction as a function of luminosity and redshift.
We will then reconstruct the GW background from SMBH binaries based on 
the observed dual-AGN fractions, together with the AGN luminosity function and theoretical modeling of the evolution of binary separation. Ultimately, we aim to reveal the SMBH coalescence history. 
In the 2030s, SKA and LISA are expected to operate \citep[][]{Carilli2004NewAR..48..979C,Colpi2024arXiv240207571C}, and will enable us to translate the emerging nHz GW signals into a physical picture of SMBH growth through galaxy and SMBH mergers in unprecedented detail. 


\subsection{Required observations and expected results}

\begin{figure}[h]
    \centering
    \includegraphics[width=0.98\columnwidth]{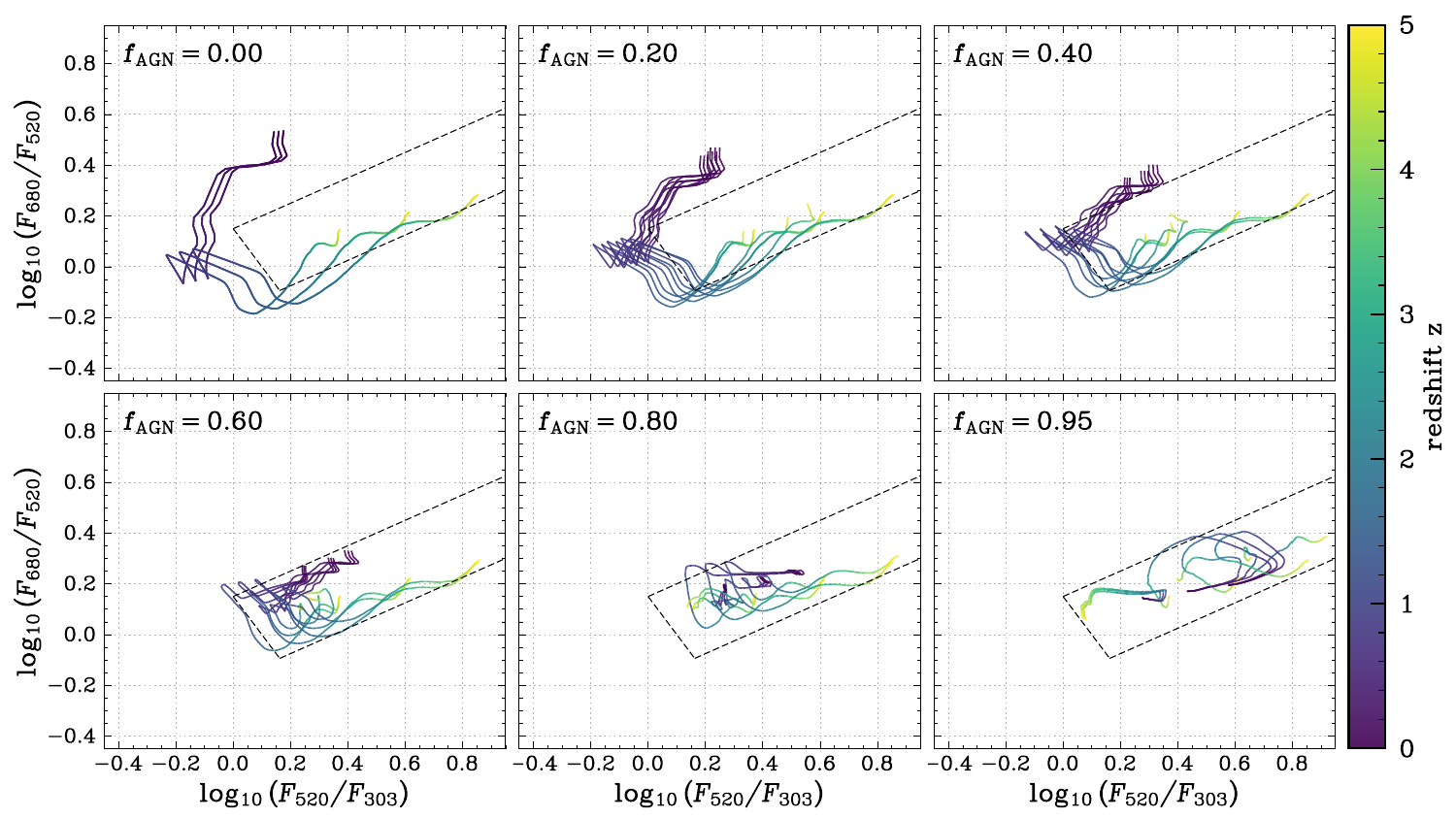}
    \caption{
    GREX-PLUS mid-IR color--color diagrams for composite AGN+star-forming galaxy SEDs.
    The six panels show models with intrinsic AGN fractions
    $f_{\rm AGN} = 0.0, 0.2, 0.4, 0.6, 0.8,$ and $0.95$ (defined over $1$--$10~\mu{\rm m}$), as
    indicated in each panel.
    Line colors represent the redshift evolution tracks from $z=0$ to $z=5$, while the
    different curves within each panel correspond to combinations of dust extinction applied
    to the AGN and host components
    ($A_V^{\rm AGN} = 0, 10, 20$; $A_V^{\rm gal} = 0, 1, 2$),
    using the Milky Way $R_V = 3.1$ law and 
    the dust extinction curve of \citet{Draine2003ARA&A..41..241D}. 
    Flux densities are band-averaged $F_\nu$ in the GREX-PLUS bands
    F303 ($2.6$--3.4 $\mu{\rm m}$), 
    F520 ($4.5$--5.9 $\mu{\rm m}$), and F680 ($5.9$--7.7 $\mu{\rm m}$),
    plotted as $\log_{10}(F_{520}/F_{303})$ versus $\log_{10}(F_{680}/F_{520})$.
    The region depicted by solid lines marks the proposed GREX-PLUS AGN selection wedge, tuned to
    encompass primarily the $f_{\rm AGN} \gtrsim 0.6$ tracks, while rejecting
    most purely star-forming ($f_{\rm AGN}=0$) loci.
    }
    \label{fig:smbhcor_color2}
\end{figure}

To achieve the goal, GREX-PLUS imaging surveys in the 2--8\,$\mu$m band are important. 
Three WFC bands spanning 2--8\,$\mu$m will be used to sample the rest-frame near-IR SED and to separate AGN-dominated from star-formation-dominated sources via color diagnostics. 
To interpret GREX-PLUS IR colors of AGNs, we constructed synthetic
color--color tracks based on composite AGN + star-forming galaxy SEDs.
As intrinsic templates, we adopted the SWIRE library \citep{Polletta2007ApJ...663...81P},
using the ``QSO1'' template to represent the AGN component and the ``M82''
template to represent a star-forming host. 
The relative AGN contribution was parameterized by the fraction of the
integrated $1$--10\,$\mu$m rest-frame flux density.
Dust attenuation was applied separately to the AGN and host components using
the Milky Way $R_V=3.1$ extinction law with $A_V^{\rm AGN} = 0$, 10, and 20 and $A_V^{\rm gal} = 0$, 1, and 2 \citep{Draine2003ARA&A..41..241D}.
For a given redshift $z$, the GREX-PLUS bands F303, F520, and F680 ($2.6$--3.4, $4.5$--5.9, and $5.9$--7.7\,$\mu$m, respectively) were mapped to rest-frame wavelength intervals via
$\lambda_{\rm rest} = \lambda_{\rm obs}/(1+z)$ and simply treated as top-hat filters. 
Finally, we defined the GREX-PLUS colors as
$\log_{10}(F_{520}/F_{303})$ versus $\log_{10}(F_{680}/F_{520})$ and
computed tracks over $0 \leq z \leq 5$ for each choice of $f_{\rm AGN}$ and
($A_V^{\rm AGN}$, $A_V^{\rm gal}$). 
These synthetic tracks form the basis of
the color--color diagrams shown in Figure~\ref{fig:smbhcor_color2} and are used to define a GREX-PLUS AGN selection wedge (areas surrounded by dashed lines in the figure); 
    \begin{equation}
      0.5\,x - 0.175 \;\le\; y \;\le\; 0.5\,x + 0.15,
      \qquad
      y \;\ge\; -1.5\,x + 0.15, 
    \end{equation}
    where 
    \[
      x \equiv \log_{10}\!\left(\frac{F_{520}}{F_{303}}\right),\qquad
      y \equiv \log_{10}\!\left(\frac{F_{680}}{F_{520}}\right). 
    \]

Dual and merging AGN candidates will be identified using the IR colors. 
We will then use morphological information from GREX-PLUS itself to identify interacting systems and galaxies with multiple nuclei. 
Pairs with angular separations larger than the WFC PSF will be recognized as resolved dual nuclei, while systems with smaller separations may still show disturbed morphologies indicative of late-stage mergers that would be candidates for follow-up observations at various wavelengths. 

We estimated the number of AGNs that GREX-PLUS can detect by adopting an X-ray luminosity function (XLF), or AGN number density, at a fixed intrinsic $2$--$10~\mathrm{keV}$ luminosity ($L_{\rm 2-10} = 3\times10^{44}$ $\mathrm{erg\,s^{-1}}$), corresponding to a bolometric luminosity of $L_{\rm bol}\sim10^{46}~\mathrm{erg\,s^{-1}}$. 
The luminosity matches a range in which past surveys for dual AGNs might be biased because of narrow survey coverage and/or obscuration. 
We then converted the X-ray luminosity to a rest-frame mid-IR luminosity at $6~\mu\mathrm{m}$ by assuming a fixed correction between the bands (i.e., $\lambda L_{\rm 6\,\mu m}/L_{\rm 2-10} \sim$ 1.1; \citealt{Mateos2015MNRAS.449.1422M}). 
Then, based on the number density, corresponding mid-IR luminosity, and design of each survey tier (Deep/Medium/Wide), we calculated 
the total number of AGNs that GREX-PLUS is expected to detect in the Deep, Medium, and Wide surveys. 
Under the conservative assumption that the dual-AGN fraction is $\sim 3\times10^{-3}$ \citep{Kusakabe2026ApJ...999..117K}, 
the numbers of detectable dual-AGN pairs for the tiers are 
$N_{\mathrm{D}}(z\lesssim5) \sim \mathrm{10}$,
$N_{\mathrm{M}}(z\lesssim5) \sim \mathrm{90}$, and
$N_{\mathrm{W}}(z\lesssim3) \sim \mathrm{440}$. 
These predicted counts demonstrate that GREX-PLUS will provide statistically meaningful samples of high-luminosity AGNs out to high redshift.

In summary, GREX-PLUS imaging surveys would be able to deliver a less biased census of dual AGNs across cosmic time up to $z \sim 5$. 
The expected results include (1) robust measurements of the dual-AGN fraction and its dependence on luminosity and redshift and (2) quantitative constraints on the SMBH merger rate density that can be directly compared with the existing PTA observations and future GW observations.

\subsection{Scientific goals}

The primary goal of this science program is to estimate how efficiently SMBHs grow through their coalescence by making a census of dual AGNs with GREX-PLUS. 
Specifically, we aim (i) to measure the dual-AGN fraction and its dependence on bolometric luminosity and redshift, while including the contribution from heavily obscured and Compton-thick systems and (ii) to reconstruct the cosmic history of SMBH coalescence by connecting these electromagnetic constraints to the population of SMBH binaries and comparing the predicted nHz SGWB with actual SGWB spectra in the 2030s.   

\begin{table}
    \begin{center}
    \caption{Required observational parameters.}\label{tab:firstgals}
    \begin{tabular}{|l|p{9cm}|l|}
    \hline
     & Requirement & Remarks \\
    \hline
    Wavelength & 2--8 $\mu$m & \multirow{1}{*}{$a$} \\
    \hline
    Spatial resolution & $<1$ arcsec & \multirow{1}{*}{$b$} \\
    \hline
    Field of view & 1000 deg$^2$ & \multirow{1}{*}{$c$}\\
    \hline
    Sensitivity & 21 AB mag (or better) in the F680 band ($5\sigma$, point source) & \multirow{1}{*}{$d$} \\
    \hline
    \end{tabular}
    \end{center}
    $^a$ To calculate two colors for identifying AGNs.\\
    $^b$ To identify dual nuclei at scales of several kpc.\\
    $^c$ To identify a rare population of dual AGNs with bolometric luminosities above $\sim 10^{46}$ erg s$^{-1}$. \\ 
    $^d$ To identify AGNs with bolometric luminosities above $\sim 10^{46}$ erg s$^{-1}$ based on the color--color diagram.\\
\end{table}

\printbibliography[heading=subbibliography]
\end{refsection}

\clearpage

\begin{refsection}[2-17_AGN_TimeDomain/AGN_TimeDomain.bib]

\section{AGN Time Domain Studies}
\label{sec:agn_timedomain}

\noindent
\begin{flushright}
Mitsuru Kokubo$^{1}$
\\
$^{1}$ Division of Science, National Astronomical Observatory of Japan
\end{flushright}
\vspace{0.5cm}

\subsection{Scientific background and motivation}

\label{sec:intro}

\begin{figure}[ht]
    \centering
    \includegraphics[width=0.69\columnwidth]{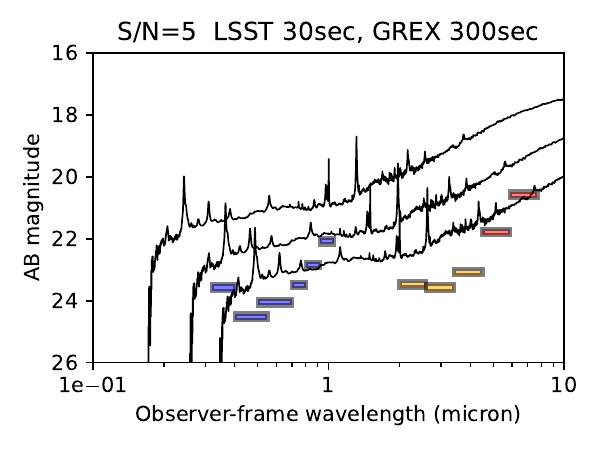}
    \caption{
    The AGN template spectrum redshifted to $z=1$ ($m_{{\rm LSST}-i}=21$~mag), $2$ (22~mag), and $3$ (23~mag). The UV--NIR spectrum at $0.086~\mu$m~$\leq$~$\lambda \leq 1.4~\mu$m is that provided by STScI \citep{2001AJ....122..549V,2006ApJ...640..579G}, and the MIR spectrum at $\lambda > 1.4~\mu$m is that provided by \cite{2016MNRAS.463.2064H}. The vertical bars indicate the $5\sigma$ limiting magnitudes of the LSST bands (30~s exposure) and of the GREX-PLUS/WFC bands (300~s exposure).
    }
    \label{fig:agn_template_at_z123}
\end{figure}

The central engine of active galactic nuclei (AGNs) is thought to be an accretion disk surrounding a supermassive black hole (SMBH). 
The AGN accretion disk emits thermal radiation primarily at UV--optical wavelengths, and a characteristic feature of this emission is temporal flux variability with amplitudes of $\sim 0.1$~mag on timescales of days to years \citep[][and references therein]{2004ApJ...601..692V, 2009ApJ...698..895K,2010ApJ...708..927K,2010ApJ...721.1014M,2021Sci...373..789B}. 
Because the broad emission lines from the broad-line region (BLR) and the IR thermal emission from the dusty torus are both re-radiated responses to the irradiating UV--optical disk emission, variations in the disk flux drive corresponding variations in the BLR and torus emission. 
As a result, AGNs exhibit variability ubiquitously across the UV-to-IR wavelengths (in unobscured AGNs, the disk component dominates at $\lambda_{\text{rest}}<2\,\mu$m, while the torus dominates at $\lambda_{\text{rest}}>2\,\mu$m; Figure~\ref{fig:agn_template_at_z123}). 
AGNs constitute one of the major classes of variable extragalactic objects, with extragalactic time-domain surveys detecting variable AGNs in numbers comparable to those of supernovae \citep[e.g.,][]{2019PASJ...71...74Y, 2025A&A...697A.204D, 2025ApJ...985..223M}.

The ubiquity of AGN variability allows for variability-based AGN selection, in which AGNs are identified through the detection of flux variability at galaxy centers \citep[][and references therein]{2025ApJ...995...24K,Burke_2026}.
The temporal behavior of the UV--optical AGN variability is known to follow a stochastic process, specifically the damped random walk \citep[DRW;][]{2009ApJ...698..895K,2010ApJ...708..927K,2010ApJ...721.1014M}, which allows it to be distinguished from one-off events such as supernovae.
Because AGN variability is empirically known to depend only weakly on wavelength \citep{2010ApJ...721.1014M}, variability-based selection is expected to identify AGNs in a more redshift-independent manner than commonly used optical color-based methods.
In addition, variability-based selection is particularly powerful for detecting low-luminosity AGNs (i.e., systems with low black hole masses and/or low accretion rates), whose emission is often overwhelmed by the more luminous stellar light of their host galaxies and can therefore be easily missed by optical color selection \citep[e.g.,][]{2022ApJ...936..104W,2023MNRAS.518.1880B,Burke_2026}.

In the SDSS/eBOSS survey, variability information from the SDSS Stripe~82 Supernova Survey and Palomar Transient Factory (PTF) optical time-domain surveys was used to select AGNs, complementing the canonical color-selected AGN samples \citep{2015ApJS..221...27M, 2016A&A...587A..41P}. 
SDSS/eBOSS demonstrated that, compared with color-based selection (which inevitably misses AGNs at specific redshifts; the so-called `redshift desert' \citealt{2015ApJS..221...27M}), variability-based selection yields AGN samples that are more uniformly distributed in redshift \citep{2016A&A...587A..41P}. 
A deep optical time-domain survey with Subaru/HSC ($m_{\text{lim}} \sim 26$~mag) revealed $\simeq 300$ variable AGNs in the COSMOS field, many of which were missed by optical color, IR color, and X-ray AGN selection methods \citep{2020ApJ...894...24K}. 
In the 2020s--2030s, the Rubin/LSST survey will enable variability-based AGN selection down to $m_{\text{lim}} \sim 24$~mag across the entire southern sky (Chapter~10.5 of \citealt{2009arXiv0912.0201L}, \citealt{panda2026agnvariabilityrubinobservatory}).

By investigating IR variability, we can expect to identify not only unobscured AGNs but also obscured AGNs through variability-based selection. 
Although several IR wide-field time-domain surveys have been conducted with ground-based telescopes \citep[up to the $K$ band;][]{2017ApJ...849..110S,2020MNRAS.493.3026E,2024MNRAS.531.2551G,2024MNRAS.531.3310L} and IR satellite missions \citep[WISE/NEOWISE and Spitzer, mostly up to the $M$ band;][]{2010ApJ...716..530K,2017ApJ...839...88K,2018MNRAS.476.1111P,2019ApJ...886...33L,2023ApJ...958..135S}, no deep and wide MIR time-domain survey has been available to date. 
Among planned IR satellite missions, the Nancy Grace Roman Space Telescope will perform an ultradeep and wide High Latitude Time Domain Survey \citep[HLTDS;][]{2025arXiv250510574Z} which will be useful for finding low-luminosity AGNs via variability detection \citep{yamada23}, but the observing wavelengths are limited to the $K$ band.
The GREX-PLUS time-domain survey, as detailed in this GREX-PLUS Science Book (Section~\ref{sec:highzsupernovae}; see also below), will be the first to provide both depth and wide-area coverage in the MIR ($2$--8~$\mu$m), offering an unprecedented opportunity to perform IR variability-based AGN selection and to construct a more complete AGN sample, including obscured AGNs.

Moreover, the temporal lags between the driving UV--optical continuum light curve and the responding BLR and dust IR emission---corresponding to the light-crossing times between the central accretion disk and the surrounding BLR and dust torus---provide a unique observational means of probing the otherwise unresolved innermost structures of AGNs \citep[referred to as AGN reverberation mapping;][]{1993PASP..105..247P}.
The BLR and dust IR reverberation lags have been measured through intensive photometric and spectroscopic monitoring, leading to empirical relationships between the AGN accretion-disk luminosity and the characteristic radii of the BLR and the innermost dust torus in the form of $R = c\tau \propto L^{1/2}$, where $R$ is the radius of the BLR or the innermost dust torus, $\tau$ is the rest-frame reverberation lag, and $L$ denotes the AGN accretion-disk luminosity \citep{2004ApJ...613..682P,2013ApJ...767..149B,2014ApJ...788..159K,2019ApJ...886..150M,2024MNRAS.531.3310L}.

As for dust reverberation mapping, the MAGNUM project observed a sample of low-$z$ AGNs ($z < 0.5$) with a dedicated 2.0 m IR telescope in the optical and IR bands (up to $K$ band) \citep{2006ApJ...639...46S,2014ApJ...788..159K,2014ApJ...784L..11Y,2019ApJ...886..150M}.
The WISE/NEOWISE satellite has surveyed the entire sky once every half year in the W1 ($\sim 3.4~\mu\mathrm{m}$) and W2 ($\sim 4.5~\mu\mathrm{m}$) bands, enabling dust reverberation mapping for luminous AGNs whose dust lags are longer than $\simeq 100$~days \citep{2019ApJ...886...33L,2020ApJ...900...58Y,2024ApJ...968...59M,2025ApJ...993..203T}.
Except for a few examples, such as several luminous Palomar-Green quasars \citep{1989AJ.....97..957N,1999AJ....118...35N,2019ApJ...886...33L} and NGC~4151 \citep{2021ApJ...912..126L}, there have been no dedicated studies of IR variability in the MIR bands at $\lambda > 5~\mu\mathrm{m}$.
The combination of $K$-, W1-, and W2-band dust-reverberation-lag measurements indicates a hint of wavelength-dependent dust lags for a given AGN luminosity, revealing a steep radial distribution of dust temperature in the innermost dust torus, $R_{\rm dust} \propto \lambda^{0.62} \propto T^{-0.62}$ \citep{2024ApJ...968...59M}. 
However, because of the limited accuracy due to the low sampling rate of NEOWISE and the narrow wavelength coverage from $K$ band to W2 band, the current constraint on the wavelength dependence of dust lags remains weak.
The GREX-PLUS time-domain survey, if conducted with sufficiently high cadence and wide wavelength coverage, will provide an unprecedented opportunity to measure wavelength-dependent dust lags for a large sample of AGNs at various redshifts up to $z \lesssim 3$, enabling us to constrain the temperature structure of the innermost dust torus.

The AGN reverberation mapping measurements require densely sampled UV--optical accretion-disk continuum light curves, obtained simultaneously with or prior to multi-epoch broad-line spectroscopy and/or dust IR photometry.
In the 2030s, the Rubin/LSST will provide such baseline optical light-curve measurements for AGNs across the entire southern sky.
For example, the 4MOST Time Domain Extragalactic Survey (TiDES) has been proposed to conduct spectroscopic monitoring observations with 4MOST in conjunction with LSST, enabling BLR reverberation-mapping measurements for $\lesssim 1000$ AGNs \citep[][]{2025ApJ...992..158F}.
Similarly, by selecting the GREX-PLUS time-domain survey field in the southern sky, we can use the LSST light curves to characterize the driving AGN accretion-disk variability, which, in combination with the lower-cadence multi-epoch GREX-PLUS photometry, will allow accurate dust reverberation-lag measurements.

One application of the dust lag-luminosity relationship is its use in turning AGNs into standard candles. 
The BLR/dust lag measurements for high-$z$ AGNs, combined with the established BLR/dust lag-luminosity relationship, can be used to infer AGN luminosities without assuming a cosmological model, thereby enabling the construction of an AGN Hubble diagram \citep{2011ApJ...740L..49W,2014ApJ...784L..11Y,2017MNRAS.464.1693H}.
Because AGNs are observable beyond $z \sim 2$, where observations of Type~Ia supernovae become increasingly difficult \citep[e.g.,][]{2025ApJ...993..116K}, it has been suggested that AGNs may serve as invaluable probes of the cosmic expansion history out to $z>2$ \citep[e.g.,][]{2014MNRAS.441.3454K,2019ApJ...886..150M}.
Probing the cosmic expansion history over a wide redshift range with a single standard candle offers strong constraining power on dynamical dark energy, which has recently been invoked to explain DESI’s measurements of baryon acoustic oscillations \citep[e.g.,][]{2025PhRvD.112h3515A}.
For example, with BLR reverberation-mapping measurements from 4MOST TiDES for $\lesssim 1000$ AGNs, it is anticipated that the AGN Hubble diagram will be tightly constrained up to $z \sim 2.5$ \citep[][]{2025ApJ...992..158F}.

Dust reverberation mapping can also be used to constrain the AGN Hubble diagram, as is the case for BLR reverberation mapping. However, its current application to observational cosmology is limited to $z \sim 0.5$ \citep{2019ApJ...886..150M}, primarily due to the restricted infrared wavelength coverage and sensitivity of ground-based telescopes.
As discussed in \citet{2019ApJ...886..150M}, reducing possible sources of systematic uncertainty in constraining cosmic expansion with AGN dust-reverberation measurements requires quantifying the wavelength dependence of dust-reverberation lags and tightly constraining the slope of the radius-luminosity relation. Both of these goals can be addressed by the GREX-PLUS transient survey proposed here.
The dust reverberation mapping observations carried out with GREX-PLUS will pave the way for next-generation reverberation mapping studies with the PRIMA mission, which will target longer wavelengths and higher-$z$ AGNs (\citealt{2025JATIS..11c1626G}, Chapter 53 of \citealt{2025arXiv251110927M}).

\subsection{Required observations and expected results}

\subsubsection{UV-optical continuum variability}

The UV--optical light curves of AGNs can be described by the damped random walk (DRW) process \citep[e.g.,][]{2010ApJ...721.1014M,2021ApJ...907...96S,2023MNRAS.518.1880B,2025arXiv250913308S}. 
We assume a DRW process as the time-series model describing the latent signal $s(t_i)$ in units of magnitude with zero mean. 
The relationship between the signals at epochs $t_i$ and $t_j$ for a regularly sampled DRW process is \citep{2009ApJ...698..895K,2010ApJ...708..927K}\footnote{The corresponding continuous autoregressive process follows the stochastic differential equation $ds(t) = -\frac{1}{\tau_{d}}s(t)dt + \hat{\sigma}\sqrt{dt}\epsilon(t)$, where $\hat{\sigma} = \sqrt{\frac{2}{\tau_{d}}}\sigma_{d} = \frac{\text{SF}_{\infty}}{\sqrt{\tau_{d}}}$, and $\epsilon(t)$ is a white noise process with zero mean and variance equal to 1 \citep{2009ApJ...698..895K}.}:
\begin{eqnarray}
s(t_i) &=& s(t_j)\exp\left(-\frac{\Delta t_{ij}}{\tau_d}\right) + G\!\left[\sigma_d^2\left(1-\exp\left(-\frac{2\Delta t_{ij}}{\tau_d}\right)\right)\right],
\label{eqn:driving_lightcurve}
\end{eqnarray}
where $\Delta t_{ij} = t_i - t_j > 0$.
Here, $\sigma_d$ and $\tau_d$ denote the wavelength-dependent asymptotic variability amplitude and the decorrelation timescale of the DRW process, respectively.

The structure function (SF) of the DRW process is
\begin{eqnarray}
\mathrm{SF}(\lambda_a,\Delta t_{ij})
&\equiv& \sqrt{\left\langle \left[s(t_i)-s(t_j)\right]^2 \right\rangle}
\nonumber\\
&=& \sqrt{2}\,\sigma_d\,\sqrt{1-\exp\left(-\frac{\Delta t_{ij}}{\tau_d}\right)} 
\nonumber\\
&=& \mathrm{SF}_{\infty}\,\sqrt{1-\exp\left(-\frac{\Delta t_{ij}}{\tau_d}\right)},
\end{eqnarray}
where the asymptotic SF at $\Delta t_{ij} \rightarrow +\infty$ is defined as $\mathrm{SF}_{\infty} = \sqrt{2} \sigma_d$ \citep{2010ApJ...721.1014M}.

$\text{SF}_{\infty}$ and $\tau_{d}$ are empirically parametrized as \citep{2021ApJ...907...96S,2023MNRAS.518.1880B}:
\begin{eqnarray}
\log\left(\frac{\text{SF}_{\infty}}{\text{mag}}\right) &=& A + B\log\left(\frac{\lambda}{4000\text{\AA}}\right) + C (M_{i}+23) + D \log \left( \frac{M_{\text{BH}}}{10^{9}~M_{\odot}} \right),
\label{eqn:sf_infty_relation}
\end{eqnarray}
where $A = -0.476 \pm 0.008$, $B = -0.479 \pm 0.005$, $C = 0.118 \pm 0.003$,
and $D = 0.118 \pm 0.008$ from Table~2 of \cite{2021ApJ...907...96S}; and,
\begin{eqnarray}
\log\left(\frac{\tau_{d}}{\text{days}}\right) &=& A + B\log\left(\frac{\lambda}{4000\text{\AA}}\right) + C(M_{i}+23) + D\log\left(\frac{M_{\text{BH}}}{10^{9}~M_{\odot}}\right),
\label{eqn:tau_d_relation}
\end{eqnarray}
where $A = 2.597 \pm 0.02$, $B = 0.17 \pm 0.02$, $C = 0.035 \pm 0.007$,
and $D = 0.141 \pm 0.02$ from Table~2 of \cite{2021ApJ...907...96S}.
We assume that the absolute $i$-band magnitude $M_{i}$ can be approximated as $M_{i} = 90 - 2.5\log\left(L_{\text{bol}}/\text{erg~s}^{-1}\right)$ \citep{2009ApJ...697.1656S,2023MNRAS.518.1880B}.
The bolometric luminosity $L_{\text{bol}}$ is related to the monochromatic luminosity at 5100\AA\ $L_{5100}$ via a bolometric correction factor $L_{\text{bol}}/L_{5100} = 9.26$ \citep{2006ApJS..166..470R}.
For simplicity, we fix the SMBH mass to $M_{\text{BH}} = 10^{8}~M_{\odot}$.

\subsubsection{Dust reverberation}
\label{sec:dust_reverberation}

We define $m(t) = s(t) + \bar{m}$ as the disk continuum light curve in units of magnitude, where $\bar{m} = -2.5\log(\bar{f}/f_{0})$ indicates the mean magnitude, $\bar{f}$ indicates the mean flux density in units of Jansky, and $f_{0}=3631~{\rm Jy}$.
We can convert $m(t)$ to flux-density units as
$f(t) = f_{0}\times 10^{-0.4m(t)} = f_{0}\times 10^{-0.4\left[s(t) + \bar{m}\right]} = \bar{f}\times 10^{-0.4s(t)}$.
The responding dust IR light curve in flux units is modeled as
\begin{eqnarray}
f_{\rm dust}(\lambda, t) - \bar{f}_{\rm dust}(\lambda) &=& \frac{\bar{f}_{\rm dust}(\lambda)}{\bar{f}}\int_{-\infty}^{\infty} d\tau\, \Psi(\lambda,\tau) \left[ f(t-\tau) - \bar{f} \right]\nonumber\\
\therefore m_{\rm dust}(\lambda, t) &=& -2.5\log\left[\frac{f_{\rm dust}(\lambda, t)}{f_{0}}\right]\nonumber\\
&=& \bar{m}_{\rm dust}(\lambda) - 2.5\log\left[ \int_{-\infty}^{\infty} d\tau\, \Psi(\lambda,\tau) \left[ 10^{-0.4\left[s(t-\tau)\right]} \right]  \right]\nonumber\\
\therefore s_{\rm dust}(\lambda, t) &=& m_{\rm dust}(\lambda, t) - \bar{m}_{\rm dust}(\lambda) \nonumber\\
&=&  - 2.5\log\left[ \int_{-\infty}^{\infty} d\tau\, \Psi(\lambda,\tau) \left[ 10^{-0.4\left[s(t-\tau)\right]} \right]  \right],
\label{eqn:responding_lightcurve}
\end{eqnarray}
where we adopt a top-hat transfer function,
\begin{eqnarray}
\Psi(\lambda, t) =
\begin{cases}
\dfrac{1}{2 t_{\rm lag}(\lambda)}, & 0 \le t \le 2 t_{\rm lag}(\lambda), \\[6pt]
0, & \text{otherwise},
\end{cases}
\end{eqnarray}
and $t_{\rm lag}(\lambda)$ denotes the wavelength-dependent dust IR reverberation lag.


This transfer function corresponds to the response of an optically thin, dust-filled sphere with radius $R_{\text{dust}}(\lambda) = c\,t_{\rm lag}(\lambda)$, which provides a reasonable approximation to a more realistic dust torus transfer function \citep{2017ApJ...843....3A}. 
The driving light curve $s(t)$ is taken to be the optical continuum light curve at
$\lambda_{\rm rest} = 5100\,\text{\AA}$.
Equation~\ref{eqn:responding_lightcurve} implies that we implicitly assume the variability amplitude of the dust IR emission (expressed in magnitudes) to be identical to that of the $\lambda_{\rm rest} = 5100\,\text{\AA}$ accretion-disk light curve.
When the width of the transfer function ($=2t_{\rm lag}(\lambda)$) is large, the IR variability amplitude decreases.
This normalization factor, however, is not well constrained observationally.

The wavelength dependence of the dust reverberation radius is not well constrained, and here we assume a simple power-law following \citet{2024ApJ...968...59M}:
\begin{eqnarray}
R_{\rm dust}(\lambda) = ct_{\text{lag}}(\lambda) = R_{\rm dust}(\lambda_{K})
\left( \frac{\lambda}{\lambda_K} \right)^{0.62},
\label{eqn:dust_lambda_radius}
\end{eqnarray}
where $\lambda_{K}=2.2\,\mu$m.
The rest-frame $K$ band reverberation radius $R_{\rm dust}(\lambda_{K}) = ct_{\text{lag}}(\lambda_{K})$ is evaluated from the lag-luminosity relation of \citet{2019ApJ...886..150M}:
\begin{eqnarray}
\log \left( \frac{R_{\rm dust}}{1~{\rm pc}} \right)
= \beta
+ \alpha \log \left( \frac{L_V}{10^{43.70}~{\rm erg~s}^{-1}} \right)
+ G(\sigma),
\label{eqn:dust_K_radius}
\end{eqnarray}
with $\beta = -1.021 \pm 0.023$, $\alpha = 0.424 \pm 0.026$, and $\sigma = 0.12$.  
For simplicity, the $V$-band monochromatic luminosity $L_V$ is assumed to be identical to the rest-frame 
$5100\,\text{\AA}$ monochromatic luminosity, $L_{5100\,\text{\AA}}$.

\subsubsection{AGN UV luminosity function}

\begin{figure}
    \centering
    \includegraphics[width=0.69\columnwidth]{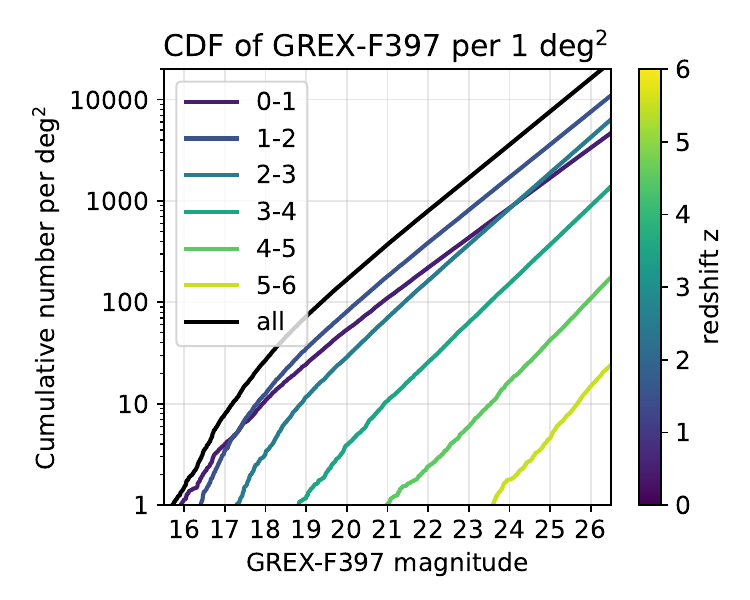}
    \caption{
    Cumulative AGN surface number density per 1~deg$^{2}$ as a function of F397-band magnitude.
    Each curve represents a different redshift interval ($z=0$--1, 1--2, 2--3, 3--4, 4--5, 5--6, and $z=0$--6 (all)), as indicated by the labels.
    Note that the number densities shown here include only unobscured AGNs and do not account for the obscured AGN population.
    The $z=0$--6 CDF can be approximated as $\log_{10}({\rm CDF}) = -4.304 + 0.327m_{\rm F397}$.
    }
    \label{fig:surface_number_density}
\end{figure}

To simulate the population of unobscured AGNs, we adopted the UV AGN luminosity function $\phi(M_{1450})$ from \citet{2019MNRAS.488.1035K} (Model 3 in their Table 3) together with the AGN spectral template shown in Figure~\ref{fig:agn_template_at_z123} \citep{2001AJ....122..549V,2006ApJ...640..579G,2016MNRAS.463.2064H}. 
The spectral template was used to compute the $K$-corrections, i.e., the conversions among the UV absolute magnitude at 1450~\AA\ ($M_{1450}$), the monochromatic luminosity at 5100~\AA\ ($L_{5100}$), and the apparent magnitudes in the LSST and GREX-PLUS filters at various redshifts. 
The $K$-corrections were calculated with \texttt{speclite}\footnote{\url{https://speclite.readthedocs.io}}, using the LSST (2016) filter transmission curves and custom top-hat GREX-PLUS transmission curves.
The observer-frame wavelength ranges (and central wavelengths) of the GREX-PLUS/WFC bands were taken from the GREX-PLUS Quick Factsheet ver.~20250702: 
$\lambda_{\rm obs} = 2.0$--2.6~$\mu$m (2.3~$\mu$m), 2.6--3.4~$\mu$m (3.0~$\mu$m), 3.4--4.5~$\mu$m (4.0~$\mu$m), 4.5--5.9~$\mu$m (5.2~$\mu$m), and 5.9--7.7~$\mu$m (6.8~$\mu$m) for the F232, F303, F397, F520, and F680 filters, respectively.

Under the above assumptions, a Monte Carlo AGN sample of $(z, M_{1450}, m_{{\rm LSST-}z})$ for a $40~\rm deg^2$ area was generated over $m_{{\rm LSST-}z} = 15$--$29$~mag and $z=0$--6.85.
We do not include the intergalactic medium (IGM) Lyman-$\alpha$ extinction; therefore, this simulation is accurate only up to $z \sim 6$ (note that the IGM affects the $z$-band photometry at $z>6$). 
Figure~\ref{fig:surface_number_density} shows the cumulative surface number density per deg${}^2$ of the AGNs as a function of the GREX-PLUS F397-band magnitude, calculated from a Monte Carlo AGN sample generated for a 40~deg${}^2$ sky region as described above.

\subsubsection{Optical-IR light-curve simulations for the GREX-PLUS time-domain surveys}

Here, we perform light-curve simulations of the GREX-PLUS time-domain surveys for a Monte Carlo AGN sample of $(z, M_{1450}, m_{{\rm LSST-}z})$ for a $10~\rm deg^2$ area generated over $m_{{\rm LSST-}z} = 15$--$29$~mag and $z=0$--6.85.
We do not include the intergalactic medium (IGM) Lyman-$\alpha$ extinction; therefore, this simulation is accurate only up to $z \sim 6$ (note that the IGM affects the $z$-band photometry at $z>6$). 
For each simulated AGN, apparent magnitudes in all LSST and GREX-PLUS filters, $L_{5100}$, randomly drawn values of $\mathrm{SF}_{\infty}$ and $\tau_d$ from Equations~\ref{eqn:sf_infty_relation} and \ref{eqn:tau_d_relation}, and randomly drawn dust radii at the observer-frame GREX-PLUS bands from Equations~\ref{eqn:dust_lambda_radius} and \ref{eqn:dust_K_radius} were calculated. 
Subsequently, an observer-frame DRW continuum light curve $s(t)$ at $\lambda_{\rm rest} = 5100$~\AA\ was simulated following Equation~\ref{eqn:driving_lightcurve}. 
Note that, when observing an AGN at redshift $z$, the observer-frame decorrelation timescale becomes $\tau_{d,\mathrm{obs}} = (1+z)\,\tau_d$ because of cosmological time dilation. 
Here, $G(\sigma^2)$ represents a Gaussian deviate with variance $\sigma^2$. 
The corresponding responding dust IR light curves $s_{\mathrm{dust}}(t)$ at the observer-frame GREX-PLUS bands were generated for each AGN following Equation~\ref{eqn:responding_lightcurve} (noting that the observer-frame dust lag is $t_{\mathrm{lag,\,obs}} = (1+z)\,t_{\mathrm{lag}}$).

To simulate the GREX-PLUS dust light curves, we adopted the $5\sigma$ (300~sec) limiting magnitudes provided in the Quick Factsheet ver.~20250702\footnote{\url{https://waseda.box.com/shared/static/859qijbpr6embs2zqicuavr0vtbarfue.pdf}} (see Figure~\ref{fig:agn_template_at_z123}).
Assuming the standard scaling relation, $m_{\rm lim}(t_{\rm exp}, {\rm SNR}) = m_{\rm lim}(300~{\rm sec}, 5\sigma) + 1.25 \log_{10}\left(t_{\rm exp}/300~{\rm sec}\right) - 2.5 \log_{10}\left(\rm SNR/5\right)$, we adopt an exposure time of 10~hours or 63 ($=10\times10^{0.8}$) hours per epoch, which yields an approximate $5\sigma$ limiting magnitude of $\simeq 26$~mag per epoch, or $\simeq 27$~mag per epoch in the F232 band, consistent with the sensitivity required for the GREX-PLUS transient survey.
The cadence of the GREX-PLUS transient survey is not yet fixed; here, we consider two cases:  
(a) a low-cadence survey with one observation per year, aimed at surveying AGNs via IR variability-based selection, and  
(b) a high-cadence survey, which enables not only IR variability-based AGN selection but also multiwavelength IR dust reverberation mapping in conjunction with the LSST observations.

\subsubsection*{Low-Cadence Time-domain Survey for IR Variability-Based AGN Selection}

\begin{figure}
    \centering
    \begin{minipage}[c]{0.45\textwidth}  
        \centering
        \includegraphics[width=1.00\linewidth]{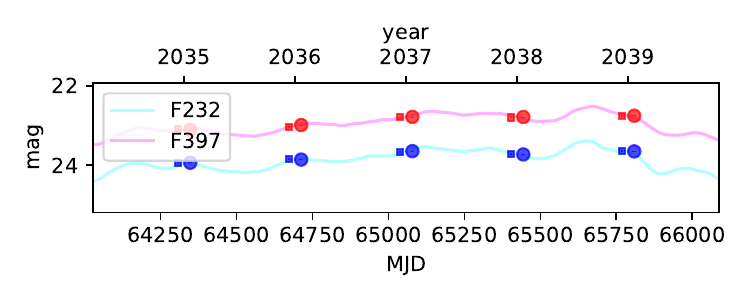}\\[-3.0mm]
        \includegraphics[width=1.00\linewidth]{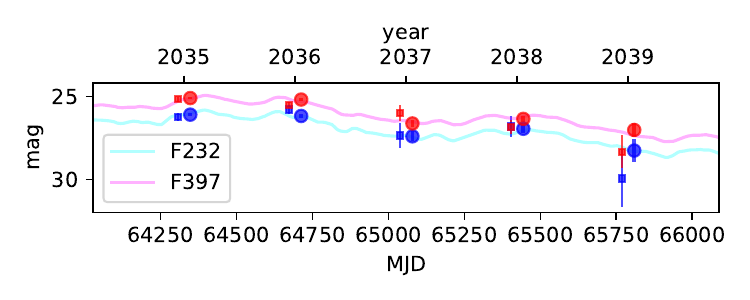}
    \end{minipage}%
    \hfill
    \begin{minipage}[c]{0.53\textwidth}   
        \centering
        \includegraphics[width=1.00\linewidth]{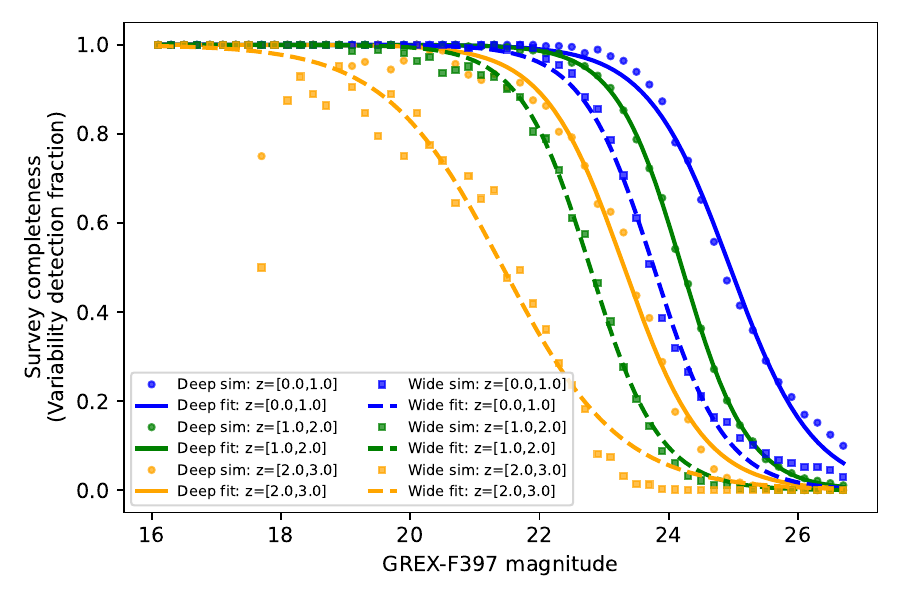}
    \end{minipage}
    \caption{
    {\it Left}: The continuous light curves indicate simulated light curves in the F232 and F397 bands for $m_{F397} = 23$~mag and $26$~mag AGNs at $z=1$.
    The squares and circles denote the low-cadence transient survey with per-epoch exposure times of 10~hours (wide survey) and 100~hours (deep survey), respectively. 
    For clarity, these symbols have been slightly offset along the horizontal axis when plotted.
    The significances of the variability detection $\sigma_{\rm XS}/\bar{\sigma}_{\rm err}$ in the F232/F397 bands are 4.42/8.59 (wide) and 11.48/21.88 (deep) for the 23~mag object, and 0.66/1.47 (wide) and 3.23/3.90 (deep) for the 26~mag object, respectively.
    {\it Right}: The variability detection fraction ($\sigma_{\rm XS}/\bar{\sigma}_{\rm err} > 10$) as a function of F397-band magnitude, computed for a Monte Carlo AGN sample at $z=0-1$ (blue), $1-2$ (green), and $2-3$ (orange) over a 10~deg${}^{2}$ sky area.
    The solid and dashed curves denote the best-fitting logistic function for the deep and wide surveys, respectively, $\eta_{0} \times (1+e^{(m-m_{50})/\tau})^{-1}$ with $\eta_{0}=1$:
    ($m_{50}$, $\tau$) = 
    (24.950, 0.641) for $z=0$--1,
    (24.205, 0.514) for $z=1$--2, and
    (23.311, 0.617) for $z=2$--3
    for the deep survey, and
    (23.767, 0.555) for $z=0$--1,
    (22.783, 0.544) for $z=1$--2, and
    (21.434, 0.909) for $z=2$--3
    for the wide survey.
    }
    \label{fig:lightcurve_sims}
\end{figure}

The GREX-PLUS transient survey for high-$z$ ($z>6$) pair-instability supernovae and superluminous supernovae proposed in this Science Book (Section~\ref{sec:highzsupernovae}) requires a low-cadence time-domain survey over the 5-year mission, with one observation per year. 
It consists of a two-tiered transient survey in the F232 and F397 bands: a wide 40~deg$^{2}$ survey reaching $m_{\rm lim}({\rm SNR}=5)=26$~mag per epoch (for a total exposure time of 10 hours), and a deep 1~deg$^{2}$ survey reaching $m_{\rm lim}({\rm SNR}=5)=27$~mag per epoch (also with a 10-hour exposure).
Although dust-reverberation measurements are not feasible with such a low cadence, the wide-field time-domain datasets produced by this survey will allow us to identify unobscured and mildly obscured AGNs across the survey fields. 
To detect variability with a typical amplitude of 0.1~mag, the single-epoch photometry must achieve a precision of $\sim 0.01$~mag, corresponding to ${\rm SNR} \sim 100$. 
Therefore, the GREX-PLUS transient survey will enable a nearly complete variability-selected AGN census down to $26~\mathrm{mag} - 2.5\log_{10}(100/5) \sim 23~\mathrm{mag}$ in the wide survey, and $27~\mathrm{mag} - 2.5\log_{10}(100/5) \sim 24~\mathrm{mag}$ in the deep survey. 
However, low-luminosity AGNs tend to exhibit larger variability amplitudes (Equation~\ref{eqn:sf_infty_relation}), which partly compensates for the survey depth required for variability-based AGN selection.

We quantitatively evaluate the survey completeness using light-curve simulations.
The simulated GREX-PLUS light curves were generated by sampling $s_{\rm dust}(t)$ with a cadence of one observation per year over the 5-year mission (assuming observations on January~1 from 2035 to 2039), and then uncorrelated Gaussian measurement noise, with variance calculated from the $5\sigma$ depth per epoch (10~hours) and the scaling relation $m_{\rm lim}(t_{\rm exp}, {\rm SNR}) = m_{\rm lim}(300~{\rm sec}, 5\sigma) + 1.25 \log_{10}\left(t_{\rm exp}/300~{\rm sec}\right) - 2.5 \log_{10}\left({\rm SNR}/5\right)$, was added to the sampled dust IR light curves.

As shown in the left panels of Figure~\ref{fig:lightcurve_sims}, the variability is clearly identifiable in the simulated light curve of a bright AGN ($m_{F397}=23$~mag) even with a sampling rate of once per year, whereas the variability signal is buried in the photometric noise for a faint AGN ($m_{F397}=26$~mag).
Following \citet[][their Equation~8]{2003MNRAS.345.1271V}, we define the excess variance as $\sigma_{\rm XS}^{2} = \sigma_{\rm obs}^{2} - \bar{\sigma}_{\rm err}^{2}$, where the observed sample variance is $\sigma_{\rm obs}^{2} = \left\langle (m-\bar{m})^{2} \right\rangle = \frac{1}{N-1}\sum_{i=1}^{N} (m_{i}-\bar{m})^{2}$,  the arithmetic mean magnitude is $\bar{m} = \frac{1}{N}\sum_{i=1}^{N} m_{i}$, and the mean square photometric error is $\bar{\sigma}_{\rm err}^{2} = \left\langle \sigma_{\rm err}^{2} \right\rangle = \frac{1}{N}\sum_{i=1}^{N} \sigma_{{\rm err},i}^{2}$.
The variability detection significance is defined as the ratio $\sigma_{\rm XS}/\bar{\sigma}_{\rm err}$, and in this study we adopt a threshold of $\sigma_{\rm XS}/\bar{\sigma}_{\rm err} > 10$ to classify a source as variable.
In the example light curves in the left panels of Figure~\ref{fig:lightcurve_sims}, $\sigma_{\rm XS}/\bar{\sigma}_{\rm err}$ in the F232/F397 bands is 4.42/8.59 (wide) and 11.48/21.88 (deep) for the 23~mag object, and 0.66/1.47 (wide) and 3.23/3.90 (deep) for the 26~mag object, respectively.
These values suggest that variability detection is feasible for objects at 23 mag but becomes challenging for those at 26 mag.

The right panel of Figure~\ref{fig:lightcurve_sims} shows the variability detection fraction ($\sigma_{\rm XS}/\bar{\sigma}_{\rm err} > 10$), i.e., the variability-based AGN selection completeness curve, as a function of F397-band magnitude, computed for a Monte Carlo AGN sample ($z=0$--6) over a 10~deg${}^{2}$ sky area. 
Under our simulation setup, the variability detection fraction exhibits only a weak dependence on redshift; therefore, we assume that it is primarily a function of the apparent magnitude.
The variability detection fraction exhibits a gradual decrease at $\gtrsim 23~\text{mag}$ for the wide survey and $\gtrsim 24~\text{mag}$ for the deep survey.
By fitting a logistic function to the simulated completeness curve as a function of F397-band magnitude, $\eta_{0} \times (1+e^{(m-m_{50})/\tau})^{-1}$ with $\eta_{0}=1$ \citep[e.g.,][]{2008ApJ...676..121M,2014AJ....148...13R}, we obtained 
($m_{50}$, $\tau$) = 
(24.950, 0.641) for $z=0$--1,
(24.205, 0.514) for $z=1$--2, and
(23.311, 0.617) for $z=2$--3
for the deep survey, and
(23.767, 0.555) for $z=0$--1,
(22.783, 0.544) for $z=1$--2, and
(21.434, 0.909) for $z=2$--3
for the wide survey.
As mentioned in Section~\ref{sec:dust_reverberation}, the higher the redshift of the AGN, the smaller the IR variability amplitude becomes, making variability detection more challenging.

This analysis and the simulated AGN surface number density (Figure~\ref{fig:surface_number_density}) indicate that the IR variability-based AGN selection with the GREX-PLUS low-cadence wide (deep) time-domain survey will enable us to establish a complete AGN sample at $z=0$--3 down to $m_{\rm F397} \sim 22$~mag with $\sim 800$ AGNs per 1~deg${}^{2}$ (down to $m_{\rm F397} \sim 23$~mag with $\sim 2000$ AGNs per 1~deg${}^{2}$).
Note that the simulation presented here, based on the AGN UV luminosity function of \citet{2019MNRAS.488.1035K}, includes only the unobscured AGN population. 
In actual GREX-PLUS observations, however, both unobscured and (mildly) obscured AGN populations will be detected. 
Because the (mildly) obscured AGN population is thought to dominate near the knee of the bolometric luminosity function \citep{2020MNRAS.495.3252S}, one of our key goals is to place observational constraints on this population using GREX-PLUS transient survey data, in conjunction with follow-up spectroscopic observations with multi-object spectroscopy instruments and multiwavelength SED analyses.

\subsubsection*{High-Cadence Time-domain Survey for Multi-Wavelength IR Dust Reverberation Mapping}

\begin{figure}
    \centering
    \includegraphics[width=0.69\columnwidth]{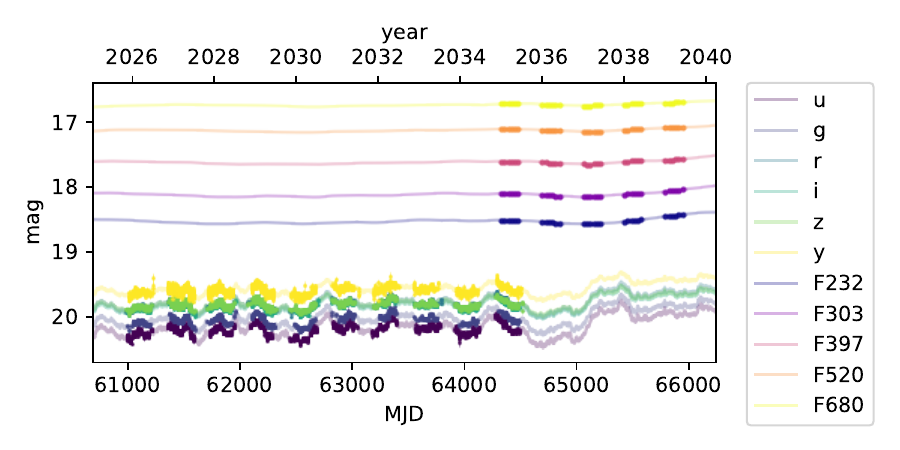}
    \caption{
    Same as the left panel of Figure~\ref{fig:lightcurve_sims} (for an AGN at $z=1.0$, $m_{i} = 19.81$~mag and $m_{\rm F397} = 17.61$~mag), but for all five GREX-PLUS bands and including LSST COSMOS DDF simulations.
    The input dust reverberation lags in the rest frame are $t_{\rm lag}=241, 284, 340, 400, 472$~days, for the five GREX-PLUS bands, respectively.
    }
    \label{fig:lightcurve_sims_highcadence}
\end{figure}

Here we consider a High-Cadence Time-domain Survey in which observations are conducted once per month, but only during a contiguous 6-month observing season each year, over a total baseline of five years. 
The per-epoch exposure time is assumed to be 10 hours, corresponding to a limiting magnitude of $m_{\rm lim}(SNR=5) = 26$~mag in the F232 band.
We assume that data are obtained in all five GREX-PLUS bands at each epoch.

As described above, the LSST optical survey is currently planned to operate from 2026 until 2035 \citep{panda2026agnvariabilityrubinobservatory}, whereas GREX-PLUS will likely begin operations after 2035. 
Although LSST observations may continue beyond that date, in this study, we follow the up-to-date LSST survey simulations ({\tt baseline\_v5.0.0\_10yrs.db})\footnote{\url{https://community.lsst.org/t/release-of-v5-0-simulations/11114}} and assume that LSST observations are available only through 2035.
Under this assumption, LSST photometry will tightly constrain the UV--optical variability of AGNs prior to 2035, and we then consider a scenario in which the GREX-PLUS light curves obtained after 2035 capture the dust reverberation signals, with observer-frame time delays of several hundred days or longer.

The simulated GREX-PLUS light curves $s_{\rm dust}(t)$ were sampled once per month for six contiguous months each year, from January~1, 2035 to June~1, 2040.

The continuum light curves at the observer-frame LSST bands $s_{x}(t)$ (where $x = u, g, r, i, z,$ or $y$) were obtained by scaling the amplitude of the $\lambda_{\rm rest} = 5100$~\AA\ light curve $s(t)$ by ${\rm SF}_{\infty}(\lambda_{{\rm LSST-}x})/{\rm SF}_{\infty}(5100~{\text{\AA}})$; i.e., $s_{x}(t) = \frac{{\rm SF}_{\infty}(\lambda_{{\rm LSST-}x})}{{\rm SF}_{\infty}(5100~{\text{\AA}})}s(t)$, without any UV-optical interband continuum lags assumed in this calculation.
Note that the UV--optical interband continuum lags are typically a few days, much shorter than the dust reverberation lags. 
These continuum lags are expected to be tightly constrained by LSST observations \citep[e.g., see simulations by][]{2025arXiv251121479Y}, which can be used to correct for their influence on the dust reverberation lag measurements if necessary.
The optical light curves were then unevenly sampled according to the COSMOS Deep Drilling Field (DDF) samplings given in the LSST survey simulation {\tt baseline\_v5.0.0\_10yrs.db}.
Uncorrelated Gaussian measurement noise, with variance calculated from the $5\sigma$ depth per epoch (30~sec) provided in the LSST survey simulation and the scaling relation $m_{\rm lim}(t_{\rm exp}, {\rm SNR}) = m_{\rm lim}(30~{\rm sec}, 5\sigma) + 1.25 \log_{10}\left(t_{\rm exp}/30~{\rm sec}\right) - 2.5 \log_{10}\left({\rm SNR}/5\right)$, was added to the sampled optical light curves.

An example of simulated GREX-PLUS and LSST COSMOS DDF light curves is shown in Figure~\ref{fig:lightcurve_sims_highcadence}.
The input observer-frame dust reverberation lags are 
$t_{\rm lag}=482, 568, 680, 800, 944$~days in the F232, F303, F397, F520, and F680 bands, respectively.
As seen in Figure~\ref{fig:lightcurve_sims_highcadence}, under the current survey plans, the temporal overlap between the LSST and GREX-PLUS observing periods is limited, making it difficult to directly identify the optical--infrared time delay within the light curves.
Nevertheless, the high-cadence observations are capable of capturing subtle flux variations on observer-frame timescales of approximately one month.
If the driving accretion-disk light curves are densely and robustly constrained by LSST observations, it is in principle possible for GREX-PLUS to detect the delayed responding light curves even in cases where the time lag exceeds one year.
A more quantitative assessment, however, will require detailed simulations once the final survey cadence is determined.


A more detailed assessment of how well these dust reverberation lag measurements can constrain the dust lag-luminosity relation over a wide luminosity range and thereby the cosmic expansion history (e.g., the Hubble diagram) will require dedicated simulations of lag detectability \citep[e.g.,][]{2014ApJ...784L...4H,2015ApJS..216....4S,2015MNRAS.453.1701K,2017MNRAS.464.1693H,2021JKAS...54...37K}. 
Such simulations will be carried out as part of our future work.

\subsection{Scientific goals}

With the wide-field IR time-domain observations of GREX-PLUS, we can construct a deep and nearly complete variability-selected AGN sample, reaching down to $m_{F397} \sim 23$--24~mag and including not only unobscured but also mildly obscured AGNs. 
Such a sample will, for the first time, allow us to directly constrain the bolometric AGN luminosity function around its knee, where the contribution of obscured systems remains poorly measured.

Furthermore, the high-cadence IR time-domain survey will enable multi-wavelength dust reverberation mapping for thousands of AGNs across the five GREX-PLUS bands. 
By measuring dust reverberation lags as a function of wavelength, we will establish a robust wavelength-dependent dust lag-luminosity relation. 
Applying this relation to the large high-$z$ AGN samples will allow us to construct an AGN Hubble diagram, providing an independent cosmological probe capable of constraining the cosmic expansion history out to $z\sim3$.

Together, these efforts will make GREX-PLUS a uniquely powerful mission for both AGN demographics and cosmology.

\printbibliography[heading=subbibliography]
\end{refsection}

\clearpage

\begin{refsection}[2-18_AGNmolecularoutflow/AGNmolecularoutflow.bib]

\section{AGN Molecular Outflow}
\label{sec:AGNmolecularoutflow}

\noindent
\begin{flushright}
Shunsuke Baba$^{1}$
\\
$^{1}$ ISAS/JAXA
\end{flushright}
\vspace{0.5cm}

\subsection{Scientific background and motivation}

Gas outflows from active galactic nuclei (AGNs) have been proposed as a mechanism regulating the coevolution of supermassive black holes and their host galaxies.
Such outflows have been observed in multiple phases and on multiple velocity scales, from fast ionized gas to slow molecular gas, but it is the molecular outflow that accounts for the majority of the mass.
In addition, stars form from gas in the molecular phase.
Therefore, understanding the kinematics of molecular outflows is essential for assessing AGN feedback effects on the host galaxy.
Observations of nearby ultra-luminous infrared galaxies (ULIRGs, $L_\mathrm{IR}>10^{12}\,{\rm L}_\odot$) have shown that the kinetic power and momentum rate of molecular outflows correlate with AGN luminosity, indicating that AGNs are the dominant source driving the outflow \citep{2014A&A...562A..21C}, although there may be a large scatter in the coupling efficiency and momentum boost \citep{2019MNRAS.483.4586F}.
Investigating what types of molecular gas outflows were driven by AGNs in distant galaxies at epochs of more active star formation will help us understand how present-day massive galaxies were formed.
The innermost spatial scale at which an AGN can directly affect the molecular gas is the region corresponding to the molecular torus that surrounds the nucleus ($\lesssim$10\,pc).
In this subsection, we propose a method to observe the kinematics of the molecular gas in this region.

\subsection{Required observations and expected results}

\begin{figure}[htbp]
\centering
\includegraphics[width=\textwidth]{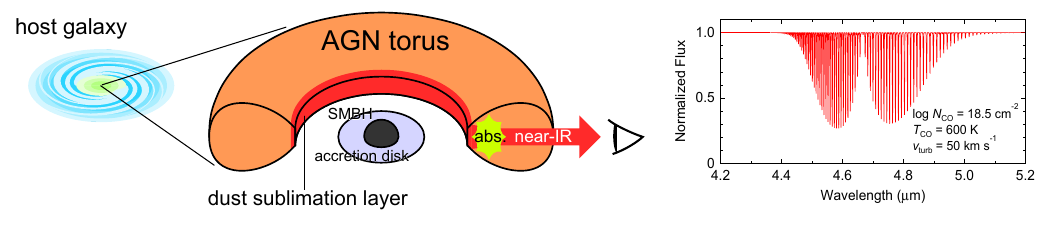}
\caption{
Left: schematic of the observation of an AGN torus region using near-infrared absorption lines.
The thermal emission from the dust sublimation layer is the dominant continuum emission source over the galaxy, and by spectroscopic observations of absorption lines against it, effectively high spatial resolution can be achieved.
Right: example of a CO vibrational-rotational absorption spectrum (column density $N(\mathrm{CO})=10^{18.5}\,\mathrm{cm^{-2}}$, temperature 600\,K, velocity width 50\,km\,s$^{-1}$).
Lines of different rotational levels appear at intervals of $\sim$0.01\,$\mu$m around the rest wavelength of 4.67\,$\mu$m.
}
\label{fig:Baba_torus}
\end{figure}

The biggest obstacle to observing the vicinity of an AGN is the small physical size of the region itself.
Its angular size is only a few hundred milliarcseconds, even for the nearest AGNs.
This hinders imaging observations of distant AGNs.
This problem can be solved by spectroscopic observations of absorption lines in the rest near-infrared region.
Figure~\ref{fig:Baba_torus} illustrates the concept of this method: the dominant near-infrared continuum in an AGN host galaxy is thermal radiation from dust heated by the nucleus, especially from a compact region where the dust is sublimating.
With absorption line spectroscopy, the effective spatial resolution is determined by the size of the background source, and the vicinity of the AGN can be selectively observed with minimal influence from the host galaxy.
In the near-infrared region, a useful absorption line band is the vibrational-rotational transition of CO ($v=1\leftarrow0$, $\Delta J=\pm1$, band center $\sim$4.67\,$\mu$m, line spacing $\sim$0.01\,$\mu$m).
An example spectrum is shown in Figure~\ref{fig:Baba_torus}.
Dozens of lines belonging to different rotational excitation levels can be observed at once, which helps us effectively constrain the gas column density, temperature, and line-of-sight velocity.
This is an advantage that CO pure rotational emission lines in the submillimeter region do not have.

Previous observations of CO ro-vibrational absorption lines have been performed for nearby AGNs;
\citet{2013PASJ...65....5S} and \citet{2021ApJ...921..141O} observed the ULIRG IRAS 08572$+$3915 NW, which has a buried AGN, with the Infrared Camera and Spectrograph (IRCS) of the Subaru telescope at wavelength resolution of $R\sim10,000$ ($\Delta v\sim 30\,\mathrm{km\,s^{-1}}$) and detected an outflow component at $-160\,\mathrm{km\,s^{-1}}$ in each line of CO.
This component is found to have a high temperature (several hundred K) and a large column density ($N(\mathrm{CO})\sim5\times10^{18}\,\mathrm{cm^{-2}}$) from its level population diagram, suggesting that it is in the vicinity of the AGN, such that X-rays heat it.
Thus, CO absorption at rest-frame near-infrared wavelengths is a unique probe sensitive to outflows near AGNs.
Unfortunately, ground-based observations with such high spectral resolution have been carried out for only a few nearby objects, let alone distant ones, due to poor atmospheric transmittance.
\citet{2023ApJ...951...87O,2024ApJ...976..106O} conducted Subaru/IRCS observations similar to those of IRAS 08572$+$3915 NW on three other (U)LIRGs and discussed their gas dynamics; however, all of these objects were in the local Universe ($z<0.05$).
There are some examples of space-based observations using ISO, Spitzer, and AKARI \citep{2004A&A...426L...5L,2004ApJS..154..184S,2018ApJ...852...83B,2022ApJ...928..184B}, but in these cases, individual rotational lines are not resolved due to poor spectral resolution.
As an example using JWST, \citet{2024A&A...682A.182G} observed CO absorption lines with NIRSpec toward a massive clump in the nearby merging galaxy VV 114 E and detected two outflow components of different blueshifts.
However, as the spectral resolution of NIRSpec is at most $R=2,700$, the velocity profiles of individual lines were just barely resolved.
Unfortunately, for JWST/MIRI, which covers 4.9--27.9\,$\mu$m and could observe CO lines in high-$z$ AGNs, the spectral resolution is not higher than that of NIRSpec.

Observing outflows through CO ro-vibrational absorption lines for distant AGNs, which is not possible with existing telescopes, is a science case that will be pioneered for the first time with GREX-PLUS.
If the spectral resolution of the high-dispersion spectrograph is better than $R=10,000$ ($\Delta v\sim 30\,\mathrm{km\,s^{-1}}$), the velocity decomposition of CO lines can be adequately achieved.
If the 10--18\,$\mu$m wavelength range is covered, the CO band (rest 4.67\,$\mu$m) can be observed at redshifts beyond $z=1.2$, an important era in galaxy evolution.
One concern is that the sensitivity of the spectrograph is a bit challenging.
The current nominal sensitivity to continuum emission is 6\,mJy at the 5$\sigma$ level for a one-hour integration (low background case).
Let us use IRAS 00397$-$1312 ($z=0.26$, $L_\mathrm{IR}=1\times10^{13}\,{\rm L}_\odot$) as an example of a luminous AGN that shows CO absorption.
If this galaxy were at $z=1.2$, the S/N reached in a 6-hour integration would be 2.7 at rest-frame 5\,$\mu$m.
Although multiple-line stacking as described below can be employed to obtain velocity profiles, it is essential first to find a promising bright target in the hyper-luminous infrared galaxy (HyLIRG, $L_\mathrm{IR}>10^{13}\,{\rm L}_\odot$) class.

\begin{figure}[htbp]
\centering
\includegraphics[width=\textwidth]{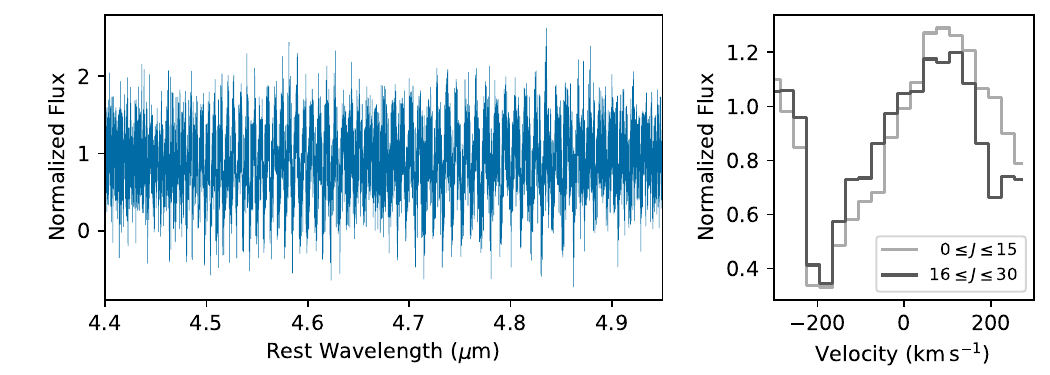}
\caption{
Left: expected $R=25,000$ spectrum of the outflow described in this text observed from the same line of sight as the propagation direction with S/N=2.7 for the continuum level.
Right: velocity profiles for the lines of different rotational levels $J$ averaged for each low and high excitation regime.
The binning is applied with a velocity width corresponding to the required resolution ($R=10,000$, i.e., $\delta v =30$\,km\,s$^{-1}$), rather than the nominal wavelength resolution of the high-dispersion spectrometer.
}
\label{fig:Baba_spec}
\end{figure}

Now, let us assume an outflow with a certain mass outflow rate as an example and show what kind of spectrum is expected.
Here, we assume that the velocity decelerates from $-200$\,km\,s$^{-1}$ to $-50$\,\,km\,s$^{-1}$ and the temperature decreases from 800\,K to 600\,K during the propagation process, and the gas density is $10^6$\,cm$^{-3}$ at the base, the opening angle is 90\,deg, and the CO-to-H$_2$ abundance ratio is $10^{-5}$.
The resulting mass outflow rate is $30\,{\rm M}_\odot\,\mathrm{yr}^{-1}$.
Figure~\ref{fig:Baba_spec} (left) shows a spectrum of the CO absorption caused by this outflow expected for S/N=2.7 for the continuum level.
The observational line of sight is assumed to be in the same direction as the outflow.
As this figure shows, it is difficult to analyze the individual lines separately.
However, since the band contains many lines of different rotational excitation levels, the velocity profile can be determined well by averaging (stacking) them.
Figure~\ref{fig:Baba_spec} (right) shows the averaged velocity profiles for low ($0\leq J\leq15$) and high ($15\leq J\leq30$) excitation levels.
Binning is applied with a velocity width of 30\,km\,s$^{-1}$.
In both cases, a P-Cygni profile consisting of absorption and emission, which is characteristic of outflows, is observed.
It can also be seen that the line profiles extend toward lower velocities at low excitation than at high excitation.
This indicates that the outflow is slowing down as the gas moves from the center, where it is warm, outward to where it is cold (i.e., decelerating outflow).
Thus, even with the current nominal sensitivity, the overall velocity profile of the gas can be studied by line stacking, and if the temperature change associated with propagation is significant, velocity changes in an outflow can also be discussed.
The required wavelength resolution, wavelength coverage, and sensitivity are summarized in Table \ref{tab:Baba_requirements}.

\subsection{Scientific goals}

By performing pointed high-dispersion spectroscopy of CO ro-vibrational absorption lines in HyLIRG-class AGNs at $z>1.2$, we will detect molecular gas outflows with an order of magnitude higher wavelength resolution than existing space telescopes.
This will provide a more detailed examination of the AGN feedback during the era of high galactic activity.

\begin{table}[ht]
    \begin{center}
    \caption{Required observational parameters.}
    \label{tab:Baba_requirements}
    \begin{tabular}{|l|p{9cm}|l|}
    \hline
     & Requirement & Remarks \\
    \hline
    Wavelength & 10--18 $\mu$m &  \\
    \hline
    Spatial resolution & N/A &  \\
    \hline
    Wavelength resolution & $\lambda/\Delta \lambda>10,000$ &  \\
    \hline
    Field of view & N/A &  \\
    \hline
    Sensitivity & 6.0 mJy (for continuum, 1 hr, $5\sigma$) &  \\
    \hline
    Observing field & N/A &  \\
    \hline
    Observing cadence & N/A &  \\
    \hline
    \end{tabular}
    \end{center}
\end{table}

\printbibliography[heading=subbibliography]
\end{refsection}

\clearpage

\begin{refsection}[2-19_IGMmoleculargas/IGMmoleculargas.bib]

\section{Molecular Gas in the Intergalactic Medium}
\label{sec:IGMmoleculargas}

\noindent
\begin{flushright}
Toru Misawa$^{1}$, Shunsuke Baba$^{2}$
\\
$^{1}$ Shinshu University, 
$^{2}$ ISAS/JAXA 
\end{flushright}
\vspace{0.5cm}

\subsection{Scientific background and motivation}

Neutral hydrogen absorbers with very large column densities of log($N_{\rm HI}$/cm$^{-2}$) $\geq$ 20.3 provide strong absorption features (Damped Ly$\alpha$ systems; DLAs) in spectra of background quasars \citep{1986ApJS...61..249W}. In DLAs, hydrogen is mainly neutral, while it is almost ionized in all other classes of quasar absorption systems with log($N_{\rm HI}$/cm$^{-2}$) $\lesssim$ 20. The neutrality of the hydrogen gas in DLAs is important, since such cold neutral clouds could be precursors of molecular clouds, corresponding to the birthplace of stars in young galaxies \citep{2003ApJ...587..278W}. Indeed, a clear chemical evolution is found in DLAs; the metallicity increases by a factor of $\sim$50--100 from $z=5$ to today \citep{2018A&A...611A..76D}.

Since the first detections in the 1960s, more than 50,000 DLAs have been discovered so far \citep{2022ApJS..258...18C}. \citet{2014A&A...566A..24N} recently discovered $\sim$100 extremely strong DLAs (ESDLAs) with log($N_{\rm HI}$/cm$^{-2}$) $\geq$ 21.7 in quasar spectra from the Baryon Oscillation Spectroscopic Survey (BOSS) of the Sloan Digital Sky Survey (SDSS), and confirmed that their column-density frequency distribution becomes steeper than what is seen at lower column densities as \citet{2005ApJ...635..123P} had already pointed out, which could be due to the conversion of neutral hydrogen atoms to molecular hydrogen (H$_2$), dust obscuration of the background quasars, and/or a low covering factor (i.e., a small size) of the absorbers. Based on a statistical analysis of the column density distribution function, ESDLAs are supposed to have column densities of up to log($N_{\rm HI}$/cm$^{-2}$) $\sim$ 22.5 \citep{2021MNRAS.507..704H}.
In such high-column-density absorbers, the presence of molecular hydrogen (H$_2$) has been theoretically expected \citep{2004ApJ...609..667S}. The molecular fraction, $f$(H$_2$) = 2N(H$_2$)/[2N(H$_2$)+N(HI)], actually increases according to the total hydrogen column density in the ISM toward Galactic stars \citep{1977ApJ...216..291S}. Extragalactic molecular hydrogen has also been discovered in high-column-density absorbers at high redshifts, such as ESDLAs, through the detection of Lyman- and Werner-band absorption lines \citep{2014A&A...566A..24N, 2019MNRAS.490.2668B}.

\begin{figure}[ht]
   \begin{center}
   \includegraphics[width=100mm,angle=270]{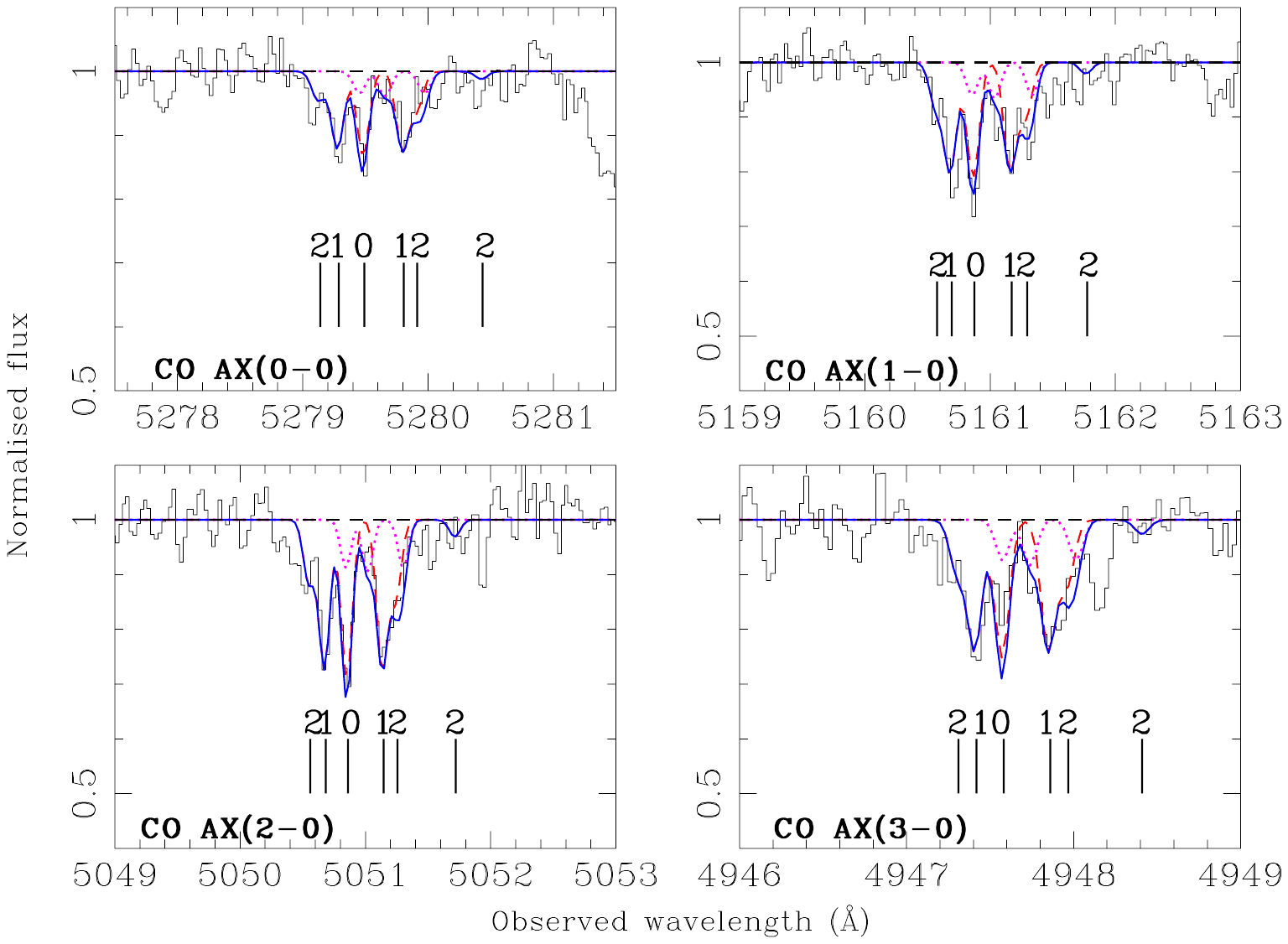}
   \end{center}
   \caption{Voigt profile fits (blue curves) to the 4.7~$\mu$m CO bands at $z_{\rm abs}=2.4185$ toward a quasar SDSS J143912.04+111740.5 (black histograms). The vertical lines with numbers are the locations of different CO transitions from different $J$ levels. The figure is taken from \citet{2008A&A...482L..39S}.}
   \label{fig:2_10_1}
\end{figure}

Recently, carbon monoxide (CO) molecular bands have also been detected in quasar spectra through the detection of A--X band absorption lines (Fig.~\ref{fig:2_10_1}), although the sample size ($\sim$10) is still small \citep{2008A&A...482L..39S, 2018A&A...612A..58N, 2019A&A...625L...9G,2024MNRAS.533.1367K}. For example, the first discovered CO-bearing DLA system at $z$ $\sim$ 2.42 has an extremely large molecular fraction ($f$ $\sim$ 0.27), a large metallicity with a dust depletion, and a column density ratio of N(CO)/N(H$_2$) = 3$\times$10$^{-6}$ \citep{2008A&A...482L..39S}. Such conditions correspond to cold, well-shielded, and chemically evolved molecular gas. Thus, CO-bearing DLAs, as well as H$_2$-bearing ones, are crucial for studying the star/galaxy formation history at high redshift.

Since column densities of CO molecular clouds can be measured with high accuracy of $\sim$0.02~dex by applying Voigt-profile fitting, we can not only estimate their physical properties like metallicity and molecular fraction, but also use them to measure the temperature of the cosmic microwave background radiation (CMBR) as follows. Assuming that the relative level populations of the CO rotational levels are thermalized by the CMBR, \citet{2008A&A...482L..39S} measured the CMBR temperature at $z$ $\sim$ 2.42 as T$_{\rm CMBR}$ = 9.15$\pm$0.72~K, while 9.315$\pm$0.007~K is expected from theory (i.e., T$_{\rm CMBR}$($z$) = T$_{\rm CMBR}$($z=0$)$\times$(1+$z$)). Among several methods of measuring T$_{\rm CMBR}$ at high redshift \citep{2022Natur.602...58R}, this is one of the most accurate measurements.

In previous studies, including \citet{2008A&A...482L..39S}, extragalactic CO molecular clouds have been detected through the A--X bands that correspond to electronic transitions in the ultraviolet region in the rest frame (optical wavelengths in the observed frame). They are also accessible through the detection of absorption lines corresponding to rotational and vibrational (ro-vib) transitions at $\lambda$ $\sim$4.7~$\mu$m in the NIR region in the rest frame (MIR region in the observed frame). 
The high-resolution MIR spectrograph with GREX-PLUS (covering 10--18$\,\mu$m) will, for the first time, enable us to start a systematic search for the 4.7~$\mu$m CO bands in ESDLAs at high redshift.

\subsection{Required observations and expected results}

\begin{figure}[ht]
   \begin{center}
   \includegraphics[width=150mm,angle=0]{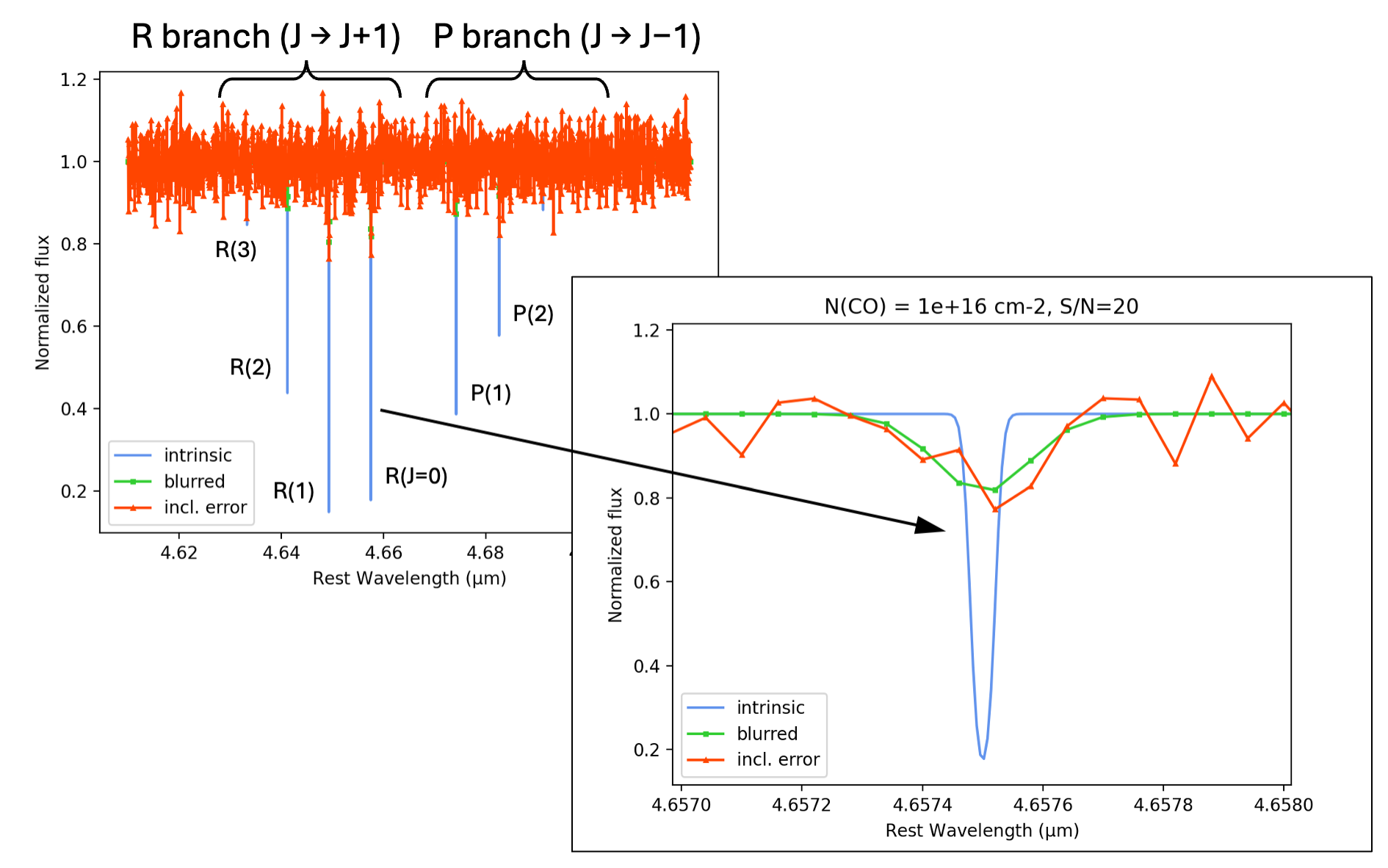}
   \end{center}
   \caption{Synthesized spectrum around the 4.7~$\mu$m CO ro-vib bands at $z=2.2$. We assume log($N_{\rm CO}$/cm$^{-2}$) = 16, T$_{\rm kin}=100$~K, and n(H$_2$) = 18 cm$^{-3}$, following \citet{2008A&A...482L..39S}, \citet{2019MNRAS.490.2668B}, and \citet{2019A&A...625L...9G}. The blue curve denotes the intrinsic spectrum with a thermally broadened line width, while green and red curves are those after convolution with $R=30,000$ and adding noise to produce an S/N = 20 pixel$^{-1}$ spectrum. Both the R and P branches are clearly detected. We iterate model fits to the synthesized spectrum using four free parameters (T$_{\rm CMBR}$, T$_{\rm kin}$, n(H$_2$), and log$N_{\rm CO}$) after changing a seed for noise. The accuracy of the T$_{\rm CMBR}$ measurement ($\delta$T) depends strongly on the S/N ratio; $\delta$T $\sim$ 1~K or 0.1~K if S/N = 20 or 1000 pixel$^{-1}$.}
   \label{fig:2_10_2}
\end{figure}

In this proposal, we aim to detect the 4.7~$\mu$m CO ro-vib bands in ESDLAs, which will be the first systematic survey of the 4.7~$\mu$m CO bands at high redshift. We also attempt to measure T$_{\rm CMBR}$ based on the relative strength of several transitions of both the R branch (J $\rightarrow$ J+1) and the P branch (J $\rightarrow$ J-1), and verify the past measurements of T$_{\rm CMBR}$. To achieve these goals, both high spectral resolution ($R\sim25,000$) and high signal-to-noise ratio (S/N $>$ 20 pixel$^{-1}$) are required. As a feasibility test, we perform a simulation as follows: 1) we assume an ESDLA system with a CO column density of log($N_{\rm CO}$/cm$^{-2}$) = 16 at $z=2.2$, with a kinetic temperature of T$_{\rm kin}=100$~K and a gas volume density of n(H$_2$) = 18 cm$^{-3}$, following previous studies \citep{2008A&A...482L..39S, 2019MNRAS.490.2668B, 2019A&A...625L...9G}, 2) we assume that CO rotational excitation is dominated by radiative pumping by the CMBR at $z=2.2$ (T$_{\rm CMBR}$ = 8.74~K) but is partially contributed by collisional excitation, 3) we run the non-LTE radiative transfer code RADEX \citep{2007A&A...468..627V} for the modeled absorption system to calculate the column densities of CO molecules in the ground and excited levels, 4) we synthesize a spectrum around the 4.7~$\mu$m CO ro-vib bands with a thermally broadened line width, convolve it assuming a spectral resolution of $R$ = 25,000, and then add noise to produce a spectrum with S/N = 20 pixel$^{-1}$ (Fig.~\ref{fig:2_10_2}), and finally 5) we repeat model fitting to the synthesized spectrum using four free parameters (T$_{\rm CMBR}$, T$_{\rm kin}$, n(H$_2$), and log$N_{\rm CO}$). The accuracy of the T$_{\rm CMBR}$ measurement ($\delta$T) depends strongly on the S/N ratio. For example, $\delta$T $\sim$ 1~K (similar to those measured in previous studies) if S/N = 20 (i.e., the minimum S/N ratio we require) and $\delta$T $\sim$ 0.1~K if S/N = 1000.

If we assume WISE Band 3 magnitude (at $\sim12\,\mu$m) of W3 = 9.5 ($\sim$1~mag fainter than one of the brightest quasars at $z$ $>$ 2, J13260399+7023462 \citep{2020ApJ...899...76J}) as a target's magnitude, a 20 hr exposure would provide a $R\sim25,000$ spectrum with S/N $\sim$ 20 pixel$^{-1}$ based on the sensitivity curve of GREX-PLUS high-resolution mode for a point source. Although such bright quasars with W3 $\lesssim$ 9.5 are still rare, a Gaia-assisted quasar survey using a novel combination of astrometry from Gaia-DR3 and photometry from optical, NIR, and MIR missions effectively continues to discover bright quasars even if they are heavily dust-reddened (e.g., \citealt{2018A&A...615L...8H}). GREX-PLUS also plans to conduct its own survey of dust-obscured AGNs. These surveys will provide a number of quasars that are bright enough for our project.

\subsection{Scientific goals}

We will conduct high-resolution spectroscopy ($R=25,000$) of several MIR-bright quasars (W3 $\lesssim$ 9.5) to search for the 4.7~$\mu$m CO ro-vib bands in high-redshift ESDLAs. This will be the first systematic search of the 4.7~$\mu$m CO band to our knowledge. If we successfully acquire MIR spectra with a sufficiently high S/N ratio, we will measure T$_{\rm CMBR}$ based on the relative strength of several transitions of both the R and P branches. We have already confirmed this is feasible if spectra with S/N $>$ 20 pixel$^{-1}$ are available. The total exposure time to acquire spectra with S/N $\sim$ 20 pixel$^{-1}$ is $\sim$20~hours for quasars with W3 $\sim$ 9.5. Here, we emphasize that it is desirable to select targets whose 4.7~$\mu$m CO ro-vib bands fall in spectral regions with relatively higher resolving power, since these bands are not fully resolved at the typical resolving power of GREX-PLUS ($R\sim25,000$). With data of this quality, we will be able to place more stringent constraints on T$_{\rm CMBR}$ at high redshift with higher-quality spectra, which will be a complementary test to future ELT/ANDES observations targeting the A--X bands in the UV region \citep{2024ExA....57....5M} and a direct crucial test of the standard $\Lambda$CDM cosmology.

\begin{table}[ht]
    \begin{center}
    \caption{Required observational parameters.}
    \label{tab:2_10}
    \begin{tabular}{|l|p{9cm}|l|}
    \hline
     & Requirement & Remarks \\
    \hline
    Wavelength & $\sim$14$\mu$m (10--18$\,\mu$m) & $a$ \\
    \hline
    Spatial resolution & $<1$ arcsec & \\
    \hline
    Wavelength resolution & $\lambda/\Delta \lambda \sim 25,000 $ & $b$ \\
    \hline
    Field of view & 8$^{\prime\prime} \times 35^{\prime\prime}$ & \\
    \hline
    Sensitivity & 6.0~mJy (1 hr, $5\sigma$, point source, low BG) & $c$\\
    \hline
    Observing field & N/A & \\
    \hline
    Observing cadence & N/A & \\
    \hline
    \end{tabular}
    \end{center}
    $^a$ Any wavelength is useful as long as the redshifted 4.7~$\mu$m CO bands are covered.\\
    $^b$ $R=25,000$ is sufficient, but higher resolution is better since 4.7~$\mu$m CO lines are probably not resolved if we assume they are thermally broadened with T$_{\rm kin}=100$~K.\\
    $^c$ At least S/N = 20 pixel$^{-1}$ at $\lambda\sim14\,\mu$m is required with a reasonable observing time of $\leq$ 20 hours.
\end{table}

\printbibliography[heading=subbibliography]
\end{refsection}

\clearpage

\begin{refsection}[2-20_magellanicclouds/magellanicclouds.bib]

\section{Magellanic Clouds}
\label{sec:magellanicclouds}

\noindent
\begin{flushright}
Takashi Shimonishi$^{1}$
\\
$^{1}$ Niigata University
\end{flushright}
\vspace{0.5cm}

\subsection{Scientific background and motivation}
The Large and Small Magellanic Clouds (LMC and SMC) are the nearest star-forming low-metallicity dwarf galaxies. 
Their proximity \citep[about 50 kpc for the LMC and 60 kpc for the SMC;][]{Pie13, Gra14} enables us to spatially resolve the distribution of individual stars at moderate angular resolution (1 arcsec $\sim$0.24 pc at the LMC). 
A nearly face-on viewing angle of the LMC ($\sim$27 degrees) and a large separation from the Galactic disk direction allow us to obtain a bird's-eye view of the distributions of stars and the interstellar medium (ISM) in a galaxy. 
Their low-metallicity environments \citep[$\sim$1/2--1/3 for the LMC and $\sim$1/5--1/10 for the SMC; e.g.,][]{Rus92} play an important role as a laboratory for investigating various astrophysical and astrochemical phenomena in environments similar to the past metal-poor Universe. 

Owing to this uniqueness, various types of survey observations have been carried out toward the Magellanic Clouds. 
Infrared surveys are especially useful for detecting and classifying embedded or low-temperature objects, such as young stellar objects (YSOs) and mass-losing late-type stars. 
Studies of extragalactic YSOs have made substantial progress with the advent of high-sensitivity space-borne infrared telescopes such as Spitzer, AKARI, and Herschel. 

$\sim$1000 YSO candidates have been identified in the LMC and SMC through whole-galaxy infrared imaging surveys \citep[e.g.,][]{Whin08,GC09,Kat12,Sew13} and follow-up spectroscopy \citep[e.g.,][]{ST,ST13,Sea09,Woo11,Oli13}. 
Infrared spectroscopic data of embedded YSOs in the LMC/SMC have been used to investigate the chemistry of ices and dust, as well as the spectral properties of PAH emission in protostellar envelopes \citep[e.g.,][]{ST,ST10,ST16,Oli09,Oli11,Sea11}. 
These data have also been used to classify the evolutionary stages of massive YSOs \citep[e.g.,][]{Sea09}.

These studies have provided unique targets for follow-up observations in the radio regime with ALMA, which can spatially resolve molecular cloud cores at the distance of the Magellanic Clouds. 
Based on the synergy between infrared telescopes and ALMA, we have begun to investigate various core-scale ($\sim$0.1~pc) phenomena related to star formation, such as protostellar outflows \citep{Fuk15,ST16b,Tok22} and the emergence of hot molecular cores \citep{ST16b,ST20,ST23,Sew18,Sew22a}. 

Clearly, studies of star formation and the ISM at the star-forming core scale are now being extended to extragalactic and low-metallicity environments. 
However, despite the large number of identified extragalactic YSOs, current observations are biased toward luminous (mostly high-mass) sources due to the sensitivity limits of existing infrared surveys. 

\begin{figure}[thp!]
\begin{center}
\includegraphics[width=11.0cm]{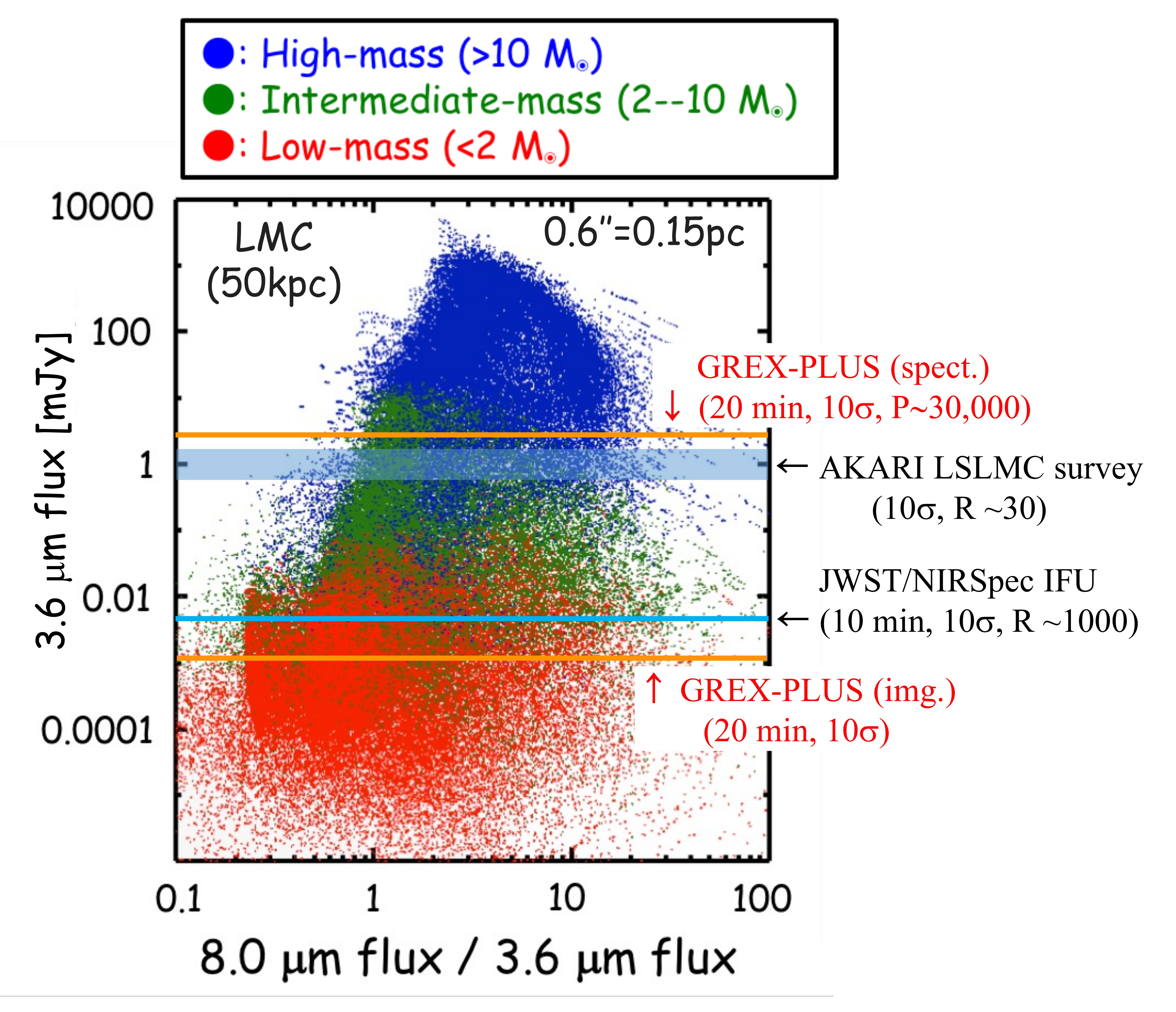}
\caption
{Theoretically predicted infrared fluxes and colors of YSOs at the distance of the LMC (50 kpc), based on \citet{Rob07}. 
Symbols are color-coded by YSO mass: high-mass ($>$10 M$_{\odot}$; blue), intermediate-mass (2--10 M$_{\odot}$; green), and low-mass ($<$2 M$_{\odot}$; red). 
The solid orange lines indicate the expected sensitivities of GREX-PLUS for spectroscopy (upper) and imaging (lower). 
The spectroscopic sensitivities of the AKARI LSLMC survey \citep{ST13} and JWST/NIRSpec/IFU (10 minutes on source) are also shown. }
\label{tab:f1_ST}
\end{center}
\end{figure}

\begin{figure}[thp!]
\begin{center}
\includegraphics[width=13.0cm]{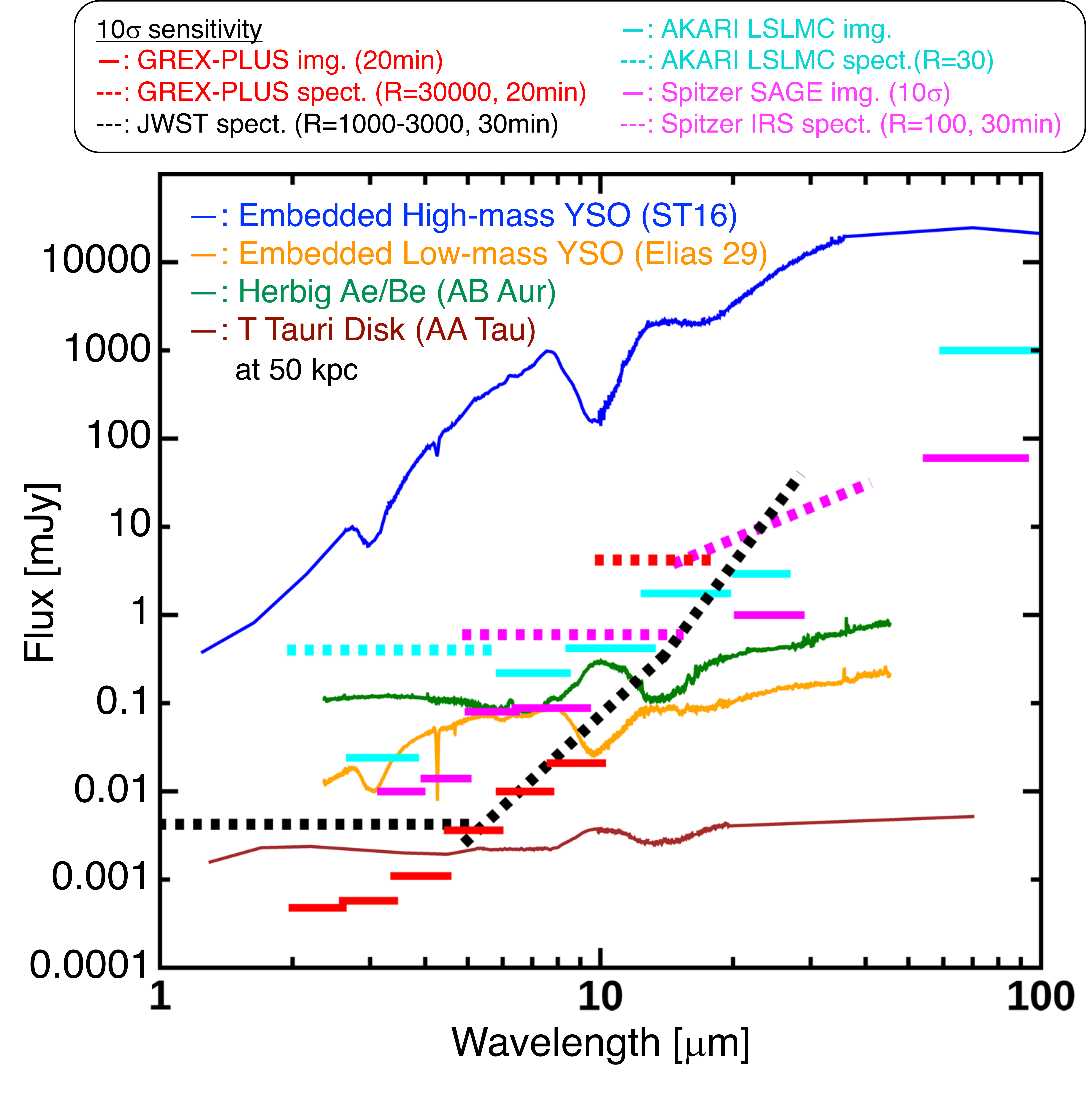}
\caption
{Spectral energy distributions of various objects related to star and planet formation, scaled to the distance of the LMC. 
(Blue) embedded high-mass LMC YSO \citep{ST16}; 
(yellow) embedded low-mass Galactic YSO (Elias 29); 
(green) Herbig Ae/Be star (AB Aur), representing an intermediate-mass protoplanetary disk; 
(brown) T Tauri star (AA Tau), representing a low-mass protoplanetary disk. 
The wavelength coverages and sensitivities of selected infrared telescopes are indicated by horizontal lines (solid: imaging; dotted: spectroscopy), with those of GREX-PLUS shown in red.
}
\label{tab:f2_ST}
\end{center}
\end{figure}

\subsection{Required observations and expected results}
GREX-PLUS is expected to extend the mass limit of extragalactic YSO surveys down to the low-mass regime with its unprecedented survey efficiency and high sensitivity in the near- and mid-infrared. 
Such a survey will enable us to investigate galaxy-scale star-formation activity over a wide range of protostellar masses. 

Figure~\ref{tab:f1_ST} shows theoretically predicted infrared fluxes and colors of YSOs at the distance of the LMC \citep[based on the SED models of][]{Rob07}. 
These results demonstrate that the imaging sensitivity of GREX-PLUS is sufficient to detect low-mass YSOs in the Magellanic Clouds. 
Figure~\ref{tab:f2_ST} shows the wavelength coverage and sensitivity of GREX-PLUS in comparison with those of previous infrared space telescopes, together with expected spectral energy distributions of star-forming objects at the distance of the LMC. 
These figures indicate that GREX-PLUS will significantly lower the mass limit of extragalactic YSO surveys. 

Imaging surveys of embedded YSOs are crucial for identifying the earliest phases of star formation, where protostars are deeply embedded in dense envelopes and remain undetectable at optical wavelengths. 
The mid-infrared wavelength coverage of GREX-PLUS is particularly important for characterizing such objects, as embedded YSOs typically show peak emission in the mid- to far-infrared, and detecting this infrared excess is key to their identification. 
Such surveys will provide a statistically significant, unbiased census of protostellar populations across a wide range of masses, enabling us to investigate how star formation proceeds in diverse environments within the LMC and SMC. 

In particular, the spatial distribution and clustering properties of embedded YSOs offer key constraints on the initial conditions of star formation across the entire galaxy. 
We can investigate how cloud fragmentation, star formation efficiency, and the relative populations of low- and high-mass YSOs, which reflect the initial mass function, vary among different regions within the LMC and SMC that are subject to different radiation-field strengths, metallicities, and degrees of interaction with the Magellanic Bridge. 
These results are essential for understanding how star formation in the Magellanic Clouds differs from that in the Milky Way and for placing constraints on star formation in environments analogous to those in the early Universe. 

For such a study, a multi-band (2--10~$\mu$m) survey covering the entire LMC ($\sim$60~deg$^2$) and SMC ($\sim$30~deg$^2$) is desirable. 
An imaging survey of the Magellanic Bridge would also be of great interest. 
High mid-infrared sensitivity is critical for detecting embedded objects. 
For example, with photometric sensitivities of 0.001~mJy at 3--4~$\mu$m and 0.01~mJy at 8--10~$\mu$m, an embedded low-mass Class~I protostar such as Elias~29 would be detectable at the distance of the LMC.

\vspace{10pt}
The high-dispersion mid-infrared spectroscopic capability of GREX-PLUS will provide a unique opportunity to probe the chemical composition of high-temperature molecular gas in the vicinity of protostars. 
Embedded YSOs are known to exhibit various infrared absorption bands arising from associated molecular gas and ice. 
High-dispersion mid-infrared spectroscopy is a powerful tool for resolving ro-vibrational transitions of high-temperature gaseous molecules \citep[e.g.,][]{Boo98,vD04,Lah06,Dun18,Ind20}. 

A variety of astrochemically important molecular species, such as H$_2$O, CO$_2$, C$_2$H$_2$, HCN, CH$_4$, and NH$_3$, are detectable in the mid-infrared. 
In particular, for "radio-inactive" molecules such as CO$_2$, CH$_4$, and C$_2$H$_2$, infrared spectroscopy is the only way to directly detect them. 
In addition, observations of gaseous H$_2$O at radio wavelengths are severely affected by Earth's atmosphere, often making them difficult to carry out from the ground. 
These molecules are expected to be abundant in high-temperature regions of protostellar envelopes. 
Investigating their abundances and chemical diversity based on a statistically significant sample of YSOs is essential for understanding the emergence of chemical complexity in a low-metallicity ISM. 

At the distance of the Magellanic Clouds, the spectroscopic sensitivity of GREX-PLUS will allow us to probe most high-mass YSOs with high-dispersion spectroscopy. 
Approximately 300 infrared sources have been identified as high-mass YSOs in the LMC and SMC through infrared spectroscopy with Spitzer and AKARI. 
A systematic high-dispersion spectroscopic survey of these targets is therefore highly desirable. 

Such observations will provide statistical data on the chemistry of forming stars in diverse galactic environments, and will reveal how the chemical compositions of protostellar envelopes vary depending on stellar luminosity, evolutionary stage, and local environmental conditions such as metallicity, radiation field strength, degree of clustering, and location within a galaxy. 
Furthermore, these spectroscopic data can be directly compared with previous high-resolution studies of Galactic massive protostars. 
Such comparisons will enable us to quantify how chemical compositions and molecular abundances vary as a function of metallicity and environment.

\subsection{Scientific goals}
In summary, GREX-PLUS observations of the Magellanic Clouds will provide an unprecedented dataset for studying ongoing star formation and chemical evolution in nearby low-metallicity galaxies. 
A whole-galaxy infrared imaging survey will reveal star-formation activity across a wide mass range, while high-dispersion spectroscopy will uncover the chemical recipe of stars and planets at low metallicity and its relation to local interstellar conditions. 
Such a galaxy-scale systematic survey is not feasible for Galactic YSOs, as a bird's-eye view of the Milky Way cannot be obtained and observations are significantly affected by foreground contamination. 
Finally, we note that the outer Galaxy is another important nearby laboratory for studying star formation and the ISM under low-metallicity conditions \citep[e.g.,][]{Bra07,Yas08,Izu17,Izu22,ST21,Ike25,Ike26}, and infrared imaging and spectroscopic surveys of such regions represent an additional promising science case for GREX-PLUS. 

In this chapter, we have discussed the science cases of GREX-PLUS related to star formation and astrochemistry in the Magellanic Clouds. 
We also note that other potential science cases, such as those involving mass-losing late-type stars, diffuse interstellar bands, polycyclic aromatic hydrocarbons, and supernova remnants, will also be important for GREX-PLUS observations of the Magellanic Clouds and should be explored in the future.

\printbibliography[heading=subbibliography]
\end{refsection}

\clearpage

\begin{refsection}[2-21_AGBdust/AGBdust.bib]

\section{AGB Dust Feedback in Nearby Galaxies}
\label{sec:AGBdust}

\noindent
\begin{flushright}
Noriyuki Matsunaga$^{1}$
\\
$^{1}$ University of Tokyo
\end{flushright}
\vspace{0.5cm}

\subsection{Scientific background and motivation}

The formation and evolution of cosmic dust is a central issue in understanding how galaxies grow and recycle their baryons. In the nearby Universe, dust plays a key role in regulating radiative transfer, cooling, chemical reactions, and star formation. Observations of both local and high-redshift galaxies reveal dust masses that cannot be explained solely by production from stars such as asymptotic giant branch (AGB) stars and supernovae \citep{Zhukovska-2008,Galliano-2018}, often referred to as the dust budget crisis \citep{Rowlands-2014}. This discrepancy has motivated significant advances in theoretical models that include grain growth in the interstellar medium \citep{Asano-2013a,Asano-2013b,RemyRuyer-2014}. In such models, stellar dust acts as the ``seed population'' on which accretion of metals in the ISM becomes efficient once the ambient metallicity exceeds a threshold. This threshold, known as the critical metallicity \citep{Asano-2013a}, marks a transition between stellar-dominated dust production at low metallicity and ISM-dominated dust growth at higher metallicity. Furthermore, the critical metallicity depends on the chemical species that form dust: ${\sim}0.05\,Z_\odot$ for iron, ${\sim}0.2\,Z_\odot$ for silicates, and ${\sim}0.5\,Z_\odot$ for carbonaceous dust \citep{Choban-2024}.

AGB stars are among the most important stellar dust factories in the Universe. Through slow, radiation-driven winds, they return material enriched with heavy elements into the ISM over timescales of $10^5$--$10^6$~years \citep{Schneider-2014,Goldman-2022}. The chemistry of the dust they produce---oxygen-rich silicates or carbon-rich amorphous carbon---is determined by the C/O ratio in their atmospheres and varies strongly with metallicity and star-formation histories \citep{Hofner-2018,Choban-2024}. Mapping how O-rich and C-rich AGB stars populate galaxies of different masses and metallicities provides a direct observational handle on the chemical pathways available for dust formation. While numerous studies in the Local Group have established empirical correlations between metallicity, C/O ratio, and AGB mass loss \citep{Boyer-2015a,Boyer-2019,Jones-2023}, very little is known about dust-enshrouded AGB populations in more distant systems beyond a few megaparsecs. Present constraints rely almost exclusively on limited-field surveys with Spitzer, such as DUSTiNGS \citep{Boyer-2015b}, that miss significant fractions of galaxy disks and cannot resolve radial variations or population gradients.

A deep, wide-field mid-infrared survey covering $2$--8\,$\mu$m offers a new opportunity to address these questions. This wavelength regime is sensitive to thermal emission from circumstellar dust shells, allowing both the detection of heavily obscured AGB stars and the characterization of their chemistry through broadband colors. When combined with theoretical dust-evolution models, such a survey can determine whether the composition of stellar dust traces the metallicity and star-formation structure of galaxies, and whether the stellar dust supply regulates the onset of rapid ISM grain growth. By extending studies of dusty AGB stars beyond the Local Group to galaxies within {$\sim$}20\,Mpc, including systems in the Virgo and Fornax clusters, the proposed survey will provide a large-scale systematic census of stellar dust sources in diverse galactic environments. Such GREX-PLUS surveys would complement and extend emerging deep surveys of AGB stars in relatively distant galaxies with Roman and JWST \citep{Boyer-2024}.

\subsection{Required observations and expected results}

A survey capable of advancing this field must achieve three key observational capabilities: sensitivity to dust-rich AGB stars, multi-band photometry across $2$--8\,$\mu$m, and sufficient area coverage to map entire galaxies. Depths of 24--26~AB mag in the near- and mid-infrared allow detection of dusty AGB stars and large-amplitude Mira variables out to distances of {$\sim$}20\,Mpc. At these distances, a typical 30-kpc galactic disk spans only a few arcminutes, which would be entirely covered by GREX-PLUS footprints. Five photometric bands across $2$--8\,$\mu$m provide sensitivity to both silicate-rich and carbon-rich dust shells, enabling classification based on color--color diagnostics analogous to those derived from Spitzer/IRAC observations in the Local Group, although adding near-IR photometric data or any spectroscopic data would be very beneficial for the classification \citep{Reiter-2015,Boyer-2024}. Importantly, repeated imaging enables time-domain analysis: measuring amplitudes of variable AGB stars, identifying extreme mass-losing Miras, and distinguishing them from red supergiants or background contaminants. Long-term monitoring over a few years (with $\sim$20 or more epochs) would be necessary to determine periods of Miras considering their long periods (from $\sim$100 to hundreds of days), but their amplitudes are uniquely large and even a few-epoch time-series data would enable us to identify good Mira candidates.

\begin{figure*}[ht!]
\begin{center}
    \includegraphics[width=15cm]{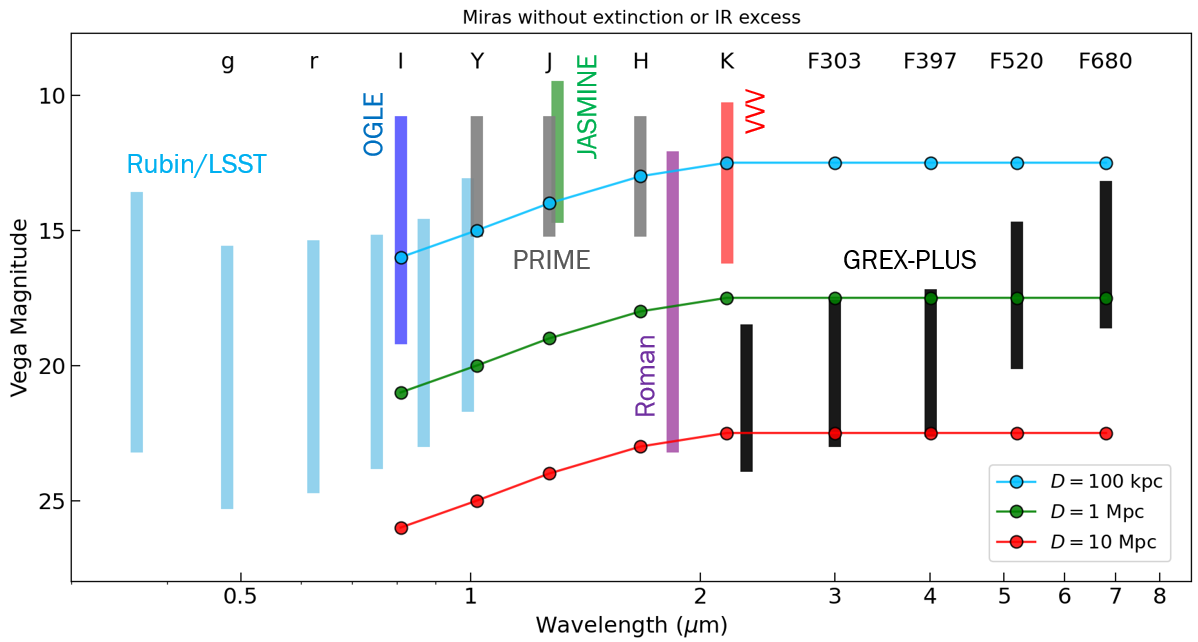}
\end{center}
\caption{Expected brightnesses of Miras ($P=300$~days, no dust excess or absorption) and coverage of various surveys indicated in the panel.}
\label{fig:AGBdust_fig1}
\end{figure*}

From such a dataset, catalogs of O-rich and C-rich AGB stars over wide regions of galactic disks can be produced for many galaxies within GREX-PLUS survey areas. At $5$--10~Mpc, the survey will detect several tens of thousands of AGB candidates in massive spirals, including the faint and heavily reddened populations that dominate dust injection (Figure~\ref{fig:AGBdust_fig1}). As is evident from this figure, GREX-PLUS is the only project capable of fully detecting faint objects as far as 10~Mpc away. Even at $15$--20~Mpc, the survey remains sensitive at $2$--4\,$\mu$m to the brightest, most dust-rich AGB stars whose luminosity and mass-loss rates make them key contributors to the total dust return budget. Multi-band colors will allow estimates of circumstellar optical depth and approximate dust mass-loss rates using radiative-transfer grids such as GRAMS \citep{Riebel-2012}. Spatial distributions of AGB chemistries can then be mapped across each galaxy, reflecting internal metallicity gradients and tracing regions where conditions favor the production of carbonaceous versus silicate dust. Despite its potential importance, however, iron dust would present gray, featureless signals that are hard to confirm with current knowledge. While some studies have reported that its presence is necessary to explain the SEDs of specific objects, mainly metal-poor AGBs with infrared excess \citep{McDonald-2010,Marini-2019,Boyer-2025}, methods for determining the iron dust content need to be established (such as improving the decomposition of various dust grains affecting the broad SEDs to address the presence/absence of featureless iron dust).

The observational outputs---AGB catalogs, dust-chemistry maps, and variability information---can be directly compared with existing multiwavelength datasets. Far-infrared dust maps from Herschel and JWST reveal the distribution of ISM dust, while ALMA observations trace molecular gas reservoirs. Optical and near-infrared imaging from HST and Roman provide stellar population context. Combining these datasets with the mid-infrared AGB survey will enable the first empirical comparison between the stellar dust supply and the ISM dust reservoir in galaxies outside the Local Group. This comparison is essential for constraining dust-evolution models and quantifying the conditions under which ISM grain growth becomes dominant.

\subsection{Scientific goals}

The overarching scientific objective is to determine how AGB stars regulate the supply of dust seeds to galaxies and to test how this supply controls the balance between stellar dust production and ISM grain growth. By extending studies to a statistically meaningful sample of galaxies within {$\sim$}20\,Mpc, the survey can explore how dust injection varies with galaxy mass, morphology, metallicity, and environment. A central goal is to quantify the O-rich and C-rich AGB populations, particularly their spatial gradients, in each system and assess how dust chemistries reflect AGB populations. These gradients reveal how internal chemical evolution shapes dust chemistry, allowing direct tests of the prediction that carbon-rich dust production becomes inefficient at higher metallicity. 
Furthermore, it would be possible to constrain the onset of rapid ISM dust growth by comparing stellar dust maps with the distribution of ISM dust. The critical metallicity predicted by theoretical models depends sensitively on dust composition: galaxies with more efficient stellar production of carbonaceous or silicate grains may cross this threshold earlier \citep{Asano-2013a,Choban-2024}. On the other hand, the C-rich star formation is even more sensitive to the metallicity and changes dramatically at around the solar metallicity \citep{Boyer-2019,Jones-2023}. By measuring the relative contributions of O-rich and C-rich dust production across galaxies of different metallicities, the survey would be able to determine whether the dust seed population by AGB stars is sufficient to trigger ISM grain growth. 

Finally, time-domain information allows identification of extreme mass-losing Miras, which dominate the instantaneous dust injection rate despite their short lifetimes. Pulsation drives dust formation and mass loss \citep{Hofner-2018}. These stars provide a direct measure of current dust feedback, while the overall AGB populations reflect long-term star-formation and chemical histories. Theoretically modeling Mira pulsation, or the interplay between pulsation and convection in general, and their mass loss is notoriously difficult \citep{Hofner-2018}. Observational data for large samples of stellar systems, capturing rare objects in short-lifetime evolutionary stages, would be crucial to explore time-varying mass-loss phenomena \citep{Whitelock-2003} and to constrain the evolutionary models.

\printbibliography[heading=subbibliography]
\end{refsection}

\chapter{Galactic and Planetary Sciences}
\label{chap:galacticplanetary}


\begin{refsection}[3-1_PS_theoreticalperspective/PS_theoreticalperspective.bib]

\section{Theoretical Perspective}
\label{sec:PS_theoreticalperspective}

\newcommand{\HK}[1]{{\color{red}#1}}

\noindent
\begin{flushright}
Hiroyuki Kurokawa$^{1}$
\\
$^{1}$ University of Tokyo
\end{flushright}
\vspace{0.5cm}

\subsection{Introduction}

\begin{figure}
    \centering
    \includegraphics{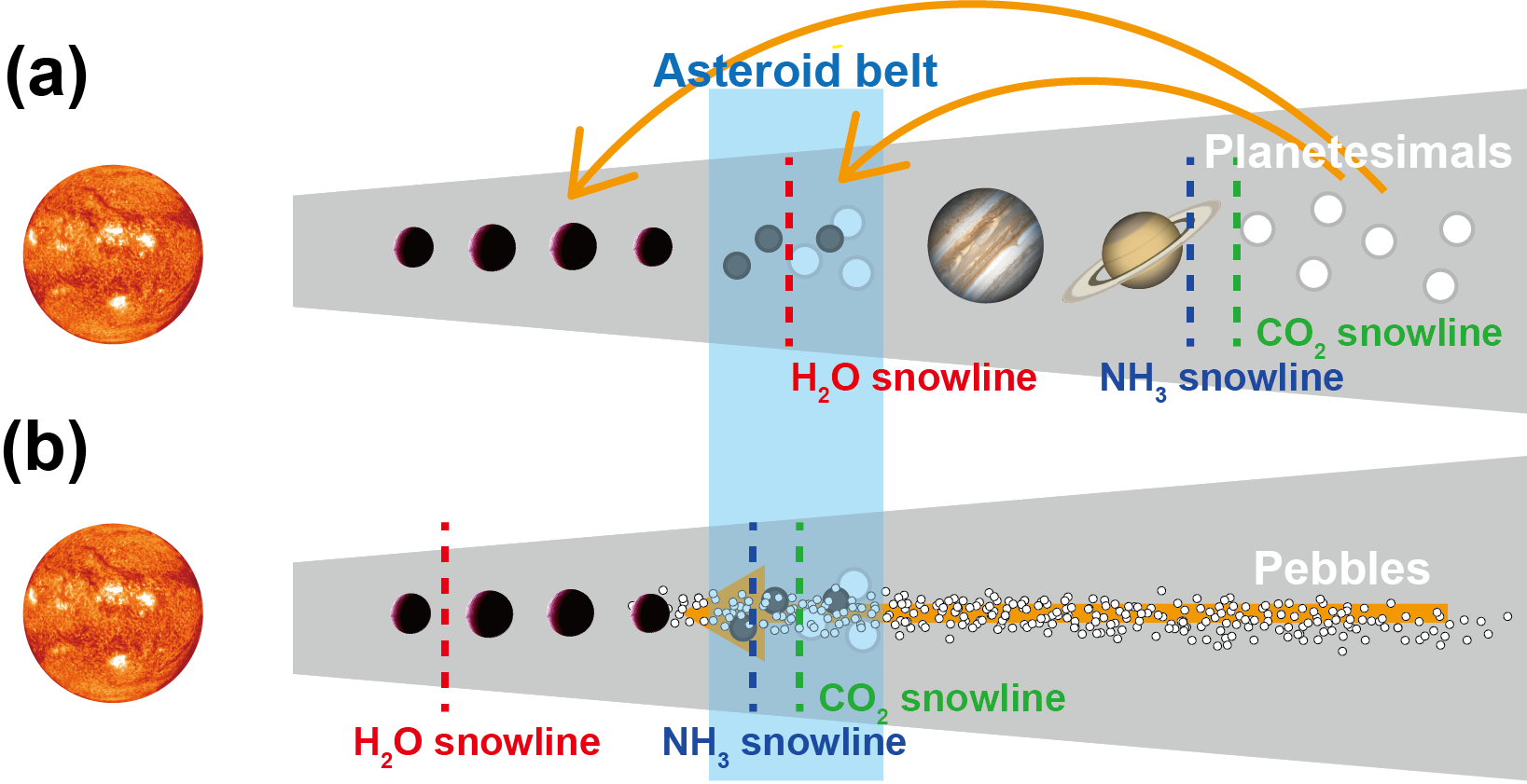}
    \caption{Schematic illustration of two different modes of volatile delivery to terrestrial planets. (a) Volatile delivery by planetesimals before and/or after the dissipation of protoplanetary disk gas. (b) Volatile delivery by pebbles following the migration of snow lines before the dissipation of protoplanetary disk gas.}
    \label{fig:Fig_3_1_1}
\end{figure}

GREX-PLUS will be able to perform key observations of protoplanetary disks, Solar System small bodies, and planets in the Solar and extrasolar systems to understand the origins of planets, their habitability, and building blocks for life. Such scientific goals fit ISAS's space-science and exploration roadmap toward exploring life in the Universe. In Section~\ref{sec:PS_theoreticalperspective}, we summarize the scientific background and theoretical perspective on how life-essential volatile elements were delivered to terrestrial planets to motivate the observations with the GREX-PLUS instruments to unveil the dynamic evolution of snow lines in protoplanetary disks, volatile-rich small bodies in the Solar System, and giant planets in the Solar and extrasolar systems (see the following subsections for details).

In addition to these topics, GREX-PLUS will provide key insights into a broader range of environments relevant to the origin and evolution of planetary systems. Observations of interstellar molecules (Section~\ref{sec:interstellarmolecules}) will constrain the chemical pathways linking the interstellar medium to planet-forming disks. Studies of star-forming regions (Section~\ref{sec:starformingregions}) will bridge disk-scale processes and their natal environments. On larger Galactic scales, surveys of the Galactic plane and the Galactic center (Sections~\ref{sec:galacticplane} and \ref{sec:galaxycenter}) will provide a comprehensive census of stars, dust, and molecular material, placing planet formation in a broader Galactic context. Observations of brown dwarfs (Section~\ref{sec:bd}) will further provide important links between giant planets and low-mass stars.

\subsection{Protoplanetary disks}
\label{subsec:ppd}

Planets form in protoplanetary disks surrounding host stars, and thus constraining their properties is important for understanding the origin of planetary systems. How dust particles, which ultimately form planets, are transported in disks and aggregate into larger bodies is not fully understood, but planet formation theory suggests that the locations of snow lines, where molecules of interest (most importantly H$_2$O, but others as well) condense, in a protoplanetary disk control the sizes and compositions of resulting planets (Figure~\ref{fig:Fig_3_1_1}). Sublimation of volatiles changes the composition of solid materials \citep[e.g.,][]{1981PThPS..70...35H} and thus planets that form \citep[e.g., rocky vs. icy planets,][]{2004ApJ...604..388I}. Moreover, changes in gas and dust surface densities associated with snow lines influence the efficiency of dust aggregation and, ultimately, the architectures of planetary systems. Classical planet formation models assume that increased dust surface density beyond the H$_2$O snow line is preferable for giant planet formation \citep[e.g.,][]{1981PThPS..70...35H,2000ApJ...537.1013I,2008ApJ...673..502K}. More recent models that take dust migration \citep[the inward drift due to aerodynamic drag induced by disk gas rotating with sub-Keplerian speed,][]{1976PThPh..56.1756A,1977MNRAS.180...57W} into account also suggest that sublimation at the snow line causes dust pile-up and thus efficient aggregation \citep[e.g.,][]{2011ApJ...733L..41S,2016ApJ...821...82O}. 

The locations of snow lines in protoplanetary disks are poorly understood. A snow line in a disk is defined as the location where the partial pressure of the molecule of interest becomes equal to its saturation vapor pressure; therefore, the temperature matters. The temperature at the midplane of the disk is determined by i) the stellar irradiation to the disk surface and re-emission to the midplane and ii) release of gravitational potential energy of gas accreting onto the host star. Theoretical models suggest that the decline of the second component causes a temperature decrease and, consequently, inward migration of snow lines with time. Notably, in the late stage of protoplanetary disk evolution (a few Myrs), the H$_2$O snow line is predicted to be located inside the inner boundary of the habitable zone, namely, even inside Earth's orbit \citep{2007ApJ...654..606G,2011ApJ...738..141O}. A recent study considering non-ideal MHD effects on disk gas accretion showed that accretion energy is released at the high altitudes of the disk and thus the temperature at the disk midplane is even lower from the earlier stages than previously thought \citep{2021ApJ...916...72M}. Such a configuration of the H$_2$O snow line with respect to the habitable zone can drastically change our understanding of how Earth and other terrestrial planets formed. Once the H$_2$O snow line passes the orbits of protoplanets, migration and accretion of mm- to cm-sized icy dust particles (icy pebbles) can easily supply water comparable to Earth's ocean mass \citep[Figure~\ref{fig:Fig_3_1_1}b,][]{2016A&A...589A..15S,2016Icar..267..368M}. Such an efficient supply of water, if it occurs, is thought to even change the question on Earth's habitability from \textquotedblleft how did Earth acquire its water?" into \textquotedblleft how did Earth avoid becoming an icy giant?" Moreover, there are uncertainties in the theoretical predictions for effective snow line locations depending on the formation of clathrates and salts \citep{2014ApJ...796L..28M,2020Sci...367.7462P}; therefore, observational constraints are needed.

The mid-infrared high-resolution spectrometer onboard GREX-PLUS will be able to determine the position of the H$_2$O snow line in the midplane regions of protoplanetary disks by observing the Doppler shift of water vapor emission lines caused by Keplerian motion (see Section~\ref{sec:protoplanetarydisks}). Previous attempts to constrain the location of the H$_2$O snow line were more sensitive to disk surfaces, where the snow line is further from the host star than in the midplane. The locations of other volatile species including NH$_3$, H$_2$S, and CO$_2$ snow lines will, if detected, further test the theoretical predictions for snow line locations and constrain how different solid-material compositions are distributed in disks and whether icy pebbles can contribute significantly to the supply of volatile elements essential for life.

\subsection{Small bodies in the solar system}

Asteroids and comets (hereafter, together called small bodies) in the Solar System record the origin of Earth and its habitability. Small bodies have been thought to be remnants of planet-formation stages and also to have delivered water and other life-essential elements to terrestrial planets. Classical planet formation theory assumed that planets form hierarchically from dust through km-sized planetesimals \citep[namely, small bodies,][]{1985prpl.conf.1100H}. The terrestrial-planet-forming region was assumed to be volatile free, and thus Earth was thought to have acquired these elements by accreting small bodies which originate from the outer Solar System \citep[Figure~\ref{fig:Fig_3_1_1}a, e.g.,][]{2000M&PS...35.1309M}. Recent theory, where migration of dust particles and snow lines is taken into account, points to another possibility: volatile delivery by icy pebbles (Figure~\ref{fig:Fig_3_1_1}b, Section \ref{subsec:ppd}).

The two modes of volatile delivery (planetesimals vs. pebbles) would leave distinct records in small-body composition and population and thus potentially be distinguished with asteroid observations. In the planetesimal migration scenario, the formation and migration of giant planets cause gravitational scattering of small bodies, which leads to bodies originally formed at and beyond the giant-planet-forming region being transported to the inner solar system including the terrestrial-planet-forming region and current asteroid belt. Therefore, some classes of volatile-rich asteroids in the asteroid belt should have compositional similarities to Trans-Neptunian objects \citep[e.g.,][]{2021ApJ...916L...6H} for various sizes. On the other hand, inward migration and accretion of icy pebbles onto terrestrial planets would also cause accretion onto small bodies; for instance, ammonia on the dwarf planet Ceres (the largest body in the asteroid belt) has been interpreted as a potential consequence of pebble accretion \citep{2015Natur.528..241D,2019ESS.....431719N}. Because the accretion efficiency is highly dependent on the mass of the target \citep{2012A&A...544A..32L,2016A&A...586A..66V}, icy pebble accretion would make larger asteroids more volatile-rich \citep{2019ESS.....431719N}.

The near-infrared camera and the mid-infrared high-resolution spectrometer onboard GREX-PLUS will be able to survey the presence/absence of important ices and minerals on small bodies down to $\sim10$ km in diameter, which is much smaller than the sizes probed with previous spectral observations in these wavelengths. If a body shows spectral features of H$_2$O ice ($\sim3.1\ \mathrm{\mu m}$) or hydrated minerals ($2.7$--$2.9\ \mathrm{\mu m}$), the body should have formed beyond the H$_2$O snow line \citep{2012Icar..219..641T,2015aste.book...65R,2019PASJ...71....1U}. Similarly, ammoniated materials ($\sim3.1\ \mathrm{\mu m}$) suggest formation beyond the NH$_3$ snow line; carbonated minerals ($3.4$ and $3.9\ \mathrm{\mu m}$) suggest the formation beyond the CO$_2$ snow line \citep{2012ApJ...752...15O,2017AJ....153...72V,2020Sci...367.7462P,2022AGUA....300568K}. These key ices and minerals for understanding the dominant mode of volatile delivery show diagnostic spectral features in near- to mid-infrared wavelengths which will be covered by the GREX-PLUS instruments (see Section~\ref{sec:icysmallbodies}). 

\subsection{Giant planets in the Solar and extrasolar systems}

Formation and migration of giant planets in planetary systems determine how volatile elements are delivered to terrestrial planets (Figure~\ref{fig:Fig_3_1_1}). A giant planet forming in a protoplanetary disk can carve a gap in the gas disk by its gravity \citep[e.g.,][]{1986ApJ...307..395L}. Because such a gas gap induces a positive pressure gradient at its edge, gas starts to rotate with super-Keplerian speed. Consequently, the direction of pebble migration changes from inward to outward, halting icy pebble accretion to terrestrial planets on inner orbits \citep{2016Icar..267..368M}. Both growth and migration of giant planets cause gravitational scattering of small bodies and subsequent accretion of volatile-rich asteroids onto terrestrial planets \citep{2011Natur.475..206W,2017Icar..297..134R}.

The chemical compositions of giant planets record their formation locations in protoplanetary disks and, consequently, migration history. Observational constraints and theoretical modeling point to the fact that giant planets form by the \textquotedblleft nucleated instability" -- a solid core first forms, and then the core starts to accrete gas rapidly once it becomes massive enough \citep[e.g.,][]{2000ApJ...537.1013I}. Compositions of gas and solid phases change across volatile snow lines, leaving diagnostic signatures in atmospheric compositions of giant planets (see Sections~\ref{sec:exoplanetatmosphere} and \ref{sec:solarsystemplanet}). Atmospheric H, C, N, and O ratios of short-period extrasolar planets have been considered to reflect their formation locations relative to the relevant snow lines \citep[H$_2$O, NH$_3$, CO$_2$, CH$_4$, CO, and N$_2$; e.g.,][]{2011ApJ...743L..16O,2016ApJ...833..203P,Notsu+2020}. The diagnosis of long-period giant planets including the Solar System ones is less straightforward than that of short-period ones because condensation in the depths of the atmosphere changes the chemical composition in the gas phase and limits the detectability of the condensable molecules. Nevertheless, uniform enrichment of C, N, O, S, P, and noble gases with respect to their solar abundances in Jupiter's atmosphere has been interpreted as a signature of its formation beyond the N$_2$ snow line \citep{2019AJ....158..194O}. Solar-like $^{14}$N/$^{15}$N ratios of Jupiter's and Saturn's atmospheres also support their formation beyond the N$_2$ snow line \citep{2014Icar..238..170F,2014ApJ...796L..28M}. Possible super-solar S/N ratios of Uranus' and Neptune's atmospheres have been discussed to inform their formation between the H$_2$S and N$_2$ snow lines \citep{2020RSPTA.37800107M}.

The mid-infrared high-resolution spectrometer onboard GREX-PLUS will be able to acquire atmospheric spectra of solar and extrasolar giant planets to constrain molecular abundances of interest (See Sections~\ref{sec:exoplanetatmosphere} and \ref{sec:solarsystemplanet}). The inferred molecular abundances including H$_2$O and NH$_3$ combined with chemical network modeling would constrain bulk elemental abundances and consequently formation locations of short-period extrasolar planets. Detection of NH$_3$ and H$_2$S in the atmospheres of the solar system ice giants would be an additional success, but GREX-PLUS will provide better estimates of atmospheric temperature structures by observing H$_2$ lines, which provide critical information for constraining molecular abundances with observations at other wavelengths \citep[e.g., near-infrared,][]{2018NatAs...2..420I,2019Icar..321..550I}. 
Brown dwarfs, which occupy the mass range between giant planets and low-mass stars, provide complementary benchmarks for atmospheric chemistry and thermal structure, and thus offer important comparative constraints on the formation and evolution of giant planets (see Section~\ref{sec:bd}).
Reconstructed atmospheric elemental compositions would enable us to constrain the formation and migration histories of giant planets and, consequently, volatile delivery to terrestrial planets.


\printbibliography[heading=subbibliography]
\end{refsection}

\clearpage

\begin{refsection}[3-2_protoplanetarydisks/protoplanetarydisks.bib]

\section{Protoplanetary Disks}
\label{sec:protoplanetarydisks}

\noindent
\begin{flushright}
Shota Notsu$^{1,2}$, 
Hideko Nomura$^{3}$
\\
$^{1}$ The University of Tokyo
$^{2}$ RIKEN
$^{3}$ NAOJ
\end{flushright}
\vspace{0.5cm}

\subsection{Scientific background and motivation}
Protoplanetary disks are rotating accretion disks surrounding young, newly formed stars (e.g., T Tauri stars, Herbig Ae/Be stars). 
They are composed of dust grains and gas, and contain all the material that will form planetary systems orbiting main-sequence stars. They are active environments for the creation of simple and complex molecules, including organic matter and H$_{2}$O. 

In the hot inner regions of disks, H$_{2}$O ice evaporates from dust-grain surfaces. In contrast, it is frozen out on dust-grain surfaces in the outer cold parts of disks. The border of these two regions is the H$_{2}$O snowline \citep{Hayashi+1981}. Outside the H$_{2}$O snowline, the solid material is enhanced with the supply of H$_{2}$O ice. In addition, dust grains covered with water ice mantles can stick even at higher collisional velocities, and efficient coagulation is promoted. Thus, the formation of gaseous planetary cores is promoted in such regions, and we can regard the H$_{2}$O snowline in the disk midplane as dividing the regions of gas-giant planet and rocky planet formation. Icy planetesimals, comets, and/or icy pebbles coming from outside the H$_{2}$O snowline may bring water to rocky planets including Earth (e.g., \citealt{Morbidelli+2016, Sato+2016}).
 Recent theoretical studies (e.g., \citealt{Oka+2011, Mori+2021}) suggested that the location of the H$_{2}$O snowline will change if we change the physical conditions such as the luminosity of the central star, the mass accretion rate, and the dust-grain size distribution in the disk.
 Thus, observationally locating the positions of the H$_{2}$O snowline in protoplanetary disks is important, since it will provide constraints on current formation theories of planetesimals and planets, and will help to clarify the origin of water on rocky planets including Earth.

Recently, high-spatial resolution observations with the Subaru Telescope and ALMA have revealed detailed substructures in the dust emission from the disks, and signs of planet formation have been found (e.g., \citealt{Andrews+2018}). Molecular emission line observations using ALMA have also spatially resolved the positions of CO snowlines (e.g., \citealt{Qi+2013, Qi+2019}). CO evaporates from the dust-grain surfaces into the gas phase with lower desorption temperature ($\sim20$ K) than that of water ($\sim100$--150 K), and thus the CO snowline is located at a larger radius, making it relatively easy to observe. Molecular emission lines of water from protoplanetary disks have been detected by the Spitzer Space Telescope and the Herschel Space Observatory in the mid- and far-infrared wavelengths, respectively (e.g., \citealt{Carr+2008, Hogerheijde+2011, Blevins+2016, vanDishoeck+2021}). However, these emission lines are mainly emitted from water molecules at the hot disk surface and at the cold outer disk outside the water snowline, and the position of the water snowline in the disk midplane has not been directly located.

Recently, spatially resolved water line emission within the water snowline has been observed for younger (Class 0-I) protostellar disks and envelopes (e.g., \citealt{Harsono+2020, Tobin+2023, Facchini+2024, Nakasone+2026}) and one Class II Herbig disk \citep{Rampinelli+2026} with, e.g., ALMA. In addition, hot and warm water gas from the innermost region of the disks has been detected with JWST, and these observations suggest that warm water line emission is stronger in compact disks than in large disks with outer gaps, suggesting that there is indeed a higher inward mass flux of icy pebbles in compact disks (e.g., \citealt{Banzatti+2023, Banzatti+2025, Gasman+2025}).

The formation processes of large organic molecules, which could lead to the origin of life, are another interesting topic. In star-forming regions, complex organic molecules have been observed spectroscopically mainly in the (sub-)millimeter wavelengths (e.g., \citealt{Booth+2021, Yang+2021, Yamato+2024a}), but the maximum number of atoms in each molecule is about 10 at most. In contrast, various large organic compounds, including amino acids, have been detected in comets and meteorites in the Solar System. It is still unclear how the large organic compounds found in the inner Solar System objects are formed from molecules found in star-forming regions. Complex organic molecules are thought to be formed on dust-grain surfaces. Grain-surface reactions behave differently in the low-temperature and warm regions (e.g., \citealt{Walsh+2014, Notsu+2022}). At low temperatures, hydrogen diffuses mainly on the dust-grain surface and reacts with other grain-surface molecules to form hydrogenated molecules. In the warmer regions, some of the hydrogenated molecules are destroyed by UV and X-ray radiation to form radicals, which react with each other to form more complex organic molecules. Investigating the distributions of organic molecules in the warm regions will help clarify the formation processes of prebiotic molecular species in the planet-forming regions.
Recent ALMA observations with high sensitivity have made it possible to detect complex organic molecules that were previously undetectable in protoplanetary disks (e.g., \citealt{Oberg+2015, Walsh+2016, Booth+2021, Booth+2025, Brunken+2022, Yamato+2024b}).
They trace the emission of organic molecular lines from the outer region in the disk (around a few tens of au), where the cold dust-grain surface reactions are active, as well as the emission from the inner hot region of the disks around Herbig Ae/Be stars.

Here, we propose detecting water snowlines in various protoplanetary disks through survey observations of water emission lines using GREX-PLUS in order to elucidate the evolution of the positions of water snowlines.
In addition, we also propose to observe complex organic molecular lines emitted from the inner region of the disks in order to constrain their formation processes in planet-forming regions.

\subsection{Required observations and expected results}
Based on detailed physical and chemical structure models of protoplanetary disks, we investigated the candidate water emission lines that trace water snowline positions in the disks \citep{Notsu+2016, Notsu+2017, Notsu+2018}. We found that emission lines with small Einstein A coefficients ($\sim10^{-6}$--$10^{-3}$) and relatively high excitation energies ($\sim1000$ K) are suitable for identifying the location of the water snowline near the equatorial plane of the disk. Such water emission lines are widely distributed from the mid-infrared to submillimeter wavelengths; in the GREX-PLUS wavelength coverage, the water emission line at 17.75 $\mu$m can be used to trace the position of the water snowline. The water snowline in a disk with a central star of one solar mass is difficult to observe with spatially resolved imaging even with ALMA. However, because of the Keplerian rotation of the disk, the emitting region can be identified by analyzing the Doppler-shifted emission profiles, even if spatially unresolved. 

\begin{figure}[ht]
\centering
\includegraphics[width=0.7\linewidth]{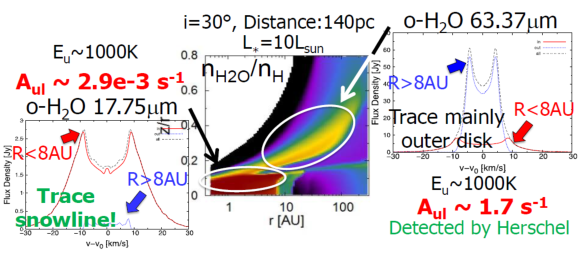}
\includegraphics[width=0.25\linewidth]{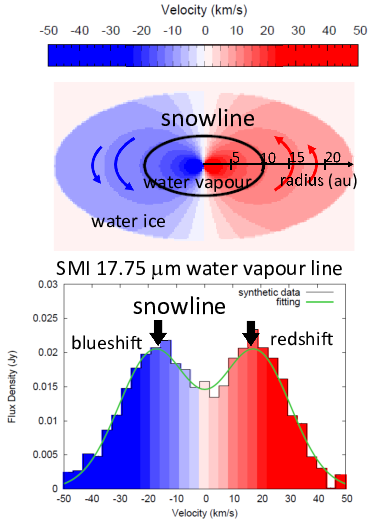}
\caption{({\it Left}) Model calculations of the distribution of gas-phase water in a protoplanetary disk and their emission-line spectra \citep{Notsu+2017}. The 63.37 $\mu$m emission line detected by Herschel mainly traces the hot surface layer of the outer disk, whereas the 17.75 $\mu$m emission line, observable by GREX-PLUS, traces the position of the water snowline. ({\it Right}) By analysing emission profiles that trace the Doppler shift due to the Keplerian rotation, the emission region, i.e. the water snowline, can be located using a high spectral resolution of $R=29,000$ even if it cannot be spatially resolved \citep{Kamp+2021}.}
\end{figure}

Simulated observations based on model calculations show that the location of the water snowline can be identified from the line profiles with a wavelength resolution of $R=29,000$. We also investigated the sensitivity required for detection using disk models with various physical quantities such as dust size, disk mass, and central star luminosity \citep{Kamp+2021}. Using a model with typical physical quantities, water line emission from a disk with a central star of about one solar mass in a star-forming region near the Solar System ($d=150$ pc) can be detected by GREX-PLUS with an integration time of 50 min. The water line emission from a disk with a central star of $2$--3 times the solar mass can be detected by GREX-PLUS with an integration time of about 30 minutes, even in star-forming regions in giant molecular clouds at a distance of $d=420$ pc from the Solar System. We expect to detect target water emission lines with JWST's medium-dispersion spectroscopic observations, and to identify water snowline positions by analyzing the emission line profiles obtained with GREX-PLUS's high-dispersion spectroscopic observations.

The GREX-PLUS wavelength bands include transition lines of complex organic molecules such as CH$_{3}$CN and CH$_{3}$OH as well as small organic molecules such as HCN and C$_{2}$H$_{2}$. Model calculations of CH$_{3}$CN molecular emission lines from a disk with a central star of one solar mass show that at the distance of TW Hya (d=60 pc), emission lines can be detected with an integration time of a few hours. The GREX-PLUS observations will help clarify the composition of organic molecules formed by grain-surface reactions in the warm region, which was difficult to detect by ALMA. Emission lines of small organic molecules such as HCN and C$_{2}$H$_{2}$ have already been detected with the Spitzer Space Telescope (e.g., \citealt{Carr+2008, Pontoppidan+2010}) and JWST (e.g., \citealt{Tabone+2023, Grant+2025}). Analysis of the emission-line profiles obtained by GREX-PLUS's high-dispersion spectroscopy is expected to provide information on the spatial distribution of the C/O and N/O elemental composition ratios in the disk. Since water is a major carrier of oxygen elements in both gas and icy phases, the C/O elemental composition ratio in the gas phase (and also the solid phase) is expected to change significantly inside and outside the water snowline. On the other hand, the C/O elemental composition ratios in the gas and solid phases are also affected by chemical reactions. Since the elemental compositions of gas-giant planets are considered to reflect the elemental compositions of protoplanetary disks, statistically investigating the spatial distribution of the elemental composition ratios in disks and comparing them with those of short-period gas-giant planets will provide constraints on where in the disks such gas-giant planets formed (e.g., \citealt{Notsu+2020, Notsu+2022}). In addition, the C/O ratios in the innermost region of the disk can increase within the refractory carbon sublimation front (carbon soot line, e.g., \citealt{Houge+2025}). Observations of C$_{2}$H$_{2}$ and water line profiles can constrain the distribution of C/O ratios in these innermost regions of the disks (e.g., \citealt{Tabone+2023, Grant+2025}).

\subsection{Scientific goals}
Line survey observations will cover about 100 protoplanetary disks in the relatively nearby solar-mass and intermediate-mass star-forming regions such as Taurus, $\rho$ Oph, Chameleon, Upper Sco, Orion, Perseus, Serpens, Aquila, etc., for a total observing time of about 300 hours. The target objects will be carefully selected based on the existing ALMA and JWST observations of dust continuum emission and molecular lines. We will study the time evolution of the positions of water snowlines and clarify the formation processes of planetary systems and how water and organic molecules are supplied to rocky planets. In addition, we will detect small organic molecules such as HCN and C$_{2}$H$_{2}$ and complex organic molecules (CH$_{3}$CN and CH$_{3}$OH) in the nearby disks. This will reveal the distribution of elemental compositions such as C/O in planet-forming regions and the complex formation process of organic molecules on warm grain surfaces, and discuss the formation process of short-period gaseous planets and molecular species related to the origin of life. Thanks to the wide wavelength coverage (10--18 $\mu$m), which includes water and various target molecular lines, we expect to achieve these science goals from the same line-survey observations.

\begin{table}[ht]
    \label{tab:protoplanetarydisks}
    \begin{center}
    \caption{Required observational parameters.}
    \begin{tabular}{|l|p{5cm}|l|}
    \hline
     & Requirement & Remarks \\
    \hline
    Wavelength & 10--18 $\mu$m & The target water line is at 17.75 $\mu$m \\
    \cline{1-2}
    \hline
    Wavelength resolution & $\lambda/\Delta \lambda>29,000$ & To resolve the double peaks of line profiles \\
    \hline
    Sensitivity & $<$ 5 mJy (5$\sigma$, 1 hr) & \\
    \hline
    Observing field & Nearby star-forming regions with $d<500$ pc &  \\
    \hline
    \end{tabular}
    \end{center}
\end{table}

\printbibliography[heading=subbibliography]
\end{refsection}

\clearpage

\begin{refsection}[3-3_interstellarmolecules/interstellarmolecules.bib]

\section{Interstellar Molecules}
\label{sec:interstellarmolecules}

\noindent
\begin{flushright}
Yasuhiro Hirahara$^{1}$
\\
$^{1}$ Nagoya University
\end{flushright}
\vspace{0.5cm}

\subsection{Scientific background and motivation}

High-resolution spectroscopic observations provide useful information about the physical and chemical conditions in the interstellar medium (ISM), such as interstellar molecular clouds and circumstellar envelopes. So far, more than 340 interstellar molecular species have been identified, showing the complex chemical processes in various ISM environments. For the identification of interstellar molecules, contributions of the pure rotational molecular transitions in the microwave region were the most prominent both in spectroscopic observations and laboratory experiments, especially for the short-lived species, such as ionic and radical species. In the mm- and submm-wavelength region, the so-called radio region, the heterodyne techniques make it quite easy to achieve high spectral resolution, which is the most important for the precise assignment of the spectra. Rotational transitions can only be observed for the gas-phase molecules with permanent electric dipole moments. A few molecules have been identified in the ISM through their electronic transitions mostly in the optical and ultraviolet regions. In general, molecules other than radical species have dissociative excited states, which makes it rare and difficult to observe their electronic transitions.

Since all molecules have vibrational transitions mostly observed in the infrared (IR) wavelength regions, IR spectroscopic observations have a significant advantage. Many infrared spectral databases are published and updated especially for the chemical analyses of complex organic molecules. In particular, the mid-infrared region is called the ``fingerprint region'' because transitions accompanying vibrations of various functional groups of organic materials appear in this region, and hence it is a useful band used in spectroscopic analysis to determine the structure of the carbon skeleton. In addition to organic compounds, Si-O skeletal vibration modes in various silicates also appear in this region, providing a wealth of information on the structures of mineral or amorphous solid silicates. When focusing on the observations of the ISM, infrared vibrational spectroscopy is a unique method for identifying non-polar gas-phase molecules, such as H${_2}$, CH${_4}$, CO${_2}$, and SiH${_4}$, being complementary to radio observations.

Spectroscopic observations, however, to date in the infrared region have detected only about 30 species of molecules, many of which were identified earlier in the rotational spectra in the radio-frequency region. Only 26 chemical species, shown in Table 3.2, were first detected  in infrared observations.  The main reason for the paucity of past studies lies in the difficulty of high-sensitivity and high-resolution spectroscopic observations at wavelengths longer than 8--13 $\mu$m (N-band) due to deep absorption by the Earth's atmosphere. In fact, as shown in Table 3.2, about half of the interstellar molecules have been discovered by space telescopes equipped with medium or low wavelength resolution spectrometers in the infrared region. Therefore, GREX-PLUS with an unprecedentedly high-wavelength-resolution spectrometer will contribute to the progress of interstellar chemistry.

\begin{table}[ht]
    \label{tab:listofmolecules}
    \begin{center}
    \caption{Interstellar molecules detected by infrared spectroscopic observations
    (underline: first detection by IR space telescopes).}
    \begin{tabular}{ll}
    \hline \hline
    Type & Species \\
    \hline
    Simple hydrides, inorganic species & H$_2$, CH$_4$, CO$_2$, SiH$_4$ \\
    Aromatic molecules & \underline{C$_6$H$_6$ (benzene) [ISO], C$_{60}$, C$_{70}$ [Spitzer]} \\
    Linear carbon chains & C$_3$, C$_5$, C$_2$H$_4$, C$_2$H$_2$, C$_4$, C$_6$, \underline{HC$_4$H, HC$_6$H [ISO]} \\
    Molecular ions & H$_3$$^+$, HeH$^+$, \\
    & \underline{H$_2$O$^+$, H$_3$O$^+$, OH$^+$, SH$^+$, ArH$^+$, HCl$^+$, H$_2$Cl$^+$ [Herschel]}\\
    & \underline{CH$_3$$^+$ [JWST]} \\
    Radicals & CH$_3(^2A_2")$ \\
    \hline
    \end{tabular}
    \end{center}
    References: [Spitzer]:\citet{2010Sci...329.1180C}, [ISO]:\citet{2001ApJ...546L.123C},
    
    [HERSHEL]:\citet{Indriolo_2015}, \citet{lis:hal-00633439}, \citet{2010A&A...521L..35B}, 
    
    [JWST]:\citet{2023Natur.621...56B}

\end{table}

The infrared vibrational transitions of the target interstellar molecular species can be observed by resolving their rotational structures, and the wavelength resolution necessary to determine the molecular-species assignment was investigated. As an example, we show the result for benzene, which was first detected by the infrared astronomical satellite ISO (Infrared Space Observatory) in Figure~3.3.

\begin{figure}
    \centering
    \includegraphics[width=14cm]{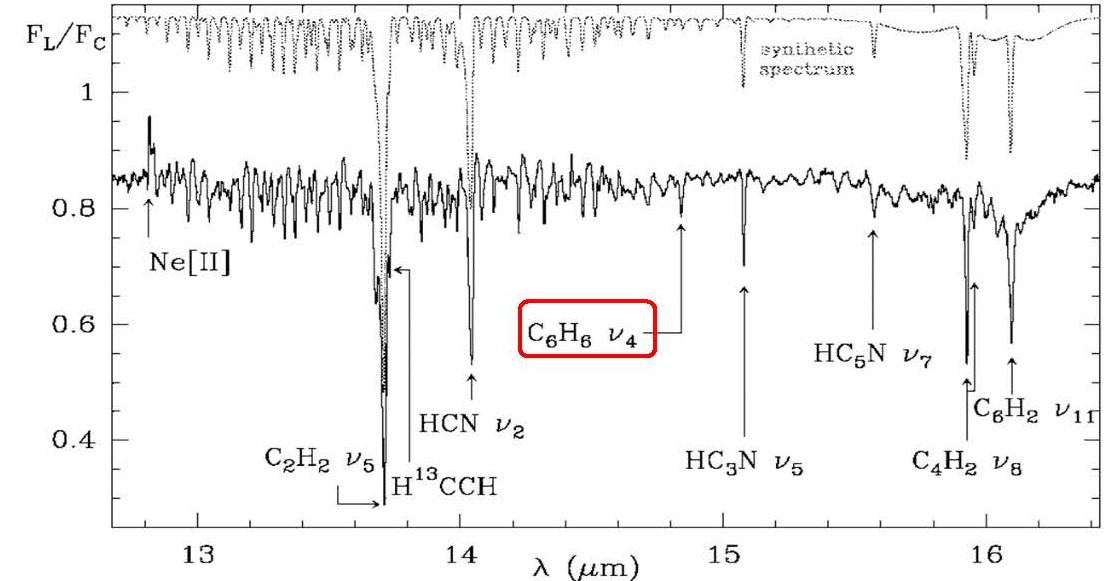}
    \caption{The infrared absorption spectrum of benzene (C$_6$H$_6$) and polyynes (C$_4$H$_2$, C$_6$H$_2$) detected for the first time, by the SWS (Short Wavelength Spectrometer) of ISO (Infrared Space Observatory) with $R\sim2,000$ toward the protoplanetary nebula CRL618 \citep{2001ApJ...546L.123C}.}
    \end{figure}

\underline{Science case I: Interstellar aromatic chemistry}

By the spectroscopic observation with the ISO Short Wavelength Spectrometer (SWS, $R\sim2,000$), only the Q--branch ($\Delta J=0$) of the $\nu$$_4$ vibrational-rotational transition of C$_6$H$_6$, an oblate symmetric top molecule, is prominent, and the molecular species may not be reliably assigned. We calculated the spectral shape of C$_6$H$_6$ for various wavelength resolutions by adopting the molecular constants for benzene \citep{DangNhu1989SpectralII}. As shown in Figure~3.4 ($\it{right}$), it is found that all P-- and R-- branches of benzene $\nu$$_4$ band can be fully resolved at a resolution of $R\geq28,000$. High-resolution observations for benzene will also contribute to increasing the intensity of the spectra by a factor of $\sim5$, allowing precise estimation of the rotational excitation temperature of the molecular species. Therefore, mid-IR high-resolution spectroscopic surveys of benzene and related small aromatic molecules as described below will contribute to the understanding of the material cycle and evolution in the Universe. In an early model calculation for the synthesis mechanism \citep{woods2002synthesis}, benzene in CRL-618 can be produced through the ``bottom-up'' scheme with successive ion-molecule reactions of acetylene (C$_2$H$_2$) and protonated acetylene (C$_2$H$_3$$^+$), yet-undetected ionic species that may be produced effectively in proto-planetary nebulae by proton-exchange reactions between (C$_2$H$_2$) and HCO$^+$. In their model, benzonitrile (C$_{6}$H$_{5}$CN) is predicted to be produced with an abundance similar to that of benzene. Interestingly, the first detection of benzonitrile in the interstellar medium was made in the low-temperature dark cloud TMC-1 ($T$$_k$$\sim10$ K) by the single-dish 100 m Robert C. Byrd Green Bank Telescope (GBT) and NRO 45 m telescope \citep{2018Sci...359..202M}.

\begin{figure}
    \centering
    \includegraphics[width=14cm]{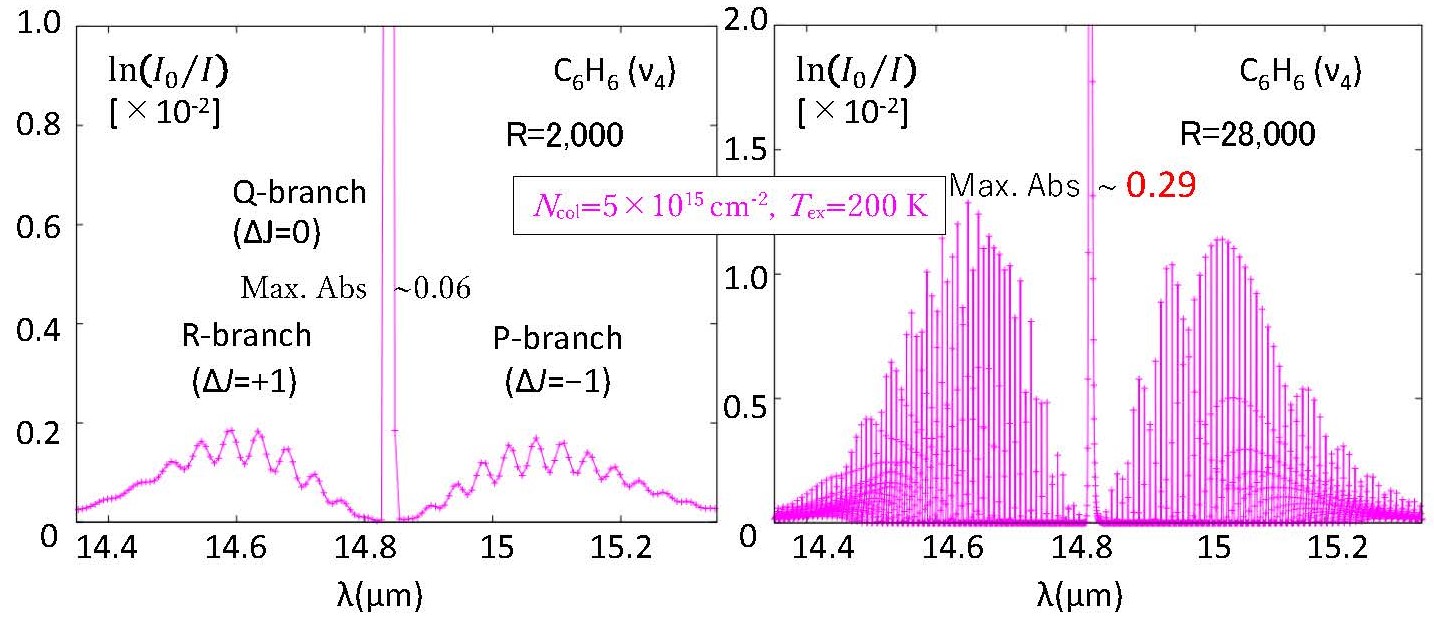}
    
    \caption{Simulated IR high-resolution spectrum of benzene (C$_6$H$_6$) $\nu$$_4$ vibration band for a column density $\it{N}$$_{col}$=5$\times$10$^{15}$ cm$^{-2}$ and $\it{T}$$_{ex}$=200 K, with the wavelength resolution $\it{R}$=2,000 ($\it{left}$) and 28,000 ($\it{right}$) \citep{2017GradThesis.Sci.Nagoya-U.}.}
    \end{figure}

Following the successful detection of benzonitrile,  the first detection of the two isomeric aromatic compounds 1-- and 2--cyanonaphthalene (C$_{10}$H$_{7}$CN) in TMC-1 was reported by submillimeter-wave spectroscopy with the GBT \citep{2021Sci...371.1265M}. Soon after the discovery, new aromatic molecules, as illustrated in Figure~3.5, ethynyl cyclopropenylidene (c-C$_{3}$HCCH), cyclopentadiene (c-C$_{5}$H$_{6}$), indene (c-C$_{9}$H$_{8}$), and ortho-benzyne (o-C$_{6}$H$_{4}$) were found in TMC-1 using the Yebes 40 m radio telescope (\citealp{2021A&A...649L..15C}, \citealp{2021A&A...652L...9C}).  Since the abundance of indene in TMC-1 is as high as 1.6 ${\times}$ 10$^{-9}$, the formation mechanism or origin of the aromatic hydrocarbon molecules should be reconsidered.
Recently, 1-cyanopyrene, a cyano-substituted derivative of the four-ring PAH pyrene, has been detected toward TMC-1 in GBT observations (\citealp{2024Sci...386.810M}), representing one of the largest PAH-related molecules identified in the cold interstellar medium to date.

\begin{figure}
    \centering
    \includegraphics[width=14cm]{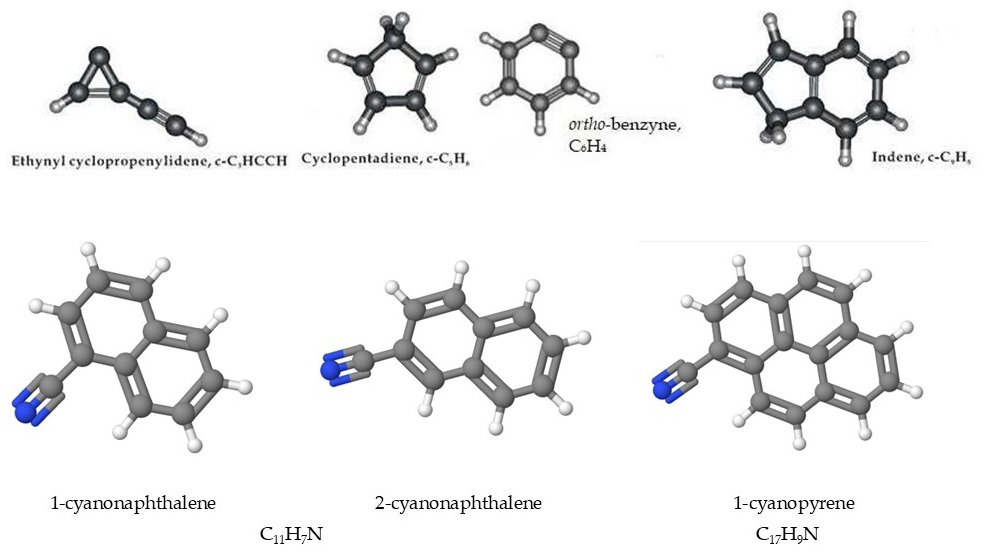}
    \caption{Molecular structures of aromatic hydrocarbons recently detected in the cold dark cloud TMC-1.}
\end{figure}

In contrast, the ``ultimate aromatic molecules'', fullerenes (C$_{60}$ and C$_{70}$), were detected by infrared spectroscopic observation with $R\sim500$ by the Spitzer Space Telescope toward the hydrogen-deficient young planetary nebula Tc 1, as shown in Figure~3.6 \citep{2010Sci...329.1180C}. Meanwhile, spectra of fullerenes have been detected toward the molecular cloud associated with the reflection nebula NGC2023 \citep{2010ApJ...722L..54S} and NGC7023 \citep{2012PNAS..109..401B}. 
Subsequent observations have identified infrared spectra of fullerenes in a wide variety of astrophysical environments, including planetary and proto-planetary nebulae in both our Galaxy and the Small Magellanic Cloud, mainly through detailed analyses of Spitzer archival data \citep{2014MNRAS.437.2577O}. More recently, JWST/MIRI observations have provided significantly improved spectra of C$_{60}$ and C$_{70}$ in UV-irradiated circumstellar and interstellar environments, revealing detailed variations in excitation conditions, spatial distributions, and relationships with aromatic hydrocarbons and carbonaceous dust, building upon the top-down fullerene formation scenario proposed for UV-irradiated PDRs (\citealp{2012PNAS..109..401B}, which is also supported by laboratory studies on graphene-to-fullerene transformation (\citealp{2010NatCh...2..450C}). These results suggest close evolutionary links among aromatic hydrocarbons, fullerene-related species, and carbonaceous grains under strong UV radiation fields.

By applying thermal emission models, \citet{2010Sci...329.1180C} first obtained low and different excitation temperatures, 330 K and 180 K for C$_{60}$ and C$_{70}$ in Tc1, respectively, suggesting that the emission of fullerenes does not originate from free molecules in the gas phase, but from molecular carriers attached to solid material. Even if the attachment to the host solid is not strong enough to hinder the rotational motion of guest fullerene molecules, it may only be possible to decompose the sharp Q-- branch features for the 17 and 19 $\mu$m bands of C$_{60}$ and 15 and 16 $\mu$m bands of C$_{70}$ by high-wavelength-resolution observations, as in the case of the benzene spectrum observed by ISO. In theory, C$_{60}$ is a symmetric top rotor with a small rotational constant of B $\approx$ 0.0028 cm$^{-1}$, with a nearly spherical shape with icosahedral ($\it{I_h}$) symmetry. Since all 60 carbon nuclei have zero spin for $^{12}$C$_{60}$, boson-exchange symmetry restrictions require that many of the rotational quantum levels disappear. These predicted ``missing'' levels result in a characteristic pattern of spectral line spacings that may be a key to observing and assigning a rotationally resolved C$_{60}$ spectrum. In spite of the small B values, the spacings between the rotational levels are thinned out for the P- and Q- branches of $^{12}$C$_{60}$. However, such simplification in rotational structure only occurs for $^{12}$C$_{60}$. The relative concentrations of $^{12}$C$_{60}$, $^{13}$C$^{12}$C$_{59}$, and $^{13}$C$_{2}^{12}$C$_{58}$ are $\approx$ 0.51, 0.34, and 0.11, respectively, even when the isotopic abundance ratio [$^{13}$C]/[$^{12}$C] is as low as 0.011 (terrestrial value).  Since nearly half of the C$_{60}$  molecules will include at least one $^{13}$C atom, the vibrational spectrum of C$_{60}$ in the low temperature gas in the ISM, if it exists, will be seriously complicated.

Since the first discovery, the formation mechanisms of fullerenes are intriguing subjects of discussion. According to the results of chemical model calculations involving the UV irradiation from the PDR region \citep{2016A&A...588C...1B} , the ``top-down'' scheme where larger carbon clusters shrink to reach C$_{60}$ may be more plausible than the traditional ``bottom-up'' approach. In the ``bottom-up'' scheme, large aromatic molecules are built up from small molecules such as benzene and other aromatic molecules shown in Figure 3.5, and also as simpler hydrocarbon ions. Inspired by the successive detections of cyano-substituted aromatics in the cold molecular clouds as described above, \citet{refId0} conducted UV and VUV photodissociation experiments of protonated benzonitrile (C$_{6}$H$_{5}$CNH$^{+}$). They found that the primary dissociation channel is the phenylium cation (C$_{6}$H$_{5}^{+}$) which is highly reactive species to produce larger polycyclic aromatic hydrocarbons. The high wavelength resolution spectroscopic observation in the 10--20 ${\mu}$m region toward PDRs with GREX-PLUS may clarify the formation mechanism, through new detections of pure non-polar aromatic hydrocarbons in the gas phase, whose sizes are between benzene and fullerenes.  $\it{ex.}$ naphthalene (C$_{10}$H$_{8}$) and its cation \citep{doi:10.1021/j100199a011}, and phenalenyl radical (C$_{13}$H$_{9}$$\cdot$) \citep{doi:10.1021/ja206322n}, an oblate symmetric top, ($\it{D}$$_{3h}$) showing simple rotational structure for the high-resolution vibrational spectra.  

Recent observations with the JWST/MIRI instrument have further expanded the observational frontier of interstellar aromatic chemistry in UV-irradiated environments. In particular, aromatic and carbon-rich molecular spectra including benzene-related species and fullerene bands have been investigated toward photodissociation regions (PDRs) with unprecedented sensitivity. Early results from JWST/MIRI programs on low- and high-mass protostars and disks have revealed unexpectedly rich warm hydrocarbon chemistry in irradiated inner planet-forming disk regions. In particular, benzene (C$_6$H$_6$) and diacetylene (C$_4$H$_2$) were first identified toward a very low-mass star disk with a high C/O ratio, suggesting active aromatic chemistry associated with terrestrial planet-forming environments (\citealp{Tabone2023NatAstronomy}; \citealp{vanDishoeck2023DiskChemistry}). These discoveries imply that the formation of aromatic hydrocarbons may proceed efficiently not only in evolved circumstellar envelopes and cold dark clouds, but also in UV-irradiated protoplanetary environments. These observations further suggest that aromatic chemistry proceeds actively not only in cold dark clouds such as TMC-1, but also in warm irradiated environments associated with star and planet formation. However, most observed spectra still remain unresolved in rotational structure, limiting our understanding of the detailed molecular excitation and formation processes.
High-resolution mid-infrared spectroscopy with GREX-PLUS will provide a unique opportunity to investigate rotationally resolved vibrational spectra of symmetric aromatic molecules and fullerene-related species. Such observations are expected to discriminate between bottom-up and top-down formation pathways through detailed analyses of excitation conditions, molecular structures, and isotopic compositions in various interstellar environments.

\begin{figure}
    \centering
    \includegraphics[width=14cm]{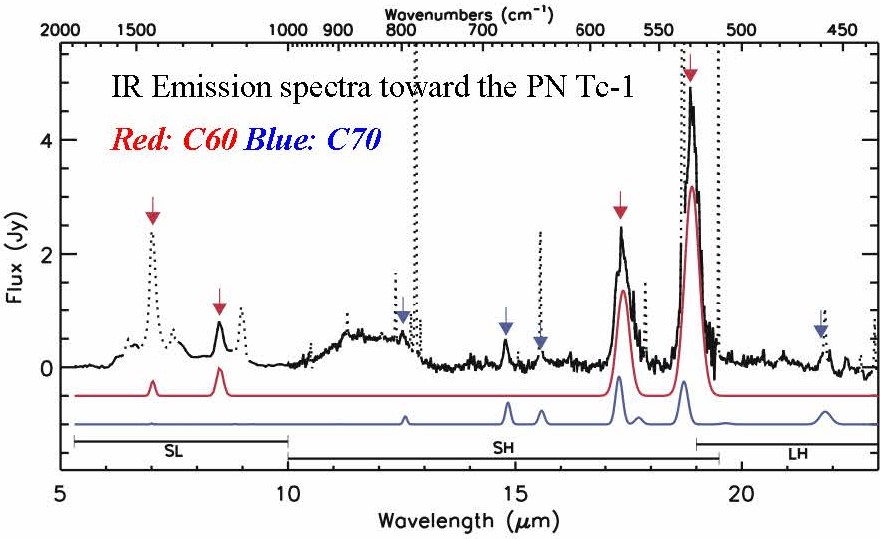}
    \caption{Infrared emission spectra of C$_{60}$ and C$_{70}$ toward the young planetary nebula Tc-1, continuum-subtracted spectrum between 5 and 23 ${\mu}$m, with the Spitzer IRS \citep{2010Sci...329.1180C}.
The red and blue curves below the data are thermal emission models for all infrared active bands of C$_{60}$ and C$_{70}$ at temperatures of 330 K and 180 K, respectively.}
\end{figure}

\underline{Science case II: Search for the ``key carbo-ions'' for interstellar chemistry}

Figure~3.7 shows an excerpt of conventional ``build-up'' ion-molecule reaction network for the formation of organic molecules starting from simple molecules in an interstellar molecular cloud at low temperatures ($T\sim10 K$), where UV photons are heavily attenuated and cosmic-ray-induced ionization dominates \citep{1992ChRv..92.1473,2013ChRv..113.8710A}. The hydrogen molecular ions (H$_2$$^+$) produced by high-energy cosmic rays quickly react with H$_2$ to form stable H$_3$$^+$, which then becomes a ``seed'' for the formation of more complex molecules via various ion-molecule reactions. Since so-called ``terminal ions'': CH$_3$$^+$, C$_2$H$_2$$^+$, C$_2$H$_3$$^+$, and ``the smallest aromatic hydrocarbon'', cyclic-C$_3$H$_3$$^+$ (cyclopropenyl cation) are very slow to react with ambient H$_2$, they react with atoms and molecules other than H$_2$ ($\it{e.g.}$, C, N, CO, H$_2$O, HCN) and electrons. In particular, ``carbo-ions'', as named by \citet{doi:10.1098/rsta.1988.0002}, located at the branch points of the reaction network are predicted to be abundant and chemically important in interstellar chemistry. Owing to their highly symmetric molecular structures and consequently very small or zero permanent dipole moments, however, most of them have remained observationally elusive in the interstellar medium. 
Comprehensive spectroscopic searches under various interstellar environments are therefore essential for validating reaction-network models of interstellar chemistry.
The carbo-ions listed in Table 3.3 are chemically stable, highly symmetric molecules with no or only very small permanent dipole moments.

It is worth pointing out that deep spectroscopic searches for carbo-ions through vibrational transitions are important for the advancement of laboratory astrophysics and astrochemistry.
In earlier studies, laboratory high-resolution absorption spectroscopy of CH$_3^+$($\nu_3$), C$_2$H$_2$$^+$($^2\Pi_u$)($\nu_3$), and C$_2$H$_3$$^+$($\nu_6$) was uniquely reported by Prof. Takeshi Oka's group using difference-frequency laser spectroscopy around the $3.2~\mu$m region (\citealp{doi:10.1063/1.454194}; \citealp{doi:10.1063/1.451929}; \citealp{doi:10.1063/1.457612}). The spectral resolutions achieved in these measurements are sufficiently high to resolve rotational structures, which is highly useful for astronomical searches.
Later, laboratory measurements of vibrational transitions at ${\lambda}>10~\mu$m for carbo-ions were conducted by \citet{2005PhRvL..94g3001A} for C$_2$H$_2$$^+$($^2\Pi_u$)($\nu_9$), and by \citet{MARIMUTHU2020111377}, using tunable mid-infrared laser photodissociation spectroscopy combined with ion-trap detection of fragment ions. 

It should be noted that these measurements achieved high sensitivity, although their spectral resolution was insufficient for resolving detailed rotational structures.
Since the molecular rotational constants of the carbo-ions listed in Table 3.3 are relatively large, GREX-PLUS high-resolution observations with ${R\geq20,000}$ are expected to provide the best and first detailed information on their intriguing molecular structures.
Recently, \citet{2023Natur.621...56B} reported the first detection of interstellar methyl cation (CH$_3^+$) using the MIRI-MRS instrument on board the James Webb Space Telescope (JWST), as shown in Fig.~3.8.
The JWST detection of CH$_3^+$ in emission from a UV-irradiated photodissociation region (PDR), rather than in absorption toward a cold starless core, suggests that carbo-ion chemistry plays an important role not only in conventional low-temperature ion-molecule chemistry, but also in warm and strongly UV-irradiated environments associated with star and planet formation. Such environments are expected to exhibit highly excited rotational and vibrational populations, complex velocity structures, and non-equilibrium chemistry. High-resolution spectroscopy by GREX-PLUS will therefore provide a unique opportunity to investigate the excitation, kinematics, and chemical evolution of carbo-ions in irradiated protoplanetary environments beyond the sensitivity-limited discovery space opened by JWST.

\begin{figure}
    \centering
    \includegraphics[width=14cm]{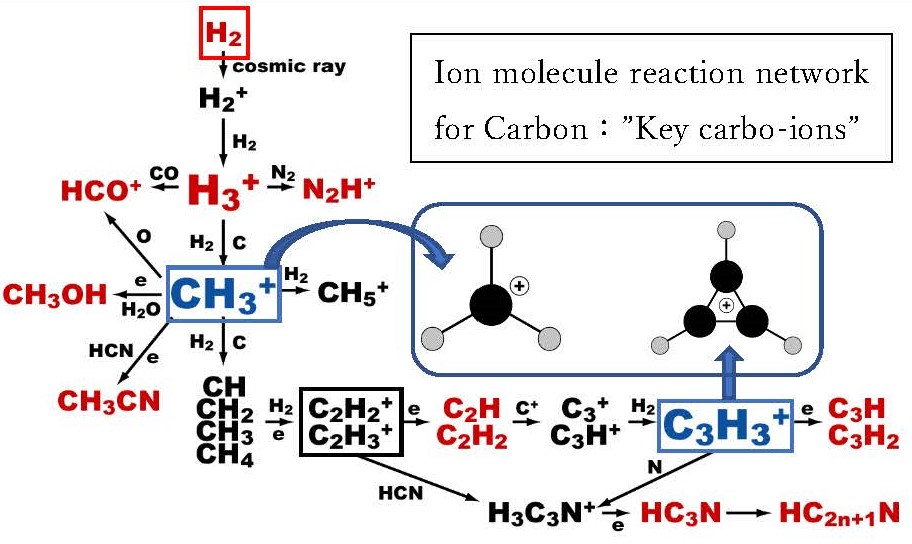}
    \caption{A chemical reaction network for the formation of organic molecules from simple molecules in an interstellar molecular cloud at low temperatures ($T\sim10K$)
    (\citealp{1992ChRv..92.1473},\citealp{2013ChRv..113.8710A}).
    }
\end{figure}

\begin{figure}
    \centering
    \includegraphics[width=14cm]{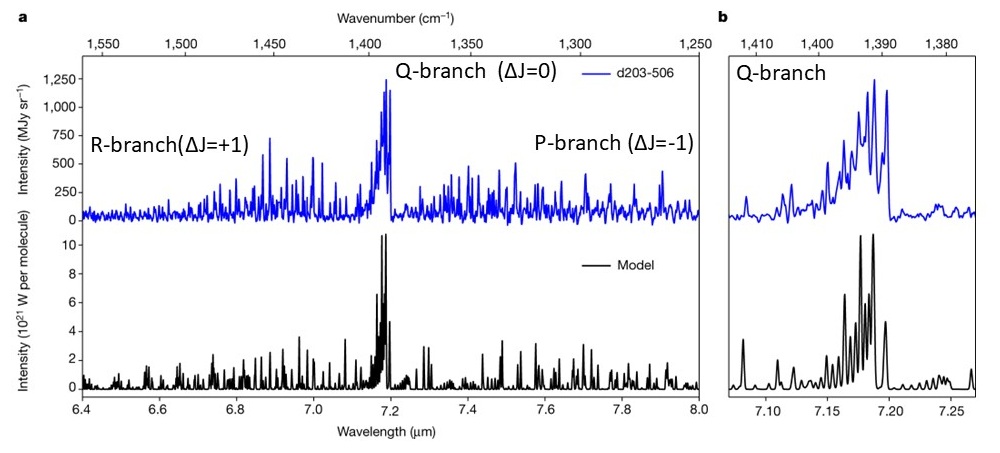}
    \caption{Comparison between the observed JWST spectrum of d203-506 and modeled CH$_3$$^+$ spectrum. a, Full spectrum. b, Enlargement of the strongest lines. The model (black curve) for the $\nu$$_2$ and $\nu$$_4$ vibrations of CH$_3$$^+$ was obtained using the detailed spectroscopic constants (\citealp{2023Natur.621...56B}).
    }
\end{figure}

\begin{table}[ht]
    \label{tab:carboions}
    \begin{center}
    \caption{List of target key carbo-ions for spectroscopic observation.}
    \small
    \setlength{\tabcolsep}{4pt}
    \begin{tabular}{p{1.3cm}cccp{7.2cm}}
    \hline \hline
    Species & Transition & $\lambda$/[$\mu$m] & Ref. & Note\\
    \hline
    CH$_3$$^+$
    & $\nu_2$, $\nu_4$ & 7.2 & ($a$)
    & \begin{tabular}[t]{@{}l@{}}
    First interstellar detection by JWST/MIRI; \\ rotationally resolved spectroscopy unavailable
    \end{tabular} \\
    C$_2$H$_2$$^+$($^2\Pi_u$) & $\nu$$_9$ & 11.8 & ($b$) & Laser-induced reaction with ion-trap detection\\
    C$_2$H$_3$$^+$ & $\nu$$_7$ & 13.2 & ($c$) & No high-resolution laboratory data\\
    C$_3$H$_3$$^+$ & $\nu$$_7$ & 11.0 & ($d$) & Partial IRPD spectroscopy \\
    \hline
    \end{tabular}
    \end{center}
    ($a$) \citet{1985JCP...82...1...333}, ($b$) \citet{2005PhRvL..94g3001A}, ($c$) \citet{1986JChPh..85.3437L}, ($d$) \citet{MARIMUTHU2020111377}
\end{table}

\begin{table}[ht]
    \label{tab:ISMrequirements}
    \begin{center}
    \caption{Required observational parameters.}
    \begin{tabular}{|l|p{9cm}|l|}
    \hline
     & Requirement & Remarks \\
    \hline
    Wavelength & 11--18 $\mu$m & \\
    \hline
    Spatial resolution & $<3$ arcsec at $\lambda=15\,\mu$m &\\
    \hline
    Wavelength resolution & $\lambda/\Delta \lambda>30,000$ & \\
    \hline
    Field of view & N/A & \\
    \hline
    Sensitivity & Line sensitivity 5$\sigma$, 1 hr & \\
    \hline
    Observing field & Our Galaxy & \\
    \hline
    Observing cadence & N/A & \\
    \hline
    \end{tabular}
    \end{center}
\end{table}

\subsection{Required observations}

\subsection{Scientific goals}

\printbibliography[heading=subbibliography]
\end{refsection}

\clearpage

\begin{refsection}[3-4_exoplanetatmosphere/exoplanetatmosphere.bib]

\section{Exoplanet Atmosphere}
\label{sec:exoplanetatmosphere}

\noindent
\begin{flushright}
Yuka Fujii$^{1}$, 
Yui Kawashima$^{2}$,
Taro Matsuo$^{3}$, 
Kazumasa Ohno$^{1,4}$
\\
$^{1}$ NAOJ,
$^{2}$ Kyoto University, 
$^{3}$ University of Osaka,
$^{4}$ University of California, Santa Cruz
\end{flushright}
\vspace{0.5cm}

\subsection{Scientific background and motivation}

Discoveries of exoplanets since the 1990s have revolutionized our view of the Universe by revealing the ubiquity of planetary systems and the wide diversity in their architectures. 
Understanding the trends and origins of this diversity is crucial for a unified picture of planetary system formation that puts the Solar System into context. 

To this end, characterization of their atmospheric properties by detailed follow-up observations is expected to provide valuable clues. 
This is because the atmospheric composition of gaseous planets reflects their formation environment. For example, since the gaseous C/O ratio in the protoplanetary disk is expected to vary within the disk owing to chemical gradients induced by snow lines \citep[e.g.,][]{2011ApJ...743L..16O, Eistrup+16, Booth&Ilee19,Notsu+2020,Notsu+2022,Schneider&Bitsch21}, we can estimate where in the disk the planet captures its atmosphere by measuring the C/O ratio of the atmosphere.
So far, C/O ratios of planetary atmospheres have been investigated mainly by constraining the abundance ratios of CO/$\mathrm{CH_4}$ to $\mathrm{H_2O}$ through low-resolution spectroscopy \citep[e.g.,][]{2011Natur.469...64M, 2016AJ....152..203L, 2020A&A...642A..28M}.
This is because those species are the main carbon- and oxygen-bearing molecules in the hydrogen-dominated atmospheres \citep[e.g.,][]{Lodders&Fegley02,Moses+13} and have prominent absorption features in near-infrared wavelengths accessible with the current observations.
While less abundant than CO, $\mathrm{CH_4}$, and $\mathrm{H_2O}$, there are also other good tracers of the C/O ratio, such as $\mathrm{CO_2}$, HCN, and $\mathrm{C_2H_2}$, which are produced photochemically in the upper atmosphere \citep[e.g.,][]{2014RSPTA.37230073M, 2019ApJ...877..109K}.
These photochemical species can provide insights not only into the C/O ratio, but also into the UV irradiation intensity from the host star and the strength of atmospheric vertical mixing. 
UV irradiation from the host star plays vital roles in both atmospheric chemistry and planetary evolution accompanied by atmospheric escape, while the strength of atmospheric vertical mixing provides a key to understanding atmospheric dynamics across the entire planet.
However, their low absolute abundances hamper detection by low-resolution spectroscopy.

In addition to the C/O ratio, recent studies suggested that the atmospheric N/O ratio also provides strong constraints on the planet formation environment \citep{Piso+16,Cridland+20,Ohno&Ueda21,Notsu+2022}.
This is because the main nitrogen reservoirs in protoplanetary disks, namely NH$_3$ and N$_2$, have an order-of-magnitude difference between their abundances \citep{Oberg&Bergin21}, which produces a drastic radial variation in the N/O ratio in the disk across the N$_2$ snowline.
Based on the atmospheric nitrogen abundance, several studies have suggested that Solar System Jupiter might have originally formed in extremely cold environments of $<30~{\rm K}$ \citep{Owen+99,Oberg&Wordsworth19,Bosman+19,Ohno&Ueda21}.
In exoplanetary atmospheres, NH$_3$ and HCN provide a way to constrain the atmospheric nitrogen abundance \citep{Macdonald&Madhusudhan17,Ohno&Fortney22a,Ohno&Fortney22b}. 
However, low-resolution spectroscopy has not conclusively detected NH$_3$ and HCN in exoplanets.
This is possibly owing to the overlap of their near-infrared spectral features with those of O- and C-bearing species and the prevalence of photochemical hazes in warm exoplanets that mute absorption features of gaseous molecules \citep[e.g.,][]{Gao+20,Dymont+21}.

To overcome the difficulty of current observations of exoplanetary atmospheres, high-resolution spectroscopy allows robust detection by distinguishing the feature of interest from other molecular features.  
Indeed, the aforementioned minor species have recently started to be found in exoplanet atmospheres by high-resolution spectroscopy with ground-based telescopes, which has advanced our understanding of the C/O ratios of exoplanetary atmospheres \citep{2021Natur.592..205G}.
The species of interest, especially C$_2$H$_2$, HCN, and NH$_3$, have stronger absorption features in the mid-infrared wavelength region, where significant telluric absorption hampers observations from the ground.
Thus, utilizing the high-resolution spectrograph (10--18~$\mu$m) mounted on GREX-PLUS, we can expect robust detection of the above species, 
which will further advance our understanding of the atmospheric C/O ratio, N/O ratio, vertical mixing, and UV environment.

Operating in the mid-infrared, GREX-PLUS high-resolution spectroscopy is inherently sensitive to spectral features in thermal emission, over a wide temperature range of planets down to $\sim $500~K. 
The coolest ones are accessible only in the mid-infrared where their thermal emission is in the Jeans domain and the planet-to-star flux ratio becomes larger. 
High-resolution spectroscopy of their thermal emission will be able to constrain the composition and the thermal structure of the atmospheres with high fidelity for the first time. 
Furthermore, the high-resolution spectroscopy allows us to resolve the Doppler shift of planetary thermal emission due to the planetary orbital motion, thereby measuring the radial velocities of planets. 
This suggests that planetary spectral features can be separated from those of the host star in a model-independent manner even for non-transiting (non-eclipsing) planets. 
Radial velocity measurements of non-transiting planets can constrain their orbital inclinations and hence their ``true'' masses. 
The ability of GREX-PLUS to probe non-transiting planets thanks to its high spectral resolution provides a significant advantage over JWST for studying relatively cool targets, since transit probabilities decrease rapidly with increasing distance from the host star.
The unique parameter space that GREX-PLUS mid-infrared high-resolution spectroscopy can probe will fill the gap between the two major categories of planets for which atmospheric observations are/will be actively pursued: close-in transiting planets and more distant directly-imaged planets. 
While challenging, the radial velocity measurements of Jupiter-like planets also serve as a unique probe of the presence of (large) satellites or binary planets. 

Potential targets also include cooling gas giants in distant orbits that are still luminous due to their intrinsic energy gained through gas accretion. Both the ``hot-start" and ``cold-start" models \citep[e.g.,][]{2007ApJ...655..541M} predict that these self-luminous gas giants with ages of less than 100 Myr have effective temperatures of more than 500~K. Such self-luminous gas giants will be excellent targets for GREX-PLUS in terms of improving the signal-to-noise ratio. Before the era of GREX-PLUS, several space telescope missions, such as JWST and Nancy Grace Roman Space Telescope (hereafter {\it Roman}) will search for self-luminous gas giants around nearby moving groups at ages of 10--100 Myr \citep{2020AJ....159..166U}. Note that the memberships of the nearby moving groups are well characterized by the high-precision proper-motion measurements of Gaia. 
For example, the Mid-InfraRed Imager (MIRI) of JWST provides several coronagraphic capabilities in the mid-infrared regime for the first time, which are sensitive to warm and temperate gas giants in the outer planetary system \citep[e.g.,][]{2023ApJ...951L..20C}. 
In addition, {\it Roman} will search for reflected light from young gas giants close to the central stars (less than 5 AU), thanks to its small inner working angle \citep{2018SPIE10698E..2IM}. 
Since both missions perform only high-contrast imaging (not high-contrast spectroscopy), GREX-PLUS will build strong synergies with these missions.


\subsection{Required observations and expected results}

\subsubsection{Transmission spectra}

There are two possible approaches to observing exoplanetary atmospheres using the high-resolution spectrograph of GREX-PLUS.
One is transmission spectroscopy of transiting planets, which measures the absorption by the planetary atmosphere during transit relative to the stellar spectrum observed outside of transit.
The cross-correlation method is often used to investigate the existence of chemical species in exoplanetary atmospheres \citep[e.g.,][]{2013MNRAS.436L..35B, 2017AJ....154..221N}.
This method calculates the cross-correlation between the observed spectrum and a theoretical template spectrum that predicts the wavelengths and strengths of absorption lines for a given chemical species. 
One can infer the presence or absence of each chemical species by looking at the goodness of the cross-correlation between the observed and template spectra.
In transmission spectroscopy, the amplitudes of spectral features are scaled by the atmospheric scale height. 
Thus, our prime targets are planets with sizes ranging from that of Jupiter to Neptune, which are likely to have hydrogen-dominated atmospheres with large scale heights. 

\begin{figure}
    \centering
    \includegraphics[width=\textwidth]{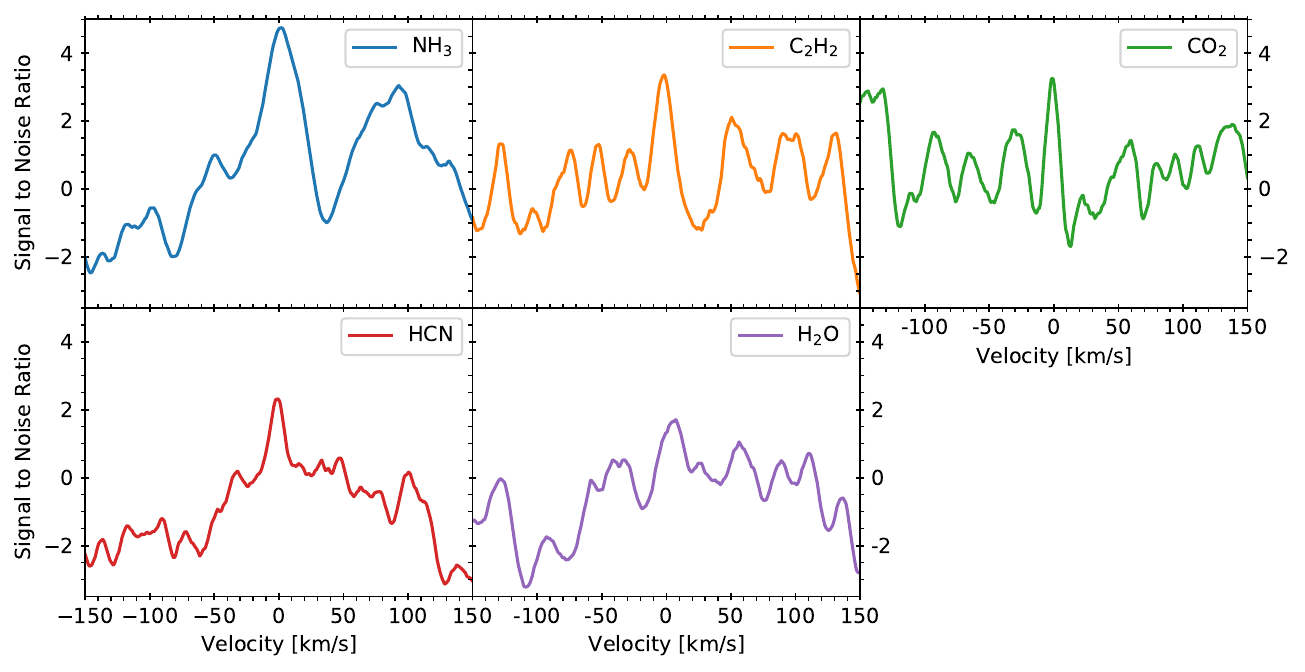}
    \caption{Results of the cross-correlation between the mock and template spectra for the case of 20 transit observations of a mini-Neptune planet GJ 1214b by GREX-PLUS.}
    \label{fig:ccf_transit}
\end{figure}

As an example, we simulate transit observations of the mini-Neptune GJ 1214b orbiting an M4.5-type host star at a distance of 14.65~pc by GREX-PLUS.
The transit duration is about one hour, and we consider a case in which 20 transits are observed.
For the spectral model, we adopt the fiducial model of \citet{2019ApJ...877..109K}. 
The abundance profiles of chemical species are calculated using the photochemical model of \citet{2018ApJ...853....7K}, assuming solar elemental abundance ratios and a constant eddy diffusivity of $10^7$~$\mathrm{cm^2/s}$ throughout the atmosphere. 
For the temperature profile, the analytical formula of \citet{2010A&A...520A..27G} is used. For further details, see \citet{2019ApJ...877..109K}.
In Figure~\ref{fig:ccf_transit}, we show the results of the cross-correlation between the mock and template spectra for $\mathrm{NH_3}$, $\mathrm{C_2H_2}$, $\mathrm{CO_2}$, $\mathrm{HCN}$, and $\mathrm{H_2O}$.
The detection significances are 4.7, 3.4, 3.3, 2.3, and 1.7$\sigma$ for $\mathrm{NH_3}$, $\mathrm{C_2H_2}$, $\mathrm{CO_2}$, $\mathrm{HCN}$, and $\mathrm{H_2O}$, respectively.
(We note that, for $\mathrm{H_2O}$, the existence of its absorption features over the wide observational wavelength range makes it somewhat challenging to detect via the cross-correlation technique despite its large abundance.)
Thus, high-resolution transmission spectroscopy with GREX-PLUS opens a new window for detecting minor chemical species, such as NH$_3$, C$_2$H$_2$, and HCN, which have not been robustly discovered by current low-resolution spectroscopy.

\subsubsection{Thermal emission spectra}

Another approach to studying exoplanet atmospheres is detecting the thermal emission spectra of the planet. 
Even if the star and planet are not spatially resolved, features in planetary spectra may be identified in the composite spectra due to the distinct Doppler shift of the planetary signal (for the application of this method to mid-infrared high-resolution spectroscopy, see \cite{2021AJ....161..180F}). 
In what follows, we examine the viability of this method for the GREX-PLUS mission using modeled thermal emission spectra and mock observations. 

\begin{figure}[tbh]
    \centering
    \includegraphics[width=0.5\textwidth]{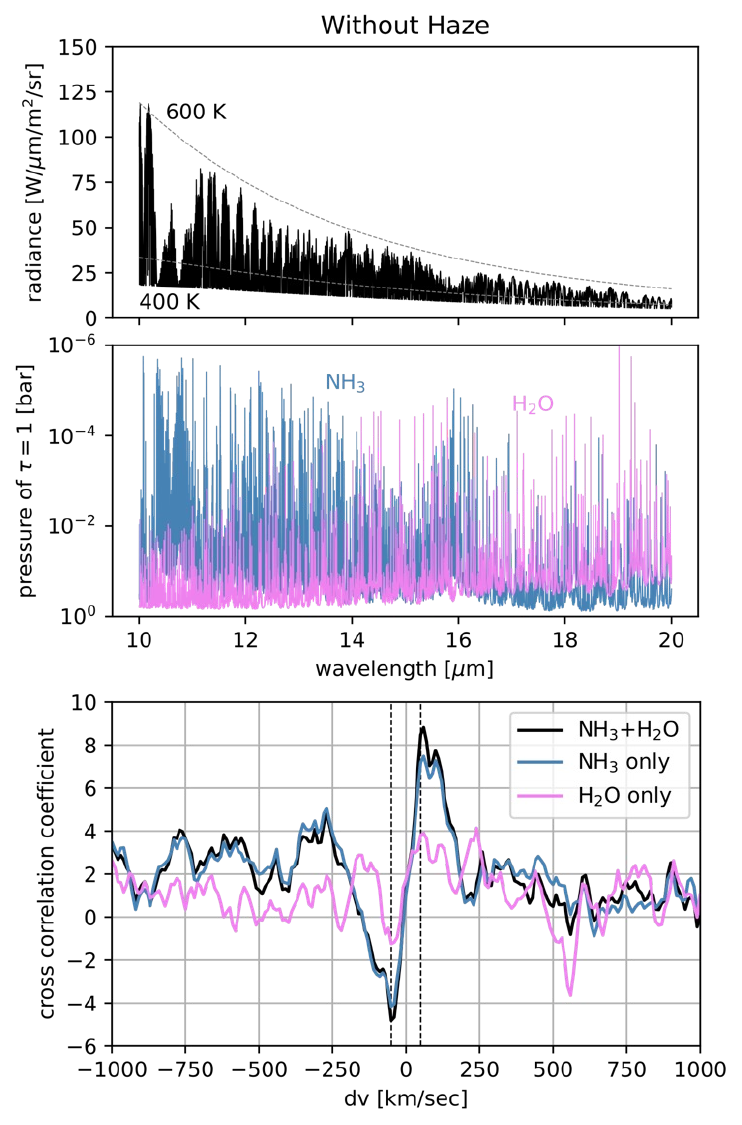}
    \caption{Top panel: model thermal emission spectrum of a Jupiter-like planet with an equilibrium temperature of 500~K. Middle panel: pressure levels at which $\tau =1$. Bottom panel: result of the cross-correlation analysis between the mock and model spectra. The mock data assume a $\sim500$~K Jupiter-sized planet around a Solar-type star at 20~pc, with observations of 1~day performed at both quadrature phases. The two vertical dashed lines represent the radial velocities at the two quadrature phases.}
    \label{fig:sp_corr_WJ500K}
\end{figure}

The top and middle panels of Figure~\ref{fig:sp_corr_WJ500K} are the modeled thermal emission spectra of a warm Jupiter-sized planet (500~K, 1$R_J$, 1$M_J$) and the pressure levels where the opacity of each molecule from the top of the atmosphere becomes unity. 
Here, we employed a 1-dimensional atmospheric profile assuming solar metallicity, where the temperature profile is taken from \citet{2014A&A...562A.133P} and the chemical profile is calculated using \citet{2018ApJ...853....7K}. 
Specifically, the mixing ratios of NH$_3$ and H$_2$O are approximately 10$^{-4}$ and 10$^{-3}$, respectively. 
Spectral features at $<15~\mu\mathrm{m}$ are dominated by NH$_3$, while H$_2$O features appear at longer wavelengths.

For the mock observations, we place this planet around a Solar-type star at a distance of 20~pc from Earth. 
The target system is observed at two quadrature phases (corresponding to the maximum and minimum radial velocities) for one day each (any combination of phases would work as long as the two radial velocities are well separated). 
We neglect the radial velocity of the central star, as it is much slower than that of the planet. 
Each spectrum is high-pass filtered and then the difference between the two is extracted. 
By cross-correlating this differential spectrum with the modeled planetary spectrum, we examine the detectability of specific features in planetary spectra and measure the radial velocity of the planet. 

The cross-correlation between the mock data and the theoretical models is shown in the bottom panel. 
The signal-to-noise ratio (SNR) of the peak of the cross-correlation function can be $>5\sigma$ if the model is correct, while a model with only NH$_3$ can also lead to a similar SNR. 
H$_2$O is harder to detect in this case, because the detectable cross-correlation signal is largely driven by NH$_3$ features below 15~$\mu$m, whereas H$_2$O contributes mainly at longer wavelengths with weaker effective line contrast.

\begin{figure}[tbh]
    \centering
    \includegraphics[width=0.5\textwidth]{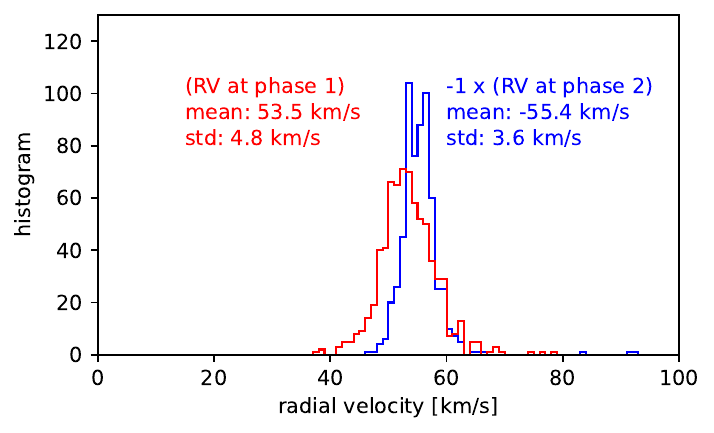}
    \caption{Probability distributions of the planetary radial velocity at the two quadrature phases based on mock observations of a $\sim500$~K Jupiter-sized planet around a solar-type star at 20~pc, where each quadrature phase is observed for 1~day.}
    \label{fig:hist_rv_WJ500K}
\end{figure}

Figure~\ref{fig:hist_rv_WJ500K} shows the distribution of the estimated radial velocity at these two phases (indicated by different colors), suggesting that the uncertainty in the radial velocity is a few km~s$^{-1}$. 
This can be compared with the typical orbital velocity ($30$--100~km s$^{-1}$), implying that the orbital inclination can be well constrained. 

Such measurements of orbital velocity can be extended to the detection of binary planets. If an Earth-sized planet or a Neptune-sized planet orbits the planet in the orbit of the Jovian moon Io, the resultant variations of radial velocity are approximately 50~m~s$^{-1}$ and 900~m~s$^{-1}$, respectively. 
The radial velocity variation of binary planets, as proposed by \cite{2014ApJ...790...92O}, can reach up to 17~km~s$^{-1}$. 
Monitoring the radial velocity over different epochs may allow us to constrain the presence of relatively large moons or binary planets.

While we did not assume a specific target in the study above, 
the analysis can be directly applied to, for example, 
Rho CrB b (17.5~pc, $M_p \sin i\sim 1.1 M_J$) and 
70 Vir b (17.9~pc, $M_p \sin i\sim 7.4 M_J$). 
Furthermore, cool Jupiter-like planets ($\gtrsim300$~K) around nearby M-type stars may also be investigated. 
For example, GJ~876~c, a Jovian planet around an M4 star at 4.7~pc and receiving an incident energy flux similar to that of Earth, exhibits a star-to-planet contrast and photon flux similar to those of a warm Jupiter around a solar-type star at 20~pc discussed above. 
Thus, atmospheric species on such planets may also be detected depending on their atmospheric properties. 

\begin{figure}
    \centering
    \includegraphics[width=0.5\textwidth]{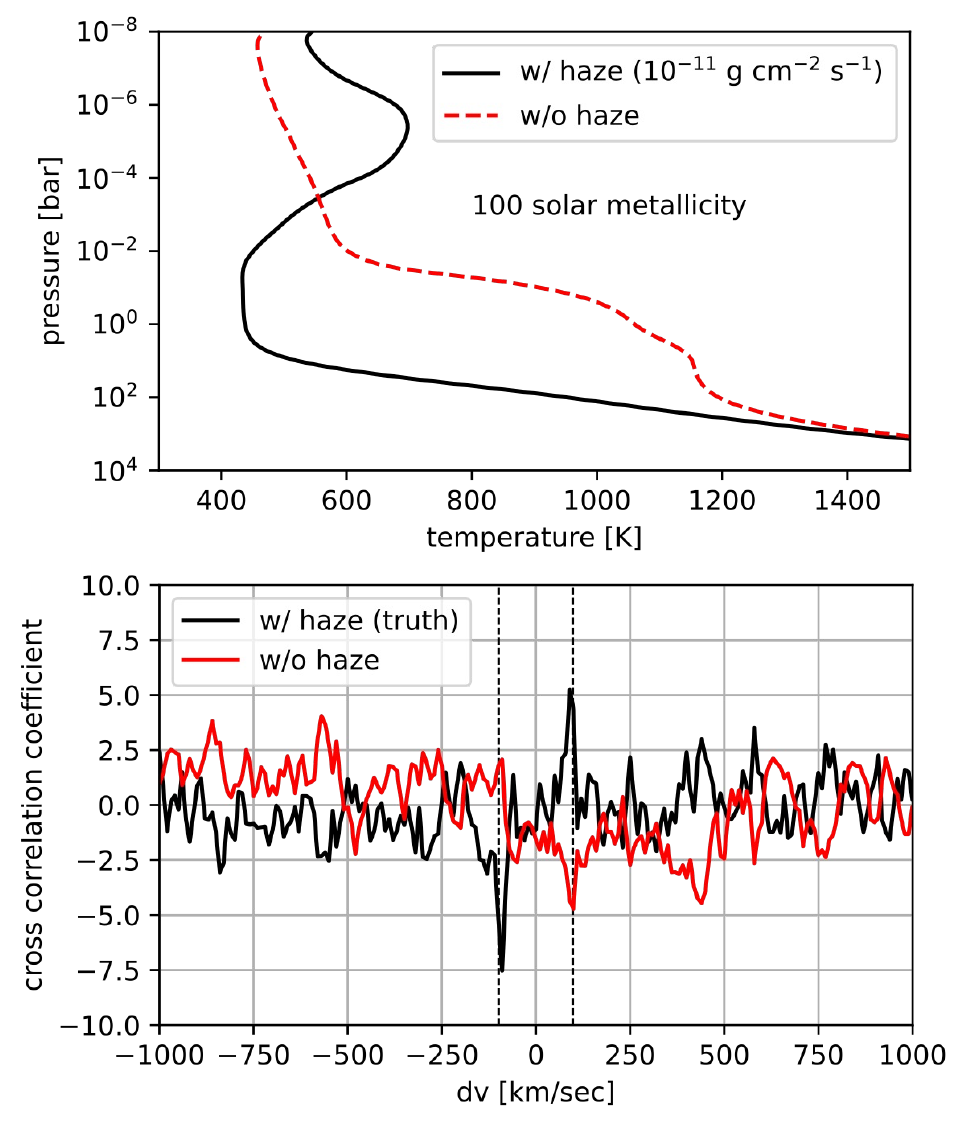}
    \caption{Top panel: modeled temperature profiles of GJ 1214b with and without a haze layer based on \citet{Ohno&Fortney22a,Ohno24}; the presence of a haze layer in GJ 1214b is suggested by the observed near-infrared transmission spectrum \citep{Kreidberg+14}. The haze layer would induce a thermal inversion in the upper atmosphere. Bottom panel: cross-correlation coefficients between the mock data of GJ 1214b and the two modeled spectra with and without a haze layer (black and red lines in the top panel, respectively). The mock data assume an atmospheric profile with a haze layer (black line in the top panel) and observations of 2.5 days at each quadrature phase. The two vertical dashed lines represent the radial velocities at the two quadrature phases. While the model with a haze layer produces signals at the expected radial velocities with SNR $>$ 5, the model without a haze layer shows less significant signals.}
    \label{fig:GJ1214b_crr_haze}
\end{figure}

A more intensive use of telescope time would be required to detect the thermal emission of Neptune-sized planets such as GJ~1214b. 
The detectability depends strongly on the actual atmospheric structure.
As a case study, we examine whether the thermal emission of a planet such as GJ~1214b would be detectable by adopting plausible atmospheric structure models. To generate the mock spectra, we computed the atmospheric pressure--temperature (PT) profile for 100$\times$ solar metallicity using a 1D radiative-convective equilibrium model \citep{McKay+89,Marley&Robinson15} and then postprocessed it with a publicly available photochemical model, VULCAN \citep{Tsai+21}.
We also consider the possible presence of photochemical hazes by iteratively computing the PT profile and vertical haze distribution using an aerosol microphysical model (\citealt{Ohno&Okuzumi18,Ohno&Kawashima20,Ohno24}).
Our simulations suggest that observations of $\gtrsim5$ days would enable the detection of NH$_3$ on GJ~1214b. 
Note that thermal emission is sensitive to both the atmospheric composition and its vertical structure. 
A cross-correlation analysis using model thermal emission spectra with and without a haze layer, the former of which has a thermal inversion in the upper atmosphere \citep[e.g.,][]{2015ApJ...815..110M, Lavvas&Aufaux21, Ohno24}, could provide evidence of the haze layer (Figure~\ref{fig:GJ1214b_crr_haze}).

\subsection{Scientific goals}

\begin{itemize}
    \item Detect minor species in exoplanet atmospheres as a probe of atmospheric elemental abundances, vertical mixing, and UV environment. 
    \item Constrain the atmospheric thermal structure of the planet to infer the presence or absence of aerosols. 
    \item Measure the radial velocity of the planet. 
\end{itemize}

\begin{table}
    \label{tab:exoplanet}
    \begin{center}
    \caption{Required observational parameters.}
    \begin{tabular}{|l|p{9cm}|l|}
    \hline
     & Requirement & Remarks \\
    \hline
    Wavelength & 10--18 $\mu$m & \multirow{2}{*}{} \\
    \cline{1-2}
    Spatial resolution & N/A & \\
    \hline
    Wavelength resolution & $\lambda/\Delta \lambda>30,000$ & \\
    \hline
    Field of view & N/A & \\
    \hline
    Sensitivity & ?? & \\
    \hline
    Observing field & N/A & \\
    \hline
    Observing cadence & N/A & \\
    \hline
    \end{tabular}
    \end{center}
\end{table}

\printbibliography[heading=subbibliography]
\end{refsection}

\clearpage

\begin{refsection}[3-5_solarsystemplanet/solarsystemplanet.bib]

\section{Solar System Planetary Atmospheres}
\label{sec:solarsystemplanet}

\noindent
\begin{flushright}
Hideo Sagawa$^{1}$
\\
$^{1}$ Kyoto Sangyo University
\end{flushright}
\vspace{0.5cm}

\subsection{Scientific background and motivation}
Observations of planetary atmospheres are important not only for studying meteorology and climate, but also for understanding the formation and evolution of planets through knowledge of their atmospheric compositions. In addition, detailed characterization of planetary atmospheres---such as their redox state and chemical stability, the presence or absence of clouds and hazes, and atmospheric escape---can provide essential information on whether a planet can harbor life, i.e., its habitability. 

For a long time, attempts were made to describe the atmospheres of other planets based on knowledge obtained from the Earth's atmosphere. However, it is now evident that Earth's atmospheric conditions are not universal; our planet cannot be regarded as a standard reference for planetary atmospheres. Each planet has a different distance from the Sun, resulting in different levels of solar radiation incident on the atmosphere. Differences in planetary mass and rotation rate lead to variations in gravitational acceleration and the Coriolis force, both of which are fundamental parameters governing atmospheric dynamics. The total atmospheric mass and its composition, including minor trace gases, also vary from planet to planet, and these parameters play key roles in radiative transfer, i.e., the distribution of energy within the atmosphere. Furthermore, non-atmospheric factors such as the presence or absence of an intrinsic magnetic field and volcanic activity also contribute to the diversity of planetary atmospheres. The more we learn about this diversity, the more important it becomes to expand comparative studies of planetary atmospheres. In this context, detailed observations of Solar System planetary atmospheres and a thorough understanding of the processes occurring within them are also crucial for interpreting the atmospheres of exoplanets, whose observational accessibility has rapidly expanded in recent years with the advent of facilities such as the James Webb Space Telescope (JWST). 

Infrared spectroscopy has long served as one of the most effective tools for observing planetary atmospheres. Thermal infrared radiation from a planet is observed as a combination of continuum emission from the atmosphere and absorption lines (or emission lines, in some cases) produced by specific molecular species. By analyzing such infrared spectra, we can derive both the temperature structure and the atmospheric composition.

The temperature structure is undoubtedly one of the most fundamental physical properties of planetary atmospheres, and its variation with latitude and time (season) drives a variety of atmospheric phenomena, including global circulation. The gaseous and icy giant planets beyond Jupiter possess internal heat sources; however, it remains unclear to what altitudes these internal heat sources influence the atmospheric temperature structure. For example, internal heat appears to have a limited effect on the weather layer of Uranus, whereas it plays a more significant role in Neptune. This implies that the atmospheric thermal structure encodes information not only about meteorological processes in the observable atmosphere but also about the transport of energy from the deep interior. Therefore, knowledge of the thermal structures of gaseous and icy giant planets is also important for constraining their internal structure and formation processes.

The temperature structures of these outer Solar System planets have been investigated in several studies, beginning with flyby observations by the Voyager spacecraft \citep[e.g.,][]{Lindal81, Lindal85, Conrath98}, followed by observations using infrared space telescopes and orbiting spacecraft \citep[e.g.,][]{Lellouch01, Orton14, Fletcher16}. However, no single observation has yet provided comprehensive coverage of the vertical temperature structure from the lower to the upper atmosphere. In addition, the temporal coverage of existing observations is sparse and limited, making them insufficient for fully understanding seasonal variations in the atmospheres of ice giant planets.

\begin{figure}[!t]
\centering
\includegraphics[width=0.88\linewidth]{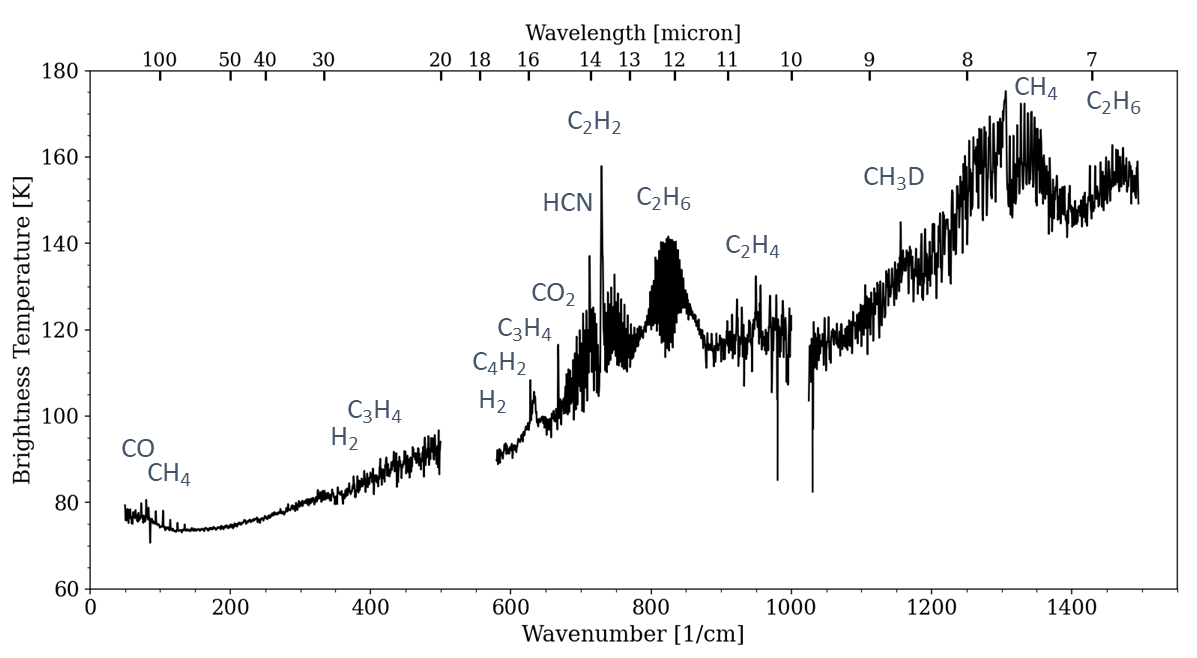}
\caption{An example of an infrared spectrum of Titan's atmosphere observed by Cassini/CIRS. 
The spectral resolution of CIRS is $\sim$0.5~cm$^{-1}$.} 
\label{fig:SolSysAtmos_Titan}
\end{figure}

Observations of atmospheric composition, including isotopologues, are essential for understanding atmospheric chemistry. In particular, Titan's atmosphere is of great interest among Solar System bodies, as its N$_2$-dominated and CH$_4$-rich reducing environment hosts one of the most complex known atmospheric chemistries \citep{Horst17}. Observations by the Composite Infrared Spectrometer (CIRS) on board the Cassini spacecraft have revealed the presence of numerous organic molecules in the stratosphere, including nitriles (HCN, HC$_3$N, CH$_3$CN, etc.) and hydrocarbons (C$_4$H$_2$, C$_2$H$_2$, C$_2$H$_6$, etc.) (Figure~\ref{fig:SolSysAtmos_Titan}). These species exhibit strong spatial and temporal variability in their abundances, which is interpreted as a result of Titan's atmospheric circulation and photochemistry \citep{Coustenis07, Coustenis10}. However, Cassini/CIRS observations covered less than half of one Titan year (approximately 30 Earth years). Continued monitoring observations are therefore required to capture the full characteristics of seasonal variations in Titan's atmosphere. 

\begin{figure}[!t]
\centering
\includegraphics[width=0.64\linewidth]{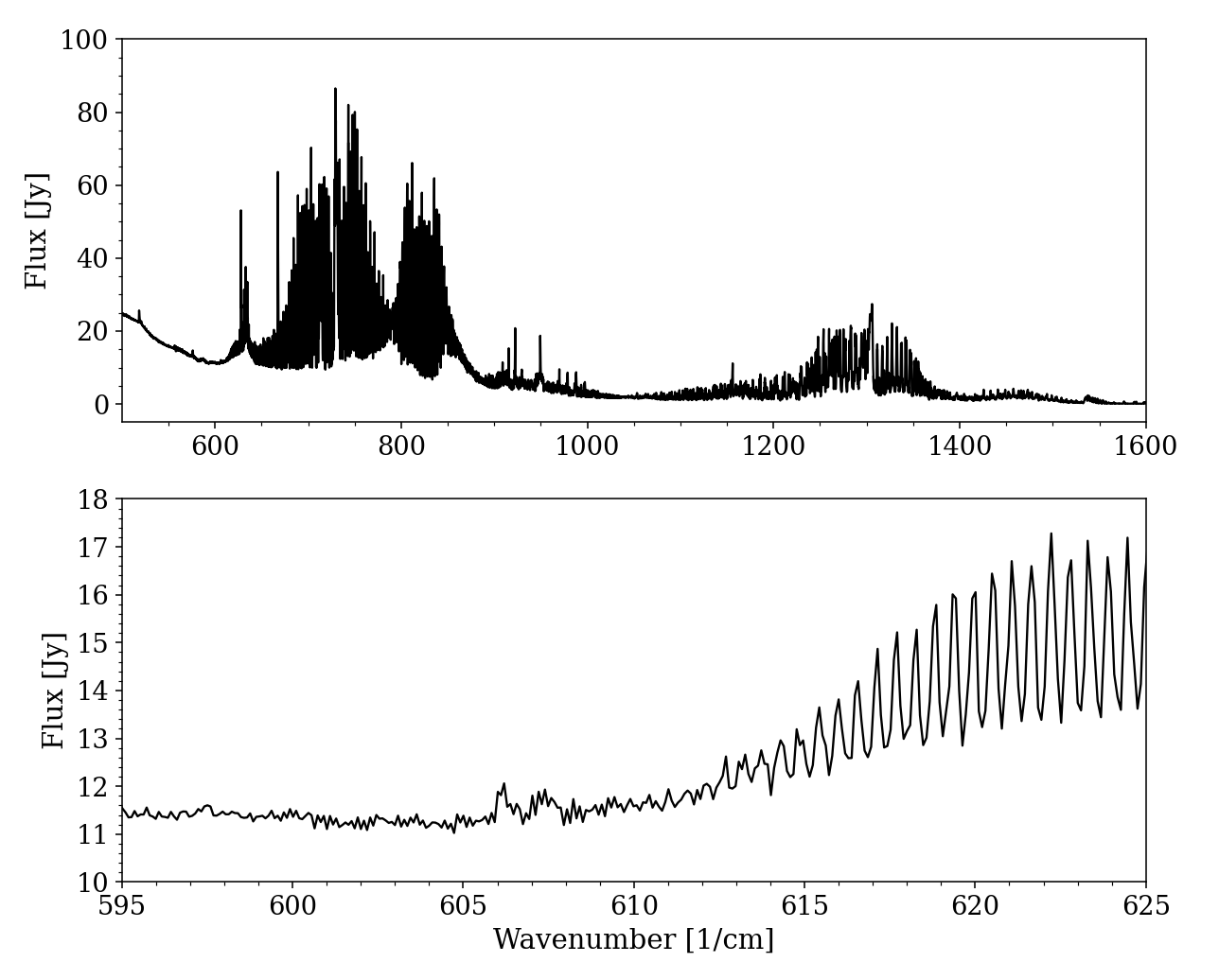}
\caption{Top: Titan JWST/MIRI spectra obtained in a Guaranteed Time Observation program GTO~1251. Data are retrieved from the data archive (Level-3 ``x1d'' product). Bottom: Close-up of the 600~cm$^{-1}$ region (16.5~$\mu$m). The weak emission feature at 606~cm$^{-1}$ is attributed to CH$_3$ lines \citep{Nixon25}.} 
\label{fig:SolSysAtmos_MIRI}
\end{figure}

In this context, spectroscopic observations with the Mid-Infrared Instrument (MIRI) on board JWST have recently begun \citep{Nixon25}, extending and complementing the earlier investigations of Titan’s seasonal variability conducted with Cassini/CIRS. While the spectral resolution of MIRI is comparable to that of CIRS, its superior sensitivity and spatial resolution represent a significant advancement. Notably, JWST/MIRI observations have successfully detected the methyl radical (CH$_3$), a key intermediate species in Titan’s atmospheric photochemistry that had long eluded direct detection (Figure~\ref{fig:SolSysAtmos_MIRI}). The ability to detect such fundamental species demonstrates the enhanced capability of JWST and highlights the importance of continued observations to further constrain the complex chemical processes in Titan’s atmosphere.

One aspect of the observational parameter space that cannot be fully explored even with JWST/MIRI is high spectral resolution. Such resolution enables more precise separation of absorption lines from different molecular species in the observed spectrum; as discussed below, this capability is essential for studies of planetary atmospheres.

\subsection{Required observations and expected results}

\subsubsection{Vertical profile of atmospheric temperature}
 
Infrared emission from gaseous and icy giant planets exhibits spectrally broad absorption features caused by collision-induced absorption (CIA) of molecular hydrogen (H$_2$) and helium, which are the two major constituents of their atmospheres. In addition, several narrow spectral features arising from the quadrupole rotational transitions of H$_2$ are observed. The intensity (opacity) of CIA and quadrupole line emission depends on both the abundance of H$_2$ and the temperature distribution along the line of sight. Since the volume mixing ratio of H$_2$ is relatively well constrained, the atmospheric temperature profile can be retrieved from infrared spectral observations. It should be noted, however, that the intensity of the quadrupole lines also depends on the ortho-to-para ratio of H$_2$ nuclear spin isomers; therefore, an assumption about this ratio is required. A value of 3:1, corresponding to room-temperature conditions, is commonly adopted.

\begin{figure}[!t]
\centering
\includegraphics[width=0.92\linewidth]{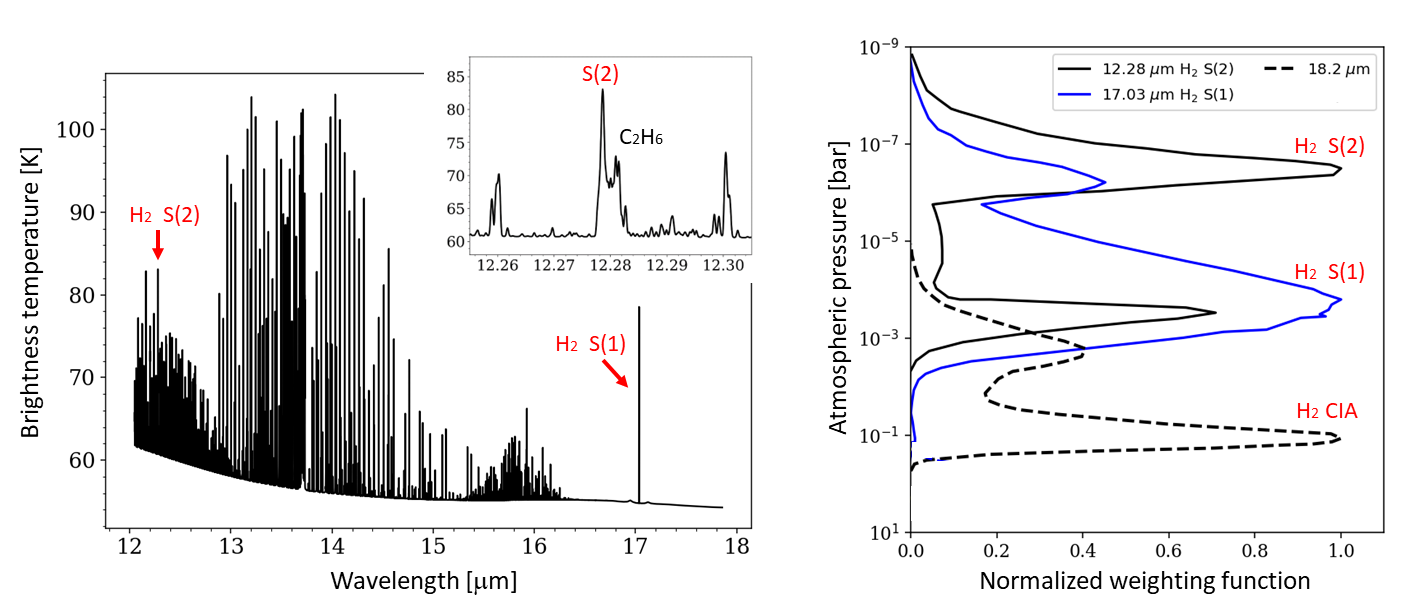}
\caption{Left: Simulated infrared spectrum of Uranus's atmosphere. A close-up to H$_2$ S(2) is shown in the upper right panel. A spectral resolving power of $\sim$30,000 is assumed. Right: Temperature weighting functions at selected wavelengths.} 
\label{fig:SolSysAtmos_Uranus}
\end{figure}

As an example, Figure~\ref{fig:SolSysAtmos_Uranus} shows a simulated infrared spectrum of Uranus, assuming a spectral resolving power of $\sim$30,000. The H$_2$ quadrupole lines S(1) and S(2) are observable at 17.04 and 12.28~$\mu$m, respectively. Absorption features of C$_4$H$_2$, C$_2$H$_2$, and C$_2$H$_6$ are also present at wavelengths shorter than 16~$\mu$m (although they are not labeled in the figure). The sensitivity of the observed spectrum to the atmospheric temperature profile is evaluated using the weighting function, defined as the functional derivative of spectral radiance with respect to temperature at each altitude. Larger values of the weighting function indicate greater sensitivity to temperature at the corresponding atmospheric level. 
The continuum emission primarily constrains the temperature in the upper troposphere (around the $\sim$0.1 bar level; see the dashed line in the right panel of Figure~\ref{fig:SolSysAtmos_Uranus}). In contrast, observations of the narrow H$_2$ quadrupole lines S(1) and S(2) extend the maximum altitude range up to the upper stratosphere (from $\sim$10$^{-3}$ to $\sim$10$^{-7}$ bar, as indicated by the solid lines). One of the main challenges is the precise measurement of the H$_2$ S(2) line intensity. This line is located at 12.28~$\mu$m, where numerous C$_2$H$_6$ lines are also present, leading to spectral blending. High spectral resolving power is therefore required to separate the H$_2$ S(2) line from interfering C$_2$H$_6$ features. 
As shown in Figure~\ref{fig:SolSysAtmos_Uranus}, more accurate retrieval of the temperature profile from the H$_2$ S(2) line would improve sensitivity to higher atmospheric layers ($\sim$10$^{-7}$ bar). In particular, recent stellar occultation observations probing these pressure levels have revealed lower temperatures than previously inferred \citep{Saunders24}, highlighting the need for additional observational constraints.

Previous observations with Spitzer/IRS ($\lambda/\Delta\lambda \sim 600$) did not provide sufficient spectral resolution to fully resolve H$_2$ S(2) and C$_2$H$_6$ lines, complicating the analysis compared to that of the H$_2$ S(1) line \citep{Orton14}. 
Even with $\lambda/\Delta\lambda \sim 10{,}000$, it remains uncertain whether the H$_2$ S(2) line can be clearly resolved in observational spectra alone; comparison with models that incorporate atmospheric radiative transfer is required. In this context, an even higher spectral resolving power (e.g., $\lambda/\Delta\lambda \sim 30{,}000$) would be preferable. 

Furthermore, infrared spectra of Uranus and Neptune have been obtained only infrequently since earlier observations by the Infrared Space Observatory (ISO) and the Spitzer Space Telescope. Given the long orbital periods of these ice giant planets, new infrared spectroscopic observations in the 2030s would provide important constraints on seasonal variations in their atmospheric temperature structures.

\subsubsection{Abundances of minor species}

In addition to temperature, the atmospheric compositions of gaseous and icy giant planets remain incompletely understood. Some minor species contribute to greenhouse warming and/or radiative cooling; therefore, accurate knowledge of their abundances is essential for constraining atmospheric thermal structures. 

As shown in Figures~\ref{fig:SolSysAtmos_Titan}--\ref{fig:SolSysAtmos_Uranus}, infrared spectroscopy is a highly effective technique for the spectral detection of minor species, particularly in the stratosphere. Because these species are often closely linked through photochemical reactions, simultaneous measurements of multiple molecular species are required to develop reliable photochemical models. Such observations can be achieved in the wavelength range of $\sim12$--18~$\mu$m, where numerous molecular lines are densely distributed. For molecules that lack suitable transitions in the infrared region, complementary observations with (sub)millimeter facilities such as ALMA are also effective. 

In addition to their chemical interconnections, many of these minor species exhibit significant spatial and temporal variability driven by atmospheric circulation and photochemistry. Titan provides a representative example of such behavior. To understand the seasonal variations in Titan’s atmosphere, it is important to continue monitoring observations over as long a period as possible, with a frequency of once or twice a year. This cadence is roughly equivalent to sampling seasonal changes on Earth every two weeks to one month. 

For both topics (1) and (2), a spatial resolution better than a few arcseconds is required. This resolution is sufficient to resolve latitudinal belts and zones, as well as some localized features (e.g., the Great Red Spot), on Jupiter, which has an apparent disk diameter of 30--50 arcsec. Saturn (15--20 arcsec in diameter) can similarly be resolved into equatorial, mid-latitude, and high-latitude regions. 
A spatial resolution of a few arcseconds is comparable to the apparent diameters of Uranus and Neptune, and is therefore insufficient to fully resolve their disks. Nevertheless, such resolution still provides an advantage by enabling the detection of planetary emission without significant dilution by the cold background sky. Titan, with an angular diameter of approximately 0.8 arcsec, is effectively observed as a point source. The required sensitivity is better than $10^{-18}$~W~m$^{-2}$, corresponding to the typical flux level of the weakest target emission lines.

\subsection{Scientific goals}
The primary objective of this study is to obtain high-resolution thermal infrared spectra of planetary atmospheres in our Solar System over the wavelength range of 12--18~$\mu$m, with a spectral resolving power of $\lambda/\Delta\lambda \gtrsim 30{,}000$. From these observations, the vertical temperature structure of the atmospheres, as well as the abundances of minor species---including their isotopologues---will be derived. Particular emphasis is placed on the gaseous and icy giant planets, for which observational constraints remain limited, and on Titan, which hosts highly complex atmospheric chemistry. To constrain temporal (seasonal) variability, the data should be acquired at multiple epochs.

\begin{table}[ht]
    \label{tab:ssplanet}
    \begin{center}
    \caption{Required observational parameters.}
    \begin{tabular}{|l|p{9cm}|l|}
    \hline
     & Requirement & Remarks \\
    \hline
    Wavelength & 12--18~$\mu$m & $a$ \\
    \hline
    Spatial resolution & $<$ a few arcsec & $b$ \\
    \hline
    Wavelength resolution & $\lambda/\Delta \lambda \gtrsim 30{,}000$ & $c$ \\
    \hline
    \multirow{2}{*}{Sensitivity} & $<$10$^{-8}$~W~m$^{-2}$~sr$^{-1}$ ($5\sigma$, extended source, 1~hr) & \multirow{2}{*}{$d$} \\
    & $<$10$^{-18}$~W~m$^{-2}$ ($5\sigma$, point source, 1~hr) & \\
    \hline
    Observing field & N/A & \\
    \hline
    Observing cadence & once every 6--12 months & $e$ \\
    \hline
    \end{tabular}
    \end{center}
    $^a$ To detect the dense spectral lines of hydrocarbons at 12--16~$\mu$m and also the H$_2$ S(1) quadrupole line at 17.03~$\mu$m.\\
    $^b$ To spatially resolve Jupiter and Saturn.\\
    $^c$ To spectrally resolve H$_2$ S(2) from surrounding C$_2$H$_6$ lines.\\ 
    $^d$ To achieve a sufficient S/N on H$_2$ S(2) observations. For Titan, a point source sensitivity is applied. \\
    $^e$ To monitor the seasonal variability of Titan's atmosphere.
\end{table}

\printbibliography[heading=subbibliography]
\end{refsection}

\clearpage

\begin{refsection}[3-6_icysmallbodies/icysmallbodies.bib]

\section{Icy Small Solar System Bodies}
\label{sec:icysmallbodies}

\noindent
\begin{flushright}
Takafumi Ootsubo$^{1}$, 
Tsuyoshi Terai$^{2}$,
Shuya Tan$^{3}$,
\\
$^{1}$~University of Occupational and Environmental Health Japan, 
$^{2}$~Subaru Telescope, NAOJ, 
$^{3}$~JAMSTEC
\end{flushright}
\vspace{0.5cm}

\subsection{Scientific background and motivation}

Small Solar System bodies---asteroids, comets, and trans-Neptunian objects (TNOs)---preserve primordial ices and volatile species from the epoch of planet formation 4.6 billion years ago. The abundances of these volatiles reflect the thermal structure of the protosolar disk---specifically, the locations of snow lines for H$_2$O, CO$_2$, CO, and other species---as well as the subsequent thermal processing within the parent bodies. Characterizing these volatile inventories is crucial for understanding where and how planetesimals formed, and how they were subsequently scattered and mixed during giant planet migration.

The key spectral diagnostics of these volatiles, including H$_2$O, CO$_2$, CO, and organic ices, fall within the 2.5--5~$\mu$m wavelength range that is largely inaccessible from the ground due to strong atmospheric absorption. Space-based infrared observations are therefore indispensable. GREX-PLUS's wide-field camera, with its five photometric bands covering 2--8~$\mu$m, offers a unique capability to survey volatile compositions across diverse small-body populations. This section presents three complementary science cases: (1) carbonate minerals as tracers of aqueous alteration on asteroids, (2) surface ice compositions of TNOs and Centaurs, and (3) CO$_2$/H$_2$O mixing ratios in comets and interstellar objects.

\subsubsection{Carbonate minerals on asteroids}
Specific types of asteroids, the so-called carbonaceous asteroids, have been investigated as the primitive remnants of early Solar System history and planet formation. These asteroids would possess volatile elements, such as water, possibly reflecting their evolution around the snow line. In GREX-PLUS Science Book v1 \citep{2023arXiv230408104G}, absorption features in the 3-$\mu$m band in reflectance spectra of asteroids were targeted as diagnostic features of hydrated minerals and/or water ice. This suggested that GREX-PLUS can detect absorption features on a number of asteroids smaller than 10 km, and that a survey of the presence and shapes of absorption features on over 100 small asteroids can provide insight into material evolution in the early Solar System.

\begin{figure}[htb]
 \begin{center}
  \includegraphics[width=100mm]{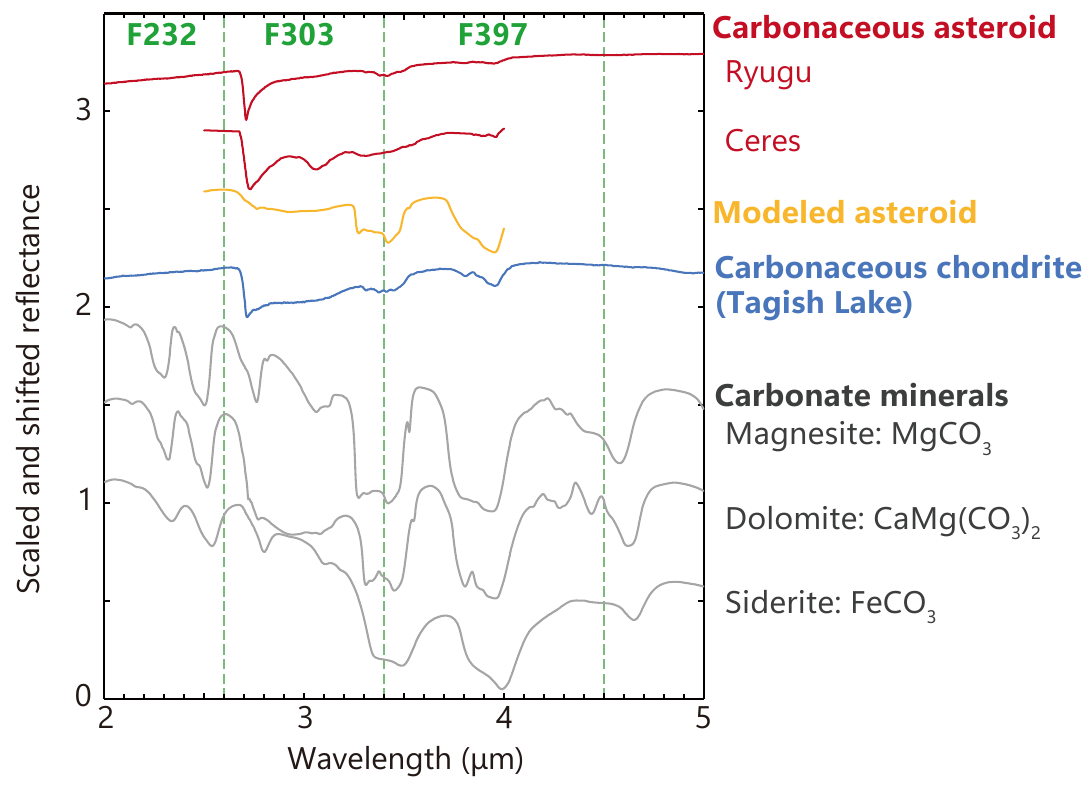}
 \end{center}
 \caption{
   Reflectance spectra of representative carbonaceous asteroids (Ryugu and Ceres) and a carbonaceous chondrite (Tagish Lake), together with those of a modeled asteroid and carbonate minerals (i.e., magnesite, dolomite, and siderite). The dashed lines show the wavelength ranges of GREX-PLUS's NIR filters (F232, F303, and F397). Sources of the data are as follows: Ryugu \citep{2023Sci...379.8671N}, Ceres and modeled asteroid \citep{2022AGUA....300568K}, and Tagish Lake and carbonate minerals \citep{2023SciA....9I3789A}.
 }
 \label{fig_tan01}
\end{figure}

In this context, recent sample returns from carbonaceous asteroids Ryugu and Bennu have provided detailed compositional information on these objects, including elemental and mineralogical evidence of past water. In addition to water, evidence of CO$_2$ has been identified. Carbonate minerals and fluid inclusions containing CO$_2$ were observed in the returned samples \citep[e.g.,][]{2023Sci...379.8671N}. These CO$_2$-rich materials are interpreted as resulting from the formation of CO$_2$-rich fluids within the parent bodies, through the melting of ice accreted beyond the CO$_2$ snowline. Moreover, geochemical models of planetesimals suggest that CO$_2$-rich fluids would have produced abundant carbonate minerals in the interior regions dominated by water \citep{2022AGUA....300568K, 2024GeCoA.374..264S}. Asteroids derived from those water-dominated regions could exhibit absorptions characteristic of carbonates at wavelengths of 3.4 and 4.0~$\mu$m in reflectance spectra \citep{2022AGUA....300568K} (Figure~\ref{fig_tan01}), providing clear evidence of planetesimals beyond the CO$_2$ snowline. Indeed, strong absorptions due to carbonates have been observed locally on Bennu \citep{2020Sci...370.3557K}, supporting the idea that carbonates are observable under space-weathering processes.

AKARI, the Japanese infrared space telescope, also obtained reflectance spectra of carbonaceous asteroids, some of which exhibit weak absorption features likely due to carbonates \citep{2019PASJ...71....1U}. However, its targets were limited to a small number of asteroids with sizes larger than 40~km. Compensating for the lack of strong carbonate absorptions on large asteroids, asteroids smaller than $\sim$10 km could potentially possess these minerals preferentially. In addition, these carbonate absorptions have not been surveyed in asteroid spectra, due to dark reflectance and strong telluric absorptions at wavelengths longer than 3~$\mu$m. GREX-PLUS's near-infrared camera covers the wavelengths of the 3.4 and 4.0~$\mu$m features. Its high sensitivity can provide spectra of asteroids at wavelengths reaching 4~$\mu$m, with the F303 and F397 filters (Figure~\ref{fig_tan01}). A survey of carbonates, in addition to hydrated minerals/water-ice, on a large number of asteroids would provide information on the evolutionary history and size distributions for planetesimals beyond H$_2$O and CO$_2$ snowlines.

\subsubsection{Volatiles on trans-Neptunian objects and small icy bodies beyond Jupiter}

TNOs, a population of small Solar System bodies beyond Neptune's orbit, 
are believed to be primitive icy bodies and remnants of planetesimals formed in the early stage of 
our Solar System.
They are located far from the Sun where it is cold ($\lesssim$~50~K) enough for volatile materials 
to condense and be retained on their surfaces.
In fact, various kinds of ices, such as H$_2$O, CH$_4$, CH$_3$OH, C$_2$H$_6$, CO, and NH$_3$, 
were detected on the objects (including outer-planet satellites and dwarf planets) in the outer Solar 
System \citep[e.g.,][]{2020tnss.book..109B}.
The composition of the surface ices provides essential information about the thermal and chemical 
histories of planetesimals, which would lead us to an understanding of the physical and chemical 
conditions in the protoplanetary disk as well as the dynamical evolution of small bodies including 
the radial mixing during planet formation/migration processes.

H$_2$O ice is the most common icy component on small bodies in the outer Solar System.
Most TNOs that are known to be covered by icy surfaces have spectra dominated by H$_2$O ice 
except for the largest objects, such as Pluto, Eris, Makemake, and Sedna that have 
CH$_4$ ice-rich surfaces \citep[e.g.,][]{2013ASSL..356..107D}.
The abundance of H$_2$O ice has been investigated for a large number of TNOs and Centaurs 
(a small-body population located between Jupiter and Neptune) primarily based on the characteristic 
absorption bands of H$_2$O ice at 1.5~$\mu$m and 2.0~$\mu$m by several near-infrared spectroscopic 
surveys performed with 8--10 meter ground-based telescopes 
\citep[e.g.,][]{2008AJ....135...55B,2011Icar..214..297B,2012AJ....143..146B}.
These studies revealed that the surface H$_2$O ice fraction varies among objects: TNOs larger than 
$\sim$800~km in diameter have H$_2$O ice-rich surfaces, while none or only small amounts of H$_2$O 
ice were detected on most of the smaller objects. 

\begin{figure}[htb]
 \begin{center}
  \includegraphics[width=160mm]{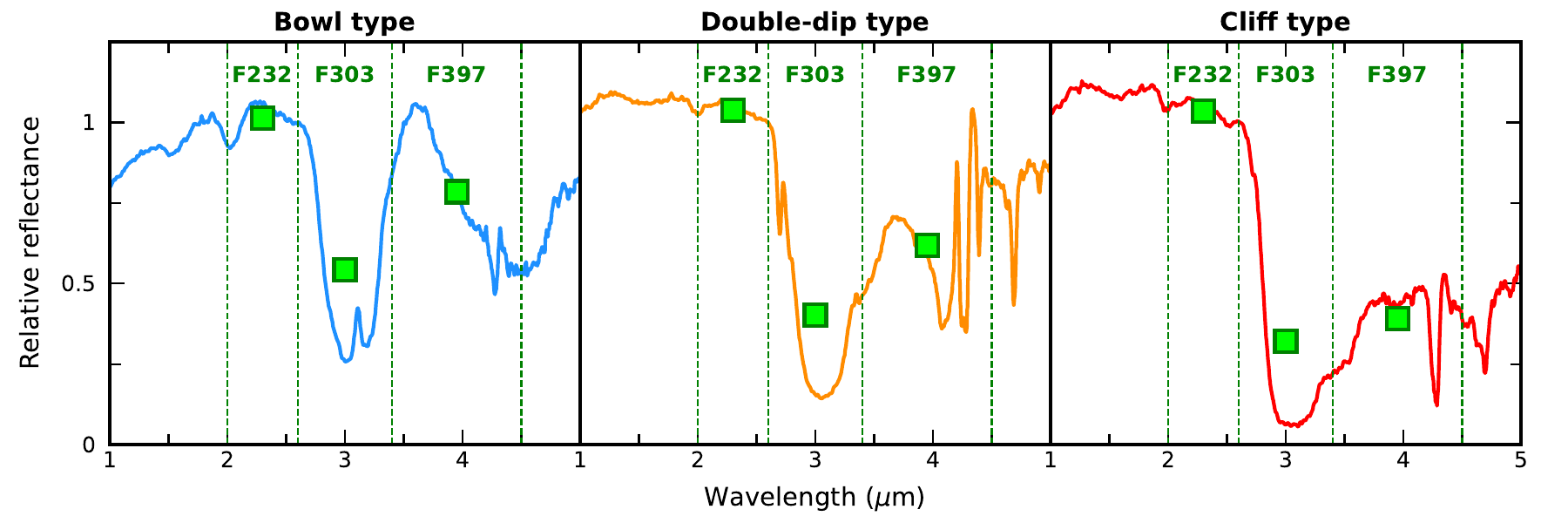}
 \end{center}
 \caption{
  Median reflectance spectra of the TNO compositional groups (left: bowl type, middle: double-dip 
  type, right: cliff type) defined by \cite{2025NatAs...9..230P} based on the JWST/DiSCo-TNOs data.
  The dashed lines show the wavelength ranges of GREX-PLUS's NIR filters (F232, F303, and F397).
  The squares show the average values of the spectra in each band-pass range (assuming a constant 
  transmittance). 
 }
 \label{fig_terai01}
\end{figure}

Although the factors behind this variety of the surface H$_2$O ice abundance remained unclear until 
recently, JWST observations completely changed the situation. 
The JWST Cycle~1 program ``DiSCo-TNOs'' \citep{2021jwst.prop.2418P} provided unprecedented 
high-sensitivity near-infrared spectra of 54~TNOs and five Centaurs in that wavelength range from 0.6 
and 5.3~$\mu$m, which detected several volatile ices such as H$_2$O, CO$_2$, CO, and CH$_3$OH, 
as well as complex molecules and refractory materials throughout the observed TNOs 
\citep{2025NatAs...9..230P}.
Cluster analyses of the acquired reflectance spectra indicated the existence of three 
compositional groups as shown in Figure~\ref{fig_terai01}: 
(1)~``Bowl type'' with spectra characterized by H$_2$O-ice absorptions (25\% of the DiSCo sample), 
(2)~``Double-dip type'' with spectra characterized by the features of CO$_2$, its isotopologue
$^{13}$CO$_2$, and CO (43\% of the sample),  
(3)~``Cliff type'' with spectra characterized by bands corresponding to O$-$H-, C$-$H-, and
N$-$H-bearing materials, as well as two minima at 2.27 and 2.34~$\mu$m attributed to CH$_3$OH 
ice (32\% of the sample).

The DiSCo dataset revealed that all of the observed spectra exhibit a broad absorption band at 
3~$\mu$m, whose shape is what the above group names were based on. 
This band is apparently attributed to H$_2$O ice in the bowl-type TNOs, which also show the other
H$_2$O-ice features at 1.5 and 2.0~$\mu$m together with the Fresnel reflection peak of crystalline 
H$_2$O ice at 3.1~$\mu$m in their spectra.
In contrast, the 3-$\mu$m band seems to be primarily due to O$-$H and possibly N$-$H 
bands of complex organics and CH$_3$OH ice in the double-dip and cliff types.

Another interesting point is that the cliff-type spectra have significantly reddish spectral slopes 
($\sim$20--40\%/100~nm) in the visible range comparable to that of cold classical TNOs 
\citep{2012A&A...546A.115H}.
Note that all five cold classical TNOs in the DiSCo sample are in the cliff-type group.
This fact implies a rich abundance of red complex organics such as tholin on the surfaces that can 
be darkened and reddened through irradiation by energetic ions and photons from the Sun, so-called 
space weathering.

CH$_3$OH ice is a promising precursor candidate for forming red organic material by irradiation
\citep[e.g.,][]{2025Icar..44116669H}.
Actually, laboratory experiments showed that irradiated CH$_3$OH ice is destroyed and induces 
reddening and darkening of visible reflectance spectra due to the formation of a refractory organic 
residue \citep{2006ApJ...644..646B}.
In addition, the diagnostic absorption at 2.27~$\mu$m of CH$_3$OH ice has been detected at the 
surface of the Centaur Pholus \citep{1998Icar..135..389C}, the plutino 2002~VE$_{95}$ 
\citep{2006A&A...455..725B}, and the cold-classical Arrokoth, target of the extended New Horizons 
mission \citep{2019Sci...364.9771S}, all of which have very red spectral slopes in the visible 
range.
The steep visible spectral slopes of the cliff types are consistent with the high abundance of 
CH$_3$OH ice and indicative of the irradiation history of their surface materials.

\citet{2025NatAs...9..230P} also found that both the bowl and double-dip objects are on more excited 
orbits (higher inclination and eccentricity) than the cliff objects are, and suggested that objects 
were formed in the order bowl, double-dip, and cliff, with increasing distance from the Sun.
This hypothesis agrees well with the model of volatile evaporation gradients on surfaces of 
planetesimals in the early Solar System, which predicts the retention lines of H$_2$O, CO$_2$, and CH$_3$OH 
in order of increasing formation distance from the Sun \citep{2011ApJ...739L..60B}.
According to this scenario, the bowl-type, double-dip-type, and cliff-type TNOs formed within 
the areas between the H$_2$O and CO$_2$ retention lines, between the CO$_2$ and CH$_3$OH retention 
lines, and beyond the CH$_3$OH retention line, respectively.
This could be an essential clue to investigating the origin and dynamical evolution processes of primordial
small bodies in the ancient outer Solar System and could also give crucial constraints on 
planetary migration models \citep[e.g.,][]{2005Natur.435..459T}.

\begin{figure}[htb]
 \begin{center}
  \includegraphics[width=90mm]{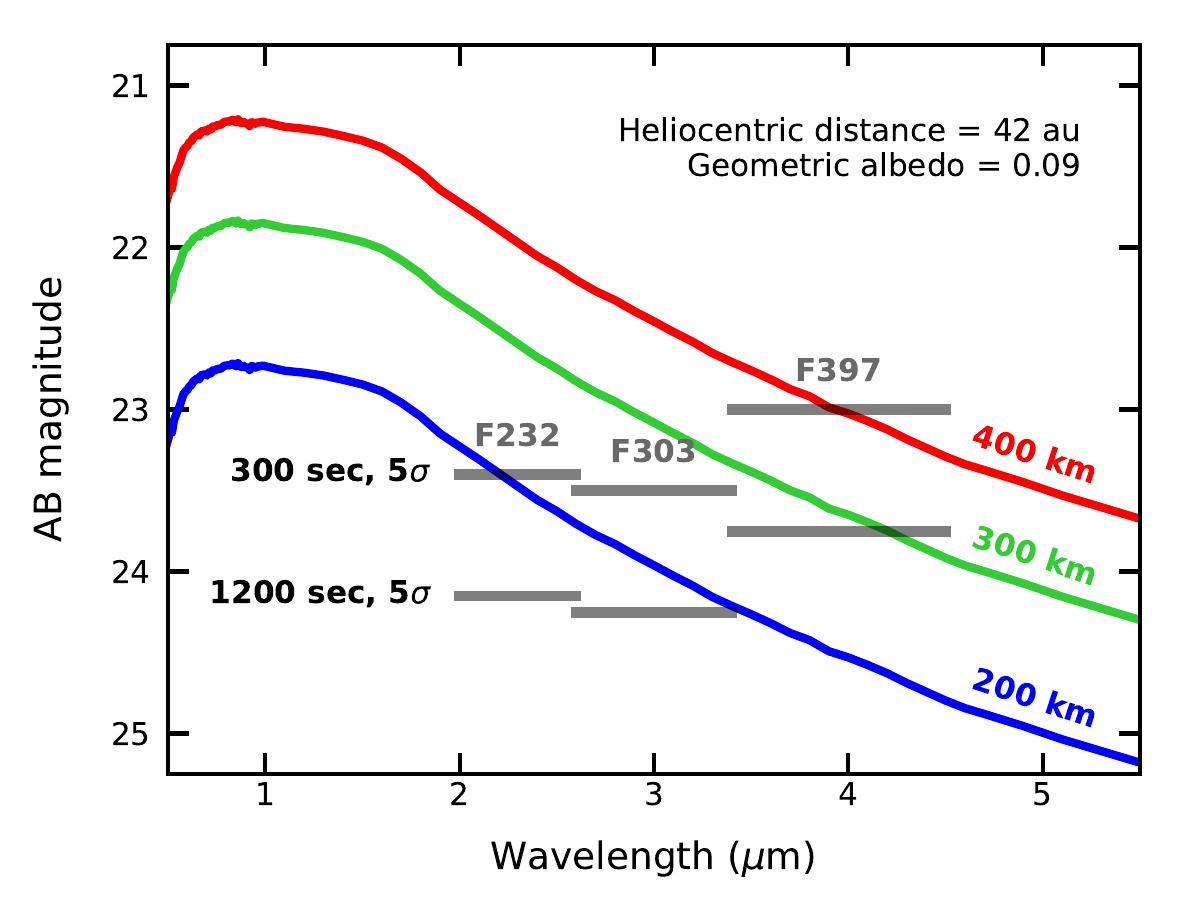}
 \end{center}
 \caption{
   AB~magnitude vs. wavelength plots of TNOs at a heliocentric distance of 42~au with diameters of 
   400~km, 300~km, and 200~km from top to bottom, assuming a constant geometric albedo of 0.09.
   The gray bars show the 5$\sigma$ limiting magnitudes of GREX-PLUS's NIR filters (F232, F303, and 
   F397) with exposure times of 300~sec (upper) and 1200~sec (lower).
 }
 \label{fig_terai02}
\end{figure}

The near-infrared wide-field camera on GREX-PLUS can provide high-quality photometric data 
covering most of the wavelength range (2--4.5~$\mu$m) of the DiSCo spectra.
Our estimates show that the color data obtained from the F232, F303, and F397 filters are useful 
to divide TNOs into the three compositional groups with photometric accuracy of $\sim$0.1~mag
(see Figure~\ref{fig_terai01}).
GREX-PLUS's sensitivity enables the detection of small TNOs down to 200~km in diameter with 1200-sec
exposures by the F232 and F303 filters (see Figure~\ref{fig_terai02}), which corresponds to the 
size range of the DiSCo sample objects.
It would allow us to conduct a detailed investigation of the orbital distributions of each 
compositional group and relationships between the near-infrared spectra and the spectral 
slope/geometric albedo in the visible range from a large TNO/Centaur sample.

In addition to the TNO survey, JWST/NIRSpec spectroscopy has revealed similarities between TNO and Centaur spectra, including absorptions due to CO$_2$ ice and organic ices \citep{2025NatAs...9..245L}. Centaurs are a category of small Solar System bodies. They orbit mainly between Jupiter and Neptune and have been known to exhibit characteristics of both asteroids and comets. The similarity between TNOs and Centaurs indicates inward migration of TNOs. Moreover, some Centaurs and other small bodies orbiting in the outer Solar System are categorized as active asteroids and comet-asteroid transition objects. They would possess fresh icy material in their subsurfaces and possibly on their surfaces, as shown for the active centaur Chiron \citep{2024A&A...692L..11P}. Thus, observations of them could provide insight into the original and/or interior volatile materials of TNOs. However, observations of these icy bodies at 2--5~$\mu$m are limited, lacking information on their volatiles. In addition, small-to-mid-sized icy moons of Saturn, Uranus, and Neptune also show non-water volatile materials, such as CO$_2$, on reflectance spectra \citep[e.g.,][]{2006Icar..184..543G}, similar to TNOs. Comparisons of volatile inventories and species on these planetary bodies would provide insight into the evolution of small icy Solar System bodies between the orbits of asteroids and TNOs.

\subsubsection{CO$_2$/H$_2$O Ice Ratio Survey of Comets and Interstellar Objects}

Cometary nuclei preserve primordial icy materials from the early Solar System. Among volatile species in cometary ice, H$_2$O, CO$_2$, and CO are the three most abundant molecules, serving as the primary carriers of oxygen atoms. Their relative abundances provide fundamental constraints on comet formation environments, because these molecules have significantly different condensation temperatures: H$_2$O at $\sim$150~K, CO$_2$ at $\sim$70--80~K, and CO at $\sim$20--25~K. The mixing ratios thus serve as thermometers for the formation region and tracers of radial mixing in the protosolar disk.

Direct observations of CO$_2$ are difficult because the 4.3~$\mu$m fundamental band is completely absorbed by the Earth's atmosphere. The AKARI infrared satellite conducted the first systematic survey, detecting CO$_2$ in 17 out of 18 comets and revealing CO$_2$/H$_2$O ratios ranging from several percent to $\sim$30\% \citep{Ootsubo2012}. Subsequent compilations expanded the sample to $\sim$25 comets \citep{HarringtonPinto2022}, showing that Jupiter-family comets preferentially have CO$_2$-dominated comae, while dynamically new Oort Cloud comets exhibit higher CO$_2$/CO ratios than thermally processed returning comets.

Recent JWST and SPHEREx observations have demonstrated remarkable compositional diversity: Main-Belt Comet 238P/Read shows CO$_2$/H$_2$O $<$ 0.7\% \citep{Kelley2023}; Centaur 39P/Oterma exhibits CO$_2$-driven activity at $\sim$6~au \citep{HarringtonPinto2023}; and the interstellar comet 3I/ATLAS displayed an extraordinary CO$_2$/H$_2$O ratio of $7.6\pm0.3$ ($\sim$760\%) \citep{Cordiner2025,Lisse2026}. Despite these advances, the current sample ($\sim$25 comets) remains insufficient to statistically address whether dynamical classes show systematically different compositions and whether 3I/ATLAS-like CO$_2$ enrichment is typical of interstellar objects.

\subsection{Required observations and expected results}

\subsubsection{Asteroids}
As proposed in GREX-PLUS Science Book v1 \citep{2023arXiv230408104G}, consecutive imaging with the F232, F303, and F397 filters in a single field within a short time ($\ll$~1~hour) is necessary to reduce the effect of brightness variation due to the object's spin. Imaging with the F303 and F397 filters is essential for the survey of absorption features due to hydrated minerals/water ice and carbonates. More than 100 asteroids smaller than 10~km are necessary for the survey of absorption features. Ideally, narrow-band filters with widths of 0.2--0.3~$\mu$m are better for detecting carbonate absorptions at 3.4 and/or 4.0~$\mu$m.

\subsubsection{Trans-Neptunian objects and small icy bodies beyond Jupiter}
For TNOs, the minimum request is consecutive imaging with the F232 and F303 filters in a field within 
a short time ($\ll$~1~hour) to reduce the effect of brightness variation due to the object's spin. 
In addition, F397 photometry with a similar configuration would be helpful to classify the type of 
the compositional groups more accurately.
According to our estimate, 1200 sec exposures with these two filters can detect TNOs as small as 
$\sim$200~km in diameter at 5$\sigma$ sensitivity (see Figure~\ref{fig_terai02}).
Wide-field survey observations at low ecliptic latitude (within $\pm$30$^\circ$) regions are also 
required as much as possible so that we can obtain photometric data of a large number of TNOs.
We plan to identify the compositional types of $\sim$100~TNOs by combining the data from the Wide survey and 
pointing observations.
This will allow us to achieve an approximately threefold increase in the number of the DiSCo's TNO sample, 
yielding more detailed insights into the orbital distributions of each spectral group and 
relationships between the surface volatile composition and the visible spectra/albedo.

For small icy bodies in the interplanetary regions, consecutive imaging with the F232, F303, and F397 filters in a field within a short time ($\ll$~1~hour) is required to reduce the effect of brightness variation due to the object's spin. According to our estimate, a 300-second integration with the F303 filter is sufficient to observe small icy bodies with diameters of 10~km in Saturnian orbit, 50~km in Uranian orbit, and 100~km in Neptunian orbit. We plan to determine surface volatile species and abundances of small icy bodies beyond Jupiter’s orbit with diameters larger than $10$--100~km.

\subsubsection{Comets}

The GREX-PLUS WFC covers the 2.0--8.0~$\mu$m wavelength range with five broad photometric bands. Three bands serve primarily for continuum characterization: F232 (2.0--2.6~$\mu$m) measures dust-scattered light, while F520 (4.5--5.9~$\mu$m) and F680 (5.9--7.7~$\mu$m) capture dust thermal emission. The remaining two bands target major volatile species: F303 (2.6--3.4~$\mu$m) is sensitive to H$_2$O gas emission at 2.7~$\mu$m and H$_2$O ice absorption at 3.0~$\mu$m, and F397 (3.4--4.5~$\mu$m) covers the CO$_2$ fundamental band at 4.3~$\mu$m. Additionally, F520 contains CO gas emission at 4.7~$\mu$m, enabling separation of all three major volatile species. The analysis follows the approach demonstrated by Spitzer/IRAC \citep{Reach2013} and SPHEREx: model the dust continuum using F232, F520, and F680; subtract from F303 and F397 to isolate gas emission; convert flux ratios to production rate ratios using appropriate fluorescence models. However, the CO$_2$ and CO fundamental bands at 4.3 and 4.7~$\mu$m occupy $<$10\% of the F397 and F520 bandwidths, respectively, limiting detection sensitivity for these narrow features. We propose the addition of narrow-band filters around 4.3~$\mu$m and 4.7~$\mu$m optimized for cometary volatile detection, if possible. These narrow-band filters, combined with adjacent continuum bands (F397, F520), enable measurement of band-to-continuum flux ratios that can be directly converted to production rate ratios.

For bright comets ($V \sim 10$--12) at 1.5--3~au, expected gas emission excess is 10--30\% above the continuum. The WFC sensitivity is sufficient for comets with $Q(\mathrm{H_2O}) > 10^{27}$~molecules/s. The wide field of view enables both targeted observations of 30--50 comets and serendipitous detections in survey fields near the ecliptic. Observations of the same comet within a short time ($<$~1~day) are desired to reduce the effect of brightness variation due to the coma activity.

\subsection{Scientific goals}

\subsubsection{Asteroids}
The survey for carbonate minerals on asteroids, in addition to hydrated minerals/water ice, aims to reveal evidence for asteroids derived from regions around or beyond the CO$_2$ snowline. Observations of $>100$ small asteroids would provide clues to the size and distribution of planetesimals in the early Solar System and to the conditions of aqueous alteration in their parent bodies, such as temperatures and the abundances of CO$_2$-rich fluids.

\subsubsection{Trans-Neptunian objects and small icy bodies beyond Jupiter}
TNO imaging with the NIR filters allows us to efficiently classify them into the three compositional 
groups \citep[bowl, double-dip, and cliff;][]{2025NatAs...9..230P} to study 
(1)~orbital distributions of each compositional group to investigate the formation regions and 
extensive dynamical evolution driven by gravitational perturbations from the giant planets that 
probably moved outward during the planet migration phase, 
(2)~relationships between the near-infrared spectra and the colors/spectral slopes as well as 
geometric albedos at visible wavelengths to understand the irradiation effects depending on 
surface volatile composition, and
(3)~comparisons of their size distributions among the compositional groups as tracers of the growth 
rate of planetesimals as a function of radial distance in the protoplanetary disk that could provide 
essential information for elucidating the planet formation process.

The survey of small icy Solar System bodies beyond Jupiter's orbit aims to determine the abundances and species of volatile materials, such as H$_2$O and CO$_2$, on their surfaces and to compare this information on volatiles with that for TNOs. Observations of centaurs and icy moons would provide clues to migration and thermal histories, and potentially the original volatiles of TNOs.

\subsubsection{Comets}

Newly derived CO$_2$/H$_2$O and CO/H$_2$O mixing ratios for $\sim$30 comets, combined with the existing AKARI database, will enable robust statistical analysis of compositional differences between Jupiter-family and Oort Cloud comets, providing insights into their birthplaces in the protosolar disk. The survey will also characterize the heliocentric distance dependence of these ratios across the H$_2$O sublimation zone ($\sim$2.5~au) and identify compositional outliers analogous to C/2016 R2 (CO-rich) or 3I/ATLAS (CO$_2$-rich). Multi-epoch observations of $\sim$10 comets at different heliocentric distances will further probe nucleus heterogeneity and evolutionary effects on volatile release. If the survey achieves $\geq$50 comets including rare populations such as dynamically new comets, Halley-type comets, and active Centaurs, it would provide robust constraints on the primordial distribution of major ice species in the protosolar disk. Observations of newly discovered interstellar objects would establish whether the extreme CO$_2$ enrichment observed in 3I/ATLAS is characteristic of extrasolar comets, offering a unique comparison between the volatile inventory of our Solar System and that of other planetary systems.

\begin{table}[bth]
    \label{tab:icybodies}
    \begin{center}
    \caption{Required observational parameters.}
    \begin{tabular}{|l|p{9cm}|l|}
    \hline
     & Requirement & Remarks \\
    \hline
    Wavelength & \underline{asteroids \& TNOs}: 2--5~$\mu$m & a \\
               & \underline{comets}:  2--8~$\mu$m & b \\
    \hline
    Spatial resolution & $<1$ arcsec & \\
    \hline
    Wavelength resolution & $\lambda/\Delta \lambda>3$ & \\
    \hline
    Field of view & \underline{TNOs}: 40 deg$^2$, 23--24 AB mag ($5\sigma$, point source) & \\
    \hline
    Observing field &\underline{asteroids \& TNOs}: ecliptic latitude $<$~30$^\circ$ & \\
    \hline
    Observing cadence & \underline{asteroids}: consecutive imaging with the three NIR filters within $\ll$~1~hour & \\
                      & \underline{TNOs}: consecutive imaging with the two or three NIR filters within $\ll$~1~hour & \\
                      & \underline{comets}: imaging with the five filters within $\ll$~1~day; multi-epoch for $\sim$10 comets at
                  different heliocentric distances & \\
    \hline
    \end{tabular}
    \begin{flushleft}
    $^{a}$ F232, F303, and F397 filters \\
    $^{b}$ Addition of narrow-band filters at 3.4, 4.0, 4.3, and 4.7~$\mu$m 
    ($\Delta\lambda \sim 0.1$--0.3~$\mu$m) would enhance detection of 
    carbonate minerals on asteroids (3.4, 4.0~$\mu$m) and CO$_2$/CO 
    volatile species in comets (4.3, 4.7~$\mu$m).    
    \end{flushleft}
    \end{center}
\end{table}

\printbibliography[heading=subbibliography]
\end{refsection}

\clearpage

\begin{refsection}[3-7_starformingregions/starformingregions.bib]

\section{Star and Planetary Forming Regions}
\label{sec:starformingregions}

\noindent
\begin{flushright}
Chikako Yasui$^{1}$, 
Michihiro Takami$^{2}$
\\
$^{1}$ NAOJ, 
$^{2}$ ASIAA
\end{flushright}
\vspace{0.5cm}

\subsection{Scientific background and motivation}



\subsubsection{Protoplanetary Disks as Sites of Planet Formation}

\begin{figure*}[ht!]
\begin{center}
    \includegraphics[width=8cm]{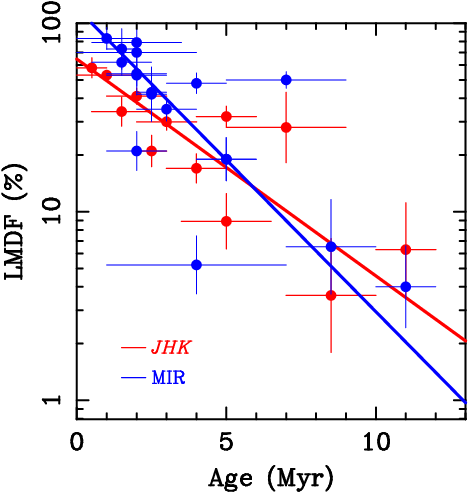}
\end{center}
\caption{Evolution of protoplanetary disks---fraction of stars
with NIR disk excess as a function of age. Figure~5 (left panel) of \citet{Yasui2014}, ``Rapid evolution of the innermost dust disc of protoplanetary discs surrounding intermediate-mass stars''.}
\label{fig:yasui1}
\end{figure*}

Because planets are formed in protoplanetary disks in the process of
star formation, the evolution and dissipation of protoplanetary disks
should be an essential key to determining planet formation.
Observational studies of protoplanetary disks have progressed rapidly
from around 2000.
Figure~\ref{fig:yasui1} is an example of one such result, obtained from
near-infrared and mid-infrared
observations, where each dot represents an individual
star-forming region, the horizontal axis indicates the age, and the
vertical axis indicates the fraction of stars that still have their
disks within each region.
Using the fact that the stars in each star-forming region are born
almost simultaneously, the lifetime of protoplanetary disks, i.e., the
timescale of planet formation, was estimated to be about 10 Myr, because
almost all stars lose their disks at about 10 Myr.
However, the figure also shows that some young stars have already lost
their disks, while others still retain their disks even at relatively
old ages, indicating that there is a large variation in the timescale
for the disk dispersal of individual stars.
In an attempt to explain the variation, various environmental
dependencies have been proposed:
e.g.,
stellar mass \citep{Ribas2015}, 
cluster density \citep{Fang2013},
and metallicity \citep{Yasui2010}. 
However, the dispersion of the disk frequency is still large, even after considering the dependence of disk lifetimes on each parameter.
Therefore, what determines the disk lifetime is a very important
question in planet formation that has been left unanswered.

Various theoretical models existed prior to about 2000 as mechanisms
that attempted to explain disk dispersal. 
Meanwhile, from around the same period, it has been observationally known
that disks disappear from the inner to outer parts almost
simultaneously \citep{Williams2011}. 
The UV-switch model, a combination of the mass accretion process from
inner disks onto the central stars and the photoevaporation process of
outer disks, was proposed as a possible explanation for disk dispersal
\citep{Clarke2001}, and is now widely accepted.
Mass accretion activities have been confirmed in many objects, mainly by
observations of hydrogen emission lines,
and the comparison of the line profiles with theoretical models has
revealed the basic properties of mass accretion quantitatively, such as
the mass accretion rate and the geometric structure of accretion
\citep{Hartmann2016}.
Meanwhile, \textit{\textbf{ the photoevaporation process has not been well understood
because there have been a very small number of observational detections ($N
\sim 20$), even though it is the main process of disk dissipation.}}
Photoevaporation is a phenomenon in which disk gas is heated by
high-energy photons from the central star or nearby stars and flows out
of the disk.
Although theoretical studies of the complex chemical network have led to
a better understanding of this phenomenon, the very basics, such as what
types of radiation (EUV, FUV, or X-rays) primarily induce
photoevaporation, are still poorly understood.


\subsubsection{The Physical Mechanism of High-Mass Star Formation}
The formation of high-mass stars has represented a tricky puzzle from both theoretical and observational points of view.
After a certain evolutionary phase, accretion should be inhibited by radiation pressure, making the formation of stars more massive than $\sim$10 ${\rm M}_\odot$ impossible \citep{Tan14}. This is not consistent with the clear existence of a significant number of more massive stars.
Disk accretion is regarded as the most promising scenario to resolve this issue, as such a geometry could significantly reduce the effects of radiation pressure on the accreting material \citep{Beltran16}. However, there is not yet observational proof that the mass accretion from the innermost region of these disks to the central protostar is ongoing. For instance, millimeter interferometry of thermal dust continuum emission (e.g., using ALMA) allows us to observe disk structures with a spatial resolution as good as $\sim$100 au, with the given angular resolution (typically $\sim$0".03) and distances to the nearest high-mass young stars (typically $\gtrsim$2 kpc) \citep{Beltran16}. Such observations do not allow us to investigate to what extent the dust(+gas) remains at $r \ll 100$ au against radiation pressure. Moreover, dust continuum observations do not directly provide kinematic information in principle; therefore, they do not allow us to investigate whether disk accretion actually occurs. The latter issue could be solved by observing molecular line emission using millimeter interferometry, but only with modest angular/spatial resolutions.

Furthermore, these stars are heavily embedded even when they reach the main sequence so that it is difficult to determine the evolutionary stages of the individual targets. We cannot use optical spectroscopy to determine the stellar properties, and extensive ground-based near-infrared spectroscopy to date has not been able to solve this issue because of the limited number of observable targets and spectral features \citep[e.g.,][]{Caratti17,Beuther10, Hsieh21}.
There is a proposed evolutionary sequence based on the presence and size of the ionized region \citep[No HII Region $\rightarrow$ Hyper Compact HII Region $\rightarrow$ Ultra Compact HII Region;][]{Beuther05}, which is easily measured in the radio. However, this phase is determined by the amount of stellar ionizing radiation, which is in principle a function of both mass and evolutionary status. The far-IR luminosity is used as an indicator of the mass of the central source \citep[e.g.,][]{Beltran16}, but may also be a function of the mass accretion rate \citep[e.g.,][]{Audard14}.

Breakthroughs in the above issues would be made with space mid-infrared spectroscopy, in particular using JWST and GREX-PLUS. These missions will allow us to observe a variety of spectral lines and features at 3--20 $\mu$m, which are tremendously useful for determining the physical properties of the target protostars and the inner disks. Such lines and features include: Br-$\alpha$ 4.05 $\mu$m, Pf-$\beta$ 4.65 $\mu$m, PAH 3.3-$\mu$m, and perhaps He I (e.g., 4.04/4.05 $\mu$m) lines, that are useful for probing the color of UV radiation, allowing the calculation of the temperature of the (proto-)stellar photosphere \citep{Osterbrock89}; and a variety of molecular lines (CO, H$_2$O, CO$_2$, HCN, C$_2$H$_2$, etc.) useful for directly probing the accretion and ejection of molecular gas \citep{Elias06} \citep[see also][]{Hsieh21} or the presence of the inner disk ($\ll$100 AU) \citep{Najita03,Pontoppidan10,Gasman2025}, thereby allowing us to determine if these protostars are still in an active accretion phase. If the stellar continuum is significantly brighter than disk emission, CO, SiO and Br-$\alpha$ may be seen in absorption in the stellar photosphere, allowing us to probe stellar evolution at the start of ionizing radiation \citep{Hosokawa10}. Otherwise, if the star is associated with a disk with a very high mass accretion rate, the spectra would be similar to M supergiants with CO and SiO bands \citep{Audard14}. Although many of the above spectral lines/features lie at $\lambda<5\,\mu$m, for which the protostars and disks are not directly seen due to extremely large circumstellar extinction, their infrared spectra are probably accessible via {\it bright} infrared reflection nebulae at $\lambda=3$--5 $\mu$m \citep[][]{Takami12}.

\begin{figure*}[ht!]
\begin{center}
    \includegraphics[width=13cm]{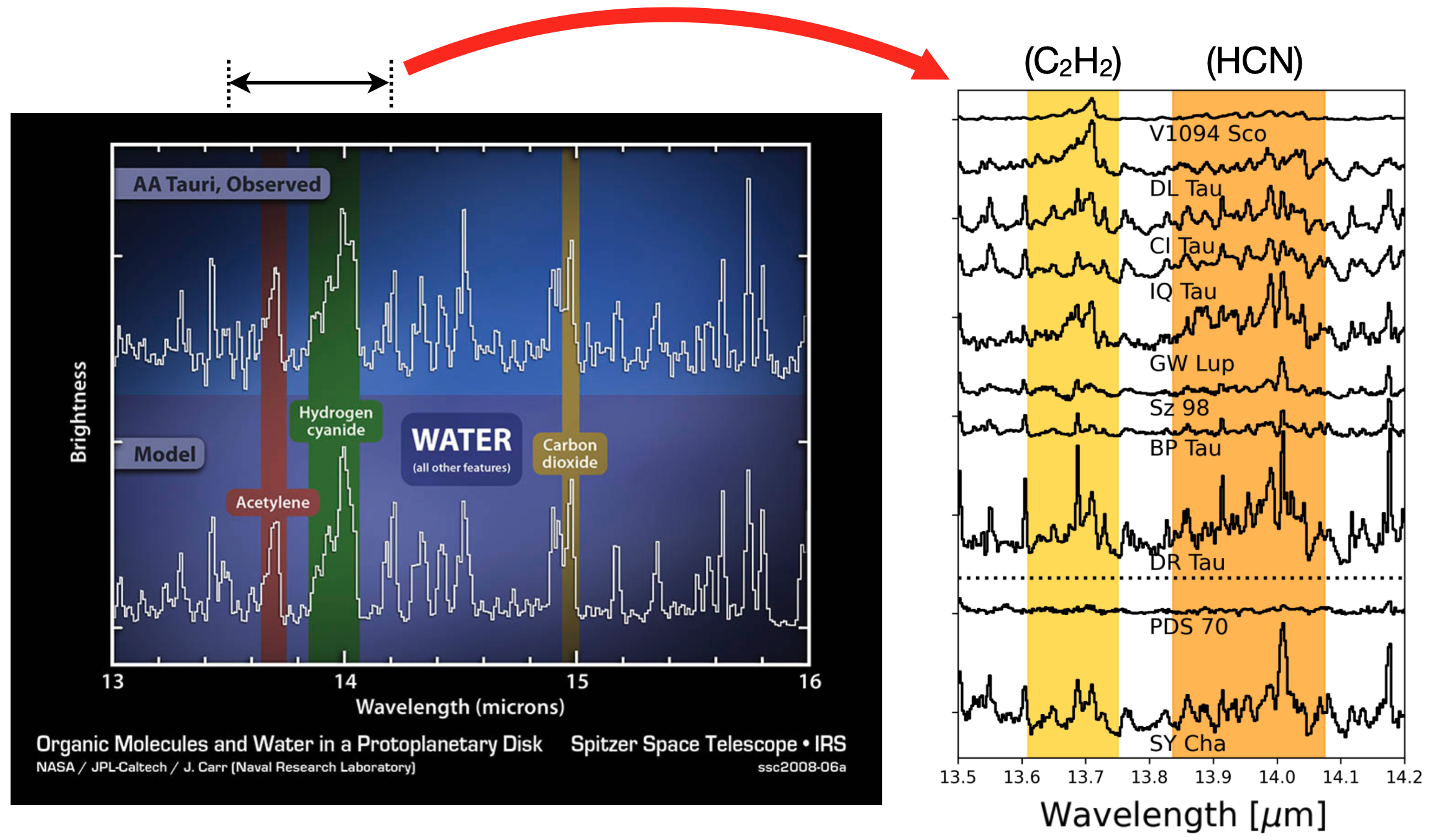}
\end{center}
\caption{(Left) A variety of molecular emission lines from the disk associated with a low-mass young star. Observed using the Spitzer Space Telescope at $R=600$ (Figure courtesy: NASA/JPL/J. Carr). (Right) The C$_2$H$_2$ and HCN bands from the disks associated with low-mass young stars observed using JWST at $R=3000$ \citep{Gasman2025}.
\label{fig:takami1}}
\end{figure*}

GREX-PLUS will offer a unique opportunity for high-resolution spectroscopy at 10--18 $\mu$m, which covers a number of molecular lines associated with H$_2$O, CO$_2$, HCN, C$_2$H$_2$, NH$_3$ (see Figure~\ref{fig:takami1}). Although the spectral coverage of its high-resolution (HR) spectrograph is significantly smaller than JWST, its contributions will be powerful and essential as described below.

\subsection{Required observations and expected results}

\subsubsection{Low-Mass Disk Evolution: The First Large Statistical Study with the Mid-Infrared [NeII] Line}

Forbidden lines with low-velocity components, such as [NeII] 12.8 $\mu$m
in the mid-infrared wavelength range and [OI] and [SII] in the optical
wavelength range, are known tracers of photoevaporation that have been
identified so far \citep{Pascucci2023}. 
For [OI] and [SII], although relatively high sensitivities can be
achieved in the optical wavelength region even from the ground-based
observations, the observed lines often show several components other than
photoevaporation, and it is sometimes difficult to separate the
photoevaporation component.
Meanwhile, the line profiles of [NeII] show only the component tracing
photoevaporation (although the profiles for some objects show only the
component tracing jets) and is currently considered a promising tracer
of photoevaporation.
However, since the sensitivity of the mid-infrared wavelength region is low
for ground-based observations, space telescopes are required, but no
previous space telescopes have had high-resolution spectrographs.

GREX-PLUS is the first space telescope capable of mid-infrared
high-resolution spectroscopy.
High sensitivity observations will dramatically increase the number of
detected samples.
In previous studies \citep{Pascucci2009, Pascucci2020}, sensitivity was
$F_{\rm cont} \le 300$ mJy using VISIR ($R\sim30,000$) on Melipal/VLT,
and at this sensitivity, the observations were selective for sources 
with strong 12 $\mu$m emission. 
Meanwhile, GREX-PLUS achieves a sensitivity of $F_{\rm cont} = 5$ mJy at
$5\sigma$ for an integration time of 1 hr, making it possible to observe
objects down to $F_{\rm cont} = 30$ mJy (1 hr, $30\sigma$).
This sensitivity allows us to detect stars down to $\simeq0.5\,{\rm M}_\odot$
in the well-known star-forming regions, such as Taurus, Cham, Lup, CrA,
etc.
This will be the first comprehensive observation of photoevaporative
activity in nearby star-forming regions.


\subsubsection{High-Mass Star Formation}

We expect that the high-resolution spectrometer on GREX-PLUS will yield powerful and essential progress toward understanding the physical mechanism of high-mass star formation using mid-infrared spectroscopy. First, it will allow us to determine the origin of these emission lines, which must originate from either the inner disk region \citep[e.g.,][]{Pontoppidan10}, an outflow, or an inflow.
If it turns out that the emission lines originate from the disk, their line profiles would be useful for investigating the structure of the inner disk regions even without direct spatial information.
We expect that the spectral resolution of the high-resolution spectrometer on GREX-PLUS ($R\sim3\times10^4$, $\Delta v=10$ km s$^{-1}$) is optimal for such studies 
\citep{Elias06,Hsieh21}.


We will investigate the following using the spectra obtained with GREX-PLUS: (1) what is responsible for the emission/absorption lines (disk/inflow/outflow) for which stellar masses and evolutionary stages; (2) if the emission/absorption lines are associated with the inner disk, how their physical conditions change with their stellar masses and evolutionary stages, and when and at which conditions those disks start to dissipate (and therefore what actually determines their final stellar masses).

\subsection{Scientific goals}


\subsubsection{Low-Mass Disk Evolution: The Relationship Between Photoevaporation and Other Physical Parameters}

[NeII] observations in nearby star-forming regions are expected
to dramatically increase the sample size of observations to more than
300.
The majority of the observed objects have been studied to date to reveal
phenomena related to disk dispersal other than the photoevaporation
process (e.g., mass accretion, disk wind, stellar winds, jets, etc.),
as well as other parameters besides age and central stellar mass. 
For the majority of target objects, previous studies have revealed
physical parameters other than age and central stellar mass, as well as parameters
for disk dispersal processes other than photoevaporation (mass
accretion, disk winds, stellar winds, jets, etc.).
We plan to handle data from more than 300 statistically sufficient
objects and discuss the evolution of stars and disks in general, with a
focus on photoevaporation.
This will allow us to clarify under what conditions photoevaporation is
efficient and to answer the fundamental question of what determines
disk dissipation and thus the planet formation process.



\subsubsection{High-Mass Star Formation}

Using the high-resolution spectrometer, we will obtain spectra for a number of high-mass protostars with a variety of masses and evolutionary stages. The required instrument specifications are tabulated below. High-mass star-forming regions are very complex, so it is not easy to predict the brightness of the emission features or the depths of the absorption features in advance. Therefore, we propose to organize a pilot survey for a fraction of targets (in particular the bright ones) to determine the best integration time to study the above emission/absorption lines and features. As our targets are bright, they would be useful for system-verification observations of the spectrograph.

\begin{table}[ht]
    \label{tab:takam}
    \begin{center}
    \caption{Required observational parameters.}
    \begin{tabular}{|l|p{9cm}|l|}
    \hline
     & Requirement & Remarks \\
    \hline
    Wavelength & 10--18 $\mu$m &  \\
    \cline{1-2}
    Spatial resolution & No critical requirement & \\
    \hline
    Wavelength resolution & $\lambda/\Delta \lambda = 3 \times 10^4$--$1 \times 10^5$ & \\
    \hline
    Sensitivity & 30 mJy, S/N $\geq$ 30 &\\
    \hline
    Observing field & None & \\
    \hline
    Observing cadence & N/A & \\
    \hline
    \end{tabular}
    \end{center}
\end{table}

\printbibliography[heading=subbibliography]
\end{refsection}

\clearpage

\begin{refsection}[3-8_browndwarf/browndwarf.bib]

\section{Brown Dwarfs}
\label{sec:bd}

\noindent
\begin{flushright}
Yui Kawashima$^{1}$, 
Shota Miyazaki$^{2}$, Hajime Kawahara$^{2, 3}$
\\
$^{1}$ Kyoto University, 
$^{2}$ Institute of Space and Astronautical Science, Japan Aerospace Exploration Agency,
$^{3}$ The University of Tokyo
\end{flushright}
\vspace{0.5cm}

\subsection{Scientific background and motivation}



\begin{figure}[htbp]
  \begin{minipage}[b]{0.5\columnwidth}
    \centering
    \includegraphics[height=5.5cm]{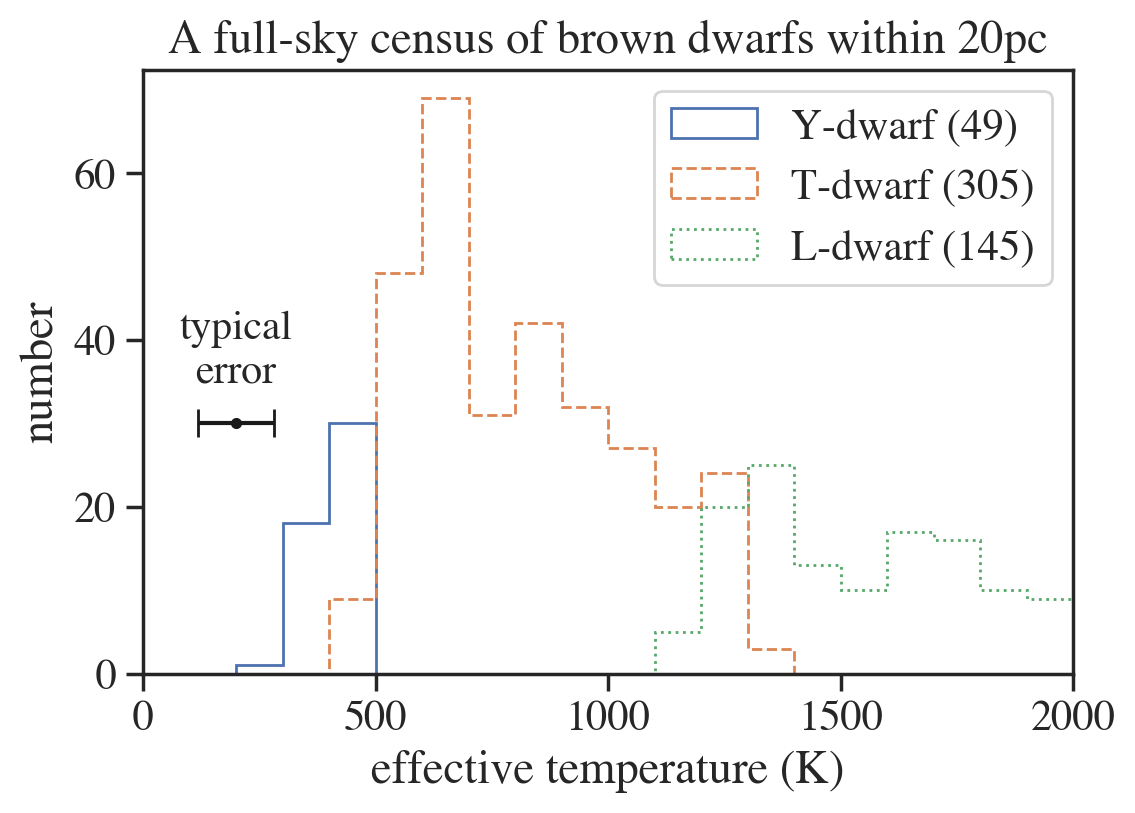}
  \end{minipage}\hfill
  \begin{minipage}[b]{0.5\columnwidth}
    \centering
    \includegraphics[height=5.5cm]{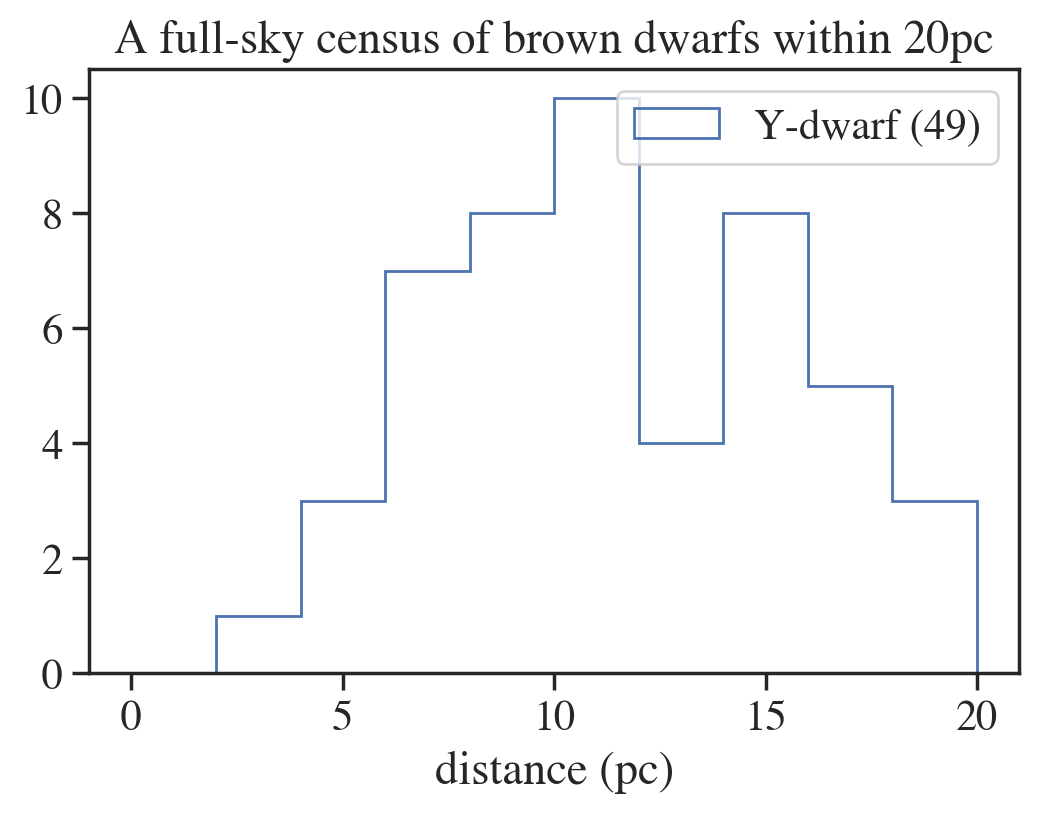}
  \end{minipage}\hfill
    \caption{(Left) Number of brown dwarfs within 20 pc of the Sun as a function of estimated effective temperature, based on the catalog of \cite{2021ApJS..253....7K}. Numbers in parentheses in the labels denote the total number of discoveries for each spectral type. (Right) Number of Y dwarfs as a function of distance.}
    \label{fig:BD_20pc}
\end{figure}

Brown dwarfs, which bridge the gap between planets and stars, are key objects for developing a unified understanding of formation, evolution, and atmospheric processes in the low-mass regime. The effective temperatures of brown dwarfs span three spectral types \citep{2004AJ....127.3516G, 2011ApJ...743...50C}: L-type ($\sim 1,500$--$2,300$~K), T-type ($\sim 500$--$1,500$~K), and Y-type ($\lesssim 500$~K). While previous surveys are considered to have achieved complete detection of the hotter L and T dwarfs within the nearby 20 pc volume, the completeness limit for Y dwarfs remains 10 pc due to their extreme faintness \citep[][see Figure~\ref{fig:BD_20pc}]{2021ApJS..253....7K}. Given the observed increase in space density 
with decreasing effective temperature down to $\sim 600$~K, these cold objects are expected to be significantly more abundant than their warmer counterparts.
However, robust constraints on their space density and luminosity function are still lacking. Therefore, improving 
the census of Y dwarfs is essential for 
understanding their formation and evolution \citep{2003PASP..115..763C}.

The temperature range of Y dwarfs ($\lesssim 500$~K) also directly overlaps with that of the primary targets of current and near-future exoplanet atmospheric characterization. In contrast to the well-studied atmospheres at $\sim 500$--$2,500$~K, atmospheric processes in this temperature regime, including chemical reactions, cloud formation, and atmospheric circulation, remain largely unexplored. For example, silicate clouds are considered to dominate atmospheres at $\sim 1,000$--$2,000$~K in both self-luminous brown dwarfs and planets \citep{2006ApJ...648..614C, 2023ApJ...946L...6M}, as well as in irradiated planets \citep{2020NatAs...4..951G}. In the cooler Y-dwarf regime, 
other cloud species are predicted to play a major role, including $\mathrm{Na_2S}$, $\mathrm{ZnS}$, and $\mathrm{KCl}$ clouds in the warmest Y dwarfs \citep{2012ApJ...756..172M}, and $\mathrm{H_2O}$ and $\mathrm{NH_3}$ clouds in the coldest Y dwarfs \citep{2014ApJ...787...78M, 2023ApJ...950....8L}, analogous to cloud processes observed on Earth and Jupiter \citep{2004AmJPh..72..767S}.
However, observational confirmation of these predictions is absent due to the current lack of a statistically sufficient sample of Y dwarfs.

Compared with planets orbiting bright host stars, whose atmospheric characterization relies on either indirect measurements through their effects on the host stars (e.g., transit or eclipse observations) or on high-contrast imaging, isolated brown dwarfs with similar temperatures typically allow substantially higher observational precision through direct observation. As such, isolated Y dwarfs serve as a unique and powerful laboratory for studying atmospheric processes in this temperature regime. Nevertheless, the number of discovered Y dwarfs remains limited to $\sim 50$, precluding statistically meaningful population studies. To date, all confirmed Y dwarfs have been discovered by the infrared all-sky survey mission WISE \citep[e.g.,][]{2011ApJ...743...50C, 2012ApJ...753..156K}.

In this context, the wide-field camera on GREX-PLUS is expected to reach a limiting magnitude approximately five magnitudes deeper than that of WISE\footnote{\url{https://irsa.ipac.caltech.edu/data/WISE/docs/release/All-Sky/expsup/sec6_3a.html}} (see Table~\ref{tab:GPsurveys} for the GREX-PLUS Deep survey plan), enabling a survey of Y dwarfs that reaches about ten times farther.
Taking into account an increase in survey volume by a factor of $10^3$ and a survey area covering 1/40 of the entire sky (Table~\ref{tab:GPsurveys}), we anticipate that the effective surveyable volume and hence the number of detectable Y dwarfs will increase by a factor of $1,000 \times 1/40 \sim 25$. Such a substantial increase in the Y-dwarf sample would, for the first time, enable comprehensive population studies of Y dwarfs.
We emphasize that this census of Y dwarfs can be carried out as a natural by-product of the wide-field imaging survey of distant galaxies, without requiring any additional dedicated observing time.

Moreover, the GREX-PLUS wide-field camera is not limited to the detection of Y dwarfs. It also enables atmospheric characterization of newly discovered Y dwarfs through the use of five photometric filters covering the wavelength range of $2$--$8$~$\mu$m.
The infrared spectra of Y dwarfs are shaped by distinct molecular absorption features, including those of $\mathrm{H_2O}$, $\mathrm{CH_4}$, $\mathrm{CO}$, $\mathrm{CO_2}$, and $\mathrm{NH_3}$ \citep[e.g.,][]{2023ApJ...951L..48B}, and even photometric measurements alone can provide valuable constraints on their atmospheric properties.
With such a large sample of newly discovered Y dwarfs, we can therefore investigate key atmospheric processes through the temperature dependence of their five photometric colors.
This prompt, initial atmospheric characterization at the time of discovery is highly valuable for planning follow-up spectroscopic observations with other facilities, such as the James Webb Space Telescope (JWST).

Furthermore, GREX-PLUS is equipped with a high-resolution spectrograph ($R \sim 30,000$) covering the wavelength range of $12$--$18$~$\mu$m.
With a spectral resolving power approximately an order of magnitude higher than that of the Mid-Infrared Instrument (MIRI) on board JWST over the same wavelength range, this instrument enables detailed and robust atmospheric characterization of brown dwarfs across all spectral types.
Owing to its wavelength coverage, the high-resolution spectrograph on GREX-PLUS is particularly well suited to studies of very low-temperature objects, such as Y dwarfs.

\subsection{Required observations and expected results}


\subsubsection{Detection of Y dwarfs}

\paragraph{Estimated number of detectable Y dwarfs}


Based on the survey volume scaling of $\sim 25$ for GREX-PLUS relative to WISE, discussed above, and the 49 Y dwarfs detected in the WISE survey, GREX-PLUS is expected to detect $\sim 49 \times 25 = 1,225$ Y dwarfs analogous to those discovered by WISE. This estimate relies on the following assumptions:
(i) The local spatial distribution of Y dwarfs is homogeneous, and any age dependence is neglected, although the age distribution is expected to correlate with their spatial distribution in the Galaxy, since brown dwarfs cool as they age.
(ii) The survey area of GREX-PLUS covers 1/40 of the entire sky (the Deep survey plan of GREX-PLUS; see Table~\ref{tab:GPsurveys}).
(iii) The maximum distance that can be probed, $d$, scales as $d \propto \sqrt{\mathrm{flux}} \propto \mathrm{SNR}$, resulting in a survey volume increase proportional to the cube of $\mathrm{SNR}$.

While this simple scaling provides a useful first-order estimate, it implicitly assumes that the population of Y dwarfs detected by GREX-PLUS is analogous to that already discovered by WISE. In reality, GREX-PLUS will probe substantially deeper into lower-temperature regimes that lie at or beyond the detection limit of WISE. If objects with effective temperatures of $\lesssim 300$~K are detected, they would enable direct comparisons between extrasolar atmospheres and planetary atmospheres in the Solar System, such as that of Jupiter, within the same temperature regime. Such comparisons would significantly advance our understanding of atmospheric physical and chemical processes operating at these extremely low temperatures.

Because the detectability and space density of brown dwarfs depend sensitively on their effective temperatures, a more physically motivated estimate requires treating different temperature ranges separately. We therefore refine the above estimate by explicitly considering brown dwarf populations in distinct effective temperature intervals. 
Based on the dependence of space density on effective temperature above 450~K, presented in \cite{2021ApJS..253....7K}, we roughly extrapolate the space densities for the colder effective temperature ranges of $300$--$450$~K and $150$--$300$~K as follows:
\begin{eqnarray}
\rho_{300-450} \sim 6 \times 10^{-3} \: \mathrm{pc}^{-3} \: [150 \: \mathrm{K}]^{-1} \\
\rho_{150-300} \sim 1 \times 10^{-2} \: \mathrm{pc}^{-3} \: [150 \: \mathrm{K}]^{-1}
\end{eqnarray}
Among the brown dwarfs within 20 pc, the current census completeness limit for Y dwarfs, the faintest object in terms of the combination of effective temperature and distance has $T_\mathrm{eff} = 377$~K and $d = 19.25$~pc. Assuming these values represent the detection limit of WISE, the limiting distances for brown dwarfs with effective temperatures of 300~K and 150~K are $d_{300 \: \mathrm{K}, \: \mathrm{WISE}} = 12.2$~pc and $d_{150 \: \mathrm{K}, \: \mathrm{WISE}} = 3.1$~pc, respectively.

Assuming 100\% survey completeness within these distances, and adopting the space densities given above, the expected numbers of brown dwarfs detectable by WISE in these temperature ranges are
\begin{eqnarray}
    N_{300-450, \: \mathrm{WISE}} \sim \frac{4\pi}{3} d^3_{300 \: \mathrm{K}, \: \mathrm{WISE}} \times \rho_{300-450} = 45.6 \\
    N_{150-300, \: \mathrm{WISE}} \sim \frac{4\pi}{3} d^3_{150 \: \mathrm{K}, \: \mathrm{WISE}} \times \rho_{150-300} = 1.2
\end{eqnarray}
Note that, since $d_{450 \: \mathrm{K}, \: \mathrm{WISE}} > d_{300 \: \mathrm{K}, \: \mathrm{WISE}}$ and $d_{300 \: \mathrm{K}, \: \mathrm{WISE}} > d_{150 \: \mathrm{K}, \: \mathrm{WISE}}$, these estimates should be regarded as conservative.
These values are consistent with the results of the WISE survey to within an order of magnitude.
Indeed, one brown dwarf cooler than 300~K was detected by WISE \citep{2014ApJ...786L..18L}.

Using the same scaling relation as mentioned above, we estimate the number of the coldest brown dwarfs detectable by GREX-PLUS as follows:
\begin{eqnarray}
    N_{300-450, \: \mathrm{GREX-PLUS}} = N_{150-300, \: \mathrm{WISE}} \times 10^3 \times 1/40 \sim 1,100 \\
    N_{150-300, \: \mathrm{GREX-PLUS}} = N_{150-300, \: \mathrm{WISE}} \times 10^3 \times 1/40 \sim 30
\end{eqnarray}

\paragraph{Stellar spatial homogeneity as a proxy for the Y-dwarf distribution}
As a proxy for assessing the spatial homogeneity of Y dwarfs, we examine the spatial distribution of stars using the TESS Input Catalog v8.0 \citep{2019yCat.4038....0S}. This proxy is valid insofar as Y dwarfs trace the local disk population similarly to low-mass stars. We first confirm that the number of stars within each distance bin increases with the square of the distance out to $\sim 100$~pc, indicating that the stellar distribution is homogeneous.
This result is consistent with the homogeneous spatial distribution of Y dwarfs discovered by the WISE survey ($\lesssim 20$~pc).

We estimate the relative gain in the number of detectable Y dwarfs for GREX-PLUS compared to WISE as
\begin{equation}
    G = \frac{ N_{<d_\mathrm{GREX-PLUS}} }{N_{<d_\mathrm{WISE}}} f_\mathrm{fov},
\end{equation}
where $f_\mathrm{fov} = 1/40$ is the ratio of the survey area to the entire sky. The quantity
\begin{equation}
    N_{<d_\mathrm{GREX-PLUS}} = \sum_{\mathrm{pixel}} N_\mathrm{pix, \: <d_\mathrm{GREX-PLUS}} 
\end{equation}
represents the integration of the surface number density of stars within a distance $d_\mathrm{GREX-PLUS}$ over the entire sky.
Here, $N_\mathrm{pix, \: <d_\mathrm{GREX-PLUS}}$ denotes the number of stars within a distance $d_\mathrm{GREX-PLUS}$ in a single pixel.
Similarly, $N_{<d_\mathrm{WISE}}$ represents the number of stars within a distance $d_\mathrm{WISE}$.

Assuming a relative sensitivity of GREX-PLUS that is 100 times greater than that of WISE, as discussed above, we adopt
\begin{equation}
    d_\mathrm{GREX-PLUS} = 10 d_\mathrm{WISE}.
\end{equation}

\begin{figure}[htbp]
  \begin{minipage}[b]{0.5\columnwidth}
    \centering
    \includegraphics[height=5cm]{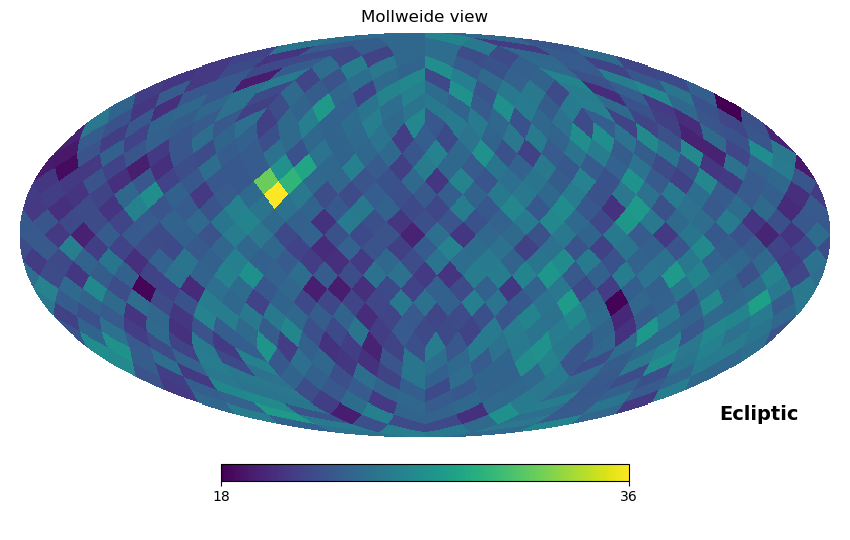}
  \end{minipage}\hfill
  \begin{minipage}[b]{0.5\columnwidth}
    \centering
    \includegraphics[height=5cm]{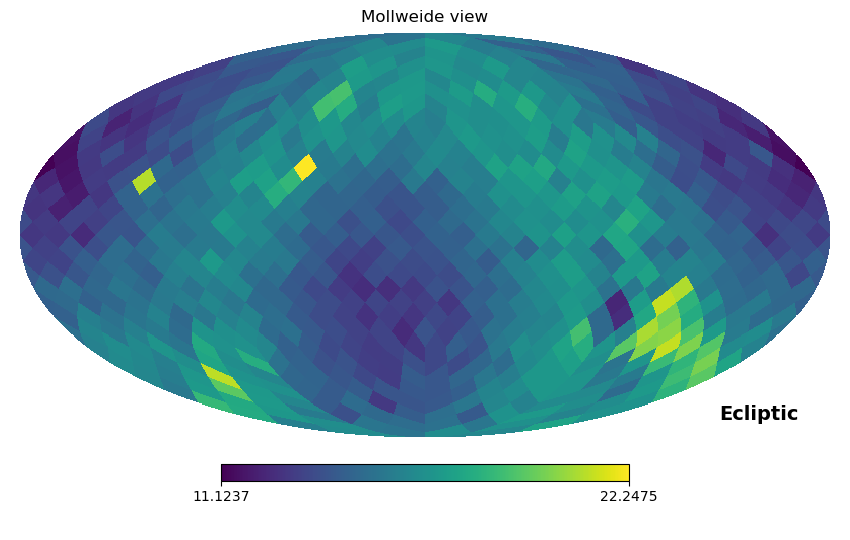}
  \end{minipage}\hfill
    \caption{Expected gain in the number of detected Y dwarfs in the GREX-PLUS survey relative to the WISE survey, using the stellar spatial distribution as a proxy for that of Y dwarfs (shown in equatorial coordinates). (Left) Case of $d_\mathrm{WISE} = 10$~pc and $d_\mathrm{GREX-PLUS} = 100$~pc. (Right) Case of $d_\mathrm{WISE} = 20$~pc and $d_\mathrm{GREX-PLUS} = 200$~pc.}
    \label{fig:BD_density}
\end{figure}

Figure~\ref{fig:BD_density} shows the relative gain of GREX-PLUS compared to WISE for two cases: $d_\mathrm{WISE} = 10$~pc and $d_\mathrm{GREX-PLUS} = 100$~pc (left), and $d_\mathrm{WISE} = 20$~pc and $d_\mathrm{GREX-PLUS} = 200$~pc (right). In regions with low stellar density, the relative gain is found to be approximately $10$--20. By multiplying this gain by the number of Y dwarfs within a distance $d_\mathrm{WISE}$, denoted as $m_{<d_\mathrm{WISE}}$, we can estimate the expected number of detections in cases where the survey field corresponds to such a low-stellar-density region. Adopting $m_{<d_\mathrm{WISE}} = 25$ for $d_\mathrm{WISE} = 10$~pc, the expected number of detections in the lowest-density regions is $\sim 500$.

\paragraph{Discrimination between Y dwarfs and distant galaxies}

For the detection of faint Y dwarfs, discrimination against similarly faint distant galaxies is crucial. Y dwarfs are bright at wavelengths of $4$--$5$~$\mu$m and faint at adjacent wavelengths because of the relative lack of molecular absorption at these wavelengths \citep[e.g.,][]{2023ApJ...951L..48B}. Capturing this spectral feature is therefore essential.
However, photometric color information alone may be insufficient for robust discrimination. We therefore propose proper motion and parallax measurements, which would yield a clear distinction between nearby Y dwarfs and distant galaxies.
The proper motions of nearby Y dwarfs are expected to be $\gtrsim 20$--$30$~mas/yr at distances of $\sim 100$~pc, based on the typical velocity dispersion in the solar neighborhood \citep[e.g.,][]{2013A&ARv..21...61R}, and are therefore much larger than the negligible proper motions and parallaxes of distant galaxies.

In addition, multi-epoch observational data enable investigations of possible binarity of the discovered Y dwarfs through measurements of their proper motions and astrometric reflex motion.
While mass estimates for isolated single brown dwarfs must rely on thermal evolution models, leading to relatively large uncertainties due to the typical lack of age information, dynamical mass measurements are feasible for binary brown dwarf systems \citep[e.g.,][]{2018A&A...618A.111L}.
Such reliable mass determinations allow much tighter constraints on their radii, as well as their atmospheric properties, including elemental and molecular abundances, thermal structure, and cloud characteristics \citep[e.g.,][]{2025ApJ...988...53K}.
Consequently, binary brown dwarfs serve as 
excellent benchmark targets for advancing our understanding of formation, evolution, and atmospheric processes through comparisons of their fundamental parameters, namely mass, radius, and luminosity, and atmospheric properties, as demonstrated by studies of the nearest brown dwarf binary system, WISE~1049AB \citep[also known as Luhman~16AB; e.g.,][]{2014Natur.505..654C, 2024MNRAS.532.2207B, 2026ApJ...997..118Y}.

\subsubsection{Characterization of Y dwarfs}

\begin{figure}[htbp]
  \begin{minipage}[b]{0.5\columnwidth}
    \centering
    \includegraphics[height=5.3cm]{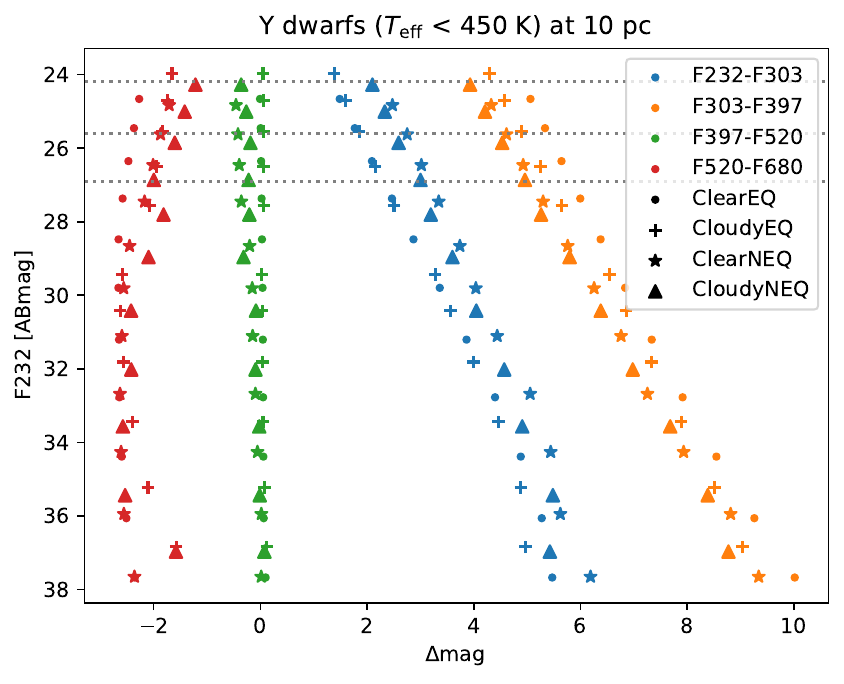}
  \end{minipage}\hfill
  \begin{minipage}[b]{0.5\columnwidth}
    \centering
    \includegraphics[height=5.3cm]{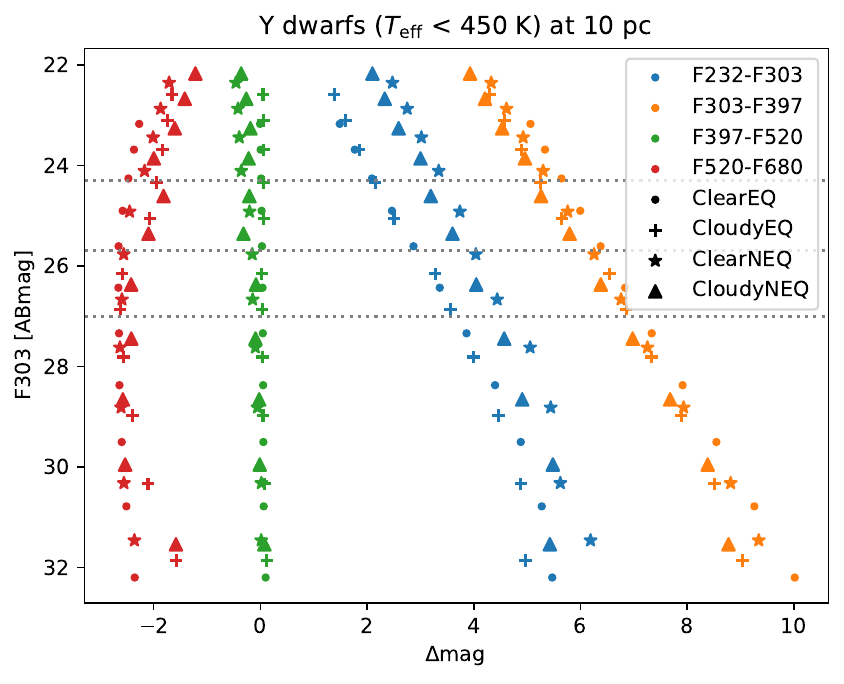}
  \end{minipage}\hfill
  \begin{minipage}[b]{0.5\columnwidth}
    \centering
    \includegraphics[height=5.3cm]{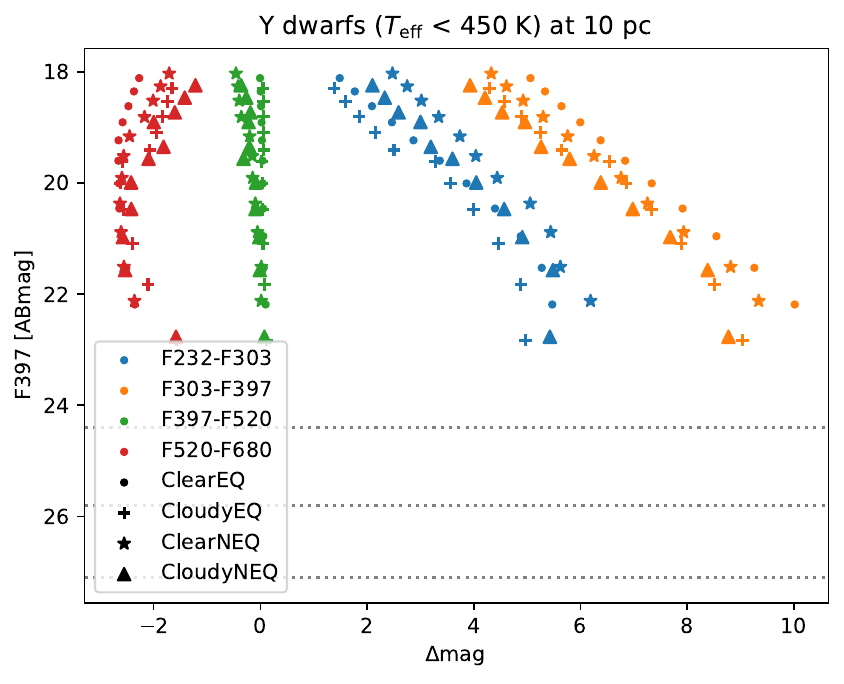}
  \end{minipage}\hfill
    \begin{minipage}[b]{0.5\columnwidth}
    \centering
    \includegraphics[height=5.3cm]{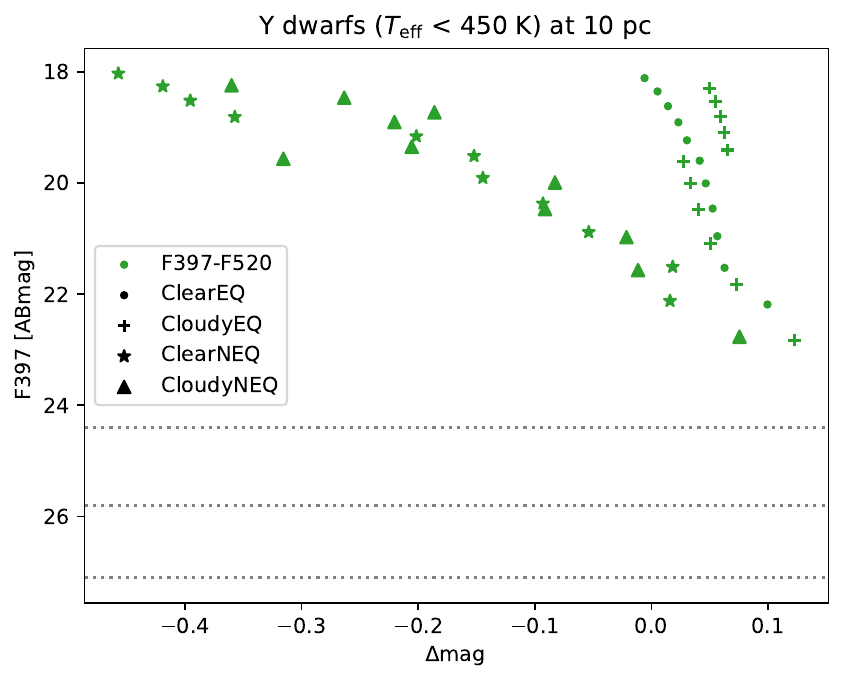}
  \end{minipage}\hfill
  \begin{minipage}[b]{0.5\columnwidth}
    \centering
    \includegraphics[height=5.3cm]{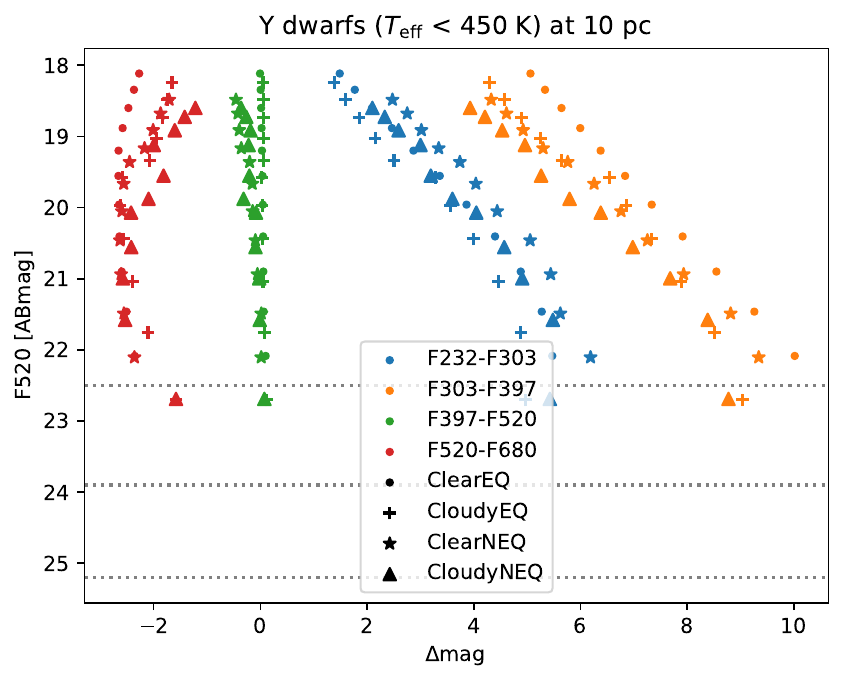}
  \end{minipage}\hfill
  \begin{minipage}[b]{0.5\columnwidth}
    \centering
    \includegraphics[height=5.3cm]{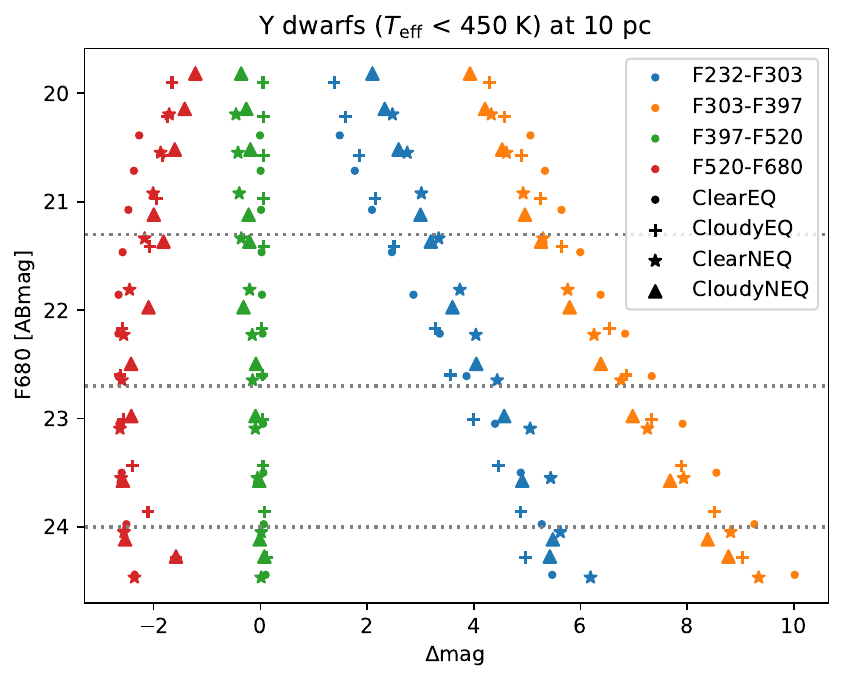}
  \end{minipage}\hfill
  \caption{Simulated trends of photometric colors observable with the GREX-PLUS wide-field camera, computed using atmospheric spectral model grids for several atmospheric scenarios developed by \citet{2023ApJ...950....8L}. Each panel shows the color differences, focusing on one of the five photometric filters (F232, F303, F397, F520, and F680; from top left to bottom right). The considered scenarios include clear equilibrium chemistry (ClearEQ; circles), cloudy equilibrium chemistry (CloudyEQ; crosses), clear nonequilibrium chemistry (ClearNEQ; stars), and cloudy nonequilibrium chemistry (CloudyNEQ; triangles). Each point represents an effective temperature ranging from $200$~K to $450$~K at 25~K intervals. The dotted lines indicate the limiting magnitudes of the GREX-PLUS wide-field survey plans (Wide, Medium, and Deep; see Table~\ref{tab:GPsurveys}). 
  Note that the distance was assumed to be 10~pc.
  The middle-right panel shows a zoomed-in view of the middle-left panel, highlighting the F397 and F520 filters, in which Y dwarfs are brightest and the different scenarios can thus be more easily distinguished.
  }
  \label{fig:bd_gp_color}
\end{figure}

\paragraph{Wide-field camera}
As mentioned above, the five photometric filters of the wide-field camera on GREX-PLUS enable first-step atmospheric characterization of newly discovered Y dwarfs. Figure~\ref{fig:bd_gp_color} shows the simulated trends of five photometric colors observable with the GREX-PLUS wide-field camera, computed using publicly available\footnote{\url{https://zenodo.org/records/7779180}} atmospheric spectral model grids for several atmospheric scenarios developed by \citet{2023ApJ...950....8L}.
We used their models for a gravity of $\log{g} = 4.00$ and a metallicity of $Z = 0$. The distance was assumed to be 10~pc.
The considered scenarios include clear equilibrium chemistry (ClearEQ), cloudy equilibrium chemistry (CloudyEQ), clear nonquilibrium chemistry (ClearNEQ), and cloudy nonequilibrium chemistry (CloudyNEQ).
For the cloudy cases, water cloud formation is included.
Nonequilibrium chemistry is treated using an eddy diffusion coefficient of $K_\mathrm{zz} = 10^6$~$\mathrm{cm}^2$~$\mathrm{s}^{-1}$.

Figure~\ref{fig:bd_gp_color} clearly demonstrates that the color--magnitude trends differ depending on the atmospheric scenario (clear/cloudy and equilibrium/nonequilibrium chemistry), as indicated by the different symbols.
Therefore, five-color photometric information from the first large, statistically significant sample of Y dwarfs expected to be discovered by GREX-PLUS will enable the effects of cloud formation and chemical disequilibrium to be identified in the previously unexplored temperature regime of Y dwarfs.

\paragraph{High-resolution spectrograph}
GREX-PLUS carries a high-resolution spectrograph ($R \sim 30,000$) operating over $12$--$18$~$\mu$m.
Since the launch of JWST in 2021, brown dwarf atmospheres have also been investigated in the mid-infrared using MIRI on board JWST through its low-resolution ($R \sim 100$) and medium-resolution ($R \sim 3,000$) spectroscopic modes \citep[e.g.,][]{2023Natur.624..263B, 2024ApJ...977L..32X, 2025A&A...695A.224K, 2025A&A...703A..70V}.

By providing a spectral resolution $\sim 10$ times higher than that of JWST/MIRI, the GREX-PLUS high-resolution spectrograph will enable significantly more detailed atmospheric characterization of brown dwarfs, including reliable constraints on gravity (and hence mass), robust identification of molecules and their isotopologues, tighter abundance constraints, and improved retrievals of temperature structures \citep[e.g.,][]{2022ApJS..258...31K, 2022MNRAS.514.3160T, 2023Natur.624..263B, 2025ApJ...988...53K, 2025AJ....170..211K, 2025arXiv251123018Y}.
Its mid-infrared wavelength coverage makes this instrument particularly powerful for studying very low-temperature objects such as Y dwarfs within 100--200~pc, which are exceedingly faint at optical and near-infrared wavelengths and thus largely inaccessible to ground-based high-resolution spectroscopy.

\subsection{Scientific goals}


Thanks to its sensitivity, improved by a factor of $\sim 100$ compared to WISE, which has been the primary and essentially only facility responsible for the detection of faint Y dwarfs, GREX-PLUS is expected to detect $\gtrsim 1,000$ new Y dwarfs with its wide-field camera, as a natural by-product of the wide-field imaging survey of distant galaxies. 
Such a large sample will, for the first time, enable a statistical exploration of the coolest substellar objects, providing critical insights into the low-mass end of star formation and evolutionary processes.

Beyond mere detection, the five photometric filters of the GREX-PLUS wide-field camera will allow initial atmospheric characterization of a large population of newly detected Y dwarfs at the time of discovery, opening a new window onto the atmospheric processes operating in the lowest-temperature regime.
The temperature range of Y dwarfs overlaps with that of the primary targets of current and upcoming exoplanet atmospheric studies, as well as with Solar System objects such as Jupiter. Therefore, statistical investigations of Y-dwarf atmospheres will have strong synergy with efforts to understand the common physical and chemical processes governing atmospheres in low-temperature environments.

Furthermore, with its unique capability to perform high-resolution spectroscopy in the mid-infrared, GREX-PLUS also enables detailed and robust atmospheric characterization of brown dwarfs, leveraging a resolving power $\sim 10$ times higher than that of JWST/MIRI.
The wavelength coverage of the high-resolution spectrograph is particularly well suited to low-temperature objects such as Y dwarfs, for which ground-based high-resolution spectroscopy is nearly impossible due to their extreme faintness in the optical and near-infrared.

\begin{table}
    \begin{center}
    \caption{Required observational parameters.}\label{tab:browndwarf}
    \begin{tabular}{|l|p{9cm}|l|}
    \hline
     & Requirement & Remarks \\
    \hline
    Wavelength & 2--8 $\mu$m/10--18 $\mu$m & \multirow{2}{*}{} \\
    \cline{1-2}
    Wavelength resolution & $\lambda/\Delta \lambda \gtrsim 30,000$ &  \\
    \hline
    Field of view & 1000 deg$^2$, $\gtrsim 24$ AB mag ($5\sigma$, point source) & \multirow{2}{*}{}\\
    \cline{1-1}
    Sensitivity &  & \\
    \hline
    Observing cadence & A few epochs over a moderate time span & $a$ \\
    \hline
    \end{tabular}
    \end{center}
    $^a$ We need proper motion and parallax measurements to discriminate Y dwarfs ($\gtrsim 20$--$30$~mas/yr at distances of $\sim 100$~pc) from similarly faint, distant galaxies and potentially to identify their binarity.\\
\end{table}

\printbibliography[heading=subbibliography]
\end{refsection}

\clearpage

\begin{refsection}[3-9_Galactic_Plane/Galactic_Plane.bib]

\section{Galactic Plane Survey}
\label{sec:galacticplane}

\noindent
\begin{flushright}
Kumiko Morihana$^{1}$

$^{1}$ Subaru Telescope
\end{flushright}
\vspace{0.5cm}

\subsection{Scientific background and motivation}

\begin{figure}[htpb]
    \centering
    \includegraphics[width=12cm]{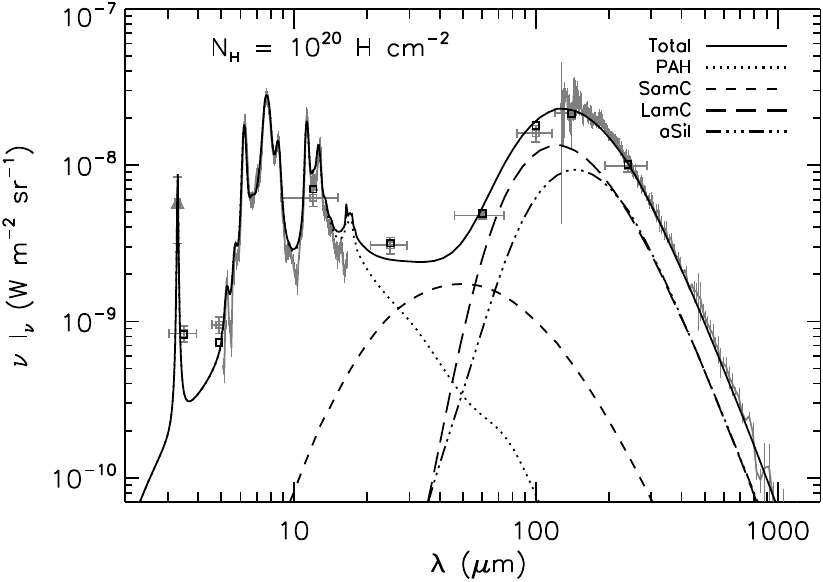}
    \caption{The typical spectral energy distribution (SED) of interstellar dust in the Milky Way. Grey symbols and lines represent observed data from ISO, COBE, and AROME. The black solid line indicates the best-fit model calculated using the DustEM code. The prominent emission peaks between 3 and 12 $\mu$m are primarily attributed to Polycyclic Aromatic Hydrocarbons (Compiegne et al. 2011).}
    \label{fig:dust}
\end{figure}

To explore the evolution of our Galaxy, it is fundamental to investigate the nature of the interstellar medium (ISM) in addition to studying individual sources that constitute the Galaxy. The ISM is a dynamic reservoir; it provides the raw materials from which new generations of stars are born and, in turn, acts as the primary medium that absorbs and thermalizes the energy and heavy elements injected by stellar winds and supernova explosions. Understanding the physical state and spatial distribution of the ISM---including gas, dust, and organic molecules such as polycyclic aromatic hydrocarbons (PAHs)---is therefore essential to quantifying the efficiency of star formation and the impact of stellar feedback. PAHs are widely distributed throughout our Galaxy and in nearby galaxies (e.g., \cite{Onaka1996}), with major emission features observed at 3.3, 6.2, 7.7, 8.6, and 11.3 $\mu$m. 
These features are observed at high galactic latitudes and within the Galactic plane, serving as a reference for dust evolution processes (Figure~\ref{fig:dust}).

\begin{figure}[htpb]
    \centering
    \includegraphics[width=12cm]{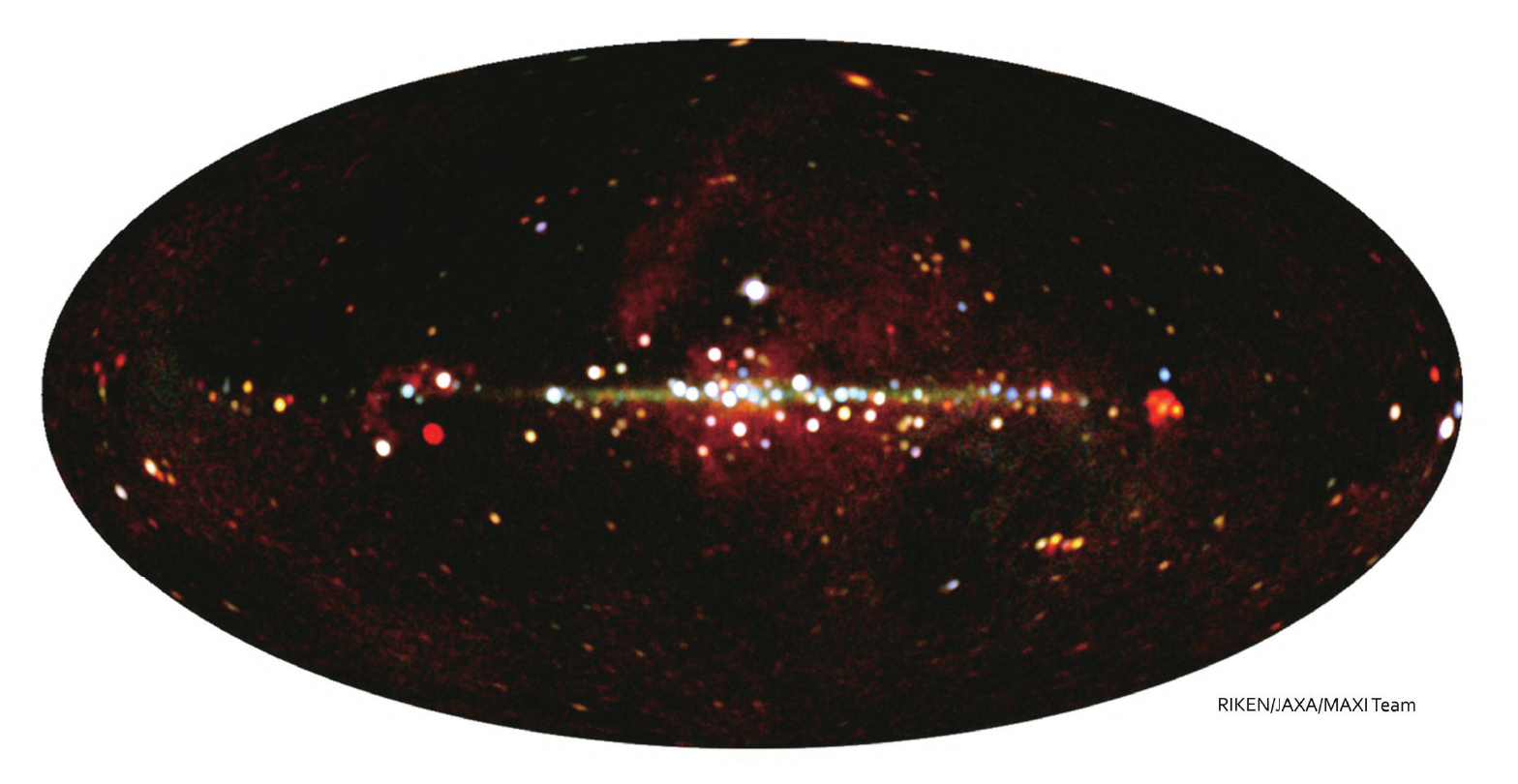}
    \caption{All-sky X-ray image (3--20 keV) with MAXI/GSC from 2009 August to 2015 October in Galactic coordinates \copyright RIKEN/JAXA. In addition to the bright stars (mainly black holes and neutron stars), such extended emission along the Galactic plane is recognized as the Galactic Diffuse X-ray Emission.}
    \label{fig:gdxe}
\end{figure}

However, the ISM is not a monolithic entity; it consists of multiple phases that are constantly exchanging energy. Although infrared observations of PAHs trace the "cool" molecular and neutral phases influenced by stellar radiation, they do not provide a complete picture of the energetic balance of the Galaxy. A significant fraction of the energy injected by stellar feedback is stored in a much hotter phase: the Galactic Diffuse X-ray Emission (GDXE; e.g., \cite{Worrall1982}, Figure~\ref{fig:gdxe}). The GDXE is an apparently extended emission of high-temperature plasma ($\sim10^8$ K) observed along the Galactic plane ($|l| < 45^{\circ}$, $|b| < 2^{\circ}$), characterized by thermal emission lines such as the Fe K-shell complex: a triad of lines at 6.40 keV (neutral Fe $\mathrm{I}$), 6.68 keV (He-like Fe $\mathrm{XXV}$), and 6.97 keV (H-like Fe $\mathrm{XXVI}$).

Since its discovery, the origin of the GDXE has been debated: is it truly diffuse plasma, or is it the integrated emission of faint X-ray sources? If the GDXE is composed of point sources, the primary candidates are specific stellar populations, such as magnetic (e.g., \cite{Hong2012}, \cite{Yuasa2012}) and non-magnetic (e.g., \cite{Nobukawa2016}; \cite{Morihana2022}) Cataclysmic Variables (CVs) and active stars (e.g., \cite{Revnivtsev2006}). However, if the GDXE originates from a truly diffuse $10^8$ K plasma, it implies a staggering rate of energy injection into the ISM that challenges our current understanding of stellar feedback.

To solve this decades-old mystery, we must compare the spatial distribution of this high-energy plasma with a high-fidelity template of the diffuse interstellar gas. However, obtaining such a template has been a long-standing challenge in infrared astronomy.

\begin{figure}[htpb]
    \centering
     \includegraphics[width=13cm]{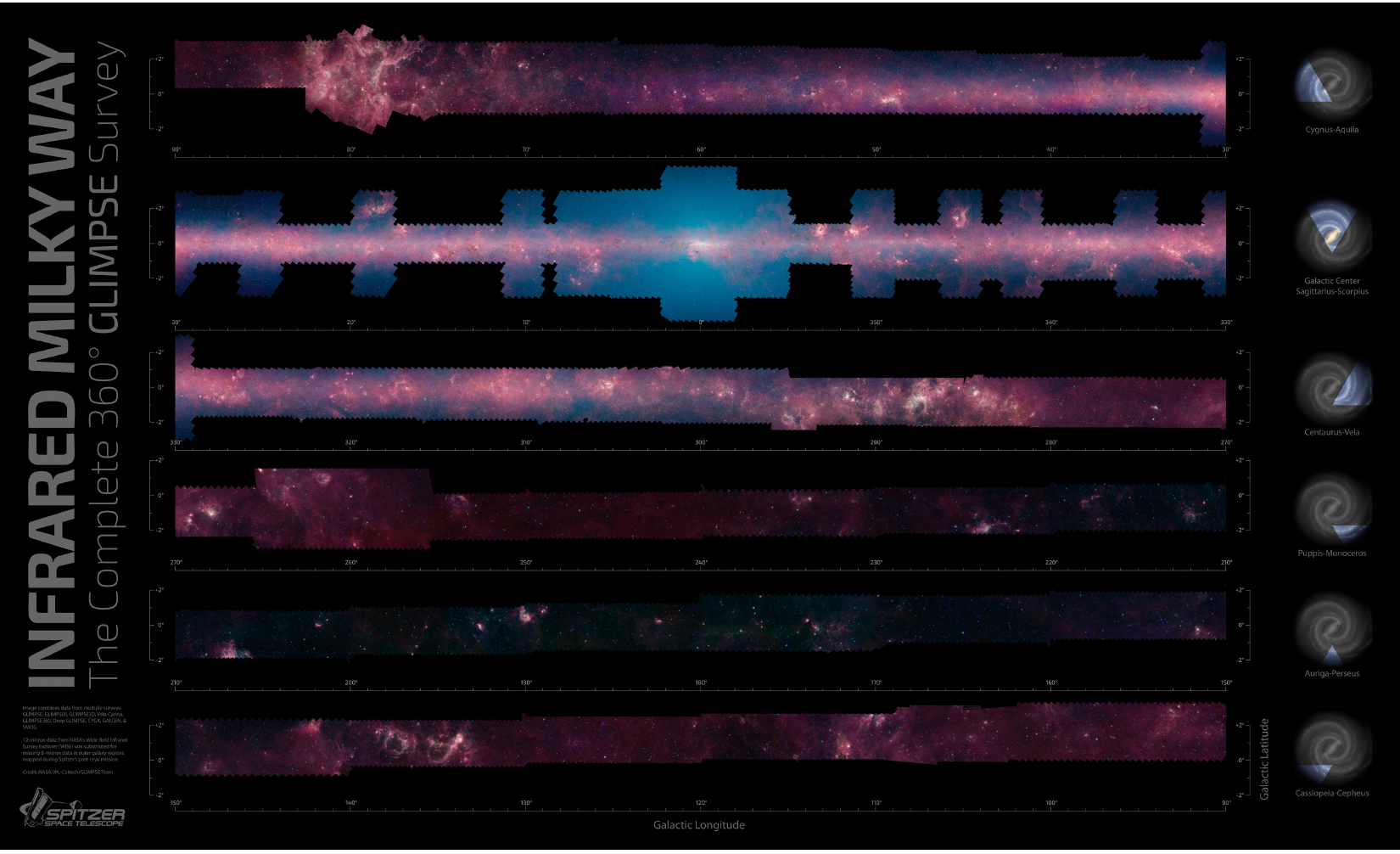}
    \caption{Spitzer/GLIMPSE 360-degree infrared panorama of the Milky Way. This mosaic, compiled from over two million snapshots, reveals the Galactic plane in a three-color composite: 3.6 $\mu$m (blue), 4.5 $\mu$m (green), and 8.0 $\mu$m (red). The pervasive red emission traces the distribution of PAHs and warm dust, highlighting active star-forming regions and stellar feedback structures throughout the Galaxy (\copyright NASA/JPL-Caltech/GLIMPSE Team.)}
    \label{fig:glimpse}
\end{figure}

Our understanding of the dust distribution along the Galactic plane has evolved through several generations of infrared sky surveys. Early missions such as the Infrared Astronomical Satellite (IRAS) provided the first all-sky maps of thermal dust emission. Subsequent missions, such as AROME (Association de Recherche Optique des Molécules Extragalactiques; \cite{Giard1994}) and COBE (Cosmological Background Explorer)/Diffuse Infrared Background Experiment (DIRBE; \cite{Bernard1994}) further expanded our view of the infrared sky. The first spectroscopic evidence of dust in the Galactic plane was provided by the IRTS (InfraRed Telescope in Space; \cite{Tanaka1996}) and the Infrared Space Observatory (ISO; \cite{Mattila1996}).
Further advancements were made by the Spitzer Space Telescope, whose GLIMPSE survey (Galactic Legacy Infrared Mid-Plane Survey Extraordinaire; \cite{Benjamin2003}; \cite{Churchwell2009}) used the 8.0 $\mu$m band to map the 7.7 and 8.6 $\mu$m PAH complexes as precise tracers of stellar feedback (Figure~\ref{fig:glimpse}). Despite these leaps in capability, a critical piece of the puzzle remains missing: the 3.3 $\mu$m PAH. This specific feature, arising from C-H stretching vibrations of small, neutral PAHs, has remained difficult to isolate due to the dominant stellar continuum in broad filters, such as Spitzer's IRAC 3.6 $\mu$m band.

This is where GREX-PLUS provides a decisive breakthrough. While previous missions were limited by spatial resolution or spectral contamination, sub-arcsecond ($0.3^{\prime\prime}$) resolution of the GREX-PLUS allows for the precise identification and subtraction of millions of individual stars. By effectively "cleaning" the stellar continuum, GREX-PLUS will provide the first high-fidelity template of the 3.3 $\mu$m PAH distribution. This unique dataset is essential for determining whether the GDXE is physically coupled with the diffuse gas or whether it originates from an underlying population of faint stellar remnants such as CVs.

\subsection{Required observations and expected results}
\subsubsection{Wide-area Survey Strategy and Spectral Coverage}

To investigate the physical origin of the Galactic high-energy environment, we propose a wide-area survey covering the region where the GDXE is most prominent ($|l| < 45^{\circ}$, $|b| < 1^{\circ}$), totaling $180 \text{ deg}^2$. This survey area is strategically selected to encompass the high-intensity ridge of the GDXE, allowing for a statistically robust comparison between the X-ray surface brightness and the underlying stellar and gaseous components. GREX-PLUS will perform deep imaging in the F232 ($2.3\,\mu\text{m}$) and F397 ($4.0\,\mu\text{m}$) bands with a sensitivity sufficient to resolve individual stars down to the faint populations at the distance of the Galactic Center.

\begin{table}
    \begin{center}
    \caption{Required observational parameters for the Galactic Plane Survey}\label{tab:galacticplane}
    \begin{tabular}{|l|p{9cm}|l|}
    \hline
     & Requirement & Remarks \\
    \hline
    Wavelength & 2.0--4.5 $\mu$m & F232, F303, F397\\
    \hline
    Field of view & $|l| < 45^{\circ}$, $|b| < 1^{\circ}$ ($\sim180$ deg$^2$) &\\
    \hline
    Sensitivity &  22 AB mag ($5\sigma$, $S/N=10$)& \\
    \hline
    \end{tabular}
    \end{center}
\end{table}

\subsubsection{Expected Results}
The analysis will leverage the sub-arcsecond resolution of GREX-PLUS to follow three primary paths, starting with the removal of stellar contamination to isolate the diffuse signals:

\begin{itemize}
\item Point-Source Census and Population Classification: We will utilize the $F397/F232$ color ratio to identify and classify specific stellar populations. While normal late-type stars such as Active Binaries (ABs) exhibit a ratio near unity, Cataclysmic Variables (CVs) are expected to show a smaller ratio. This is because the high-density environment of accretion disks ($n_e \sim 10^8$--$10^{14}$ cm$^{-3}$) causes hydrogen recombination lines to deviate from standard "Case B" theory, which typically predicts a $Br\alpha/Br\gamma$ ratio of $0.44$--0.55. By quantifying the spatial density of these infrared-identified point sources, we can directly test the "point-source origin" hypothesis of the GDXE.

\item PAH Depletion Analysis: The simultaneous mapping of the $3.3\,\mu\text{m}$ PAH feature ($F335$) will provide the first sub-arcsecond, "cleaned" template of the neutral ISM. Since high-temperature plasma is known to destroy PAHs, an anti-correlation between PAH intensity and X-ray brightness would serve as strong evidence for the existence of a truly diffuse, hot plasma ($10^8$ K).

\item Plasma Cooling through Recombination Lines: If a hot diffuse plasma exists, its cooling process is expected to emit hydrogen recombination lines such as $Br\alpha$ ($4.05\,\mu\text{m}$). By comparing the $Br\alpha$ intensity map with the X-ray surface brightness (after removing resolved X-ray point sources), we can trace the distribution of cooling gas and further constrain the diffuse plasma model.
\end{itemize}

\subsection{Scientific goals}


The goal of this survey is to establish a comprehensive energy and matter budget for the Milky Way Galaxy by bridging the gap between the infrared and X-ray windows.

First, we aim to solve the decades-old mystery of the origin of the GDXE. While X-ray observatories such as Chandra provide high-resolution data, they alone struggle to resolve the millions of faint stellar populations—such as Cataclysmic Variables (CVs) and active stars—that are deeply embedded in the crowded and obscured Galactic plane. By integrating GREX-PLUS’s sub-arcsecond infrared stellar census with Chandra’s X-ray data, we will definitively determine whether the Galactic plane is filled with a staggering volume of hot plasma or whether the emission is simply the collective glow of these faint, unresolved stellar remnants. This determination is fundamental to our understanding of the Galactic energy budget, as it dictates the required rate of energy injection from supernova explosions and stellar winds.
Second, this study seeks to characterize the chemical lifecycle of the ISM across the entire Galactic disk. The resulting high-resolution map of the $3.3\,\mu$m PAH feature, isolated for the first time by removing stellar contamination, will serve as a high-fidelity proxy for the metallicity gradient and the state of chemical enrichment from the Galactic Center to the outer disk.
Ultimately, by connecting the distribution of neutral organic molecules to the most energetic phenomena in the Galaxy, this program will provide a new, multi-phase framework for understanding how the Milky Way evolves as a dynamic, interconnected system of stars and gas.

\printbibliography[heading=subbibliography]
\end{refsection}

\clearpage

\begin{refsection}[3-10_galaxycenter/galaxycenter.bib]

\section{The Galactic Center and Disk}
\label{sec:galaxycenter}

\noindent
\begin{flushright}
Naoteru Gouda$^{1}$, Taihei Yano$^{1}$
\\
$^{1}$ NAOJ, 
\end{flushright}
\vspace{0.5cm}
\subsection{Scientific background and motivation}

{\bf ~~~~~~\normalsize $\sim$ Scientific significance of the Milky Way Galaxy (the Galaxy) $\sim$}

The Galaxy has been considered to have undergone various evolutionary processes, including collisions and mergers with several other dwarf galaxies so far. In order to understand the evolutionary process of our Galaxy (``Galactic Archaeology''), it is very important to understand the dynamical structure and its history in the Galaxy. In addition, the Galaxy is a typical disk galaxy in the present Universe, and detailed observations of the Galaxy provide clues to the formation and evolution of many other galaxies in the Universe.
For external galaxies, in contrast, it is difficult to accurately reveal in detail the physical characteristics of individual stars by observations because of their large distances. On the other hand, the Galaxy, where we live, is the only one for which we can reveal its physical characteristics, spatial distribution, and motion in an accurate and detailed manner for individual stars. In other words, it is possible to use methods that cannot be applied to other galaxies in the study of the formation and evolution of galaxies, and the Galaxy is a very good ``test bed'' from this point of view. In fact, the spatial distribution and motions of stars are already being revealed in increasing detail by Gaia.
 
Furthermore, stars, planets, and life including humanity in the Galaxy are born and evolve in the Galaxy, which is a collection of over 100 billion stars, so they are complexly affected by the Galaxy we live in (the effects depend on the locations where in the Galaxy the stars and planets were born, the orbits the stars 
with planets follow, and the influence from the surrounding space environment, etc.). Therefore, ``knowing'' the Galaxy is also important for solving the mystery of the existence of humanity.
So, the exploration of the Galaxy, which is the most familiar galaxy, but still a mystery, is very significant.

\begin{figure}[tbh]
  \begin{center}
   \includegraphics[width=110mm]{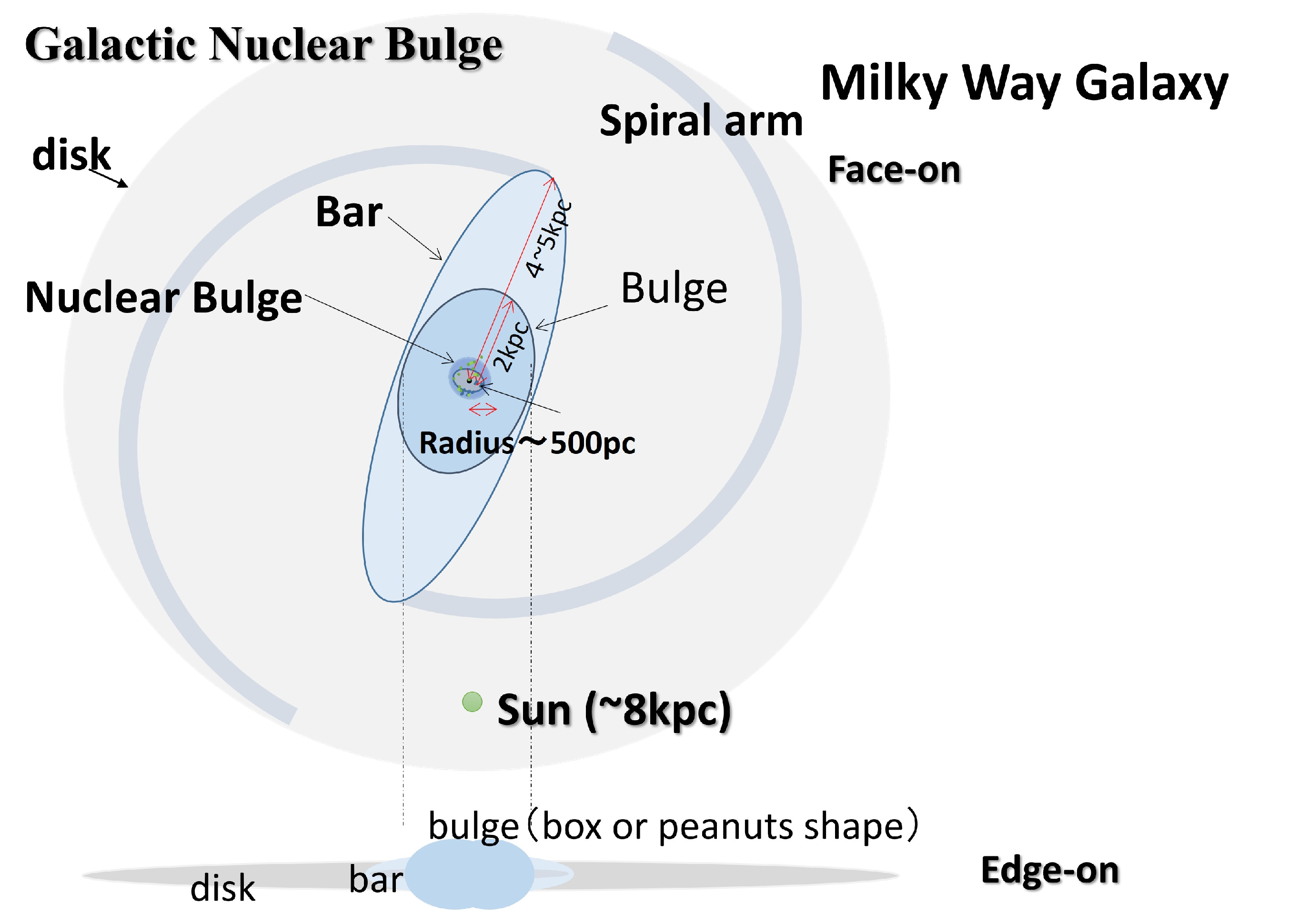}
  \end{center}
  \caption{Schematic picture of sub-structures of the Galaxy}
  \label{fig-MW}
\end{figure}
 
\subsection{Required observations and expected results}

{\bf ~~~~~~\normalsize $\sim$ Exploration of the Galactic nuclear bulge $\sim$}

The structure of the Galaxy is roughly divided into the disk containing the Solar System, the bulge structure at the center of the Galaxy, and a large structure called the halo surrounding the disk and the bulge \citep{2016ARA&A...54..529}. In addition, there are spiral patterns called spiral arms and a bar-like structure (bar). The bulge structure is a bulging structural element (about 2 kpc in size) around the Galactic center, where stars are densely gathered (see Fig.~\ref{fig-MW}). In addition, as a structure in close contact with the bulge, there is a non-axisymmetric structure called a bar structure (the length from the center is about 5 kpc) in the Galaxy, which is considered to be classified as a ``barred spiral galaxy''. Non-axisymmetric structures such as the bar structure strongly influence the dynamical structure of a wide range of the Galaxy including the vicinity of the solar system (such as the orbits and velocity distribution of stars). The velocities and orbits of stars are strongly influenced by the non-axisymmetric structures. Therefore, for example, information on the motions of stars in the disk alone does not allow us to understand all the physical information on the disk and spiral arms. It should be understood together with the physical information of the bulge and bar structures including the formation epoch of the bar structure.
Bulges are roughly classified into two types. One of them is a ``classical bulge'' and the other is a ``pseudo-bulge''. The classical bulge exhibits a brightness profile similar to that of an elliptical galaxy, while the pseudo-bulge exhibits a profile similar to that of a disk structure, and these are considered to have different origins. Furthermore, the pseudo-bulge is categorized into two types, a box/peanut type and a disk-like bulge. There is a possibility that the Galaxy has these three types of bulges (\citealt{2016ASSL...418..199G}). The Milky Way certainly has a box/peanut bulge, but the other two types, namely a disk-like bulge and a classical bulge, might also be located in the Galactic nucleus region.
 The Galactic nucleus region, whose radius is about 500 pc (or less than 500 pc), is called the Galactic nuclear bulge (see Fig.~\ref{fig-MW}) (\citealt{2016ARA&A...54..529}). The nuclear bulge has different physical characteristics from those of the outer bulge. The nuclear bulge is the place where the material elements of the Galaxy, such as stars, gas, and dark matter, are the most concentrated, and it is the place where the history from the early Galaxy formation to the present is intensively hidden. 
In the Galactic nuclear bulge, stars of various ages have different spatial distributions and motions according to their ages. It is like having different strata in geology. 
Although the Galactic nuclear bulge covers only a small part of the sky, it is a particularly important region for understanding our Galaxy's structure and its formation and evolution. Thus, revealing the Milky Way's central core structure and its formation history is very interesting and important for understanding the Galaxy as a whole. We call it ``the Galactic Center Archeology''. It can be said to be the core of exploring the Galaxy's history. 
In addition, the nuclear bulge region is an important area that links the physical relationship among the entire outer bulge, the bar structure, and the supermassive black hole at the Galactic center (e.g., \citealt{2017ApJL...844..15D}).

Furthermore, the recent development of powerful, multi-wavelength observational capabilities has enabled the era of Galactic archaeology (e.g., \citealt{2002ARA&A...40..487}). 
For example, X-ray observations of the Galactic center region have revealed periods of enhanced accretion activity just a few hundred years ago (e.g., \citealt{2013A&A...558A..32C}). On larger scales, ionized gas emission points to a period of AGN activity several
million years ago \citep{2013ApJ...778..58B}. This result is independently supported by X-ray observations of hot gas \citep{2016ApJL...828..12N}. AGN activity on possibly even longer timescales is revealed by gamma-ray and ultraviolet spectroscopic studies of large-scale biconical plasma that exists above and below the Galactic plane.
In addition, the Galactic central region includes nuclear star clusters. These clusters exhibit ongoing episodes of star formation, resulting in a generally younger stellar population \citep{2016ARA&A...54..529}. As described above, the Galactic central region shows very intriguing phenomena.

GREX-PLUS can carry out imaging surveys in the Galactic central region 
in near-infrared and mid-infrared bands. This may allow us to find many intriguing new phenomena and/or celestial objects and clarify them in collaboration
with X-ray and radio observations.
Furthermore, as described below, GREX-PLUS can provide not only images of celestial objects in the Galactic central region but also astrometric information of stars, which is very important for elucidating the Galactic central region, including the Galactic nuclear bulge.

An example of astrometric observations with GREX-PLUS is presented. Observations are carried out at a wavelength of $2.2~\mu\mathrm{m}$. This wavelength is advantageous because it allows us to see through to the Galactic center, while it is also necessary that the wavelength is not too long in order to determine the centroid positions of stellar images; thus, $2.2~\mu\mathrm{m}$ is a suitable choice.
The survey region is defined as $-1.5^\circ < \ell < 1.5^\circ$ and $-0.3^\circ < b < 0.3^\circ$. This corresponds to a region large enough to cover the nuclear disk within the bulge, enabling comparison with observational data from JASMINE.
The mission duration is set to 7 days in spring and 7 days in autumn every year for 5 years. Alternatively, for a 3-year mission, 12 days of observations in both spring and autumn each year would also be acceptable.
For the ratio of PSF size to pixel size, in order to perform astrometric observations efficiently, it is required to satisfy $(\lambda / D) \cdot (f / w) \sim 1.5$--2.0,
where $\lambda$ is the wavelength, $D$ is the primary mirror diameter, $f$ is the focal length, and $w$ is the pixel size. In fact, at a wavelength of $2.2~\mu\mathrm{m}$, this condition is satisfied for GREX-PLUS.
For pointing stability, in order to ensure sufficient positional accuracy and to keep it smaller than the stellar image size, a requirement of $<100\,\mathrm{mas}/20\,\mathrm{s}$ is imposed. However, the currently proposed specification of the pointing stability is $100\,\mathrm{mas}/300\,\mathrm{s}$, and thus this requirement is satisfied.
In astrometry, thermal stability is important, and the thermal stability of the instrument should be controlled so that variations on the detector plane are kept below $1\,\mathrm{nm}/5\,\mathrm{hr}$. This depends on the telescope structure and other parameters, and the degree to which it can be satisfied requires detailed investigation. This condition refers to temporal variations in the centroids of stellar images on the focal plane.
In this observing case, an annual parallax accuracy of $25\,\mu\mathrm{as}$ can be achieved for stars brighter than $K<14$ mag. Of course, various observational configurations are possible depending on the survey region and mission duration.

\subsection{Scientific goals}

{\bf ~~~~~~\normalsize $\sim$ Infrared astrometry in the Galactic nuclear bulge $\sim$}

The key to deciphering the history of the central region of the Galaxy, which is about 26 thousand light-years ($\sim8$ kpc) away from us, is the positional distribution and movement of stars of various ages that still remain in the nuclear bulge. Stars born at different times in the history of the Galaxy's formation, like fossils, exist as witnesses of history in the present Galaxy. Information on the current positional distribution and motion of those stars reflects the history of the central region of the Galaxy.
Furthermore, the past of the nuclear bulge can lead to the elucidation of the whole Galactic history, such as determining the formation time of the outer long bar structure surrounding the nuclear bulge \citep{2020MN...492..4500}. The history of the formation and evolution of the Galactic center region can be clarified by precisely observing astrometric information, such as the positions and velocities of stars in the nuclear bulge, to also resolve the outer bar and bulge. 
We are located at the edge of the Galactic disk, and because the light from stars is blocked by the thick gas and dust between us and the Galactic center, the Galactic central region is a place where astronomical observation is difficult. Therefore, infrared astrometric measurements are necessary.
JASMINE is the first satellite mission in the world to perform high-precision astrometry from space with a stable stellar image captured by a space telescope using infrared wavelengths that are transparent to dust and gas, and JASMINE provides information on the positional distribution and movement of stars in the central region of the Galaxy, which are not yet understood \citep{2021ASPC...528..163G}. 
The mission objective of JASMINE is to use an optical telescope with a primary mirror aperture of around 36 cm to perform infrared astrometric observations (Hw band: $0.9$--1.6 $\mu$m). JASMINE can measure, at the highest precision, annual parallaxes at a precision of less than or equal to 25 $\mu$as and proper motions, or transverse angular velocities across the celestial sphere, at a precision of less than or equal to 25 $\mu$as/year in the direction of an area of a few square degrees of the Galactic nuclear bulge in order to create a catalog of the positions and movements of stars within the region. This mission will help to achieve revolutionary breakthroughs in astronomy, including the formation history of the Galactic nuclear bulge (Galactic Center Archeology); Galacto-seismology; the supermassive black hole at the Galactic Center; the gravitational field in the Galactic nuclear bulge; the activity around the Galactic Center; formation of star clusters; the orbital elements of X-ray binary stars and the identification of the compact object in an X-ray binary; the physics of fixed stars; star formation; planetary systems; and gravitational lensing. Such data will allow the compilation of a more meaningful catalog when combined with data from ground-based observations of the line-of-sight velocities and chemical compositions of stars in the bulge.
However, a relatively small telescope provides astrometric information only for bright stars (brighter than Hw$\sim$14.5 mag).
On the other hand, GREX-PLUS may provide astrometric information for stars including fainter stars while the precision of the astrometric measurements may not be very good if we use only data provided by GREX-PLUS\footnote{If GREX-PLUS satisfies the conditions of specifications of the instruments which are suggested by Table~\ref{tb:astrometry}, and also if it can carry out the survey of stellar images at the Galactic nuclear bulge region in the near-infrared band, then GREX-PLUS may have possibility to provide highly precise ($\sim25$ $\mu$as) astrometric measurements for stars brighter than $K\sim14$ mag.}. However, for stars observed by both JASMINE and GREX-PLUS, if JASMINE can provide highly accurate astrometric information for these stars, GREX-PLUS can use the astrometric information for these stars to obtain precise astrometric information even for fainter stars that JASMINE cannot measure. 
Furthermore, if it becomes possible to observe fainter and redder stars than those accessible with JASMINE, for example, the kinematics and history of the nuclear star cluster (NSC) in the Galactic center and the structure of the nuclear stellar disk (NSD), which are among the main science targets of JASMINE \citep{2024PASJ...76..386K,}, could be investigated more precisely. Such observations would reduce sample biases and enable more detailed information, such as the age structure, to be obtained.
This strengthens the scientific performance
of kinematic analysis at the Galactic nuclear bulge, in particular,
leading to investigations of new objects and new phenomena, such as
the discovery of intermediate-mass black holes and clarification of physical characteristics of ultra-light dark matter, etc.
In this way, many objects and phenomena that are scientifically interesting and important still remain for astrometric observations in the Galactic nuclear bulge, and GREX-PLUS has the possibility to contribute to this exploration.

\begin{table}
    \begin{center}
        \caption{Required observational parameters.}
        \label{tb:astrometry}
        \begin{tabular}{|l|l|l|} \hline
        &Requirements&Remarks \\ \hline
        Wavelength& $\sim 2.2~\mu {\rm m}$&a \\ \hline
        Observation region& $-1.5^{\circ}<{\ell}<1.5^{\circ}, -0.3^{\circ}<b<0.3^{\circ}$ &b \\ \hline
        Observation periods& 7 days in spring, 7 days in autumn every year for 5 years&c \\ \hline
        Ratio of PSF size to pixel size& $(\lambda/D)\cdot (f/w) \sim 1.5$--2.0 & d \\ \hline
        Pointing stability& $<100~{\rm mas}/20~{\rm s}$ & e \\ \hline
        Thermal~stability of instruments& $<1~{\rm nm}/5~{\rm hr}$& f \\ \hline
        \end{tabular}
    \end{center}
    ${}^a$ The K band is suitable for astrometric measurements of stars in the Galactic central region.\\
    ${}^b$ This region corresponds to the Galactic nuclear stellar disk region in the Galactic bulge.\\
    ${}^c$ For 3 years of operation instead of 5 years of operation, 12 days in spring and 12 days in autumn every year are necessary.\\
    ${}^d$ The required ratio of the PSF size to the pixel size is suitable for precise estimation of the centroids of stellar images $(\lambda:{\rm wavelength}, D:{\rm the~diameter~of~the~primary~mirror}, f:{\rm focal~length}, w:{\rm pixel ~size})$ The present proposed specification of GREX-PLUS for this ratio is 2 and thus this requirement is satisfied.\\
    ${}^e$ The present proposed specification of the pointing stability is 100 mas/300 s and thus this requirement is satisfied.\\
    ${}^f$ This condition is for temporal variations of the centroids of stellar images on the focal plane.
\end{table}

\printbibliography[heading=subbibliography]
\end{refsection}

\chapter{Synergy with Other Projects}
\label{chap:synergywithotherprojects}

\begin{refsection}[4_synergywithotherprojects/synergywithotherprojects.bib]

In this chapter, we will review expected synergies between GREX-PLUS and other telescope projects.


\subsection*{Subaru Telescope}

Wide and deep optical imaging survey fields produced with the Hyper Suprime-Cam (HSC; \citealt{2018PASJ...70S...1M}) on the Subaru Telescope are natural target fields for any infrared wide-field imaging survey, including GREX-PLUS.
This is a clear synergy case.
In the late 2020s, the Subaru Telescope will have a next-generation adaptive optics (AO) system, Ground-Layer AO (GLAO), that enables near diffraction-limited imaging quality across a wide area up to the $K$ band (i.e. wavelength 2.2 $\mu$m).
This will be achieved by the instrument called ULTIMATE-Subaru \citep{2022SPIE12185E..21M}.
The expected angular resolution of $0.2''$ in the $K$ band is several times better than that of GREX-PLUS.
For the scientific theme of EGS2, galaxy mass assembly, this higher spatial resolution is very useful for resolving the internal structure of galaxies and examining the stellar mass assembly process.
Therefore, it is another excellent synergy case to make follow-up observations of galaxies detected in GREX-PLUS imaging surveys using ULTIMATE-Subaru to spatially resolve their internal structure.

\subsection*{Rubin Observatory}

The Legacy Survey of Space and Time (LSST) released its first image on June 23, 2025 and will obtain a ``movie'' of about 20,000 square degrees in the southern sky over 10 years.
The imaging depth will be as deep as 27.5 AB to 24.9 AB in $u,g,r,i,z,y$ after the 10-year survey.
These wide-area and deep optical images will be the best fields for near-infrared imaging surveys including GREX-PLUS.
This is a natural synergistic point between Rubin Observatory and GREX-PLUS.

\subsection*{Extremely Large Telescopes}

The next-generation ground-based 30--40 m extremely large telescopes (ELTs) will have unprecedented sensitivity and angular resolution in near-infrared wavelengths with AO systems.
On the other hand, the field-of-view is too small to make any wide-field survey and ELTs themselves cannot find interesting objects to be observed in great detail.
Therefore, ELTs inevitably need other telescopes that provide excellent targets to be followed up with ELTs.
GREX-PLUS is one of such target providers for ELTs.
Since GREX-PLUS wide-field imaging surveys are the only $>1$ degree-scale surveys in the 2--8 $\mu$m wavelength range in the coming decades, the unique role of GREX-PLUS as a target provider for ELTs stands unrivaled.
The spectroscopic capability of ELTs is also extremely high and can reach spectral resolutions $>100,000$.
However, the atmospheric absorption is unavoidable and the sensitivity and wavelength coverage are limited at wavelengths longer than 2.5 $\mu$m.
Therefore, the GREX-PLUS high-resolution spectroscopic capability is very unique even compared to ELTs.

\subsection*{JWST}

James Webb Space Telescope (JWST) was successfully launched on December 25, 2021 and has started its science observations in 2022.
Thanks to the highly successful launch and cruise to the Sun-Earth L2, fuel consumption was minimal, and the fuel is expected to last over 20 years.
JWST observes in the 0.6--28 $\mu$m wavelength range, which includes that of GREX-PLUS.
The primary mirror aperture of 6.5 m is overwhelmingly large as a space telescope, yielding ultra-high sensitivity and high spatial resolution.
On the other hand, although its field-of-view is larger than that of the Hubble Space Telescope, it is limited to less than 10 arcmin$^2$.
Hence, it is difficult to conduct square-degree-scale imaging surveys.
Since GREX-PLUS will achieve such ultra-wide-field imaging surveys, they are complementary to each other.
Follow-up observations using JWST for unique and interesting objects detected in GREX-PLUS surveys provide straightforward synergy.
The wavelength resolution of the JWST spectrographs is limited to 3,000 at most, which is 10 times coarser than that of GREX-PLUS.
For example, JWST cannot resolve the Keplerian motion of water molecules around the ``snowline'', while it can detect the water 18 $\mu$m line in protoplanetary disks very easily.
Therefore, we can select the disks to be targeted with GREX-PLUS after JWST water-line detections, which constitutes another very nice synergy case between GREX-PLUS and JWST.

\subsection*{Euclid}

Euclid is a wide-field survey space telescope developed by the European Space Agency (ESA) and was successfully launched on July 1, 2023.
The primary mirror aperture is 1.2 m, similar to that of GREX-PLUS.
It has a field-of-view of approximately 2,000 arcmin$^2$, which is also similar to that of GREX-PLUS.
The observed wavelength range extends from optical wavelengths to the near-infrared at 2 $\mu$m, and there is no overlap with GREX-PLUS. 
It also has a wide-field spectroscopic capability, but with a low wavelength resolution of 250.
The main scientific goal is to resolve the origin of the accelerated expansion of the Universe by measuring gravitational weak-lensing effects and baryonic acoustic oscillations, and to conduct an ultra-wide-field survey of up to 15,000 square degrees.
The ultra-wide imaging survey will be shallow (24 AB), and Euclid will also conduct a narrower survey (40 square degrees) to reach a deeper depth (26 AB).
Compared to GREX-PLUS, there is a difference in the combination of the width and depth of the surveys, in addition to the difference in the wavelength band.
As synergy between GREX-PLUS and Euclid, GREX-PLUS survey fields will be selected from among Euclid survey fields, especially from its Deep surveys, to have the imaging data at wavelengths shorter than 2 $\mu$m.
The GREX-PLUS imaging data at wavelengths longer than 2 $\mu$m are also extremely useful to trace the longer wavelength radiation from objects detected in the Euclid imaging data.

\subsection*{Roman}

Nancy Grace Roman Space Telescope (hereafter Roman) is NASA's next-generation flagship space telescope, scheduled for launch in September 2026. 
The primary mirror aperture is 2.4 m, more than twice as large as that of GREX-PLUS, and its observational wavelength range is 0.5--2.3 $\mu$m. 
GREX-PLUS covers the wavelength range of 2 $\mu$m or longer, which is complementary to Roman's wavelength range. 
In particular, the High Latitude Wide Area Survey (HLWAS) conducted by Roman provides a deep survey at wavelengths shorter than 2 $\mu$m.
Among the planned surveys, the Deep tier will have an area of 19.2 square degrees with depths of 27.7 AB in $ZY$, 27.6 AB in $J$, 27.5 AB in $H$, 27.0 AB in $F$ and 25.9 AB in $K$ bands \citep{2025arXiv250510574O}, which will be the best target fields for the Deep survey of GREX-PLUS.
The Medium tier of Roman HLWAS will have a super-wide area of 2415 square degrees with depths of 26.5 AB in $Y$ and 26.4 AB in $JH$ \citep{2025arXiv250510574O}, which will be the basis of the GREX-PLUS Medium and Wide surveys.
Roman has a prism and grism spectrograph. 
The wavelength resolution is low, which is completely different from the GREX-PLUS high-resolution spectrometer in terms of both wavelength band and wavelength resolution.
Roman will also conduct microlensing surveys near the Galactic center, which is very powerful for identifying exoplanets with orbital radii of 1 au or larger.
Roman also has a Coronagraph Instrument to realize direct observations of exoplanets, including polarimetric spectroscopy as well as imaging.

Extremely high synergy between Roman and GREX-PLUS is expected in a wide range of scientific themes.
In fact, most of the science themes with the GREX-PLUS wide-field camera assume wide-field deep imaging data at wavelengths shorter than 2 $\mu$m by Roman.
Therefore, GREX-PLUS imaging surveys must be conducted in the Roman imaging survey fields.
There is also potential synergy in observations of the Galaxy center (Section~\ref{sec:galaxycenter}) between Roman microlensing surveys and GREX-PLUS.
Synergy between the Roman Coronagraph Instrument for direct observations of exoplanets and GREX-PLUS high-resolution spectrometer observations is also possible and under investigation.

\subsection*{JASMINE}

Japan Astrometry Satellite Mission for INfrared Exploration (JASMINE; \citealt{2021ASPC..528..163G}) is scheduled for launch in 2032 by an Epsilon Launch Vehicle of ISAS/JAXA.
With a primary mirror of 36 cm, JASMINE will conduct astrometric observations in the near-infrared wavelength range of 1.0--1.6 $\mu$m.
The target field is the Galactic nuclear region to understand the nuclear bulge formation history in the Milky Way.
JASMINE will also conduct exoplanet transit observations when the Galaxy center is not visible from the orbit.
As described in Section~\ref{sec:galaxycenter}, GREX-PLUS can realize Galaxy center observations at wavelengths longer than those of JASMINE.
Since the dust extinction is less severe at the wavelengths covered by GREX-PLUS than at those covered by JASMINE, the observations will be highly complementary.
For the astrometric measurements in GREX-PLUS imaging data, it will be necessary to calibrate the astrometry with JASMINE data which have higher astrometric accuracy.
There is another synergy in exoplanet observations.
JASMINE will discover many transit exoplanet systems, especially, Earth-type planets in the habitable zone around M-type dwarf stars.
These exoplanets will be excellent follow-up targets with the GREX-PLUS high-resolution spectrometer to examine rare molecules in planetary atmospheres.

\subsection*{HiZ-GUNDAM}

High-z Gamma-ray bursts for Unraveling the Dark Ages Mission (HiZ-GUNDAM; \citealt{2020SPIE11444E..2ZY}) is a mission candidate for launch in the 2030s by an Epsilon Launch Vehicle of ISAS/JAXA.
The scientific aim is to explore the very early Universe by using gamma-ray bursts (GRBs).
To do so, HiZ-GUNDAM will carry wide-field gamma-ray detectors and an optical-to-near-infrared small telescope for the immediate follow-up of afterglows of GRBs.
The infrared wavelength coverage of the telescope is up to 2.5 $\mu$m.
The primary mirror aperture is 30 cm.
While the scientific theme is partly overlapping with those of the GREX-PLUS wide-field camera, the sensitivity (i.e. mirror aperture) and the wavelength coverage are very different.
Therefore, HiZ-GUNDAM cannot achieve the scientific goals of GREX-PLUS.
At the same time, GREX-PLUS cannot replace HiZ-GUNDAM because GREX-PLUS does not carry any gamma-ray detectors.
On the other hand, if HiZ-GUNDAM and GREX-PLUS are in orbit simultaneously, the most distant GRBs found by HiZ-GUNDAM will be excellent follow-up targets for GREX-PLUS.




%

\subsection*{ALMA}

GREX-PLUS also has strong synergy with world-leading radio telescopes, for example, 
the Atacama Large Millimeter/submillimeter Array (ALMA).
It has unprecedentedly high sensitivity, high angular resolution, and high spectral resolution in a wide wavelength range from 300 $\mu$m to 3.6 mm.
Since its field-of-view is very tiny, however, ALMA generally needs targets to be observed {\it a priori}.
GREX-PLUS wide-field surveys will supply a number of excellent targets to ALMA.
Especially, ALMA follow-up observations of [O~{\sc iii}] 88 $\mu$m and [C~{\sc ii}] 158 $\mu$m lines for very high-$z$ galaxy candidates detected with GREX-PLUS are a natural synergy.
Another excellent synergy is dust continuum observations of these high-$z$ galaxies with ALMA.
GREX-PLUS high-resolution spectroscopy also has strong synergy with ALMA in molecular chemistry in the interstellar medium, star-forming regions, and protoplanetary disks.

By 2030, ALMA will enhance its capability by broadening the receiver bandwidth by a factor of two.
This development will also enhance the synergy with GREX-PLUS considerably.
For example, the speed of line searches for distant galaxy candidates will simply become faster by a factor of two, and the continuum sensitivity will also improve.



\printbibliography[heading=subbibliography]

@ARTICLE{2024PASJ...76..386K,
       author = {{Kawata}, Daisuke and {Kawahara}, Hajime and {Gouda}, Naoteru and {Secrest}, Nathan J. and {Kano}, Ryouhei and {Kataza}, Hirokazu and {Isobe}, Naoki and {Ohsawa}, Ryou and {Usui}, Fumihiko and {Yamada}, Yoshiyuki and {Graham}, Alister W. and {Pettitt}, Alex R. and {Asada}, Hideki and {Baba}, Junichi and {Bekki}, Kenji and {Dorland}, Bryan N. and {Fujii}, Michiko and {Fukui}, Akihiko and {Hattori}, Kohei and {Hirano}, Teruyuki and {Kamizuka}, Takafumi and {Kashima}, Shingo and {Kawanaka}, Norita and {Kawashima}, Yui and {Klioner}, Sergei A. and {Kodama}, Takanori and {Koshimoto}, Naoki and {Kotani}, Takayuki and {Kuzuhara}, Masayuki and {Levine}, Stephen E. and {Majewski}, Steven R. and {Masuda}, Kento and {Matsunaga}, Noriyuki and {Miyakawa}, Kohei and {Miyoshi}, Makoko and {Morihana}, Kumiko and {Nishi}, Ryoichi and {Notsu}, Yuta and {Omiya}, Masashi and {Sanders}, Jason and {Tanikawa}, Ataru and {Tsujimoto}, Masahiro and {Yano}, Taihei and {Aizawa}, Masataka and {Arimatsu}, Ko and {Biermann}, Michael and {Boehm}, Celine and {Chiba}, Masashi and {Debattista}, Victor P. and {Gerhard}, Ortwin and {Hirabayashi}, Masayuki and {Hobbs}, David and {Ikenoue}, Bungo and {Izumiura}, Hideyuki and {Jordi}, Carme and {Kohara}, Naoki and {L{\"o}ffler}, Wolfgang and {Luri}, Xavier and {Mase}, Ichiro and {Miglio}, Andrea and {Mitsuda}, Kazuhisa and {Newswander}, Trent and {Nishiyama}, Shogo and {Obuchi}, Yoshiyuki and {Ootsubo}, Takafumi and {Ouchi}, Masami and {Ozaki}, Masanobu and {Perryman}, Michael and {Prusti}, Timo and {Ramos}, Pau and {Read}, Justin I. and {Rich}, R. Michael and {Sch{\"o}nrich}, Ralph and {Shikauchi}, Minori and {Shimizu}, Risa and {Suematsu}, Yoshinori and {Tada}, Shotaro and {Takahashi}, Aoi and {Tatekawa}, Takayuki and {Tatsumi}, Daisuke and {Tsujimoto}, Takuji and {Tsuzuki}, Toshihiro and {Urakawa}, Seitaro and {Uraguchi}, Fumihiro and {Utsunomiya}, Shin and {Van Eylen}, Vincent and {van Leeuwen}, Floor and {Wada}, Takehiko and {Walton}, Nicholas A.},
        title = "{JASMINE: Near-infrared astrometry and time-series photometry science}",
      journal = {\pasj},
         year = 2024,
        month = jun,
       volume = {76},
       number = {3},
        pages = {386-425},
          doi = {10.1093/pasj/psae020},
archivePrefix = {arXiv},
       eprint = {2307.05666},
 primaryClass = {astro-ph.IM},
       adsurl = {https://ui.adsabs.harvard.edu/abs/2024PASJ...76..386K}
}

@INPROCEEDINGS{2018cwla.conf...49M,
       author = {{Honda}, Mitsuhiko and {Packham}, Chris and {Chun}, M. and {Imanishi}, Masatoshi and {Ichikawa}, Kohei and {Marois}, C. and {Birkby}, J. and {Crossfield}, I. and {Herczeg}, G. and {Greathouse}, T.~K. and {Richter}, M. and {Sakon}, Itsuki and {Katazaokamoto}, Yoshiko and {Kataza}, Hirokazu and {MICHI Science and Instrument Team}},
        title = "{TMT/MICHI current concept}",
    booktitle = {The Cosmic Wheel and the Legacy of the AKARI Archive: From Galaxies and Stars to Planets and Life},
         year = 2018,
       editor = {{Ootsubo}, Takafumi and {Yamamura}, Issei and {Murata}, Kazumi and {Onaka}, Takashi},
        month = mar,
        pages = {49-52},
       adsurl = {https://ui.adsabs.harvard.edu/abs/2018cwla.conf...49M}
}

@INPROCEEDINGS{2012SPIE.8442E..1AY,
       author = {{Yamada}, Toru and {Iwata}, Ikuru and {Ando}, Makiko and {Doi}, Mamoru and {Goto}, Tomotsugu and {Ikeda}, Yuji and {Imanishi}, Masatoshi and {Inoue}, Akio K. and {Iwamura}, Satoru and {Kawai}, Nobuyuki and {Kobayashi}, Masakazu A.~R. and {Kodama}, Tadayuki and {Komiyama}, Yutaka and {Kubo}, Mariko and {Matsuhara}, Hideo and {Mawatari}, Ken and {Matsuoka}, Yoshiki and {Morokuma}, Tomoki and {Nakaya}, Hidehiko and {Ohta}, Koji and {Okamoto}, Atsushi and {Oyabu}, Shinki and {Sato}, Yohichi and {Sugita}, Hiroyuki and {Tsutsui}, Ryo and {Tokoku}, Chihiro and {Toshikawa}, Jun and {Tsuneta}, Saku and {Wada}, Takehiko and {Yabe}, Kiyoto and {Yasuda}, Naoki and {Yonetoku}, Daisuke},
        title = "{WISH for deep and wide NIR surveys}",
    booktitle = {Space Telescopes and Instrumentation 2012: Optical, Infrared, and Millimeter Wave},
         year = 2012,
       editor = {{Clampin}, Mark C. and {Fazio}, Giovanni G. and {MacEwen}, Howard A. and {Oschmann}, Jacobus M., Jr.},
       series = {Society of Photo-Optical Instrumentation Engineers (SPIE) Conference Series},
       volume = {8442},
        month = sep,
          eid = {84421A},
        pages = {84421A},
          doi = {10.1117/12.925632},
       adsurl = {https://ui.adsabs.harvard.edu/abs/2012SPIE.8442E..1AY}
}

@ARTICLE{2018PASA...35...30R,
       author = {{Roelfsema}, P.~R. and {Shibai}, H. and {Armus}, L. and {Arrazola}, D. and {Audard}, M. and {Audley}, M.~D. and {Bradford}, C.~M. and {Charles}, I. and {Dieleman}, P. and {Doi}, Y. and {Duband}, L. and {Eggens}, M. and {Evers}, J. and {Funaki}, I. and {Gao}, J.~R. and {Giard}, M. and {di Giorgio}, A. and {Gonz{\'a}lez Fern{\'a}ndez}, L.~M. and {Griffin}, M. and {Helmich}, F.~P. and {Hijmering}, R. and {Huisman}, R. and {Ishihara}, D. and {Isobe}, N. and {Jackson}, B. and {Jacobs}, H. and {Jellema}, W. and {Kamp}, I. and {Kaneda}, H. and {Kawada}, M. and {Kemper}, F. and {Kerschbaum}, F. and {Khosropanah}, P. and {Kohno}, K. and {Kooijman}, P.~P. and {Krause}, O. and {van der Kuur}, J. and {Kwon}, J. and {Laauwen}, W.~M. and {de Lange}, G. and {Larsson}, B. and {van Loon}, D. and {Madden}, S.~C. and {Matsuhara}, H. and {Najarro}, F. and {Nakagawa}, T. and {Naylor}, D. and {Ogawa}, H. and {Onaka}, T. and {Oyabu}, S. and {Poglitsch}, A. and {Reveret}, V. and {Rodriguez}, L. and {Spinoglio}, L. and {Sakon}, I. and {Sato}, Y. and {Shinozaki}, K. and {Shipman}, R. and {Sugita}, H. and {Suzuki}, T. and {van der Tak}, F.~F.~S. and {Torres Redondo}, J. and {Wada}, T. and {Wang}, S.~Y. and {Wafelbakker}, C.~K. and {van Weers}, H. and {Withington}, S. and {Vandenbussche}, B. and {Yamada}, T. and {Yamamura}, I.},
        title = "{SPICA-A Large Cryogenic Infrared Space Telescope: Unveiling the Obscured Universe}",
      journal = {Publications of the Astronomical Society of Australia},
         year = 2018,
        month = aug,
       volume = {35},
          eid = {e030},
        pages = {e030},
          doi = {10.1017/pasa.2018.15},
archivePrefix = {arXiv},
       eprint = {1803.10438},
 primaryClass = {astro-ph.IM},
       adsurl = {https://ui.adsabs.harvard.edu/abs/2018PASA...35...30R}
}

@ARTICLE{2020MNRAS.493.2059B,
       author = {{Bowler}, R.~A.~A. and {Jarvis}, M.~J. and {Dunlop}, J.~S. and {McLure}, R.~J. and {McLeod}, D.~J. and {Adams}, N.~J. and {Milvang-Jensen}, B. and {McCracken}, H.~J.},
        title = {A lack of evolution in the very bright end of the galaxy luminosity function from $z=8$ to 10},
      journal = {Monthly Notices of the Royal Astronomical Society},
         year = 2020,
        month = apr,
       volume = {493},
       number = {2},
        pages = {2059-2084}
}

@ARTICLE{GopiraRoadmap2025, 
	author = "{GOPIRA Roadmap 2025 Creation Committee}", 
	title  = "GOPIRA Optical and Infrared Astronomy Roadmap 2025", 
	journal = "\url{http://gopira.jp/siryo/GOPIRA_Roadmap_2025.pdf}",
	year   = 2026	
}

@ARTICLE{york00,
       author = {{York}, Donald G. and {Adelman}, J. and {Anderson}, Jr., John E. and {Anderson}, Scott F. and {Annis}, James and {Bahcall}, Neta A. and {Bakken}, J.~A. and {Barkhouser}, Robert and {Bastian}, Steven and {Berman}, Eileen and et al.},
        title = "{The Sloan Digital Sky Survey: Technical Summary}",
      journal = {\aj},
         year = 2000,
        month = sep,
       volume = {120},
       number = {3},
        pages = {1579-1587},
          doi = {10.1086/301513},
archivePrefix = {arXiv},
       eprint = {astro-ph/0006396},
 primaryClass = {astro-ph},
       adsurl = {https://ui.adsabs.harvard.edu/abs/2000AJ....120.1579Y}
}

@ARTICLE{sdss_DR7,
       author = {{Abazajian}, Kevork N. and {Adelman-McCarthy}, Jennifer K. and {Ag{\"u}eros}, Marcel A. and {Allam}, Sahar S. and {Allende Prieto}, Carlos and {An}, Deokkeun and {Anderson}, Kurt S.~J. and {Anderson}, Scott F. and {Annis}, James and {Bahcall}, Neta A. and et al.},
        title = "{The Seventh Data Release of the Sloan Digital Sky Survey}",
      journal = {\apjs},
         year = 2009,
        month = jun,
       volume = {182},
       number = {2},
        pages = {543-558},
          doi = {10.1088/0067-0049/182/2/543},
archivePrefix = {arXiv},
       eprint = {0812.0649},
 primaryClass = {astro-ph},
       adsurl = {https://ui.adsabs.harvard.edu/abs/2009ApJS..182..543A}
}

@article{2023A&A...677A.184W,
	adsurl = {https://ui.adsabs.harvard.edu/abs/2023A&A...677A.184W},
	archiveprefix = {arXiv},
	author = {{Weaver}, J.~R. and {Davidzon}, I. and {Toft}, S. and {Ilbert}, O. and {McCracken}, H.~J. and {Gould}, K.~M.~L. and {Jespersen}, C.~K. and {Steinhardt}, C. and {Lagos}, C.~D.~P. and {Capak}, P.~L. and et al.},
	doi = {10.1051/0004-6361/202245581},
	eid = {A184},
	eprint = {2212.02512},
	journal = {\aap},
	month = sep,
	pages = {A184},
	primaryclass = {astro-ph.GA},
	title = {{COSMOS2020: The galaxy stellar mass function. The assembly and star formation cessation of galaxies at 0.2< z {\ensuremath{\leq}} 7.5}},
	volume = {677},
	year = 2023}

@ARTICLE{2025A&A...703A...5M,
       author = {{Mattolini}, D. and {Zibetti}, S. and {Gallazzi}, A.~R. and {Scholz-D{\'\i}az}, L. and {Pratesi}, J.},
        title = "{Re-assessing the stellar population scaling relations of the galaxies in the Local Universe}",
      journal = {\aap},
         year = 2025,
        month = oct,
       volume = {703},
          eid = {A5},
        pages = {A5},
          doi = {10.1051/0004-6361/202554972},
archivePrefix = {arXiv},
       eprint = {2509.04570},
 primaryClass = {astro-ph.GA},
       adsurl = {https://ui.adsabs.harvard.edu/abs/2025A&A...703A...5M}
}

@ARTICLE{2004ApJ...613..898T,
       author = {{Tremonti}, Christy A. and {Heckman}, Timothy M. and {Kauffmann}, Guinevere and {Brinchmann}, Jarle and {Charlot}, St{\'e}phane and {White}, Simon D.~M. and {Seibert}, Mark and {Peng}, Eric W. and {Schlegel}, David J. and {Uomoto}, Alan and et al.},
        title = "{The Origin of the Mass-Metallicity Relation: Insights from 53,000 Star-forming Galaxies in the Sloan Digital Sky Survey}",
      journal = {\apj},
         year = 2004,
        month = oct,
       volume = {613},
       number = {2},
        pages = {898-913},
          doi = {10.1086/423264},
archivePrefix = {arXiv},
       eprint = {astro-ph/0405537},
 primaryClass = {astro-ph},
       adsurl = {https://ui.adsabs.harvard.edu/abs/2004ApJ...613..898T}
}

@ARTICLE{2022MNRAS.512.1415Z,
       author = {{Zibetti}, Stefano and {Gallazzi}, Anna R.},
        title = "{Stellar mass as the 'glocal' driver of galaxies' stellar population properties}",
      journal = {\mnras},
         year = 2022,
        month = may,
       volume = {512},
       number = {1},
        pages = {1415-1429},
          doi = {10.1093/mnras/stac370},
archivePrefix = {arXiv},
       eprint = {2202.03975},
 primaryClass = {astro-ph.GA},
       adsurl = {https://ui.adsabs.harvard.edu/abs/2022MNRAS.512.1415Z}
}

@ARTICLE{2009MNRAS.400.1181Z,
       author = {{Zibetti}, Stefano and {Charlot}, St{\'e}phane and {Rix}, Hans-Walter},
        title = "{Resolved stellar mass maps of galaxies - I. Method and implications for global mass estimates}",
      journal = {\mnras},
         year = 2009,
        month = dec,
       volume = {400},
       number = {3},
        pages = {1181-1198},
          doi = {10.1111/j.1365-2966.2009.15528.x},
archivePrefix = {arXiv},
       eprint = {0904.4252},
 primaryClass = {astro-ph.CO},
       adsurl = {https://ui.adsabs.harvard.edu/abs/2009MNRAS.400.1181Z}
}

@ARTICLE{2001ApJ...550..212B,
       author = {{Bell}, Eric F. and {de Jong}, Roelof S.},
        title = "{Stellar Mass-to-Light Ratios and the Tully-Fisher Relation}",
      journal = {\apj},
         year = 2001,
        month = mar,
       volume = {550},
       number = {1},
        pages = {212-229},
          doi = {10.1086/319728},
archivePrefix = {arXiv},
       eprint = {astro-ph/0011493},
 primaryClass = {astro-ph},
       adsurl = {https://ui.adsabs.harvard.edu/abs/2001ApJ...550..212B}
}

@ARTICLE{2012ApJ...744...17M,
       author = {{Meidt}, Sharon E. and {Schinnerer}, Eva and {Knapen}, Johan H. and {Bosma}, Albert and {Athanassoula}, E. and {Sheth}, Kartik and {Buta}, Ronald J. and {Zaritsky}, Dennis and {Laurikainen}, Eija and {Elmegreen}, Debra and et al.},
        title = "{Reconstructing the Stellar Mass Distributions of Galaxies Using S$^{4}$G IRAC 3.6 and 4.5 {\ensuremath{\mu}}m Images. I. Correcting for Contamination by Polycyclic Aromatic Hydrocarbons, Hot Dust, and Intermediate-age Stars}",
      journal = {\apj},
         year = 2012,
        month = jan,
       volume = {744},
       number = {1},
          eid = {17},
        pages = {17},
          doi = {10.1088/0004-637X/744/1/17},
archivePrefix = {arXiv},
       eprint = {1110.2683},
 primaryClass = {astro-ph.CO},
       adsurl = {https://ui.adsabs.harvard.edu/abs/2012ApJ...744...17M}
}

@ARTICLE{2011MNRAS.417..812Z,
       author = {{Zibetti}, Stefano and {Groves}, Brent},
        title = "{Resolved optical-infrared spectral energy distributions of galaxies: universal relations and their break-down on local scales}",
      journal = {\mnras},
         year = 2011,
        month = oct,
       volume = {417},
       number = {2},
        pages = {812-834},
          doi = {10.1111/j.1365-2966.2011.19286.x},
archivePrefix = {arXiv},
       eprint = {1106.2165},
 primaryClass = {astro-ph.CO},
       adsurl = {https://ui.adsabs.harvard.edu/abs/2011MNRAS.417..812Z}
}

@ARTICLE{2008MNRAS.388.1595D,
       author = {{da Cunha}, Elisabete and {Charlot}, St{\'e}phane and {Elbaz}, David},
        title = "{A simple model to interpret the ultraviolet, optical and infrared emission from galaxies}",
      journal = {\mnras},
         year = 2008,
        month = aug,
       volume = {388},
       number = {4},
        pages = {1595-1617},
          doi = {10.1111/j.1365-2966.2008.13535.x},
archivePrefix = {arXiv},
       eprint = {0806.1020},
 primaryClass = {astro-ph},
       adsurl = {https://ui.adsabs.harvard.edu/abs/2008MNRAS.388.1595D}
}

@ARTICLE{2005MNRAS.362..799M,
       author = {{Maraston}, Claudia},
        title = "{Evolutionary population synthesis: models, analysis of the ingredients and application to high-z galaxies}",
      journal = {\mnras},
         year = 2005,
        month = sep,
       volume = {362},
       number = {3},
        pages = {799-825},
          doi = {10.1111/j.1365-2966.2005.09270.x},
archivePrefix = {arXiv},
       eprint = {astro-ph/0410207},
 primaryClass = {astro-ph},
       adsurl = {https://ui.adsabs.harvard.edu/abs/2005MNRAS.362..799M}
}

@ARTICLE{2010ApJ...722L..64K,
       author = {{Kriek}, Mariska and {Labb{\'e}}, Ivo and {Conroy}, Charlie and {Whitaker}, Katherine E. and {van Dokkum}, Pieter G. and {Brammer}, Gabriel B. and {Franx}, Marijn and {Illingworth}, Garth D. and {Marchesini}, Danilo and {Muzzin}, Adam and et al.},
        title = "{The Spectral Energy Distribution of Post-starburst Galaxies in the NEWFIRM Medium-band Survey: A Low Contribution from TP-AGB Stars}",
      journal = {\apjl},
         year = 2010,
        month = oct,
       volume = {722},
       number = {1},
        pages = {L64-L69},
          doi = {10.1088/2041-8205/722/1/L64},
archivePrefix = {arXiv},
       eprint = {1008.4357},
 primaryClass = {astro-ph.CO},
       adsurl = {https://ui.adsabs.harvard.edu/abs/2010ApJ...722L..64K}
}

@ARTICLE{2013MNRAS.428.1479Z,
       author = {{Zibetti}, Stefano and {Gallazzi}, Anna and {Charlot}, St{\'e}phane and {Pierini}, Daniele and {Pasquali}, Anna},
        title = "{Near-infrared spectroscopy of post-starburst galaxies: a limited impact of TP-AGB stars on galaxy spectral energy distributions}",
      journal = {\mnras},
         year = 2013,
        month = jan,
       volume = {428},
       number = {2},
        pages = {1479-1497},
          doi = {10.1093/mnras/sts126},
archivePrefix = {arXiv},
       eprint = {1205.4717},
 primaryClass = {astro-ph.CO},
       adsurl = {https://ui.adsabs.harvard.edu/abs/2013MNRAS.428.1479Z}
}

@ARTICLE{2007ApJ...666..870C,
       author = {{Calzetti}, D. and {Kennicutt}, R.~C. and {Engelbracht}, C.~W. and {Leitherer}, C. and {Draine}, B.~T. and {Kewley}, L. and {Moustakas}, J. and {Sosey}, M. and {Dale}, D.~A. and {Gordon}, K.~D. and et al.},
        title = "{The Calibration of Mid-Infrared Star Formation Rate Indicators}",
      journal = {\apj},
         year = 2007,
        month = sep,
       volume = {666},
       number = {2},
        pages = {870-895},
          doi = {10.1086/520082},
archivePrefix = {arXiv},
       eprint = {0705.3377},
 primaryClass = {astro-ph},
       adsurl = {https://ui.adsabs.harvard.edu/abs/2007ApJ...666..870C}
}

@ARTICLE{2010MNRAS.407..163F,
       author = {{Foyle}, Kelly and {Rix}, Hans-Walter and {Zibetti}, Stefano},
        title = "{An observational estimate for the mean secular evolution rate in spiral galaxies}",
      journal = {\mnras},
         year = 2010,
        month = sep,
       volume = {407},
       number = {1},
        pages = {163-180},
          doi = {10.1111/j.1365-2966.2010.16931.x},
archivePrefix = {arXiv},
       eprint = {0911.2231},
 primaryClass = {astro-ph.GA},
       adsurl = {https://ui.adsabs.harvard.edu/abs/2010MNRAS.407..163F}
}

@ARTICLE{2026arXiv260104156C,
       author = {{Coulter}, David A. and {Larison}, Conor and {Pierel}, Justin D.~R. and {Fujimoto}, Seiji and {Kokorev}, Vasily and {Allingham}, Joseph F.~V. and {Moriya}, Takashi J. and {Siebert}, Matthew and {Asada}, Yoshihisa and {Bezanson}, Rachel and {Brada{\v{c}}}, Maru{\v{s}}a and {Brammer}, Gabriel and {Chisholm}, John and {Coe}, Dan and {Dayal}, Pratika and {Engesser}, Michael and {Finkelstein}, Steven L. and {Fox}, Ori D. and {Furtak}, Lukas J. and {Koekemoer}, Anton M. and {Moore}, Thomas and {Nakane}, Minami and {Ouchi}, Masami and {Pan}, Richard and {Quimby}, Robert and {Rest}, Armin and {Richard}, Johan and {Robbins}, Luke and {Strolger}, Louis-Gregory and {Sun}, Fengwu and {Treu}, Tommaso and {Yanagisawa}, Hiroto and {Abdurro'uf} and {Agrawal}, Aadya and {Amor{\'\i}n}, Ricardo and {Anderson}, Joseph P. and {Angulo}, Rodrigo and {Atek}, Hakim and {Bauer}, Franz E. and {Bradley}, Larry D. and {Bromm}, Volker and {Bronikowski}, Mateusz and {Conselice}, Christopher J. and {DeCoursey}, Christa and {DerKacy}, James M. and {Desprez}, Guillaume and {Dhawan}, Suhail and {Diego}, Jose M. and {Egami}, Eiichi and {Faisst}, Andreas and {Frye}, Brenda and {Gomez}, Sebastian and {Gonz{\'a}lez-Otero}, Mauro and {Griggio}, Massimo and {Harikane}, Yuichi and {Inayoshi}, Kohei and {Jha}, Saurabh W. and {Jim{\'e}nez-Teja}, Yolanda and {Kartaltepe}, Jeyhan S. and {Kelly}, Patrick L. and {Kwok}, Lindsey A. and {Lane}, Zachary G. and {Li}, Xiaolong and {Lobbe}, Ivo and {Lucas}, Ray A. and {Magdis}, Georgios E. and {Martis}, Nicholas S. and {Matthee}, Jorryt and {Meena}, Ashish K. and {Naidu}, Rohan P. and {Noirot}, Ga{\"e}l and {Oguri}, Masamune and {Padilla Gonzalez}, Estefania and {Pascale}, Massimo and {Petrushevska}, Tanja and {Ricotti}, Massimo and {Schaerer}, Daniel and {Schuldt}, Stefan and {Shahbandeh}, Melissa and {Sheu}, William and {Shukawa}, Koji and {Tsujita}, Akiyoshi and {Vanzella}, Eros and {Wang}, Qinan and {Weaver}, John and {Windhorst}, Rogier and {Xu}, Yi and {Zenati}, Yossef and {Zitrin}, Adi},
        title = "{A spectroscopically confirmed, strongly lensed, metal-poor Type II supernova at z = 5.13}",
      journal = {arXiv e-prints},
         year = 2026,
        month = jan,
          eid = {arXiv:2601.04156},
        pages = {arXiv:2601.04156},
          doi = {10.48550/arXiv.2601.04156},
archivePrefix = {arXiv},
       eprint = {2601.04156},
 primaryClass = {astro-ph.HE},
       adsurl = {https://ui.adsabs.harvard.edu/abs/2026arXiv260104156C}
}

@ARTICLE{2022ApJ...940L...1W,
       author = {{Welch}, Brian and {Coe}, Dan and {Zackrisson}, Erik and {de Mink}, S.~E. and {Ravindranath}, Swara and {Anderson}, Jay and {Brammer}, Gabriel and {Bradley}, Larry and {Yoon}, Jinmi and {Kelly}, Patrick and {Diego}, Jose M. and {Windhorst}, Rogier and {Zitrin}, Adi and {Dimauro}, Paola and {Jim{\'e}nez-Teja}, Yolanda and {Abdurro'uf} and {Nonino}, Mario and {Acebron}, Ana and {Andrade-Santos}, Felipe and {Avila}, Roberto J. and {Bayliss}, Matthew B. and {Ben{\'\i}tez}, Alex and {Broadhurst}, Tom and {Bhatawdekar}, Rachana and {Brada{\v{c}}}, Maru{\v{s}}a and {Caminha}, Gabriel B. and {Chen}, Wenlei and {Eldridge}, Jan and {Farag}, Ebraheem and {Florian}, Michael and {Frye}, Brenda and {Fujimoto}, Seiji and {Gomez}, Sebastian and {Henry}, Alaina and {Hsiao}, Tiger Y. -Y. and {Hutchison}, Taylor A. and {James}, Bethan L. and {Joyce}, Meridith and {Jung}, Intae and {Khullar}, Gourav and {Larson}, Rebecca L. and {Mahler}, Guillaume and {Mandelker}, Nir and {McCandliss}, Stephan and {Morishita}, Takahiro and {Newshore}, Rosa and {Norman}, Colin and {O'Connor}, Kyle and {Oesch}, Pascal A. and {Oguri}, Masamune and {Ouchi}, Masami and {Postman}, Marc and {Rigby}, Jane R. and {Ryan}, Russell E., Jr. and {Sharma}, Soniya and {Sharon}, Keren and {Strait}, Victoria and {Strolger}, Louis-Gregory and {Timmes}, F.~X. and {Toft}, Sune and {Trenti}, Michele and {Vanzella}, Eros and {Vikaeus}, Anton},
        title = "{JWST Imaging of Earendel, the Extremely Magnified Star at Redshift z = 6.2}",
      journal = {\apjl},
         year = 2022,
        month = nov,
       volume = {940},
       number = {1},
          eid = {L1},
        pages = {L1},
          doi = {10.3847/2041-8213/ac9d39},
archivePrefix = {arXiv},
       eprint = {2208.09007},
 primaryClass = {astro-ph.GA},
       adsurl = {https://ui.adsabs.harvard.edu/abs/2022ApJ...940L...1W}
}

@ARTICLE{2002ApJ...567..532H,
       author = {{Heger}, A. and {Woosley}, S.~E.},
        title = "{The Nucleosynthetic Signature of Population III}",
      journal = {\apj},
         year = 2002,
        month = mar,
       volume = {567},
       number = {1},
        pages = {532-543},
          doi = {10.1086/338487},
archivePrefix = {arXiv},
       eprint = {astro-ph/0107037},
 primaryClass = {astro-ph},
       adsurl = {https://ui.adsabs.harvard.edu/abs/2002ApJ...567..532H}
}

@ARTICLE{2013MNRAS.431..912Q,
       author = {{Quimby}, Robert M. and {Yuan}, Fang and {Akerlof}, Carl and {Wheeler}, J. Craig},
        title = "{Rates of superluminous supernovae at z {\ensuremath{\sim}} 0.2}",
      journal = {\mnras},
         year = 2013,
        month = may,
       volume = {431},
       number = {1},
        pages = {912-922},
          doi = {10.1093/mnras/stt213},
archivePrefix = {arXiv},
       eprint = {1302.0911},
 primaryClass = {astro-ph.CO},
       adsurl = {https://ui.adsabs.harvard.edu/abs/2013MNRAS.431..912Q}
}

@ARTICLE{2007ApJ...663.1187H,
       author = {{Hsiao}, E.~Y. and {Conley}, A. and {Howell}, D.~A. and {Sullivan}, M. and {Pritchet}, C.~J. and {Carlberg}, R.~G. and {Nugent}, P.~E. and {Phillips}, M.~M.},
        title = "{K-Corrections and Spectral Templates of Type Ia Supernovae}",
      journal = {\apj},
         year = 2007,
        month = jul,
       volume = {663},
       number = {2},
        pages = {1187-1200},
          doi = {10.1086/518232},
archivePrefix = {arXiv},
       eprint = {astro-ph/0703529},
 primaryClass = {astro-ph},
       adsurl = {https://ui.adsabs.harvard.edu/abs/2007ApJ...663.1187H}
}

@ARTICLE{2015ApJ...802L..19R,
       author = {{Robertson}, Brant E. and {Ellis}, Richard S. and {Furlanetto}, Steven R. and {Dunlop}, James S.},
        title = "{Cosmic Reionization and Early Star-forming Galaxies: A Joint Analysis of New Constraints from Planck and the Hubble Space Telescope}",
      journal = {\apjl},
         year = 2015,
        month = apr,
       volume = {802},
       number = {2},
          eid = {L19},
        pages = {L19},
          doi = {10.1088/2041-8205/802/2/L19},
archivePrefix = {arXiv},
       eprint = {1502.02024},
 primaryClass = {astro-ph.CO},
       adsurl = {https://ui.adsabs.harvard.edu/abs/2015ApJ...802L..19R}
}

@ARTICLE{2022ApJ...925..211M,
       author = {{Moriya}, Takashi J. and {Quimby}, Robert M. and {Robertson}, Brant E.},
        title = "{Discovering Supernovae at the Epoch of Reionization with the Nancy Grace Roman Space Telescope}",
      journal = {\apj},
         year = 2022,
        month = feb,
       volume = {925},
       number = {2},
          eid = {211},
        pages = {211},
          doi = {10.3847/1538-4357/ac415e},
archivePrefix = {arXiv},
       eprint = {2108.01801},
 primaryClass = {astro-ph.HE},
       adsurl = {https://ui.adsabs.harvard.edu/abs/2022ApJ...925..211M}
}

@ARTICLE{2021MNRAS.503.1206M,
       author = {{Moriya}, Takashi J. and {Chen}, Ke-Jung and {Nakajima}, Kimihiko and {Tominaga}, Nozomu and {Blinnikov}, Sergei I.},
        title = "{Observational properties of a general relativistic instability supernova from a primordial supermassive star}",
      journal = {\mnras},
         year = 2021,
        month = may,
       volume = {503},
       number = {1},
        pages = {1206-1213},
          doi = {10.1093/mnras/stab622},
archivePrefix = {arXiv},
       eprint = {2103.01336},
 primaryClass = {astro-ph.HE},
       adsurl = {https://ui.adsabs.harvard.edu/abs/2021MNRAS.503.1206M}
}

@ARTICLE{2011ApJ...734..102K,
       author = {{Kasen}, Daniel and {Woosley}, S.~E. and {Heger}, Alexander},
        title = "{Pair Instability Supernovae: Light Curves, Spectra, and Shock Breakout}",
      journal = {\apj},
         year = 2011,
        month = jun,
       volume = {734},
       number = {2},
          eid = {102},
        pages = {102},
          doi = {10.1088/0004-637X/734/2/102},
archivePrefix = {arXiv},
       eprint = {1101.3336},
 primaryClass = {astro-ph.HE},
       adsurl = {https://ui.adsabs.harvard.edu/abs/2011ApJ...734..102K}
}

@ARTICLE{2015MNRAS.448..568H,
       author = {{Hirano}, S. and {Hosokawa}, T. and {Yoshida}, N. and {Omukai}, K. and {Yorke}, H.~W.},
        title = "{Primordial star formation under the influence of far ultraviolet radiation: 1540 cosmological haloes and the stellar mass distribution}",
      journal = {\mnras},
         year = 2015,
        month = mar,
       volume = {448},
       number = {1},
        pages = {568-587},
          doi = {10.1093/mnras/stv044},
archivePrefix = {arXiv},
       eprint = {1501.01630},
 primaryClass = {astro-ph.GA},
       adsurl = {https://ui.adsabs.harvard.edu/abs/2015MNRAS.448..568H}
}

@ARTICLE{1998A&AS..127....1L,
       author = {{Leinert}, Ch. and {Bowyer}, S. and {Haikala}, L.~K. and {Hanner}, M.~S. and {Hauser}, M.~G. and {Levasseur-Regourd}, A.-Ch. and {Mann}, I. and {Mattila}, K. and {Reach}, W.~T. and {Schlosser}, W. and {Staude}, H.~J. and {Toller}, G.~N. and {Weiland}, J.~L. and {Weinberg}, J.~L. and {Witt}, A.~N.},
        title = "{The 1997 reference of diffuse night sky brightness}",
      journal = {\aaps},
         year = 1998,
        month = jan,
       volume = {127},
        pages = {1-99},
          doi = {10.1051/aas:1998105},
       adsurl = {https://ui.adsabs.harvard.edu/abs/1998A&AS..127....1L}
}

@ARTICLE{2004MNRAS.351L..71C,
       author = {{Cooray}, Asantha and {Yoshida}, Naoki},
        title = "{First sources in infrared light: stars, supernovae and miniquasars}",
      journal = {\mnras},
         year = 2004,
        month = jul,
       volume = {351},
       number = {3},
        pages = {L71-L77},
          doi = {10.1111/j.1365-2966.2004.08047.x},
archivePrefix = {arXiv},
       eprint = {astro-ph/0404109},
 primaryClass = {astro-ph},
       adsurl = {https://ui.adsabs.harvard.edu/abs/2004MNRAS.351L..71C}
}

@INPROCEEDINGS{2024SPIE13092E..0VM,
       author = {{Matsuura}, Shuji and {Bock}, James J. and {Cooray}, Asantha and {Fazar}, Candice and {Feder}, Richard M. and {Hashimoto}, Ryo and {Heaton}, Grigory and {Hristov}, Viktor and {Kawano}, Yuya and {Korngut}, Phillip and {Lee}, Dae-Hee and {Matsumi}, Chika and {Mercado}, Dale and {Nakagawa}, Shunsuke and {Nakagawa}, Tomoya and {Nakahata}, Shuta and {Nguyen}, Chi H. and {Noda}, Kazuma and {Patru}, Dorin and {Park}, Won-Kee and {Sano}, Kei and {Takahashi}, Aoi and {Takimoto}, Kohji and {Tamai}, Momoko and {Tsumura}, Kohji and {Zemcov}, Michael},
        title = "{The cosmic infrared background experiment 2 (CIBER-2): status of recent flights}",
    booktitle = {Space Telescopes and Instrumentation 2024: Optical, Infrared, and Millimeter Wave},
         year = 2024,
       editor = {{Coyle}, Laura E. and {Matsuura}, Shuji and {Perrin}, Marshall D.},
       series = {Society of Photo-Optical Instrumentation Engineers (SPIE) Conference Series},
       volume = {13092},
        month = aug,
          eid = {130920V},
        pages = {130920V},
          doi = {10.1117/12.3017526},
       adsurl = {https://ui.adsabs.harvard.edu/abs/2024SPIE13092E..0VM}
}

@ARTICLE{2025ApJS..280...66Z,
       author = {{Zemcov}, Michael and {Bock}, James J. and {Cooray}, Asantha and {Matsuura}, Shuji and {Lee}, Dae-Hee and {Fazar}, Candice and {Feder}, Richard M. and {Heaton}, Grigory and {Hashimoto}, Ryo and {Korngut}, Phillip and {Matsumoto}, Toshio and {Nguyen}, Chi H. and {Noda}, Kazuma and {Park}, Won-Kee and {Sano}, Kei and {Takimoto}, Kohji and {Arai}, Toshiaki and {Bang}, Seung-Cheol and {Bangale}, Priyadarshini and {Furutani}, Masaki and {Hristov}, Viktor and {Kawano}, Yuya and {Kida}, Arisa and {Kojima}, Tomoya and {Lanz}, Alicia and {Matsumi}, Chika and {Mercado}, Dale and {Nakagawa}, Shunsuke and {Nakagawa}, Tomoya and {Nakahata}, Shuta and {Ohta}, Ryo and {Patru}, Dorin and {Shirahata}, Mai and {Suzuki}, Hiroko and {Takahashi}, Aoi and {Tamai}, Momoko and {Tramm}, Serena and {Tsumura}, Kohji and {Yamada}, Yasuhiro and {Wang}, Shiang-Yu},
        title = "{The Cosmic Infrared Background Experiment-2: An Intensity-mapping Optimized Sounding-rocket Payload to Understand the Near-IR Extragalactic Background Light}",
      journal = {\apjs},
         year = 2025,
        month = oct,
       volume = {280},
       number = {2},
          eid = {66},
        pages = {66},
          doi = {10.3847/1538-4365/adfc6e},
archivePrefix = {arXiv},
       eprint = {2510.05210},
 primaryClass = {astro-ph.IM},
       adsurl = {https://ui.adsabs.harvard.edu/abs/2025ApJS..280...66Z}
}

@ARTICLE{2005Natur.438...45K,
       author = {{Kashlinsky}, A. and {Arendt}, R.~G. and {Mather}, J. and {Moseley}, S.~H.},
        title = "{Tracing the first stars with fluctuations of the cosmic infrared background}",
      journal = {\nat},
         year = 2005,
        month = nov,
       volume = {438},
       number = {7064},
        pages = {45-50},
          doi = {10.1038/nature04143},
archivePrefix = {arXiv},
       eprint = {astro-ph/0511105},
 primaryClass = {astro-ph},
       adsurl = {https://ui.adsabs.harvard.edu/abs/2005Natur.438...45K}
}

@ARTICLE{2018RvMP...90b5006K,
       author = {{Kashlinsky}, A. and {Arendt}, R.~G. and {Atrio-Barandela}, F. and {Cappelluti}, N. and {Ferrara}, A. and {Hasinger}, G.},
        title = "{Looking at cosmic near-infrared background radiation anisotropies}",
      journal = {Reviews of Modern Physics},
         year = 2018,
        month = apr,
       volume = {90},
       number = {2},
          eid = {025006},
        pages = {025006},
          doi = {10.1103/RevModPhys.90.025006},
archivePrefix = {arXiv},
       eprint = {1802.07774},
 primaryClass = {astro-ph.CO},
       adsurl = {https://ui.adsabs.harvard.edu/abs/2018RvMP...90b5006K}
}

@ARTICLE{2017ApJ...839....7M,
       author = {{Matsuura}, Shuji and {Arai}, Toshiaki and {Bock}, James J. and {Cooray}, Asantha and {Korngut}, Phillip M. and {Kim}, Min Gyu and {Lee}, Hyung Mok and {Lee}, Dae Hee and {Levenson}, Louis R. and {Matsumoto}, Toshio and {Onishi}, Yosuke and {Shirahata}, Mai and {Tsumura}, Kohji and {Wada}, Takehiko and {Zemcov}, Michael},
        title = "{New Spectral Evidence of an Unaccounted Component of the Near-infrared Extragalactic Background Light from the CIBER}",
      journal = {\apj},
         year = 2017,
        month = apr,
       volume = {839},
       number = {1},
          eid = {7},
        pages = {7},
          doi = {10.3847/1538-4357/aa6843},
archivePrefix = {arXiv},
       eprint = {1704.07166},
 primaryClass = {astro-ph.GA},
       adsurl = {https://ui.adsabs.harvard.edu/abs/2017ApJ...839....7M}
}

@ARTICLE{2013ApJ...769...68C,
       author = {{Cappelluti}, N. and {Kashlinsky}, A. and {Arendt}, R.~G. and {Comastri}, A. and {Fazio}, G.~G. and {Finoguenov}, A. and {Hasinger}, G. and {Mather}, J.~C. and {Miyaji}, T. and {Moseley}, S.~H.},
        title = "{Cross-correlating Cosmic Infrared and X-Ray Background Fluctuations: Evidence of Significant Black Hole Populations among the CIB Sources}",
      journal = {\apj},
         year = 2013,
        month = may,
       volume = {769},
       number = {1},
          eid = {68},
        pages = {68},
          doi = {10.1088/0004-637X/769/1/68},
archivePrefix = {arXiv},
       eprint = {1210.5302},
 primaryClass = {astro-ph.CO},
       adsurl = {https://ui.adsabs.harvard.edu/abs/2013ApJ...769...68C}
}

@ARTICLE{2012Natur.490..514C,
       author = {{Cooray}, Asantha and {Smidt}, Joseph and {de Bernardis}, Francesco and {Gong}, Yan and {Stern}, Daniel and {Ashby}, Matthew L.~N. and {Eisenhardt}, Peter R. and {Frazer}, Christopher C. and {Gonzalez}, Anthony H. and {Kochanek}, Christopher S. and {Koz{\l}owski}, Szymon and {Wright}, Edward L.},
        title = "{Near-infrared background anisotropies from diffuse intrahalo light of galaxies}",
      journal = {\nat},
         year = 2012,
        month = oct,
       volume = {490},
       number = {7421},
        pages = {514-516},
          doi = {10.1038/nature11474},
       adsurl = {https://ui.adsabs.harvard.edu/abs/2012Natur.490..514C}
}

@ARTICLE{2014ApJ...781...60H,
       author = {{Hirano}, Shingo and {Hosokawa}, Takashi and {Yoshida}, Naoki and {Umeda}, Hideyuki and {Omukai}, Kazuyuki and {Chiaki}, Gen and {Yorke}, Harold W.},
        title = "{One Hundred First Stars: Protostellar Evolution and the Final Masses}",
      journal = {\apj},
         year = 2014,
        month = feb,
       volume = {781},
       number = {2},
          eid = {60},
        pages = {60},
          doi = {10.1088/0004-637X/781/2/60},
archivePrefix = {arXiv},
       eprint = {1308.4456},
 primaryClass = {astro-ph.CO},
       adsurl = {https://ui.adsabs.harvard.edu/abs/2014ApJ...781...60H}
}

@ARTICLE{1971MNRAS.152...75H,
       author = {{Hawking}, Stephen},
        title = "{Gravitationally collapsed objects of very low mass}",
      journal = {\mnras},
         year = 1971,
        month = jan,
       volume = {152},
        pages = {75},
          doi = {10.1093/mnras/152.1.75},
       adsurl = {https://ui.adsabs.harvard.edu/abs/1971MNRAS.152...75H}
}

@ARTICLE{2013A&A...550A...4H,
       author = {{H.~E.~S.~S. Collaboration} and {Abramowski}, A. and {Acero}, F. and {Aharonian}, F. and {Akhperjanian}, A.~G. and {Anton}, G. and {Balenderan}, S. and {Balzer}, A. and {Barnacka}, A. and {Becherini}, Y. and {Becker Tjus}, J. and {Bernl{\"o}hr}, K. and {Birsin}, E. and {Biteau}, J. and {Bochow}, A. and {Boisson}, C. and {Bolmont}, J. and {Bordas}, P. and {Brucker}, J. and {Brun}, F. and {Brun}, P. and {Bulik}, T. and {Carrigan}, S. and {Casanova}, S. and {Cerruti}, M. and {Chadwick}, P.~M. and {Charbonnier}, A. and {Chaves}, R.~C.~G. and {Cheesebrough}, A. and {Cologna}, G. and {Conrad}, J. and {Couturier}, C. and {Dalton}, M. and {Daniel}, M.~K. and {Davids}, I.~D. and {Degrange}, B. and {Deil}, C. and {deWilt}, P. and {Dickinson}, H.~J. and {Djannati-Ata{\"\i}}, A. and {Domainko}, W. and {O'C. Drury}, L. and {Dubus}, G. and {Dutson}, K. and {Dyks}, J. and {Dyrda}, M. and {Egberts}, K. and {Eger}, P. and {Espigat}, P. and {Fallon}, L. and {Farnier}, C. and {Fegan}, S. and {Feinstein}, F. and {Fernandes}, M.~V. and {Fernandez}, D. and {Fiasson}, A. and {Fontaine}, G. and {F{\"o}rster}, A. and {F{\"u}{\ss}ling}, M. and {Gajdus}, M. and {Gallant}, Y.~A. and {Garrigoux}, T. and {Gast}, H. and {Giebels}, B. and {Glicenstein}, J.~F. and {Gl{\"u}ck}, B. and {G{\"o}ring}, D. and {Grondin}, M. -H. and {H{\"a}ffner}, S. and {Hague}, J.~D. and {Hahn}, J. and {Hampf}, D. and {Harris}, J. and {Heinz}, S. and {Heinzelmann}, G. and {Henri}, G. and {Hermann}, G. and {Hillert}, A. and {Hinton}, J.~A. and {Hofmann}, W. and {Hofverberg}, P. and {Holler}, M. and {Horns}, D. and {Jacholkowska}, A. and {Jahn}, C. and {Jamrozy}, M. and {Jung}, I. and {Kastendieck}, M.~A. and {Katarzy{\'n}ski}, K. and {Katz}, U. and {Kaufmann}, S. and {Kh{\'e}lifi}, B. and {Klochkov}, D. and {Klu{\'z}niak}, W. and {Kneiske}, T. and {Komin}, Nu. and {Kosack}, K. and {Kossakowski}, R. and {Krayzel}, F. and {Laffon}, H. and {Lamanna}, G. and {Lenain}, J. -P. and {Lennarz}, D. and {Lohse}, T. and {Lopatin}, A. and {Lu}, C. -C. and {Marandon}, V. and {Marcowith}, A. and {Masbou}, J. and {Maurin}, G. and {Maxted}, N. and {Mayer}, M. and {McComb}, T.~J.~L. and {Medina}, M.~C. and {M{\'e}hault}, J. and {Menzler}, U. and {Moderski}, R. and {Mohamed}, M. and {Moulin}, E. and {Naumann}, C.~L. and {Naumann-Godo}, M. and {de Naurois}, M. and {Nedbal}, D. and {Nguyen}, N. and {Niemiec}, J. and {Nolan}, S.~J. and {Ohm}, S. and {de O{\~n}a Wilhelmi}, E. and {Opitz}, B. and {Ostrowski}, M. and {Oya}, I. and {Panter}, M. and {Parsons}, D. and {Paz Arribas}, M. and {Pekeur}, N.~W. and {Pelletier}, G. and {Perez}, J. and {Petrucci}, P. -O. and {Peyaud}, B. and {Pita}, S. and {P{\"u}hlhofer}, G. and {Punch}, M. and {Quirrenbach}, A. and {Raue}, M. and {Reimer}, A. and {Reimer}, O. and {Renaud}, M. and {de los Reyes}, R. and {Rieger}, F. and {Ripken}, J. and {Rob}, L. and {Rosier-Lees}, S. and {Rowell}, G. and {Rudak}, B. and {Rulten}, C.~B. and {Sahakian}, V. and {Sanchez}, D.~A. and {Santangelo}, A. and {Schlickeiser}, R. and {Schulz}, A. and {Schwanke}, U. and {Schwarzburg}, S. and {Schwemmer}, S. and {Sheidaei}, F. and {Skilton}, J.~L. and {Sol}, H. and {Spengler}, G. and {Stawarz}, {\L}. and {Steenkamp}, R. and {Stegmann}, C. and {Stinzing}, F. and {Stycz}, K. and {Sushch}, I. and {Szostek}, A. and {Tavernet}, J. -P. and {Terrier}, R. and {Tluczykont}, M. and {Valerius}, K. and {van Eldik}, C. and {Vasileiadis}, G. and {Venter}, C. and {Viana}, A. and {Vincent}, P. and {V{\"o}lk}, H.~J. and {Volpe}, F. and {Vorobiov}, S. and {Vorster}, M. and {Wagner}, S.~J. and {Ward}, M. and {White}, R. and {Wierzcholska}, A. and {Wouters}, D. and {Zacharias}, M. and {Zajczyk}, A. and {Zdziarski}, A.~A. and {Zech}, A. and {Zechlin}, H. -S.},
        title = "{Measurement of the extragalactic background light imprint on the spectra of the brightest blazars observed with H.E.S.S.}",
      journal = {\aap},
         year = 2013,
        month = feb,
       volume = {550},
          eid = {A4},
        pages = {A4},
          doi = {10.1051/0004-6361/201220355},
archivePrefix = {arXiv},
       eprint = {1212.3409},
 primaryClass = {astro-ph.HE},
       adsurl = {https://ui.adsabs.harvard.edu/abs/2013A&A...550A...4H}
}

@ARTICLE{2011ApJ...742..124M,
       author = {{Matsumoto}, T. and {Seo}, H.~J. and {Jeong}, W. -S. and {Lee}, H.~M. and {Matsuura}, S. and {Matsuhara}, H. and {Oyabu}, S. and {Pyo}, J. and {Wada}, T.},
        title = "{AKARI Observation of the Fluctuation of the Near-infrared Background}",
      journal = {\apj},
         year = 2011,
        month = dec,
       volume = {742},
       number = {2},
          eid = {124},
        pages = {124},
          doi = {10.1088/0004-637X/742/2/124},
archivePrefix = {arXiv},
       eprint = {1010.0491},
 primaryClass = {astro-ph.CO},
       adsurl = {https://ui.adsabs.harvard.edu/abs/2011ApJ...742..124M}
}

@ARTICLE{2024ApJ...963..129M,
       author = {{Matthee}, Jorryt and {Naidu}, Rohan P. and {Brammer}, Gabriel and {Chisholm}, John and {Eilers}, Anna-Christina and {Goulding}, Andy and {Greene}, Jenny and {Kashino}, Daichi and {Labbe}, Ivo and {Lilly}, Simon J. and {Mackenzie}, Ruari and {Oesch}, Pascal A. and {Weibel}, Andrea and {Wuyts}, Stijn and {Xiao}, Mengyuan and {Bordoloi}, Rongmon and {Bouwens}, Rychard and {van Dokkum}, Pieter and {Illingworth}, Garth and {Kramarenko}, Ivan and {Maseda}, Michael V. and {Mason}, Charlotte and {Meyer}, Romain A. and {Nelson}, Erica J. and {Reddy}, Naveen A. and {Shivaei}, Irene and {Simcoe}, Robert A. and {Yue}, Minghao},
        title = "{Little Red Dots: An Abundant Population of Faint Active Galactic Nuclei at z {\ensuremath{\sim}} 5 Revealed by the EIGER and FRESCO JWST Surveys}",
      journal = {\apj},
         year = 2024,
        month = mar,
       volume = {963},
       number = {2},
          eid = {129},
        pages = {129},
          doi = {10.3847/1538-4357/ad2345},
archivePrefix = {arXiv},
       eprint = {2306.05448},
 primaryClass = {astro-ph.GA},
       adsurl = {https://ui.adsabs.harvard.edu/abs/2024ApJ...963..129M}
}

@ARTICLE{2013ARA&A..51..511K,
       author = {{Kormendy}, John and {Ho}, Luis C.},
        title = "{Coevolution (Or Not) of Supermassive Black Holes and Host Galaxies}",
      journal = {\araa},
         year = 2013,
        month = aug,
       volume = {51},
       number = {1},
        pages = {511-653},
          doi = {10.1146/annurev-astro-082708-101811},
archivePrefix = {arXiv},
       eprint = {1304.7762},
 primaryClass = {astro-ph.CO},
       adsurl = {https://ui.adsabs.harvard.edu/abs/2013ARA&A..51..511K}
}

@ARTICLE{2021ApJ...907L...1W,
       author = {{Wang}, Feige and {Yang}, Jinyi and {Fan}, Xiaohui and {Hennawi}, Joseph F. and {Barth}, Aaron J. and {Banados}, Eduardo and {Bian}, Fuyan and {Boutsia}, Konstantina and {Connor}, Thomas and {Davies}, Frederick B. and {Decarli}, Roberto and {Eilers}, Anna-Christina and {Farina}, Emanuele Paolo and {Green}, Richard and {Jiang}, Linhua and {Li}, Jiang-Tao and {Mazzucchelli}, Chiara and {Nanni}, Riccardo and {Schindler}, Jan-Torge and {Venemans}, Bram and {Walter}, Fabian and {Wu}, Xue-Bing and {Yue}, Minghao},
        title = "{A Luminous Quasar at Redshift 7.642}",
      journal = {\apjl},
         year = 2021,
        month = jan,
       volume = {907},
       number = {1},
          eid = {L1},
        pages = {L1},
          doi = {10.3847/2041-8213/abd8c6},
archivePrefix = {arXiv},
       eprint = {2101.03179},
 primaryClass = {astro-ph.GA},
       adsurl = {https://ui.adsabs.harvard.edu/abs/2021ApJ...907L...1W}
}

@ARTICLE{2018ApJ...869..150M,
       author = {{Matsuoka}, Yoshiki and {Strauss}, Michael A. and {Kashikawa}, Nobunari and {Onoue}, Masafusa and {Iwasawa}, Kazushi and {Tang}, Ji-Jia and {Lee}, Chien-Hsiu and {Imanishi}, Masatoshi and {Nagao}, Tohru and {Akiyama}, Masayuki and {Asami}, Naoko and {Bosch}, James and {Furusawa}, Hisanori and {Goto}, Tomotsugu and {Gunn}, James E. and {Harikane}, Yuichi and {Ikeda}, Hiroyuki and {Izumi}, Takuma and {Kawaguchi}, Toshihiro and {Kato}, Nanako and {Kikuta}, Satoshi and {Kohno}, Kotaro and {Komiyama}, Yutaka and {Lupton}, Robert H. and {Minezaki}, Takeo and {Miyazaki}, Satoshi and {Murayama}, Hitoshi and {Niida}, Mana and {Nishizawa}, Atsushi J. and {Noboriguchi}, Akatoki and {Oguri}, Masamune and {Ono}, Yoshiaki and {Ouchi}, Masami and {Price}, Paul A. and {Sameshima}, Hiroaki and {Schulze}, Andreas and {Shirakata}, Hikari and {Silverman}, John D. and {Sugiyama}, Naoshi and {Tait}, Philip J. and {Takada}, Masahiro and {Takata}, Tadafumi and {Tanaka}, Masayuki and {Toba}, Yoshiki and {Utsumi}, Yousuke and {Wang}, Shiang-Yu and {Yamashita}, Takuji},
        title = "{Subaru High-z  Exploration of Low-luminosity Quasars (SHELLQs). V. Quasar Luminosity Function and Contribution to Cosmic Reionization at z = 6}",
      journal = {\apj},
         year = 2018,
        month = dec,
       volume = {869},
       number = {2},
          eid = {150},
        pages = {150},
          doi = {10.3847/1538-4357/aaee7a},
archivePrefix = {arXiv},
       eprint = {1811.01963},
 primaryClass = {astro-ph.GA},
       adsurl = {https://ui.adsabs.harvard.edu/abs/2018ApJ...869..150M}
}

@ARTICLE{2016ApJ...833..222J,
       author = {{Jiang}, Linhua and {McGreer}, Ian D. and {Fan}, Xiaohui and {Strauss}, Michael A. and {Ba{\~n}ados}, Eduardo and {Becker}, Robert H. and {Bian}, Fuyan and {Farnsworth}, Kara and {Shen}, Yue and {Wang}, Feige and {Wang}, Ran and {Wang}, Shu and {White}, Richard L. and {Wu}, Jin and {Wu}, Xue-Bing and {Yang}, Jinyi and {Yang}, Qian},
        title = "{The Final SDSS High-redshift Quasar Sample of 52 Quasars at z>5.7}",
      journal = {\apj},
         year = 2016,
        month = dec,
       volume = {833},
       number = {2},
          eid = {222},
        pages = {222},
          doi = {10.3847/1538-4357/833/2/222},
archivePrefix = {arXiv},
       eprint = {1610.05369},
 primaryClass = {astro-ph.GA},
       adsurl = {https://ui.adsabs.harvard.edu/abs/2016ApJ...833..222J}
}

@ARTICLE{schindler2025,
       author = {{Schindler}, Jan-Torge and {Hennawi}, Joseph F. and {Davies}, Frederick B. and {Bosman}, Sarah E.~I. and {Endsley}, Ryan and {Wang}, Feige and {Yang}, Jinyi and {Barth}, Aaron J. and {Eilers}, Anna-Christina and {Fan}, Xiaohui and {Kakiichi}, Koki and {Maseda}, Michael and {Pizzati}, Elia and {Nanni}, Riccardo},
        title = "{A little red dot at z = 7.3 within a large galaxy overdensity}",
      journal = {Nature Astronomy},
         year = 2025,
        month = nov,
       volume = {9},
       number = {11},
        pages = {1732-1744},
          doi = {10.1038/s41550-025-02660-1},
archivePrefix = {arXiv},
       eprint = {2411.11534},
 primaryClass = {astro-ph.GA},
       adsurl = {https://ui.adsabs.harvard.edu/abs/2025NatAs...9.1732S}
}

@ARTICLE{escudero2025,
       author = {{Carranza-Escudero}, Mar{\'\i}a and {Conselice}, Christopher J. and {Adams}, Nathan and {Harvey}, Thomas and {Austin}, Duncan and {Behroozi}, Peter and {Ferreira}, Leonardo and {Ormerod}, Katherine and {Duan}, Qiao and {Trussler}, James and {Li}, Qiong and {Westcott}, Lewi and {Windhorst}, Rogier A. and {Coe}, Dan and {Cohen}, Seth H. and {Cheng}, Cheng and {Driver}, Simon P. and {Frye}, Brenda and {Furtak}, Lukas J. and {Grogin}, Norman A. and {Hathi}, Nimish P. and {Jansen}, Rolf A. and {Koekemoer}, Anton M. and {Marshall}, Madeline A. and {O'Brien}, Rosalia and {Pirzkal}, Norbert and {Polletta}, Maria and {Robotham}, Aaron and {Rutkowski}, Michael J. and {Summers}, Jake and {Wilkins}, Stephen M. and {Willmer}, Christopher N.~A. and {Yan}, Haojing and {Zitrin}, Adi},
        title = "{Lonely Little Red Dots: Challenges to the Active Galactic Nucleus Nature of Little Red Dots through Their Clustering and Spectral Energy Distributions}",
      journal = {\apjl},
         year = 2025,
        month = aug,
       volume = {989},
       number = {2},
          eid = {L50},
        pages = {L50},
          doi = {10.3847/2041-8213/adf73d},
archivePrefix = {arXiv},
       eprint = {2506.04004},
 primaryClass = {astro-ph.GA},
       adsurl = {https://ui.adsabs.harvard.edu/abs/2025ApJ...989L..50C}
}

@ARTICLE{2023arXiv230408104G,
       author = {{GREX-PLUS Science Team} and {:} and {Inoue}, Akio K. and {Harikane}, Yuichi and {Moriya}, Takashi and {Nomura}, Hideko and {Baba}, Shunsuke and {Fujii}, Yuka and {Gouda}, Naoteru and {Hirahara}, Yasuhiro and {Kawashima}, Yui and {Kodama}, Tadayuki and {Koyama}, Yusei and {Kurokawa}, Hiroyuki and {Matsuo}, Taro and {Matsuoka}, Yoshiki and {Matsuura}, Shuji and {Mawatari}, Ken and {Misawa}, Toru and {Nagamine}, Kentaro and {Nakajima}, Kimihiko and {Notsu}, Shota and {Ootsubo}, Takafumi and {Ohno}, Kazumasa and {Sagawa}, Hideo and {Shimonishi}, Takashi and {Tadaki}, Ken-ichi and {Takami}, Michihiro and {Terai}, Tsuyoshi and {Toba}, Yoshiki and {Yamashita}, Takuji and {Yasui}, Chikako},
        title = "{GREX-PLUS Science Book}",
      journal = {arXiv e-prints},
         year = 2023,
        month = apr,
          eid = {arXiv:2304.08104},
        pages = {arXiv:2304.08104},
          doi = {10.48550/arXiv.2304.08104},
archivePrefix = {arXiv},
       eprint = {2304.08104},
 primaryClass = {astro-ph.CO},
       adsurl = {https://ui.adsabs.harvard.edu/abs/2023arXiv230408104G}
}

@ARTICLE{Harikane2023_AGN,
       author = {{Harikane}, Yuichi and {Zhang}, Yechi and {Nakajima}, Kimihiko and {Ouchi}, Masami and {Isobe}, Yuki and {Ono}, Yoshiaki and {Hatano}, Shun and {Xu}, Yi and {Umeda}, Hiroya},
        title = "{A JWST/NIRSpec First Census of Broad-line AGNs at z = 4-7: Detection of 10 Faint AGNs with M $_{BH}$ {}10$^{6}$-{}10$^{8}$ M $_{{\ensuremath{\odot}}}$ and Their Host Galaxy Properties}",
      journal = {\apj},
         year = 2023,
        month = dec,
       volume = {959},
       number = {1},
          eid = {39},
        pages = {39},
          doi = {10.3847/1538-4357/ad029e},
archivePrefix = {arXiv},
       eprint = {2303.11946},
 primaryClass = {astro-ph.GA},
       adsurl = {https://ui.adsabs.harvard.edu/abs/2023ApJ...959...39H}
}

@ARTICLE{Matthee2024,
       author = {{Matthee}, Jorryt and {Naidu}, Rohan P. and {Brammer}, Gabriel and {Chisholm}, John and {Eilers}, Anna-Christina and {Goulding}, Andy and {Greene}, Jenny and {Kashino}, Daichi and {Labbe}, Ivo and {Lilly}, Simon J. and {Mackenzie}, Ruari and {Oesch}, Pascal A. and {Weibel}, Andrea and {Wuyts}, Stijn and {Xiao}, Mengyuan and {Bordoloi}, Rongmon and {Bouwens}, Rychard and {van Dokkum}, Pieter and {Illingworth}, Garth and {Kramarenko}, Ivan and {Maseda}, Michael V. and {Mason}, Charlotte and {Meyer}, Romain A. and {Nelson}, Erica J. and {Reddy}, Naveen A. and {Shivaei}, Irene and {Simcoe}, Robert A. and {Yue}, Minghao},
        title = "{Little Red Dots: An Abundant Population of Faint Active Galactic Nuclei at z {\ensuremath{\sim}} 5 Revealed by the EIGER and FRESCO JWST Surveys}",
      journal = {\apj},
         year = 2024,
        month = mar,
       volume = {963},
       number = {2},
          eid = {129},
        pages = {129},
          doi = {10.3847/1538-4357/ad2345},
archivePrefix = {arXiv},
       eprint = {2306.05448},
 primaryClass = {astro-ph.GA},
       adsurl = {https://ui.adsabs.harvard.edu/abs/2024ApJ...963..129M}
}

@ARTICLE{Kocevski2025_LRD,
       author = {{Kocevski}, Dale D. and {Finkelstein}, Steven L. and {Barro}, Guillermo and {Taylor}, Anthony J. and {Calabr{\`o}}, Antonello and {Laloux}, Brivael and {Buchner}, Johannes and {Trump}, Jonathan R. and {Leung}, Gene C.~K. and {Yang}, Guang and {Dickinson}, Mark and {P{\'e}rez-Gonz{\'a}lez}, Pablo G. and {Pacucci}, Fabio and {Inayoshi}, Kohei and {Somerville}, Rachel S. and {McGrath}, Elizabeth J. and {Akins}, Hollis B. and {Bagley}, Micaela B. and {Bowler}, Rebecca A.~A. and {Bisigello}, Laura and {Carnall}, Adam and {Casey}, Caitlin M. and {Cheng}, Yingjie and {Cleri}, Nikko J. and {Costantin}, Luca and {Cullen}, Fergus and {Davis}, Kelcey and {Donnan}, Callum T. and {Dunlop}, James S. and {Ellis}, Richard S. and {Ferguson}, Henry C. and {Fujimoto}, Seiji and {Fontana}, Adriano and {Giavalisco}, Mauro and {Grazian}, Andrea and {Grogin}, Norman A. and {Hathi}, Nimish P. and {Hirschmann}, Michaela and {Huertas-Company}, Marc and {Holwerda}, Benne W. and {Illingworth}, Garth and {Juneau}, St{\'e}phanie and {Kartaltepe}, Jeyhan S. and {Koekemoer}, Anton M. and {Li}, Wenxiu and {Lucas}, Ray A. and {Magee}, Dan and {Mason}, Charlotte and {McLeod}, Derek J. and {McLure}, Ross J. and {Napolitano}, Lorenzo and {Papovich}, Casey and {Pirzkal}, Nor and {Rodighiero}, Giulia and {Santini}, Paola and {Wilkins}, Stephen M. and {Yung}, L.~Y. Aaron},
        title = "{The Rise of Faint, Red Active Galactic Nuclei at z > 4: A Sample of Little Red Dots in the JWST Extragalactic Legacy Fields}",
      journal = {\apj},
         year = 2025,
        month = jun,
       volume = {986},
       number = {2},
          eid = {126},
        pages = {126},
          doi = {10.3847/1538-4357/adbc7d},
archivePrefix = {arXiv},
       eprint = {2404.03576},
 primaryClass = {astro-ph.GA},
       adsurl = {https://ui.adsabs.harvard.edu/abs/2025ApJ...986..126K}
}

@ARTICLE{Akins2025,
       author = {{Akins}, Hollis B. and {Casey}, Caitlin M. and {Lambrides}, Erini and {Allen}, Natalie and {Andika}, Irham T. and {Brinch}, Malte and {Champagne}, Jaclyn B. and {Cooper}, Olivia and {Ding}, Xuheng and {Drakos}, Nicole E. and {Faisst}, Andreas and {Finkelstein}, Steven L. and {Franco}, Maximilien and {Fujimoto}, Seiji and {Gentile}, Fabrizio and {Gillman}, Steven and {Gozaliasl}, Ghassem and {Harish}, Santosh and {Hayward}, Christopher C. and {Hirschmann}, Michaela and {Ilbert}, Olivier and {Kartaltepe}, Jeyhan S. and {Kocevski}, Dale D. and {Koekemoer}, Anton M. and {Kokorev}, Vasily and {Liu}, Daizhong and {Long}, Arianna S. and {McCracken}, Henry Joy and {McKinney}, Jed and {Onoue}, Masafusa and {Paquereau}, Louise and {Renzini}, Alvio and {Rhodes}, Jason and {Robertson}, Brant E. and {Shuntov}, Marko and {Silverman}, John D. and {Tanaka}, Takumi S. and {Toft}, Sune and {Trakhtenbrot}, Benny and {Valentino}, Francesco and {Zavala}, Jorge},
        title = "{COSMOS-Web: The Overabundance and Physical Nature of ``Little Red Dots''{\textemdash}Implications for Early Galaxy and SMBH Assembly}",
      journal = {\apj},
         year = 2025,
        month = sep,
       volume = {991},
       number = {1},
          eid = {37},
        pages = {37},
          doi = {10.3847/1538-4357/ade984},
archivePrefix = {arXiv},
       eprint = {2406.10341},
 primaryClass = {astro-ph.GA},
       adsurl = {https://ui.adsabs.harvard.edu/abs/2025ApJ...991...37A}
}

@ARTICLE{Delvecchio2025,
       author = {{Delvecchio}, I. and {Daddi}, E. and {Magnelli}, B. and {Elbaz}, D. and {Giavalisco}, M. and {Traina}, A. and {Lanzuisi}, G. and {Akins}, H.~B. and {Belli}, S. and {Casey}, C.~M. and {Gentile}, F. and {Gruppioni}, C. and {Pozzi}, F. and {Zamorani}, G.},
        title = "{Active galactic nuclei-heated dust revealed in ``little red dots''}",
      journal = {\aap},
         year = 2025,
        month = dec,
       volume = {704},
          eid = {A313},
        pages = {A313},
          doi = {10.1051/0004-6361/202557164},
archivePrefix = {arXiv},
       eprint = {2509.07100},
 primaryClass = {astro-ph.GA},
       adsurl = {https://ui.adsabs.harvard.edu/abs/2025A&A...704A.313D}
}

@ARTICLE{Ananna2024,
       author = {{Ananna}, Tonima Tasnim and {Bogd{\'a}n}, {\'A}kos and {Kov{\'a}cs}, Orsolya E. and {Natarajan}, Priyamvada and {Hickox}, Ryan C.},
        title = "{X-Ray View of Little Red Dots: Do They Host Supermassive Black Holes?}",
      journal = {\apjl},
         year = 2024,
        month = jul,
       volume = {969},
       number = {1},
          eid = {L18},
        pages = {L18},
          doi = {10.3847/2041-8213/ad5669},
archivePrefix = {arXiv},
       eprint = {2404.19010},
 primaryClass = {astro-ph.GA},
       adsurl = {https://ui.adsabs.harvard.edu/abs/2024ApJ...969L..18A}
}

@ARTICLE{Casey2025,
       author = {{Casey}, Caitlin M. and {Akins}, Hollis B. and {Finkelstein}, Steven L. and {Franco}, Maximilien and {Fujimoto}, Seiji and {Liu}, Daizhong and {Long}, Arianna S. and {Magdis}, Georgios and {Manning}, Sinclaire M. and {McKinney}, Jed and {Shuntov}, Marko and {Tanaka}, Takumi S.},
        title = "{An Upper Limit of {}10$^{6}$ M$_{{\ensuremath{\odot}}}$ in Dust from ALMA Observations in 60 Little Red Dots}",
      journal = {\apjl},
         year = 2025,
        month = sep,
       volume = {990},
       number = {2},
          eid = {L61},
        pages = {L61},
          doi = {10.3847/2041-8213/adfa91},
archivePrefix = {arXiv},
       eprint = {2505.18873},
 primaryClass = {astro-ph.GA},
       adsurl = {https://ui.adsabs.harvard.edu/abs/2025ApJ...990L..61C}
}

@ARTICLE{Inayoshi2025_superEdd,
       author = {{Inayoshi}, Kohei and {Kimura}, Shigeo S. and {Noda}, Hirofumi},
        title = "{Weakness of X-rays and variability in high-redshift active galactic nuclei with super-Eddington accretion}",
      journal = {\pasj},
         year = 2025,
        month = aug,
       volume = {77},
       number = {4},
        pages = {811-822},
          doi = {10.1093/pasj/psaf050},
archivePrefix = {arXiv},
       eprint = {2412.03653},
 primaryClass = {astro-ph.HE},
       adsurl = {https://ui.adsabs.harvard.edu/abs/2025PASJ...77..811I}
}

@ARTICLE{Greene2024,
       author = {{Greene}, Jenny E. and {Labbe}, Ivo and {Goulding}, Andy D. and {Furtak}, Lukas J. and {Chemerynska}, Iryna and {Kokorev}, Vasily and {Dayal}, Pratika and {Volonteri}, Marta and {Williams}, Christina C. and {Wang}, Bingjie and {Setton}, David J. and {Burgasser}, Adam J. and {Bezanson}, Rachel and {Atek}, Hakim and {Brammer}, Gabriel and {Cutler}, Sam E. and {Feldmann}, Robert and {Fujimoto}, Seiji and {Glazebrook}, Karl and {de Graaff}, Anna and {Khullar}, Gourav and {Leja}, Joel and {Marchesini}, Danilo and {Maseda}, Michael V. and {Matthee}, Jorryt and {Miller}, Tim B. and {Naidu}, Rohan P. and {Nanayakkara}, Themiya and {Oesch}, Pascal A. and {Pan}, Richard and {Papovich}, Casey and {Price}, Sedona H. and {van Dokkum}, Pieter and {Weaver}, John R. and {Whitaker}, Katherine E. and {Zitrin}, Adi},
        title = "{UNCOVER Spectroscopy Confirms the Surprising Ubiquity of Active Galactic Nuclei in Red Sources at z > 5}",
      journal = {\apj},
         year = 2024,
        month = mar,
       volume = {964},
       number = {1},
          eid = {39},
        pages = {39},
          doi = {10.3847/1538-4357/ad1e5f},
archivePrefix = {arXiv},
       eprint = {2309.05714},
 primaryClass = {astro-ph.GA},
       adsurl = {https://ui.adsabs.harvard.edu/abs/2024ApJ...964...39G}
}

@ARTICLE{Inayoshi2025_num,
       author = {{Inayoshi}, Kohei},
        title = "{Little Red Dots as the Very First Activity of Black Hole Growth}",
      journal = {\apjl},
         year = 2025,
        month = jul,
       volume = {988},
       number = {1},
          eid = {L22},
        pages = {L22},
          doi = {10.3847/2041-8213/adea66},
archivePrefix = {arXiv},
       eprint = {2503.05537},
 primaryClass = {astro-ph.GA},
       adsurl = {https://ui.adsabs.harvard.edu/abs/2025ApJ...988L..22I}
}

@ARTICLE{Ma2025_lowz,
       author = {{Ma}, Yilun and {Greene}, Jenny E. and {Setton}, David J. and {Goulding}, Andy D. and {Annunziatella}, Marianna and {Fan}, Xiaohui and {Kokorev}, Vasily and {Labbe}, Ivo and {Li}, Jiaxuan and {Lin}, Xiaojing and {Marchesini}, Danilo and {Matthee}, Jorryt and {Robbins}, Luke and {Sajina}, Anna and {Sawicki}, Marcin and {Telford}, O. Grace},
        title = "{Counting Little Red Dots at z < 4 with Ground-based Surveys and Spectroscopic Follow-up}",
      journal = {\apj},
         year = 2026,
        month = mar,
       volume = {1000},
       number = {1},
          eid = {59},
        pages = {59},
          doi = {10.3847/1538-4357/ae4596},
archivePrefix = {arXiv},
       eprint = {2504.08032},
 primaryClass = {astro-ph.GA},
       adsurl = {https://ui.adsabs.harvard.edu/abs/2026ApJ..1000...59M}
}

@ARTICLE{Ma2025_LFcut,
       author = {{Ma}, Yilun and {Greene}, Jenny E. and {Volonteri}, Marta and {Goulding}, Andy D. and {Setton}, David J. and {Annunziatella}, Marianna and {Egami}, Eiichi and {Fan}, Xiaohui and {Kokorev}, Vasily and {Labbe}, Ivo and {Lin}, Xiaojing and {Marchesini}, Danilo and {Matthee}, Jorryt and {Nanayakkara}, Themiya and {Robbins}, Luke and {Sajina}, Anna and {Sawicki}, Marcin},
        title = "{No Luminous Little Red Dots: A Sharp Cutoff in Their Luminosity Function}",
      journal = {arXiv e-prints},
         year = 2025,
        month = sep,
          eid = {arXiv:2509.02662},
        pages = {arXiv:2509.02662},
          doi = {10.48550/arXiv.2509.02662},
archivePrefix = {arXiv},
       eprint = {2509.02662},
 primaryClass = {astro-ph.GA},
       adsurl = {https://ui.adsabs.harvard.edu/abs/2025arXiv250902662M}
}

@ARTICLE{Inayoshi2025_uv_opt,
       author = {{Inayoshi}, Kohei and {Murase}, Kohta and {Kashiyama}, Kazumi},
        title = "{Spectral Uniformity of Little Red Dots: A Natural Outcome of Coevolving Seed Black Holes and Nascent Starbursts}",
      journal = {\apj},
         year = 2026,
        month = mar,
       volume = {1000},
       number = {1},
          eid = {90},
        pages = {90},
          doi = {10.3847/1538-4357/ae42ce},
archivePrefix = {arXiv},
       eprint = {2509.19422},
 primaryClass = {astro-ph.GA},
       adsurl = {https://ui.adsabs.harvard.edu/abs/2026ApJ..1000...90I}
}

@ARTICLE{InayoshiMaiolino2025,
       author = {{Inayoshi}, Kohei and {Maiolino}, Roberto},
        title = "{Extremely Dense Gas around Little Red Dots and High-redshift Active Galactic Nuclei: A Nonstellar Origin of the Balmer Break and Absorption Features}",
      journal = {\apjl},
         year = 2025,
        month = feb,
       volume = {980},
       number = {2},
          eid = {L27},
        pages = {L27},
          doi = {10.3847/2041-8213/adaebd},
archivePrefix = {arXiv},
       eprint = {2409.07805},
 primaryClass = {astro-ph.GA},
       adsurl = {https://ui.adsabs.harvard.edu/abs/2025ApJ...980L..27I}
}

@ARTICLE{Naidu2025_BHstar,
       author = {{Naidu}, Rohan P. and {Matthee}, Jorryt and {Katz}, Harley and {de Graaff}, Anna and {Oesch}, Pascal and {Smith}, Aaron and {Greene}, Jenny E. and {Brammer}, Gabriel and {Weibel}, Andrea and {Hviding}, Raphael and {Chisholm}, John and {Labb\textbackslash'e}, Ivo and {Simcoe}, Robert A. and {Witten}, Callum and {Atek}, Hakim and {Baggen}, Josephine F.~W. and {Belli}, Sirio and {Bezanson}, Rachel and {Boogaard}, Leindert A. and {Bose}, Sownak and {Covelo-Paz}, Alba and {Dayal}, Pratika and {Fudamoto}, Yoshinobu and {Furtak}, Lukas J. and {Giovinazzo}, Emma and {Goulding}, Andy and {Gronke}, Max and {Heintz}, Kasper E. and {Hirschmann}, Michaela and {Illingworth}, Garth and {Inoue}, Akio K. and {Johnson}, Benjamin D. and {Leja}, Joel and {Leonova}, Ecaterina and {McConachie}, Ian and {Maseda}, Michael V. and {Natarajan}, Priyamvada and {Nelson}, Erica and {Setton}, David J. and {Shivaei}, Irene and {Sobral}, David and {Stefanon}, Mauro and {Tacchella}, Sandro and {Toft}, Sune and {Torralba}, Alberto and {van Dokkum}, Pieter and {van der Wel}, Arjen and {Volonteri}, Marta and {Walter}, Fabian and {Wang}, Bingjie and {Watson}, Darach},
        title = "{A ``Black Hole Star'' Reveals the Remarkable Gas-Enshrouded Hearts of the Little Red Dots}",
      journal = {arXiv e-prints},
         year = 2025,
        month = mar,
          eid = {arXiv:2503.16596},
        pages = {arXiv:2503.16596},
          doi = {10.48550/arXiv.2503.16596},
archivePrefix = {arXiv},
       eprint = {2503.16596},
 primaryClass = {astro-ph.GA},
       adsurl = {https://ui.adsabs.harvard.edu/abs/2025arXiv250316596N}
}

@ARTICLE{Liu2025_break,
       author = {{Liu}, Hanpu and {Jiang}, Yan-Fei and {Quataert}, Eliot and {Greene}, Jenny E. and {Ma}, Yilun},
        title = "{The Balmer Break and Optical Continuum of Little Red Dots from Super-Eddington Accretion}",
      journal = {\apj},
         year = 2025,
        month = nov,
       volume = {994},
       number = {1},
          eid = {113},
        pages = {113},
          doi = {10.3847/1538-4357/ae0c19},
archivePrefix = {arXiv},
       eprint = {2507.07190},
 primaryClass = {astro-ph.GA},
       adsurl = {https://ui.adsabs.harvard.edu/abs/2025ApJ...994..113L}
}

@ARTICLE{Pacucci2025_first,
       author = {{Pacucci}, Fabio and {Hernquist}, Lars and {Fujii}, Michiko},
        title = "{Little Red Dots are Nurseries of Massive Black Holes}",
      journal = {\apj},
         year = 2025,
        month = nov,
       volume = {994},
       number = {1},
          eid = {40},
        pages = {40},
          doi = {10.3847/1538-4357/ae1619},
archivePrefix = {arXiv},
       eprint = {2509.02664},
 primaryClass = {astro-ph.GA},
       adsurl = {https://ui.adsabs.harvard.edu/abs/2025ApJ...994...40P}
}

@ARTICLE{Pacucci2024_superEdd,
       author = {{Pacucci}, Fabio and {Narayan}, Ramesh},
        title = "{Mildly Super-Eddington Accretion onto Slowly Spinning Black Holes Explains the X-Ray Weakness of the Little Red Dots}",
      journal = {\apj},
         year = 2024,
        month = nov,
       volume = {976},
       number = {1},
          eid = {96},
        pages = {96},
          doi = {10.3847/1538-4357/ad84f7},
archivePrefix = {arXiv},
       eprint = {2407.15915},
 primaryClass = {astro-ph.HE},
       adsurl = {https://ui.adsabs.harvard.edu/abs/2024ApJ...976...96P}
}

@ARTICLE{Tanaka2025_z10,
       author = {{Tanaka}, Takumi S. and {Akins}, Hollis B. and {Harikane}, Yuichi and {Silverman}, John D. and {Casey}, Caitlin M. and {Inayoshi}, Kohei and {Schindler}, Jan-Torge and {Shimasaku}, Kazuhiro and {Kocevski}, Dale D. and {Onoue}, Masafusa and {Faisst}, Andreas L. and {Robertson}, Brant E. and {Kokorev}, Vasily and {Shuntov}, Marko and {Koekemoer}, Anton M. and {Franco}, Maximilien and {Egami}, Eiichi and {Liu}, Daizhong and {Taylor}, Anthony J. and {Kartaltepe}, Jeyhan S. and {Bosman}, Sarah E.~I. and {Champagne}, Jaclyn B. and {Kakiichi}, Koki and {Harish}, Santosh and {Zhang}, Zijian and {Newman}, Sophie L. and {Kakkad}, Darshan and {Fei}, Qinyue and {Fujimoto}, Seiji and {Li}, Mingyu and {Finkelstein}, Steven L. and {Li}, Zi-Jian and {Lambrides}, Erini and {Sommovigo}, Laura and {Zavala}, Jorge A. and {Ito}, Kei and {Liu}, Zhaoxuan and {Treister}, Ezequiel and {Aravena}, Manuel and {Gozaliasl}, Ghassem and {Zhang}, Haowen and {Hatamnia}, Hossein and {Umeda}, Hiroya and {Inoue}, Akio K. and {Yang}, Jinyi and {Ando}, Makoto and {Arita}, Junya and {Ding}, Xuheng and {Matsui}, Suin and {Shibanuma}, Yuki and {Magdis}, Georgios and {Zhuang}, Mingyang and {Fan}, Xiaohui and {Li}, Zihao and {Liu}, Weizhe and {Lyu}, Jianwei and {Rhodes}, Jason and {Toft}, Sune and {Wang}, Feige and {Zou}, Siwei and {Arango-Toro}, Rafael C. and {Battisti}, Andrew J. and {Gillman}, Steven and {Khostovan}, Ali Ahmad and {Long}, Arianna S. and {Mobasher}, Bahram and {Sanders}, David B.},
        title = "{Discovery of a Little Red Dot Candidate at z {\ensuremath{\gtrsim}} 10 in COSMOS-web Based on MIRI-NIRCam Selection}",
      journal = {\apj},
         year = 2025,
        month = dec,
       volume = {995},
       number = {1},
          eid = {21},
        pages = {21},
          doi = {10.3847/1538-4357/ae145f},
archivePrefix = {arXiv},
       eprint = {2508.00057},
 primaryClass = {astro-ph.GA},
       adsurl = {https://ui.adsabs.harvard.edu/abs/2025ApJ...995...21T}
}

@ARTICLE{Lin2025_local,
       author = {{Lin}, Xiaojing and {Fan}, Xiaohui and {Cai}, Zheng and {Bian}, Fuyan and {Liu}, Hanpu and {Sun}, Fengwu and {Ma}, Yilun and {Greene}, Jenny E. and {Strauss}, Michael A. and {Green}, Richard and {Lyu}, Jianwei and {Champagne}, Jaclyn B. and {Goulding}, Andy D. and {Inayoshi}, Kohei and {Jin}, Xiangyu and {Leung}, Gene C.~K. and {Li}, Mingyu and {Liu}, Weizhe and {Liu}, Yichen and {Mao}, Junjie and {Pudoka}, Maria Anne and {Tee}, Wei Leong and {Wang}, Ben and {Wang}, Feige and {Wu}, Yunjing and {Yang}, Jinyi and {Zhang}, Haowen and {Zhu}, Yongda},
        title = "{The Discovery of Little Red Dots in the Local Universe: Signatures of Cool Gas Envelopes}",
      journal = {\apj},
         year = 2026,
        month = feb,
       volume = {997},
       number = {2},
          eid = {364},
        pages = {364},
          doi = {10.3847/1538-4357/ae2bdf},
archivePrefix = {arXiv},
       eprint = {2507.10659},
 primaryClass = {astro-ph.GA},
       adsurl = {https://ui.adsabs.harvard.edu/abs/2026ApJ...997..364L}
}

@ARTICLE{Taylor2025_z9,
       author = {{Taylor}, Anthony J. and {Kokorev}, Vasily and {Kocevski}, Dale D. and {Akins}, Hollis B. and {Cullen}, Fergus and {Dickinson}, Mark and {Finkelstein}, Steven L. and {Arrabal Haro}, Pablo and {Bromm}, Volker and {Giavalisco}, Mauro and {Inayoshi}, Kohei and {Juneau}, St{\'e}phanie and {Leung}, Gene C.~K. and {P{\'e}rez-Gonz{\'a}lez}, Pablo G. and {Somerville}, Rachel S. and {Trump}, Jonathan R. and {Amor{\'\i}n}, Ricardo O. and {Barro}, Guillermo and {Burgarella}, Denis and {Brooks}, Madisyn and {Carnall}, Adam C. and {Casey}, Caitlin M. and {Cheng}, Yingjie and {Chisholm}, John and {Chworowsky}, Katherine and {Davis}, Kelcey and {Donnan}, Callum T. and {Dunlop}, James S. and {Ellis}, Richard S. and {Fern{\'a}ndez}, Vital and {Fujimoto}, Seiji and {Grogin}, Norman A. and {Gupta}, Ansh R. and {Hathi}, Nimish P. and {Jung}, Intae and {Hirschmann}, Michaela and {Kartaltepe}, Jeyhan S. and {Koekemoer}, Anton M. and {Larson}, Rebecca L. and {Leung}, Ho-Hin and {Llerena}, Mario and {Lucas}, Ray A. and {McLeod}, Derek J. and {McLure}, Ross and {Napolitano}, Lorenzo and {Papovich}, Casey and {Stanton}, Thomas M. and {Tripodi}, Roberta and {Wang}, Xin and {Wilkins}, Stephen M. and {Yung}, L.~Y. Aaron and {Zavala}, Jorge A.},
        title = "{CAPERS-LRD-z9: A Gas-enshrouded Little Red Dot Hosting a Broad-line Active Galactic Nucleus at z = 9.288}",
      journal = {\apjl},
         year = 2025,
        month = aug,
       volume = {989},
       number = {1},
          eid = {L7},
        pages = {L7},
          doi = {10.3847/2041-8213/ade789},
archivePrefix = {arXiv},
       eprint = {2505.04609},
 primaryClass = {astro-ph.GA},
       adsurl = {https://ui.adsabs.harvard.edu/abs/2025ApJ...989L...7T}
}

@ARTICLE{Lin2025_C3D,
       author = {{Lin}, Xiaojing and {Fan}, Xiaohui and {Wang}, Feige and {Sun}, Fengwu and {Champagne}, Jaclyn B. and {Egami}, Eiichi and {Kakiichi}, Koki and {Lyu}, Jianwei and {Tee}, Wei Leong and {Yang}, Jinyi and {Bian}, Fuyan and {Bosman}, Sarah E.~I. and {Cai}, Zheng and {Casey}, Caitlin M. and {Decarli}, Roberto and {Faisst}, Andreas L. and {Finkelstein}, Steven L. and {Fujimoto}, Seiji and {Harish}, Santosh and {Ilbert}, Olivier and {Inoue}, Akio K. and {Jin}, Xiangyu and {Kartaltepe}, Jeyhan S. and {Kocevski}, Dale D. and {Li}, Mingyu and {Liu}, Weizhe and {Liu}, Yichen and {Schindler}, Jan-Torge and {Shuntov}, Marko and {Tanaka}, Takumi S. and {Vestergaard}, Marianne and {Wu}, Yunjing and {Zhang}, Haowen and {Zhang}, Zijian},
        title = "{Bridging Quasars and Little Red Dots: Insights into Broad-line Active Galactic Nuclei at z = 5─8 from the First JWST COSMOS-3D Dataset}",
      journal = {\apj},
         year = 2026,
        month = jan,
       volume = {996},
       number = {1},
          eid = {93},
        pages = {93},
          doi = {10.3847/1538-4357/ae1b9b},
archivePrefix = {arXiv},
       eprint = {2504.08039},
 primaryClass = {astro-ph.GA},
       adsurl = {https://ui.adsabs.harvard.edu/abs/2026ApJ...996...93L}
}

@ARTICLE{KormendyHo2013,
       author = {{Kormendy}, John and {Ho}, Luis C.},
        title = "{Coevolution (Or Not) of Supermassive Black Holes and Host Galaxies}",
      journal = {\araa},
         year = 2013,
        month = aug,
       volume = {51},
       number = {1},
        pages = {511-653},
          doi = {10.1146/annurev-astro-082708-101811},
archivePrefix = {arXiv},
       eprint = {1304.7762},
 primaryClass = {astro-ph.CO},
       adsurl = {https://ui.adsabs.harvard.edu/abs/2013ARA&A..51..511K}
}

@ARTICLE{Volonteri2010,
       author = {{Volonteri}, Marta},
        title = "{Formation of supermassive black holes}",
      journal = {\aapr},
         year = 2010,
        month = jul,
       volume = {18},
       number = {3},
        pages = {279-315},
          doi = {10.1007/s00159-010-0029-x},
archivePrefix = {arXiv},
       eprint = {1003.4404},
 primaryClass = {astro-ph.CO},
       adsurl = {https://ui.adsabs.harvard.edu/abs/2010A&ARv..18..279V}
}

@ARTICLE{Kido2025,
       author = {{Kido}, Daisaburo and {Ioka}, Kunihito and {Hotokezaka}, Kenta and {Inayoshi}, Kohei and {Irwin}, Christopher M.},
        title = "{Black hole envelopes in Little Red Dots}",
      journal = {\mnras},
         year = 2025,
        month = dec,
       volume = {544},
       number = {4},
        pages = {3407-3416},
          doi = {10.1093/mnras/staf1898},
archivePrefix = {arXiv},
       eprint = {2505.06965},
 primaryClass = {astro-ph.HE},
       adsurl = {https://ui.adsabs.harvard.edu/abs/2025MNRAS.544.3407K}
}

@ARTICLE{deGraaff2025_Cliff,
       author = {{de Graaff}, Anna and {Rix}, Hans-Walter and {Naidu}, Rohan P. and {Labb{\'e}}, Ivo and {Wang}, Bingjie and {Leja}, Joel and {Matthee}, Jorryt and {Katz}, Harley and {Greene}, Jenny E. and {Hviding}, Raphael E. and {Baggen}, Josephine and {Bezanson}, Rachel and {Boogaard}, Leindert A. and {Brammer}, Gabriel and {Dayal}, Pratika and {van Dokkum}, Pieter and {Goulding}, Andy D. and {Hirschmann}, Michaela and {Maseda}, Michael V. and {McConachie}, Ian and {Miller}, Tim B. and {Nelson}, Erica and {Oesch}, Pascal A. and {Setton}, David J. and {Shivaei}, Irene and {Weibel}, Andrea and {Whitaker}, Katherine E. and {Williams}, Christina C.},
        title = "{A remarkable ruby: Absorption in dense gas, rather than evolved stars, drives the extreme Balmer break of a little red dot at z = 3.5}",
      journal = {\aap},
         year = 2025,
        month = sep,
       volume = {701},
          eid = {A168},
        pages = {A168},
          doi = {10.1051/0004-6361/202554681},
archivePrefix = {arXiv},
       eprint = {2503.16600},
 primaryClass = {astro-ph.GA},
       adsurl = {https://ui.adsabs.harvard.edu/abs/2025A&A...701A.168D}
}

@ARTICLE{2025arXiv251110927M,
       author = {{Moullet}, A. and {Burgarella}, D. and {Kataria}, T. and {Beuther}, H. and {Battersby}, C. and {Cheng}, M. and {Essinger-Hileman}, T. and {Inami}, H. and {Mills}, E. and {Nagao}, T. and {Unwin}, S.},
        title = "{PRIMA General Observer Science Book Volume 2}",
      journal = {arXiv e-prints},
         year = 2025,
        month = nov,
          eid = {arXiv:2511.10927},
        pages = {arXiv:2511.10927},
          doi = {10.48550/arXiv.2511.10927},
archivePrefix = {arXiv},
       eprint = {2511.10927},
 primaryClass = {astro-ph.IM},
       adsurl = {https://ui.adsabs.harvard.edu/abs/2025arXiv251110927M}
}

@ARTICLE{2010MNRAS.407.1701N,
       author = {{Narayanan}, Desika and {Dey}, Arjun and {Hayward}, Christopher C. and {Cox}, Thomas J. and {Bussmann}, R. Shane and {Brodwin}, Mark and {Jonsson}, Patrik and {Hopkins}, Philip F. and {Groves}, Brent and {Younger}, Joshua D. and {Hernquist}, Lars},
        title = "{A physical model for z \raisebox{-0.5ex}\textasciitilde 2 dust-obscured galaxies}",
      journal = {\mnras},
         year = 2010,
        month = sep,
       volume = {407},
       number = {3},
        pages = {1701-1720},
          doi = {10.1111/j.1365-2966.2010.16997.x},
archivePrefix = {arXiv},
       eprint = {0910.2234},
 primaryClass = {astro-ph.CO},
       adsurl = {https://ui.adsabs.harvard.edu/abs/2010MNRAS.407.1701N}
}

@ARTICLE{2015PASJ...67...86T,
       author = {{Toba}, Yoshiki and {Nagao}, Tohru and {Strauss}, Michael A. and {Aoki}, Kentaro and {Goto}, Tomotsugu and {Imanishi}, Masatoshi and {Kawaguchi}, Toshihiro and {Terashima}, Yuichi and {Ueda}, Yoshihiro and {Bosch}, James and {Bundy}, Kevin and {Doi}, Yoshiyuki and {Inami}, Hanae and {Komiyama}, Yutaka and {Lupton}, Robert H. and {Matsuhara}, Hideo and {Matsuoka}, Yoshiki and {Miyazaki}, Satoshi and {Morokuma}, Tomoki and {Nakata}, Fumiaki and {Oi}, Nagisa and {Onoue}, Masafusa and {Oyabu}, Shinki and {Price}, Paul and {Tait}, Philip J. and {Takata}, Tadafumi and {Tanaka}, Manobu M. and {Terai}, Tsuyoshi and {Turner}, Edwin L. and {Uchida}, Tomohisa and {Usuda}, Tomonori and {Utsumi}, Yousuke and {Yamada}, Yoshihiko and {Wang}, Shiang-Yu},
        title = "{Hyper-luminous dust-obscured galaxies discovered by the Hyper Suprime-Cam on Subaru and WISE}",
      journal = {\pasj},
         year = 2015,
        month = oct,
       volume = {67},
       number = {5},
          eid = {86},
        pages = {86},
          doi = {10.1093/pasj/psv057},
archivePrefix = {arXiv},
       eprint = {1506.00320},
 primaryClass = {astro-ph.GA},
       adsurl = {https://ui.adsabs.harvard.edu/abs/2015PASJ...67...86T}
}

@ARTICLE{2018MNRAS.478.3056B,
       author = {{Blecha}, Laura and {Snyder}, Gregory F. and {Satyapal}, Shobita and {Ellison}, Sara L.},
        title = "{The power of infrared AGN selection in mergers: a theoretical study}",
      journal = {\mnras},
         year = 2018,
        month = aug,
       volume = {478},
       number = {3},
        pages = {3056-3071},
          doi = {10.1093/mnras/sty1274},
archivePrefix = {arXiv},
       eprint = {1711.02094},
 primaryClass = {astro-ph.GA},
       adsurl = {https://ui.adsabs.harvard.edu/abs/2018MNRAS.478.3056B}
}

@ARTICLE{2018ARA&A..56..625H,
       author = {{Hickox}, Ryan C. and {Alexander}, David M.},
        title = "{Obscured Active Galactic Nuclei}",
      journal = {\araa},
         year = 2018,
        month = sep,
       volume = {56},
        pages = {625-671},
          doi = {10.1146/annurev-astro-081817-051803},
archivePrefix = {arXiv},
       eprint = {1806.04680},
 primaryClass = {astro-ph.GA},
       adsurl = {https://ui.adsabs.harvard.edu/abs/2018ARA&A..56..625H}
}

@ARTICLE{2019A&A...622A.103B,
       author = {{Boquien}, M. and {Burgarella}, D. and {Roehlly}, Y. and {Buat}, V. and {Ciesla}, L. and {Corre}, D. and {Inoue}, A.~K. and {Salas}, H.},
        title = "{CIGALE: a python Code Investigating GALaxy Emission}",
      journal = {\aap},
         year = 2019,
        month = feb,
       volume = {622},
          eid = {A103},
        pages = {A103},
          doi = {10.1051/0004-6361/201834156},
archivePrefix = {arXiv},
       eprint = {1811.03094},
 primaryClass = {astro-ph.GA},
       adsurl = {https://ui.adsabs.harvard.edu/abs/2019A&A...622A.103B}
}

@ARTICLE{2020A&A...640L...8U,
       author = {{Umehata}, Hideki and {Smail}, Ian and {Swinbank}, A.~M. and {Kohno}, Kotaro and {Tamura}, Yoichi and {Wang}, Tao and {Ao}, Yiping and {Hatsukade}, Bunyo and {Kubo}, Mariko and {Nakanishi}, Kouchiro and {Hayatsu}, Natsuki N.},
        title = "{ALMA Deep Field in SSA22. A near-infrared-dark submillimeter galaxy at z = 4.0}",
      journal = {\aap},
         year = 2020,
        month = aug,
       volume = {640},
          eid = {L8},
        pages = {L8},
          doi = {10.1051/0004-6361/202038146},
archivePrefix = {arXiv},
       eprint = {2007.09887},
 primaryClass = {astro-ph.GA},
       adsurl = {https://ui.adsabs.harvard.edu/abs/2020A&A...640L...8U}
}

@ARTICLE{2020ApJ...899...35T,
       author = {{Toba}, Yoshiki and {Goto}, Tomotsugu and {Oi}, Nagisa and {Wang}, Ting-Wen and {Kim}, Seong Jin and {Ho}, Simon C. -C. and {Burgarella}, Denis and {Hashimoto}, Tetsuya and {Hsieh}, Bau-Ching and {Huang}, Ting-Chi and {Hwang}, Ho Seong and {Ikeda}, Hiroyuki and {Kim}, Helen K. and {Kim}, Seongjae and {Lee}, Dongseob and {Malkan}, Matthew A. and {Matsuhara}, Hideo and {Miyaji}, Takamitsu and {Momose}, Rieko and {Ohyama}, Youichi and {Oyabu}, Shinki and {Pearson}, Chris and {Santos}, Daryl Joe D. and {Shim}, Hyunjin and {Takagi}, Toshinobu and {Ueda}, Yoshihiro and {Utsumi}, Yousuke and {Wada}, Takehiko},
        title = "{Search for Optically Dark Infrared Galaxies without Counterparts of Subaru Hyper Suprime-Cam in the AKARI North Ecliptic Pole Wide Survey Field}",
      journal = {\apj},
         year = 2020,
        month = aug,
       volume = {899},
       number = {1},
          eid = {35},
        pages = {35},
          doi = {10.3847/1538-4357/ab9cb7},
archivePrefix = {arXiv},
       eprint = {2006.07577},
 primaryClass = {astro-ph.GA},
       adsurl = {https://ui.adsabs.harvard.edu/abs/2020ApJ...899...35T}
}

@ARTICLE{2021Natur.597..489F,
       author = {{Fudamoto}, Y. and {Oesch}, P.~A. and {Schouws}, S. and {Stefanon}, M. and {Smit}, R. and {Bouwens}, R.~J. and {Bowler}, R.~A.~A. and {Endsley}, R. and {Gonzalez}, V. and {Inami}, H. and {Labbe}, I. and {Stark}, D. and {Aravena}, M. and {Barrufet}, L. and {da Cunha}, E. and {Dayal}, P. and {Ferrara}, A. and {Graziani}, L. and {Hodge}, J. and {Hutter}, A. and {Li}, Y. and {De Looze}, I. and {Nanayakkara}, T. and {Pallottini}, A. and {Riechers}, D. and {Schneider}, R. and {Ucci}, G. and {van der Werf}, P. and {White}, C.},
        title = "{Normal, dust-obscured galaxies in the epoch of reionization}",
      journal = {\nat},
         year = 2021,
        month = sep,
       volume = {597},
       number = {7877},
        pages = {489-492},
          doi = {10.1038/s41586-021-03846-z},
archivePrefix = {arXiv},
       eprint = {2109.10378},
 primaryClass = {astro-ph.GA},
       adsurl = {https://ui.adsabs.harvard.edu/abs/2021Natur.597..489F}
}

@ARTICLE{2022ApJ...926..155S,
       author = {{Shu}, Xinwen and {Yang}, Lei and {Liu}, Daizhong and {Wang}, Wei-Hao and {Wang}, Tao and {Han}, Yunkun and {Huang}, Xingxing and {Lim}, Chen-Fatt and {Chang}, Yu-Yen and {Zheng}, Wei and {Zheng}, XianZhong and {Wang}, Junxian and {Kong}, Xu},
        title = "{A Census of Optically Dark Massive Galaxies in the Early Universe from Magnification by Lensing Galaxy Clusters}",
      journal = {\apj},
         year = 2022,
        month = feb,
       volume = {926},
       number = {2},
          eid = {155},
        pages = {155},
          doi = {10.3847/1538-4357/ac3de5},
archivePrefix = {arXiv},
       eprint = {2112.03709},
 primaryClass = {astro-ph.GA},
       adsurl = {https://ui.adsabs.harvard.edu/abs/2022ApJ...926..155S}
}

@ARTICLE{2023MNRAS.522..449B,
       author = {{Barrufet}, L. and {Oesch}, P.~A. and {Weibel}, A. and {Brammer}, G. and {Bezanson}, R. and {Bouwens}, R. and {Fudamoto}, Y. and {Gonzalez}, V. and {Gottumukkala}, R. and {Illingworth}, G. and {Heintz}, K.~E. and {Holden}, B. and {Labbe}, I. and {Magee}, D. and {Naidu}, R.~P. and {Nelson}, E. and {Stefanon}, M. and {Smit}, R. and {van Dokkum}, P. and {Weaver}, J.~R. and {Williams}, C.~C.},
        title = "{Unveiling the nature of infrared bright, optically dark galaxies with early JWST data}",
      journal = {\mnras},
         year = 2023,
        month = jun,
       volume = {522},
       number = {1},
        pages = {449-456},
          doi = {10.1093/mnras/stad947},
archivePrefix = {arXiv},
       eprint = {2207.14733},
 primaryClass = {astro-ph.GA},
       adsurl = {https://ui.adsabs.harvard.edu/abs/2023MNRAS.522..449B}
}

@ARTICLE{2023ApJ...946L..16P,
       author = {{P{\'e}rez-Gonz{\'a}lez}, Pablo G. and {Barro}, Guillermo and {Annunziatella}, Marianna and {Costantin}, Luca and {Garc{\'\i}a-Argum{\'a}nez}, {\'A}ngela and {McGrath}, Elizabeth J. and {M{\'e}rida}, Rosa M. and {Zavala}, Jorge A. and {Arrabal Haro}, Pablo and {Bagley}, Micaela B. and {Backhaus}, Bren E. and {Behroozi}, Peter and {Bell}, Eric F. and {Bisigello}, Laura and {Buat}, V{\'e}ronique and {Calabr{\`o}}, Antonello and {Casey}, Caitlin M. and {Cleri}, Nikko J. and {Coogan}, Rosemary T. and {Cooper}, M.~C. and {Cooray}, Asantha R. and {Dekel}, Avishai and {Dickinson}, Mark and {Elbaz}, David and {Ferguson}, Henry C. and {Finkelstein}, Steven L. and {Fontana}, Adriano and {Franco}, Maximilien and {Gardner}, Jonathan P. and {Giavalisco}, Mauro and {G{\'o}mez-Guijarro}, Carlos and {Grazian}, Andrea and {Grogin}, Norman A. and {Guo}, Yuchen and {Huertas-Company}, Marc and {Jogee}, Shardha and {Kartaltepe}, Jeyhan S. and {Kewley}, Lisa J. and {Kirkpatrick}, Allison and {Kocevski}, Dale D. and {Koekemoer}, Anton M. and {Long}, Arianna S. and {Lotz}, Jennifer M. and {Lucas}, Ray A. and {Papovich}, Casey and {Pirzkal}, Nor and {Ravindranath}, Swara and {Somerville}, Rachel S. and {Tacchella}, Sandro and {Trump}, Jonathan R. and {Wang}, Weichen and {Wilkins}, Stephen M. and {Wuyts}, Stijn and {Yang}, Guang and {Yung}, L.~Y. Aaron},
        title = "{CEERS Key Paper. IV. A Triality in the Nature of HST-dark Galaxies}",
      journal = {\apjl},
         year = 2023,
        month = mar,
       volume = {946},
       number = {1},
          eid = {L16},
        pages = {L16},
          doi = {10.3847/2041-8213/acb3a5},
archivePrefix = {arXiv},
       eprint = {2211.00045},
 primaryClass = {astro-ph.GA},
       adsurl = {https://ui.adsabs.harvard.edu/abs/2023ApJ...946L..16P}
}

@ARTICLE{2022ApJ...936..118Y,
       author = {{Yutani}, Naomichi and {Toba}, Yoshiki and {Baba}, Shunsuke and {Wada}, Keiichi},
        title = "{Origin and Evolution of Dust-obscured Galaxies in Galaxy Mergers}",
      journal = {\apj},
         year = 2022,
        month = sep,
       volume = {936},
       number = {2},
          eid = {118},
        pages = {118},
          doi = {10.3847/1538-4357/ac87a2},
archivePrefix = {arXiv},
       eprint = {2205.00567},
 primaryClass = {astro-ph.GA},
       adsurl = {https://ui.adsabs.harvard.edu/abs/2022ApJ...936..118Y}
}

@ARTICLE{2006SSRv..123..485G,
       author = {{Gardner}, Jonathan P. and {Mather}, John C. and {Clampin}, Mark and {Doyon}, Rene and {Greenhouse}, Matthew A. and {Hammel}, Heidi B. and {Hutchings}, John B. and {Jakobsen}, Peter and {Lilly}, Simon J. and {Long}, Knox S. and {Lunine}, Jonathan I. and {McCaughrean}, Mark J. and {Mountain}, Matt and {Nella}, John and {Rieke}, George H. and {Rieke}, Marcia J. and {Rix}, Hans-Walter and {Smith}, Eric P. and {Sonneborn}, George and {Stiavelli}, Massimo and {Stockman}, H.~S. and {Windhorst}, Rogier A. and {Wright}, Gillian S.},
        title = "{The James Webb Space Telescope}",
      journal = {\ssr},
         year = 2006,
        month = apr,
       volume = {123},
       number = {4},
        pages = {485-606},
          doi = {10.1007/s11214-006-8315-7},
archivePrefix = {arXiv},
       eprint = {astro-ph/0606175},
 primaryClass = {astro-ph},
       adsurl = {https://ui.adsabs.harvard.edu/abs/2006SSRv..123..485G}
}

@ARTICLE{2022A&A...662A.112E,
       author = {{Euclid Collaboration} and {Scaramella}, R. and {Amiaux}, J. and {Mellier}, Y. and {Burigana}, C. and {Carvalho}, C.~S. and {Cuillandre}, J. -C. and {Da Silva}, A. and {Derosa}, A. and {Dinis}, J. and {Maiorano}, E. and {Maris}, M. and {Tereno}, I. and {Laureijs}, R. and {Boenke}, T. and {Buenadicha}, G. and {Dupac}, X. and {Gaspar Venancio}, L.~M. and {G{\'o}mez-{\'A}lvarez}, P. and {Hoar}, J. and {Lorenzo Alvarez}, J. and {Racca}, G.~D. and {Saavedra-Criado}, G. and {Schwartz}, J. and {Vavrek}, R. and {Schirmer}, M. and {Aussel}, H. and {Azzollini}, R. and {Cardone}, V.~F. and {Cropper}, M. and {Ealet}, A. and {Garilli}, B. and {Gillard}, W. and {Granett}, B.~R. and {Guzzo}, L. and {Hoekstra}, H. and {Jahnke}, K. and {Kitching}, T. and {Maciaszek}, T. and {Meneghetti}, M. and {Miller}, L. and {Nakajima}, R. and {Niemi}, S.~M. and {Pasian}, F. and {Percival}, W.~J. and {Pottinger}, S. and {Sauvage}, M. and {Scodeggio}, M. and {Wachter}, S. and {Zacchei}, A. and {Aghanim}, N. and {Amara}, A. and {Auphan}, T. and {Auricchio}, N. and {Awan}, S. and {Balestra}, A. and {Bender}, R. and {Bodendorf}, C. and {Bonino}, D. and {Branchini}, E. and {Brau-Nogue}, S. and {Brescia}, M. and {Candini}, G.~P. and {Capobianco}, V. and {Carbone}, C. and {Carlberg}, R.~G. and {Carretero}, J. and {Casas}, R. and {Castander}, F.~J. and {Castellano}, M. and {Cavuoti}, S. and {Cimatti}, A. and {Cledassou}, R. and {Congedo}, G. and {Conselice}, C.~J. and {Conversi}, L. and {Copin}, Y. and {Corcione}, L. and {Costille}, A. and {Courbin}, F. and {Degaudenzi}, H. and {Douspis}, M. and {Dubath}, F. and {Duncan}, C.~A.~J. and {Dusini}, S. and {Farrens}, S. and {Ferriol}, S. and {Fosalba}, P. and {Fourmanoit}, N. and {Frailis}, M. and {Franceschi}, E. and {Franzetti}, P. and {Fumana}, M. and {Gillis}, B. and {Giocoli}, C. and {Grazian}, A. and {Grupp}, F. and {Haugan}, S.~V.~H. and {Holmes}, W. and {Hormuth}, F. and {Hudelot}, P. and {Kermiche}, S. and {Kiessling}, A. and {Kilbinger}, M. and {Kohley}, R. and {Kubik}, B. and {K{\"u}mmel}, M. and {Kunz}, M. and {Kurki-Suonio}, H. and {Lahav}, O. and {Ligori}, S. and {Lilje}, P.~B. and {Lloro}, I. and {Mansutti}, O. and {Marggraf}, O. and {Markovic}, K. and {Marulli}, F. and {Massey}, R. and {Maurogordato}, S. and {Melchior}, M. and {Merlin}, E. and {Meylan}, G. and {Mohr}, J.~J. and {Moresco}, M. and {Morin}, B. and {Moscardini}, L. and {Munari}, E. and {Nichol}, R.~C. and {Padilla}, C. and {Paltani}, S. and {Peacock}, J. and {Pedersen}, K. and {Pettorino}, V. and {Pires}, S. and {Poncet}, M. and {Popa}, L. and {Pozzetti}, L. and {Raison}, F. and {Rebolo}, R. and {Rhodes}, J. and {Rix}, H. -W. and {Roncarelli}, M. and {Rossetti}, E. and {Saglia}, R. and {Schneider}, P. and {Schrabback}, T. and {Secroun}, A. and {Seidel}, G. and {Serrano}, S. and {Sirignano}, C. and {Sirri}, G. and {Skottfelt}, J. and {Stanco}, L. and {Starck}, J.~L. and {Tallada-Cresp{\'\i}}, P. and {Tavagnacco}, D. and {Taylor}, A.~N. and {Teplitz}, H.~I. and {Toledo-Moreo}, R. and {Torradeflot}, F. and {Trifoglio}, M. and {Valentijn}, E.~A. and {Valenziano}, L. and {Verdoes Kleijn}, G.~A. and {Wang}, Y. and {Welikala}, N. and {Weller}, J. and {Wetzstein}, M. and {Zamorani}, G. and {Zoubian}, J. and {Andreon}, S. and {Baldi}, M. and {Bardelli}, S. and {Boucaud}, A. and {Camera}, S. and {Di Ferdinando}, D. and {Fabbian}, G. and {Farinelli}, R. and {Galeotta}, S. and {Graci{\'a}-Carpio}, J. and {Maino}, D. and {Medinaceli}, E. and {Mei}, S. and {Neissner}, C. and {Polenta}, G. and {Renzi}, A. and {Romelli}, E. and {Rosset}, C. and {Sureau}, F. and {Tenti}, M. and {Vassallo}, T. and {Zucca}, E. and {Baccigalupi}, C. and {Balaguera-Antol{\'\i}nez}, A. and {Battaglia}, P. and {Biviano}, A. and {Borgani}, S. and {Bozzo}, E. and {Cabanac}, R. and {Cappi}, A. and {Casas}, S. and {Castignani}, G. and {Colodro-Conde}, C. and {Coupon}, J. and {Courtois}, H.~M. and {Cuby}, J. and {de la Torre}, S. and {Desai}, S. and {Dole}, H. and {Fabricius}, M. and {Farina}, M. and {Ferreira}, P.~G. and {Finelli}, F. and {Flose-Reimberg}, P. and {Fotopoulou}, S. and {Ganga}, K. and {Gozaliasl}, G. and {Hook}, I.~M. and {Keihanen}, E. and {Kirkpatrick}, C.~C. and {Liebing}, P. and {Lindholm}, V. and {Mainetti}, G. and {Martinelli}, M. and {Martinet}, N. and {Maturi}, M. and {McCracken}, H.~J. and {Metcalf}, R.~B. and {Morgante}, G. and {Nightingale}, J. and {Nucita}, A. and {Patrizii}, L. and {Potter}, D. and {Riccio}, G. and {S{\'a}nchez}, A.~G. and {Sapone}, D. and {Schewtschenko}, J.~A. and {Schultheis}, M. and {Scottez}, V. and {Teyssier}, R. and {Tutusaus}, I. and {Valiviita}, J. and {Viel}, M. and {Vriend}, W. and {Whittaker}, L.},
        title = "{Euclid preparation. I. The Euclid Wide Survey}",
      journal = {\aap},
         year = 2022,
        month = jun,
       volume = {662},
          eid = {A112},
        pages = {A112},
          doi = {10.1051/0004-6361/202141938},
archivePrefix = {arXiv},
       eprint = {2108.01201},
 primaryClass = {astro-ph.CO},
       adsurl = {https://ui.adsabs.harvard.edu/abs/2022A&A...662A.112E}
}

@ARTICLE{2019arXiv190205569A,
       author = {{Akeson}, Rachel and {Armus}, Lee and {Bachelet}, Etienne and {Bailey}, Vanessa and {Bartusek}, Lisa and {Bellini}, Andrea and {Benford}, Dominic and {Bennett}, David and {Bhattacharya}, Aparna and {Bohlin}, Ralph and {Boyer}, Martha and {Bozza}, Valerio and {Bryden}, Geoffrey and {Calchi Novati}, Sebastiano and {Carpenter}, Kenneth and {Casertano}, Stefano and {Choi}, Ami and {Content}, David and {Dayal}, Pratika and {Dressler}, Alan and {Dor{\'e}}, Olivier and {Fall}, S. Michael and {Fan}, Xiaohui and {Fang}, Xiao and {Filippenko}, Alexei and {Finkelstein}, Steven and {Foley}, Ryan and {Furlanetto}, Steven and {Kalirai}, Jason and {Gaudi}, B. Scott and {Gilbert}, Karoline and {Girard}, Julien and {Grady}, Kevin and {Greene}, Jenny and {Guhathakurta}, Puragra and {Heinrich}, Chen and {Hemmati}, Shoubaneh and {Hendel}, David and {Henderson}, Calen and {Henning}, Thomas and {Hirata}, Christopher and {Ho}, Shirley and {Huff}, Eric and {Hutter}, Anne and {Jansen}, Rolf and {Jha}, Saurabh and {Johnson}, Samson and {Jones}, David and {Kasdin}, Jeremy and {Kelly}, Patrick and {Kirshner}, Robert and {Koekemoer}, Anton and {Kruk}, Jeffrey and {Lewis}, Nikole and {Macintosh}, Bruce and {Madau}, Piero and {Malhotra}, Sangeeta and {Mandel}, Kaisey and {Massara}, Elena and {Masters}, Daniel and {McEnery}, Julie and {McQuinn}, Kristen and {Melchior}, Peter and {Melton}, Mark and {Mennesson}, Bertrand and {Peeples}, Molly and {Penny}, Matthew and {Perlmutter}, Saul and {Pisani}, Alice and {Plazas}, Andr{\'e}s and {Poleski}, Radek and {Postman}, Marc and {Ranc}, Cl{\'e}ment and {Rauscher}, Bernard and {Rest}, Armin and {Roberge}, Aki and {Robertson}, Brant and {Rodney}, Steven and {Rhoads}, James and {Rhodes}, Jason and {Ryan}, Russell, Jr. and {Sahu}, Kailash and {Sand}, David and {Scolnic}, Dan and {Seth}, Anil and {Shvartzvald}, Yossi and {Siellez}, Karelle and {Smith}, Arfon and {Spergel}, David and {Stassun}, Keivan and {Street}, Rachel and {Strolger}, Louis-Gregory and {Szalay}, Alexander and {Trauger}, John and {Troxel}, M.~A. and {Turnbull}, Margaret and {van der Marel}, Roeland and {von der Linden}, Anja and {Wang}, Yun and {Weinberg}, David and {Williams}, Benjamin and {Windhorst}, Rogier and {Wollack}, Edward and {Wu}, Hao-Yi and {Yee}, Jennifer and {Zimmerman}, Neil},
        title = "{The Wide Field Infrared Survey Telescope: 100 Hubbles for the 2020s}",
      journal = {arXiv e-prints},
         year = 2019,
        month = feb,
          eid = {arXiv:1902.05569},
        pages = {arXiv:1902.05569},
archivePrefix = {arXiv},
       eprint = {1902.05569},
 primaryClass = {astro-ph.IM},
       adsurl = {https://ui.adsabs.harvard.edu/abs/2019arXiv190205569A}
}

@ARTICLE{2022MNRAS.512.4248E,
       author = {{Endsley}, Ryan and {Stark}, Daniel P. and {Fan}, Xiaohui and {Smit}, Renske and {Wang}, Feige and {Yang}, Jinyi and {Hainline}, Kevin and {Lyu}, Jianwei and {Bouwens}, Rychard and {Schouws}, Sander},
        title = "{Radio and far-IR emission associated with a massive star-forming galaxy candidate at z ≃ 6.8: a radio-loud AGN in the reionization era?}",
      journal = {\mnras},
         year = 2022,
        month = may,
       volume = {512},
       number = {3},
        pages = {4248-4261},
          doi = {10.1093/mnras/stac737},
archivePrefix = {arXiv},
       eprint = {2108.01084},
 primaryClass = {astro-ph.GA},
       adsurl = {https://ui.adsabs.harvard.edu/abs/2022MNRAS.512.4248E}
}

@ARTICLE{2019ApJS..243...15T,
       author = {{Toba}, Yoshiki and {Yamashita}, Takuji and {Nagao}, Tohru and {Wang}, Wei-Hao and {Ueda}, Yoshihiro and {Ichikawa}, Kohei and {Kawaguchi}, Toshihiro and {Akiyama}, Masayuki and {Hsieh}, Bau-Ching and {Kajisawa}, Masaru and {Lee}, Chien-Hsiu and {Matsuoka}, Yoshiki and {Noboriguchi}, Akatoki and {Onoue}, Masafusa and {Schramm}, Malte and {Tanaka}, Masayuki and {Komiyama}, Yutaka},
        title = "{A Wide and Deep Exploration of Radio Galaxies with Subaru HSC (WERGS). II. Physical Properties Derived from the SED Fitting with Optical, Infrared, and Radio Data}",
      journal = {\apjs},
         year = 2019,
        month = jul,
       volume = {243},
       number = {1},
          eid = {15},
        pages = {15},
          doi = {10.3847/1538-4365/ab238d},
archivePrefix = {arXiv},
       eprint = {1905.01419},
 primaryClass = {astro-ph.GA},
       adsurl = {https://ui.adsabs.harvard.edu/abs/2019ApJS..243...15T}
}

@ARTICLE{2020AJ....160...60Y,
       author = {{Yamashita}, Takuji and {Nagao}, Tohru and {Ikeda}, Hiroyuki and {Toba}, Yoshiki and {Kajisawa}, Masaru and {Ono}, Yoshiaki and {Tanaka}, Masayuki and {Akiyama}, Masayuki and {Harikane}, Yuichi and {Ichikawa}, Kohei and {Kawaguchi}, Toshihiro and {Kawamuro}, Taiki and {Kohno}, Kotaro and {Lee}, Chien-Hsiu and {Lee}, Kianhong and {Matsuoka}, Yoshiki and {Niida}, Mana and {Ogura}, Kazuyuki and {Onoue}, Masafusa and {Uchiyama}, Hisakazu},
        title = "{A Wide and Deep Exploration of Radio Galaxies with Subaru HSC (WERGS). III. Discovery of a z = 4.72 Radio Galaxy with the Lyman Break Technique}",
      journal = {\aj},
         year = 2020,
        month = aug,
       volume = {160},
       number = {2},
          eid = {60},
        pages = {60},
          doi = {10.3847/1538-3881/ab98fe},
archivePrefix = {arXiv},
       eprint = {2006.04844},
 primaryClass = {astro-ph.GA},
       adsurl = {https://ui.adsabs.harvard.edu/abs/2020AJ....160...60Y}
}

@ARTICLE{2019Natur.572..211W,
       author = {{Wang}, T. and {Schreiber}, C. and {Elbaz}, D. and {Yoshimura}, Y. and {Kohno}, K. and {Shu}, X. and {Yamaguchi}, Y. and {Pannella}, M. and {Franco}, M. and {Huang}, J. and {Lim}, C.-F. and {Wang}, W.-H.},
        title = "{A dominant population of optically invisible massive galaxies in the early Universe}",
      journal = {\nat},
         year = 2019,
        month = aug,
       volume = {572},
       number = {7768},
        pages = {211-214},
          doi = {10.1038/s41586-019-1452-4},
archivePrefix = {arXiv},
       eprint = {1908.02372},
 primaryClass = {astro-ph.GA},
       adsurl = {https://ui.adsabs.harvard.edu/abs/2019Natur.572..211W}
}

@ARTICLE{Kormendy2013ARA&A..51..511K,
       author = {{Kormendy}, John and {Ho}, Luis C.},
        title = "{Coevolution (Or Not) of Supermassive Black Holes and Host Galaxies}",
      journal = {\araa},
         year = 2013,
        month = aug,
       volume = {51},
       number = {1},
        pages = {511-653},
          doi = {10.1146/annurev-astro-082708-101811},
archivePrefix = {arXiv},
       eprint = {1304.7762},
 primaryClass = {astro-ph.CO},
       adsurl = {https://ui.adsabs.harvard.edu/abs/2013ARA&A..51..511K}
}

@ARTICLE{Alexander2012NewAR..56...93A,
       author = {{Alexander}, D.~M. and {Hickox}, R.~C.},
        title = "{What drives the growth of black holes?}",
      journal = {\nar},
         year = 2012,
        month = jun,
       volume = {56},
       number = {4},
        pages = {93-121},
          doi = {10.1016/j.newar.2011.11.003},
archivePrefix = {arXiv},
       eprint = {1112.1949},
 primaryClass = {astro-ph.GA},
       adsurl = {https://ui.adsabs.harvard.edu/abs/2012NewAR..56...93A}
}

@ARTICLE{Blecha2018MNRAS.478.3056B,
       author = {{Blecha}, Laura and {Snyder}, Gregory F. and {Satyapal}, Shobita and {Ellison}, Sara L.},
        title = "{The power of infrared AGN selection in mergers: a theoretical study}",
      journal = {\mnras},
         year = 2018,
        month = aug,
       volume = {478},
       number = {3},
        pages = {3056-3071},
          doi = {10.1093/mnras/sty1274},
archivePrefix = {arXiv},
       eprint = {1711.02094},
 primaryClass = {astro-ph.GA},
       adsurl = {https://ui.adsabs.harvard.edu/abs/2018MNRAS.478.3056B}
}

@ARTICLE{Ricci2017MNRAS.468.1273R,
       author = {{Ricci}, C. and {Bauer}, F.~E. and {Treister}, E. and {Schawinski}, K. and {Privon}, G.~C. and {Blecha}, L. and {Arevalo}, P. and {Armus}, L. and {Harrison}, F. and {Ho}, L.~C. and et al.},
        title = "{Growing supermassive black holes in the late stages of galaxy mergers are heavily obscured}",
      journal = {\mnras},
         year = 2017,
        month = jun,
       volume = {468},
       number = {2},
        pages = {1273-1299},
          doi = {10.1093/mnras/stx173},
archivePrefix = {arXiv},
       eprint = {1701.04825},
 primaryClass = {astro-ph.HE},
       adsurl = {https://ui.adsabs.harvard.edu/abs/2017MNRAS.468.1273R}
}

@ARTICLE{Koss2012ApJ...746L..22K,
       author = {{Koss}, Michael and {Mushotzky}, Richard and {Treister}, Ezequiel and {Veilleux}, Sylvain and {Vasudevan}, Ranjan and {Trippe}, Margaret},
        title = "{Understanding Dual Active Galactic Nucleus Activation in the nearby Universe}",
      journal = {\apjl},
         year = 2012,
        month = feb,
       volume = {746},
       number = {2},
          eid = {L22},
        pages = {L22},
          doi = {10.1088/2041-8205/746/2/L22},
archivePrefix = {arXiv},
       eprint = {1201.2944},
 primaryClass = {astro-ph.HE},
       adsurl = {https://ui.adsabs.harvard.edu/abs/2012ApJ...746L..22K}
}

@ARTICLE{Silverman2020ApJ...899..154S,
       author = {{Silverman}, John D. and {Tang}, Shenli and {Lee}, Khee-Gan and {Hartwig}, Tilman and {Goulding}, Andy and {Strauss}, Michael A. and {Schramm}, Malte and {Ding}, Xuheng and {Riffel}, Rogemar A. and {Fujimoto}, Seiji and et al.},
        title = "{Dual Supermassive Black Holes at Close Separation Revealed by the Hyper Suprime-Cam Subaru Strategic Program}",
      journal = {\apj},
         year = 2020,
        month = aug,
       volume = {899},
       number = {2},
          eid = {154},
        pages = {154},
          doi = {10.3847/1538-4357/aba4a3},
archivePrefix = {arXiv},
       eprint = {2007.05581},
 primaryClass = {astro-ph.GA},
       adsurl = {https://ui.adsabs.harvard.edu/abs/2020ApJ...899..154S}
}

@ARTICLE{Li2025ApJ...986..101L,
       author = {{Li}, Junyao and {Zhuang}, Ming-Yang and {Shen}, Yue and {Volonteri}, Marta and {Chen}, Nianyi and {Matteo}, Tiziana Di},
        title = "{Active Galactic Nuclei and Host Galaxies in COSMOS-Web. II. First Look At the Kiloparsec-scale Dual and Offset AGN Population}",
      journal = {\apj},
         year = 2025,
        month = jun,
       volume = {986},
       number = {1},
          eid = {101},
        pages = {101},
          doi = {10.3847/1538-4357/adbae2},
archivePrefix = {arXiv},
       eprint = {2405.14980},
 primaryClass = {astro-ph.GA},
       adsurl = {https://ui.adsabs.harvard.edu/abs/2025ApJ...986..101L}
}

@ARTICLE{Agazie2023ApJ...951L...8A,
       author = {{Agazie}, Gabriella and {Anumarlapudi}, Akash and {Archibald}, Anne M. and {Arzoumanian}, Zaven and {Baker}, Paul T. and {B{\'e}csy}, Bence and {Blecha}, Laura and {Brazier}, Adam and {Brook}, Paul R. and {Burke-Spolaor}, Sarah and et al.},
        title = "{The NANOGrav 15 yr Data Set: Evidence for a Gravitational-wave Background}",
      journal = {\apjl},
         year = 2023,
        month = jul,
       volume = {951},
       number = {1},
          eid = {L8},
        pages = {L8},
          doi = {10.3847/2041-8213/acdac6},
archivePrefix = {arXiv},
       eprint = {2306.16213},
 primaryClass = {astro-ph.HE},
       adsurl = {https://ui.adsabs.harvard.edu/abs/2023ApJ...951L...8A}
}

@ARTICLE{Draine2003ARA&A..41..241D,
       author = {{Draine}, B.~T.},
        title = "{Interstellar Dust Grains}",
      journal = {\araa},
         year = 2003,
        month = jan,
       volume = {41},
        pages = {241-289},
          doi = {10.1146/annurev.astro.41.011802.094840},
archivePrefix = {arXiv},
       eprint = {astro-ph/0304489},
 primaryClass = {astro-ph},
       adsurl = {https://ui.adsabs.harvard.edu/abs/2003ARA&A..41..241D}
}

@ARTICLE{Lonsdale2004ApJS..154...54L,
       author = {{Lonsdale}, Carol and {Polletta}, Maria del Carmen and {Surace}, Jason and {Shupe}, Dave and {Fang}, Fan and {Xu}, C. Kevin and {Smith}, Harding E. and {Siana}, Brian and {Rowan-Robinson}, Michael and {Babbedge}, Tom and {Oliver}, Seb and {Pozzi}, Francesca and {Davoodi}, Payam and {Owen}, Frazer and {Padgett}, Deborah and {Frayer}, Dave and {Jarrett}, Tom and {Masci}, Frank and {O'Linger}, JoAnne and {Conrow}, Tim and {Farrah}, Duncan and {Morrison}, Glenn and {Gautier}, Nick and {Franceschini}, Alberto and {Berta}, Stefano and {Perez-Fournon}, Ismael and {Hatziminaoglou}, Evanthia and {Afonso-Luis}, Alejandro and {Dole}, Herve and {Stacey}, Gordon and {Serjeant}, Steve and {Pierre}, Marguerite and {Griffin}, Matt and {Puetter}, Rick},
        title = "{First Insights into the Spitzer Wide-Area Infrared Extragalactic Legacy Survey (SWIRE) Galaxy Populations}",
      journal = {\apjs},
         year = 2004,
        month = sep,
       volume = {154},
       number = {1},
        pages = {54-59},
          doi = {10.1086/423206},
archivePrefix = {arXiv},
       eprint = {astro-ph/0406209},
 primaryClass = {astro-ph},
       adsurl = {https://ui.adsabs.harvard.edu/abs/2004ApJS..154...54L}
}

@ARTICLE{Carilli2004NewAR..48..979C,
       author = {{Carilli}, C.~L. and {Rawlings}, S.},
        title = "{Motivation, key science projects, standards and assumptions}",
      journal = {\nar},
         year = 2004,
        month = dec,
       volume = {48},
       number = {11-12},
        pages = {979-984},
          doi = {10.1016/j.newar.2004.09.001},
archivePrefix = {arXiv},
       eprint = {astro-ph/0409274},
 primaryClass = {astro-ph},
       adsurl = {https://ui.adsabs.harvard.edu/abs/2004NewAR..48..979C}
}

@ARTICLE{Colpi2024arXiv240207571C,
       author = {{Colpi}, Monica and {Danzmann}, Karsten and {Hewitson}, Martin and {Holley-Bockelmann}, Kelly and {Jetzer}, Philippe and {Nelemans}, Gijs and {Petiteau}, Antoine and {Shoemaker}, David and {Sopuerta}, Carlos and {Stebbins}, Robin and {Tanvir}, Nial and {Ward}, Henry and {Weber}, William Joseph and {Thorpe}, Ira and {Daurskikh}, Anna and {Deep}, Atul and {Fern{\'a}ndez N{\'u}{\~n}ez}, Ignacio and {Garc{\'\i}a Marirrodriga}, C{\'e}sar and {Gehler}, Martin and {Halain}, Jean-Philippe and {Jennrich}, Oliver and {Lammers}, Uwe and {Larra{\~n}aga}, Jonan and {Lieser}, Maike and {L{\"u}tzgendorf}, Nora and {Martens}, Waldemar and {Mondin}, Linda and {Piris Ni{\~n}o}, Ana and {Amaro-Seoane}, Pau and {Arca Sedda}, Manuel and {Auclair}, Pierre and {Babak}, Stanislav and {Baghi}, Quentin and {Baibhav}, Vishal and {Baker}, Tessa and {Bayle}, Jean-Baptiste and {Berry}, Christopher and {Berti}, Emanuele and {Boileau}, Guillaume and {Bonetti}, Matteo and {Brito}, Richard and {Buscicchio}, Riccardo and {Calcagni}, Gianluca and {Capelo}, Pedro R. and {Caprini}, Chiara and {Caputo}, Andrea and {Castelli}, Eleonora and {Chen}, Hsin-Yu and {Chen}, Xian and {Chua}, Alvin and {Davies}, Gareth and {Derdzinski}, Andrea and {Domcke}, Valerie Fiona and {Doneva}, Daniela and {Dvorkin}, Irna and {Mar{\'\i}a Ezquiaga}, Jose and {Gair}, Jonathan and {Haiman}, Zoltan and {Harry}, Ian and {Hartwig}, Olaf and {Hees}, Aurelien and {Heffernan}, Anna and {Husa}, Sascha and {Izquierdo-Villalba}, David and {Karnesis}, Nikolaos and {Klein}, Antoine and {Korol}, Valeriya and {Korsakova}, Natalia and {Kupfer}, Thomas and {Laghi}, Danny and {Lamberts}, Astrid and {Larson}, Shane and {Le Jeune}, Maude and {Lewicki}, Marek and {Littenberg}, Tyson and {Madge}, Eric and {Mangiagli}, Alberto and {Marsat}, Sylvain and {Vilchez}, Ivan Martin and {Maselli}, Andrea and {Mathews}, Josh and {van de Meent}, Maarten and {Muratore}, Martina and {Nardini}, Germano and {Pani}, Paolo and {Peloso}, Marco and {Pieroni}, Mauro and {Pound}, Adam and {Quelquejay-Leclere}, Hippolyte and {Ricciardone}, Angelo and {Rossi}, Elena Maria and {Sartirana}, Andrea and {Savalle}, Etienne and {Sberna}, Laura and {Sesana}, Alberto and {Shoemaker}, Deirdre and {Slutsky}, Jacob and {Sotiriou}, Thomas and {Speri}, Lorenzo and {Staab}, Martin and {Steer}, Dani{\`e}le and {Tamanini}, Nicola and {Tasinato}, Gianmassimo and {Torrado}, Jesus and {Torres-Orjuela}, Alejandro and {Toubiana}, Alexandre and {Vallisneri}, Michele and {Vecchio}, Alberto and {Volonteri}, Marta and {Yagi}, Kent and {Zwick}, Lorenz},
        title = "{LISA Definition Study Report}",
      journal = {arXiv e-prints},
         year = 2024,
        month = feb,
          eid = {arXiv:2402.07571},
        pages = {arXiv:2402.07571},
          doi = {10.48550/arXiv.2402.07571},
archivePrefix = {arXiv},
       eprint = {2402.07571},
 primaryClass = {astro-ph.CO},
       adsurl = {https://ui.adsabs.harvard.edu/abs/2024arXiv240207571C}
}

@ARTICLE{Polletta2007ApJ...663...81P,
       author = {{Polletta}, M. and {Tajer}, M. and {Maraschi}, L. and {Trinchieri}, G. and {Lonsdale}, C.~J. and {Chiappetti}, L. and {Andreon}, S. and {Pierre}, M. and {Le F{\`e}vre}, O. and {Zamorani}, G. and {Maccagni}, D. and {Garcet}, O. and {Surdej}, J. and {Franceschini}, A. and {Alloin}, D. and {Shupe}, D.~L. and {Surace}, J.~A. and {Fang}, F. and {Rowan-Robinson}, M. and {Smith}, H.~E. and {Tresse}, L.},
        title = "{Spectral Energy Distributions of Hard X-Ray Selected Active Galactic Nuclei in the XMM-Newton Medium Deep Survey}",
      journal = {\apj},
         year = 2007,
        month = jul,
       volume = {663},
       number = {1},
        pages = {81-102},
          doi = {10.1086/518113},
archivePrefix = {arXiv},
       eprint = {astro-ph/0703255},
 primaryClass = {astro-ph},
       adsurl = {https://ui.adsabs.harvard.edu/abs/2007ApJ...663...81P}
}

@ARTICLE{Kusakabe2026ApJ...999..117K,
       author = {{Kusakabe}, Katsunori and {Inoue}, Yoshiyuki and {Toyouchi}, Daisuke},
        title = "{Coherence of Supermassive Black Hole Binary Demographics with the nHz Stochastic Gravitational-wave Background}",
      journal = {\apj},
         year = 2026,
        month = mar,
       volume = {999},
       number = {1},
          eid = {117},
        pages = {117},
          doi = {10.3847/1538-4357/ae3962},
archivePrefix = {arXiv},
       eprint = {2510.10548},
 primaryClass = {astro-ph.HE},
       adsurl = {https://ui.adsabs.harvard.edu/abs/2026ApJ...999..117K}
}

@ARTICLE{Mateos2015MNRAS.449.1422M,
       author = {{Mateos}, S. and {Carrera}, F.~J. and {Alonso-Herrero}, A. and {Rovilos}, E. and {Hern{\'a}n-Caballero}, A. and {Barcons}, X. and {Blain}, A. and {Caccianiga}, A. and {Della Ceca}, R. and {Severgnini}, P.},
        title = "{Revisiting the relationship between 6 {\ensuremath{\mu}}m and 2-10 keV continuum luminosities of AGN}",
      journal = {\mnras},
         year = 2015,
        month = may,
       volume = {449},
       number = {2},
        pages = {1422-1440},
          doi = {10.1093/mnras/stv299},
archivePrefix = {arXiv},
       eprint = {1501.04335},
 primaryClass = {astro-ph.GA},
       adsurl = {https://ui.adsabs.harvard.edu/abs/2015MNRAS.449.1422M}
}

@ARTICLE{1989AJ.....97..957N,
       author = {{Neugebauer}, G. and {Soifer}, B.~T. and {Matthews}, K. and {Elias}, J.~H.},
        title = "{The Near-Infrared Variability of a Sample of Optically Selected Quasars}",
      journal = {\aj},
         year = 1989,
        month = apr,
       volume = {97},
        pages = {957},
          doi = {10.1086/115040},
       adsurl = {https://ui.adsabs.harvard.edu/abs/1989AJ.....97..957N}
}

@ARTICLE{1993PASP..105..247P,
       author = {{Peterson}, Bradley M.},
        title = "{Reverberation Mapping of Active Galactic Nuclei}",
      journal = {\pasp},
         year = 1993,
        month = mar,
       volume = {105},
        pages = {247},
          doi = {10.1086/133140},
       adsurl = {https://ui.adsabs.harvard.edu/abs/1993PASP..105..247P}
}

@ARTICLE{1999AJ....118...35N,
       author = {{Neugebauer}, G. and {Matthews}, K.},
        title = "{Variability of Quasars at 10 Microns}",
      journal = {\aj},
         year = 1999,
        month = jul,
       volume = {118},
       number = {1},
        pages = {35-45},
          doi = {10.1086/300945},
archivePrefix = {arXiv},
       eprint = {astro-ph/9903363},
 primaryClass = {astro-ph},
       adsurl = {https://ui.adsabs.harvard.edu/abs/1999AJ....118...35N}
}

@ARTICLE{2001AJ....122..549V,
       author = {{Vanden Berk}, Daniel E. and {Richards}, Gordon T. and {Bauer}, Amanda and {Strauss}, Michael A. and {Schneider}, Donald P. and {Heckman}, Timothy M. and {York}, Donald G. and {Hall}, Patrick B. and {Fan}, Xiaohui and {Knapp}, G.~R. and {Anderson}, Scott F. and {Annis}, James and {Bahcall}, Neta A. and {Bernardi}, Mariangela and {Briggs}, John W. and {Brinkmann}, J. and {Brunner}, Robert and {Burles}, Scott and {Carey}, Larry and {Castander}, Francisco J. and {Connolly}, A.~J. and {Crocker}, J.~H. and {Csabai}, Istv{\'a}n and {Doi}, Mamoru and {Finkbeiner}, Douglas and {Friedman}, Scott and {Frieman}, Joshua A. and {Fukugita}, Masataka and {Gunn}, James E. and {Hennessy}, G.~S. and {Ivezi{\'c}}, {\v{Z}}eljko and {Kent}, Stephen and {Kunszt}, Peter Z. and {Lamb}, D.~Q. and {Leger}, R. French and {Long}, Daniel C. and {Loveday}, Jon and {Lupton}, Robert H. and {Meiksin}, Avery and {Merelli}, Aronne and {Munn}, Jeffrey A. and {Newberg}, Heidi Jo and {Newcomb}, Matt and {Nichol}, R.~C. and {Owen}, Russell and {Pier}, Jeffrey R. and {Pope}, Adrian and {Rockosi}, Constance M. and {Schlegel}, David J. and {Siegmund}, Walter A. and {Smee}, Stephen and {Snir}, Yehuda and {Stoughton}, Chris and {Stubbs}, Christopher and {SubbaRao}, Mark and {Szalay}, Alexander S. and {Szokoly}, Gyula P. and {Tremonti}, Christy and {Uomoto}, Alan and {Waddell}, Patrick and {Yanny}, Brian and {Zheng}, Wei},
        title = "{Composite Quasar Spectra from the Sloan Digital Sky Survey}",
      journal = {\aj},
         year = 2001,
        month = aug,
       volume = {122},
       number = {2},
        pages = {549-564},
          doi = {10.1086/321167},
archivePrefix = {arXiv},
       eprint = {astro-ph/0105231},
 primaryClass = {astro-ph},
       adsurl = {https://ui.adsabs.harvard.edu/abs/2001AJ....122..549V}
}

@ARTICLE{2003MNRAS.345.1271V,
       author = {{Vaughan}, S. and {Edelson}, R. and {Warwick}, R.~S. and {Uttley}, P.},
        title = "{On characterizing the variability properties of X-ray light curves from active galaxies}",
      journal = {\mnras},
         year = 2003,
        month = nov,
       volume = {345},
       number = {4},
        pages = {1271-1284},
          doi = {10.1046/j.1365-2966.2003.07042.x},
archivePrefix = {arXiv},
       eprint = {astro-ph/0307420},
 primaryClass = {astro-ph},
       adsurl = {https://ui.adsabs.harvard.edu/abs/2003MNRAS.345.1271V}
}

@ARTICLE{2004ApJ...601..692V,
       author = {{Vanden Berk}, Daniel E. and {Wilhite}, Brian C. and {Kron}, Richard G. and {Anderson}, Scott F. and {Brunner}, Robert J. and {Hall}, Patrick B. and {Ivezi{\'c}}, {\v{Z}}eljko and {Richards}, Gordon T. and {Schneider}, Donald P. and {York}, Donald G. and {Brinkmann}, Jonathan V. and {Lamb}, Don Q. and {Nichol}, Robert C. and {Schlegel}, David J.},
        title = "{The Ensemble Photometric Variability of \raisebox{-0.5ex}\textasciitilde25,000 Quasars in the Sloan Digital Sky Survey}",
      journal = {\apj},
         year = 2004,
        month = feb,
       volume = {601},
       number = {2},
        pages = {692-714},
          doi = {10.1086/380563},
archivePrefix = {arXiv},
       eprint = {astro-ph/0310336},
 primaryClass = {astro-ph},
       adsurl = {https://ui.adsabs.harvard.edu/abs/2004ApJ...601..692V}
}

@ARTICLE{2004ApJ...613..682P,
       author = {{Peterson}, B.~M. and {Ferrarese}, L. and {Gilbert}, K.~M. and {Kaspi}, S. and {Malkan}, M.~A. and {Maoz}, D. and {Merritt}, D. and {Netzer}, H. and {Onken}, C.~A. and {Pogge}, R.~W. and {Vestergaard}, M. and {Wandel}, A.},
        title = "{Central Masses and Broad-Line Region Sizes of Active Galactic Nuclei. II. A Homogeneous Analysis of a Large Reverberation-Mapping Database}",
      journal = {\apj},
         year = 2004,
        month = oct,
       volume = {613},
       number = {2},
        pages = {682-699},
          doi = {10.1086/423269},
archivePrefix = {arXiv},
       eprint = {astro-ph/0407299},
 primaryClass = {astro-ph},
       adsurl = {https://ui.adsabs.harvard.edu/abs/2004ApJ...613..682P}
}

@ARTICLE{2006ApJS..166..470R,
       author = {{Richards}, Gordon T. and {Lacy}, Mark and {Storrie-Lombardi}, Lisa J. and {Hall}, Patrick B. and {Gallagher}, S.~C. and {Hines}, Dean C. and {Fan}, Xiaohui and {Papovich}, Casey and {Vanden Berk}, Daniel E. and {Trammell}, George B. and {Schneider}, Donald P. and {Vestergaard}, Marianne and {York}, Donald G. and {Jester}, Sebastian and {Anderson}, Scott F. and {Budav{\'a}ri}, Tam{\'a}s and {Szalay}, Alexander S.},
        title = "{Spectral Energy Distributions and Multiwavelength Selection of Type 1 Quasars}",
      journal = {\apjs},
         year = 2006,
        month = oct,
       volume = {166},
       number = {2},
        pages = {470-497},
          doi = {10.1086/506525},
archivePrefix = {arXiv},
       eprint = {astro-ph/0601558},
 primaryClass = {astro-ph},
       adsurl = {https://ui.adsabs.harvard.edu/abs/2006ApJS..166..470R}
}

@ARTICLE{2006ApJ...640..579G,
       author = {{Glikman}, Eilat and {Helfand}, David J. and {White}, Richard L.},
        title = "{A Near-Infrared Spectral Template for Quasars}",
      journal = {\apj},
         year = 2006,
        month = apr,
       volume = {640},
       number = {2},
        pages = {579-591},
          doi = {10.1086/500098},
archivePrefix = {arXiv},
       eprint = {astro-ph/0511640},
 primaryClass = {astro-ph},
       adsurl = {https://ui.adsabs.harvard.edu/abs/2006ApJ...640..579G}
}

@ARTICLE{2006ApJ...639...46S,
       author = {{Suganuma}, Masahiro and {Yoshii}, Yuzuru and {Kobayashi}, Yukiyasu and {Minezaki}, Takeo and {Enya}, Keigo and {Tomita}, Hiroyuki and {Aoki}, Tsutomu and {Koshida}, Shintaro and {Peterson}, Bruce A.},
        title = "{Reverberation Measurements of the Inner Radius of the Dust Torus in Nearby Seyfert 1 Galaxies}",
      journal = {\apj},
         year = 2006,
        month = mar,
       volume = {639},
       number = {1},
        pages = {46-63},
          doi = {10.1086/499326},
archivePrefix = {arXiv},
       eprint = {astro-ph/0511697},
 primaryClass = {astro-ph},
       adsurl = {https://ui.adsabs.harvard.edu/abs/2006ApJ...639...46S}
}

@ARTICLE{2008ApJ...676..121M,
       author = {{Morokuma}, Tomoki and {Doi}, Mamoru and {Yasuda}, Naoki and {Akiyama}, Masayuki and {Sekiguchi}, Kazuhiro and {Furusawa}, Hisanori and {Ueda}, Yoshihiro and {Totani}, Tomonori and {Oda}, Takeshi and {Nagao}, Tohru and {Kashikawa}, Nobunari and {Murayama}, Takashi and {Ouchi}, Masami and {Watson}, Mike G.},
        title = "{The Subaru/XMM-Newton Deep Survey (SXDS). VI. Properties of Active Galactic Nuclei Selected by Optical Variability}",
      journal = {\apj},
         year = 2008,
        month = mar,
       volume = {676},
       number = {1},
        pages = {121-130},
          doi = {10.1086/528788},
archivePrefix = {arXiv},
       eprint = {0712.3106},
 primaryClass = {astro-ph},
       adsurl = {https://ui.adsabs.harvard.edu/abs/2008ApJ...676..121M}
}

@ARTICLE{2009ApJ...697.1656S,
       author = {{Shen}, Yue and {Strauss}, Michael A. and {Ross}, Nicholas P. and {Hall}, Patrick B. and {Lin}, Yen-Ting and {Richards}, Gordon T. and {Schneider}, Donald P. and {Weinberg}, David H. and {Connolly}, Andrew J. and {Fan}, Xiaohui and {Hennawi}, Joseph F. and {Shankar}, Francesco and {Vanden Berk}, Daniel E. and {Bahcall}, Neta A. and {Brunner}, Robert J.},
        title = "{Quasar Clustering from SDSS DR5: Dependences on Physical Properties}",
      journal = {\apj},
         year = 2009,
        month = jun,
       volume = {697},
       number = {2},
        pages = {1656-1673},
          doi = {10.1088/0004-637X/697/2/1656},
archivePrefix = {arXiv},
       eprint = {0810.4144},
 primaryClass = {astro-ph},
       adsurl = {https://ui.adsabs.harvard.edu/abs/2009ApJ...697.1656S}
}

@ARTICLE{2009ApJ...698..895K,
       author = {{Kelly}, Brandon C. and {Bechtold}, Jill and {Siemiginowska}, Aneta},
        title = "{Are the Variations in Quasar Optical Flux Driven by Thermal Fluctuations?}",
      journal = {\apj},
         year = 2009,
        month = jun,
       volume = {698},
       number = {1},
        pages = {895-910},
          doi = {10.1088/0004-637X/698/1/895},
archivePrefix = {arXiv},
       eprint = {0903.5315},
 primaryClass = {astro-ph.CO},
       adsurl = {https://ui.adsabs.harvard.edu/abs/2009ApJ...698..895K}
}

@ARTICLE{2009arXiv0912.0201L,
       author = {{LSST Science Collaboration} and {Abell}, Paul A. and {Allison}, Julius and {Anderson}, Scott F. and {Andrew}, John R. and {Angel}, J. Roger P. and {Armus}, Lee and {Arnett}, David and {Asztalos}, S.~J. and {Axelrod}, Tim S. and {Bailey}, Stephen and {Ballantyne}, D.~R. and {Bankert}, Justin R. and {Barkhouse}, Wayne A. and {Barr}, Jeffrey D. and {Barrientos}, L. Felipe and {Barth}, Aaron J. and {Bartlett}, James G. and {Becker}, Andrew C. and {Becla}, Jacek and {Beers}, Timothy C. and {Bernstein}, Joseph P. and {Biswas}, Rahul and {Blanton}, Michael R. and {Bloom}, Joshua S. and {Bochanski}, John J. and {Boeshaar}, Pat and {Borne}, Kirk D. and {Bradac}, Marusa and {Brandt}, W.~N. and {Bridge}, Carrie R. and {Brown}, Michael E. and {Brunner}, Robert J. and {Bullock}, James S. and {Burgasser}, Adam J. and {Burge}, James H. and {Burke}, David L. and {Cargile}, Phillip A. and {Chandrasekharan}, Srinivasan and {Chartas}, George and {Chesley}, Steven R. and {Chu}, You-Hua and {Cinabro}, David and {Claire}, Mark W. and {Claver}, Charles F. and {Clowe}, Douglas and {Connolly}, A.~J. and {Cook}, Kem H. and {Cooke}, Jeff and {Cooray}, Asantha and {Covey}, Kevin R. and {Culliton}, Christopher S. and {de Jong}, Roelof and {de Vries}, Willem H. and {Debattista}, Victor P. and {Delgado}, Francisco and {Dell'Antonio}, Ian P. and {Dhital}, Saurav and {Di Stefano}, Rosanne and {Dickinson}, Mark and {Dilday}, Benjamin and {Djorgovski}, S.~G. and {Dobler}, Gregory and {Donalek}, Ciro and {Dubois-Felsmann}, Gregory and {Durech}, Josef and {Eliasdottir}, Ardis and {Eracleous}, Michael and {Eyer}, Laurent and {Falco}, Emilio E. and {Fan}, Xiaohui and {Fassnacht}, Christopher D. and {Ferguson}, Harry C. and {Fernandez}, Yanga R. and {Fields}, Brian D. and {Finkbeiner}, Douglas and {Figueroa}, Eduardo E. and {Fox}, Derek B. and {Francke}, Harold and {Frank}, James S. and {Frieman}, Josh and {Fromenteau}, Sebastien and {Furqan}, Muhammad and {Galaz}, Gaspar and {Gal-Yam}, A. and {Garnavich}, Peter and {Gawiser}, Eric and {Geary}, John and {Gee}, Perry and {Gibson}, Robert R. and {Gilmore}, Kirk and {Grace}, Emily A. and {Green}, Richard F. and {Gressler}, William J. and {Grillmair}, Carl J. and {Habib}, Salman and {Haggerty}, J.~S. and {Hamuy}, Mario and {Harris}, Alan W. and {Hawley}, Suzanne L. and {Heavens}, Alan F. and {Hebb}, Leslie and {Henry}, Todd J. and {Hileman}, Edward and {Hilton}, Eric J. and {Hoadley}, Keri and {Holberg}, J.~B. and {Holman}, Matt J. and {Howell}, Steve B. and {Infante}, Leopoldo and {Ivezic}, Zeljko and {Jacoby}, Suzanne H. and {Jain}, Bhuvnesh and {R} and {Jedicke} and {Jee}, M. James and {Garrett Jernigan}, J. and {Jha}, Saurabh W. and {Johnston}, Kathryn V. and {Jones}, R. Lynne and {Juric}, Mario and {Kaasalainen}, Mikko and {Styliani} and {Kafka} and {Kahn}, Steven M. and {Kaib}, Nathan A. and {Kalirai}, Jason and {Kantor}, Jeff and {Kasliwal}, Mansi M. and {Keeton}, Charles R. and {Kessler}, Richard and {Knezevic}, Zoran and {Kowalski}, Adam and {Krabbendam}, Victor L. and {Krughoff}, K. Simon and {Kulkarni}, Shrinivas and {Kuhlman}, Stephen and {Lacy}, Mark and {Lepine}, Sebastien and {Liang}, Ming and {Lien}, Amy and {Lira}, Paulina and {Long}, Knox S. and {Lorenz}, Suzanne and {Lotz}, Jennifer M. and {Lupton}, R.~H. and {Lutz}, Julie and {Macri}, Lucas M. and {Mahabal}, Ashish A. and {Mandelbaum}, Rachel and {Marshall}, Phil and {May}, Morgan and {McGehee}, Peregrine M. and {Meadows}, Brian T. and {Meert}, Alan and {Milani}, Andrea and {Miller}, Christopher J. and {Miller}, Michelle and {Mills}, David and {Minniti}, Dante and {Monet}, David and {Mukadam}, Anjum S. and {Nakar}, Ehud and {Neill}, Douglas R. and {Newman}, Jeffrey A. and {Nikolaev}, Sergei and {Nordby}, Martin and {O'Connor}, Paul and {Oguri}, Masamune and {Oliver}, John and {Olivier}, Scot S. and {Olsen}, Julia K. and {Olsen}, Knut and {Olszewski}, Edward W. and {Oluseyi}, Hakeem and {Padilla}, Nelson D. and {Parker}, Alex and {Pepper}, Joshua and {Peterson}, John R. and {Petry}, Catherine and {Pinto}, Philip A. and {Pizagno}, James L. and {Popescu}, Bogdan and {Prsa}, Andrej and {Radcka}, Veljko and {Raddick}, M. Jordan and {Rasmussen}, Andrew and {Rau}, Arne and {Rho}, Jeonghee and {Rhoads}, James E. and {Richards}, Gordon T. and {Ridgway}, Stephen T. and {Robertson}, Brant E. and {Roskar}, Rok and {Saha}, Abhijit and {Sarajedini}, Ata and {Scannapieco}, Evan and {Schalk}, Terry and {Schindler}, Rafe and {Schmidt}, Samuel},
        title = "{LSST Science Book, Version 2.0}",
      journal = {arXiv e-prints},
         year = 2009,
        month = dec,
          eid = {arXiv:0912.0201},
        pages = {arXiv:0912.0201},
          doi = {10.48550/arXiv.0912.0201},
archivePrefix = {arXiv},
       eprint = {0912.0201},
 primaryClass = {astro-ph.IM},
       adsurl = {https://ui.adsabs.harvard.edu/abs/2009arXiv0912.0201L}
}

@ARTICLE{2010ApJ...708..927K,
       author = {{Koz{\l}owski}, Szymon and {Kochanek}, Christopher S. and {Udalski}, A. and {Wyrzykowski}, {\L}. and {Soszy{\'n}ski}, I. and {Szyma{\'n}ski}, M.~K. and {Kubiak}, M. and {Pietrzy{\'n}ski}, G. and {Szewczyk}, O. and {Ulaczyk}, K. and {Poleski}, R. and {OGLE Collaboration}},
        title = "{Quantifying Quasar Variability as Part of a General Approach to Classifying Continuously Varying Sources}",
      journal = {\apj},
         year = 2010,
        month = jan,
       volume = {708},
       number = {2},
        pages = {927-945},
          doi = {10.1088/0004-637X/708/2/927},
archivePrefix = {arXiv},
       eprint = {0909.1326},
 primaryClass = {astro-ph.CO},
       adsurl = {https://ui.adsabs.harvard.edu/abs/2010ApJ...708..927K}
}

@ARTICLE{2010ApJ...716..530K,
       author = {{Koz{\l}owski}, Szymon and {Kochanek}, Christopher S. and {Stern}, Daniel and {Ashby}, Matthew L.~N. and {Assef}, Roberto J. and {Bock}, J.~J. and {Borys}, C. and {Brand}, K. and {Brodwin}, M. and {Brown}, M.~J.~I. and {Cool}, R. and {Cooray}, A. and {Croft}, S. and {Dey}, Arjun and {Eisenhardt}, P.~R. and {Gonzalez}, A. and {Gorjian}, V. and {Griffith}, R. and {Grogin}, N. and {Ivison}, R. and {Jacob}, J. and {Jannuzi}, B.~T. and {Mainzer}, A. and {Moustakas}, L. and {R{\"o}ttgering}, H. and {Seymour}, N. and {Smith}, H.~A. and {Stanford}, S.~A. and {Stauffer}, J.~R. and {Sullivan}, I.~S. and {van Breugel}, W. and {Willner}, S.~P. and {Wright}, E.~L.},
        title = "{Mid-infrared Variability from the Spitzer Deep Wide-field Survey}",
      journal = {\apj},
         year = 2010,
        month = jun,
       volume = {716},
       number = {1},
        pages = {530-543},
          doi = {10.1088/0004-637X/716/1/530},
archivePrefix = {arXiv},
       eprint = {1002.3365},
 primaryClass = {astro-ph.CO},
       adsurl = {https://ui.adsabs.harvard.edu/abs/2010ApJ...716..530K}
}

@ARTICLE{2010ApJ...721.1014M,
       author = {{MacLeod}, C.~L. and {Ivezi{\'c}}, {\v{Z}}. and {Kochanek}, C.~S. and {Koz{\l}owski}, S. and {Kelly}, B. and {Bullock}, E. and {Kimball}, A. and {Sesar}, B. and {Westman}, D. and {Brooks}, K. and {Gibson}, R. and {Becker}, A.~C. and {de Vries}, W.~H.},
        title = "{Modeling the Time Variability of SDSS Stripe 82 Quasars as a Damped Random Walk}",
      journal = {\apj},
         year = 2010,
        month = oct,
       volume = {721},
       number = {2},
        pages = {1014-1033},
          doi = {10.1088/0004-637X/721/2/1014},
archivePrefix = {arXiv},
       eprint = {1004.0276},
 primaryClass = {astro-ph.CO},
       adsurl = {https://ui.adsabs.harvard.edu/abs/2010ApJ...721.1014M}
}

@ARTICLE{2011ApJ...740L..49W,
       author = {{Watson}, D. and {Denney}, K.~D. and {Vestergaard}, M. and {Davis}, T.~M.},
        title = "{A New Cosmological Distance Measure Using Active Galactic Nuclei}",
      journal = {\apjl},
         year = 2011,
        month = oct,
       volume = {740},
       number = {2},
          eid = {L49},
        pages = {L49},
          doi = {10.1088/2041-8205/740/2/L49},
archivePrefix = {arXiv},
       eprint = {1109.4632},
 primaryClass = {astro-ph.CO},
       adsurl = {https://ui.adsabs.harvard.edu/abs/2011ApJ...740L..49W}
}

@ARTICLE{2013ApJ...767..149B,
       author = {{Bentz}, Misty C. and {Denney}, Kelly D. and {Grier}, Catherine J. and {Barth}, Aaron J. and {Peterson}, Bradley M. and {Vestergaard}, Marianne and {Bennert}, Vardha N. and {Canalizo}, Gabriela and {De Rosa}, Gisella and {Filippenko}, Alexei V. and {Gates}, Elinor L. and {Greene}, Jenny E. and {Li}, Weidong and {Malkan}, Matthew A. and {Pogge}, Richard W. and {Stern}, Daniel and {Treu}, Tommaso and {Woo}, Jong-Hak},
        title = "{The Low-luminosity End of the Radius-Luminosity Relationship for Active Galactic Nuclei}",
      journal = {\apj},
         year = 2013,
        month = apr,
       volume = {767},
       number = {2},
          eid = {149},
        pages = {149},
          doi = {10.1088/0004-637X/767/2/149},
archivePrefix = {arXiv},
       eprint = {1303.1742},
 primaryClass = {astro-ph.CO},
       adsurl = {https://ui.adsabs.harvard.edu/abs/2013ApJ...767..149B}
}

@ARTICLE{2014ApJ...788..159K,
       author = {{Koshida}, Shintaro and {Minezaki}, Takeo and {Yoshii}, Yuzuru and {Kobayashi}, Yukiyasu and {Sakata}, Yu and {Sugawara}, Shota and {Enya}, Keigo and {Suganuma}, Masahiro and {Tomita}, Hiroyuki and {Aoki}, Tsutomu and {Peterson}, Bruce A.},
        title = "{Reverberation Measurements of the Inner Radius of the Dust Torus in 17 Seyfert Galaxies}",
      journal = {\apj},
         year = 2014,
        month = jun,
       volume = {788},
       number = {2},
          eid = {159},
        pages = {159},
          doi = {10.1088/0004-637X/788/2/159},
archivePrefix = {arXiv},
       eprint = {1406.2078},
 primaryClass = {astro-ph.GA},
       adsurl = {https://ui.adsabs.harvard.edu/abs/2014ApJ...788..159K}
}

@ARTICLE{2014MNRAS.441.3454K,
       author = {{King}, Anthea L. and {Davis}, Tamara M. and {Denney}, K.~D. and {Vestergaard}, M. and {Watson}, D.},
        title = "{High-redshift standard candles: predicted cosmological constraints}",
      journal = {\mnras},
         year = 2014,
        month = jul,
       volume = {441},
       number = {4},
        pages = {3454-3476},
          doi = {10.1093/mnras/stu793},
archivePrefix = {arXiv},
       eprint = {1311.2356},
 primaryClass = {astro-ph.CO},
       adsurl = {https://ui.adsabs.harvard.edu/abs/2014MNRAS.441.3454K}
}

@ARTICLE{2014ApJ...784L...4H,
       author = {{H{\"o}nig}, S.~F.},
        title = "{Dust Reverberation Mapping in the Era of Big Optical Surveys and its Cosmological Application}",
      journal = {\apjl},
         year = 2014,
        month = mar,
       volume = {784},
       number = {1},
          eid = {L4},
        pages = {L4},
          doi = {10.1088/2041-8205/784/1/L4},
archivePrefix = {arXiv},
       eprint = {1401.2999},
 primaryClass = {astro-ph.GA},
       adsurl = {https://ui.adsabs.harvard.edu/abs/2014ApJ...784L...4H}
}

@ARTICLE{2014ApJ...784L..11Y,
       author = {{Yoshii}, Yuzuru and {Kobayashi}, Yukiyasu and {Minezaki}, Takeo and {Koshida}, Shintaro and {Peterson}, Bruce A.},
        title = "{A New Method for Measuring Extragalactic Distances}",
      journal = {\apjl},
         year = 2014,
        month = mar,
       volume = {784},
       number = {1},
          eid = {L11},
        pages = {L11},
          doi = {10.1088/2041-8205/784/1/L11},
archivePrefix = {arXiv},
       eprint = {1403.1693},
 primaryClass = {astro-ph.CO},
       adsurl = {https://ui.adsabs.harvard.edu/abs/2014ApJ...784L..11Y}
}

@ARTICLE{2014AJ....148...13R,
       author = {{Rodney}, Steven A. and {Riess}, Adam G. and {Strolger}, Louis-Gregory and {Dahlen}, Tomas and {Graur}, Or and {Casertano}, Stefano and {Dickinson}, Mark E. and {Ferguson}, Henry C. and {Garnavich}, Peter and {Hayden}, Brian and {Jha}, Saurabh W. and {Jones}, David O. and {Kirshner}, Robert P. and {Koekemoer}, Anton M. and {McCully}, Curtis and {Mobasher}, Bahram and {Patel}, Brandon and {Weiner}, Benjamin J. and {Cenko}, S. Bradley and {Clubb}, Kelsey I. and {Cooper}, Michael and {Filippenko}, Alexei V. and {Frederiksen}, Teddy F. and {Hjorth}, Jens and {Leibundgut}, Bruno and {Matheson}, Thomas and {Nayyeri}, Hooshang and {Penner}, Kyle and {Trump}, Jonathan and {Silverman}, Jeffrey M. and {U}, Vivian and {Azalee Bostroem}, K. and {Challis}, Peter and {Rajan}, Abhijith and {Wolff}, Schuyler and {Faber}, S.~M. and {Grogin}, Norman A. and {Kocevski}, Dale},
        title = "{Type Ia Supernova Rate Measurements to Redshift 2.5 from CANDELS: Searching for Prompt Explosions in the Early Universe}",
      journal = {\aj},
         year = 2014,
        month = jul,
       volume = {148},
       number = {1},
          eid = {13},
        pages = {13},
          doi = {10.1088/0004-6256/148/1/13},
archivePrefix = {arXiv},
       eprint = {1401.7978},
 primaryClass = {astro-ph.CO},
       adsurl = {https://ui.adsabs.harvard.edu/abs/2014AJ....148...13R}
}

@ARTICLE{2015ApJS..221...27M,
       author = {{Myers}, Adam D. and {Palanque-Delabrouille}, Nathalie and {Prakash}, Abhishek and {P{\^a}ris}, Isabelle and {Yeche}, Christophe and {Dawson}, Kyle S. and {Bovy}, Jo and {Lang}, Dustin and {Schlegel}, David J. and {Newman}, Jeffrey A. and {Petitjean}, Patrick and {Kneib}, Jean-Paul and {Laurent}, Pierre and {Percival}, Will J. and {Ross}, Ashley J. and {Seo}, Hee-Jong and {Tinker}, Jeremy L. and {Armengaud}, Eric and {Brownstein}, Joel and {Burtin}, Etienne and {Cai}, Zheng and {Comparat}, Johan and {Kasliwal}, Mansi and {Kulkarni}, Shrinivas R. and {Laher}, Russ and {Levitan}, David and {McBride}, Cameron K. and {McGreer}, Ian D. and {Miller}, Adam A. and {Nugent}, Peter and {Ofek}, Eran and {Rossi}, Graziano and {Ruan}, John and {Schneider}, Donald P. and {Sesar}, Branimir and {Streblyanska}, Alina and {Surace}, Jason},
        title = "{The SDSS-IV Extended Baryon Oscillation Spectroscopic Survey: Quasar Target Selection}",
      journal = {\apjs},
         year = 2015,
        month = dec,
       volume = {221},
       number = {2},
          eid = {27},
        pages = {27},
          doi = {10.1088/0067-0049/221/2/27},
archivePrefix = {arXiv},
       eprint = {1508.04472},
 primaryClass = {astro-ph.CO},
       adsurl = {https://ui.adsabs.harvard.edu/abs/2015ApJS..221...27M}
}

@ARTICLE{2015ApJS..216....4S,
       author = {{Shen}, Yue and {Brandt}, W.~N. and {Dawson}, Kyle S. and {Hall}, Patrick B. and {McGreer}, Ian D. and {Anderson}, Scott F. and {Chen}, Yuguang and {Denney}, Kelly D. and {Eftekharzadeh}, Sarah and {Fan}, Xiaohui and {Gao}, Yang and {Green}, Paul J. and {Greene}, Jenny E. and {Ho}, Luis C. and {Horne}, Keith and {Jiang}, Linhua and {Kelly}, Brandon C. and {Kinemuchi}, Karen and {Kochanek}, Christopher S. and {P{\^a}ris}, Isabelle and {Peters}, Christina M. and {Peterson}, Bradley M. and {Petitjean}, Patrick and {Ponder}, Kara and {Richards}, Gordon T. and {Schneider}, Donald P. and {Seth}, Anil and {Smith}, Robyn N. and {Strauss}, Michael A. and {Tao}, Charling and {Trump}, Jonathan R. and {Wood-Vasey}, W.~M. and {Zu}, Ying and {Eisenstein}, Daniel J. and {Pan}, Kaike and {Bizyaev}, Dmitry and {Malanushenko}, Viktor and {Malanushenko}, Elena and {Oravetz}, Daniel},
        title = "{The Sloan Digital Sky Survey Reverberation Mapping Project: Technical Overview}",
      journal = {\apjs},
         year = 2015,
        month = jan,
       volume = {216},
       number = {1},
          eid = {4},
        pages = {4},
          doi = {10.1088/0067-0049/216/1/4},
archivePrefix = {arXiv},
       eprint = {1408.5970},
 primaryClass = {astro-ph.IM},
       adsurl = {https://ui.adsabs.harvard.edu/abs/2015ApJS..216....4S}
}

@ARTICLE{2015MNRAS.453.1701K,
       author = {{King}, Anthea L. and {Martini}, Paul and {Davis}, Tamara M. and {Denney}, K.~D. and {Kochanek}, C.~S. and {Peterson}, Bradley M. and {Skielboe}, Andreas and {Vestergaard}, Marianne and {Huff}, Eric and {Watson}, Darach and {Banerji}, Manda and {McMahon}, Richard and {Sharp}, Rob and {Lidman}, C.},
        title = "{Simulations of the OzDES AGN reverberation mapping project}",
      journal = {\mnras},
         year = 2015,
        month = oct,
       volume = {453},
       number = {2},
        pages = {1701-1726},
          doi = {10.1093/mnras/stv1718},
archivePrefix = {arXiv},
       eprint = {1504.03031},
 primaryClass = {astro-ph.GA},
       adsurl = {https://ui.adsabs.harvard.edu/abs/2015MNRAS.453.1701K}
}

@ARTICLE{2016MNRAS.463.2064H,
       author = {{Hern{\'a}n-Caballero}, Antonio and {Hatziminaoglou}, Evanthia and {Alonso-Herrero}, Almudena and {Mateos}, Silvia},
        title = "{The near-to-mid infrared spectrum of quasars}",
      journal = {\mnras},
         year = 2016,
        month = dec,
       volume = {463},
       number = {2},
        pages = {2064-2078},
          doi = {10.1093/mnras/stw2107},
archivePrefix = {arXiv},
       eprint = {1605.04867},
 primaryClass = {astro-ph.GA},
       adsurl = {https://ui.adsabs.harvard.edu/abs/2016MNRAS.463.2064H}
}

@ARTICLE{2016A&A...587A..41P,
       author = {{Palanque-Delabrouille}, N. and {Magneville}, Ch. and {Y{\`e}che}, Ch. and {P{\^a}ris}, I. and {Petitjean}, P. and {Burtin}, E. and {Dawson}, K. and {McGreer}, I. and {Myers}, A.~D. and {Rossi}, G. and {Schlegel}, D. and {Schneider}, D. and {Streblyanska}, A. and {Tinker}, J.},
        title = "{The extended Baryon Oscillation Spectroscopic Survey: Variability selection and quasar luminosity function}",
      journal = {\aap},
         year = 2016,
        month = mar,
       volume = {587},
          eid = {A41},
        pages = {A41},
          doi = {10.1051/0004-6361/201527392},
archivePrefix = {arXiv},
       eprint = {1509.05607},
 primaryClass = {astro-ph.CO},
       adsurl = {https://ui.adsabs.harvard.edu/abs/2016A&A...587A..41P}
}

@ARTICLE{2017MNRAS.464.1693H,
       author = {{H{\"o}nig}, S.~F. and {Watson}, D. and {Kishimoto}, M. and {Gandhi}, P. and {Goad}, M. and {Horne}, K. and {Shankar}, F. and {Banerji}, M. and {Boulderstone}, B. and {Jarvis}, M. and {Smith}, M. and {Sullivan}, M.},
        title = "{Cosmology with AGN dust time lags-simulating the new VEILS survey}",
      journal = {\mnras},
         year = 2017,
        month = jan,
       volume = {464},
       number = {2},
        pages = {1693-1703},
          doi = {10.1093/mnras/stw2484},
archivePrefix = {arXiv},
       eprint = {1609.09091},
 primaryClass = {astro-ph.CO},
       adsurl = {https://ui.adsabs.harvard.edu/abs/2017MNRAS.464.1693H}
}

@ARTICLE{2017ApJ...843....3A,
       author = {{Almeyda}, Triana and {Robinson}, Andrew and {Richmond}, Michael and {Vazquez}, Billy and {Nikutta}, Robert},
        title = "{Modeling the Infrared Reverberation Response of the Circumnuclear Dusty Torus in AGNs: The Effects of Cloud Orientation and Anisotropic Illumination}",
      journal = {\apj},
         year = 2017,
        month = jul,
       volume = {843},
       number = {1},
          eid = {3},
        pages = {3},
          doi = {10.3847/1538-4357/aa7687},
archivePrefix = {arXiv},
       eprint = {1709.07011},
 primaryClass = {astro-ph.GA},
       adsurl = {https://ui.adsabs.harvard.edu/abs/2017ApJ...843....3A}
}

@ARTICLE{2017ApJ...839...88K,
       author = {{Kasliwal}, Mansi M. and {Bally}, John and {Masci}, Frank and {Cody}, Ann Marie and {Bond}, Howard E. and {Jencson}, Jacob E. and {Tinyanont}, Samaporn and {Cao}, Yi and {Contreras}, Carlos and {Dykhoff}, Devin A. and {Amodeo}, Samuel and {Armus}, Lee and {Boyer}, Martha and {Cantiello}, Matteo and {Carlon}, Robert L. and {Cass}, Alexander C. and {Cook}, David and {Corgan}, David T. and {Faella}, Joseph and {Fox}, Ori D. and {Green}, Wayne and {Gehrz}, R.~D. and {Helou}, George and {Hsiao}, Eric and {Johansson}, Joel and {Khan}, Rubab M. and {Lau}, Ryan M. and {Langer}, Norbert and {Levesque}, Emily and {Milne}, Peter and {Mohamed}, Shazrene and {Morrell}, Nidia and {Monson}, Andy and {Moore}, Anna and {Ofek}, Eran O. and {O' Sullivan}, Donal and {Parthasarathy}, Mudumba and {Perez}, Andres and {Perley}, Daniel A. and {Phillips}, Mark and {Prince}, Thomas A. and {Shenoy}, Dinesh and {Smith}, Nathan and {Surace}, Jason and {Van Dyk}, Schuyler D. and {Whitelock}, Patricia A. and {Williams}, Robert},
        title = "{SPIRITS: Uncovering Unusual Infrared Transients with Spitzer}",
      journal = {\apj},
         year = 2017,
        month = apr,
       volume = {839},
       number = {2},
          eid = {88},
        pages = {88},
          doi = {10.3847/1538-4357/aa6978},
archivePrefix = {arXiv},
       eprint = {1701.01151},
 primaryClass = {astro-ph.HE},
       adsurl = {https://ui.adsabs.harvard.edu/abs/2017ApJ...839...88K}
}

@ARTICLE{2017ApJ...849..110S,
       author = {{S{\'a}nchez}, P. and {Lira}, P. and {Cartier}, R. and {P{\'e}rez}, V. and {Miranda}, N. and {Yovaniniz}, C. and {Ar{\'e}valo}, P. and {Milvang-Jensen}, B. and {Fynbo}, J. and {Dunlop}, J. and {Coppi}, P. and {Marchesi}, S.},
        title = "{Near-infrared Variability of Obscured and Unobscured X-Ray-selected AGNs in the COSMOS Field}",
      journal = {\apj},
         year = 2017,
        month = nov,
       volume = {849},
       number = {2},
          eid = {110},
        pages = {110},
          doi = {10.3847/1538-4357/aa9188},
archivePrefix = {arXiv},
       eprint = {1710.01306},
 primaryClass = {astro-ph.GA},
       adsurl = {https://ui.adsabs.harvard.edu/abs/2017ApJ...849..110S}
}

@ARTICLE{2018MNRAS.476.1111P,
       author = {{Polimera}, Mugdha and {Sarajedini}, Vicki and {Ashby}, Matthew L.~N. and {Willner}, S.~P. and {Fazio}, Giovanni G.},
        title = "{Morphologies of mid-IR variability-selected AGN host galaxies}",
      journal = {\mnras},
         year = 2018,
        month = may,
       volume = {476},
       number = {1},
        pages = {1111-1119},
          doi = {10.1093/mnras/sty164},
       adsurl = {https://ui.adsabs.harvard.edu/abs/2018MNRAS.476.1111P}
}

@ARTICLE{2019MNRAS.488.1035K,
       author = {{Kulkarni}, Girish and {Worseck}, G{\'a}bor and {Hennawi}, Joseph F.},
        title = "{Evolution of the AGN UV luminosity function from redshift 7.5}",
      journal = {\mnras},
         year = 2019,
        month = sep,
       volume = {488},
       number = {1},
        pages = {1035-1065},
          doi = {10.1093/mnras/stz1493},
archivePrefix = {arXiv},
       eprint = {1807.09774},
 primaryClass = {astro-ph.GA},
       adsurl = {https://ui.adsabs.harvard.edu/abs/2019MNRAS.488.1035K}
}

@ARTICLE{2019ApJ...886..150M,
       author = {{Minezaki}, Takeo and {Yoshii}, Yuzuru and {Kobayashi}, Yukiyasu and {Sugawara}, Shota and {Sakata}, Yu and {Enya}, Keigo and {Koshida}, Shintaro and {Tomita}, Hiroyuki and {Suganuma}, Masahiro and {Aoki}, Tsutomu and {Peterson}, Bruce A.},
        title = "{Reverberation Measurements of the Inner Radii of the Dust Tori in Quasars}",
      journal = {\apj},
         year = 2019,
        month = dec,
       volume = {886},
       number = {2},
          eid = {150},
        pages = {150},
          doi = {10.3847/1538-4357/ab4f7b},
archivePrefix = {arXiv},
       eprint = {1910.08722},
 primaryClass = {astro-ph.GA},
       adsurl = {https://ui.adsabs.harvard.edu/abs/2019ApJ...886..150M}
}

@ARTICLE{2019ApJ...886...33L,
       author = {{Lyu}, Jianwei and {Rieke}, George H. and {Smith}, Paul S.},
        title = "{Mid-IR Variability and Dust Reverberation Mapping of Low-z Quasars. I. Data, Methods, and Basic Results}",
      journal = {\apj},
         year = 2019,
        month = nov,
       volume = {886},
       number = {1},
          eid = {33},
        pages = {33},
          doi = {10.3847/1538-4357/ab481d},
archivePrefix = {arXiv},
       eprint = {1909.11101},
 primaryClass = {astro-ph.GA},
       adsurl = {https://ui.adsabs.harvard.edu/abs/2019ApJ...886...33L}
}

@ARTICLE{2019PASJ...71...74Y,
       author = {{Yasuda}, Naoki and {Tanaka}, Masaomi and {Tominaga}, Nozomu and {Jiang}, Ji-an and {Moriya}, Takashi J. and {Morokuma}, Tomoki and {Suzuki}, Nao and {Takahashi}, Ichiro and {Yamaguchi}, Masaki S. and {Maeda}, Keiichi and {Sako}, Masao and {Ikeda}, Shiro and {Kimura}, Akisato and {Morii}, Mikio and {Ueda}, Naonori and {Yoshida}, Naoki and {Lee}, Chien-Hsiu and {Suyu}, Sherry H. and {Komiyama}, Yutaka and {Regnault}, Nicolas and {Rubin}, David},
        title = "{The Hyper Suprime-Cam SSP transient survey in COSMOS: Overview}",
      journal = {\pasj},
         year = 2019,
        month = aug,
       volume = {71},
       number = {4},
          eid = {74},
        pages = {74},
          doi = {10.1093/pasj/psz050},
archivePrefix = {arXiv},
       eprint = {1904.09697},
 primaryClass = {astro-ph.GA},
       adsurl = {https://ui.adsabs.harvard.edu/abs/2019PASJ...71...74Y}
}

@ARTICLE{2020ApJ...894...24K,
       author = {{Kimura}, Yuki and {Yamada}, Toru and {Kokubo}, Mitsuru and {Yasuda}, Naoki and {Morokuma}, Tomoki and {Nagao}, Tohru and {Matsuoka}, Yoshiki},
        title = "{Properties of AGN Multiband Optical Variability in the HSC SSP Transient Survey}",
      journal = {\apj},
         year = 2020,
        month = may,
       volume = {894},
       number = {1},
          eid = {24},
        pages = {24},
          doi = {10.3847/1538-4357/ab83f3},
archivePrefix = {arXiv},
       eprint = {2004.08835},
 primaryClass = {astro-ph.GA},
       adsurl = {https://ui.adsabs.harvard.edu/abs/2020ApJ...894...24K}
}

@ARTICLE{2020MNRAS.495.3252S,
       author = {{Shen}, Xuejian and {Hopkins}, Philip F. and {Faucher-Gigu{\`e}re}, Claude-Andr{\'e} and {Alexander}, D.~M. and {Richards}, Gordon T. and {Ross}, Nicholas P. and {Hickox}, R.~C.},
        title = "{The bolometric quasar luminosity function at z = 0-7}",
      journal = {\mnras},
         year = 2020,
        month = jan,
       volume = {495},
       number = {3},
        pages = {3252-3275},
          doi = {10.1093/mnras/staa1381},
archivePrefix = {arXiv},
       eprint = {2001.02696},
 primaryClass = {astro-ph.GA},
       adsurl = {https://ui.adsabs.harvard.edu/abs/2020MNRAS.495.3252S}
}

@ARTICLE{2020MNRAS.493.3026E,
       author = {{Elmer}, E. and {Almaini}, O. and {Merrifield}, M. and {Hartley}, W.~G. and {Maltby}, D.~T. and {Lawrence}, A. and {Botti}, I. and {Hirst}, P.},
        title = "{Long-term NIR variability in the UKIDSS Ultra Deep Survey: a new probe of AGN activity at high redshift}",
      journal = {\mnras},
         year = 2020,
        month = apr,
       volume = {493},
       number = {2},
        pages = {3026-3035},
          doi = {10.1093/mnras/staa381},
archivePrefix = {arXiv},
       eprint = {2002.02468},
 primaryClass = {astro-ph.GA},
       adsurl = {https://ui.adsabs.harvard.edu/abs/2020MNRAS.493.3026E}
}

@ARTICLE{2020ApJ...900...58Y,
       author = {{Yang}, Qian and {Shen}, Yue and {Liu}, Xin and {Aguena}, Michel and {Annis}, James and {Avila}, Santiago and {Banerji}, Manda and {Bertin}, Emmanuel and {Brooks}, David and {Burke}, David and {Carnero Rosell}, Aurelio and {Carrasco Kind}, Matias and {da Costa}, Luiz and {De Vicente}, Juan and {Desai}, Shantanu and {Diehl}, H. Thomas and {Doel}, Peter and {Flaugher}, Brenna and {Fosalba}, Pablo and {Frieman}, Josh and {Garcia-Bellido}, Juan and {Gerdes}, David and {Gruen}, Daniel and {Gruendl}, Robert and {Gschwend}, Julia and {Gutierrez}, Gaston and {Hinton}, Samuel and {Hollowood}, Devon L. and {Honscheid}, Klaus and {Kuropatkin}, Nikolay and {Maia}, Marcio and {March}, Marisa and {Marshall}, Jennifer and {Martini}, Paul and {Melchior}, Peter and {Menanteau}, Felipe and {Miquel}, Ramon and {Paz-Chinchon}, Francisco and {Malag{\'o}n}, Andr{\'e}s Plazas and {Romer}, Kathy and {Sanchez}, Eusebio and {Scarpine}, Vic and {Schubnell}, Michael and {Serrano}, Santiago and {Sevilla}, Ignacio and {Smith}, Mathew and {Suchyta}, Eric and {Tarle}, Gregory and {Varga}, Tamas Norbert and {Wilkinson}, Reese},
        title = "{Dust Reverberation Mapping in Distant Quasars from Optical and Mid-infrared Imaging Surveys}",
      journal = {\apj},
         year = 2020,
        month = sep,
       volume = {900},
       number = {1},
          eid = {58},
        pages = {58},
          doi = {10.3847/1538-4357/aba59b},
archivePrefix = {arXiv},
       eprint = {2007.02402},
 primaryClass = {astro-ph.GA},
       adsurl = {https://ui.adsabs.harvard.edu/abs/2020ApJ...900...58Y}
}

@ARTICLE{2021ApJ...912..126L,
       author = {{Lyu}, Jianwei and {Rieke}, George H.},
        title = "{The Dusty Heart of NGC 4151 Revealed by {\ensuremath{\lambda}} {\ensuremath{\sim}} 1-40 {\ensuremath{\mu}}m Reverberation Mapping and Variability: A Challenge to Current Clumpy Torus Models}",
      journal = {\apj},
         year = 2021,
        month = may,
       volume = {912},
       number = {2},
          eid = {126},
        pages = {126},
          doi = {10.3847/1538-4357/abee14},
archivePrefix = {arXiv},
       eprint = {2011.07638},
 primaryClass = {astro-ph.GA},
       adsurl = {https://ui.adsabs.harvard.edu/abs/2021ApJ...912..126L}
}

@ARTICLE{2021Sci...373..789B,
       author = {{Burke}, Colin J. and {Shen}, Yue and {Blaes}, Omer and {Gammie}, Charles F. and {Horne}, Keith and {Jiang}, Yan-Fei and {Liu}, Xin and {McHardy}, Ian M. and {Morgan}, Christopher W. and {Scaringi}, Simone and {Yang}, Qian},
        title = "{A characteristic optical variability time scale in astrophysical accretion disks}",
      journal = {Science},
         year = 2021,
        month = aug,
       volume = {373},
       number = {6556},
        pages = {789-792},
          doi = {10.1126/science.abg9933},
archivePrefix = {arXiv},
       eprint = {2108.05389},
 primaryClass = {astro-ph.GA},
       adsurl = {https://ui.adsabs.harvard.edu/abs/2021Sci...373..789B}
}

@ARTICLE{2021JKAS...54...37K,
       author = {{Kim}, Minjin and {Jeong}, Woong-Seob and {Yang}, Yujin and {Son}, Jiwon and {Ho}, Luis C. and {Woo}, Jong-Hak and {Im}, Myungshin and {Byun}, Woowon},
        title = "{Simulations of Torus Reverberation Mapping Experiments with SPHEREx}",
      journal = {Journal of Korean Astronomical Society},
         year = 2021,
        month = apr,
       volume = {54},
        pages = {37-47},
          doi = {10.5303/JKAS.2021.54.2.37},
archivePrefix = {arXiv},
       eprint = {2107.04219},
 primaryClass = {astro-ph.GA},
       adsurl = {https://ui.adsabs.harvard.edu/abs/2021JKAS...54...37K}
}

@ARTICLE{2021ApJ...907...96S,
       author = {{Suberlak}, Krzysztof L. and {Ivezi{\'c}}, {\v{Z}}eljko and {MacLeod}, Chelsea},
        title = "{Improving Damped Random Walk Parameters for SDSS Stripe 82 Quasars with Pan-STARRS1}",
      journal = {\apj},
         year = 2021,
        month = feb,
       volume = {907},
       number = {2},
          eid = {96},
        pages = {96},
          doi = {10.3847/1538-4357/abc698},
archivePrefix = {arXiv},
       eprint = {2012.12907},
 primaryClass = {astro-ph.GA},
       adsurl = {https://ui.adsabs.harvard.edu/abs/2021ApJ...907...96S}
}

@ARTICLE{2022ApJ...936..104W,
       author = {{Ward}, Charlotte and {Gezari}, Suvi and {Nugent}, Peter and {Bellm}, Eric C. and {Dekany}, Richard and {Drake}, Andrew and {Duev}, Dmitry A. and {Graham}, Matthew J. and {Kasliwal}, Mansi M. and {Kool}, Erik C. and {Masci}, Frank J. and {Riddle}, Reed L.},
        title = "{Variability-selected Intermediate-mass Black Hole Candidates in Dwarf Galaxies from ZTF and WISE}",
      journal = {\apj},
         year = 2022,
        month = sep,
       volume = {936},
       number = {2},
          eid = {104},
        pages = {104},
          doi = {10.3847/1538-4357/ac8666},
archivePrefix = {arXiv},
       eprint = {2110.13098},
 primaryClass = {astro-ph.GA},
       adsurl = {https://ui.adsabs.harvard.edu/abs/2022ApJ...936..104W}
}

@ARTICLE{2023ApJ...958..135S,
       author = {{Son}, Suyeon and {Kim}, Minjin and {Ho}, Luis C.},
        title = "{The Structure Function of Mid-infrared Variability in Low-redshift Active Galactic Nuclei}",
      journal = {\apj},
         year = 2023,
        month = dec,
       volume = {958},
       number = {2},
          eid = {135},
        pages = {135},
          doi = {10.3847/1538-4357/ad01bc},
archivePrefix = {arXiv},
       eprint = {2310.05389},
 primaryClass = {astro-ph.GA},
       adsurl = {https://ui.adsabs.harvard.edu/abs/2023ApJ...958..135S}
}

@ARTICLE{yamada23,
       author = {{Yamada}, Toru},
        title = "{Roman CCS White Paper: Low-mass SMBH at High Redshift: Deepest variability search for low-luminosity AGN}",
      journal = {Roman CCS White Paper},
         year = 2023,
        month = jun,
          eid = {},
        pages = {65},
       adsurl = {https://asd.gsfc.nasa.gov/roman/white_papers/2023/075_Yamada_HLTDS.pdf},
}

@ARTICLE{2023MNRAS.518.1880B,
       author = {{Burke}, Colin J. and {Shen}, Yue and {Liu}, Xin and {Natarajan}, Priyamvada and {Caplar}, Neven and {Bellovary}, Jillian M. and {Wang}, Z. Franklin},
        title = "{Dwarf AGNs from variability for the origins of seeds (DAVOS): Intermediate-mass black hole demographics from optical synoptic surveys}",
      journal = {\mnras},
         year = 2023,
        month = jan,
       volume = {518},
       number = {2},
        pages = {1880-1904},
          doi = {10.1093/mnras/stac2478},
archivePrefix = {arXiv},
       eprint = {2207.04092},
 primaryClass = {astro-ph.GA},
       adsurl = {https://ui.adsabs.harvard.edu/abs/2023MNRAS.518.1880B}
}

@ARTICLE{2024MNRAS.531.2551G,
       author = {{Green}, K. and {Elmer}, E. and {Maltby}, D.~T. and {Almaini}, O. and {Merrifield}, M. and {Hartley}, W.~G.},
        title = "{Increasing AGN sample completeness using long-term near-infrared variability}",
      journal = {\mnras},
         year = 2024,
        month = jun,
       volume = {531},
       number = {2},
        pages = {2551-2562},
          doi = {10.1093/mnras/stae1322},
archivePrefix = {arXiv},
       eprint = {2405.14809},
 primaryClass = {astro-ph.GA},
       adsurl = {https://ui.adsabs.harvard.edu/abs/2024MNRAS.531.2551G}
}

@ARTICLE{2024MNRAS.531.3310L,
       author = {{Lira}, P. and {S{\'a}nchez-S{\'a}ez}, P. and {Ar{\'e}valo}, P. and {Tristram}, K. and {De Cicco}, D. and {Milvang-Jensen}, B. and {Dunlop}, J.~S.},
        title = "{Dust reverberation mapping of 0.3 {\ensuremath{\leq}} z {\ensuremath{\leq}} 0.8 AGN with UltraVISTA: lessons for future large surveys}",
      journal = {\mnras},
         year = 2024,
        month = jul,
       volume = {531},
       number = {3},
        pages = {3310-3325},
          doi = {10.1093/mnras/stae1095},
       adsurl = {https://ui.adsabs.harvard.edu/abs/2024MNRAS.531.3310L}
}

@ARTICLE{2024ApJ...968...59M,
       author = {{Mandal}, Amit Kumar and {Woo}, Jong-Hak and {Wang}, Shu and {Rakshit}, Suvendu and {Cho}, Hojin and {Son}, Donghoon and {Stalin}, C.~S.},
        title = "{Revisiting the Dust Torus Size{\textendash}Luminosity Relation Based on a Uniform Reverberation-mapping Analysis}",
      journal = {\apj},
         year = 2024,
        month = jun,
       volume = {968},
       number = {2},
          eid = {59},
        pages = {59},
          doi = {10.3847/1538-4357/ad414d},
archivePrefix = {arXiv},
       eprint = {2403.01885},
 primaryClass = {astro-ph.GA},
       adsurl = {https://ui.adsabs.harvard.edu/abs/2024ApJ...968...59M}
}

@ARTICLE{2025arXiv250510574Z,
       author = {{Zasowski}, Gail and {Jha}, Saurabh W. and {Chomiuk}, Laura and {Fan}, Xiaohui and {Hickox}, Ryan and {Huber}, Dan and {Kerins}, Eamonn and {Kobulnicky}, Chip and {Lauer}, Tod and {Sako}, Masao and {Shapley}, Alice and {Stephens}, Denise and {Weinberg}, David and {Williams}, Ben},
        title = "{Roman Observations Time Allocation Committee: Final Report and Recommendations}",
      journal = {arXiv e-prints},
         year = 2025,
        month = may,
          eid = {arXiv:2505.10574},
        pages = {arXiv:2505.10574},
          doi = {10.48550/arXiv.2505.10574},
archivePrefix = {arXiv},
       eprint = {2505.10574},
 primaryClass = {astro-ph.IM},
       adsurl = {https://ui.adsabs.harvard.edu/abs/2025arXiv250510574Z}
}

@ARTICLE{2025ApJ...992..158F,
       author = {{Frohmaier}, C. and {Vincenzi}, M. and {Sullivan}, M. and {H{\"o}nig}, S.~F. and {Smith}, M. and {Addison}, H. and {Collett}, T. and {Dimitriadis}, G. and {Ellis}, R.~S. and {Gandhi}, P. and {Graur}, O. and {Hook}, I. and {Kelsey}, L. and {Kim}, Y.-L. and {Lidman}, C. and {Maguire}, K. and {Makrygianni}, L. and {Martin}, B. and {M{\"o}ller}, A. and {Nichol}, R.~C. and {Nicholl}, M. and {Schady}, P. and {Simmons}, B.~D. and {Smartt}, S.~J. and {Tempel}, E. and {Wiseman}, P. and {the LSST Dark Energy Science Collaboration}},
        title = "{TiDES: The 4MOST Time Domain Extragalactic Survey}",
      journal = {\apj},
         year = 2025,
        month = oct,
       volume = {992},
       number = {1},
          eid = {158},
        pages = {158},
          doi = {10.3847/1538-4357/adff4e},
archivePrefix = {arXiv},
       eprint = {2501.16311},
 primaryClass = {astro-ph.HE},
       adsurl = {https://ui.adsabs.harvard.edu/abs/2025ApJ...992..158F}
}

@ARTICLE{2025A&A...697A.204D,
       author = {{De Cicco}, D. and {Zazzaro}, G. and {Cavuoti}, S. and {Paolillo}, M. and {Longo}, G. and {Petrecca}, V. and {Saccheo}, I. and {S{\'a}nchez-S{\'a}ez}, P.},
        title = "{Selection of optically variable active galactic nuclei via a random forest algorithm}",
      journal = {\aap},
         year = 2025,
        month = may,
       volume = {697},
          eid = {A204},
        pages = {A204},
          doi = {10.1051/0004-6361/202453630},
archivePrefix = {arXiv},
       eprint = {2505.15819},
 primaryClass = {astro-ph.GA},
       adsurl = {https://ui.adsabs.harvard.edu/abs/2025A&A...697A.204D}
}

@ARTICLE{2025ApJ...985..223M,
       author = {{Messick}, Alexander and {Baldassare}, Vivienne and {Jones}, David O. and {French}, K. Decker and {Raimundo}, Sandra I. and {Earl}, Nicholas and {Auchettl}, Katie and {Coulter}, David A. and {Huber}, Mark E. and {Verrico}, Margaret E. and {de Boer}, Thomas and {Chambers}, Kenneth C. and {Gao}, Hua and {Lin}, Chien-Cheng and {Wainscoat}, Richard J.},
        title = "{A Large-scale Search for Photometrically Variable Active Galactic Nuclei in Dwarf Galaxies Using the Young Supernova Experiment}",
      journal = {\apj},
         year = 2025,
        month = jun,
       volume = {985},
       number = {2},
          eid = {223},
        pages = {223},
          doi = {10.3847/1538-4357/adcdff},
archivePrefix = {arXiv},
       eprint = {2504.00971},
 primaryClass = {astro-ph.GA},
       adsurl = {https://ui.adsabs.harvard.edu/abs/2025ApJ...985..223M}
}

@ARTICLE{2025arXiv250913308S,
       author = {{Satheesh-Sheeba}, S. and {Assef}, R.~J. and {Anguita}, T. and {S{\'a}nchez-S{\'a}ez}, P. and {Shirley}, R. and {Ananna}, T.~T. and {Bauer}, F.~E. and {Bobrick}, A. and {Bornancini}, C.~G. and {Bosman}, S.~E.~I. and {Brandt}, W.~N. and {De Cicco}, D. and {Czerny}, B. and {Fatovi{\'c}}, M. and {Ichikawa}, K. and {Ili{\'c}}, D. and {Kova{\v{c}}evi{\'c}}, A.~B. and {Li}, G. and {Liao}, M. and {Rojas-Lilay{\'u}}, A. and {Marculewicz}, M. and {Marsango}, D. and {Mazzucchelli}, C. and {Mkrtchyan}, T. and {Panda}, S. and {Peca}, A. and {Rani}, B. and {Ricci}, C. and {Richards}, G.~T. and {Salvato}, M. and {Schneider}, D.~P. and {Temple}, M.~J. and {Tombesi}, F. and {Yu}, W. and {Yoon}, I. and {Zou}, F.},
        title = "{VAR-PZ: Constraining the Photometric Redshifts of Quasars using Variability}",
      journal = {arXiv e-prints},
         year = 2025,
        month = sep,
          eid = {arXiv:2509.13308},
        pages = {arXiv:2509.13308},
          doi = {10.48550/arXiv.2509.13308},
archivePrefix = {arXiv},
       eprint = {2509.13308},
 primaryClass = {astro-ph.GA},
       adsurl = {https://ui.adsabs.harvard.edu/abs/2025arXiv250913308S}
}

@ARTICLE{2025ApJ...993..203T,
       author = {{Tomar}, Ashutosh and {Rakshit}, Suvendu and {Mandal}, Amit Kumar and {Pandey}, Shivangi},
        title = "{Probing the Dust Torus Radius─Luminosity Relation: A WISE View}",
      journal = {\apj},
         year = 2025,
        month = nov,
       volume = {993},
       number = {2},
          eid = {203},
        pages = {203},
          doi = {10.3847/1538-4357/ae0f96},
archivePrefix = {arXiv},
       eprint = {2510.04517},
 primaryClass = {astro-ph.GA},
       adsurl = {https://ui.adsabs.harvard.edu/abs/2025ApJ...993..203T}
}

@ARTICLE{2025JATIS..11c1626G,
       author = {{Gorjian}, Varoujan and {Werner}, Michael W.},
        title = "{Studying the dust distribution around accreting black holes with reverberation mapping using PRIMA}",
      journal = {Journal of Astronomical Telescopes, Instruments, and Systems},
         year = 2025,
        month = jul,
       volume = {11},
          eid = {031626},
        pages = {031626},
          doi = {10.1117/1.JATIS.11.3.031626},
archivePrefix = {arXiv},
       eprint = {2510.16977},
 primaryClass = {astro-ph.IM},
       adsurl = {https://ui.adsabs.harvard.edu/abs/2025JATIS..11c1626G}
}

@ARTICLE{2025ApJ...993..116K,
       author = {{Kessler}, Richard and {Hounsell}, Rebekah and {Joshi}, Bhavin and {Rubin}, David and {Sako}, Masao and {Chen}, Rebecca and {Miranda}, Vivian and {Rose}, Benjamin M.},
        title = "{Cosmology Constraints from Type Ia Supernova Simulations of the Nancy Grace Roman Space Telescope Strategy Recommended by the High-Latitude Time Domain Survey Definition Committee}",
      journal = {\apj},
         year = 2025,
        month = nov,
       volume = {993},
       number = {1},
          eid = {116},
        pages = {116},
          doi = {10.3847/1538-4357/ae07c9},
archivePrefix = {arXiv},
       eprint = {2506.04402},
 primaryClass = {astro-ph.CO},
       adsurl = {https://ui.adsabs.harvard.edu/abs/2025ApJ...993..116K}
}

@ARTICLE{2025PhRvD.112h3515A,
       author = {{Abdul Karim}, M. and {Aguilar}, J. and {Ahlen}, S. and {Alam}, S. and {Allen}, L. and {Prieto}, C. Allende and {Alves}, O. and {Anand}, A. and {Andrade}, U. and {Armengaud}, E. and {Aviles}, A. and {Bailey}, S. and {Baltay}, C. and {Bansal}, P. and {Bault}, A. and {Behera}, J. and {BenZvi}, S. and {Bianchi}, D. and {Blake}, C. and {Brieden}, S. and {Brodzeller}, A. and {Brooks}, D. and {Buckley-Geer}, E. and {Burtin}, E. and {Calderon}, R. and {Canning}, R. and {Rosell}, A. Carnero and {Carrilho}, P. and {Casas}, L. and {Castander}, F.~J. and {Charles}, M. and {Chaussidon}, E. and {Chaves-Montero}, J. and {Chebat}, D. and {Chen}, X. and {Claybaugh}, T. and {Cole}, S. and {Cooper}, A.~P. and {Cuceu}, A. and {Dawson}, K.~S. and {de la Macorra}, A. and {de Mattia}, A. and {Deiosso}, N. and {Della Costa}, J. and {Demina}, R. and {Dey}, A. and {Dey}, B. and {Ding}, Z. and {Doel}, P. and {Edelstein}, J. and {Eisenstein}, D.~J. and {Elbers}, W. and {Fagrelius}, P. and {Fanning}, K. and {Fern{\'a}ndez-Garc{\'\i}a}, E. and {Ferraro}, S. and {Font-Ribera}, A. and {Forero-Romero}, J.~E. and {Frenk}, C.~S. and {Garcia-Quintero}, C. and {Garrison}, L.~H. and {Gazta{\~n}aga}, E. and {Gil-Mar{\'\i}n}, H. and {Gontcho A Gontcho}, S. and {Gonzalez}, D. and {Gonzalez-Morales}, A.~X. and {Gordon}, C. and {Green}, D. and {Gutierrez}, G. and {Guy}, J. and {Hadzhiyska}, B. and {Hahn}, C. and {He}, S. and {Herbold}, M. and {Herrera-Alcantar}, H.~K. and {Ho}, M.-F. and {Honscheid}, K. and {Howlett}, C. and {Huterer}, D. and {Ishak}, M. and {Juneau}, S. and {Kamble}, N.~V. and {Kara{\c{c}}ayl{\i}}, N.~G. and {Kehoe}, R. and {Kent}, S. and {Kim}, A.~G. and {Kirkby}, D. and {Kisner}, T. and {Koposov}, S.~E. and {Kremin}, A. and {Krolewski}, A. and {Lahav}, O. and {Lamman}, C. and {Landriau}, M. and {Lang}, D. and {Lasker}, J. and {Le Goff}, J.~M. and {Le Guillou}, L. and {Leauthaud}, A. and {Levi}, M.~E. and {Li}, Q. and {Li}, T.~S. and {Lodha}, K. and {Lokken}, M. and {Lozano-Rodr{\'\i}guez}, F. and {Magneville}, C. and {Manera}, M. and {Martini}, P. and {Matthewson}, W.~L. and {Meisner}, A. and {Mena-Fern{\'a}ndez}, J. and {Menegas}, A. and {Mergulh{\~a}o}, T. and {Miquel}, R. and {Moustakas}, J. and {Mu{\~n}oz-Guti{\'e}rrez}, A. and {Mu{\~n}oz-Santos}, D. and {Myers}, A.~D. and {Nadathur}, S. and {Naidoo}, K. and {Napolitano}, L. and {Newman}, J.~A. and {Niz}, G. and {Noriega}, H.~E. and {Paillas}, E. and {Palanque-Delabrouille}, N. and {Pan}, J. and {Peacock}, J.~A. and {Pellejero Ibanez}, M. and {Percival}, W.~J. and {P{\'e}rez-Fern{\'a}ndez}, A. and {P{\'e}rez-R{\`a}fols}, I. and {Pieri}, M.~M. and {Poppett}, C. and {Prada}, F. and {Rabinowitz}, D. and {Raichoor}, A. and {Ram{\'\i}rez-P{\'e}rez}, C. and {Rashkovetskyi}, M. and {Ravoux}, C. and {Rich}, J. and {Rocher}, A. and {Rockosi}, C. and {Rohlf}, J. and {Rom{\'a}n-Herrera}, J.~O. and {Ross}, A.~J. and {Rossi}, G. and {Ruggeri}, R. and {Ruhlmann-Kleider}, V. and {Samushia}, L. and {Sanchez}, E. and {Sanders}, N. and {Schlegel}, D. and {Schubnell}, M. and {Seo}, H. and {Shafieloo}, A. and {Sharples}, R. and {Silber}, J. and {Sinigaglia}, F. and {Sprayberry}, D. and {Tan}, T. and {Tarl{\'e}}, G. and {Taylor}, P. and {Turner}, W. and {Ure{\~n}a-L{\'o}pez}, L.~A. and {Vaisakh}, R. and {Valdes}, F. and {Valogiannis}, G. and {Vargas-Maga{\~n}a}, M. and {Verde}, L. and {Walther}, M. and {Weaver}, B.~A. and {Weinberg}, D.~H. and {White}, M. and {Wolfson}, M. and {Y{\`e}che}, C. and {Yu}, J. and {Zaborowski}, E.~A. and {Zarrouk}, P. and {Zhai}, Z. and {Zhang}, H. and {Zhao}, C. and {Zhao}, G.~B. and {Zhou}, R. and {Zou}, H. and {DESI Collaboration}},
        title = "{DESI DR2 results. II. Measurements of baryon acoustic oscillations and cosmological constraints}",
      journal = {\prd},
         year = 2025,
        month = oct,
       volume = {112},
       number = {8},
          eid = {083515},
        pages = {083515},
          doi = {10.1103/tr6y-kpc6},
archivePrefix = {arXiv},
       eprint = {2503.14738},
 primaryClass = {astro-ph.CO},
       adsurl = {https://ui.adsabs.harvard.edu/abs/2025PhRvD.112h3515A}
}

@ARTICLE{2025arXiv251121479Y,
       author = {{Yu}, Weixiang and {Ruan}, John J. and {Burke}, Colin J. and {Assef}, Roberto J. and {Ananna}, Tonima T. and {Bauer}, Franz E. and {De Cicco}, Demetra and {Horne}, Keith and {Hern{\'a}ndez-Garc{\'\i}a}, Lorena and {Ili{\'c}}, Dragana and {Kova{\v{c}}evi{\'c}}, Andjelka B. and {Marculewicz}, Marcin and {Panda}, Swayamtrupta and {Ricci}, Claudio and {Richards}, Gordon T. and {Riffel}, Rogemar A. and {Schneider}, Donald P. and {S{\'a}nchez-S{\'a}ez}, Paula and {Satheesh Sheeba}, Sarath and {Tombesi}, Francesco and {Temple}, Matthew J. and {Vogeley}, Michael S. and {Yoon}, Ilsang and {Zou}, Fan},
        title = "{Scalable and Robust Multiband Modeling of AGN Light Curves in Rubin-LSST}",
      journal = {arXiv e-prints},
         year = 2025,
        month = nov,
          eid = {arXiv:2511.21479},
        pages = {arXiv:2511.21479},
          doi = {10.48550/arXiv.2511.21479},
archivePrefix = {arXiv},
       eprint = {2511.21479},
 primaryClass = {astro-ph.GA},
       adsurl = {https://ui.adsabs.harvard.edu/abs/2025arXiv251121479Y}
}

@ARTICLE{2025ApJ...995...24K,
       author = {{Kokubo}, Mitsuru and {Harikane}, Yuichi},
        title = "{Challenging the Active Galactic Nucleus Scenario for JWST/NIRSpec Little Red Dot and Non─Little Red Dot Broad H{\ensuremath{\alpha}} Emitters in Light of Nondetection of NIRCam Photometric Variability and X-Ray}",
      journal = {\apj},
         year = 2025,
        month = dec,
       volume = {995},
       number = {1},
          eid = {24},
        pages = {24},
          doi = {10.3847/1538-4357/ae119e},
archivePrefix = {arXiv},
       eprint = {2407.04777},
 primaryClass = {astro-ph.GA},
       adsurl = {https://ui.adsabs.harvard.edu/abs/2025ApJ...995...24K}
}

@article{Burke_2026,
   title={Variability as a new discovery channel for intermediate-mass black holes in the time-domain era},
   ISSN={2397-3366},
   url={http://dx.doi.org/10.1038/s41550-025-02759-5},
   DOI={10.1038/s41550-025-02759-5},
   journal={Nature Astronomy},
   publisher={Springer Science and Business Media LLC},
   author={Burke, Colin J. and Natarajan, Priyamvada},
   year={2026},
   month=jan }

@misc{panda2026agnvariabilityrubinobservatory,
      title={AGN Variability with Rubin Observatory in the 2030s}, 
      author={Swayamtrupta Panda and Francisco Pozo Nuñez and Hygor Benati Gonçalves and Guodong Li and Bożena Czerny and Paola Marziani and Thaisa Storchi-Bergmann},
      year={2026},
      eprint={2601.21769},
      archivePrefix={arXiv},
      primaryClass={astro-ph.GA},
      url={https://arxiv.org/abs/2601.21769}, 
}

@ARTICLE{2014A&A...562A..21C,
       author = {{Cicone}, C. and {Maiolino}, R. and {Sturm}, E. and {Graci{\'a}-Carpio}, J. and {Feruglio}, C. and {Neri}, R. and {Aalto}, S. and {Davies}, R. and {Fiore}, F. and {Fischer}, J. and {Garc{\'\i}a-Burillo}, S. and {Gonz{\'a}lez-Alfonso}, E. and {Hailey-Dunsheath}, S. and {Piconcelli}, E. and {Veilleux}, S.},
        title = "{Massive molecular outflows and evidence for AGN feedback from CO observations}",
      journal = {\aap},
         year = 2014,
        month = feb,
       volume = {562},
          eid = {A21},
        pages = {A21},
          doi = {10.1051/0004-6361/201322464},
archivePrefix = {arXiv},
       eprint = {1311.2595},
 primaryClass = {astro-ph.CO},
       adsurl = {https://ui.adsabs.harvard.edu/abs/2014A&A...562A..21C}
}

@ARTICLE{2019MNRAS.483.4586F,
       author = {{Fluetsch}, A. and {Maiolino}, R. and {Carniani}, S. and {Marconi}, A. and {Cicone}, C. and {Bourne}, M.~A. and {Costa}, T. and {Fabian}, A.~C. and {Ishibashi}, W. and {Venturi}, G.},
        title = "{Cold molecular outflows in the local Universe and their feedback effect on galaxies}",
      journal = {\mnras},
         year = 2019,
        month = mar,
       volume = {483},
       number = {4},
        pages = {4586-4614},
          doi = {10.1093/mnras/sty3449},
archivePrefix = {arXiv},
       eprint = {1805.05352},
 primaryClass = {astro-ph.GA},
       adsurl = {https://ui.adsabs.harvard.edu/abs/2019MNRAS.483.4586F}
}

@ARTICLE{2013PASJ...65....5S,
       author = {{Shirahata}, Mai and {Nakagawa}, Takao and {Usuda}, Tomonori and {Goto}, Miwa and {Suto}, Hiroshi and {Geballe}, Thomas R.},
        title = "{Infrared Spectroscopy of CO Ro-Vibrational Absorption Lines toward the Obscured AGN IRAS 08572+3915}",
      journal = {\pasj},
         year = 2013,
        month = feb,
       volume = {65},
          eid = {5},
        pages = {5},
          doi = {10.1093/pasj/65.1.5},
archivePrefix = {arXiv},
       eprint = {1208.4663},
 primaryClass = {astro-ph.GA},
       adsurl = {https://ui.adsabs.harvard.edu/abs/2013PASJ...65....5S}
}

@ARTICLE{2021ApJ...921..141O,
       author = {{Onishi}, Shusuke and {Nakagawa}, Takao and {Baba}, Shunsuke and {Matsumoto}, Kosei and {Isobe}, Naoki and {Shirahata}, Mai and {Terada}, Hiroshi and {Usuda}, Tomonori and {Oyabu}, Shinki},
        title = "{Study of the Inner Structure of the Molecular Torus in IRAS 08572+3915 NW with Velocity Decomposition of CO Rovibrational Absorption Lines}",
      journal = {\apj},
         year = 2021,
        month = nov,
       volume = {921},
       number = {2},
          eid = {141},
        pages = {141},
          doi = {10.3847/1538-4357/ac1c6d},
archivePrefix = {arXiv},
       eprint = {2108.04838},
 primaryClass = {astro-ph.GA},
       adsurl = {https://ui.adsabs.harvard.edu/abs/2021ApJ...921..141O}
}

@ARTICLE{2024ApJ...976..106O,
       author = {{Onishi}, Shusuke and {Nakagawa}, Takao and {Baba}, Shunsuke and {Matsumoto}, Kosei and {Isobe}, Naoki and {Shirahata}, Mai and {Terada}, Hiroshi and {Usuda}, Tomonori and {Oyabu}, Shinki},
        title = "{Systematic Study of the Inner Structure of Molecular Tori in Nearby U/LIRGs Using Velocity Decomposition of CO Rovibrational Absorption Lines}",
      journal = {\apj},
         year = 2024,
        month = nov,
       volume = {976},
       number = {1},
          eid = {106},
        pages = {106},
          doi = {10.3847/1538-4357/ad84f4},
archivePrefix = {arXiv},
       eprint = {2410.07328},
 primaryClass = {astro-ph.GA},
       adsurl = {https://ui.adsabs.harvard.edu/abs/2024ApJ...976..106O}
}

@ARTICLE{2023ApJ...951...87O,
       author = {{Ohyama}, Youichi and {Onishi}, Shusuke and {Nakagawa}, Takao and {Matsumoto}, Kosei and {Isobe}, Naoki and {Shirahata}, Mai and {Baba}, Shunsuke and {Sakamoto}, Kazushi},
        title = "{Warm Molecular Gas in the Central Parsecs of the Buried Nucleus of NGC 4418 Traced with the Fundamental CO Rovibrational Absorptions}",
      journal = {\apj},
         year = 2023,
        month = jul,
       volume = {951},
       number = {2},
          eid = {87},
        pages = {87},
          doi = {10.3847/1538-4357/acd692},
archivePrefix = {arXiv},
       eprint = {2305.09959},
 primaryClass = {astro-ph.GA},
       adsurl = {https://ui.adsabs.harvard.edu/abs/2023ApJ...951...87O}
}

@ARTICLE{2004A&A...426L...5L,
       author = {{Lutz}, D. and {Sturm}, E. and {Genzel}, R. and {Spoon}, H.~W.~W. and {Stacey}, G.~J.},
        title = "{Gas near active galactic nuclei: A search for the 4.7 {\ensuremath{\mu}}m CO band}",
      journal = {\aap},
         year = 2004,
        month = oct,
       volume = {426},
        pages = {L5-L8},
          doi = {10.1051/0004-6361:200400072},
archivePrefix = {arXiv},
       eprint = {astro-ph/0409123},
 primaryClass = {astro-ph},
       adsurl = {https://ui.adsabs.harvard.edu/abs/2004A&A...426L...5L}
}

@ARTICLE{2004ApJS..154..184S,
       author = {{Spoon}, H.~W.~W. and {Armus}, L. and {Cami}, J. and {Tielens}, A.~G.~G.~M. and {Chiar}, J.~E. and {Peeters}, E. and {Keane}, J.~V. and {Charmandaris}, V. and {Appleton}, P.~N. and {Teplitz}, H.~I. and {Burgdorf}, M.~J.},
        title = "{Fire and Ice: Spitzer Infrared Spectrograph (IRS) Mid-Infrared Spectroscopy of IRAS F00183-7111}",
      journal = {\apjs},
         year = 2004,
        month = sep,
       volume = {154},
       number = {1},
        pages = {184-187},
          doi = {10.1086/422813},
       adsurl = {https://ui.adsabs.harvard.edu/abs/2004ApJS..154..184S}
}

@ARTICLE{2018ApJ...852...83B,
       author = {{Baba}, Shunsuke and {Nakagawa}, Takao and {Isobe}, Naoki and {Shirahata}, Mai},
        title = "{The Near-infrared CO Absorption Band as a Probe to the Innermost Part of an AGN-obscuring Material}",
      journal = {\apj},
         year = 2018,
        month = jan,
       volume = {852},
       number = {2},
          eid = {83},
        pages = {83},
          doi = {10.3847/1538-4357/aa9f25},
archivePrefix = {arXiv},
       eprint = {1712.01287},
 primaryClass = {astro-ph.GA},
       adsurl = {https://ui.adsabs.harvard.edu/abs/2018ApJ...852...83B}
}

@ARTICLE{2022ApJ...928..184B,
       author = {{Baba}, Shunsuke and {Imanishi}, Masatoshi and {Izumi}, Takuma and {Kawamuro}, Taiki and {Nguyen}, Dieu D. and {Nakagawa}, Takao and {Isobe}, Naoki and {Onishi}, Shusuke and {Matsumoto}, Kosei},
        title = "{The Extremely Buried Nucleus of IRAS 17208-0014 Observed at Submillimeter and Near-infrared Wavelengths}",
      journal = {\apj},
         year = 2022,
        month = apr,
       volume = {928},
       number = {2},
          eid = {184},
        pages = {184},
          doi = {10.3847/1538-4357/ac57c2},
archivePrefix = {arXiv},
       eprint = {2202.11105},
 primaryClass = {astro-ph.GA},
       adsurl = {https://ui.adsabs.harvard.edu/abs/2022ApJ...928..184B}
}

@ARTICLE{2024A&A...682A.182G,
       author = {{Gonz{\'a}lez-Alfonso}, Eduardo and {Garc{\'\i}a-Bernete}, Ismael and {Pereira-Santaella}, Miguel and {Neufeld}, David A. and {Fischer}, Jacqueline and {Donnan}, Fergus R.},
        title = "{JWST detection of extremely excited outflowing CO and H$_{2}$O in VV 114 E SW: A possible rapidly accreting IMBH}",
      journal = {\aap},
         year = 2024,
        month = feb,
       volume = {682},
          eid = {A182},
        pages = {A182},
          doi = {10.1051/0004-6361/202348469},
archivePrefix = {arXiv},
       eprint = {2312.04914},
 primaryClass = {astro-ph.GA},
       adsurl = {https://ui.adsabs.harvard.edu/abs/2024A&A...682A.182G}
}

@ARTICLE{1986ApJS...61..249W,
       author = {{Wolfe}, A.~M. and {Turnshek}, D.~A. and {Smith}, H.~E. and {Cohen}, R.~D.},
        title = "{Damped Lyman-Alpha Absorption by Disk Galaxies with Large Redshifts. I. The Lick Survey}",
      journal = {\apjs},
         year = 1986,
        month = jun,
       volume = {61},
        pages = {249},
          doi = {10.1086/191114},
       adsurl = {https://ui.adsabs.harvard.edu/abs/1986ApJS...61..249W}
}

@ARTICLE{2003ApJ...587..278W,
       author = {{Wolfire}, Mark G. and {McKee}, Christopher F. and {Hollenbach}, David and {Tielens}, A.~G.~G.~M.},
        title = "{Neutral Atomic Phases of the Interstellar Medium in the Galaxy}",
      journal = {\apj},
         year = 2003,
        month = apr,
       volume = {587},
       number = {1},
        pages = {278-311},
          doi = {10.1086/368016},
archivePrefix = {arXiv},
       eprint = {astro-ph/0207098},
 primaryClass = {astro-ph},
       adsurl = {https://ui.adsabs.harvard.edu/abs/2003ApJ...587..278W}
}

@ARTICLE{2018A&A...611A..76D,
       author = {{De Cia}, Annalisa and {Ledoux}, C{\'e}dric and {Petitjean}, Patrick and {Savaglio}, Sandra},
        title = "{The cosmic evolution of dust-corrected metallicity in the neutral gas}",
      journal = {\aap},
         year = 2018,
        month = apr,
       volume = {611},
          eid = {A76},
        pages = {A76},
          doi = {10.1051/0004-6361/201731970},
archivePrefix = {arXiv},
       eprint = {1709.06578},
 primaryClass = {astro-ph.GA},
       adsurl = {https://ui.adsabs.harvard.edu/abs/2018A&A...611A..76D}
}

@ARTICLE{2014A&A...566A..24N,
       author = {{Noterdaeme}, P. and {Petitjean}, P. and {P{\^a}ris}, I. and {Cai}, Z. and {Finley}, H. and {Ge}, J. and {Pieri}, M.~M. and {York}, D.~G.},
        title = "{A connection between extremely strong damped Lyman-{\ensuremath{\alpha}} systems and Lyman-{\ensuremath{\alpha}} emitting galaxies at small impact parameters}",
      journal = {\aap},
         year = 2014,
        month = jun,
       volume = {566},
          eid = {A24},
        pages = {A24},
          doi = {10.1051/0004-6361/201322809},
archivePrefix = {arXiv},
       eprint = {1403.4115},
 primaryClass = {astro-ph.GA},
       adsurl = {https://ui.adsabs.harvard.edu/abs/2014A&A...566A..24N}
}

@ARTICLE{2005ApJ...635..123P,
       author = {{Prochaska}, Jason X. and {Herbert-Fort}, St{\'e}phane and {Wolfe}, Arthur M.},
        title = "{The SDSS Damped Ly{\ensuremath{\alpha}} Survey: Data Release 3}",
      journal = {\apj},
         year = 2005,
        month = dec,
       volume = {635},
       number = {1},
        pages = {123-142},
          doi = {10.1086/497287},
archivePrefix = {arXiv},
       eprint = {astro-ph/0508361},
 primaryClass = {astro-ph},
       adsurl = {https://ui.adsabs.harvard.edu/abs/2005ApJ...635..123P}
}

@ARTICLE{2021MNRAS.507..704H,
       author = {{Ho}, Ming-Feng and {Bird}, Simeon and {Garnett}, Roman},
        title = "{Damped Lyman-{\ensuremath{\alpha}} absorbers from Sloan digital sky survey DR16Q with Gaussian processes}",
      journal = {\mnras},
         year = 2021,
        month = oct,
       volume = {507},
       number = {1},
        pages = {704-719},
          doi = {10.1093/mnras/stab2169},
archivePrefix = {arXiv},
       eprint = {2103.10964},
 primaryClass = {astro-ph.GA},
       adsurl = {https://ui.adsabs.harvard.edu/abs/2021MNRAS.507..704H}
}

@ARTICLE{2004ApJ...609..667S,
       author = {{Schaye}, Joop},
        title = "{Star Formation Thresholds and Galaxy Edges: Why and Where}",
      journal = {\apj},
         year = 2004,
        month = jul,
       volume = {609},
       number = {2},
        pages = {667-682},
          doi = {10.1086/421232},
archivePrefix = {arXiv},
       eprint = {astro-ph/0205125},
 primaryClass = {astro-ph},
       adsurl = {https://ui.adsabs.harvard.edu/abs/2004ApJ...609..667S}
}

@ARTICLE{1977ApJ...216..291S,
       author = {{Savage}, B.~D. and {Bohlin}, R.~C. and {Drake}, J.~F. and {Budich}, W.},
        title = "{A survey of interstellar molecular hydrogen. I.}",
      journal = {\apj},
         year = 1977,
        month = aug,
       volume = {216},
        pages = {291-307},
          doi = {10.1086/155471},
       adsurl = {https://ui.adsabs.harvard.edu/abs/1977ApJ...216..291S}
}

@ARTICLE{2019MNRAS.490.2668B,
       author = {{Balashev}, S.~A. and {Klimenko}, V.~V. and {Noterdaeme}, P. and {Krogager}, J. -K. and {Varshalovich}, D.~A. and {Ivanchik}, A.~V. and {Petitjean}, P. and {Srianand}, R. and {Ledoux}, C.},
        title = "{X-shooter observations of strong H$_{2}$-bearing DLAs at high redshift}",
      journal = {\mnras},
         year = 2019,
        month = dec,
       volume = {490},
       number = {2},
        pages = {2668-2678},
          doi = {10.1093/mnras/stz2707},
archivePrefix = {arXiv},
       eprint = {1909.11064},
 primaryClass = {astro-ph.GA},
       adsurl = {https://ui.adsabs.harvard.edu/abs/2019MNRAS.490.2668B}
}

@ARTICLE{2008A&A...482L..39S,
       author = {{Srianand}, R. and {Noterdaeme}, P. and {Ledoux}, C. and {Petitjean}, P.},
        title = "{First detection of CO in a high-redshift damped Lyman-{\ensuremath{\alpha}} system}",
      journal = {\aap},
         year = 2008,
        month = may,
       volume = {482},
       number = {3},
        pages = {L39-L42},
          doi = {10.1051/0004-6361:200809727},
archivePrefix = {arXiv},
       eprint = {0804.0116},
 primaryClass = {astro-ph},
       adsurl = {https://ui.adsabs.harvard.edu/abs/2008A&A...482L..39S}
}

@ARTICLE{2018A&A...612A..58N,
       author = {{Noterdaeme}, P. and {Ledoux}, C. and {Zou}, S. and {Petitjean}, P. and {Srianand}, R. and {Balashev}, S. and {L{\'o}pez}, S.},
        title = "{Spotting high-z molecular absorbers using neutral carbon. Results from a complete spectroscopic survey with the VLT}",
      journal = {\aap},
         year = 2018,
        month = apr,
       volume = {612},
          eid = {A58},
        pages = {A58},
          doi = {10.1051/0004-6361/201732266},
archivePrefix = {arXiv},
       eprint = {1801.08357},
 primaryClass = {astro-ph.GA},
       adsurl = {https://ui.adsabs.harvard.edu/abs/2018A&A...612A..58N}
}

@ARTICLE{2019A&A...625L...9G,
       author = {{Geier}, S.~J. and {Heintz}, K.~E. and {Fynbo}, J.~P.~U. and {Ledoux}, C. and {Christensen}, L. and {Jakobsson}, P. and {Krogager}, J. -K. and {Milvang-Jensen}, B. and {M{\o}ller}, P. and {Noterdaeme}, P.},
        title = "{Gaia-assisted selection of a quasar reddened by dust in an extremely strong damped Lyman-{\ensuremath{\alpha}} absorber at z = 2.226}",
      journal = {\aap},
         year = 2019,
        month = may,
       volume = {625},
          eid = {L9},
        pages = {L9},
          doi = {10.1051/0004-6361/201935108},
archivePrefix = {arXiv},
       eprint = {1904.01686},
 primaryClass = {astro-ph.GA},
       adsurl = {https://ui.adsabs.harvard.edu/abs/2019A&A...625L...9G}
}

@ARTICLE{2022Natur.602...58R,
       author = {{Riechers}, Dominik A. and {Weiss}, Axel and {Walter}, Fabian and {Carilli}, Christopher L. and {Cox}, Pierre and {Decarli}, Roberto and {Neri}, Roberto},
        title = "{Microwave background temperature at a redshift of 6.34 from H$_{2}$O absorption}",
      journal = {\nat},
         year = 2022,
        month = feb,
       volume = {602},
       number = {7895},
        pages = {58-62},
          doi = {10.1038/s41586-021-04294-5},
archivePrefix = {arXiv},
       eprint = {2202.00693},
 primaryClass = {astro-ph.GA},
       adsurl = {https://ui.adsabs.harvard.edu/abs/2022Natur.602...58R}
}

@ARTICLE{2007A&A...468..627V,
       author = {{van der Tak}, F.~F.~S. and {Black}, J.~H. and {Sch{\"o}ier}, F.~L. and {Jansen}, D.~J. and {van Dishoeck}, E.~F.},
        title = "{A computer program for fast non-LTE analysis of interstellar line spectra. With diagnostic plots to interpret observed line intensity ratios}",
      journal = {\aap},
         year = 2007,
        month = jun,
       volume = {468},
       number = {2},
        pages = {627-635},
          doi = {10.1051/0004-6361:20066820},
archivePrefix = {arXiv},
       eprint = {0704.0155},
 primaryClass = {astro-ph},
       adsurl = {https://ui.adsabs.harvard.edu/abs/2007A&A...468..627V}
}

@ARTICLE{2020ApJ...899...76J,
       author = {{Jeram}, Sarik and {Gonzalez}, Anthony and {Eikenberry}, Stephen and {Stern}, Daniel and {Mendes de Oliveira}, Claudia Lucia and {Izuti Nakazono}, Lilianne Mariko and {Ackley}, Kendall},
        title = "{An Extremely Bright QSO at z = 2.89}",
      journal = {\apj},
         year = 2020,
        month = aug,
       volume = {899},
       number = {1},
          eid = {76},
        pages = {76},
          doi = {10.3847/1538-4357/ab9c95},
archivePrefix = {arXiv},
       eprint = {2006.11915},
 primaryClass = {astro-ph.GA},
       adsurl = {https://ui.adsabs.harvard.edu/abs/2020ApJ...899...76J}
}

@ARTICLE{2018A&A...615L...8H,
       author = {{Heintz}, K.~E. and {Fynbo}, J.~P.~U. and {H{\o}g}, E. and {M{\o}ller}, P. and {Krogager}, J. -K. and {Geier}, S. and {Jakobsson}, P. and {Christensen}, L.},
        title = "{Unidentified quasars among stationary objects from Gaia DR2}",
      journal = {\aap},
         year = 2018,
        month = jul,
       volume = {615},
          eid = {L8},
        pages = {L8},
          doi = {10.1051/0004-6361/201833396},
archivePrefix = {arXiv},
       eprint = {1805.03394},
 primaryClass = {astro-ph.GA},
       adsurl = {https://ui.adsabs.harvard.edu/abs/2018A&A...615L...8H}
}

@ARTICLE{2024ExA....57....5M,
       author = {{Martins}, C.~J.~A.~P. and {Cooke}, R. and {Liske}, J. and {Murphy}, M.~T. and {Noterdaeme}, P. and {Schmidt}, T.~M. and {Alcaniz}, J.~S. and {Alves}, C.~S. and {Balashev}, S. and {Cristiani}, S. and {Di Marcantonio}, P. and {G{\'e}nova Santos}, R. and {Gon{\c{c}}alves}, R.~S. and {Gonz{\'a}lez Hern{\'a}ndez}, J.~I. and {Maiolino}, R. and {Marconi}, A. and {Marques}, C.~M.~J. and {Melo e Sousa}, M.~A.~F. and {Nunes}, N.~J. and {Origlia}, L. and {P{\'e}roux}, C. and {Vinzl}, S. and {Zanutta}, A.},
        title = "{Cosmology and fundamental physics with the ELT-ANDES spectrograph}",
      journal = {Experimental Astronomy},
         year = 2024,
        month = feb,
       volume = {57},
       number = {1},
          eid = {5},
        pages = {5},
          doi = {10.1007/s10686-024-09928-w},
archivePrefix = {arXiv},
       eprint = {2311.16274},
 primaryClass = {astro-ph.CO},
       adsurl = {https://ui.adsabs.harvard.edu/abs/2024ExA....57....5M}
}

@ARTICLE{2024MNRAS.533.1367K,
       author = {{Klimenko}, V.~V. and {Balashev}, S.~A. and {Noterdaeme}, P. and {Srianand}, R. and {Ivanchik}, A.~V.},
        title = "{Excitation of CO molecules in diffuse gas over cosmic history}",
      journal = {\mnras},
         year = 2024,
        month = sep,
       volume = {533},
       number = {2},
        pages = {1367-1393},
          doi = {10.1093/mnras/stae1863},
       adsurl = {https://ui.adsabs.harvard.edu/abs/2024MNRAS.533.1367K}
}

@ARTICLE{2022ApJS..258...18C,
       author = {{Chabanier}, Sol{\`e}ne and {Etourneau}, Thomas and {Le Goff}, Jean-Marc and {Rich}, James and {Stermer}, Julianna and {Abolfathi}, Bela and {Font-Ribera}, Andreu and {Gonzalez-Morales}, Alma X. and {de la Macorra}, Axel and {P{\'e}rez-R{\`a}fols}, Ignasi and {Petitjean}, Patrick and {Pieri}, Matthew M. and {Ravoux}, Corentin and {Rossi}, Graziano and {Schneider}, Donald P.},
        title = "{The Completed Sloan Digital Sky Survey IV Extended Baryon Oscillation Spectroscopic Survey: The Damped Ly{\ensuremath{\alpha}} Systems Catalog}",
      journal = {\apjs},
         year = 2022,
        month = jan,
       volume = {258},
       number = {1},
          eid = {18},
        pages = {18},
          doi = {10.3847/1538-4365/ac366e},
archivePrefix = {arXiv},
       eprint = {2107.09612},
 primaryClass = {astro-ph.CO},
       adsurl = {https://ui.adsabs.harvard.edu/abs/2022ApJS..258...18C}
}

@article{Witstok2025,
	adsurl = {https://ui.adsabs.harvard.edu/abs/2025Natur.639..897W},
	archiveprefix = {arXiv},
	author = {{Witstok}, Joris and {Jakobsen}, Peter and {Maiolino}, Roberto and {Helton}, Jakob M. and {Johnson}, Benjamin D. and {Robertson}, Brant E. and {Tacchella}, Sandro and {Cameron}, Alex J. and {Smit}, Renske and {Bunker}, Andrew J. and {Saxena}, Aayush and {Sun}, Fengwu and {Alberts}, Stacey and {Arribas}, Santiago and {Baker}, William M. and {Bhatawdekar}, Rachana and {Boyett}, Kristan and {Cargile}, Phillip A. and {Carniani}, Stefano and {Charlot}, St{\'e}phane and {Chevallard}, Jacopo and {Curti}, Mirko and {Curtis-Lake}, Emma and {D'Eugenio}, Francesco and {Eisenstein}, Daniel J. and {Hainline}, Kevin N. and {Jones}, Gareth C. and {Kumari}, Nimisha and {Maseda}, Michael V. and {P{\'e}rez-Gonz{\'a}lez}, Pablo G. and {Rinaldi}, Pierluigi and {Scholtz}, Jan and {{\"U}bler}, Hannah and {Williams}, Christina C. and {Willmer}, Christopher N.~A. and {Willott}, Chris and {Zhu}, Yongda},
	doi = {10.1038/s41586-025-08779-5},
	eprint = {2408.16608},
	journal = {\nat},
	month = mar,
	number = {8056},
	pages = {897-901},
	primaryclass = {astro-ph.GA},
	title = {{Witnessing the onset of reionization through Lyman-{\ensuremath{\alpha}} emission at redshift 13}},
	volume = {639},
	year = 2025}

@article{Hirai2025,
	adsurl = {https://ui.adsabs.harvard.edu/abs/2025ApJ...980L..25H},
	archiveprefix = {arXiv},
	author = {{Hirai}, Yutaka and {Saitoh}, Takayuki R. and {Fujii}, Michiko S. and {Kaneko}, Katsuhiro and {Beers}, Timothy C.},
	doi = {10.3847/2041-8213/adaf95},
	eid = {L25},
	eprint = {2411.18680},
	journal = {\apjl},
	month = feb,
	number = {2},
	pages = {L25},
	primaryclass = {astro-ph.GA},
	title = {{SIRIUS: Identifying Metal-poor Stars Enriched by a Single Supernova in a Dwarf Galaxy Cosmological Zoom-in Simulation Resolving Individual Massive Stars}},
	volume = {980},
	year = 2025}

@article{Grazian2024,
	adsurl = {https://ui.adsabs.harvard.edu/abs/2024ApJ...974...84G},
	archiveprefix = {arXiv},
	author = {{Grazian}, Andrea and {Giallongo}, Emanuele and {Boutsia}, Konstantina and {Cristiani}, Stefano and {Fontanot}, Fabio and {Bischetti}, Manuela and {Bisigello}, Laura and {Bongiorno}, Angela and {Calderone}, Giorgio and {Chiti Tegli}, Francesco and {Cupani}, Guido and {De Lucia}, Gabriella and {D'Odorico}, Valentina and {Feruglio}, Chiara and {Fiore}, Fabrizio and {Gandolfi}, Giovanni and {Girardi}, Giorgia and {Guarneri}, Francesco and {Hirschmann}, Michaela and {Porru}, Matteo and {Rodighiero}, Giulia and {Saccheo}, Ivano and {Simioni}, Matteo and {Trost}, Andrea and {Viitanen}, Akke},
	doi = {10.3847/1538-4357/ad6980},
	eid = {84},
	eprint = {2407.20861},
	journal = {\apj},
	month = oct,
	number = {1},
	pages = {84},
	primaryclass = {astro-ph.CO},
	title = {{What Are the Pillars of Reionization? Revising the AGN Luminosity Function at z {\ensuremath{\sim}} 5}},
	volume = {974},
	year = 2024}

@article{Lin2024,
	adsurl = {https://ui.adsabs.harvard.edu/abs/2024RAA....24a5009L},
	archiveprefix = {arXiv},
	author = {{Lin}, Hengjie and {Gong}, Yan and {Yue}, Bin and {Chen}, Xuelei},
	doi = {10.1088/1674-4527/ad0864},
	eid = {015009},
	eprint = {2306.05648},
	journal = {Research in Astronomy and Astrophysics},
	month = jan,
	number = {1},
	pages = {015009},
	primaryclass = {astro-ph.CO},
	title = {{Implications of the Stellar Mass Density of High-z Massive Galaxies from JWST on Warm Dark Matter}},
	volume = {24},
	year = 2024}

@article{Mason2026,
	adsurl = {https://ui.adsabs.harvard.edu/abs/2026A&A...705A.114M},
	archiveprefix = {arXiv},
	author = {{Mason}, Charlotte A. and {Chen}, Zuyi and {Stark}, Daniel P. and {Yi Lu}, Ting and {Topping}, Michael and {Tang}, Mengtao},
	doi = {10.1051/0004-6361/202553820},
	eid = {A114},
	eprint = {2501.11702},
	journal = {\aap},
	month = jan,
	pages = {A114},
	primaryclass = {astro-ph.GA},
	title = {{Constraints on the z {\ensuremath{\sim}} 6{\ensuremath{-}}13 intergalactic medium from JWST spectroscopy of Lyman-alpha damping wings in galaxies}},
	volume = {705},
	year = 2026}

@article{Kageura2025,
	adsurl = {https://ui.adsabs.harvard.edu/abs/2025ApJS..278...33K},
	archiveprefix = {arXiv},
	author = {{Kageura}, Yuta and {Ouchi}, Masami and {Nakane}, Minami and {Umeda}, Hiroya and {Harikane}, Yuichi and {Yoshiura}, Shintaro and {Nakajima}, Kimihiko and {Yajima}, Hidenobu and {Thai}, Tran Thi},
	doi = {10.3847/1538-4365/adc690},
	eid = {33},
	eprint = {2501.05834},
	journal = {\apjs},
	month = jun,
	number = {2},
	pages = {33},
	primaryclass = {astro-ph.GA},
	title = {{Census of Ly{\ensuremath{\alpha}} Emission from {\ensuremath{\sim}}600 Galaxies at z = 5{\textendash}14: Evolution of the Ly{\ensuremath{\alpha}} Luminosity Function and a Late Sharp Cosmic Reionization}},
	volume = {278},
	year = 2025}

@article{Cain2025,
	adsurl = {https://ui.adsabs.harvard.edu/abs/2025ApJ...980...83C},
	archiveprefix = {arXiv},
	author = {{Cain}, Christopher and {Lopez}, Garett and {D'Aloisio}, Anson and {Mu{\~n}oz}, Julian B. and {Jansen}, Rolf A. and {Windhorst}, Rogier A. and {Gangolli}, Nakul},
	doi = {10.3847/1538-4357/ada152},
	eid = {83},
	eprint = {2409.02989},
	journal = {\apj},
	month = feb,
	number = {1},
	pages = {83},
	primaryclass = {astro-ph.CO},
	title = {{Chasing the Beginning of Reionization in the JWST Era}},
	volume = {980},
	year = 2025}

@article{Nakane2024,
	adsurl = {https://ui.adsabs.harvard.edu/abs/2024ApJ...967...28N},
	archiveprefix = {arXiv},
	author = {{Nakane}, Minami and {Ouchi}, Masami and {Nakajima}, Kimihiko and {Harikane}, Yuichi and {Ono}, Yoshiaki and {Umeda}, Hiroya and {Isobe}, Yuki and {Zhang}, Yechi and {Xu}, Yi},
	doi = {10.3847/1538-4357/ad38c2},
	eid = {28},
	eprint = {2312.06804},
	journal = {\apj},
	month = may,
	number = {1},
	pages = {28},
	primaryclass = {astro-ph.GA},
	title = {{Ly{\ensuremath{\alpha}} Emission at z = 7{\textendash}13: Clear Evolution of Ly{\ensuremath{\alpha}} Equivalent Width Indicating a Late Cosmic Reionization History}},
	volume = {967},
	year = 2024}

@article{Kokorev2025,
	adsurl = {https://ui.adsabs.harvard.edu/abs/2025ApJ...988L..10K},
	archiveprefix = {arXiv},
	author = {{Kokorev}, Vasily and {Ch{\'a}vez Ortiz}, {\'O}scar A. and {Taylor}, Anthony J. and {Finkelstein}, Steven L. and {Arrabal Haro}, Pablo and {Dickinson}, Mark and {Chisholm}, John and {Fujimoto}, Seiji and {noz}, Julian B. Mu and {Endsley}, Ryan and et al.},
	doi = {10.3847/2041-8213/ade8f5},
	eid = {L10},
	eprint = {2504.12504},
	journal = {\apjl},
	month = jul,
	number = {1},
	pages = {L10},
	primaryclass = {astro-ph.GA},
	title = {{CAPERS Observations of Two UV-bright Galaxies at z > 10. More Evidence for Bursting Star Formation in the Early Universe}},
	volume = {988},
	year = 2025}

@article{Yung2025,
	adsurl = {https://ui.adsabs.harvard.edu/abs/2025MNRAS.543.3802Y},
	archiveprefix = {arXiv},
	author = {{Yung}, L.~Y. Aaron and {Somerville}, Rachel S. and {Iyer}, Kartheik G.},
	doi = {10.1093/mnras/staf1699},
	eprint = {2504.18618},
	journal = {\mnras},
	month = nov,
	number = {4},
	pages = {3802-3813},
	primaryclass = {astro-ph.GA},
	title = {{{\ensuremath{\Lambda}}CDM is still not broken: empirical constraints on the star formation efficiency at z {\ensuremath{\sim}}12─30}},
	volume = {543},
	year = 2025}

@article{Feldmann2025,
	adsurl = {https://ui.adsabs.harvard.edu/abs/2025MNRAS.536..988F},
	archiveprefix = {arXiv},
	author = {{Feldmann}, Robert and {Boylan-Kolchin}, Michael and {Bullock}, James S. and {{\c{C}}atmabacak}, Onur and {Faucher-Gigu{\`e}re}, Claude-Andr{\'e} and {Hayward}, Christopher C. and {Kere{\v{s}}}, Du{\v{s}}an and {Lazar}, Alexandres and {Liang}, Lichen and {Moreno}, Jorge and et al.},
	doi = {10.1093/mnras/stae2633},
	eprint = {2407.02674},
	journal = {\mnras},
	month = jan,
	number = {1},
	pages = {988-1016},
	primaryclass = {astro-ph.CO},
	title = {{Elevated UV luminosity density at Cosmic Dawn explained by non-evolving, weakly mass-dependent star formation efficiency}},
	volume = {536},
	year = 2025}

@article{Napolitano2024,
	adsurl = {https://ui.adsabs.harvard.edu/abs/2024A&A...688A.106N},
	archiveprefix = {arXiv},
	author = {{Napolitano}, L. and {Pentericci}, L. and {Santini}, P. and {Calabr{\`o}}, A. and {Mascia}, S. and {Llerena}, M. and {Castellano}, M. and {Dickinson}, M. and {Finkelstein}, S.~L. and {Amor{\'\i}n}, R. and et al.},
	doi = {10.1051/0004-6361/202449644},
	eid = {A106},
	eprint = {2402.11220},
	journal = {\aap},
	month = aug,
	pages = {A106},
	primaryclass = {astro-ph.GA},
	title = {{Peering into cosmic reionization: Ly{\ensuremath{\alpha}} visibility evolution from galaxies at z = 4.5-8.5 with JWST}},
	volume = {688},
	year = 2024}

@article{Chakraborty2024,
	adsurl = {https://ui.adsabs.harvard.edu/abs/2024JCAP...07..078C},
	archiveprefix = {arXiv},
	author = {{Chakraborty}, Anirban and {Choudhury}, Tirthankar Roy},
	doi = {10.1088/1475-7516/2024/07/078},
	eid = {078},
	eprint = {2404.02879},
	journal = {\jcap},
	month = jul,
	number = {7},
	pages = {078},
	primaryclass = {astro-ph.GA},
	title = {{Modelling the star-formation activity and ionizing properties of high-redshift galaxies}},
	volume = {2024},
	year = 2024}

@ARTICLE{Chakraborty2026,
       author = {{Chakraborty}, Anirban and {Choudhury}, Tirthankar Roy},
        title = "{Probing reionization-era galaxies with JWST UV luminosity functions and large-scale clustering}",
      journal = {\jcap},
         year = 2026,
        month = jan,
       volume = {2026},
       number = {1},
          eid = {008},
        pages = {008},
          doi = {10.1088/1475-7516/2026/01/008},
archivePrefix = {arXiv},
       eprint = {2503.07590},
 primaryClass = {astro-ph.GA},
       adsurl = {https://ui.adsabs.harvard.edu/abs/2026JCAP...01..008C}
}

@article{Kannan2022,
	adsurl = {https://ui.adsabs.harvard.edu/abs/2022MNRAS.511.4005K},
	archiveprefix = {arXiv},
	author = {{Kannan}, R. and {Garaldi}, E. and {Smith}, A. and {Pakmor}, R. and {Springel}, V. and {Vogelsberger}, M. and {Hernquist}, L.},
	doi = {10.1093/mnras/stab3710},
	eprint = {2110.00584},
	journal = {\mnras},
	month = apr,
	number = {3},
	pages = {4005-4030},
	primaryclass = {astro-ph.GA},
	title = {{Introducing the THESAN project: radiation-magnetohydrodynamic simulations of the epoch of reionization}},
	volume = {511},
	year = 2022}

@article{Rosdahl2022,
	adsurl = {https://ui.adsabs.harvard.edu/abs/2022MNRAS.515.2386R},
	archiveprefix = {arXiv},
	author = {{Rosdahl}, Joakim and {Blaizot}, J{\'e}r{\'e}my and {Katz}, Harley and {Kimm}, Taysun and {Garel}, Thibault and {Haehnelt}, Martin and {Keating}, Laura C. and {Martin-Alvarez}, Sergio and {Michel-Dansac}, L{\'e}o and {Ocvirk}, Pierre},
	doi = {10.1093/mnras/stac1942},
	eprint = {2207.03232},
	journal = {\mnras},
	month = sep,
	number = {2},
	pages = {2386-2414},
	primaryclass = {astro-ph.GA},
	title = {{LyC escape from SPHINX galaxies in the Epoch of Reionization}},
	volume = {515},
	year = 2022}

@article{Wilkins2023,
	adsurl = {https://ui.adsabs.harvard.edu/abs/2023MNRAS.519.3118W},
	archiveprefix = {arXiv},
	author = {{Wilkins}, Stephen M. and {Vijayan}, Aswin P. and {Lovell}, Christopher C. and {Roper}, William J. and {Irodotou}, Dimitrios and {Caruana}, Joseph and {Seeyave}, Louise T.~C. and {Kuusisto}, Jussi K. and {Thomas}, Peter A. and {Parris}, Shedeur A.~K.},
	doi = {10.1093/mnras/stac3280},
	eprint = {2204.09431},
	journal = {\mnras},
	month = feb,
	number = {2},
	pages = {3118-3128},
	primaryclass = {astro-ph.GA},
	title = {{First light and reionization epoch simulations (FLARES) V: the redshift frontier}},
	volume = {519},
	year = 2023}
\end{refsection}

%
%
%

\end{document}